\documentclass[11pt]{article}
\usepackage[margin=0.7in]{geometry}

\newcommand{\theltitle}{Feedback Control Methods for a Single Machine Infinite Bus System}

\usepackage{fancyhdr, lastpage, extramarks}
\usepackage{amssymb,amsfonts,amsmath,amscd,mathrsfs}
\usepackage{graphicx}
\usepackage{caption}
\usepackage[usenames,dvipsnames]{color}
\usepackage{listings}
\usepackage{csquotes}
\usepackage{pict2e}
\usepackage[square,sort,comma,numbers]{natbib}
\usepackage{pstricks, pst-node, pst-circ, pst-coil}
\usepackage{auto-pst-pdf}

\usepackage{booktabs}
\usepackage{amsmath}
\usepackage{amsfonts}
\usepackage{amssymb}
\usepackage{mathtools, amssymb, amsfonts}
\usepackage{rotating}
\usepackage{authblk}

\numberwithin{table}{section}
\numberwithin{figure}{section}
\numberwithin{equation}{section}

\usepackage{varioref}
\labelformat{equation}{(#1)}

\usepackage[colorlinks,citecolor=blue,linkcolor=blue]{hyperref}

\newenvironment{centergroup}{\trivlist\item\begin{minipage}{\columnwidth}\centering}{\end{minipage}\endtrivlist}
\newcommand{\bbm}{\begin{bmatrix}}
\newcommand{\ebm}{\end{bmatrix}}
\newcommand{\bpm}{\begin{pmatrix}}
\newcommand{\epm}{\end{pmatrix}}
\newcommand{\bal}{\begin{aligned}}
\newcommand{\eal}{\end{aligned}}
\newcommand{\beq}{\begin{equation}}
\newcommand{\eeq}{\end{equation}}

\title{\textbf{\theltitle}}
\author{Pratik Vernekar, Zhongkui Wang, Andrea Serrani, and Kevin Passino}
\affil{Department of Electrical and Computer Engineering, The Ohio State University\\
       2015 Neil Avenue, Columbus, OH 43210\\ 
\vspace{5mm}
Email: pratik.vernekar@gmail.com,\\wang.1231@osu.edu,\\serrani.1@osu.edu,\\passino.1@osu.edu\\}
\date{}

\clearpage
\begin{document}

\maketitle


\begin{abstract}
In this manuscript, we present a high-fidelity physics-based truth model of a Single Machine Infinite Bus (SMIB) system. We also present reduced-order 
control-oriented nonlinear and linear models of a synchronous generator-turbine system connected to a power grid. The reduced-order control-oriented models are next used to design various 
control strategies such as: proportional-integral-derivative (PID), linear-quadratic regulator (LQR), pole placement-based state feedback, observer-based output feedback, loop transfer 
recovery (LTR)-based linear-quadratic-Gaussian (LQG), and nonlinear feedback-linearizing control for the SMIB system. The controllers developed are then validated on the high-fidelity 
physics-based truth model of the SMIB system. Finally, a comparison is made of the performance of the controllers at different operating points of the SMIB system. The material presented in this 
manuscript is part of a course on \enquote{Control and Optimization for the Smart Grid} that was developed in the Electrical and Computer Engineering Department at the Ohio State University in 2011-2012. 
This project was funded by the U.S. Department of Energy.
\end{abstract}

\newpage

\tableofcontents
\clearpage                      

\section{A Single Generation Unit}

Fossil fuels such as coal, oil, and natural gas have been the main resources of electrical
energy for many years. However in recent years, there has been a gradual increase in the use of renewable
energy resources for electricity generation, such as hydro, biogas, solar, wind, and geothermal energy. Electricity generation is basically the process of generating electric energy from other forms of energy. An electromechanical
device called \emph{synchronous generator} driven by a prime mover, usually a turbine or a
diesel engine, converts the mechanical energy into alternating current (AC) electrical energy.
\begin{centergroup}
	\begin{pspicture}(-10,-4)(8,3.5)
		\psframe(-1.5,-1.8)(2.5,-0.2)
		\rput(0.5,-0.75){Synchronous}
		\rput(0.5,-1.25){generator}
		\psframe(-0.7,0.5)(1.7,1.3)
		\rput(0.5,1.05){Measuring}
		\rput(0.5,0.75){element}
		\psline[linewidth=2pt](0,-0.2)(0,0.5)
		\psline[linewidth=2pt](0.5,-0.2)(0.5,0.5)
		\psline[linewidth=2pt](1,-0.2)(1,0.5)
		\psline[linewidth=2pt]{->}(0,1.3)(0,2.2)
		\psline[linewidth=2pt]{->}(0.5,1.3)(0.5,2.2)
		\psline[linewidth=2pt]{->}(1,1.3)(1,2.2)
		\rput(0.5,2.5){To grid through breakers and transformers}

		\psframe(-5.7,0.45)(-3.7,1.35)
		\rput(-4.7,0.95){Governor}
		\psline[linestyle=dashed]{->}(-0.7,1.05)(-3.7,1.05)
		\psline[linestyle=dashed]{->}(-0.7,0.85)(-3.7,0.85)
		\rput(-2,1.3){$P$}
		\rput(-2,0.6){$f$}
		\psline{->}(-3.5,1.8)(-3.5,1.25)(-3.7,1.25)
		\psline{->}(-2.9,-0.8)(-2.9,0.65)(-3.7,0.65)
		\psline(-3.1,-0.8)(-2.7,-0.8)
		\rput(-3,1.6){$P_\mathrm{ref}$}
		\rput(-3.1,0.2){$\omega$}

		\psframe(-7,0.6)(-6.6,1.2)
		\psline(-6.9,1.2)(-6.9,1.6)
		\psline(-6.7,1.2)(-6.7,1.6)
		\psline{->}(-6.8,3.2)(-6.8,1.7)
		\psline{->}(-5.7,0.9)(-6.6,0.9)
		\psline(-6.9,0.6)(-6.9,-1.8)(-3.5,-1.8)(-3.5,-2.2)
		\psline(-6.7,0.6)(-6.7,-0.2)(-3.3,-0.2)(-3.3,-2.2)
		\rput(-5.1,-1){Turbine}
		\rput{90}(-7.1,2.4){Working}
		\rput{90}(-6.5,2.4){fluid in}
		\rput{90}(-7.3,0.8){Valves}
		\psline{->}(-3.4,-2.2)(-3.4,-3.8)
		\rput{90}(-3.7,-3){Working}
		\rput{90}(-3.1,-3){fluid out}

		\psline[linewidth=3pt](-6.9,-0.9)(-7.4,-0.9)(-7.4,-1.1)(-6.9,-1.1)
		\psline[linewidth=3pt](-3.3,-0.9)(-1.5,-0.9)
		\psline[linewidth=3pt](-3.3,-1.1)(-1.5,-1.1)
		\psline[linewidth=3pt](2.5,-0.9)(3,-0.9)(3,-1.1)(2.5,-1.1)
		\pscurve{->}(-1.95,-1.4)(-2.2,-1.55)(-2.5,-1)(-2.2,-0.45)(-1.95,-0.6)
		\rput(-2.2,-1.8){Shaft}

		\psframe(-0.5,-3.5)(1.5,-2.5)
		\rput(0.5,-3){Exciter}
		\psframe(2,-3.8)(4,-2.2)
		\rput(3,-2.6){Automatic}
		\rput(3,-3){voltage}
		\rput(3,-3.4){regulator}
		\psline{->}(1.7,1.1)(5,1.1)(5,-3)(4,-3)
		\psline{->}(1.7,0.7)(4.6,0.7)(4.6,-2.6)(4,-2.6)
		\rput(2.8,1.4){$V_t$}\rput(2.8,0.4){$I_t$}
		\psline{->}(6,-3.4)(4,-3.4)
		\rput(5.6,-3.1){$V_\mathrm{ref}$}
		\psline{->}(2,-3)(1.5,-3)
		\psline[linewidth=2pt](0.2,-2.5)(0.2,-1.8)
		\psline[linewidth=2pt](0.8,-2.5)(0.8,-1.8)

	\end{pspicture}
	
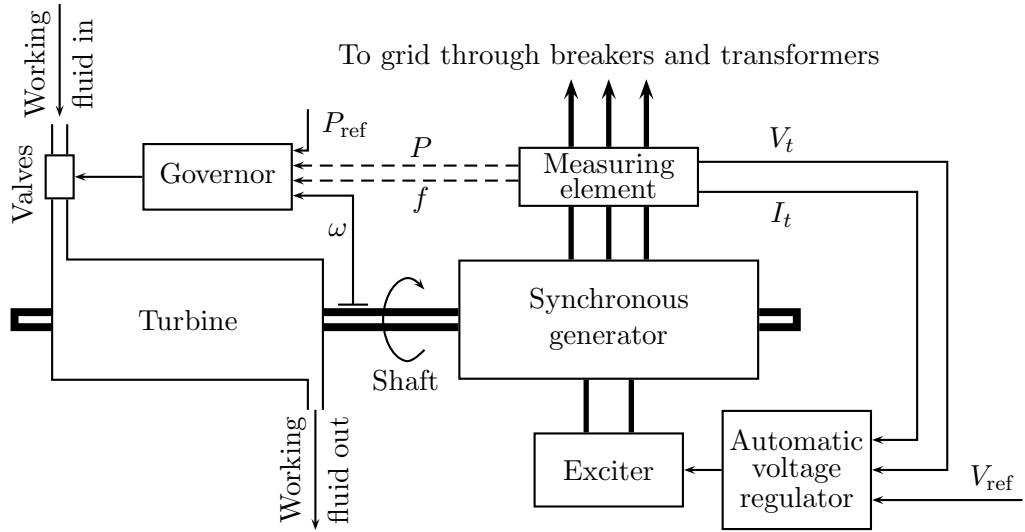
\captionof{figure}{Structure of a generation unit.}
	\label{fig:generation}
\end{centergroup}
The system shown in \autoref{fig:generation} is a general structure of a single generation
unit \cite{MBB08}. The \emph{turbine} extracts the energy from the \emph{working fluid}
flowing into the turbine through \emph{valves}. Typical working fluids are gas, steam,
and water. The \emph{shaft} is the rotary part of the turbine
on which the synchronous generator is mounted. The opening and closing of the turbine
valves or the frequency at which the turbine valves operate is regulated to a reference
frequency of $f_\mathrm{ref}$, by a \emph{turbine governor}. The frequency of the
grid $f$ which is measured by the  \emph{measuring element} is directly related to the
output power $P$. Thus, the output of the synchronous generator $P$ and the angular
frequency of the shaft $\omega $ are measured and fed back to the governor by the
measuring element. Meanwhile, the measuring element also provides information about the
output terminal voltage $V_t$ and output current $I_t$ of the synchronous generator to the
\emph{automatic voltage regulator} (AVR), which is able to control the terminal
voltage of the synchronous generator to a reference voltage $V_\mathrm{ref}$ through the
\emph{exciter}. The excitation current generated by the exciter produces the magnetic field
inside the generator.

Thus, from the above figure we can see that in an interconnected power system, where a
synchronous generator is connected to a grid, load frequency control (LFC) and automatic
voltage regulator (AVR) equipment is installed for each generator.
\autoref{fig:generation} shows two control loops, namely the load frequency
control (LFC) loop and the automatic voltage regulator (AVR) loop. The controllers are set
for a particular operating condition and accommodate small changes in load demand to
maintain the frequency and voltage magnitude within the specified limits. Small changes in
real power are mainly dependent on changes in rotor angle $\delta $, and thus the
frequency $\omega $. The reactive power is mainly dependent on the voltage magnitude (i.e., on the
generator excitation). The excitation system time constant which is an indication of how fast the transients of the AVR loop
decay exponential to zero, is much smaller than the prime
mover time constant. Thus the transients of the excitation system and thus the AVR loop decay much faster than the transients
of the LFC loop, hence it does not affect the LFC
dynamics. Thus, the cross-coupling between the LFC loop and the AVR loop is negligible.
Hence, load frequency control and excitation voltage control are usually analyzed independently \cite{RTWQ}.

The operation objectives of the LFC are to maintain reasonably uniform frequency, and to
divide the load between generators \cite{RTWQ}. The change in frequency is sensed, which is a measure
of the change in rotor angle $\delta $, i.e., the error $\Delta \delta$  to be corrected.
The error signal i.e., $\Delta f=f_\mathrm{ref}-f$ is amplified, mixed, and transformed
into a real power command signal $\Delta P_V=P_\mathrm{ref}-P$, which is sent to the prime
mover to call for an increment in the torque. The prime mover, therefore, brings about a
change in the generator output which will change the value of $\Delta f$ within the
specified tolerance.

The generator excitation system maintains the generator terminal voltage and controls the reactive
power flow. The generator excitation of older systems may be provided through slip rings
and brushes by means of DC generators mounted on the same shaft as the rotor of the
synchronous machine. However, modern excitation systems also known as brush-less excitation systems, usually use AC generators with
rotating rectifiers. The sources of reactive power
are generators, capacitors, and reactors. The generator reactive power is controlled by
field excitation using the AVR. The role of an AVR is to hold the terminal voltage
magnitude of a synchronous generator at a specified level. An increase in the reactive
power load of the generator is accompanied by a drop in the terminal voltage magnitude.
The voltage magnitude is sensed through a potential transformer on one phase. This voltage
is rectified and compared to a DC set point signal. The amplified error signal controls
the exciter field and increases the exciter terminal voltage. Thus, the generator field
current is increased, which results in an increase in the generated electromotiveforce
(emf). The reactive power generation is increased to a new equilibrium, raising the
terminal voltage to the desired value.

\section{Truth Model of the Synchronous Generator}

In the previous section we saw the basic working of a single generation unit and the
respective roles of the load frequency control and the automatic voltage regulator. In
this section we will derive the truth model of a synchronous generator. Before
we proceed to the derivation of the truth model, we present some preliminaries about the
synchronous generator.

The two main parts of a synchronous generator can be described in either electrical or mechanical
terms:
\begin{itemize}
\item Electrical:
\begin{itemize}
\item Armature: The power-producing component of an electrical machine. In a synchronous generator, 
the armature windings generate the electric current. The armature can be on either the rotor or the stator.
\item Field: The magnetic field component of an electrical machine. The magnetic field of the synchronous generator
can be provided by either electromagnets or permanent magnets mounted on either the 
rotor or the stator.
\end{itemize}
\item Mechanical:
\begin{itemize}
\item Rotor: The rotating part of the synchronous generator.
\item Stator: The stationary part of the synchronous generator.
\end{itemize}
\end{itemize}
Because power transferred into the field circuit is much smaller than in the armature
circuit, AC generators always have the field winding on the rotor and the stator has the
armature winding. Thus, a classical synchronous generator has two main magnetic parts: the
stator and the rotor, as shown in \autoref{fig:machine_structure}. The windings are
represented by one-turn coils, specifically, the small circles ``$\bigcirc$'' in the
figure. The black dots and the crosses inside the small circles indicate the directions of
the currents flowing in the windings, i.e., ``$\times$'' means the current flowing in the
direction from the outside of the paper vertically into the paper and ``$\bullet$'' means
the current flowing from the inside of the paper to the outside.
	\psset{framesep=1.5pt}
	\begin{centergroup}
		\begin{pspicture}(-5,-4)(5,4)
		\pscircle(0,0){3.8}

		\multido{\i=4+60,\n=56+60}{6}{\psarc(0,0){2.5}{\i}{\n}}

		\SpecialCoor
		\multido{\i=4+60,\n=-4+60}{6}{\psline(2.5;\i)(2.8;\i)(2.8;\n)(2.5;\n)}
		
		\multido{\i=0+60}{6}{\pscircle(2.65;\i){0.15}}
		\multido{\i=60+120}{3}{\pscircle[fillstyle=solid,
		    fillcolor=black](2.65;\i){0.05}}
		\uput{0}[0](2.5;0){$\times$} 
		\uput{0}[120](2.5;120){$\times$}
		\uput{0}[240](2.5;240){$\times$}

		\psarc(0,0){2.3}{80}{190}\psarc(0,0){2.3}{260}{370}
		\psline(2.3;80)(1.73;95)
		\psline(2.3;190)(1.73;175)
		\psline(2.3;260)(1.73;275)
		\psline(2.3;370)(1.73;355)
		\psline(1.73;95)(1.73;355)\psline(1.73;175)(1.73;275)

		\pscircle(2.1;135){0.15}\pscircle(2.1;315){0.15}
		\pscircle[fillstyle=solid,fillcolor=black](2.1;135){0.05}
		\uput{0}[315](1.90;315){$\times$}

		\pscircle(1.4;45){0.15}\pscircle(1.8;45){0.15}
		\pscircle(1.4;225){0.15}\pscircle(1.8;225){0.15}
		\pscircle[fillstyle=solid,fillcolor=black](1.4;225){0.05}
		\pscircle[fillstyle=solid,fillcolor=black](1.8;225){0.05}
		\uput{0}[45](1.2;45){$\times$}\uput{0}[45](1.6;45){$\times$}

		\uput{0}[0](2.6;-10){$a'$}
		\uput{0}[0](2.7;70){$c$}
		\uput{0}[0](2.8;110){$b'$}
		\uput{0}[0](2.8;190){$a$}
		\uput{0}[0](3.1;240){$c'$}
		\uput{0}[0](2.8;310){$b$}
		\uput{0.1}[0](1.8;40){$D$}
		\uput{0.1}[0](1.3;38){$F$}
		\uput{0.1}[0](2.1;230){$D'$}
		\uput{0.1}[0](1.6;230){$F'$}

		\uput{0.1}[0](1.8;325){$Q$}
		\uput{0}[0](1.5;310){$S$}
		\uput{0.1}[0](2.2;125){$Q'$}
        \uput{0}[0](1.8;135){$N$}
        
		\uput{0}[0](-0.4,0){Rotor}
		\uput{0.1}[0](3.2;98){Stator}

                \end{pspicture}
		
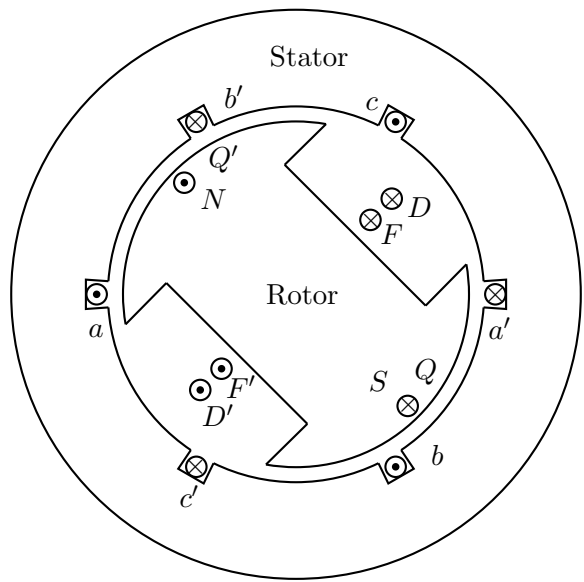
\captionof{figure}{Schematic structure of the synchronous generator.}
		\label{fig:machine_structure}
	\end{centergroup}
The \emph{armature winding}, which carries the load current $I_t$ and supplies power to the
grid, is placed in equidistant slots on the inner surface of the stator and consists of
three identical phase windings, namely, $aa'$, $bb'$ and $cc'$. The rotor is mounted on
the shaft through which the synchronous generator is driven by the prime mover, for
instance, a hydro turbine. The rotor consists of two poles, $N$ pole and $S$ pole, as seen in \autoref{fig:machine_structure}. 
The direct current (DC) \emph{excitation winding} represented by $FF'$ is
wrapped around the rotor. From basic physics, we know that the DC flowing in the
excitation winding generates a magnetic flux. Magnetic flux is a measure of the amount of
magnetic field (also called \emph {magnetic flux density}) passing through a given surface
(such as a conducting coil). The SI unit of magnetic flux is the weber (in derived units:
volt-seconds). The strength of the magnetic flux generated is proportional to the
excitation current and its direction is known by using the right-hand rule. As the rotor
rotates, the magnetic flux generated by the excitation winding wrapped on the rotor
changes spatially. Thus, there are magnetic flux changes in the armature windings as a
result of which an emf is induced in each phase of the three-phase stator armature
winding. By connecting armature windings to the grid, a closed-loop circuit is formed
which allows the AC to flow from the synchronous generator  to the grid. The AC armature
currents produce their own \emph{armature reaction} magnetic flux which is of constant
magnitude but rotates at the same speed as the rotor. The excitation flux and the armature
reaction flux then produce a resultant flux that is stationary with respect to the rotor.
Two other windings represented by $DD'$ and $QQ'$ are the two short-circuit \emph{damper}
(or, \emph{amortisseur}) \emph{windings} which help to damp the mechanical oscillations of
the rotor \cite{BV00}.  Hence, two dynamics will characterize the generator, i.e., electrical dynamics and mechanical dynamics.  

\subsection{Electrical Dynamics}
In this section we present the equations governing the electrical dynamics of a
synchronous generator which are described in \cite{BV00}.
We first present the voltage equations of a synchronous generator in the static frame, and
then use Park's transformation to convert these to the rotating frame.

\subsubsection{Voltage Equation in the Static Frame}
In this subsection we present the voltage equations in the static frame. The static frame
contains three reference axes $a$, $b$, and $c$ which correspond to the three armature
windings on the stator. Before presenting the details of the voltage equation of a
synchronous generator, we start by considering the general case of a set of coupled coils
in which one or more of the coils is mounted on a shaft and can rotate. The situation is
shown schematically in \autoref{fig:coupled_coils}.
				\begin{centergroup}
					\psset{xunit=0.8cm}
					\psset{yunit=0.8cm}
					\begin{pspicture}(-7,-5.5)(6,7.5)
						\psframe(-3,-2)(3,4)

						\pnode(-1,3){L1A}
						\pnode(1,3){L1B}
						\coil[dipolestyle=curved](L1A)(L1B){$L_1$}
						\pnode(-1,4){R1A}
						\pnode(-1,6){R1B}
						\resistor[dipolestyle=zigzag,tensionlabel
						= $i_1$,tensionlabeloffset=-1,
						tensionoffset=-0.7](R1B)(R1A){$R_1$}
						\pnode(-1,7){V1P}
						\pnode(1,7){V1N}
						\psline(R1A)(L1A)
						\psline(R1B)(V1P)
						\psline(L1B)(V1N)
						\pscircle[fillstyle=solid](V1P){0.075}
						\pscircle[fillstyle=solid](V1N){0.075}
						\pnode(2,2){L2A}
						\pnode(2,0){L2B}
						\coil[dipolestyle=curved](L2A)(L2B){$L_2$}
						\pnode(3,2){R2A}
						\pnode(5,2){R2B}
						\resistor[dipolestyle=zigzag,tensionlabel
						= $i_2$,tensionlabeloffset=-1,
						tensionoffset=-0.7](R2B)(R2A){$R_2$}
						\pnode(6,2){V2P}
						\pnode(6,0){V2N}
						\psline(L2A)(R2A)
						\psline(R2B)(V2P)
						\psline(L2B)(V2N)
						\pscircle[fillstyle=solid](V2P){0.075}
						\pscircle[fillstyle=solid](V2N){0.075}
						\pnode(-1,-1){L3A}
						\pnode(1,-1){L3B}
						\coil[dipolestyle=curved](L3A)(L3B){$L_3$}
						\pnode(1,-2){R3A}
						\pnode(1,-4){R3B}
						\resistor[dipolestyle=zigzag,tensionlabel
						= $i_3$,tensionlabeloffset=-1,
						tensionoffset=-0.7](R3B)(R3A){$R_3$}
						\pnode(1,-5){V3N}
						\pnode(-1,-5){V3P}
						\psline(R3B)(V3N)\psline(L3B)(R3A)
						\psline(L3A)(V3P)
						\pscircle[fillstyle=solid](V3P){0.075}
						\pscircle[fillstyle=solid](V3N){0.075}
						\pnode(1,2){L4A}
						\pnode(-1,0){L4B}
						\coil[dipolestyle=curved](L4A)(L4B){$L_4$}
						\pnode(-4,-2){R4A}\pnode(-4,-4){R4B}
						\resistor[dipolestyle=zigzag,tensionlabel
						= $i_4$,tensionlabeloffset=-1,
						tensionoffset=-0.7](R4B)(R4A){$R_4$}
						\pnode(-4,-5){V4P}
						\pnode(-6,-5){V4N}
						\psline(R4B)(V4P)
						\psline(L4B)(-4,0)(R4A)
						\psline(L4A)(-6,2)(V4N)
						\pscircle[fillstyle=solid](V4P){0.075}
						\pscircle[fillstyle=solid](V4N){0.075}
						\pscustom[linecolor=white,linewidth=1.5pt]{
						\psline(-0.3,1.2)(-7,4)
						\stroke[linewidth=5\pslinewidth] 
						\stroke[linewidth=3\pslinewidth,
						linecolor=black]}

						\rput(-4.8,3.6){Shaft}
						\uput{0.3}[90](V1P){$+$}
						\uput{0.3}[90](V1N){$-$}
						\uput{0.3}[90](0,7){$v_1$}
						\uput{0.3}[0](V2P){$+$}
						\uput{0.3}[0](V2N){$-$}
						\uput{0.3}[0](6,1){$v_2$}
						\uput{0.3}[270](V3N){$+$}
						\uput{0.3}[270](V3P){$-$}
						\uput{0.3}[270](0,-5){$v_3$}
						\uput{0.3}[270](V4P){$+$}
						\uput{0.3}[270](V4N){$-$}
						\uput{0.3}[270](-5,-5){$v_4$}
					\end{pspicture}
					
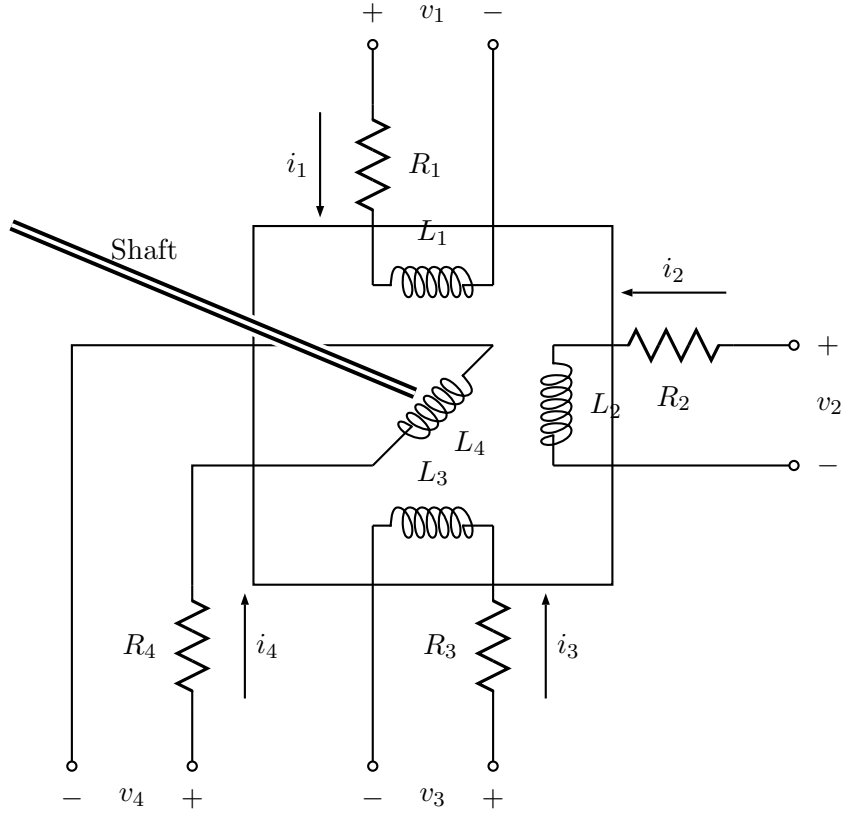
\captionof{figure}{Coupled coils.}
					\label{fig:coupled_coils}
				\end{centergroup}
Assume that for any fixed shaft angle $\theta$ there is a linear
relationship between the flux linkage $\lambda$ and current $i$. Flux linkage is defined
as the total flux passing through a surface (i.e. normal to that surface) formed by a
closed conducting loop. Thus we get the relationship $\mbox{\boldmath$\lambda$} =
\mathbf{L}(\theta)\mathbf{i}$, where, in the case of \autoref{fig:coupled_coils},
$\mathbf{i}$ and $\mbox{\boldmath$\lambda$}$ are $4\times 1$ vectors, and $\mathbf{L}$ is
a $4\times 4$ matrix. By applying Kirchhoff's voltage law (KVL) to the circuit in
\autoref{fig:coupled_coils}, we have
\begin{equation}
	\mathbf{v} = \mathbf{R}\mathbf{i} + \frac{d\mbox{\boldmath$\lambda$}}{dt}
	\label{eq:coil_kvl}
\end{equation}
where $\mathbf{R}$ is a $4\times4$ matrix. \autoref{eq:coil_kvl} indicates that the terminal voltage of each coil equals the sum of
the voltage drop on the resistance and the derivative of the flux linkage.
\psset{framesep=1.5pt}
	\begin{centergroup}
		\begin{pspicture}(-5,-5)(5,5)
		\pscircle(0,0){3.8}

		\multido{\i=4+60,\n=56+60}{6}{\psarc(0,0){2.5}{\i}{\n}}

		\SpecialCoor
		\multido{\i=4+60,\n=-4+60}{6}{\psline(2.5;\i)(2.8;\i)(2.8;\n)(2.5;\n)}
		
		\multido{\i=0+60}{6}{\pscircle(2.65;\i){0.15}}
		\multido{\i=60+120}{3}{\pscircle[fillstyle=solid,
		    fillcolor=black](2.65;\i){0.05}}
		\uput{0}[0](2.5;0){$\times$} 
		\uput{0}[120](2.5;120){$\times$}
		\uput{0}[240](2.5;240){$\times$}

		\psarc(0,0){2.3}{80}{190}\psarc(0,0){2.3}{260}{370}
		\psline(2.3;80)(1.73;95)
		\psline(2.3;190)(1.73;175)
		\psline(2.3;260)(1.73;275)
		\psline(2.3;370)(1.73;355)
		\psline(1.73;95)(1.73;355)\psline(1.73;175)(1.73;275)

		\pscircle(2.1;135){0.15}\pscircle(2.1;315){0.15}
		\pscircle[fillstyle=solid,fillcolor=black](2.1;135){0.05}
		\uput{0}[315](1.90;315){$\times$}

		\pscircle(1.4;45){0.15}\pscircle(1.8;45){0.15}
		\pscircle(1.4;225){0.15}\pscircle(1.8;225){0.15}
		\pscircle[fillstyle=solid,fillcolor=black](1.4;225){0.05}
		\pscircle[fillstyle=solid,fillcolor=black](1.8;225){0.05}
		\uput{0}[45](1.2;45){$\times$}\uput{0}[45](1.6;45){$\times$}

		\psline[linestyle=dashed]{->}(0;0)(5;45)
		\psline[linestyle=dashed]{->}(0;0)(5;90)
		\psline[linestyle=dashed]{->}(0;0)(5;135)
		\psline[linestyle=dashed]{->}(0;0)(5;210)
		\psline[linestyle=dashed]{->}(0;0)(5;330)
		\psline[linestyle=dashed](0;0)(2.4;0)
		\psline[linestyle=dashed](0;0)(2.4;180)
		\psarc{->}{4.5}{90}{135}

		\uput{0}[112.5](4.6;112.5){$\theta$}
		\uput{0.1}[0](4.7;90){Reference}\uput{0.5}[0](4.3;90){axis}
	        \uput{0.1}[180](4.7;135){Direct}\uput{0.7}[180](4.2;135){axis}
		\uput{0.2}[0](4.7;45){Quadrature}\uput{0.8}[0](4.2;45){axis} 

		\uput{0}[0](2.6;-10){$a'$}\uput{0}[0](2.6;10){$i_a$}
		\uput{0}[0](2.7;70){$c$}\uput{0}[0](3;70){$i_c$}
		\uput{0}[120](2.9;120){$i_b$}\uput{0}[0](2.8;110){$b'$}
                \uput{0}[0](2.85;170){$i_a$}\uput{0}[0](2.8;190){$a$}
		\uput{0}[0](3.1;240){$c'$}
		\uput{0}[0](2.8;290){$i_b$}\uput{0}[300](2.9;300){$b$}
		\uput{0.1}[0](1.8;40){$D$}\uput{0.1}[0](1.3;38){$F$}
		\uput{0.1}[0](1.8;65){$i_D$}\uput{0.1}[0](1.3;70){$i_F$}
		\uput{0.1}[0](2.1;230){$D'$}\uput{0.1}[0](1.6;230){$F'$}

		\uput{0.1}[0](1.8;295){$i_Q$}\uput{0.1}[0](1.8;325){$Q$}
		\uput{0.1}[0](2.2;125){$Q'$}
		
		\pscircle[linestyle=dashed](2.85;45){0.15}\pscircle[linestyle=dashed](2.85;225){0.15}
		\pscircle[fillstyle=solid,fillcolor=black](2.85;225){0.05}
		\uput{0}[45](2.65;45){$\times$}
		\uput{0.3}[0](2.85;45){$d$}
		\uput{0.3}[0](3.05;225){$d'$}
		\uput{0.3}[90](2.85;45){$i_d$}
		
		\pscircle[linestyle=dashed](3.05;135){0.15}\pscircle[linestyle=dashed](3.05;315){0.15}
		\pscircle[fillstyle=solid,fillcolor=black](3.05;135){0.05}
		\uput{0}[315](2.85;315){$\times$}
		\uput{0.3}[90](3.05;135){$q'$}
		\uput{0.3}[0](3.05;315){$q$}
		\uput{0.3}[180](3.05;135){$i_q$}
		
                \end{pspicture}
		
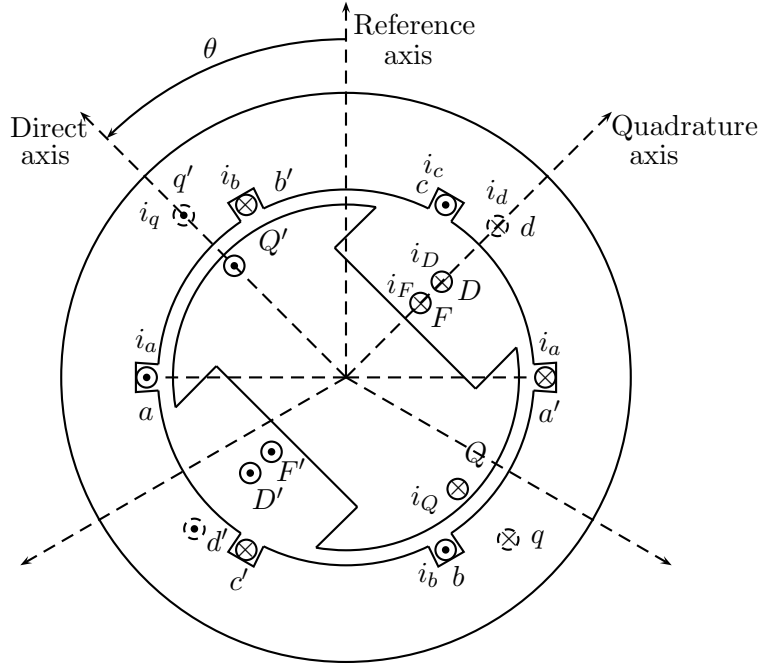
\captionof{figure}{Machine schematic.}
		\label{fig:electrical_dynamics}
		\end{centergroup}
	
By applying \autoref{eq:coil_kvl} and using the circuits
convention on the associated reference directions in \autoref{fig:electrical_dynamics}, we
get the relationship between voltages, currents, and flux linkages \cite{BV00}.
\begin{equation}
	\begin{aligned}
	\bbm v_{a'a} \\ v_{b'b} \\ v_{c'c} \\ v_{FF'} \\ v_{DD'} \\ v_{QQ'} \ebm &= 
	\bbm r & & & & & \\ & r & & & 0 &  \\ & & r & & & \\ & & & r_F & & \\ & 0 & & & r_D
	& \\ & & & & & r_Q \ebm \bbm i_a \\ i_b \\ i_c \\ i_F \\ i_D \\ i_Q \ebm +
	\frac{d}{dt} \bbm \lambda_{aa'} \\ \lambda_{bb'} \\ \lambda_{cc'} \\ \lambda_{FF'}
	\\ \lambda_{DD'} \\ \lambda_{QQ'} \ebm \\
	&= \mathbf{Ri} + \frac{d\mbox{\boldmath$\lambda$}}{dt} \\
        \end{aligned}
	\label{eq:electrical_dynamics}
\end{equation}
We simplify the equation above by using a single-subscript notation, i.e., $v_a \triangleq
v_{aa'} = -v_{a'a}$, $v_b \triangleq v_{bb'}= -v_{b'b}$, $v_c \triangleq v_{cc'}$,
$v_{F}\triangleq v_{FF'}$, $v_{D}\triangleq v_{DD'}$, and $v_{Q}\triangleq v_{QQ'}$.
Here, we define $\mathbf{v} \triangleq [v_a, v_b, v_c, -v_F, -v_D, -v_Q]^\mathrm{T}$ to be
the voltage vector consisting of the three phase terminal voltages ($v_a$, $v_b$, $v_c$), and the voltage of the field
winding ($v_F$) and two damper windings ($v_D$, $v_Q$). The corresponding current vector is defined as
$\mathbf{i} \triangleq [i_a, i_b, i_c, i_F, i_D, i_Q]^\mathrm{T}$. Then
\autoref{eq:electrical_dynamics} can be written as follows:
\begin{equation}
	\mathbf{v} = -\mathbf{R}\mathbf{i} - \frac{d\mbox{\boldmath$\lambda$}}{dt}
	\label{eq:basic}
\end{equation}
   
\subsubsection{Voltage Equation in the Synchronously Rotating Frame}
The electrical dynamics as given in \autoref{eq:electrical_dynamics} are derived in the
static $abc$ frame. The flux linkage in \autoref{eq:electrical_dynamics} is dependent on
the self and mutual inductances which are not constant, but are time varying. In the
voltage equation as given in \autoref{eq:basic} the $\dot{\lambda}$ term must be computed 
as $\dot{\lambda}=L\dot{i}+\dot{L}i$. Thus, to simplify
the equations we make a coordinate transformation which transforms
variables from the $abc$ static frame to a synchronously rotating frame (which
is also called $dq$ frame, see \autoref{fig:electrical_dynamics}). As a result of this
transformation we introduce two fictitious windings $dd'$ and $qq'$, as shown in
\autoref{fig:electrical_dynamics}. Thus we get
\begin{equation}
	\begin{aligned}
		v_d &= -ri_d - \omega\lambda_q - \frac{d\lambda_d}{dt} \\
		v_q &= -ri_q + \omega\lambda_d - \frac{d\lambda_q}{dt} \\
		v_F &= r_Fi_F + \frac{d\lambda_F}{dt} \\
		v_D &= r_Di_D + \frac{d\lambda_D}{dt} \\
		v_Q &= r_Qi_Q + \frac{d\lambda_Q}{dt} \\
	\end{aligned}
	\label{eq:dq_eq}
\end{equation}
 The details of the derivation of this transformation are given in the
 \hyperref[sec:appendix]{Appendix}. The extra terms $-\omega\lambda_q$ and
 $\omega\lambda_d$ are introduced by the transformation.

We can rearrange \autoref{eq:dq_eq} to put the quantities on the direct axis
together and the quantities on the  quadrature axis together. Hence, \autoref{eq:dq_eq} is
rewritten as follows:
\begin{equation}
	\begin{aligned}
		v_d &= -ri_d - \omega\lambda_q - \frac{d\lambda_d}{dt} \\
		v_F &= r_Fi_F + \frac{d\lambda_F}{dt} \\
		v_D &= r_Di_D + \frac{d\lambda_D}{dt} \\
		v_q &= -ri_q + \omega\lambda_d - \frac{d\lambda_q}{dt} \\
		v_Q &= r_Qi_Q + \frac{d\lambda_Q}{dt} \\
	\end{aligned}
	\label{eq:dq_eq1}
\end{equation}

As the damper windings are short-circuited, the terminal voltages are both zero.
As shown in \autoref{fig:electrical_dynamics} the direct axis is perpendicular to the
windings $dd'$, $FF'$, and $DD'$; the quadrature axis is perpendicular to the windings
$qq'$ and $QQ'$. Using the right hand thumb rule we can see that the flux linkage due to
the currents $i_d$, $i_F$, and $i_D$ is along the direct axis and the flux linkage due to
the currents $i_q$ and $i_Q$ is along the quadrature axis. Thus, the flux linkage
$\lambda_d$ along the $dd'$ winding depends on the currents $i_d$, $i_F$, and $i_D$ and is
given by $\lambda_d=L_di_d+kM_Fi_F+kM_Di_D$, where $L_d$ is the self inductance of $dd'$
winding, $M_F$ is the mutual inductance between $dd'$ and $FF'$ windings, and $M_D$ is the
mutual inductance between $dd'$ and $DD'$ windings, respectively. We can derive equations
for $\lambda_F$ and $\lambda_D$ in a similar fashion. Also, the flux linkage $\lambda_q$
along the $qq'$ winding depends on the currents $i_q$, and $i_Q$ and is given by
$\lambda_q=L_qi_q+kM_Qi_Q$, where $L_q$ is the self inductance of the $qq'$ winding, and
$M_Q$ is the mutual inductance between $qq'$ and $QQ'$ windings, respectively. We can
derive $\lambda_Q$ using the same approach. Also note that in  \autoref{eq:direct_axis}
and \autoref{eq:quadrature_axis} which gives a relationship between the flux and the
current in each winding, the mutual inductance between the windings $FF'$ and $DD'$ is
denoted by $M_R$, and self-inductances of the windings are denoted by $L_d$, $L_F$, $L_D$,
$L_q$, and $L_Q$, respectively. Thus, the connection between the flux and the current is
given by
\begin{equation}
	\bbm \lambda_d \\ \lambda_F \\ \lambda_D \ebm = 
	    \bbm L_d & kM_F & kM_D \\ kM_F & L_F & M_R \\ kM_D & M_R & L_D\\ \ebm 
	    \bbm i_d \\ i_F \\ i_D \ebm
	\label{eq:direct_axis}
\end{equation}
and 
\begin{equation}
	\bbm \lambda_q \\ \lambda_Q \ebm = \bbm L_q & kM_Q \\ kM_Q & L_Q \ebm 
	\bbm i_q \\ i_Q \ebm
	\label{eq:quadrature_axis}
\end{equation}
where $k = \sqrt{3/2}$. If we substitute \autoref{eq:direct_axis} and
\autoref{eq:quadrature_axis} into \autoref{eq:dq_eq1} and put it in matrix form, we obtain
\begin{equation}
	\bbm v_d \\ v_F \\ 0 \\ v_q \\ 0 \ebm = 
	\bbm -r & 0 & 0 & -\omega L_q & -\omega kM_Q \\
	     0 & r_F & 0 & 0 & 0\\
	     0 & 0 & r_D & 0 & 0\\
	     \omega L_d & \omega kM_F & \omega kM_D & -r & 0 \\
	     0 & 0 & 0 & 0 & r_Q \ebm \bbm i_d \\ i_F \\ i_D \\ i_q \\ i_Q \ebm + 
	\bbm -L_d & -kM_F & -kM_D & 0 & 0 \\
	     kM_F & L_F & M_R & 0 & 0 \\
	     kM_D & M_R & L_D & 0 & 0 \\
	     0 & 0 & 0 & -L_q & -kM_Q \\
	     0 & 0 & 0 & kM_Q & L_Q \ebm \bbm \dot{i_d} \\ \dot{i_F} \\
	     \dot{i_D} \\ \dot{i_q} \\ \dot{i_Q} \ebm
	\label{eq:ss_eq}
\end{equation}
Moving the derivative of the current to the left-hand side, we obtain
\begin{equation}
	\bbm L_d & kM_F & kM_D & 0 & 0 \\
	     -kM_F & -L_F & -M_R & 0 & 0 \\
	     -kM_D & -M_R & -L_D & 0 & 0 \\
	     0 & 0 & 0 & L_q & kM_Q \\
	     0 & 0 & 0 & -kM_Q & -L_Q \ebm \bbm \dot{i_d} \\ \dot{i_F} \\
	     \dot{i_D} \\ \dot{i_q} \\ \dot{i_Q} \ebm = 
        \bbm -r & 0 & 0 & -\omega L_q & -\omega kM_Q \\
	     0 & r_F & 0 & 0 & 0\\
	     0 & 0 & r_D & 0 & 0\\
	     \omega L_d & \omega kM_F & \omega kM_D & -r & 0 \\
	     0 & 0 & 0 & 0 & r_Q \ebm \bbm i_d \\ i_F \\ i_D \\ i_q \\ i_Q \ebm - 
        \bbm v_d  \\ v_F \\ 0 \\ v_q \\ 0 \ebm
	\label{eq:ss_eqq}
\end{equation}

\subsubsection{Voltage Equation of the Synchronous Generator in Per Unit System}

A normalization of variables called the per unit normalization is
always desirable. The idea is to pick base values for quantities such as voltages,
currents, impedances, power, and so on, and to define the quantity in per unit as
\begin{equation}
	\mathrm{quantity\ in\ per\ unit} = \frac{\mathrm{actual\ quantity}}{\mathrm{base\
	value\ of\ quantity}}
	\label{eq:per_unit}
\end{equation}

By carefully choosing the base quantities for both stator and rotor variables, the
electrical dynamics expressed by \autoref{eq:ss_eqq} can be expressed
in the p.u.\ system as
\begin{equation}
	\bbm L_d & kM_F & kM_D & 0 & 0 \\
	     -kM_F & -L_F & -M_R & 0 & 0 \\
	     -kM_D & -M_R & -L_D & 0 & 0 \\
	     0 & 0 & 0 & L_q & kM_Q \\
	     0 & 0 & 0 & -kM_Q & -L_Q \ebm \bbm \dot{i_d} \\ \dot{i_F} \\
	     \dot{i_D} \\ \dot{i_q} \\ \dot{i_Q} \ebm = 
        \bbm -r & 0 & 0 & -\omega L_q & -\omega kM_Q \\
	     0 & r_F & 0 & 0 & 0\\
	     0 & 0 & r_D & 0 & 0\\
	     \omega L_d & \omega kM_F & \omega kM_D & -r & 0 \\
	     0 & 0 & 0 & 0 & r_Q \ebm \bbm i_d \\ i_F \\ i_D \\ i_q \\ i_Q \ebm - 
	     \bbm v_d  \\ v_F \\ 0 \\ v_q \\ 0 \ebm \ \mathrm{p.u.}
	\label{eq:ss_eqq_pu}
\end{equation}
It is obvious that \autoref{eq:ss_eqq} and \autoref{eq:ss_eqq_pu} are identical. This is
always possible if base quantities are carefully chosen. The derivation of
\autoref{eq:ss_eqq_pu} can be found in \cite{AF03}.

\subsection{A Synchronous Generator Connected to an Infinite Bus}

A typical configuration of a generation system model is a synchronous generator
connected to an infinite bus as shown in \autoref{fig:generator_infinite_bus}. The figure
shows a synchronous generator connected to an infinite bus through a transmission line having
resistance $R_e$ and inductance $L_e$. Only the voltages and currents for phase $a$ are
shown, where $v_a$ is the phase voltage, $i_a$ is the phase current, and $V_{\infty}$ is the infinite bus voltage. An infinite bus is
an approximation of a large interconnected power system, where the action of a single
generator will not affect the operation of the power grid. In an infinite bus, the system
frequency is constant, independent of power flow, and the system voltage is constant,
independent of reactive power consumed or supplied.
\begin{centergroup}
		\begin{pspicture}(-5,-4)(5,1)
                \coil[dipolestyle=curved](0,0)(2,0){$L_e$}
		\resistor[dipolestyle=zigzag](-2,0)(0,0){$R_e$}
		\psset{arrowscale=2}
		\psline{->}(-1,-0.5)(1,-0.5)
		\rput(0,-0.8){\psframebox*{$i_a$}}
		\pscircle(-4,-2){0.5}\pscircle(4,-2){0.5}
		\psline(-4,-1.5)(-4,0)(-2,0)\psline(2,0)(4,0)(4,-1.5)
		\psline(-4,-2.5)(-4,-3.5)\psline(4,-2.5)(4,-3.5)
		\psline(-5,-3.5)(5,-3.5)
		\rput(-3.3,-1.2){\psframebox*{$+$}}\rput(-3.3,-2.8){\psframebox*{$-$}}
		\rput(-3,-2){\psframebox*{$v_a$}}
		\rput(3.3,-1.2){\psframebox*{$+$}}\rput(3.3,-2.8){\psframebox*{$-$}}
		\rput(2.8,-2){\psframebox*{$V_\infty,\ \alpha$}}
                \end{pspicture}
		
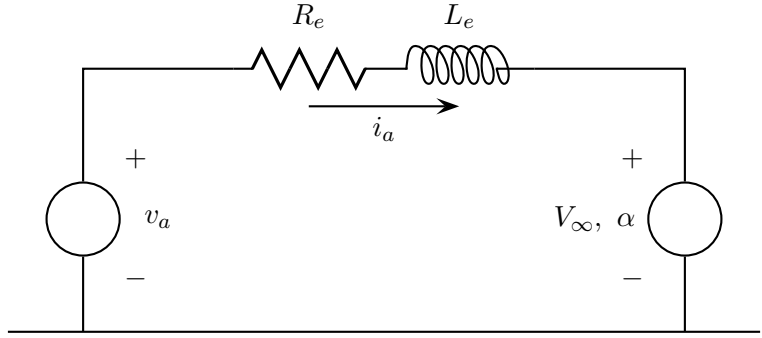
\captionof{figure}{Synchronous generator loaded by an infinite bus.}
		\label{fig:generator_infinite_bus}
\end{centergroup}
The constraints of the infinite bus are given by
\begin{equation}
	\bbm v_d \\ v_q \ebm = R_e \bbm i_d \\ i_q \ebm + L_e\bbm \dot{i_d} \\ \dot{i_q}
	\ebm - \omega L_e\bbm -i_q \\ i_d \ebm + \sqrt{3}V_{\infty}\bbm -\sin{(\delta - \alpha)} \\
	 \cos{(\delta - \alpha)}\ebm
	\label{eq:inf_bus}
\end{equation}
By including \autoref{eq:inf_bus}, we can rewrite \autoref{eq:ss_eqq_pu} as 
\begin{equation}
	\begin{aligned}
	\bbm L_d+L_e & kM_F & kM_D & 0 & 0 \\
	     -kM_F & -L_F & -M_R & 0 & 0 \\
	     -kM_D & -M_R & -L_D & 0 & 0 \\
	     0 & 0 & 0 & L_q+L_e & kM_Q \\
	     0 & 0 & 0 & -kM_Q & -L_Q \ebm \bbm \dot{i_d} \\ \dot{i_F} \\
	     \dot{i_D} \\ \dot{i_q} \\ \dot{i_Q} \ebm &= 
	     \bbm -r-R_e & 0 & 0 & -\omega (L_q+L_e) & -\omega kM_Q \\
	     0 & r_F & 0 & 0 & 0\\
	     0 & 0 & r_D & 0 & 0\\
	     \omega (L_d+L_e) & \omega kM_F & \omega kM_D & -r-R_e & 0 \\
	     0 & 0 & 0 & 0 & r_Q \ebm \bbm i_d \\ i_F \\ i_D \\ i_q \\ i_Q \ebm \\ 
	     &- \bbm 0 \\ v_F \\ 0 \\ 0 \\ 0 \ebm - \sqrt{3}V_\infty\bbm -\sin{(\delta - \alpha)} \\ 0 \\ 0 \\ 
        \cos{(\delta - \alpha)} \\ 0 \ebm \\
     \end{aligned}
	\label{eq:ss_inf}
\end{equation}
Thus, \autoref{eq:ss_inf} describes the electrical dynamics of a synchronous generator
connected to an infinite bus. 

\subsection{Mechanical Dynamics: Swing Equation}

In this subsection, we present the mechanical
dynamics of the synchronous generator.
Under normal operating conditions, the relative position of the rotor axis and the
resultant magnetic field axis is fixed. The angle between the two is known as the
\emph{power angle} or \emph{torque angle}. During any disturbance, the rotor will
decelerate or accelerate with respect to the synchronously rotating air gap magneto-motive
force (mmf), which is any physical driving (motive) force that produces magnetic flux, and
a relative motion begins. In this context, the expression 'driving force' is used in a
general sense of work potential, and is analogous, but distinct, from force measured in
Newton's. In magnetic circuits the magneto-motive
force (mmf) plays a role analogous to the role emf (voltage) plays in
electric circuits. 
The equation describing this relative motion is known as the
\emph{swing equation} \cite{AF03}. If, after this oscillatory period, the rotor locks back
into synchronous speed, the generator will maintain its stability. If the disturbance does
not involve any net change in power, the rotor returns to its original position. If the
disturbance is created by a change in generation, load, or in network conditions, the
rotor comes to a new operating power angle relative to the synchronously revolving field.

The swing equation thus governs the motion of the machine rotor relating the moment of
inertia (also referred to as the rotational inertia of the rotor) to the resultant of the
mechanical and electrical torques on the rotor, i.e., $J\ddot{\theta} = T_a\
\mathrm{N\cdot m}$, where $J$ is the moment of
inertia of all rotating masses attached to the shaft, $\theta $ is the mechanical angle of
the shaft with
respect to a fixed reference, and $T_a$ is the accelerating torque acting on the shaft. The torque is given by 
$T_a = T_m - T_e - T_d$, where $T_m$, $T_e$, and $T_d$ are mechanical,
electrical, and damping torques, respectively. The mechanical torque $T_m$ is the driving
torque provided by the prime mover. The electrical torque $T_e$ is generated by the load
currents of the armature windings on the stator. The damping torque $T_d$ is produced by
the damper windings on the rotor. The angular
reference may be chosen relative to a synchronously rotating reference frame moving with
constant angular velocity $\omega_{mR}$. The rotor angle in the static frame is
given by $\theta(t) = (\omega_{mR} t + \beta) + \delta_m$, where
$\beta$ is a constant and $\delta_m$ is the rotor position also referred to as the
\emph{mechanical torque angle}, measured from the synchronously rotating
reference frame. 
Let us denote the \emph{shaft angular velocity} in the static frame as $\omega_m$ in rad/sec, thus we
have $\omega_m =\dot{\theta} =  \omega_{mR} + \dot{\delta}_m$. By taking the
derivative of $\omega_m$ and second derivative of $\theta$ we obtain
$\dot{\omega}_m=\ddot{\theta}=\ddot{\delta}_m$, if we substitute
this in $J\ddot{\theta} = T_a$ we have 
\begin{equation}				
J\ddot{\theta} = J\ddot{\delta }_m=J\dot{\omega }_m=T_a=T_m - T_e - T_d\ [\mathrm{N\cdot m}]
\end{equation}				
The product of torque and
angular velocity is the shaft power in watts, thus we have
\begin{equation}
	J\ddot{\delta}_m\omega_m = P_m - P_e - P_d \ \ [\mathrm{W}]
\end{equation}
The quantity $J\omega_m$ is called the inertia constant and is denoted by $M$. It is
related to the kinetic energy of the rotating masses $W_k$, where $W_k=\frac{1}{2}J\omega
^2_m$. $M$ is computed as
\begin{equation}
	M=J\omega_m=\frac{2W_k}{\omega_m}\ [\mathrm{J\cdot s}]
	\label{eq:inertia_constant}
\end{equation}
Although $M$ is called an inertia constant, it is not really constant when the rotor speed
deviates from the synchronous speed $\omega_{mR}$. However, since $\omega_m$ does not
change by a large amount before stability is lost, $M$ is evaluated at the synchronous
speed $\omega_{mR}$ and is considered to remain constant, i.e.,
\begin{equation}
	M=J\omega_m\cong \frac{2W_k}{\omega_{mR}}\ [\mathrm{J\cdot s}]
	\label{eq:inertia_approximate}
\end{equation}
The swing equation in terms of the inertia constant becomes
\begin{equation}
	M\ddot{\delta }_m=M\dot{\omega }_m=P_m-P_e-P_d
	\label{eq:swing1}
\end{equation}
In relating the machine inertial performance to the network, it would be more useful to
write \autoref{eq:swing1} in terms of an electrical angle that can be conveniently related
to the position of the rotor. Such an angle is the torque angle $\delta $, which is the
angle between the magneto-motive force (mmf) and the resultant magneto-motive force (mmf) in the air gap, both rotating at synchronous
speed. It is also the electrical angle between the generated emf and the resultant stator
voltage phasors. The torque angle $\delta $, which is the same as the
electrical angle, is related to the rotor mechanical angle $\delta _m$, (measured from a
synchronously rotating frame) by
                \begin{equation}
                           \delta =\frac{p}{2}\delta _m
                 \label{eq:swing2}
                \end{equation} 
                 where $p$ is the number of poles of the synchronous generator. \autoref{fig:machine_structure} shows a 
           schematic of a synchronous generator with two poles. Also, the synchronous speed $\omega _{mR}$ used in the previous equations
         is actually the mechanical synchronous speed or the mechanical angular velocity at the synchronous reference value.
         It is related to the electrical synchronous speed $\omega _ R$ by
         \begin{equation}
                           \omega _R =\frac{p}{2}\omega  _{mR}
                 \label{eq:swing2point1}
         \end{equation}
By taking the derivative of
		 \autoref{eq:swing2} on both sides, we get
         \begin{equation}
                           \dot{\delta }=\frac{p}{2}\dot{\delta }_m
                 \label{eq:swing2point2}
         \end{equation}		 
		 Adding \autoref{eq:swing2point1} and \autoref{eq:swing2point2} we get
		 \begin{equation}
                           \dot{\delta }+\omega _R = \frac{p}{2}\dot{\delta }_m+\frac{p}{2}\omega  _{mR}\\
                 \label{eq:swing2point3}
         \end{equation}	
 Thus, the electrical angular velocity
		 $\omega$ is related to the mechanical angular velocity $\omega_m$ by
		 \begin{equation}
		 \omega = \dot{\delta} + \omega _R =
		 \frac{p}{2}(\dot{\delta}_m+\omega_{mR}) = \frac{p}{2}\omega_m
        \label{eq:swing2point4}
         \end{equation}
 Combining
		 \autoref{eq:swing1} and \autoref{eq:swing2} we get
                 \begin{equation}
                      \frac{2M}{p}\ddot{\delta }=\frac{2M}{p}\dot{\omega }= P_m-P_e-P_d
                 \label{eq:swing3}
                \end{equation}
                 Thus, we can rewrite \autoref{eq:swing3} as follows:
                \begin{equation}
                     \ddot{\delta }=\dot{\omega }=-\frac{p}{2M}P_d+\frac{p}{2M}(P_m-P_e)
                 \label{eq:swing4}
                \end{equation}
Since power system analysis is done in p.u.\ system, the swing equation is usually
expressed in per unit. Dividing \autoref{eq:swing3} by the base power $S_B$, and substituting
for $M$ results in               
 \begin{equation}
           \frac{2}{p}\frac{2W_k}{\omega _{mR}S_B}\ddot{\delta }=
          \frac{2}{p}\frac{2W_k}{\omega _{mR}S_B}\dot{\omega }= \frac{P_m}{S_B}-\frac{P_e}{S_B}-\frac{P_d}{S_B}
 \label{eq:swing5}
 \end{equation}
 We now define an important quantity known as the p.u.\ inertia constant $H$\cite{RTWQ}.
 \begin{equation}
        H=\frac{W_k}{S_B}
 \label{eq:swing6}
 \end{equation}
The unit of $H$ is seconds. The value of $H$ ranges from 1 to 10 seconds, depending
on the size and type of machine. The per unit accelerating power is related to the per unit
accelerating torque by $P_{a\mathrm{(p.u.)}}=T_{a\mathrm{(p.u.)}}\frac{\omega}{\omega _R}$. Recognizing that the electrical angular speed $\omega $
is nearly constant, and equal to $\omega_ R$, we have the p.u.\ accelerating
power $P_a$ to be numerically nearly equal to the p.u.\ accelerating torque $T_a$, i.e.
$P_{a\mathrm{(p.u.)}}\cong T_{a\mathrm{(p.u.)}}$.  
Substituting
for $H$, and $P_{a\mathrm{(p.u.)}} \cong T_{a\mathrm{(p.u.)}}$ in \autoref{eq:swing5}, we get      
\begin{equation}
           \frac{2}{p}\frac{2H}{\omega_{mR}}\ddot{\delta}=
	   \frac{2}{p}\frac{2H}{\omega_{mR}}\dot{\omega}=P_{a\mathrm{(p.u.)}}=
	   P_{m\mathrm{(p.u.)}}-P_{e\mathrm{(p.u.)}}-P_{d\mathrm{(p.u.)}}
	   \cong T_{a\mathrm{(p.u.)}}=T_{m\mathrm{(p.u.)}}-T_{e\mathrm{(p.u.)}}-T_{d\mathrm{(p.u.)}}
\label{eq:swing7}
\end{equation}
where $P_{m\mathrm{(p.u.)}}$, $P_{e\mathrm{(p.u.)}}$, and $P_{d\mathrm{(p.u.)}}$ are the
per unit mechanical power, electrical power, and damping power respectively.
Substituting $\omega _ R =\frac{p}{2}\omega  _{mR}$ in \autoref{eq:swing7} we get
\begin{equation}
          \frac{2H}{\omega_ R}\ddot{\delta }=
           \frac{2H}{\omega_ R}\dot{\omega }= P_{m\mathrm{(p.u.})}-
	   P_{e\mathrm{(p.u.)}}-P_{d\mathrm{(p.u.)}}\cong
	   T_{m\mathrm{(p.u.)}}-T_{e\mathrm{(p.u.)}}-T_{d\mathrm{(p.u.)}}
\label{eq:swing8}
\end{equation}  
In \autoref{eq:swing8}, while the torque is normalized, the angular speed $\omega $ and the time $t$ are not in per unit.
Thus the equation is not completely in per unit. We know that the angular speed $\omega $ and time $t$ in per unit are given by
\begin{equation}
\begin{aligned}
               \omega _{(p.u.)} &=\frac{\omega }{\omega _R}\\
               t_{(p.u.)} &=\omega _Rt\\
               \end{aligned}
 \label{eq:swing9}
\end{equation}
where the base angular velocity $\omega _B=\omega _R$.
Substituting \autoref{eq:swing9} in \autoref{eq:swing8} the normalized swing equation can be written as
 \begin{equation}
          \tau _j\frac {d\omega_{(p.u.)}}{dt_{(p.u.)}}=
	   T_{m\mathrm{(p.u.)}}-T_{e\mathrm{(p.u.)}}-T_{d\mathrm{(p.u.)}}
\label{eq:swing10}
\end{equation} 
where $\tau _j=2H\omega _R$.                
The damping torque is calculated as $T_{d\mathrm{(p.u.)}} = D\omega$, where
$D$ is the damping constant. The electrical torque $T_{e\phi }$ is calculated as
\begin{equation}
	T_{e\phi } = i_q\lambda_d - i_d\lambda_q 
	\label{eq:electrical_torque}
\end{equation}
Also $T_e=\frac{T_{e\phi }}{3}$, where $T_e$ is the per unit electromagnetic torque defined on a three phase $(3\phi )$ VA base, and
$T_{e\phi }$ is the per unit electromagnetic torque defined on a per phase VA base.
Substituting \autoref{eq:direct_axis} and \autoref{eq:quadrature_axis} into
\autoref{eq:electrical_torque} and writing $T_e$ in the p.u.\ system, we obtain
\begin{equation}
	T_{e\phi \mathrm{(p.u.)}} =3T_{e(p.u.)}= L_d i_d i_q + kM_F i_F i_q + kM_Di_Di_q - L_qi_di_q -
	kM_Qi_di_Q	
	\label{eq:te_pu}
\end{equation}
From \autoref{eq:swing2point4} we have $\omega = \dot{\delta} + \omega_ R$. If we choose $\omega_ R$ as the frequency base and divide
both sides of this equation by $\omega_ R$ we have 
\begin{equation}
       \frac{\omega }{\omega_ R}=\frac{\dot{\delta}}{\omega_ R}+\frac{\omega_ R}{\omega_ R}
   \label{eq:omega_pu}
\end{equation}
Since, $\omega_{\mathrm{(p.u.)}}=\frac{\omega }{\omega _ R}$ and $\dot{\delta }_{\mathrm{(p.u.)}}=\frac{\dot{\delta}}{\omega_ R}$
we can write \autoref{eq:omega_pu} as    
$\dot{\delta}_{(pu)} = \omega _{(pu)}- 1$. Thus, from \autoref{eq:swing8}, \autoref{eq:te_pu}, and \autoref{eq:omega_pu}
we can write the mechanical dynamics in the p.u.\ system as
\begin{equation}
	\begin{aligned}
		\dot{\omega} &= -\frac{1}{\tau _j}\frac{(L_d i_d i_q + kM_F i_F i_q +
		kM_Di_Di_q - L_qi_di_q - kM_Qi_di_Q)}{3} - \frac{1}{\tau _j}D\omega +
		\frac{1}{\tau _j}T_m \\
		\dot{\delta} &= \omega - 1 \\
	\end{aligned}
	\label{eq:mechanical_pu}
\end{equation}

\subsection{Truth Model of the Synchronous Generator}
By combining the electrical dynamics and mechanical dynamics, we obtain the 
truth model of the synchronous generator which is highly nonlinear. Let us define 
\begin{equation*}
	L = \bbm L_d+L_e & kM_F & kM_D & 0 & 0 \\
	     -kM_F & -L_F & -M_R & 0 & 0 \\
	     -kM_D & -M_R & -L_D & 0 & 0 \\
	     0 & 0 & 0 & L_q+L_e & kM_Q \\
	     0 & 0 & 0 & -kM_Q & -L_Q \ebm
\end{equation*}
Also denote $\mu = (L_d + L_e)M_R^2  - L_DL_F(L_d + L_e) + k^2(L_DM_F^2 + L_FM_D^2 -
2M_DM_FM_R)$ and $\nu = -k^2M_Q^2 + L_Q(L_e+L_q)$, we can derive the inverse matrix of
$L$ as
\begin{equation}
	\begin{aligned}
	L^{-1} &= \bbm \frac{1}{\mu }(M_R^2 - L_DL_F) & \frac{k}{\mu }(M_DM_R -
	              L_DM_F) & \frac{k}{\mu }(M_FM_R - L_FM_D) & 0 & 0 \\
	              -\frac{k}{\mu }(M_DM_R - L_DM_F) &
		      -\frac{1}{\mu }(M_D^2k^2-L_D(L_d+L_e)) & -\frac{1}{\mu }((L_d +
		      L_e)M_R - M_DM_Fk^2) & 0 & 0 \\
		      -\frac{k}{\mu }(M_FM_R - L_FM_D) & -\frac{1}{\mu }((L_d +
		      L_e)M_R - M_DM_Fk^2) & -\frac{1}{\mu }(M_F^2k^2-L_F(L_d+L_e)) & 0
		      & 0 \\
		      0 & 0 & 0 & \frac{L_Q}{\nu } & \frac{kM_Q}{\nu } \\
		      0 & 0 & 0 & -\frac{kM_Q}{\nu } & -\frac{L_e+L_q}{\nu } \ebm \\
	       &= \bbm L_{d1} & kM_{F1} & kM_{D1} & 0 & 0 \\
	           -kM_{F1} & -L_{F1} & -M_{R1} & 0 & 0 \\
	           -kM_{D1} & -M_{R1} & -L_{D1} & 0 & 0 \\
	           0 & 0 & 0 & L_{q1} & kM_{Q1} \\
	           0 & 0 & 0 & -kM_{Q1} & -L_{Q1} \ebm \\
	 \end{aligned}
	 \label{eq:lmatrix}
\end{equation}
where $L_{d1} = \frac{1}{\mu }(M_R^2 - L_DL_F)$, $L_{F1} =
\frac{1}{\mu }(M_D^2k^2-L_D(L_d+L_e))$, $L_{D1} =
\frac{1}{\mu }(M_F^2k^2-L_F(L_d+L_e))$, $M_{F1} = \frac{1}{\mu }(M_DM_R -
	              L_DM_F)$, $M_{D1} = \frac{1}{\mu }(M_FM_R - L_FM_D)$, $M_{R1} =
		      \frac{1}{\mu }((L_d + L_e)M_R - M_DM_Fk^2)$, $L_{q1} =
		      \frac{L_Q}{\nu }$, $L_{Q1} = \frac{L_e+L_q}{\nu }$, and
		      $M_{Q1}=\frac{M_Q}{\nu }$.
Using \autoref{eq:lmatrix} and
\autoref{eq:ss_inf} we can write
\begin{equation}
	\begin{aligned}
		&\dot{i}_d = -L_{d1}(r+R_e)i_d + kM_{F1}r_Fi_F+kM_{D1}r_Di_D-
		(L_q+L_e)L_{d1}i_q\omega - kM_QL_{d1}i_Q\omega  \\
         & \ \ \ \ \ \ \ + \sqrt{3}V_\infty L_{d1}\sin(\delta -\alpha )- kM_{F1}v_F \\
		&\dot{i}_F = kM_{F1}(r+R_e)i_d- L_{F1}r_Fi_F - M_{R1}r_Di_D + 
		kM_{F1}(L_q+L_e)i_q\omega  + k^2M_{F1}M_Qi_Q\omega \\
        & \ \ \ \ \ \ \ - \sqrt{3}V_\infty kM_{F1}\sin(\delta -\alpha )+L_{F1}v_F \\
		&\dot{i}_D = kM_{D1}(r+R_e)i_d-M_{R1}r_Fi_F-L_{D1}r_Di_D+
		kM_{D1}(L_q+L_e)i_q\omega + k^2M_{D1}M_Qi_Q\omega \\
        & \ \ \ \ \ \ \ - \sqrt{3}V_\infty kM_{D1}\sin(\delta -\alpha )+ M_{R1}v_F \\
		&\dot{i}_q = L_{q1}(L_d+L_e)i_d\omega + kM_FL_{q1}i_F\omega +
		kM_DL_{q1}i_D\omega  - L_{q1}(r+R_e)i_q + kM_{Q1}r_Qi_Q \\ 
        & \ \ \ \ \ \ \ - \sqrt{3}V_\infty L_{q1}\cos(\delta -\alpha) \\
         &\dot{i}_Q = -kM_{Q1}(L_d+L_e)i_d\omega  - k^2M_{Q1}M_Fi_F\omega  -
		k^2M_{Q1}M_Di_D\omega  + kM_{Q1}(r+R_e)i_q - L_{Q1}r_Qi_Q\\ 
        & \ \ \ \ \ \ \ + \sqrt{3}V_\infty kM_{Q1}\cos(\delta -\alpha ) \\
        \end{aligned}
	\label{eq:nonlinear_model}
\end{equation}
Dividing both LHS and RHS of \autoref{eq:nonlinear_model} by $\sqrt{3}$ we get
\begin{equation}
	\begin{aligned}
&\frac{\dot{i}_d}{\sqrt{3}} = -L_{d1}(r+R_e)\frac{i_d}{\sqrt{3}} + kM_{F1}r_F\frac{i_F}{\sqrt{3}}+kM_{D1}r_D\frac{i_D}{\sqrt{3}}-
		(L_q+L_e)L_{d1}\frac{i_q}{\sqrt{3}}\omega - kM_QL_{d1}\frac{i_Q}{\sqrt{3}}\omega \\ 
        & \ \ \ \ \ \ \ + V_\infty L_{d1}\sin(\delta -\alpha)
        - kM_{F1}\frac{v_F}{\sqrt{3}} \\
		&\frac{\dot{i}_F}{\sqrt{3}} = kM_{F1}(r+R_e)\frac{i_d}{\sqrt{3}}- L_{F1}r_F\frac{i_F}{\sqrt{3}}- 
        M_{R1}r_D\frac{i_D}{\sqrt{3}}+ kM_{F1}(L_q+L_e)\frac{i_q}{\sqrt{3}}\omega  + k^2M_{F1}M_Q\frac{i_Q}{\sqrt{3}}\omega \\ 
        & \ \ \ \ \ \ \- V_\infty kM_{F1}\sin(\delta -\alpha)+L_{F1}\frac{v_F}{\sqrt{3}} \\
		&\frac{\dot{i}_D}{\sqrt{3}} = kM_{D1}(r+R_e)\frac{i_d}{\sqrt{3}}-M_{R1}r_F\frac{i_F}{\sqrt{3}}-L_{D1}r_D\frac{i_D}{\sqrt{3}}+
		kM_{D1}(L_q+L_e)\frac{i_q}{\sqrt{3}}\omega + k^2M_{D1}M_Q\frac{i_Q}{\sqrt{3}}\omega \\ 
       & \ \ \ \ \ \ \ - V_\infty kM_{D1}\sin(\delta -\alpha )+ M_{R1}\frac{v_F}{\sqrt{3}} \\
		&\frac{\dot{i}_q}{\sqrt{3}} = L_{q1}(L_d+L_e)\frac{i_d}{\sqrt{3}}\omega + kM_FL_{q1}\frac{i_F}{\sqrt{3}}\omega +
		kM_DL_{q1}\frac{i_D}{\sqrt{3}}\omega  - L_{q1}(r+R_e)\frac{i_q}{\sqrt{3}}+ kM_{Q1}r_Q\frac{i_Q}{\sqrt{3}}\\
        & \ \ \ \ \ \ \ - V_\infty L_{q1}\cos(\delta -\alpha)\\ 
         &\frac{\dot{i}_Q}{\sqrt{3}} = -kM_{Q1}(L_d+L_e)\frac{i_d}{\sqrt{3}}\omega  - k^2M_{Q1}M_F\frac{i_F}{\sqrt{3}}\omega  -
		k^2M_{Q1}M_D\frac{i_D}{\sqrt{3}}\omega  + kM_{Q1}(r+R_e)\frac{i_q}{\sqrt{3}} - L_{Q1}r_Q\frac{i_Q}{\sqrt{3}}\\ 
       & \ \ \ \ \ \ \ + V_\infty kM_{Q1}\cos(\delta -\alpha ) \\
\end{aligned}
	\label{eq:nonlinear_model1}
\end{equation}
Converting the state variables $i_d$, $i_F$, $i_D$, $i_q$, $i_Q$, and control input $v_F$ to their corresponding RMS quantities 
$I_d$, $I_F$, $I_D$, $I_q$, $I_Q$, and $V_F$ by substituting $\frac{i_d}{\sqrt{3}}=I_d$, $\frac{i_F}{\sqrt{3}}=I_F$, $\frac{i_D}{\sqrt{3}}=I_D$, $\frac{i_q}{\sqrt{3}}=I_q$, $\frac{i_Q}{\sqrt{3}}=I_Q$, and $\frac{v_F}{\sqrt{3}}=V_F$ in 
\autoref{eq:nonlinear_model1} we get
\begin{equation}
	\begin{aligned}
		&\dot{I}_d = -L_{d1}(r+R_e)I_d + kM_{F1}r_FI_F+kM_{D1}r_DI_D-
		(L_q+L_e)L_{d1}I_q\omega - kM_QL_{d1}I_Q\omega  + V_\infty L_{d1}\sin(\delta -\alpha)\\
        & \ \ \ \ \ \ \ - kM_{F1}V_F \\
		&\dot{I}_F = kM_{F1}(r+R_e)I_d- L_{F1}r_FI_F - M_{R1}r_DI_D + 
		kM_{F1}(L_q+L_e)I_q\omega  + k^2M_{F1}M_QI_Q\omega  - V_\infty kM_{F1}\sin(\delta -\alpha)\\
		& \ \ \ \ \ \ \ +L_{F1}V_F \\
		&\dot{I}_D = kM_{D1}(r+R_e)I_d-M_{R1}r_FI_F-L_{D1}r_DI_D+
		kM_{D1}(L_q+L_e)I_q\omega + k^2M_{D1}M_QI_Q\omega  - V_\infty kM_{D1}\sin(\delta -\alpha )\\ 
		& \ \ \ \ \ \ \ + M_{R1}V_F \\
		&\dot{I}_q = L_{q1}(L_d+L_e)I_d\omega + kM_FL_{q1}I_F\omega +
		kM_DL_{q1}I_D\omega  - L_{q1}(r+R_e)I_q + kM_{Q1}r_QI_Q  
         - V_\infty L_{q1}\cos(\delta -\alpha) \\
         &\dot{I}_Q = -kM_{Q1}(L_d+L_e)I_d\omega  - k^2M_{Q1}M_FI_F\omega  -
		k^2M_{Q1}M_DI_D\omega  + kM_{Q1}(r+R_e)I_q - L_{Q1}r_QI_Q\\ 
        & \ \ \ \ \ \ \ + V_\infty kM_{Q1}\cos(\delta -\alpha ) \\
        \end{aligned}
	\label{eq:nonlinear_model2}
\end{equation}
\autoref{eq:mechanical_pu} can be written as
 \begin{equation}
	\begin{aligned}		
		&\dot{\omega } = -\frac{1}{\tau _j}\bigg{(}L_d\frac{i_d}{\sqrt{3}}\frac{i_q}{\sqrt{3}} + kM_F\frac{i_F}{\sqrt{3}}
         \frac{i_q}{\sqrt{3}} + 
         kM_D\frac{i_D}{\sqrt{3}}\frac{i_q}{\sqrt{3}} -
		L_q\frac{i_d}{\sqrt{3}}\frac{i_q}{\sqrt{3}} - kM_Q\frac{i_d}{\sqrt{3}}\frac{i_Q}{\sqrt{3}}\bigg{)} 
        -\frac{1}{\tau _j} D\omega  + \frac{1}{\tau _j}T_m \\
		&\dot{\delta } = \omega - 1 \\
	\end{aligned}
	\label{eq:nonlinear_model3}
\end{equation}
Substituting $\frac{i_d}{\sqrt{3}}=I_d$, $\frac{i_F}{\sqrt{3}}=I_F$, $\frac{i_D}{\sqrt{3}}=I_D$, $\frac{i_q}{\sqrt{3}}=I_q$, $\frac{i_Q}{\sqrt{3}}=I_Q$ in \autoref{eq:nonlinear_model3}
\begin{equation}
	\begin{aligned}		
		&\dot{\omega } = -\frac{1}{\tau _j}\bigg{(}(L_d-L_q)I_dI_q + kM_FI_FI_q + 
         kM_DI_DI_q - kM_QI_dI_Q\bigg{)} 
        -\frac{1}{\tau _j} D\omega  + \frac{1}{\tau _j}T_m \\
		&\dot{\delta } = \omega - 1 \\
	\end{aligned}
	\label{eq:nonlinear_model4}
\end{equation} 
\autoref{eq:nonlinear_model2} and
 \autoref{eq:nonlinear_model4} can be combined to get the truth model of the synchronous generator
 \begin{equation}
	\begin{aligned}
		&\dot{I}_d = -L_{d1}(r+R_e)I_d + kM_{F1}r_FI_F+kM_{D1}r_DI_D-
		(L_q+L_e)L_{d1}I_q\omega - kM_QL_{d1}I_Q\omega  + V_\infty L_{d1}\sin(\delta -\alpha)\\
        & \ \ \ \ \ \ \ - kM_{F1}V_F \\
		&\dot{I}_F = kM_{F1}(r+R_e)I_d- L_{F1}r_FI_F - M_{R1}r_DI_D + 
		kM_{F1}(L_q+L_e)I_q\omega  + k^2M_{F1}M_QI_Q\omega  - V_\infty kM_{F1}\sin(\delta -\alpha)\\
		& \ \ \ \ \ \ \ +L_{F1}V_F \\
		&\dot{I}_D = kM_{D1}(r+R_e)I_d-M_{R1}r_FI_F-L_{D1}r_DI_D+
		kM_{D1}(L_q+L_e)I_q\omega + k^2M_{D1}M_QI_Q\omega  - V_\infty kM_{D1}\sin(\delta -\alpha )\\ 
		& \ \ \ \ \ \ \ + M_{R1}V_F \\
		&\dot{I}_q = L_{q1}(L_d+L_e)I_d\omega + kM_FL_{q1}I_F\omega +
		kM_DL_{q1}I_D\omega  - L_{q1}(r+R_e)I_q + kM_{Q1}r_QI_Q  
         - V_\infty L_{q1}\cos(\delta -\alpha) \\
         &\dot{I}_Q = -kM_{Q1}(L_d+L_e)I_d\omega  - k^2M_{Q1}M_FI_F\omega  -
		k^2M_{Q1}M_DI_D\omega  + kM_{Q1}(r+R_e)I_q - L_{Q1}r_QI_Q\\ 
        & \ \ \ \ \ \ \ + V_\infty kM_{Q1}\cos(\delta -\alpha ) \\
        &\dot{\omega } = -\frac{1}{\tau _j}\bigg{(}(L_d-L_q)I_dI_q + kM_FI_FI_q + 
         kM_DI_DI_q - kM_QI_dI_Q\bigg{)} 
        -\frac{1}{\tau _j} D\omega  + \frac{1}{\tau _j}T_m \\
		&\dot{\delta } = \omega - 1 \\
	\end{aligned}
	\label{eq:nonlinear_model5}
\end{equation}
For simplification of the above expression let us denote:
\begin{equation}
\begin{aligned}
       &F_{11}=-L_{d1}(r+R_e), \ \ F_{12}=kM_{F1}r_F, \ \ F_{13}=kM_{D1}r_D,\ \ F_{14}=-(L_q+L_e)L_{d1}, 
        \ \ F_{15}=-    kM_QL_{d1},\\
       &F_{16}= V_\infty L_{d1}, \ \ G_{11}=- kM_{F1}, \ \ F_{21}=kM_{F1}(r+R_e), \ \ F_{22}=- L_{F1}r_F,
       \ \ F_{23}=- M_{R1}r_D,\\
       &F_{24}=kM_{F1}(L_q+L_e), \ \ F_{25}=k^2M_{F1}M_Q,\ \ F_{26}=-V_\infty kM_{F1}, \ \ G_{21}=L_{F1}, 
        \ \ F_{31}=kM_{D1}(r+R_e),\\
       &F_{32}=-M_{R1}r_F, \ \ F_{33}=-L_{D1}r_D, \ \ F_{34}=kM_{D1}(L_q+L_e),\ \ F_{35}=k^2M_{D1}M_Q, \ \ F_{36}= -V_\infty kM_{D1},\\
        &G_{31}=M_{R1},\ \ F_{41}=L_{q1}(L_d+L_e), \ \ F_{42}=kM_FL_{q1}, \ \ F_{43}=kM_DL_{q1},\ \ F_{44}=- L_{q1}(r+R_e),\\
        &F_{45}= kM_{Q1}r_Q, \ \ F_{46}=-V_\infty L_{q1},\ \ F_{51}=-kM_{Q1}(L_d+L_e), \ \ F_{52}=- k^2M_{Q1}M_F, 
        \ \ F_{53}=-k^2M_{Q1}M_D,\\ 
       &F_{54}=kM_{Q1}(r+R_e), \ \ F_{55}=- L_{Q1}r_Q, \ \ F_{56}= V_\infty kM_{Q1},\ \ F_{61}=-\frac{1}{\tau _j}(L_d-L_q),\\
      &F_{62}=-\frac{1}{\tau _j}kM_F, \ \ F_{63}=-\frac{1}{\tau _j}kM_D,\ \ F_{64}=\frac{1}{\tau _j}kM_Q, 
      \ \ F_{65}=-\frac{1}{\tau _j}D.\\
 \end{aligned}
	\label{eq:nonlinear_model6}
\end{equation}      
Thus, the $7^{th}$ order truth model of the synchronous generator connected to an infinite bus in per unit can be written in the
nonlinear state-space form
\begin{equation}
\begin{aligned}
    &\dot{I}_d=F_{11}I_d + F_{12}I_F+F_{13}I_D+F_{14}I_q\omega+F_{15}I_Q\omega +F_{16}\sin(\delta -\alpha)+G_{11}V_F\\
	&\dot{I}_F=F_{21}I_d+F_{22}I_F+F_{23}I_D + F_{24}I_q\omega  +F_{25}I_Q\omega  + F_{26}\sin(\delta -\alpha )+G_{21}V_F\\
	&\dot{I}_D=F_{31}I_d+F_{32}I_F+F_{33}I_D+ F_{34}I_q\omega +F_{35}I_Q\omega  + F_{36}\sin(\delta -\alpha )+G_{31}V_F\\
	&\dot{I}_q=F_{41}I_d\omega + F_{42}I_F\omega +F_{43}I_D\omega +F_{44}I_q +F_{45}I_Q + F_{46}\cos(\delta -\alpha )\\
	&\dot{I}_Q=F_{51}I_d\omega  +F_{52}I_F\omega +F_{53}I_D\omega  +F_{54}I_q+F_{55}I_Q +F_{56}\cos(\delta -\alpha ) \\
	&\dot{\omega }=F_{61}I_dI_q + F_{62}I_FI_q + F_{63}I_DI_q +F_{64}I_dI_Q+ F_{65}\omega +F_{66}T_m \\
	&\dot{\delta }=\omega - 1\\
\end{aligned}
	\label{eq:nonlinear_model7}
\end{equation} 

\subsection{Model of the Turbine-Governor System}

In this section, we present the dynamics of the turbine-governor system. For the sake of simplicity 
we assume a linear model of the turbine-governor system \cite{DSWQ}.
\begin{itemize}
\item Turbine dynamics: The dynamics of the turbine are modeled by
\end{itemize}
\begin{equation}
             \dot{P}_m=-\frac{1}{\tau _T}P_m+\frac{K_T}{\tau _T}G_V
 \label{eq:turbine1}
\end{equation}              
where $P_m$ is the mechanical power output of the turbine, $G_V$ is the gate opening of the turbine,
$\tau _T$ is the time constant of the turbine, and $K_T$ is the gain of the turbine.
As done in \autoref{eq:swing7} we have the per unit mechanical power numerically 
equal to the per unit mechanical torque, i.e. $P_{m(p.u.)}=T_{m(p.u.)}$.
Therefore, the per unit turbine dynamics are
\begin{equation}
             \dot{T}_{m(p.u.)}=-\frac{1}{\tau _T}T_{m(p.u.)}+\frac{K_T}{\tau _T}G_{V(p.u.)}
 \label{eq:turbine2}
\end{equation}  
\begin{itemize}
\item Governor dynamics: The dynamics of the governor in per unit are
\end{itemize}
\begin{equation}
             \dot{G}_{V(p.u.)}=-\frac{1}{\tau _G}G_{V(p.u.)}+\frac{K_G}{\tau _G}\bigg{(}u_T-\frac{\omega _{(p.u.)}}{R_T}\bigg{)}
\label{eq:turbine3}
\end{equation}
where $u_T$ is the turbine valve control, $\tau _G$ is the time constant of the speed governor, $K_G$ is the gain
of the speed governor, and $R_T$ is the regulation constant in per unit.\\ 
Parameters of the turbine-governor system are
\begin{equation}
      K_T = 1, \ \ K_G = 1, \ \ \tau _T = 0.5, \ \ \tau _G = 0.2, \ \ R_T = 20\\
\label{eq:turbinepara}
\end{equation}
\subsection{Truth model of the combined Synchronous Generator and Turbine-Governor System connected to an infinite bus}

In this section, we present the truth model of the combined synchronous generator and turbine-governor system connected
to an infinite bus. This model consists of $7$ nonlinear differential equations of the synchronous generator and $2$ 
linear differential equations of the turbine-governor system. Thus, the combined system consists of $9$  
differential equations. Combining \autoref{eq:nonlinear_model7}, \autoref{eq:turbine2}, and \autoref{eq:turbine3} the 
truth model of the combined 
synchronous generator and turbine-governor system connected to an infinite bus, can be written as
\begin{equation}
\begin{aligned}
    &\dot{I}_d=F_{11}I_d + F_{12}I_F+F_{13}I_D+F_{14}I_q\omega+F_{15}I_Q\omega +F_{16}\sin(\delta -\alpha)+G_{11}V_F\\
	&\dot{I}_F=F_{21}I_d+F_{22}I_F+F_{23}I_D + F_{24}I_q\omega  +F_{25}I_Q\omega  + F_{26}\sin(\delta -\alpha )+G_{21}V_F\\
	&\dot{I}_D=F_{31}I_d+F_{32}I_F+F_{33}I_D+ F_{34}I_q\omega +F_{35}I_Q\omega  + F_{36}\sin(\delta -\alpha )+G_{31}V_F\\
	&\dot{I}_q=F_{41}I_d\omega + F_{42}I_F\omega +F_{43}I_D\omega +F_{44}I_q +F_{45}I_Q + F_{46}\cos(\delta -\alpha )\\
	&\dot{I}_Q=F_{51}I_d\omega  +F_{52}I_F\omega +F_{53}I_D\omega  +F_{54}I_q+F_{55}I_Q +F_{56}\cos(\delta -\alpha ) \\
	&\dot{\omega }=F_{61}I_dI_q + F_{62}I_FI_q + F_{63}I_DI_q +F_{64}I_dI_Q+ F_{65}\omega +F_{66}T_m \\
	&\dot{\delta }=\omega - 1\\
	&\dot{T}_m=F_{81}T_m+F_{82}G_V\\
	&\dot{G}_V=F_{91}\omega +F_{92}G_V+G_{92}u_T\\
	\end{aligned}
	\label{eq:ninthorder}
\end{equation}
where $F_{81}=-\frac{1}{\tau _T}$, $F_{82}=\frac{K_T}{\tau _T}$, $F_{91}=-\frac{K_G}{\tau _GR_T}$, $F_{92}=-\frac{1}{\tau _G}$,
$G_{92}=\frac{K_G}{\tau _G}$.
Let $x = [I_d, I_F, I_D, I_q, I_Q, \omega , \delta , T_m, G_V]^\mathrm{T}$ be the vector of state variables, 
$u = [ V_F, u_T]^\mathrm{T}$ the vector of control inputs, and $y = 
[ V_t, \omega ]^\mathrm{T}$ the vector of outputs, then \autoref{eq:ninthorder}
can be written in the usual state-space form 
  \begin{equation}
	\begin{aligned}
		\dot{x} &= F(x) + G(x)u \\
		y &= h(x) \\
	\end{aligned}
	\label{eq:ninthorder1}
\end{equation}  
where
\begin{equation}
	\begin{aligned}
	F(x) &= \bpm
	F_{11}x_1 + F_{12}x_2+F_{13}x_3+F_{14}x_4x_6+F_{15}x_5x_6 +F_{16}\sin(x_7-\alpha )\\
	F_{21}x_1+F_{22}x_2+F_{23}x_3 + F_{24}x_4x_6 +F_{25}x_5x_6 + F_{26}\sin(x_7-\alpha )\\
	F_{31}x_1+F_{32}x_2+F_{33}x_3+ F_{34}x_4x_6+F_{35}x_5x_6 + F_{36}\sin(x_7-\alpha )\\
	F_{41}x_1x_6+ F_{42}x_2x_6+F_{43}x_3x_6+F_{44}x_4 +F_{45}x_5 + F_{46}\cos(x_7-\alpha )\\
	F_{51}x_1x_6 +F_{52}x_2x_6+F_{53}x_3x_6 +F_{54}x_4+F_{55}x_5 +F_{56}\cos(x_7-\alpha ) \\
	F_{61}x_1x_4 + F_{62}x_2x_4 + F_{63}x_3x_4 +F_{64}x_1x_5+ F_{65}x_6+F_{66}x_8 \\
	x_6 - 1\\
	F_{81}x_8+F_{82}x_9\\
    F_{91}x_6+F_{92}x_9\\
	\epm \\
	G(x) &= \bpm G_{11} & 0 \\ G_{21} & 0 \\ G_{31} & 0 \\0 & 0 \\ 0 & 0\\ 0 & 0\\ 0 & G_{92} \epm   \\
        \end{aligned}
	\label{eq:ninthorder2}
\end{equation}

\subsection{Derivation of the Output Generator Terminal Voltage for the Truth Model}
The synchronous generator and turbine-governor system connected to an infinite bus is a MIMO system with two inputs: excitation 
field voltage $V_F$ and turbine valve control $u_T$, i.e. $u = [u_1, u_2]^\mathrm{T} = [ V_F, u_T]^\mathrm{T}$,  and two 
regulated outputs: generator terminal voltage $V_t$ and rotor angle $\delta $, i.e.
$y = [y_1, y_2]^\mathrm{T} = [ V_t, \delta ]^\mathrm{T}$. Since the rotor angle is difficult to measure in a practical system,
we use the angular frequency $\omega $ instead of the rotor angle as the second output
in all our work, as the frequency $\omega $ can be easily measured. In this section we derive an expression for the generator terminal voltage $V_t$.
From \autoref{eq:inf_bus} we have
\begin{equation}
 \begin{aligned}
          &v_d=R_ei_d+L_e\dot{i}_d+\omega L_ei_q-\sqrt{3}V_\infty \sin(\delta -\alpha )\\
          &v_q=R_ei_q+L_e\dot{i}_q-\omega L_ei_d+\sqrt{3}V_\infty \cos(\delta -\alpha )\\
   \end{aligned}
\label{eq:truthvt1}
\end{equation}
Dividing \autoref{eq:truthvt1} by $\sqrt{3}$ and substituting $\frac{v_d}{\sqrt{3}}=V_d$, $\frac{v_q}{\sqrt{3}}=V_q$,
 $\frac{i_d}{\sqrt{3}}=I_d$, and $\frac{i_q}{\sqrt{3}}=I_q$, where $V_d$, $V_q$, $I_d$, and $I_q$ are the corresponding RMS
values, we get 
\begin{equation}
 \begin{aligned}
          &V_d=R_eI_d+L_e\dot{I}_d+\omega L_eI_q-V_\infty \sin(\delta -\alpha )\\
          &V_q=R_eI_q+L_e\dot{I}_q-\omega L_eI_d+V_\infty \cos(\delta -\alpha )\\
   \end{aligned}
\label{eq:truthvt2}
\end{equation}
Substituting $\dot{I}_d$ and $\dot{I}_q$ from \autoref{eq:ninthorder} in \autoref{eq:truthvt2} 
\begin{equation}
 \begin{aligned}
          V_d &= R_eI_d+L_e(F_{11}I_d + F_{12}I_F+F_{13}I_D+F_{14}I_q\omega +F_{15}I_Q\omega +F_{16}\sin(\delta -\alpha)
               +G_{11}v_F)+\omega L_eI_q\\
              & \ \ \ -V_\infty \sin(\delta -\alpha )\\
          V_q &=R_eI_q+L_e(F_{41}I_d\omega + F_{42}I_F\omega +F_{43}I_D\omega +F_{44}I_q +F_{45}I_Q + F_{46}\cos(\delta -\alpha ))
          -\omega L_eI_d+V_\infty \cos(\delta -\alpha )\\
 \end{aligned}
\label{eq:truthvt3}
\end{equation}    
Simplifying and rearranging \autoref{eq:truthvt3} we get
\begin{equation}
 \begin{aligned}
          &V_d=(R_e+L_eF_{11})I_d+L_eF_{12}I_F+L_eF_{13}I_D+(L_eF_{14}+L_e)I_q\omega +L_eF_{15}I_Q\omega 
          +(L_eF_{16}-V_\infty )\sin(\delta -\alpha )\\
          & \ \ \ \ \ +L_eG_{11}V_F\\
          &V_q=(L_eF_{41}-L_e)I_d\omega +L_eF_{42}I_F\omega +L_eF_{43}I_D\omega +(R_e+L_eF_{44})I_q+L_eF_{45}I_Q+(L_eF_{46}
          +V_\infty )\cos(\delta -\alpha )\\
 \end{aligned}
\label{eq:truthvt4}
\end{equation} 
For simplification of \autoref{eq:truthvt4} let us denote $R_e+L_eF_{11}=y_{11}$, $L_eF_{12}=y_{12}$,
$L_eF_{13}=y_{13}$, $L_eF_{14}+L_e=y_{14}$, $L_eF_{15}=y_{15}$, $L_eF_{16}-V_\infty =y_{16}$, $L_eG_{11}=i_{11}$,       $L_eF_{41}-L_e=y_{21}$, $L_eF_{42}=y_{22}$, $L_eF_{43}=y_{23}$, $R_e+L_eF_{44}=y_{24}$, $L_eF_{45}=y_{25}$, and
$L_eF_{46}+V_\infty =y_{26}$.
Thus, \autoref{eq:truthvt4} can be simplified to
\begin{equation}
 \begin{aligned}
          &V_d=y_{11}I_d+y_{12}I_F+y_{13}I_D+y_{14}I_q\omega +y_{15}I_Q\omega 
          +y_{16}\sin(\delta -\alpha )+i_{11}V_F\\
          &V_q=y_{21}I_d\omega +y_{22}I_F\omega +y_{23}I_D\omega +y_{24}I_q+y_{25}I_Q+y_{26}
          \cos(\delta -\alpha )\\
 \end{aligned}
\label{eq:truthvt4point1}
\end{equation} 
The generator terminal voltage $V_t$ is computed as 
\begin{equation}
         V_t=\sqrt{V^2_d+V^2_q}
 \label{eq:truthvt5}
\end{equation}
where $V_d$ and $V_q$ are as given in \autoref{eq:truthvt4}.
The output generator terminal voltage, $y_1=V_t$, as a function of the states $x$, and control inputs $u$, is
\begin{equation}
\begin{aligned}
               &V_d=y_{11}x_1+y_{12}x_2+y_{13}x_3+y_{14}x_4x_6 +y_{15}x_5x_6 
                    +y_{16}\sin(x_7 -\alpha )+i_{11}u_1\\
               &V_q=y_{21}x_1x_6 +y_{22}x_2x_6 +y_{23}x_3x_6 +y_{24}x_4+y_{25}x_5+y_{26}
                    \cos(x_7 -\alpha )\\
               &y_1=V_t=\sqrt{V^2_d+V^2_q}\\
\end{aligned}               
\label{eq:truthvt5point11}
\end{equation}
Therefore, the output equation is given by
\begin{equation}
         y=h(x)=\bbm V_t \\ x_6 \ebm
 \label{eq:truthvt6point0}
\end{equation}
where $V_t$ is as given in \autoref{eq:truthvt5}.

\section{The Reduced Order Simplified Model}

A detailed derivation of the reduced order simplified model of a synchronous generator connected to an infinite bus
is given in the \hyperref[sec:appendix]{Appendix}. 
\subsection{Reduced order model of the combined Synchronous Generator and Turbine-Governor system connected to an infinite bus}
The final system equations for the reduced order simplified model of the synchronous generator are summarized below  
\begin{equation}
\begin{aligned}
            \dot{E}'_q &= \frac{1}{\tau '_{d0}}(E_{FD}-E'_q+(L_d-L'_d)I_d) \\
            \dot{\omega } &= \frac{1}{\tau _j}[T_m-D\omega -(E'_qI_q-(L_q-L'_d)I_dI_q)]\\
            \dot{\delta } &= \omega -1\\
\end{aligned}
\label{eq:simplified1}
\end{equation}
where $E'_d$ is the $d$ axis voltage behind the transient reactance $L'_q$, and $E'_q$ is the $q$ axis voltage behind the transient reactance $L'_d$, where $L'_d=L_d-\frac{(kM_F)^2}{L_F}$. $\tau '_{d0}$ is the $d$ axis transient open circuit time constant and is 
given by the relation $\tau '_{d0}=\frac{L_F}{r_F}$. $E_{FD}$ is the excitation field emf, and $\tau _j=2H\omega _R$. Also $I_d$ and $I_q$ are the direct axis and quadrature axis currents respectively. $E'_d$ is given by an algebraic constraint
\begin{equation}
            E'_d=-(L_q-L'_q)I_q
\label{eq:algebraic}
\end{equation}            
Note that all the variables in the third axis model of \autoref{eq:simplified1} are RMS quantities.\\
By applying Kirchhoff's voltage law (KVL) to the $d$ axis and $q$ axis stator circuits, the $d$ axis and $q$ axis stator voltage equations of a synchronous generator in per unit are
\begin{equation}
\begin{aligned}
               V_d &= -rI_d-L'_qI_q+E'_d\\
               V_q &= -rI_q+L'_dI_d+E'_q\\
\end{aligned}
\label{eq:simplified2}
\end{equation}                
On substituting $E'_d$ as given in \autoref{eq:algebraic} in \autoref{eq:simplified2} we get
\begin{equation}
\begin{aligned}
               V_d &= -rI_d-L_qI_q\\
               V_q &= -rI_q+L'_dI_d+E'_q\\
\end{aligned}
\label{eq:simplified3}
\end{equation}
By applying KVL to a synchronous generator connected to an infinite bus the stator voltage equations can be written as 
\begin{equation}
\begin{aligned}
               V_{\infty d} &= V_d-R_eI_d-L_eI_q\\
                            &= -(r+R_e)I_d-(L_q+L_e)I_q\\               
               V_{\infty q} &= V_q-R_eI_q+L_eI_d\\
                            &= -(r+R_e)I_q+(L'_d+L_e)I_d+E'_q\\
\end{aligned}
\label{eq:simplified4}
\end{equation} 
where $V_{\infty d}$ and $V_{\infty d}$ are direct axis and quadrature axis infinite bus voltages respectively, $R_e$ is the
resistance and $L_e$ is the inductance of the infinite bus.
We now solve the two simultaneous equations given in \autoref{eq:simplified4} to determine the two unknowns $I_d$ and $I_q$. 
On dividing $V_{\infty d}$ in \autoref{eq:simplified4} by $-(L_q+L_e)$ 
\begin{equation}
               \frac{V_{\infty d}}{-(L_q+L_e)} = \frac{-(r+R_e)}{-(L_q+L_e)}I_d+I_q      
\label{eq:simplified5}
\end{equation}
Similarly dividing $V_{\infty q}$ in \autoref{eq:simplified4} by $r+R_e$ 
\begin{equation}
               \frac{V_{\infty q}}{(r+R_e)} = -I_q+\frac{(L'_d+L_e)}{(r+R_e)}I_d+\frac{E'_q}{(r+R_e)}       
\label{eq:simplified6}
\end{equation} 
 Now we add \autoref{eq:simplified5} and \autoref{eq:simplified6} and compute $I_d$ 
\begin{equation}
        I_d=\frac{-(E'_q-V_{\infty q})(L_q+L_e)-V_{\infty d}(r+R_e)}{(r+R_e)^2+(L'_d+L_e)(L_q+L_e)}
\label{eq:simplified7}
\end{equation}
$I_q$ is determined in a similar fashion. 
Dividing $V_{\infty d}$ in \autoref{eq:simplified4} by $(r+R_e)$ 
\begin{equation}
               \frac{V_{\infty d}}{(r+R_e)} = -I_d-\frac{(L_q+L_e)}{(r+R_e)}I_q      
\label{eq:simplified8}
\end{equation}
Also dividing $V_{\infty q}$ in \autoref{eq:simplified4} by $L'_d+L_e$ 
\begin{equation}
               \frac{V_{\infty q}}{(L'_d+L_e)} = -\frac{(r+R_e)}{(L'_d+L_e)}I_q+I_d+\frac{E'_q}{(L'_d+L_e)}       
\label{eq:simplified9}
\end{equation}
Now we add \autoref{eq:simplified8} and \autoref{eq:simplified9} and compute $I_q$ 
\begin{equation}
          I_q=\frac{(E'_q-V_{\infty q})(r+R_e)- V_{\infty d}(L'_d+L_e)}{(r+R_e)^2+(L'_d+L_e)(L_q+L_e)}
\label{eq:simplified10}
\end{equation}
We substitute $I_d$ and $I_q$ as given in \autoref{eq:simplified7} and \autoref{eq:simplified10} 
in the reduced order simplified model of the synchronous generator as given in \autoref{eq:simplified1} to get
\begin{equation}
             \dot{E}'_q = \frac{1}{\tau '_{d0}}\bigg{(}E_{FD}-E'_q+(L_d-L'_d)
             \bigg{(}\frac{-(E'_q-V_{\infty q})(L_q+L_e)-V_{\infty  d}(r+R_e)}{(r+R_e)^2+(L'_d+L_e)(L_q+L_e)}\bigg{)}\bigg{)}
\label{eq:simplified11}
\end{equation}
For simplification of the above equation we make the following substitutions:\\
$L_q+L_e=L_1, r+R_e=R_1, L_d-L'_d=L_2$ and $(r+R_e)^2+(L'_d+L_e)(L_q+L_e)=M_1$\\
Thus, \autoref{eq:simplified11} can be rewritten as
\begin{equation}
         \dot{E}'_q = \frac{1}{\tau '_{d0}}\bigg{(}E_{FD}-E'_q-\frac{L_2L_1}{M_1}E'_q+\frac{L_2L_1}{M_1}V_{\infty q}
         -\frac{L_2R_1}{M_1}V_{\infty d}\bigg{)}
\label{eq:simplified12}
\end{equation}
Substituting $V_{\infty d}=-V_{\infty }\sin(\delta -\alpha )$ and $V_{\infty q}=V_{\infty }\cos(\delta -\alpha )$ in 
\autoref{eq:simplified12} and rearranging it we get
\begin{equation}
            \dot{E}'_q=\frac{-(1+\frac{L_2L_1}{M_1})}{\tau '_{d0}}E'_q+\frac{L_2L_1V_{\infty}}{M_1\tau '_{d0}}\cos(\delta -\alpha)
             +\frac{L_2R_1V_{\infty}}{M_1\tau '_{d0}}\sin(\delta -\alpha)+\frac{1}{\tau '_{d0}}E_{FD} 
\label{eq:simplified13}
\end{equation}
Let us denote $\frac{-(1+\frac{L_2L_1}{M_1})}{\tau '_{d0}}=f_{11}$, $\frac{L_2L_1V_{\infty}}{M_1\tau '_{d0}}=f_{12}$,
$\frac{L_2R_1V_{\infty}}{M_1\tau '_{d0}}=f_{13}$ and $\frac{1}{\tau '_{d0}}=g_{11}$. Thus, \autoref{eq:simplified13} 
can be rewritten as
\begin{equation}
            \dot{E}'_q=f_{11}E'_q+f_{12}\cos(\delta -\alpha)
             +f_{13}\sin(\delta -\alpha)+g_{11}E_{FD} 
\label{eq:simplified14}
\end{equation}
Substituting $I_d$ and $I_q$ in the equation for $\dot{\omega } $ in the reduced order simplified model of 
the synchronous generator as given in \autoref{eq:simplified1} we get
\begin{equation}
\begin{aligned}
              \dot{\omega } = &\frac{1}{\tau _j}\bigg{[}T_m-D\omega- \bigg{(}E'_q\bigg{(}\frac{(E'_q-V_{\infty q})(r+R_e)- V_{\infty d}(L'_d+L_e)}{(r+R_e)^2+(L'_d+L_e)(L_q+L_e)}\bigg{)}\\
&-(L_q-L'_d)\bigg{(}\frac{-(E'_q-V_{\infty q})(L_q+L_e)-V_{\infty d}(r+R_e)}{(r+R_e)^2+(L'_d+L_e)(L_q+L_e)}\bigg{)}\bigg{(}\frac{(E'_q-V_{\infty q})(r+R_e)- V_{\infty d}(L'_d+L_e)}{(r+R_e)^2+(L'_d+L_e)(L_q+L_e)}\bigg{)}\bigg{)}\bigg{]}\\ 
\end{aligned}
\label{eq:simplified15}
\end{equation}
Let us denote $L'_d+L_e=L_3$ and $L_q-L'_d=L_4$. Substituting $V_{\infty d}=-V_{\infty }\sin(\delta -\alpha )$ and $V_{\infty q}=V_{\infty }\cos(\delta -\alpha )$ in \autoref{eq:simplified15} and on simplifying we get
\begin{equation}
\begin{aligned}
      \dot{\omega } = &-\bigg{(}\frac{R_1}{M_1\tau _j}+\frac{L_4L_1R_1}{M^2_1\tau _j}\bigg{)}E'^2_q+
      \bigg{(}\frac{R_1}{M_1\tau _j}+\frac{2L_4L_1R_1}{M^2_1\tau _j}\bigg{)}V_\infty E'_q\cos(\delta -\alpha)\\
      &- \bigg{(}\frac{L_3}{M_1\tau _j}+\frac{L_4L_1L_3}{M^2_1\tau _j}-
      \frac{L_4R^2_1}{M^2_1\tau _j}\bigg{)}V_\infty E'_q\sin(\delta -\alpha)\\
      &- \bigg{(}\frac{L_4R^2_1}{M^2_1\tau _j}-\frac{L_4L_1L_3}{M^2_1\tau _j}\bigg{)}V^2_\infty \sin(\delta -\alpha)\cos(\delta -\alpha)
      -\bigg{(}\frac{L_4L_1R_1V^2_\infty }{M^2_1\tau _j}\bigg{)}\cos^2(\delta -\alpha)\\
      &+ \bigg{(}\frac{L_4L_3R_1V^2_\infty }{M^2_1\tau _j}\bigg{)}\sin^2(\delta -\alpha)-\frac{D}{\tau _j}\omega +\frac{1}{\tau _j}T_m\\ 
\end{aligned}
\label{eq:simplified16}
\end{equation}
Let us denote
\begin{equation}
\begin{aligned}
          &-\bigg{(}\frac{R_1}{M_1\tau _j}+\frac{L_4L_1R_1}{M^2_1\tau _j}\bigg{)}=f_{21}, \  
          \bigg{(}\frac{R_1}{M_1\tau _j}+\frac{2L_4L_1R_1}{M^2_1\tau _j}\bigg{)}V_\infty=f_{22},\\
          &-\bigg{(}\frac{L_3}{M_1\tau _j}+\frac{L_4L_1L_3}{M^2_1\tau _j}-\frac{L_4R^2_1}{M^2_1\tau _j}\bigg{)}V_\infty=f_{23}, \ 
          -\bigg{(}\frac{L_4R^2_1}{M^2_1\tau _j}-\frac{L_4L_1L_3}{M^2_1\tau _j}\bigg{)}V^2_\infty=f_{24},\\
          &-\bigg{(}\frac{L_4L_1R_1V^2_\infty }{M^2_1\tau _j}\bigg{)}=f_{25}, \ 
          \bigg{(}\frac{L_4L_3R_1V^2_\infty }{M^2_1\tau _j}\bigg{)}=f_{26}, \ 
          -\frac{D}{\tau _j}=f_{27}, \ \mathrm{and} \ 
          \frac{1}{\tau _j}=f_{28}\\
\end{aligned}
\label{eq:simplified17}
\end{equation}          
Thus, \autoref{eq:simplified16} can be rewritten as 
\begin{equation}
\begin{aligned}
     \dot{\omega } =& f_{21}E'^2_q+
       f_{22}E'_q\cos(\delta -\alpha )
      + f_{23}E'_q\sin(\delta -\alpha )
      + f_{24}\sin(\delta -\alpha )\cos(\delta -\alpha )\\
      &+f_{25}\cos^2(\delta -\alpha )
      + f_{26}\sin^2(\delta -\alpha )+f_{27}\omega +f_{28}T_m\\
\end{aligned}
\label{eq:simplified18}
\end{equation} 
We can express the reduced order nonlinear model of the synchronous generator connected to an infinite bus
in the usual state space form
\begin{equation}
\begin{aligned}
            &\dot{E}'_q = f_{11}E'_q+f_{12}\cos(\delta -\alpha )
             +f_{13}\sin(\delta -\alpha )+g_{11}E_{FD}\\ 
            &\dot{\omega } = f_{21}E'^2_q+ f_{22}E'_q\cos(\delta -\alpha )+ f_{23}E'_q\sin(\delta -\alpha )
            + f_{24}\sin(\delta  -\alpha )\cos(\delta -\alpha )\\
            & \ \ \ \ \ +f_{25}\cos^2(\delta -\alpha )
            + f_{26}\sin^2(\delta -\alpha )+f_{27}\omega +f_{28}T_m\\
            &\dot{\delta } = \omega -1
 \end{aligned}           
\label{eq:simplified19}
\end{equation}
For the reduced order generator-turbine system we use the same turbine-governor model that we used for the truth model.
The model of the turbine-governor system as given in \autoref{eq:turbine2} and \autoref{eq:turbine3} in per unit is
\begin{equation}
\begin{aligned}
             &\dot{T}_{m}=-\frac{1}{\tau _T}T_{m}+\frac{K_T}{\tau _T}G_{V}\\
             &\dot{G}_{V}=-\frac{K_G}{\tau _GR_T}\omega -\frac{1}{\tau _G}G_{V}+\frac{K_G}{\tau _G}u_T\\
             \end{aligned}
\label{eq:tu1}
\end{equation}
The reduced order model of the generator-turbine system connected to an infinite bus consists of $3$ nonlinear 
differential equations of the synchronous generator and $2$ linear differential equations of the turbine-governor system. 
Thus, the combined system consists of $5$  
differential equations.
Combining \autoref{eq:simplified19} and \autoref{eq:tu1} the reduced order model of the synchronous generator and turbine connected to an infinite bus can be written as
\begin{equation}
\begin{aligned}
      &\dot{E}'_q = f_{11}E'_q+f_{12}\cos(\delta -\alpha)
             +f_{13}\sin(\delta -\alpha)+g_{11}E_{FD}\\
       &\dot{\omega } = f_{21}E'^2_q+
       f_{22}E'_q\cos(\delta -\alpha)
      + f_{23}E'_q\sin(\delta -\alpha)
      + f_{24}\sin(\delta -\alpha)\cos(\delta -\alpha)\\
      & \ \ \ \ \ +f_{25}\cos^2(\delta -\alpha)
      + f_{26}\sin^2(\delta -\alpha)+f_{27}\omega +f_{28}T_m\\      
       &\dot{\delta } = \omega -1\\
       &\dot{T}_{m} = f_{41}T_{m}+f_{42}G_{V}\\
       &\dot{G}_V = f_{51}\omega +f_{52}G_V+g_{55}u_T\\
       \end{aligned}
 \label{eq:ss_n6}
\end{equation}
where, $-\frac{1}{\tau _T}=f_{41}$, $\frac{K_T}{\tau _T}=f_{42}$, $-\frac{K_G}{\tau _GR_T}=f_{51}$,
$-\frac{1}{\tau _G}=f_{52}$, and $\frac{K_G}{\tau _G}=g_{55}$.\\
Let us define the state variables as $x = [E'_q, \omega, \delta, T_m, G_V]^\mathrm{T}$, and the two control inputs as $u = [u_1, u_2]^\mathrm{T} = [ E_{FD}, u_T]^\mathrm{T}$.
We can then express the simplified fifth order nonlinear model of the synchronous generator and turbine connected
to an infinite bus as
\begin{equation}
\begin{aligned}
            \dot{x}_1 &= f_{11}x_1+f_{12}\cos(x_3 -\alpha)+f_{13}\sin(x_3 -\alpha)+g_{11}u_1\\
            \dot{x}_2 &= f_{21}x^2_1+ f_{22}x_1\cos(x_3 -\alpha)+ f_{23}x_1\sin(x_3 -\alpha)
            + f_{24}\sin(x_3 -\alpha)\cos(x_3 -\alpha)\\
            & \ \ \ +f_{25}\cos^2(x_3 -\alpha)
            + f_{26}\sin^2(x_3 -\alpha)+f_{27}x_2 +f_{28}x_4\\
            \dot{x}_3 &= x_2-1\\
            \dot{x}_4 &= f_{41}x_4+f_{42}x_5\\
            \dot{x}_5 &= f_{51}x_2+f_{52}x_5+g_{55}u_2\\
 \end{aligned}           
\label{eq:ss_n7}
\end{equation}
Thus, we can put the simplified fifth order nonlinear model of the synchronous generator and turbine
connected to an infinite bus in the usual state-space
form
\begin{equation}
	\begin{aligned}
		\dot{x} &= f(x) + g(x)u \\
		y &= h(x) \\
	\end{aligned}
	\label{eq:ss_n8}
\end{equation}
where 
\begin{equation}
	\begin{aligned}
     f(x) &= \bpm f_{11}x_1+f_{12}\cos(x_3 -\alpha )+f_{13}\sin(x_3 -\alpha )\\
            f_{21}x^2_1+ f_{22}x_1\cos(x_3 -\alpha )+ f_{23}x_1\sin(x_3 -\alpha )
            + f_{24}\sin(x_3 -\alpha )\cos(x_3 -\alpha )\cdot \cdot \cdot \\
            \cdot \cdot \cdot +f_{25}\cos^2(x_3 -\alpha )
            + f_{26}\sin^2(x_3 -\alpha )+f_{27}x_2+f_{28}x_4\\
            x_2-1\\
            f_{41}x_4+f_{42}x_5\\
            f_{51}x_2+f_{52}x_5 \epm \\
     g(x) &=  \bpm g_{11} & 0\\ 0 & 0\\ 0 & 0 \\ 0 & 0\\ 0 & g_{55}\epm
\end{aligned}
	\label{eq:ss_n9}
\end{equation}
In \autoref{eq:ss_n8} and \autoref{eq:ss_n9}, the state variables are $x = [E'_q, \omega, \delta, T_m, G_V]^\mathrm{T}$, the two control inputs are $u =[ E_{FD}, u_T]^\mathrm{T}$, and the two regulated outputs are $y=[V_t, \omega ]^\mathrm{T}$.
The expression for the generator terminal voltage $V_t$ will be derived in the next subsection.

\subsection{Derivation of the Output Generator Terminal Voltage for the Reduced Order Model}

The reduced order model of the synchronous generator and turbine connected to an infinite bus is a MIMO system with two inputs:
excitation field EMF $E_{FD}$, and turbine valve control $u_T$, and two outputs: generator terminal
voltage $V_t$ and rotor angle $\delta $.
In this section we derive an expression for the generator terminal voltage $V_t$ which is the first output of the MIMO system.
From \autoref{eq:simplified3} the direct axis and quadrature axis stator voltage equations of a synchronous generator are given by
\begin{equation}
\begin{aligned}
               V_d &= -rI_d-L_qI_q\\
               V_q &= -rI_q+L'_dI_d+E'_q\\
\end{aligned}
\label{eq:gt1}
\end{equation}
Since the stator resistance $r\approx 0$ we can write
\begin{equation}
\begin{aligned}
               V_d &= -L_qI_q\\
               V_q &= L'_dI_d+E'_q\\
\end{aligned}
\label{eq:gt2}
\end{equation}
Substituting $I_d$ and $I_q$ as given in \autoref{eq:simplified7} and \autoref{eq:simplified10} in \autoref{eq:gt2} we get
\begin{equation}
\begin{aligned}
               V_d &= -L_qI_q\\
                   &=  -\frac{L_q(r+R_e)}{(r+R_e)^2+(L'_d+L_e)(L_q+L_e)}E'_q\\
                   & \ \  +\frac{V_\infty L_q(r+R_e)}{(r+R_e)^2+(L'_d+L_e)(L_q+L_e)}\cos(\delta -\alpha )\\
                   &  \ \ -\frac{V_\infty L_q(L'_d+L_e)}{(r+R_e)^2+(L'_d+L_e)(L_q+L_e)}\sin(\delta -\alpha )\\
\end{aligned}
\label{eq:gt3}
\end{equation}
Let us denote 
\begin{equation}
\begin{aligned}
          &-\frac{L_q(r+R_e)}{(r+R_e)^2+(L'_d+L_e)(L_q+L_e)}=V_{d1}\\ 
          &\frac{V_\infty L_q(r+R_e)}{(r+R_e)^2+(L'_d+L_e)(L_q+L_e)}=V_{d2}\\  
          &-\frac{V_\infty L_q(L'_d+L_e)}{(r+R_e)^2+(L'_d+L_e)(L_q+L_e)}=V_{d3}\\
\end{aligned}
\label{eq:gt4}
\end{equation}              
Thus, \autoref{eq:gt3} can be rewritten as
\begin{equation}
               V_d = V_{d1}E'_q+V_{d2}\cos(\delta -\alpha )+V_{d3}\sin(\delta -\alpha )
\label{eq:gt5}
\end{equation}
Now let us derive the expression for $V_q$
\begin{equation}
\begin{aligned}
               V_q &= L'_dI_d+E'_q\\
                   &=  -\frac{L'_d(L_q+L_e)}{(r+R_e)^2+(L'_d+L_e)(L_q+L_e)}E'_q\\
                   & \ \    +\frac{V_\infty L'_d(L_q+L_e)}{(r+R_e)^2+(L'_d+L_e)(L_q+L_e)}\cos(\delta -\alpha )\\
                   & \ \    +\frac{V_\infty L'_d(r+R_e)}{(r+R_e)^2+(L'_d+L_e)(L_q+L_e)}\sin(\delta -\alpha )+E'_q\\
\end{aligned}
\label{eq:gt6} 
\end{equation}
Let us denote 
\begin{equation}
\begin{aligned}
                &-\frac{L'_d(L_q+L_e)}{(r+R_e)^2+(L'_d+L_e)(L_q+L_e)}=V_{q1}\\
                &\frac{V_\infty L'_d(L_q+L_e)}{(r+R_e)^2+(L'_d+L_e)(L_q+L_e)}=V_{q2}\\
                &\frac{V_\infty L'_d(r+R_e)}{(r+R_e)^2+(L'_d+L_e)(L_q+L_e)}=V_{q3}\\
\end{aligned}
\label{eq:gt7}
\end{equation}              
Thus, \autoref{eq:gt6} can be rewritten as
\begin{equation}
               V_q = V_{q1}E'_q+V_{q2}\cos(\delta -\alpha )+V_{q3}\sin(\delta -\alpha )+E'_q
\label{eq:gt8}
\end{equation}
The generator terminal voltage $V_t$ is given by
\begin{equation}
            V_t=\sqrt{V^2_d+V^2_q}
\label{eq:gt9}
\end{equation}
The output generator terminal voltage, $y_1=V_t$, as a function of the states $x$, is
\begin{equation}
\begin{aligned}
               &V_d = V_{d1}x_1+V_{d2}\cos(x_3 -\alpha )+V_{d3}\sin(x_3 -\alpha )\\
               &V_q = V_{q1}x_1+V_{q2}\cos(x_3 -\alpha )+V_{q3}\sin(x_3 -\alpha )+x_1\\
               &V_t=\sqrt{V^2_d+V^2_q}\\
\end{aligned}               
\label{eq:gt9point00}
\end{equation}
Thus, the output equation consists of a nonlinear equation for $V_t$ as given in \autoref{eq:gt9} and \autoref{eq:gt9point00}, and a simple linear equation for $\omega $
\begin{equation}
         y=h(x)=\bbm V_t \\ x_2 \ebm
 \label{eq:gt9point0}
\end{equation}

\subsection{Linearization of the Reduced Order Model by Taylor series approximation}

In this section we linearize the fifth-order nonlinear model of the synchronous generator and turbine
connected to an infinite bus by using the Taylor series approximation about a nominal operating point ($x_0$, $u_0$).
The operating condition is a steady state equilibrium of the system. The steady state equilibrium condition is attained by the system after all the transients die out or decay to zero. The equilibrium point ($x_0$, $u_0$) is computed by 
solving the differential equation, $\dot{x} = f(x_0) + g(x_0)u_0=0$.
From \autoref{eq:ss_n9} we can write
\begin{equation}
    f_1(x)= f_{11}x_1+f_{12}\cos(x_3 -\alpha )+f_{13}\sin(x_3 -\alpha )
    \label{eq:linear1}
\end{equation}
Therefore,
\begin{equation}
\begin{aligned}
          \frac{\partial f_1}{\partial x_1}\bigg{| }_{x_0} &= f_{11}\\
          \frac{\partial f_1}{\partial x_2}\bigg{| }_{x_0} &= 0\\
          \frac{\partial f_1}{\partial x_3}\bigg{| }_{x_0} &= -f_{12}\sin(x_{30}-\alpha )+f_{13}\cos(x_{30}-\alpha )\\
          \frac{\partial f_1}{\partial x_4}\bigg{| }_{x_0} &= 0\\
          \frac{\partial f_1}{\partial x_5}\bigg{| }_{x_0} &= 0\\
\end{aligned}
	\label{eq:linear2}
\end{equation} 
\\
\begin{equation}
\begin{aligned}
    f_2(x) =& \ f_{21}x^2_1+ f_{22}x_1\cos(x_3 -\alpha )+ f_{23}x_1\sin(x_3 -\alpha )
            + f_{24}\sin(x_3 -\alpha )\cos(x_3 -\alpha )\\
            &+f_{25}\cos^2(x_3 -\alpha )
            + f_{26}\sin^2(x_3 -\alpha )+f_{27}x_2+f_{28}x_4\\
            \end{aligned}
\label{eq:linear3}
\end{equation}
Therefore,
\begin{equation}
\begin{aligned}
         \frac{\partial f_2}{\partial x_1}\bigg{| }_{x_0} &= 2f_{21}x_{10}+f_{22}\cos(x_{30}-\alpha )+f_{23}\sin(x_{30} -\alpha )\\
         \frac{\partial f_2}{\partial x_2}\bigg{| }_{x_0} &= f_{27}\\
         \frac{\partial f_2}{\partial x_3}\bigg{| }_{x_0} &= -f_{22}x_{10}\sin(x_{30} -\alpha )+f_{23}x_{10}\cos(x_{30}-\alpha )+f_{24}\cos 2(x_{30}-\alpha )
         -f_{25}\sin 2(x_{30} -\alpha )+f_{26}\sin 2(x_{30} -\alpha )\\
         \frac{\partial f_2}{\partial x_40}\bigg{| }_{x_0} &= f_{28}\\
         \frac{\partial f_2}{\partial x_5}\bigg{| }_{x_0} &= 0\\
         \end{aligned}
	\label{eq:linear4}
\end{equation}
\\
\begin{equation}
        f_3(x)=x_2-1
        \label{eq:linear5}
 \end{equation}
Therefore,
\begin{equation}
\begin{aligned}
         \frac{\partial f_3}{\partial x_1}\bigg{| }_{x_0} &= 0\\
         \frac{\partial f_3}{\partial x_2}\bigg{| }_{x_0} &= 1\\
         \frac{\partial f_3}{\partial x_3}\bigg{| }_{x_0} &= 0\\
         \frac{\partial f_3}{\partial x_4}\bigg{| }_{x_0} &= 0\\
         \frac{\partial f_3}{\partial x_5}\bigg{| }_{x_0} &= 0\\
 \end{aligned}
	\label{eq:linear6}
\end{equation}
\\
\begin{equation}
        f_4(x)=f_{41}x_4+f_{42}x_5
        \label{eq:linear7}
 \end{equation}
Therefore,
\begin{equation}
\begin{aligned}
         \frac{\partial f_4}{\partial x_1}\bigg{| }_{x_0} &= 0\\
         \frac{\partial f_4}{\partial x_2}\bigg{| }_{x_0} &= 0\\
         \frac{\partial f_4}{\partial x_3}\bigg{| }_{x_0} &= 0\\
         \frac{\partial f_4}{\partial x_4}\bigg{| }_{x_0} &= f_{41}\\
         \frac{\partial f_4}{\partial x_5}\bigg{| }_{x_0} &= f_{42}\\
 \end{aligned}
	\label{eq:linear8}
\end{equation}
\\
\begin{equation}
        f_5(x)=f_{51}x_2+f_{52}x_5
        \label{eq:linear9}
 \end{equation}
Therefore,
\begin{equation}
\begin{aligned}
         \frac{\partial f_5}{\partial x_1}\bigg{| }_{x_0} &= 0\\
         \frac{\partial f_5}{\partial x_2}\bigg{| }_{x_0} &= f_{51}\\
         \frac{\partial f_5}{\partial x_3}\bigg{| }_{x_0} &= 0\\
         \frac{\partial f_5}{\partial x_4}\bigg{| }_{x_0} &= 0\\
         \frac{\partial f_5}{\partial x_5}\bigg{| }_{x_0} &= f_{52}\\
 \end{aligned}
	\label{eq:linear10}
\end{equation}
\\
\begin{equation}
g_1(x)=g_{11}u_1
\label{eq:linear11}
 \end{equation}
Therefore,
\begin{equation}
\begin{aligned}
         \frac{\partial g_1}{\partial u_1}\bigg{| }_{u_0} &= g_{11}\\
         \frac{\partial g_1}{\partial u_2}\bigg{| }_{u_0} &= 0\\
 \end{aligned}
	\label{eq:linear12}
\end{equation}
\\
\begin{equation}
g_2(x)=0
\label{eq:linear13}
 \end{equation}
Therefore,
\begin{equation}
\begin{aligned}
         \frac{\partial g_2}{\partial u_1}\bigg{| }_{u_0} &= 0\\
         \frac{\partial g_2}{\partial u_2}\bigg{| }_{u_0} &= 0\\
 \end{aligned}
	\label{eq:linear14}
\end{equation}
\\
\begin{equation}
g_3(x)=0
\label{eq:linear15}
 \end{equation}
Therefore,
\begin{equation}
\begin{aligned}
         \frac{\partial g_3}{\partial u_1}\bigg{| }_{u_0} &= 0\\
         \frac{\partial g_3}{\partial u_2}\bigg{| }_{u_0} &= 0\\
 \end{aligned}
	\label{eq:linear16}
\end{equation}
\\
\begin{equation}
g_4(x)=0
\label{eq:linear17}
 \end{equation}
Therefore,
\begin{equation}
\begin{aligned}
         \frac{\partial g_4}{\partial u_1}\bigg{| }_{u_0} &= 0\\
         \frac{\partial g_4}{\partial u_2}\bigg{| }_{u_0} &= 0\\
 \end{aligned}
	\label{eq:linear18}
\end{equation}
\\
\begin{equation}
g_5(x)=g_{55}u_2
\label{eq:linear19}
 \end{equation}
Therefore,
\begin{equation}
\begin{aligned}
         \frac{\partial g_5}{\partial u_1}\bigg{| }_{u_0} &= 0\\
         \frac{\partial g_5}{\partial u_2}\bigg{| }_{u_0} &= g_{55}\\
 \end{aligned}
	\label{eq:linear20}
\end{equation}
Therefore, the linear reduced order model of the synchronous generator and turbine-governor system connected to an infinite bus is 
\begin{equation}
      \dot{\mathbf{x}}=A\mathbf{x}+ B\mathbf{u}
\end{equation}
where
\begin{equation}
 \begin{aligned}
      \mathbf{x}^T &= \bbm \Delta E'_q & \Delta \omega  & \Delta \delta & \Delta T_m & \Delta G_V\ebm\\ 
      \mathbf{u} &= \bbm \Delta E_{FD} \\ \Delta u_T \ebm\\
\end{aligned}
	\label{eq:linear21}
\end{equation} 
In the above equation $\Delta $ is the deviation from the nominal operating condition, i.e. $E'_q-E'_{q0}=\Delta E'_q$, $\omega -\omega _0=\Delta \omega $, $\delta -\delta _0=\Delta \delta $, $T_m-T_{m0}=\Delta T_m$, $G_V-G_{V0}=\Delta G_V$, $E_{FD}-E_{FD0}=\Delta E_{FD}$,
and $u_T-u_{T0}=\Delta u_T$, and
\begin{equation}
\begin{aligned}
        A &= \bbm \frac{\partial f_1}{\partial x_1} & \frac{\partial f_1}{\partial x_2} &  \frac{\partial f_1}{\partial x_3} &
 \frac{\partial f_1}{\partial x_4} & \frac{\partial f_1}{\partial x_5}\\ \\
\frac{\partial f_2}{\partial x_1} & \frac{\partial f_2}{\partial x_2} &  \frac{\partial f_2}{\partial x_3} &
 \frac{\partial f_2}{\partial x_4} & \frac{\partial f_2}{\partial x_5}\\ \\
 \frac{\partial f_3}{\partial x_1} & \frac{\partial f_3}{\partial x_2} &  \frac{\partial f_3}{\partial x_3} &
 \frac{\partial f_3}{\partial x_4} & \frac{\partial f_3}{\partial x_5}\\ \\
\frac{\partial f_4}{\partial x_1} & \frac{\partial f_4}{\partial x_2} &  \frac{\partial f_4}{\partial x_3} &
 \frac{\partial f_4}{\partial x_4} & \frac{\partial f_4}{\partial x_5}\\ \\
 \frac{\partial f_5}{\partial x_1} & \frac{\partial f_5}{\partial x_2} &  \frac{\partial f_5}{\partial x_3} &
 \frac{\partial f_5}{\partial x_4} & \frac{\partial f_5}{\partial x_5}\\ \\
\ebm_{x_0}\\
                 &= \bbm f_{11} & 0 & \frac{\partial f_1}{\partial x_3} & 0 & 0\\ \\
                    \frac{\partial f_2}{\partial x_1} & f_{27} & \frac{\partial f_2}{\partial x_3} & f_{28} & 0\\ \\
0 & 1 & 0 & 0 & 0\\ \\0 & 0 & 0 & f_{41} & f_{42}\\ \\0 & f_{51} & 0 & 0 & f_{52}\\
\ebm_{x_0}\\
\end{aligned}
 \label{eq:linear22}
\end{equation} 
\begin{equation}
\begin{aligned}
B &= \left[ \begin{array}{cc}
         \frac{\partial g_1}{\partial u_1} & \frac{\partial g_1}{\partial u_2} \\ \\
         \frac{\partial g_2}{\partial u_1} & \frac{\partial g_2}{\partial u_2} \\ \\
         \frac{\partial g_3}{\partial u_1} & \frac{\partial g_3}{\partial u_2} \\ \\
         \frac{\partial g_4}{\partial u_1} & \frac{\partial g_4}{\partial u_2} \\ \\
         \frac{\partial g_5}{\partial u_1} & \frac{\partial g_5}{\partial u_2} \\ \\
\end{array}\right]_{u_0}\\
           &= \left[ \begin{array}{cc} g_{11} & 0\\ 0 & 0 \\ 0 & 0 \\ 0 & 0 \\ 0 & g_{55} \\
           \end{array}\right]_{u_0}\\
\end{aligned}
\label{eq:linear23}
\end{equation}
In order to design a linear controller for the linearized model obtained by using Taylor series approximation,
we need to find an expression for the first output $\Delta V_t$.
By using Taylor series approximation, we have

\begin{equation}
     V_t(x)=V_t(x_0)+\frac{\mathrm{d}V_t}{\mathrm{d}x}\bigg{| }_{x_0}(x-x_0)+ .....
\label{eq:taylorlinear1}
\end{equation}     
Where, $x = [E'_q, \omega, \delta, T_m, G_V]^\mathrm{T}$, are the states of the reduced order nonlinear model.
Let us denote $V_t(x)-V_t(x_0)=\Delta V_t$, and $V_t(x_0)=V_{t0}$. By neglecting the second and higher order 
derivatives in \autoref{eq:taylorlinear1} we have 
\begin{equation}
     \Delta V_t=\frac{\mathrm{d}V_t}{\mathrm{d}x}\bigg{| }_{x_0}(x-x_0)
\label{eq:taylorlinear2}
\end{equation}
From \autoref{eq:gt9} and \autoref{eq:gt9point00} the generator terminal voltage $V_t$ is given by
\begin{equation}
            V_t=\sqrt{V^2_d+V^2_q}
\label{eq:taylorlinear3}
\end{equation}
where
\begin{equation}
\begin{aligned}
               &V_d = V_{d1}E'_q+V_{d2}\cos(\delta -\alpha )+V_{d3}\sin(\delta -\alpha )\\
               &V_q = V_{q1}E'_q+V_{q2}\cos(\delta -\alpha )+V_{q3}\sin(\delta -\alpha )+E'_q\\
\end{aligned}             
\label{eq:taylorlinear4}
\end{equation}
We can write 
\begin{equation}
    \Delta V_t=\frac{\mathrm{d}V_t}{\mathrm{d}x}\bigg{| }_{x_0}(x-x_0)=
\bigg{(}\frac{\partial V_t}{\partial V_d}\frac{\mathrm{d}V_d}{\mathrm{d}x}\bigg{)}\bigg{| }_{x_0}(x-x_0)
     +\bigg{(}\frac{\partial V_t}{\partial V_q}\frac{\mathrm{d}V_q}{\mathrm{d}x}\bigg{)}\bigg{| }_{x_0}(x-x_0)
\label{eq:taylorlinear5}
\end{equation}
Differentiating \autoref{eq:taylorlinear3} with respect to $V_d$ and $V_q$ respectively
\begin{equation}
\begin{aligned}
            &\frac{\partial V_t}{\partial V_d}\bigg{| }_{x_0}=\frac{V_{d0}}{\sqrt{V^2_{d0}+V^2_{q0}}}=\frac{V_{d0}}{V_{t0}}\\
            &\frac{\partial V_t}{\partial V_q}\bigg{| }_{x_0}=\frac{V_{q0}}{\sqrt{V^2_{d0}+V^2_{q0}}}=\frac{V_{q0}}{V_{t0}}\\
 \end{aligned}           
\label{eq:taylorlinear6}
\end{equation}
Also 
\begin{equation}
\begin{aligned}
            &\frac{\mathrm{d}V_d}{\mathrm{d}x}\bigg{| }_{x_0}(x-x_0)=(E'_q-E'_{q0})\frac{\partial V_d}{\partial E'_q}\bigg{| }_{x_0}
            +(\omega -\omega _0)\frac{\partial V_d}{\partial \omega }\bigg{| }_{x_0}
            +(\delta -\delta _0)\frac{\partial V_d}{\partial \delta }\bigg{| }_{x_0}
            +(T_m-T_{m0})\frac{\partial V_d}{\partial T_m }\bigg{| }_{x_0}
            +(G_V-G_{V0})\frac{\partial V_d}{\partial G_V }\bigg{| }_{x_0}\\
            &\frac{\mathrm{d}V_q}{\mathrm{d}x}\bigg{| }_{x_0}(x-x_0)=(E'_q-E'_{q0})\frac{\partial V_q}{\partial E'_q}\bigg{| }_{x_0}
            +(\omega -\omega _0)\frac{\partial V_q}{\partial \omega }\bigg{| }_{x_0}
            +(\delta -\delta _0)\frac{\partial V_q}{\partial \delta }\bigg{| }_{x_0}
            +(T_m-T_{m0})\frac{\partial V_q}{\partial T_m }\bigg{| }_{x_0}
            +(G_V-G_{V0})\frac{\partial V_q}{\partial G_V }\bigg{| }_{x_0}\\
 \end{aligned}           
\label{eq:taylorlinear7}
\end{equation}
Let us recall that the deviations of the state variables form their nominal operating condition are given by:
$E'_q-E'_{q0}=\Delta E'_q$, $\omega -\omega _0=\Delta \omega $, $\delta -\delta _0=\Delta \delta $,
$T_m-T_{m0}=\Delta T_m$, $G_V-G_{V0}=\Delta G_V$. Simplifying \autoref{eq:taylorlinear7} we get
\begin{equation}
\begin{aligned}
            &\frac{\mathrm{d}V_d}{\mathrm{d}x}\bigg{| }_{x_0}(x-x_0)=V_{d1}\Delta E'_q-V_{d2}\sin(\delta _\circ -\alpha )\Delta \delta +
              V_{d3}\cos(\delta _\circ -\alpha )\Delta \delta \\
            &\frac{\mathrm{d}V_q}{\mathrm{d}x}\bigg{| }_{x_0}(x-x_0)=V_{q1}\Delta E'_q-V_{q2}\sin(\delta _\circ -\alpha )\Delta \delta +
              V_{q3}\cos(\delta _\circ -\alpha )\Delta \delta+\triangle E'_q\\
 \end{aligned}           
\label{eq:taylorlinear8}
\end{equation}
Substituting \autoref{eq:taylorlinear6} and \autoref{eq:taylorlinear8} in \autoref{eq:taylorlinear5} and rearranging we get
\begin{equation}
\begin{aligned}
                \Delta V_t &= \bigg{(}\frac{V_{d0}}{V_{t0}}V_{d1}+\frac{V_{q0}}{V_{t0}}V_{q1}+\frac{V_{q0}}{V_{t0}}\bigg{)}
                                 \Delta E'_q\\
                              & \ \ +\bigg{(}-\frac{V_{d0}}{V_{t0}}V_{d2}\sin(\delta _\circ -\alpha )+
                           \frac{V_{d0}}{V_{t0}}V_{d3}\cos(\delta _\circ -\alpha )-\frac{V_{q0}}{V_{t0}}V_{q2}\sin(\delta _\circ -\alpha )
                           +\frac{V_{q0}}{V_{t0}}V_{q3}\cos(\delta _\circ -\alpha )\bigg{)}\Delta \delta
\end{aligned}
\label{eq:gt15}
\end{equation} 
Let us denote 
\begin{equation}
\begin{aligned} 
          &\bigg{(}\frac{V_{d0}}{V_{t0}}V_{d1}+\frac{V_{q0}}{V_{t0}}V_{q1}+\frac{V_{q0}}{V_{t0}}\bigg{)}=T_1\\
          &\bigg{(}-\frac{V_{d0}}{V_{t0}}V_{d2}\sin(\delta _\circ -\alpha )+
                           \frac{V_{d0}}{V_{t0}}V_{d3}\cos(\delta _\circ -\alpha )-
                           \frac{V_{q0}}{V_{t0}}V_{q2}\sin(\delta _\circ -\alpha )
                           +\frac{V_{q0}}{V_{t0}}V_{q3}\cos(\delta _\circ -\alpha )\bigg{)}=T_2\\
\end{aligned}
\label{eq:gt16}
\end{equation} 
Thus, \autoref{eq:gt15} can be simplified to get
\begin{equation}
                 \Delta V_t=T_1\Delta E'_q+T_2\Delta \delta
\label{eq:gt17}
\end{equation}
Thus, the linearized output equation of the reduced order model is given by
\begin{equation} 
               \mathbf{y}=C\mathbf{x} 
\label{eq:gt18}
\end{equation}
where 
\begin{equation}
           \mathbf{y} = \bbm \Delta V_t\\\Delta \omega \ebm
\label{eq:gt19}
\end{equation}
and the output matrix $C$ is 
\begin{equation}
               C=\bbm T_1 & 0 & T_2 & 0 & 0\\ 0 & 1 & 0 & 0 & 0\ebm
\label{eq:gt20}
\end{equation} 

\subsection{Example}

In this section we calculate the $(A, B, C)$ system matrices of the truth model
and the reduced order model linearized about a given nominal operating point. The power, voltage, and current ratings 
as well as the parameters of the synchronous generator 
are given in \autoref{tab:para_values1}, which contains values for an actual synchronous generator 
with some quantities, denoted by an asterisk, being estimated for our study \cite{AF03}. The power rating of a synchronous generator is equal to the product of the voltage per phase, the current per phase, and the number of phases. It is normally stated in megavolt-amperes (MVA) for large generators. The parameters given in \autoref{tab:para_values1} are not in the per unit system. 
The quantities in \autoref{tab:para_values1} are converted to the per unit system, which are given in \autoref{tab:para_values2} 
\cite{AF03}. \autoref{tab:nonlinear_para} gives the parameters of the reduced order nonlinear model in the per unit system. 

Before we proceed to the calculation of the system matrices, we briefly explain a basic power system terminology, that is used in 
determining a suitable operating condition. Real power (also known as active power) ($P$), measured in watts (W); apparent power ($S$), measured in volt-amperes (VA); and reactive power ($Q$), measured in reactive volt-amperes (var) are
related by the expression, $S=P+jQ$ in vector form. If $\phi $ is the phase angle between the current and voltage, then the power factor (PF) is equal to the cosine of the angle, $|\cos \phi |$ , and $|P|=|S|\cos \phi $ and $|Q|=|S|\sin \phi $. Power factors are usually stated as "leading" or "lagging" to show the sign of the phase angle $\phi $. If a purely resistive load is connected to a power supply, current and voltage will change polarity in step, the power factor will be unity, and the electrical energy flows in a single direction across the network in each cycle. Inductive loads such as transformers and motors (any type of wound coil) consume reactive power with current waveform lagging the voltage. Capacitive loads such as capacitor banks or buried cable generate reactive power with current phase leading the voltage. Both types of loads will absorb energy during part of the AC cycle, which is stored in the device's magnetic or electric field, only to return this energy back to the source during the rest of the cycle. Since a majority of the loads
in a power grid are inductive, we assume power factor lagging conditions, where the generator armature current lags the generator
terminal voltage. Also a high power factor is generally desirable in a transmission system to reduce transmission losses and improve voltage regulation at the load. A power factor of $0.85$ is assumed, which is within the stable operating limits of a synchronous
generator. 

The synchronous generator is connected to an infinite bus through a transmission line having $R_e=0.02$ p.u., and $L_e=0.4$ p.u. The infinite bus voltage is $1.0$ p.u. The machine loading is given by, real power $P=1.0$ p.u. at $0.85$ PF lagging conditions. The steady state operating conditions of a synchronous generator turbine system connected to an infinite bus depend on the synchronous generator turbine system parameters, transmission line parameters, and the machine loading.  For the synchronous generator with parameters and loading conditions given in \autoref{tab:para_values1} and \autoref{tab:para_values2}, a steady state operating point for the truth model is evaluated which is as given in \cite{AF03}:
\begin{equation}
\begin{aligned}
          &I_{d0}=-0.9185 \\ 
          &I_{F0}=1.6315\\ 
          &I_{D0}=-4.6204\times 10^{-6}\\
          &I_{q0}= 0.4047\\
          &I_{Q0}=5.9539\times 10^{-5}\\
          &\omega _0=1\\
          &\delta _0=1 \\
          &T_{m0}=1.0012\\
          &G_{V0}=1.0012\\
          &V_{q0}=0.9670\\
          &V_{d0}=-0.6628\\  
          &V_{t0}= 1.172\\ 
          &\delta _0-\alpha = 53.736^\circ\\ 
\end{aligned}
\label{eq:linear24}
\end{equation}

 \begin{centergroup}
	\captionof{table}{Ratings and Parameters of the Synchronous
	Generator}
	\begin{tabular}{|c||c|c|c|c|}
	    \hline
		Variables  & Rated MVA & Rated voltage & Excitation voltage & Stator current \\
		\hline
	        Values & 160 MVA & 15 kV, Y connected & 375 V & 6158.40 A \\ 
		\hline
		\hline
		Variables  & Field current & Power factor & $L_d$ & $L_F$ \\
		\hline
	        Values & 926 A & 0.85 & $6.341\times 10^{-3}$ H & 2.189 H \\ 
		\hline
		\hline
		Variables & $L_D$ & $L_q$ & $L_Q$ & $kM_F$ \\
		\hline
	        Values & $5.989\times 10^{-3} \ \mathrm{H}^*$ & $6.118\times 10^{-3}$ H & $1.423\times 10^{-3} \ \mathrm{H}^*$ 
        &  $109.01\times 10^{-3}\ \mathrm{H}^*$\\ 
		\hline
		\hline
		Variables  & $kM_D$ & $M_R$ & $kM_Q$ & $r(125^\circ C)$ \\
		\hline
	        Values  & $5.782\times 10^{-3} \ \mathrm{H}^*$ & $109.01\times 10^{-3} \ \mathrm{H}^*$ & 
        $2.779\times 10^{-3} \ \mathrm{H}^*$ & $1.542\times 10^{-3} \ \Omega $\\ 
		\hline
		\hline
		Variables & $r_F(125^\circ C)$ & $r_D$ & $r_Q$ & Inertia constant $H$\\
		\hline
	        Values & $0.371 \ \Omega $ & $18.421\times 10^{-3} \ \Omega ^*$ & 
       $18.969\times 10^{-3} \ \Omega ^*$  & $1.765 \ \frac{kW\cdot s}{hp}$\\
       \hline
		\hline
		Variables & $R_e$ & $L_e$ & $D$ & $\omega_0$ (in rads)\\
		\hline
	        Values & 0.02 p.u. & 0.4 p.u. & 0 & 376.99\\
	    \hline
	\end{tabular}
	\label{tab:para_values1}
\end{centergroup}

\begin{centergroup}
	\captionof{table}{Parameters of the Truth model of the Synchronous
	Generator-Turbine System in p.u.}
\begin{tabular}{|c||c|c|c|c|c|c|}
		\hline
		Variables (in p.u.) & $L_d$ & $L_F$ & $L_D$ & $L_q$ & $L_Q$ & $kM_F$ \\
		\hline
	        Values & 1.70 & 1.65 & 1.605 & 1.64 & 1.526 & 1.55 \\ 
		\hline
		\hline
		Variables (in p.u.)  & $kM_D$ & $M_R$ & $kM_Q$ & $r$ & $r_F$ & $r_D$ \\
		\hline
	        Values & 1.55 & 1.55 & 1.49 & 0.001096 & 0.000742 & 0.0131 \\ 
		\hline
		\hline
		Variables (in p.u.) & $r_Q$ & $H$ (in s) & $R_e$ & $L_e$ & $D$ &
		$\omega_0$ (in rads) \\
		\hline
	        Values  & 0.0540 & 2.37 & 0.02 & 0.4 & 0 & 376.99 \\ 
		\hline
		Variables (in p.u.) & $K_T$ & $K_G$ & $\tau _T$ & $\tau _G$ & $R_T$ & $k$\\
		\hline
	        Values  & 1 & 1 & 0.5 & 0.2 & 20 & $\sqrt{3/2}$\\ 
	    \hline 
         Variables (in p.u.) & $L'_d$ & $\tau '_{d0}$ & $\tau _j$ & $V_\infty $ & $\alpha $ & \\
        \hline
	        Values  & 0.245 & 5.9 & 4.74 & 1.00 & $3.5598^\circ $ & \\ 
	    \hline 
	\end{tabular}
	\label{tab:para_values2}
\end{centergroup}
\begin{centergroup}
	\captionof{table}{Parameters of the Reduced Order Nonlinear Model of the SMIB in p.u.}
\begin{tabular}{|c||c|c|c|c|c|c|}
        \hline
		Variables (in p.u.) & $V_{d1}$ & $V_{d2}$ & $V_{d3}$ & $V_{q1}$ & $V_{q2}$ & $V_{q3}$ \\
		\hline
	        Values & -0.0249 &  0.0249 & -0.8037 & -0.3797 & 0.3797 & 0.0037 \\ 
		\hline
		\hline
		Variables (in p.u.) & $f_{11}$ & $f_{12}$ & $f_{13}$ & $f_{21}$ & $f_{22}$ & $f_{23}$ \\
		\hline
	        Values & -0.5517 & 0.3822 & 0.0037 & -0.0101 & 0.0171 & -0.3269 \\ 
		\hline
		\hline
		Variables (in p.u.)  & $f_{24}$ & $f_{25}$ & $f_{26}$ & $f_{27}$ & $f_{28}$ & $f_{41}$ \\
		\hline
	        Values & 0.2235 & -0.0069 & 0.0022 & 0 & 0.2110 & -2 \\ 
		\hline
		\hline
		Variables (in p.u.) & $f_{42}$ & $f_{51}$ & $f_{52}$ & $g_{11}$ & $g_{55}$ & $E_{FDmax}$\\
		\hline
	        Values  & 2 & -0.2500 & -5 & 0.1695 & 5 & 5 \\ 
		\hline	
		\hline
	    Variables (in p.u.) & $E_{FDmin}$ & $G_{Vmax}$ & $G_{Vmin}$ &  &  & \\		
		\hline
	        Values  & -5 & 1.2 & 0 &  &  &  \\ 
		\hline	
	\end{tabular}
	\label{tab:nonlinear_para}
\end{centergroup} 
The $\mathbf{A}$ matrix of the truth model linearized about the operating point is
\begin{equation}
       \mathbf{A}=\bbm -0.0361 & 0.000437 & 0.0142 & -3.4883  & -2.5478  & -1.4119 & 1.0115 & 0 & 0\\
       0.0124 & -0.0050 & 0.0772 & 1.2022 & 0.8781 & 0.4866 & -0.3486 & 0 & 0\\
       0.0228 & 0.0044 & -0.0964  & 2.2077  & 1.6125 & 0.8936 & -0.6401 & 0 & 0\\
       3.5888 & 2.6489  & 2.6489  & -0.0361  & 0.0901 & 1.0254 & 1.3779 & 0  & 0\\
       -3.5042 & -2.5864 & -2.5864  & 0.0352 & -0.1234 & -1.0012  & -1.3454 & 0 & 0\\
       -0.000013 & -0.00035 & -0.00035  & -0.0014 & -0.00076 
         & 0  & 0  & 0.00056 & 0\\
         0 &  0 & 0 &  0  &  0  & 1.0000  & 0  &  0  &  0\\
         0 & 0 & 0 & 0 & 0 & 0 & 0  & -2.0000  &  2.0000\\
         0 & 0 & 0 & 0 & 0 & -0.2500 & 0 & 0 & -5.0000\ebm
\label{eq:linear25}
\end{equation} 
The $\mathbf{B}$ matrix of the truth model linearized about the operating point is
\begin{equation}
       \mathbf{B}= \bbm -0.5893 & 0\\6.6918 & 0\\-5.8933 & 0\\0 & 0\\0 & 0\\0 & 0\\0 & 0\\0 & 0\\ 0 & 5.0000\ebm
\label{eq:linear26}
\end{equation}
the $\mathbf{C}$ matrix of the truth model linearized about the operating point is
\begin{equation}
        \mathbf{C}=\bbm 0.8510 & 0.8739 & 0.8708 & 0.5673 & 0.6059 & 0.8691 & -0.1048 & 0 & 0\\ 
         0 & 0 & 0 & 0 & 0 & 1.0000 & 0 & 0 & 0\ebm
 \label{eq:linear27}
\end{equation} 
and the $\mathbf{D}$ matrix of the truth model linearized about the operating point is 
\begin{equation}
        \mathbf{D}=\bbm 0.1333 & 0\\0 & 0\ebm
 \label{eq:linear28}
\end{equation}
The open loop eigenvalues of the $\mathbf{A}$ matrix of the truth model are
\begin{equation}
\begin{aligned}
  &-5.0000\\          
  &-0.0359 + 0.9983i\\
  &-0.0359 - 0.9983i\\
  &-2.0000\\          
  &-0.0016 + 0.0289i\\
  &-0.0016 - 0.0289i\\
  &-0.0007\\          
  &-0.0995 \\         
  &-0.1217\\ 
  \end{aligned}
 \label{eq:linear29}
\end{equation} 
Also a steady state operating point for the reduced order nonlinear model is 
\begin{equation}
\begin{aligned}         
          &x_{10}=E'_{q0} = 1.1925 \\ 
          &x_{20}=\omega _0=1\\
          &x_{30}=\delta _0=1\\
          &x_{40}=T_{m0}=1.0012\\
          &x_{50}=G_{V0}=1.0012\\          
\end{aligned}
\label{eq:hw1_equation13}
\end{equation}      
The $A$ matrix of the reduced order model linearized about the operating point is
\begin{equation}
       A=\bbm -0.5517 & 0 & -0.3060 & 0 & 0\\ -0.2776 & 0 & -0.3054 & 0.2110 & 0\\ 
       0 & 1 & 0 & 0 & 0\\ 0 & 0 & 0 & -2 & 2\\0 & -0.25 & 0 & 0 & -5\ebm
\label{eq:linear30}
\end{equation}
The $B$ matrix of the reduced order model linearized about the operating point
 is
\begin{equation}
       B= \bbm 0.1695 & 0\\0 & 0\\0 & 0\\0 & 0\\ 0 & 5\ebm
\label{eq:linear31}
\end{equation} 
the $C$ matrix of the reduced order model linearized about the operating point
 is
\begin{equation}
        C=\bbm 0.5258 & 0 & 0.0294 & 0 & 0\\ 0 & 1 & 0 & 0 & 0\ebm
 \label{eq:linear32}
\end{equation} 
and the $D$ matrix of the reduced order model linearized about the operating point is 
\begin{equation}
        D=\bbm 0 & 0\\0 & 0\ebm
 \label{eq:linear33}
\end{equation}
The open loop eigenvalues of the $A$ matrix of the reduced order model are
\begin{equation}
\begin{aligned}
  &-5.0069\\          
  &-0.1048 + 0.4778i\\
  &-0.1048 - 0.4778i\\
  &-0.3514 \\                  
  &-1.9839\\ 
  \end{aligned}
 \label{eq:linear34}
\end{equation} 
From \autoref{eq:linear29} and \autoref{eq:linear34} we can see that the eigenvalues of the open loop system lie on the left
half plane.

\section{Open Loop Input-Output Behavior of the Synchronous Generator and Turbine-Governor System}

\begin{figure}
          \centering
          \includegraphics[trim=0cm 0cm 0cm 0cm, clip=true, totalheight=0.29\textheight, width=0.58
           \textwidth]  {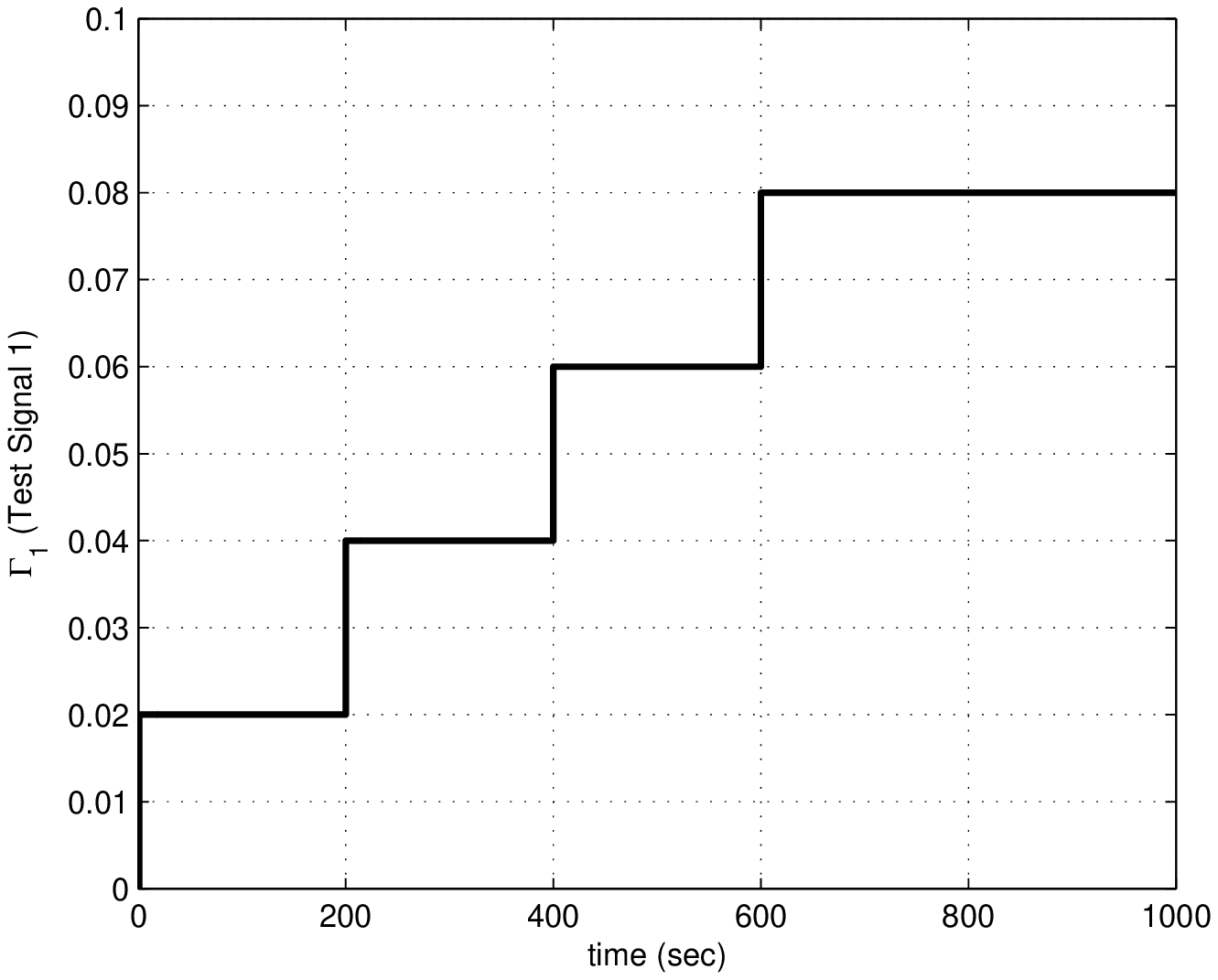}
          \caption{Plot of $\Gamma _1$ (Test Signal 1) vs time}
          \label{fig:Test_Signal1}          
          \includegraphics[trim=0cm 0cm 0cm 0cm, clip=true, totalheight=0.29\textheight, width=0.58\textwidth]{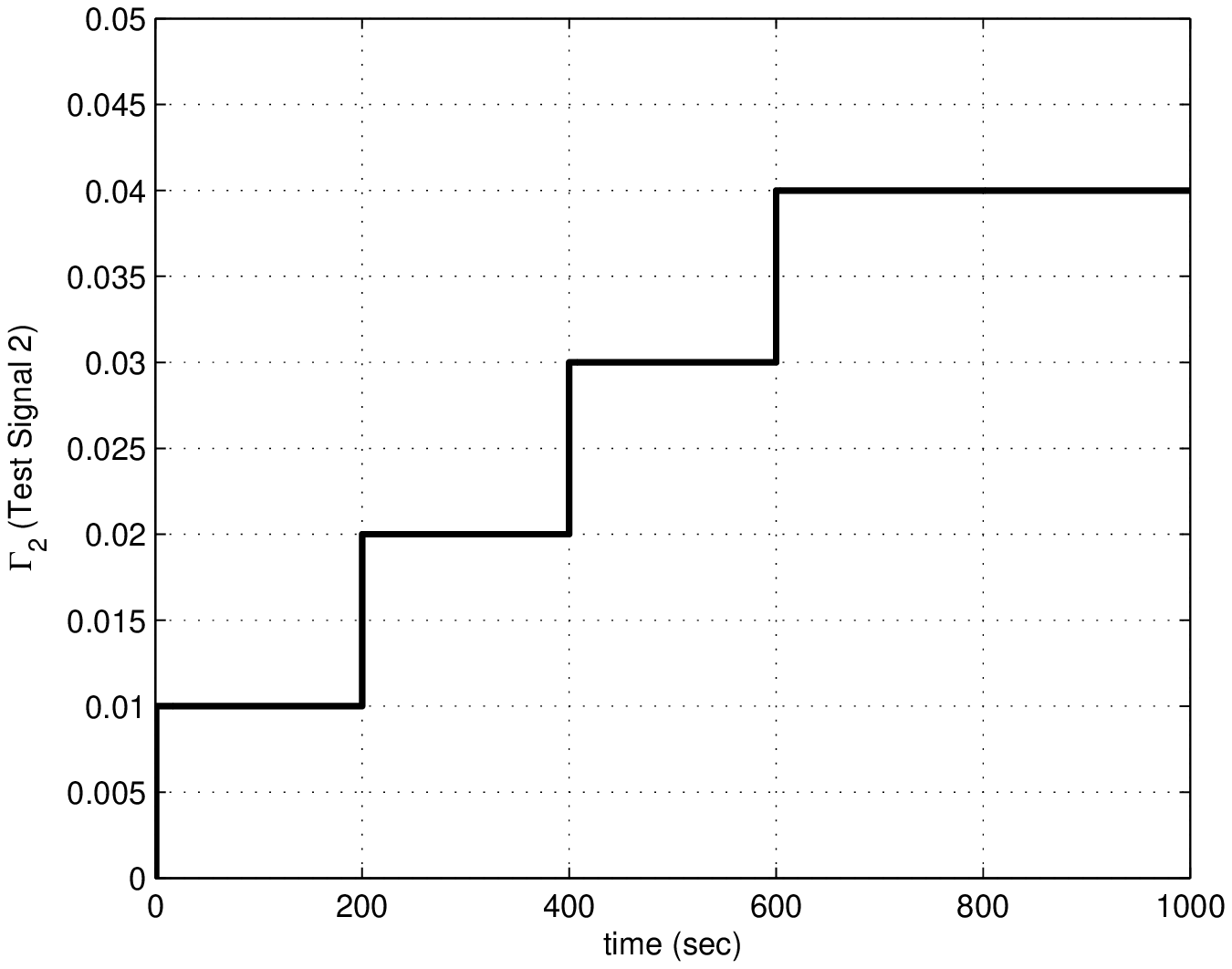}
          \caption{Plot of $\Gamma _2$ (Test Signal 2) vs time}
          \label{fig:Test_Signal2}          
\end{figure}

In this section we observe the open loop input-output behavior of the truth model and the reduced order nonlinear model 
of the SMIB for test signals given in \autoref{fig:Test_Signal1} and \autoref{fig:Test_Signal2}. 
The first control input $V_F$ of the truth model and the first control input $E_{FD}$ of the reduced order model are different. As given in
\autoref{eq:pu6} the field voltage, $V_F$, is
related to excitation field emf, $E_{FD}$, by the following expression 
\begin{equation}
\begin{aligned}
         V_F=&\bigg{(}\frac{r_F}{\omega _RkM_F}\bigg{)}E_{FD}\\
             =&e_{15}E_{FD}\\
\end{aligned}             
\label{eq:EFD_VF0}
\end{equation}
In the above expression, $\omega _R=1$ p.u., and $e_{15}=(\frac{r_F}{\omega _RkM_F})$.

We first test the truth model and the reduced order nonlinear model of the SMIB
for $\Gamma _1$ (Test signal 1) given in \autoref{fig:Test_Signal1}. 
In this case the two inputs for the truth 
model are, $u = [V_F, u_T]^\mathrm{T}=[e_{15}\Gamma _1, \Gamma _1]^\mathrm{T}$, and the two inputs for the reduced order nonlinear model are, $u = [E_{FD}, u_T]^\mathrm{T}=[\Gamma _1, \Gamma _1]^\mathrm{T}$. 

\begin{figure}
          \centering    
          \includegraphics[trim=0cm 0cm 0cm 0cm, clip=true, totalheight=0.27\textheight, 
           width=0.54\textwidth]   {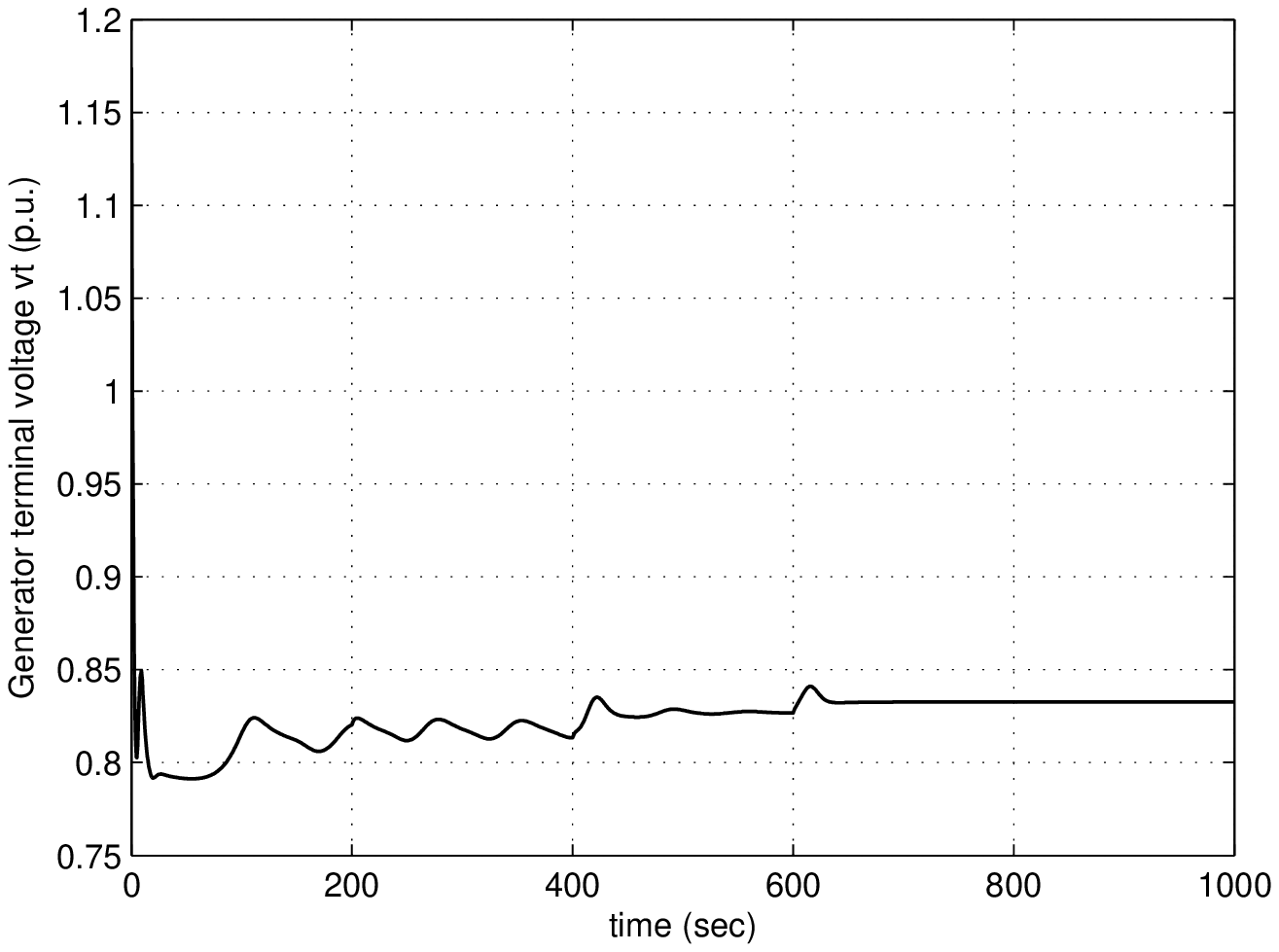}
          \caption{Plot of the generator terminal voltage $V_t$ vs time for $\Gamma _1$ (Test Signal 1) applied to 
           the reduced order nonlinear model}
          \label{fig:step_testnonlinear_vt1}
          \includegraphics[trim=0cm 0cm 0cm 0cm, clip=true, totalheight=0.27\textheight, 
           width=0.54\textwidth]{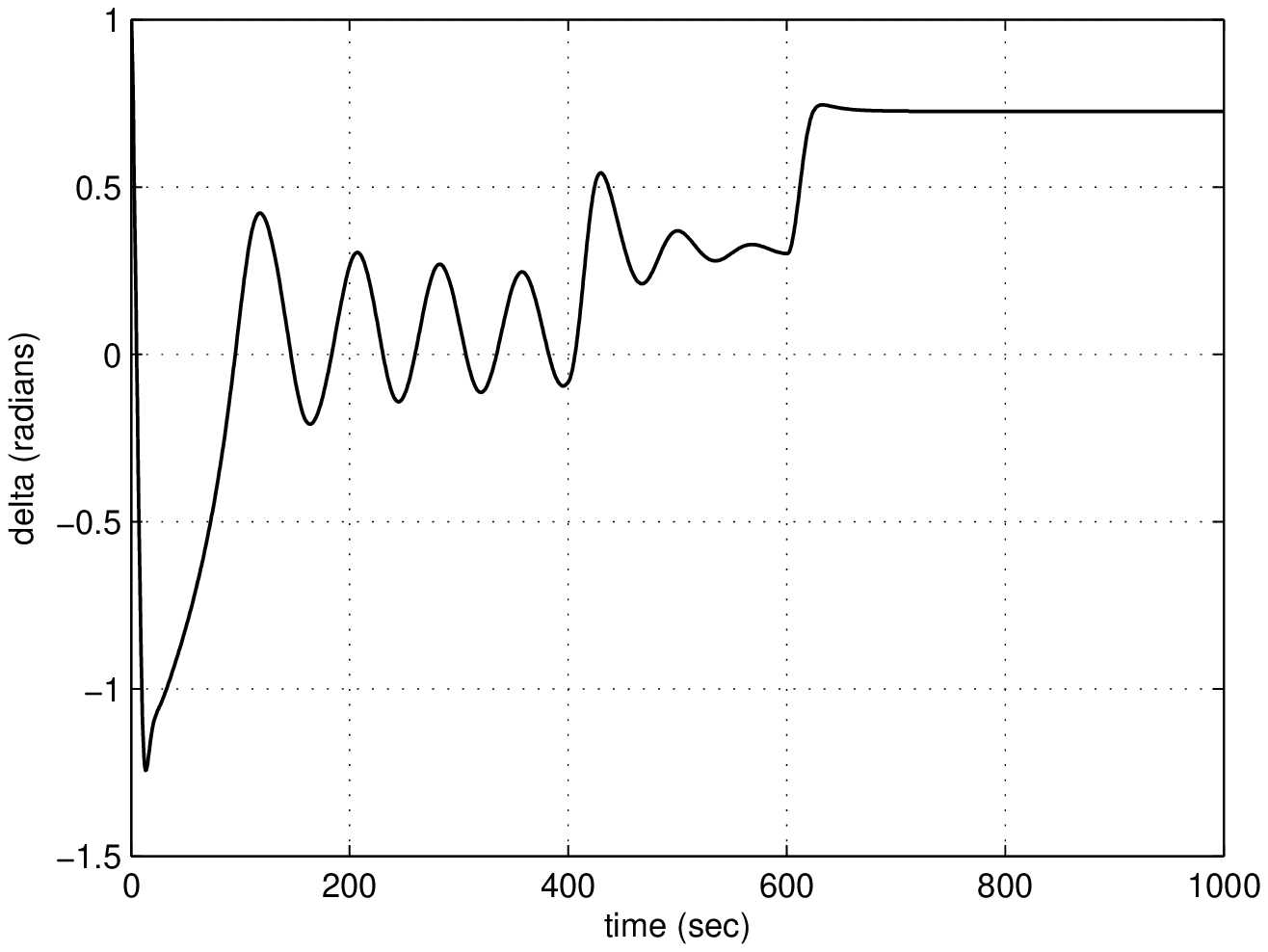}
          \caption{Plot of the rotor angle $\delta $ vs time for $\Gamma _1$ (Test Signal 1) applied to the reduced order nonlinear model}
          \label{fig:step_testnonlinear_delta1}
          \includegraphics[trim=0cm 0cm 0cm 0cm, clip=true, totalheight=0.27\textheight, 
          width=0.54\textwidth]{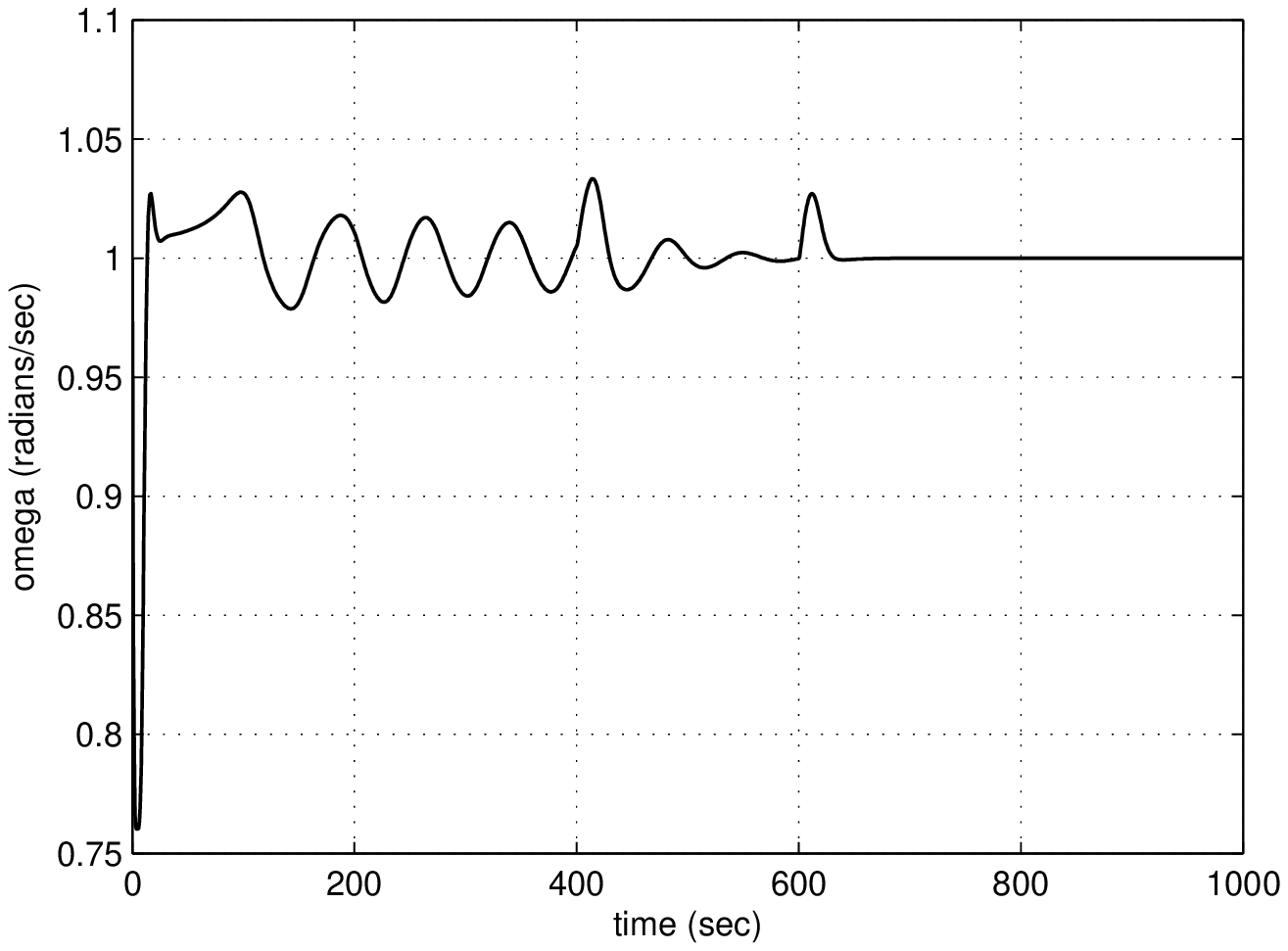}
          \caption{Plot of the frequency $\omega  $ vs time for $\Gamma _1$ (Test Signal 1) applied to the reduced order nonlinear model}
          \label{fig:step_testnonlinear_omega1}          
\end{figure}

\autoref{fig:step_testnonlinear_vt1}, \autoref{fig:step_testnonlinear_delta1}, and \autoref{fig:step_testnonlinear_omega1}
show generator terminal voltage $V_t$, rotor angle $\delta $, and frequency $\omega  $ vs time plots for $\Gamma _1$ (Test Signal 1) applied to the reduced order nonlinear model. From \autoref{fig:step_testnonlinear_vt1} we can see that after the addition of the fourth step of $\Gamma _1$ (Test Signal 1) at 600 seconds, the generator terminal voltage $V_t$ settles to a new steady state value of 0.83 p.u.. 
From \autoref{fig:step_testnonlinear_delta1} we can see that the rotor angle $\delta $ first undergoes a large undershoot, followed by oscillations about 0.1 p.u. after the application of the second incremental step of $\Gamma _1$ at 200 seconds, which is further
followed by reduced oscillations about 0.3 p.u.
after the application of the third step of $\Gamma _1$ at 400 seconds, and finally after the application of the fourth and final step at 600 seconds the oscillations almost die out and the rotor angle $\delta $ settles to a value of 0.725 p.u.. \autoref{fig:step_testnonlinear_omega1} shows that the frequency $\omega  $ settles
to its steady state value of 1 p.u. after the application of the fourth step at 600 seconds. Like the rotor angle $\delta $, the oscillations of the frequency $\omega  $ also decrease with the application of each incremental step. 

\begin{figure}
          \centering
          \includegraphics[trim=0cm 0cm 0cm 0cm, clip=true, totalheight=0.27\textheight, width=0.54
           \textwidth]  {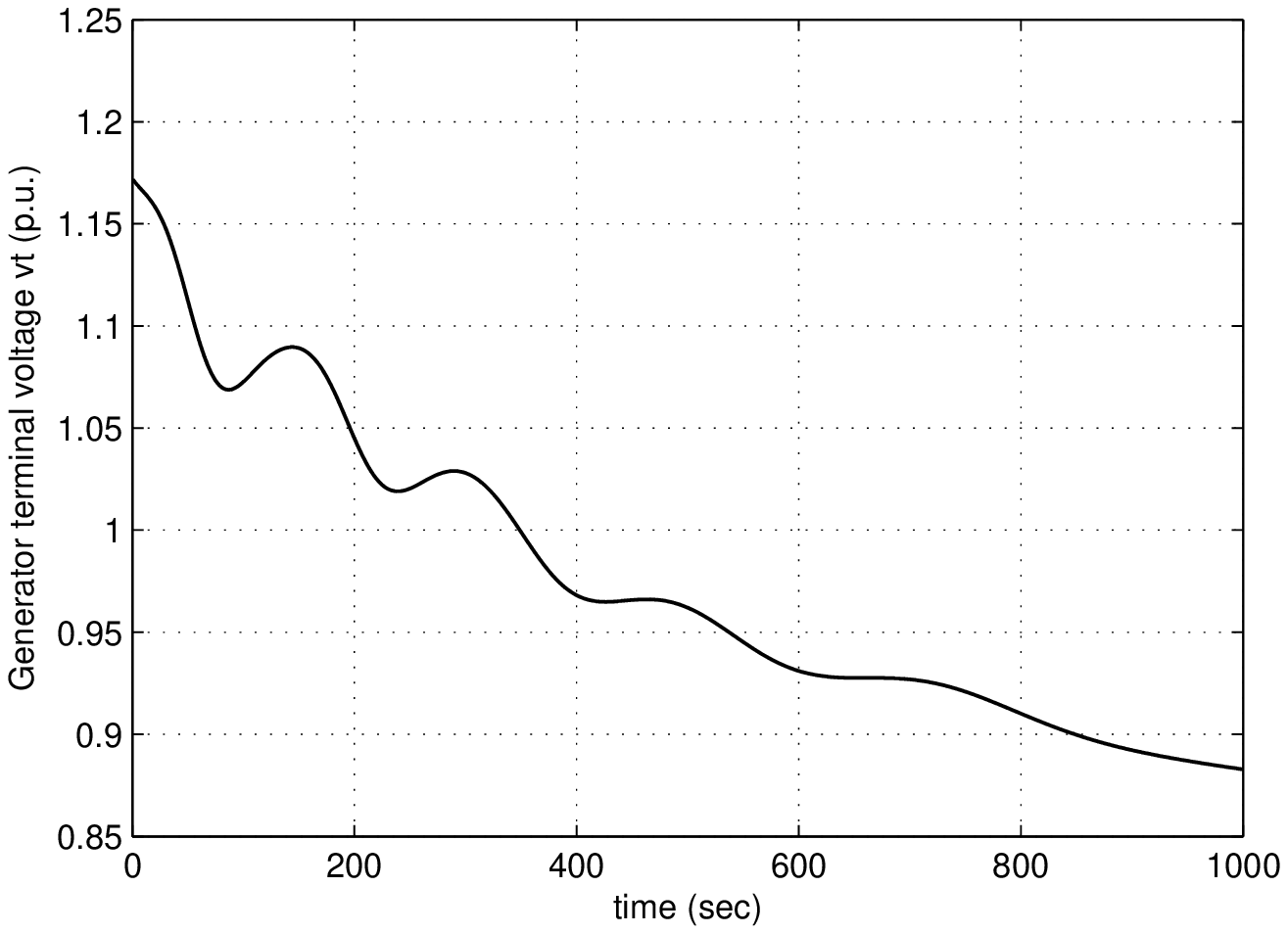}
          \caption{Plot of the generator terminal voltage $V_t$ vs time for $\Gamma _1$ (Test Signal 1) applied to the truth model}
          \label{fig:step_truth_vt1}          
          \includegraphics[trim=0cm 0cm 0cm 0cm, clip=true, totalheight=0.27\textheight, width=0.54\textwidth]{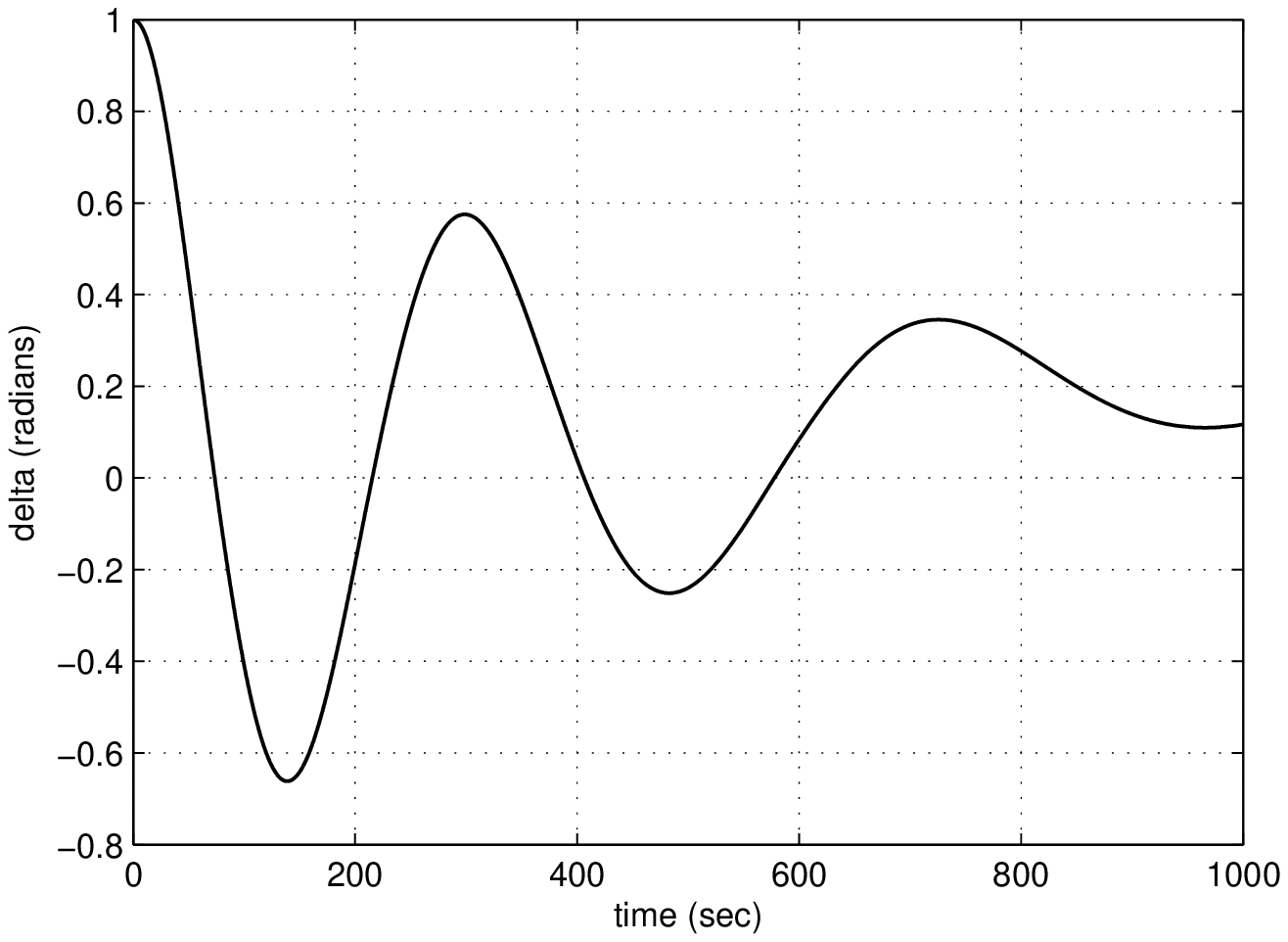}
          \caption{Plot of the rotor angle $\delta $ vs time for $\Gamma _1$ (Test Signal 1) applied to the truth model}
          \label{fig:step_truth_delta1}
          \includegraphics[trim=0cm 0cm 0cm 0cm, clip=true, totalheight=0.27\textheight, width=0.54\textwidth]{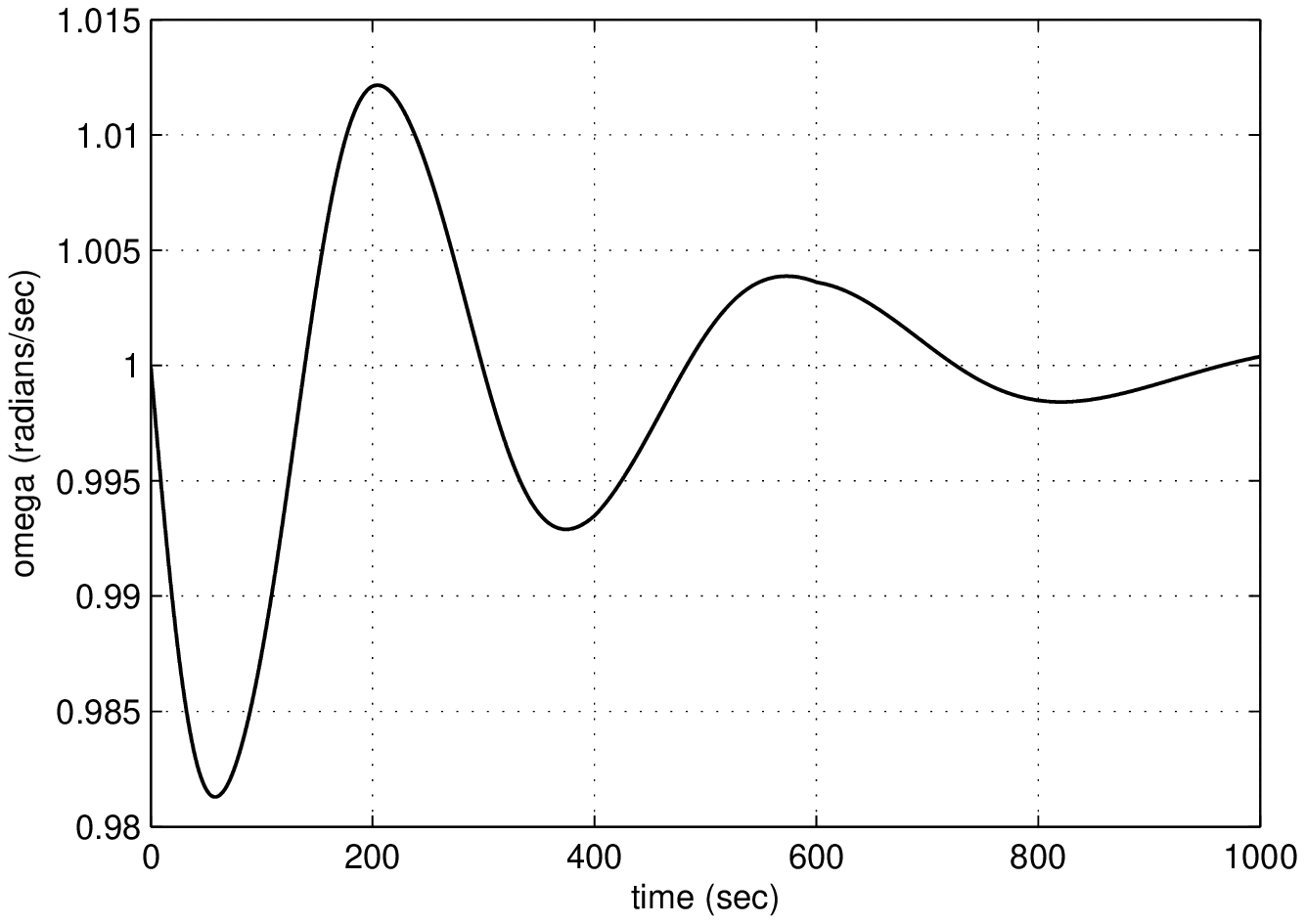}
          \caption{Plot of the angular velocity $\omega $ vs time for $\Gamma _1$ (Test Signal 1) applied to the truth model}
          \label{fig:step_truth_omega1}
\end{figure}

\autoref{fig:step_truth_vt1}, \autoref{fig:step_truth_delta1}, and \autoref{fig:step_truth_omega1}
show generator terminal voltage $V_t$, rotor angle $\delta $, and frequency $\omega  $ vs time plots for $\Gamma _1$ (Test Signal 1) applied to the truth model. From \autoref{fig:step_truth_vt1} we can see that the generator terminal voltage $V_t$ converges
very slowly to the new steady state value of 0.83 p.u.. From \autoref{fig:step_truth_delta1} we can see that the rotor angle $\delta $ first undergoes a large undershoot after which it oscillates about 0.1 p.u., where the oscillations decay with time. \autoref{fig:step_truth_omega1} shows that the frequency $\omega  $ oscillates about its steady state value of 1 p.u.. We observe a
peculiar difference between the performance of the reduced order nonlinear model and the truth model, for $\Gamma _1$ (Test Signal 1).
The addition of each incremental step of $\Gamma _1$ after a fixed interval has a distinct effect on the amplitude and damping of the output oscillations for the reduced order nonlinear model, where a rapid decrease in the output oscillations is observed with the application of each step. Thus, we observe transitions in the output response at time instants when incremental steps of $\Gamma _1$  are applied to the reduced order nonlinear model. However, for the truth model, the effect of adding incremental steps at regular intervals is negligible on the output response, compared to the reduced order nonlinear model. We do not see transitions in the output behavior at time instants when incremental steps of $\Gamma _1$  are applied to the truth model, and the oscillations decay more 
uniformly in this case.

\newpage
We now test the truth model and the reduced order nonlinear model of the SMIB
for $\Gamma _2$ (Test signal 2) given in \autoref{fig:Test_Signal2}. 
In this case the two inputs for the truth 
model are, $u = [V_F, u_T]^\mathrm{T}=[e_{15}\Gamma _2, \Gamma _2]^\mathrm{T}$, and the two inputs for the reduced order nonlinear model are, $u = [E_{FD}, u_T]^\mathrm{T}=[\Gamma _2, \Gamma _2]^\mathrm{T}$. 

\begin{figure}
          \centering    
          \includegraphics[trim=0cm 0cm 0cm 0cm, clip=true, totalheight=0.28\textheight, 
           width=0.58\textwidth]   {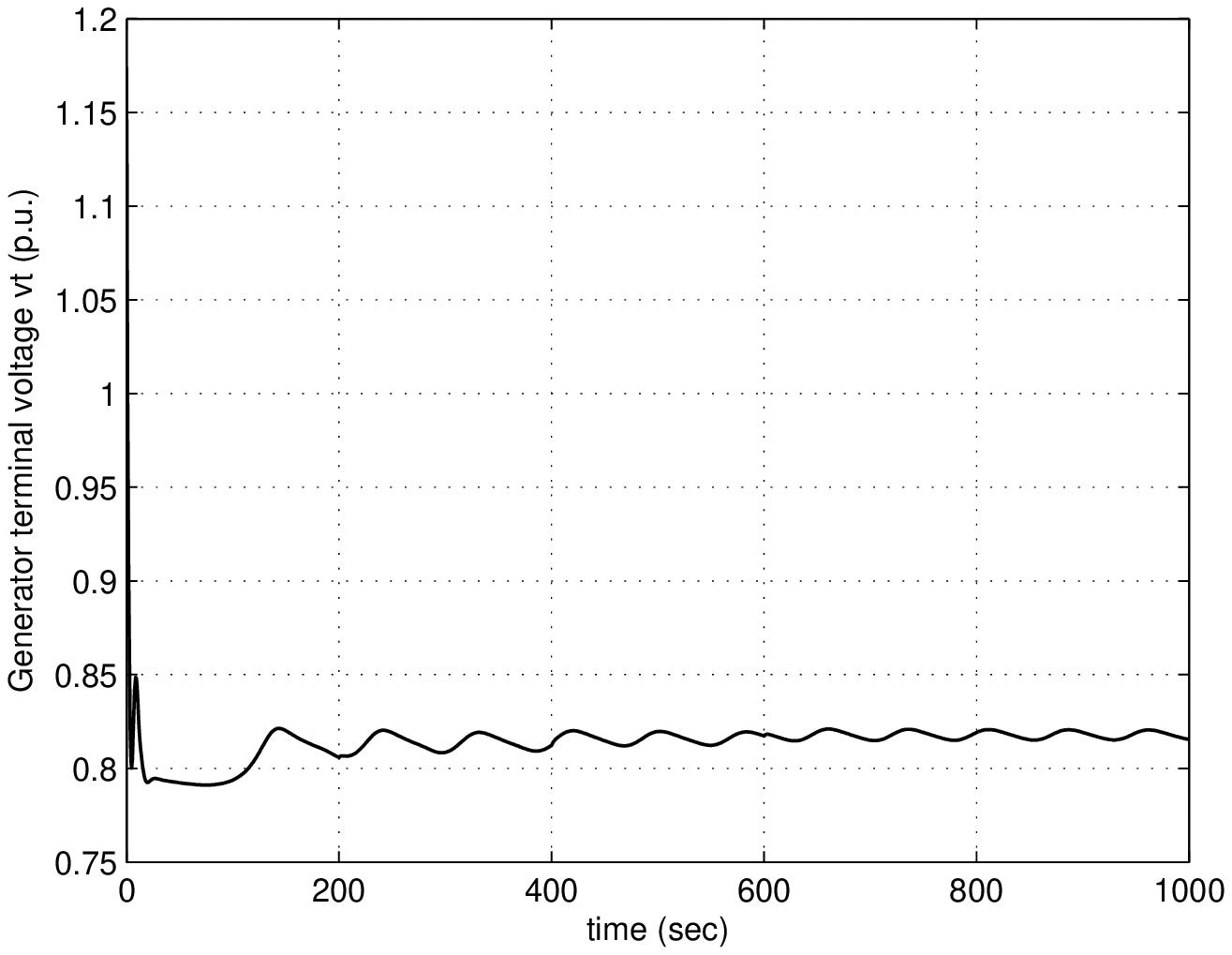}
          \caption{Plot of the generator terminal voltage $V_t$ vs time for $\Gamma _2$ (Test Signal 2) 
           applied to the reduced order nonlinear model}
          \label{fig:step_testnonlinear_vt2}
          \includegraphics[trim=0cm 0cm 0cm 0cm, clip=true, totalheight=0.28\textheight, 
           width=0.58\textwidth]{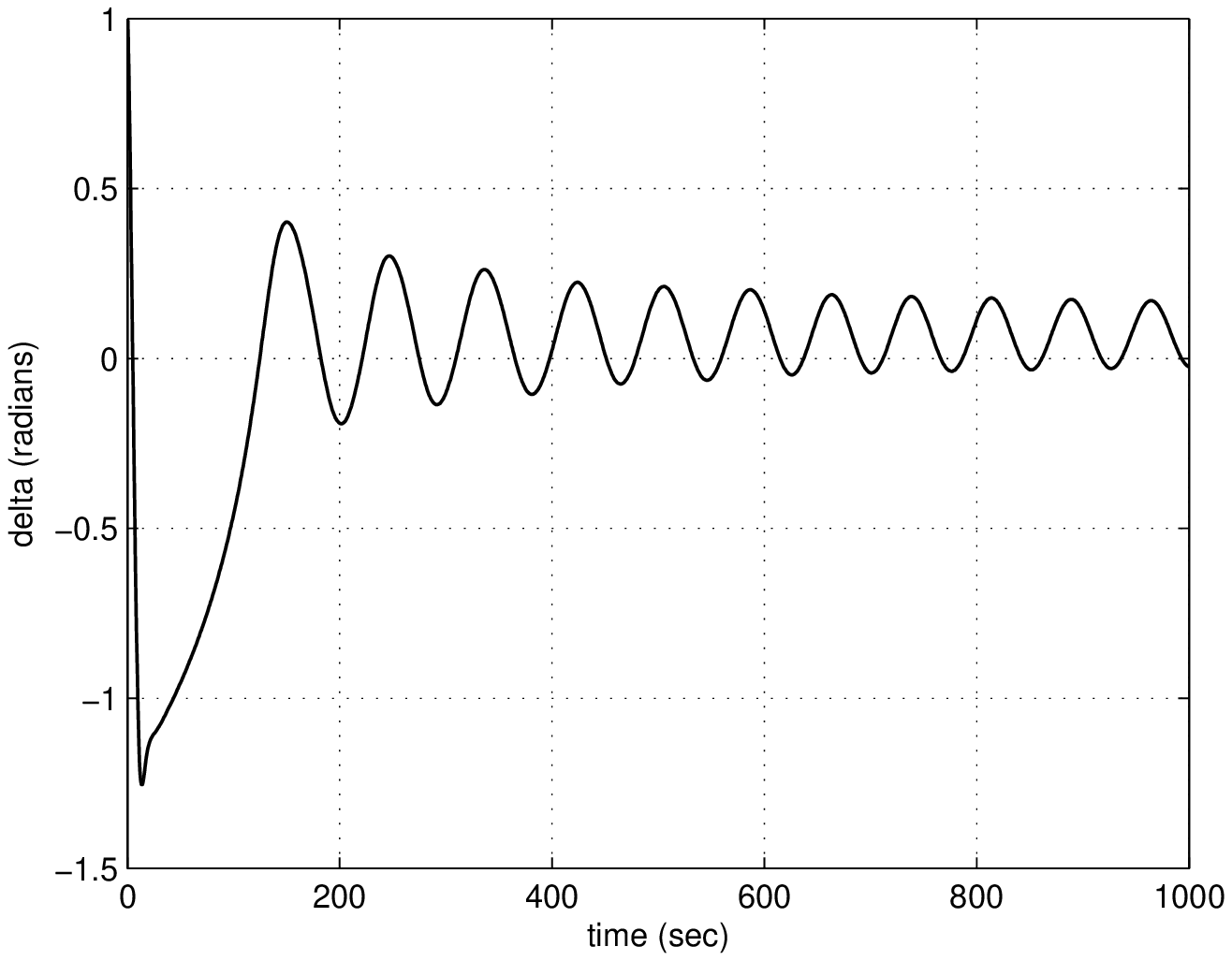}
          \caption{Plot of the rotor angle $\delta $ vs time for $\Gamma _2$ (Test Signal 2) applied to the reduced order nonlinear model}
          \label{fig:step_testnonlinear_delta2}
          \includegraphics[trim=0cm 0cm 0cm 0cm, clip=true, totalheight=0.28\textheight, 
          width=0.58\textwidth]{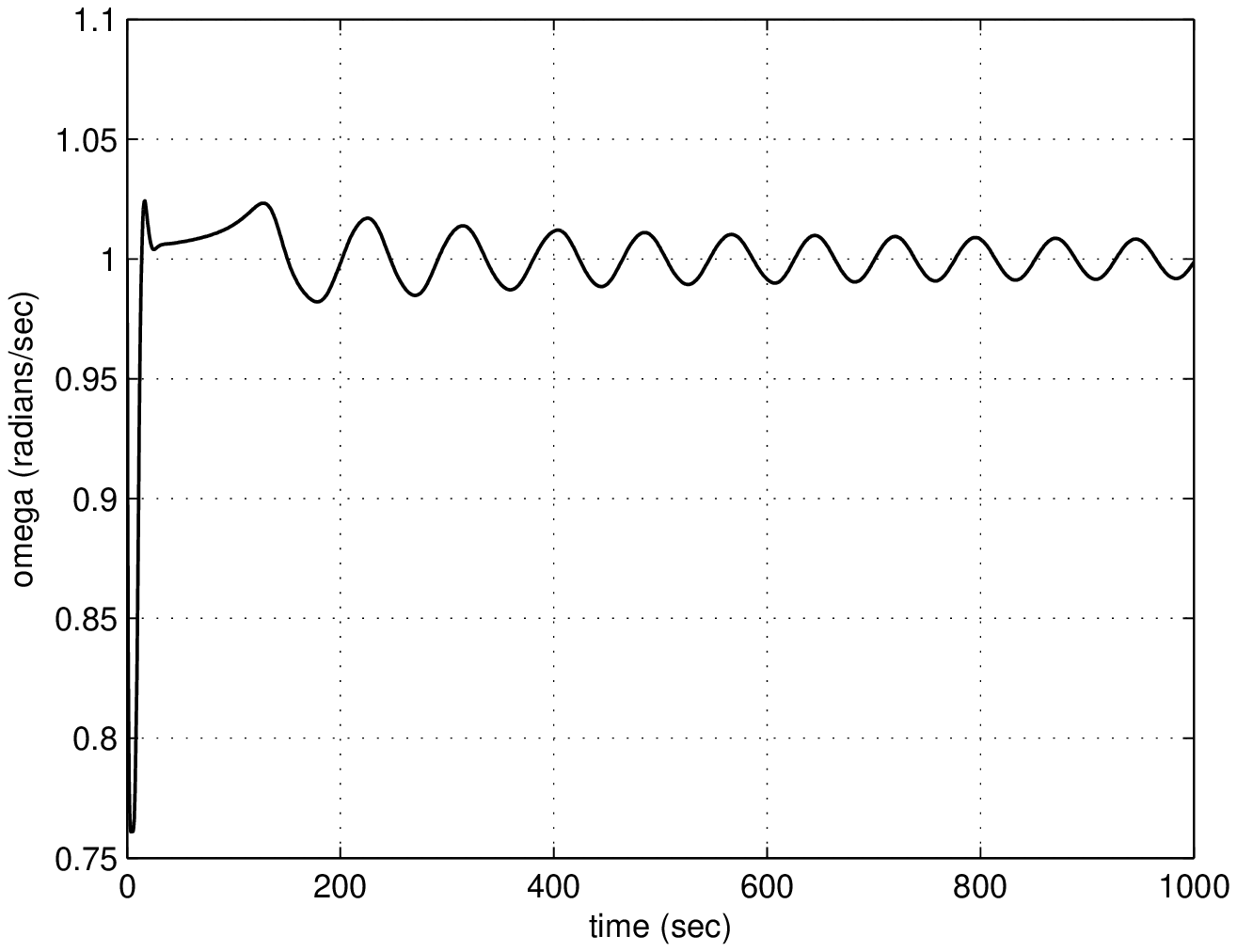}
          \caption{Plot of the frequency $\omega  $ vs time for $\Gamma _2$ (Test Signal 2) applied to the 
          reduced order nonlinear model}
          \label{fig:step_testnonlinear_omega2}          
\end{figure}

\autoref{fig:step_testnonlinear_vt2}, \autoref{fig:step_testnonlinear_delta2}, and \autoref{fig:step_testnonlinear_omega2}
show generator terminal voltage $V_t$, rotor angle $\delta $, and frequency $\omega  $ vs time plots for $\Gamma _2$ (Test Signal 2) applied to the reduced order nonlinear model. From \autoref{fig:step_testnonlinear_vt2} we can see that the generator terminal voltage $V_t$ settles to a new steady value of 0.82 p.u.. 
From \autoref{fig:step_testnonlinear_delta2} we can see that the rotor angle $\delta $ first undergoes a large undershoot after
which it oscillates about 0.1 p.u.. 
 \autoref{fig:step_testnonlinear_omega2} shows that the frequency $\omega  $ oscillates about
its steady state value of 1 p.u.. The oscillations decay with time. For $\Gamma _2$ (Test Signal 2) where the magnitudes of the incremental steps are reduced to half of $\Gamma _1$ (Test Signal 1), the effect of adding incremental steps of $\Gamma _2$ after regular 
intervals to the reduced order nonlinear model, is not as significant as compared to the effect of $\Gamma _1$ (Test Signal 1). The oscillations decay uniformly with time, and we do not see the sharp transitions that we
saw when $\Gamma _1$ was applied to the reduced order nonlinear model. 

\begin{figure}
          \centering
          \includegraphics[trim=0cm 0cm 0cm 0cm, clip=true, totalheight=0.28\textheight, width=0.58
           \textwidth]  {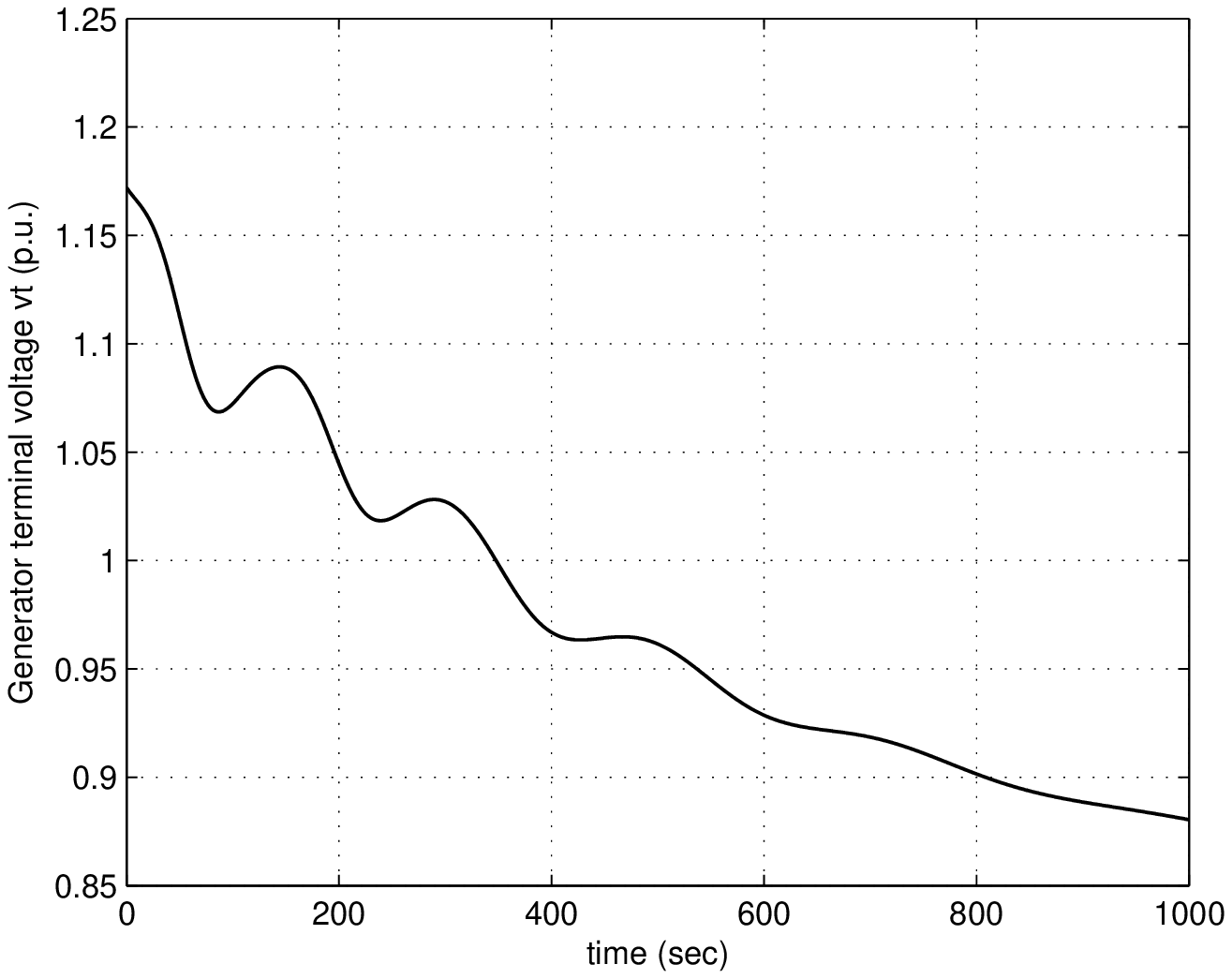}
          \caption{Plot of the generator terminal voltage $V_t$ vs time for $\Gamma _2$ (Test Signal 2)  applied to the truth model}
          \label{fig:step_truth_vt2}          
          \includegraphics[trim=0cm 0cm 0cm 0cm, clip=true, totalheight=0.28\textheight, width=0.58\textwidth]{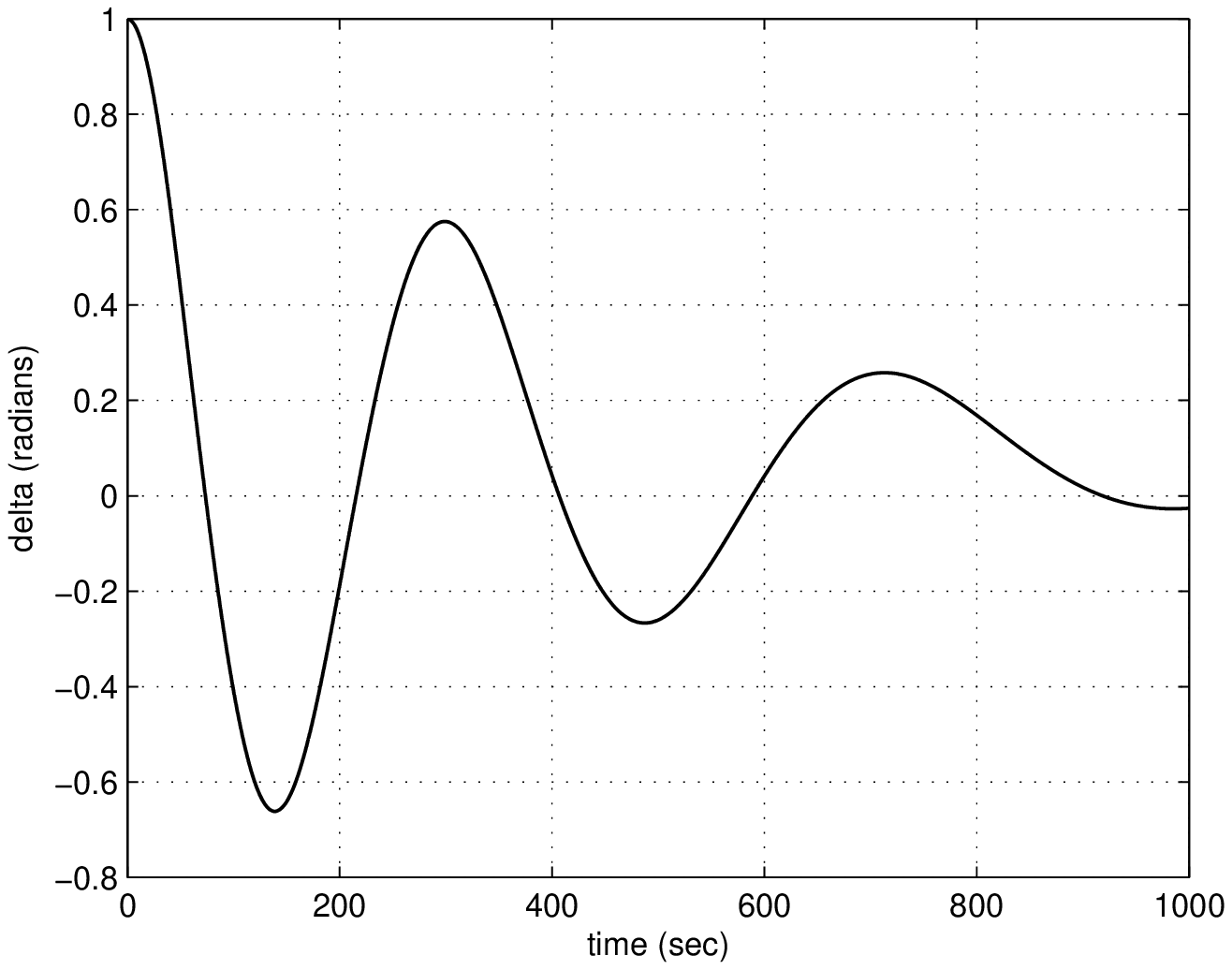}
          \caption{Plot of the rotor angle $\delta $ vs time for $\Gamma _2$ (Test Signal 2) applied to the truth model}
          \label{fig:step_truth_delta2}
          \includegraphics[trim=0cm 0cm 0cm 0cm, clip=true, totalheight=0.28\textheight, width=0.58\textwidth]{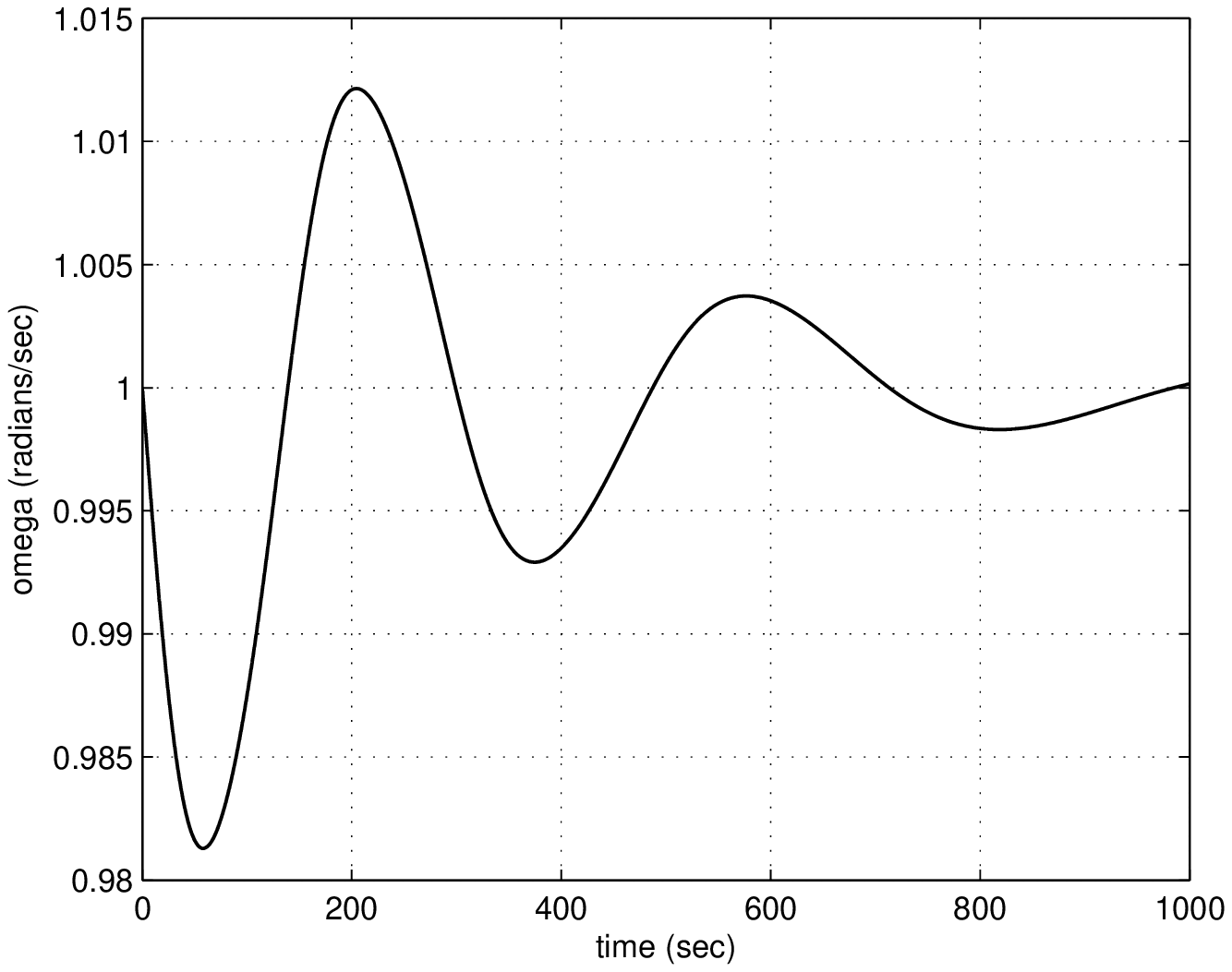}
          \caption{Plot of the angular velocity $\omega $ vs time for $\Gamma _2$ (Test Signal 2) applied to the truth model}
          \label{fig:step_truth_omega2}
\end{figure}

\autoref{fig:step_truth_vt2}, \autoref{fig:step_truth_delta2}, and \autoref{fig:step_truth_omega2}
show generator terminal voltage $V_t$, rotor angle $\delta $, and frequency $\omega  $ vs time plots for $\Gamma _2$ (Test Signal 2) applied to the truth model. These results are similar to the results obtained when $\Gamma _1$ (Test Signal 1) is applied to the truth model. 

\newpage
\section{The Decoupled Reduced Order Model}

        Let us recall from the first section that small changes in real power are mainly dependent on changes in 
        rotor angle $\delta $, and thus
        the frequency (i.e. on the prime mover valve control), whereas the reactive power is mainly dependent 
        on the voltage magnitude (i.e. on the generator excitation).
        The excitation system time constant is much smaller than the prime mover time constant and its transient decay 
        much faster and does not affect the load frequency control (LFC) dynamics. Thus, the cross-coupling between the
        LFC loop and the automatic voltage regulator (AVR) loop is negligible.
        Hence, load frequency control and excitation voltage control can be analyzed independently. Therefore, we decouple
        the two-input two-output (MIMO) reduced order system into two single input single output (SISO) subsystems. The first 
        SISO subsystem consists of the LFC loop, where the turbine valve control $u_T$ controls the rotor angle
        $\delta $ and thus the frequency $\omega $ of the synchronous generator. The second SISO subsystem consists of the AVR loop, where
        the generator excitation $E_{FD}$ controls the terminal voltage $V_t$ of the synchronous generator.
        From \autoref{eq:linear22} and \autoref{eq:linear23} the linear model of the synchronous generator and 
        turbine connected to an infinite bus is 
        \begin{equation}
        \begin{aligned}
               &\Delta \dot{E}'_q = f_{11}\Delta E'_q+\frac{\partial f_1}{\partial x_3}\bigg{| }_{x_0}
               \Delta \delta +g_{11}\Delta E_{FD}\\
               &\Delta \dot{\omega } = \frac{\partial f_2}{\partial x_1}\bigg{| }_{x_0}\Delta E'_q+f_{27}\Delta \omega +
               \frac{\partial f_2}{\partial x_3}\bigg{| }_{x_0}\Delta \delta+f_{28}\Delta T_m\\
               &\Delta \dot{\delta } = \Delta \omega \\
               &\Delta \dot{T}_m = f_{41}\Delta T_m+f_{42}\Delta G_V\\
               &\Delta \dot{G}_V = f_{51}\Delta \omega +f_{52}\Delta G_V+g_{55}\Delta u_T\\
               \end{aligned}
               \label{eq:decouple1}
        \end{equation}       
       Note that all the coefficients in the above expression are evaluated at the nominal operating point
       given in the previous section. 
Let us denote
        \begin{equation}
        \begin{aligned}
         &\frac{\partial f_1}{\partial x_3}\bigg{| }_{x_0}=A_{13}\\
         &\frac{\partial f_2}{\partial x_1}\bigg{| }_{x_0}=A_{21}\\
         &\frac{\partial f_2}{\partial x_3}\bigg{| }_{x_0}=A_{23}\\
         \end{aligned}
               \label{eq:decouple1point1}
        \end{equation} 
The MIMO system with control inputs $\Delta E_{FD}$ and $\Delta u_T$ can be decoupled into the LFC loop 
with control input $\Delta u_T$ as 
        \begin{equation}
        \begin{aligned}
               &\Delta \dot{\omega } = A_{21}\Delta E'_q+f_{27}\Delta \omega +
               A_{23}\Delta \delta+f_{28}\Delta T_m\\
               &\Delta \dot{\delta } = \Delta \omega \\
               &\Delta \dot{T}_m = f_{41}\Delta T_m+f_{42}\Delta G_v\\
               &\Delta \dot{G}_V = f_{51}\Delta \omega +f_{52}\Delta G_V+g_{55}\Delta u_T\\
               \end{aligned}
               \label{eq:decouple2point0}
        \end{equation} 
        and the AVR loop with control input $\Delta E_{FD}$ as 
        \begin{equation}
         \Delta \dot{E}'_q = f_{11}\Delta E'_q+A_{13}\Delta \delta +g_{11}\Delta E_{FD}\\
         \label{eq:decouple3point0}
        \end{equation}
The weak coupling $A_{21}\Delta E'_q$ between the LFC and the AVR loop can be neglected in the LFC loop in \autoref{eq:decouple2point0},
and the weak coupling $A_{13}\Delta \delta$ between the LFC and the AVR loop can be neglected in the AVR loop in 
\autoref{eq:decouple3point0}.
Neglecting theses terms, the LFC loop with control input $\Delta u_T$ is approximated as 
\begin{equation}
        \begin{aligned}
               &\Delta \dot{\omega } = f_{27}\Delta \omega +
               A_{23}\Delta \delta+f_{28}\Delta T_m\\
               &\Delta \dot{\delta } = \Delta \omega \\
               &\Delta \dot{T}_m = f_{41}\Delta T_m+f_{42}\Delta G_v\\
               &\Delta \dot{G}_V = f_{51}\Delta \omega +f_{52}\Delta G_V+g_{55}\Delta u_T\\
               \end{aligned}
               \label{eq:decouple2}
        \end{equation} 
and the AVR loop with control input $\Delta E_{FD}$ is approximated as
\begin{equation}
         \Delta \dot{E}'_q = f_{11}\Delta E'_q+g_{11}\Delta E_{FD}\\
         \label{eq:decouple3}
\end{equation}

\subsection{LFC Dynamics}

\begin{figure}
          \centering
          \includegraphics[trim=0cm 0cm 0cm 0cm, clip=true, totalheight=0.26\textheight, width=0.54\textwidth]{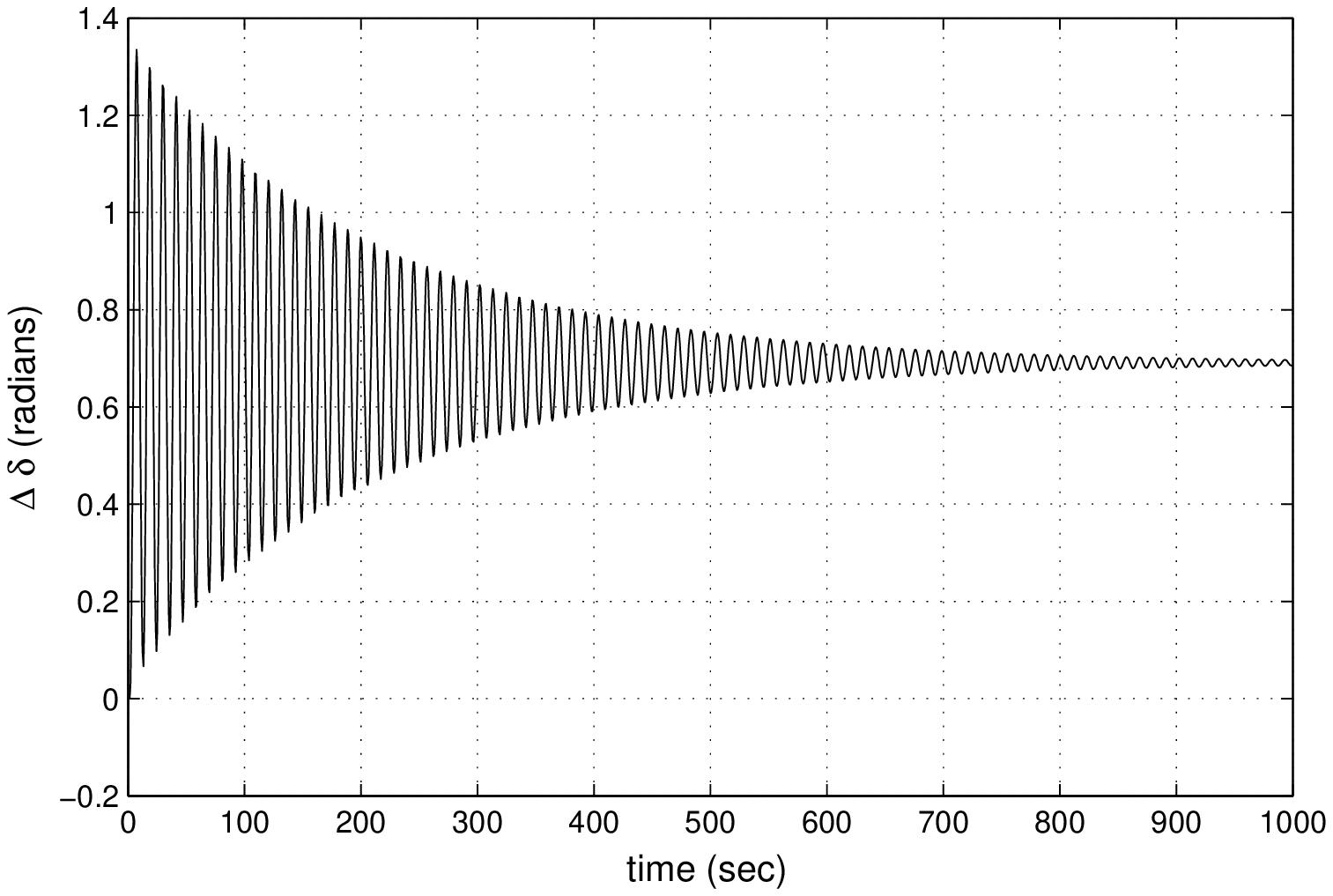}
          \caption{Plot of $\Delta \delta $(s) vs time for the uncompensated (LFC) SISO system for a step input}
          \label{fig:lfc4}
          \includegraphics[trim=0cm 0cm 0cm 0cm, clip=true, totalheight=0.26\textheight, width=0.54\textwidth]{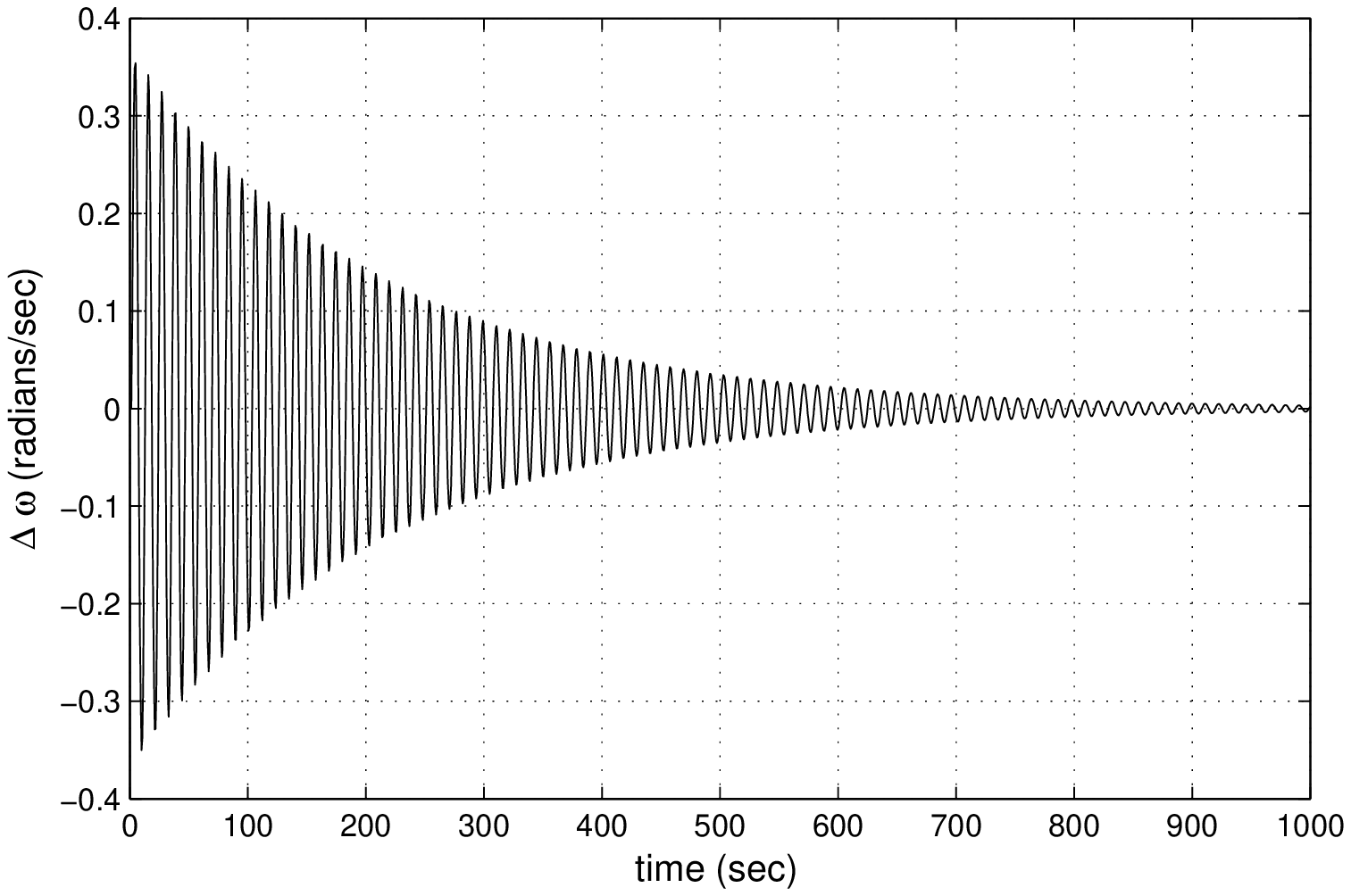}
          \caption{Plot of $\Delta \omega $(s) vs time for the uncompensated (LFC) SISO system for a step input}
          \label{fig:lfc5}
\end{figure}

\begin{figure}         
          \centering
          \includegraphics[trim=0cm 0cm 0cm 0cm, clip=true, totalheight=0.26\textheight, width=0.54\textwidth]{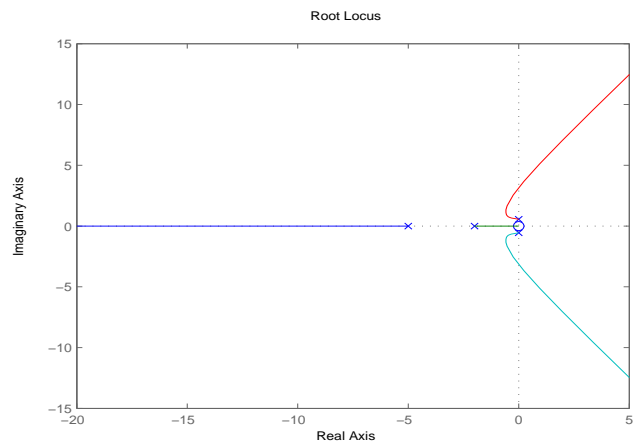}
          \caption{Root locus plot for the uncompensated (LFC) SISO system}
          \label{fig:lfc6}
\end{figure}

In this section we present the dynamics of the (LFC) SISO system with control input $\Delta u_T$. We then perform root locus analysis
of the uncompensated (LFC) SISO system. A PID controller is designed next to stabilize the (LFC) SISO system which is originally 
unstable. 
   From \autoref{eq:decouple2} the LFC dynamics are
   \begin{equation}
        \begin{aligned}
               &\Delta \dot{\omega } = f_{27}\Delta \omega +
               A_{23}\Delta \delta+f_{28}\Delta T_m\\
               &\Delta \dot{\delta } = \Delta \omega \\
               &\Delta \dot{T}_m = f_{41}\Delta T_m+f_{42}\Delta G_v\\
               &\Delta \dot{G}_V = f_{51}\Delta \omega +f_{52}\Delta G_V+g_{55}\Delta u_T\\
               \end{aligned}
               \label{eq:decouple4}
        \end{equation} 
 Differentiating $\Delta \dot{\omega }$ and substituting $\Delta \dot{\delta } = \Delta \omega $ we have
 \begin{equation}
              \Delta \ddot{\omega }=f_{27}\Delta \dot{\omega }+A_{23}\Delta \omega 
              +f_{28}\Delta \dot{T}_m
            \label{eq:decouple5}  
  \end{equation}
Taking the Laplace transform of \autoref{eq:decouple5} with zero initial conditions we get
\begin{equation}
            s^2\Delta \omega (s)=f_{27}s\Delta \omega (s)+A_{23}\Delta \omega (s)
              +f_{28}s\Delta T_m(s)
            \label{eq:decouple6}  
  \end{equation}  
On rearranging \autoref{eq:decouple6} and taking the Laplace transform of $\Delta \dot{\delta } = \Delta \omega $ 
   \begin{equation}
   \begin{aligned}
          \frac{\Delta \omega (s)}{\Delta T_m(s)} &= \frac{f_{28}s}{s^2-f_{27}s-A_{23}}\\
          \frac{\Delta \delta (s)}{\Delta \omega (s)} &= \frac{1}{s}\\
          \end{aligned}
  \label{eq:decouple7}  
  \end{equation} 
  \autoref{eq:decouple7} evaluated at the nominal operating point gives
     \begin{equation}
   \begin{aligned}
    \frac{\Delta \omega (s)}{\Delta T_m(s)} &= \frac{0.211s}{s^2+0.3054}\\
   \frac{\Delta \delta (s)}{\Delta \omega (s)} &= \frac{1}{s}\\
          \end{aligned}
  \label{eq:decouple8}  
  \end{equation}
   The dynamics of the turbine from \autoref{eq:decouple4} can be written as
   \begin{equation}
               \Delta \dot{T}_m = f_{41}\Delta T_m+f_{42}\Delta G_v
               \label{eq:decouple9}
   \end{equation} 
 Taking the Laplace transform of \autoref{eq:decouple9} with zero initial conditions 
 \begin{equation}
        s\Delta T_m(s)=f_{41}\Delta T_m(s)+f_{42}\Delta G_V(s)
 \label{eq:decouple10}
 \end{equation}    
  Rearranging and expressing \autoref{eq:decouple10} as a transfer function
 \begin{equation}
          \frac{\Delta T_m(s)}{\Delta G_V(s)}=\frac{f_{42}}{s-f_{41}}
 \label{eq:decouple11}
 \end{equation}    
 \autoref{eq:decouple11} evaluated at the nominal operating point gives
     \begin{equation}
          \frac{\Delta T_m(s)}{\Delta G_V(s)}=\frac{2}{s+2}
     \label{eq:decouple12}
 \end{equation} 
  From \autoref{eq:decouple4} the governor dynamics can be written as
\begin{equation}    
   \Delta \dot{G}_V = f_{51}\Delta \omega +f_{52}\Delta G_V+g_{55}\Delta u_T
  \label{eq:decouple13}
 \end{equation}
Taking the Laplace transform of \autoref{eq:decouple13} with zero initial conditions 
 \begin{equation}
       s\Delta G_V(s)=f_{51}\Delta \omega (s)+f_{52}\Delta G_V(s)+g_{55}\Delta u_T(s)
\label{eq:decouple14}
 \end{equation}
Rearranging and expressing \autoref{eq:decouple14} as a transfer function 
\begin{equation} 
               \Delta G_V(s)=\frac{f_{51}}{s-f_{52}}\Delta \omega (s)+\frac{g_{55}}{s-f_{52}}\Delta u_T(s)
\label{eq:decouple15}
\end{equation}
 \autoref{eq:decouple15} evaluated at the nominal operating point gives  
\begin{equation} 
               \Delta G_V(s)=\frac{-0.25}{s+5}\Delta \omega (s)+\frac{5}{s+5}\Delta u_T(s)
\label{eq:decouple15point1}
\end{equation}
\begin{figure}
          \centering
          \includegraphics[trim=8.5cm 4cm 8.5cm 4cm, clip=true, totalheight=0.7\textheight, width=0.2\textwidth, angle=90]{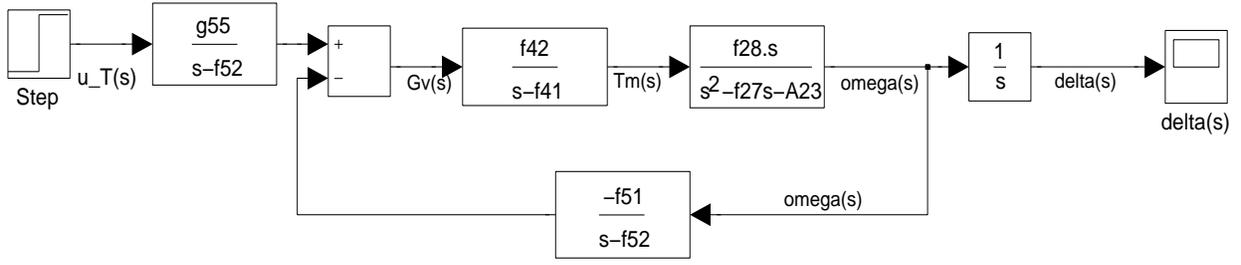}
          \caption{Simulink model of the (LFC) SISO system for a step input}
          \label{fig:lfc1}
\end{figure}
\autoref{fig:lfc1} shows the simulink block diagram of the LFC loop for a step input. In all simulink
models the subscript $\Delta $ is omitted for convenience. 
\begin{figure}
          \centering
          \includegraphics[trim=8.5cm 4cm 8.5cm 4cm, clip=true, totalheight=0.7\textheight, width=0.2\textwidth, angle=90]{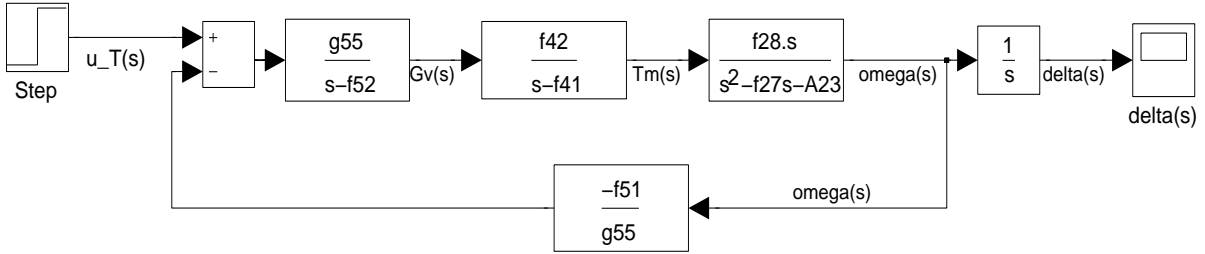}
          \caption{Simulink model of the (LFC) SISO system for a step input}
          \label{fig:lfc2}
\end{figure}
By moving the summing point ahead of the block $\frac{g_{55}}{s-f_{52}}$, the simulink model in \autoref{fig:lfc1} can be simplified to get the simulink model shown in \autoref{fig:lfc2}.
In \autoref{fig:lfc2} let 
\begin{equation}
\begin{aligned}
       G(s) &= \bigg{(}\frac{g_{55}}{s-f_{52}}\bigg{)}\bigg{(}\frac{f_{42}}{s-f_{41}}\bigg{)}
               \bigg{(}\frac{f_{28}s}{s^2-f_{27}s-A_{23}}\bigg{)}\\
            &= \frac{g_{55}f_{42}f_{28}s}{(s-f_{52})(s-f_{41})(s^2-f_{27}s-A_{23})}\\
            \end{aligned}
\label{eq:decouple16}
\end{equation}
and
\begin{equation}
       H(s)=-\frac{f_{51}}{g_{55}}
\label{eq:decouple17}
\end{equation}
\begin{figure}
          \centering
          \includegraphics[trim=0cm 0cm 0cm 0cm, clip=true, totalheight=0.15\textheight, width=0.8\textwidth, angle=0]{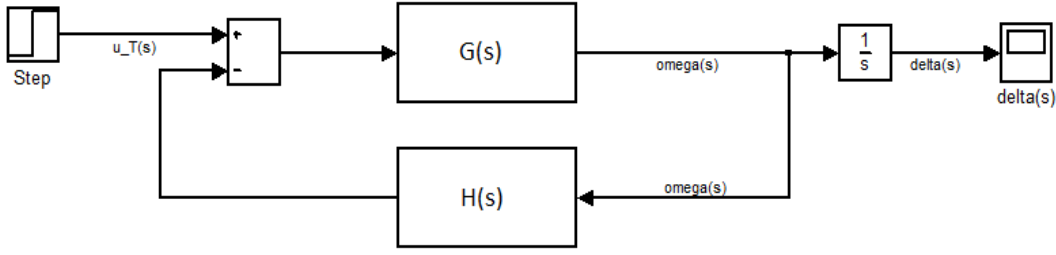}
          \caption{Simulink model of the (LFC) SISO system for a step input}
          \label{fig:lfc3}
\end{figure}
Using \autoref{eq:decouple16} and \autoref{eq:decouple17}, the simulink model in \autoref{fig:lfc2} can be simplified to 
the simulink model shown in \autoref{fig:lfc3}.
The open loop transfer function of the SISO system shown in \autoref{fig:lfc3} with output $\omega (s)$ and step input $u_T(s)$ is
\begin{equation}
\begin{aligned}
     G(s)H(s) &= -\frac{f_{42}f_{28}f_{51}s}{(s-f_{52})(s-f_{41})(s^2-f_{27}s-A_{23})}\\                
   \end{aligned}
\label{eq:decouple20}
\end{equation} 
The closed loop transfer function of the SISO system between the output $\Delta \delta (s)$ and control input
$\Delta u_T(s)$ is
\begin{equation}
\begin{aligned}
               \frac{\Delta \delta (s)}{\Delta u_T(s)} &=  \frac{G(s)}{1+G(s)H(s)}\bigg{(}\frac{1}{s}\bigg{)}= 
               \frac{\frac{g_{55}f_{42}f_{28}}{(s-f_{52})(s-f_{41})(s^2-f_{27}s-A_{23})}}
               {1 -\frac {f_{42}f_{28}f_{51}s}{(s-f_{52})(s-f_{41})(s^2-f_{27}s-A_{23})}}\\
               &= \frac{g_{55}f_{42}f_{28}}{(s-f_{52})(s-f_{41})(s^2-f_{27}s-A_{23})-f_{51}f_{42}f_{28}s}\\
 \end{aligned}
 \label{eq:decouple22}
\end{equation}
For a step input, $\Delta u_T(s)=\frac{1}{s}$. From the final value theorem, the steady state value of $\Delta \delta (s)$
is 
\begin{equation}
\begin{aligned}
                \Delta \delta _{ss} &=  \lim_{s \to 0} s\Delta \delta (s)\\
                                    &= \lim_{s \to 0} s\Delta u_T(s)
                 \frac{g_{55}f_{42}f_{28}}{(s-f_{52})(s-f_{41})(s^2-f_{27}s-A_{23})-f_{51}f_{42}f_{28}s}\\
                 &= -\frac{g_{55}f_{42}f_{28}}{f_{52}f_{41}A_{23}}\\
\end{aligned}
 \label{eq:decouple23}
\end{equation}
Substituting numerical values for the constant parameters in \autoref{eq:decouple23} we get
\begin{equation}
        \Delta \delta _{ss}=0.6909
 \label{eq:decouple24}
\end{equation} 
Also
\begin{equation}
\begin{aligned}
               \frac{\Delta \omega  (s)}{\Delta u_T(s)} &=  \frac{G(s)}{1+G(s)H(s)}= 
               \frac{\frac{g_{55}f_{42}f_{28}s}{(s-f_{52})(s-f_{41})(s^2-f_{27}s-A_{23})}}
               {1 -\frac {f_{42}f_{28}f_{51}s}{(s-f_{52})(s-f_{41})(s^2-f_{27}s-A_{23})}}\\
               &= \frac{g_{55}f_{42}f_{28}s}{(s-f_{52})(s-f_{41})(s^2-f_{27}s-A_{23})-f_{51}f_{42}f_{28}s}\\
 \end{aligned}
 \label{eq:decouple25}
\end{equation}      
From the final value theorem, the steady state value of $\Delta \omega  (s)$ for a step input is 
\begin{equation}
\begin{aligned}
                \Delta \omega  _{ss} &=  \lim_{s \to 0} s\Delta \omega  (s)\\
                                    &= \lim_{s \to 0} s\Delta u_T(s)
                 \frac{g_{55}f_{42}f_{28}s}{(s-f_{52})(s-f_{41})(s^2-f_{27}s-A_{23})-f_{51}f_{42}f_{28}s}\\
                 &= -\lim_{s \to 0}\frac{g_{55}f_{42}f_{28}}{f_{52}f_{41}A_{23}}s\\
                 &= 0
\end{aligned}
 \label{eq:decouple26}
\end{equation}
\autoref{fig:lfc4} and \autoref{fig:lfc5} show the plots of $\Delta \delta $(s) and $\Delta \omega $(s) vs time for the uncompensated (LFC) SISO system for a step input, respectively. From these plots we can see that $\Delta \delta $(s) settles to a steady state value of 0.6909 and $\Delta \omega $(s)
settles to a steady state value of 0 which are in agreement with the respective steady state values calculated in \autoref{eq:decouple24} and 
\autoref{eq:decouple26}. Also the root locus plot in \autoref{fig:lfc6} clearly shows that the uncompensated (LFC) SISO system 
with step input $u_T(s)$ is unstable.

\newpage
\subsubsection{PID controller design for the LFC loop}

\begin{figure}
          \centering
          \includegraphics[trim=8cm 2cm 8cm 2cm, clip=true, totalheight=0.7\textheight, width=0.25\textwidth, angle=90]{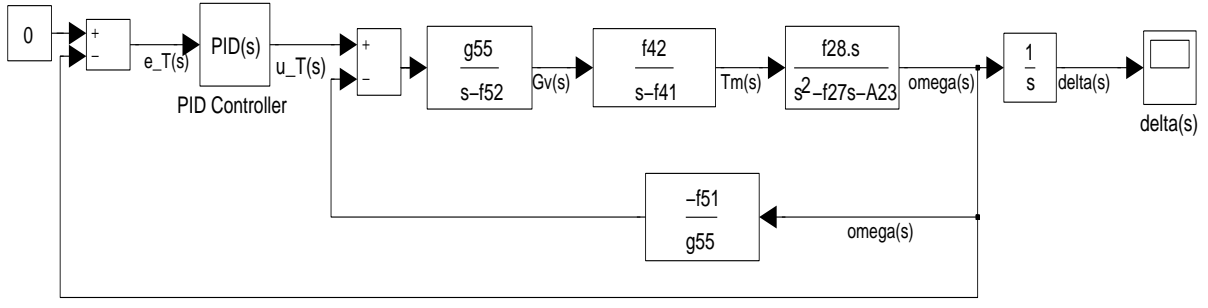}
          \caption{Simulink model of the PID compensated (LFC) SISO system}
          \label{fig:lfc7}
\end{figure}

\begin{figure}
          \centering   
          \includegraphics[trim=0cm 0cm 0cm 0cm, clip=true, totalheight=0.26\textheight, width=0.52\textwidth]{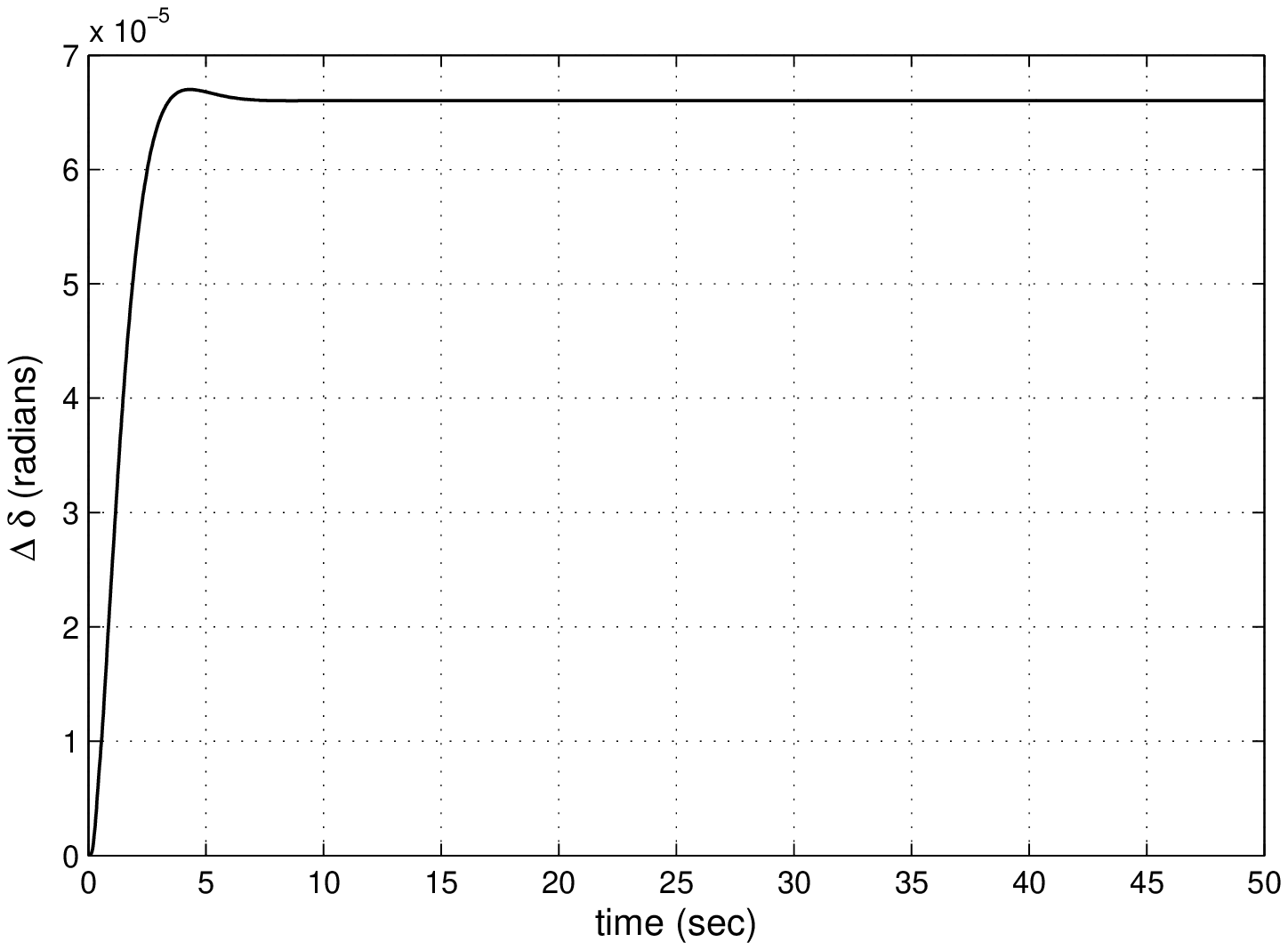}
          \caption{Plot of $\Delta \delta $(s) vs time for the PID compensated (LFC) SISO system}
          \label{fig:lfc8}          
          \includegraphics[trim=0cm 0cm 0cm 0cm, clip=true, totalheight=0.26\textheight, width=0.52\textwidth]{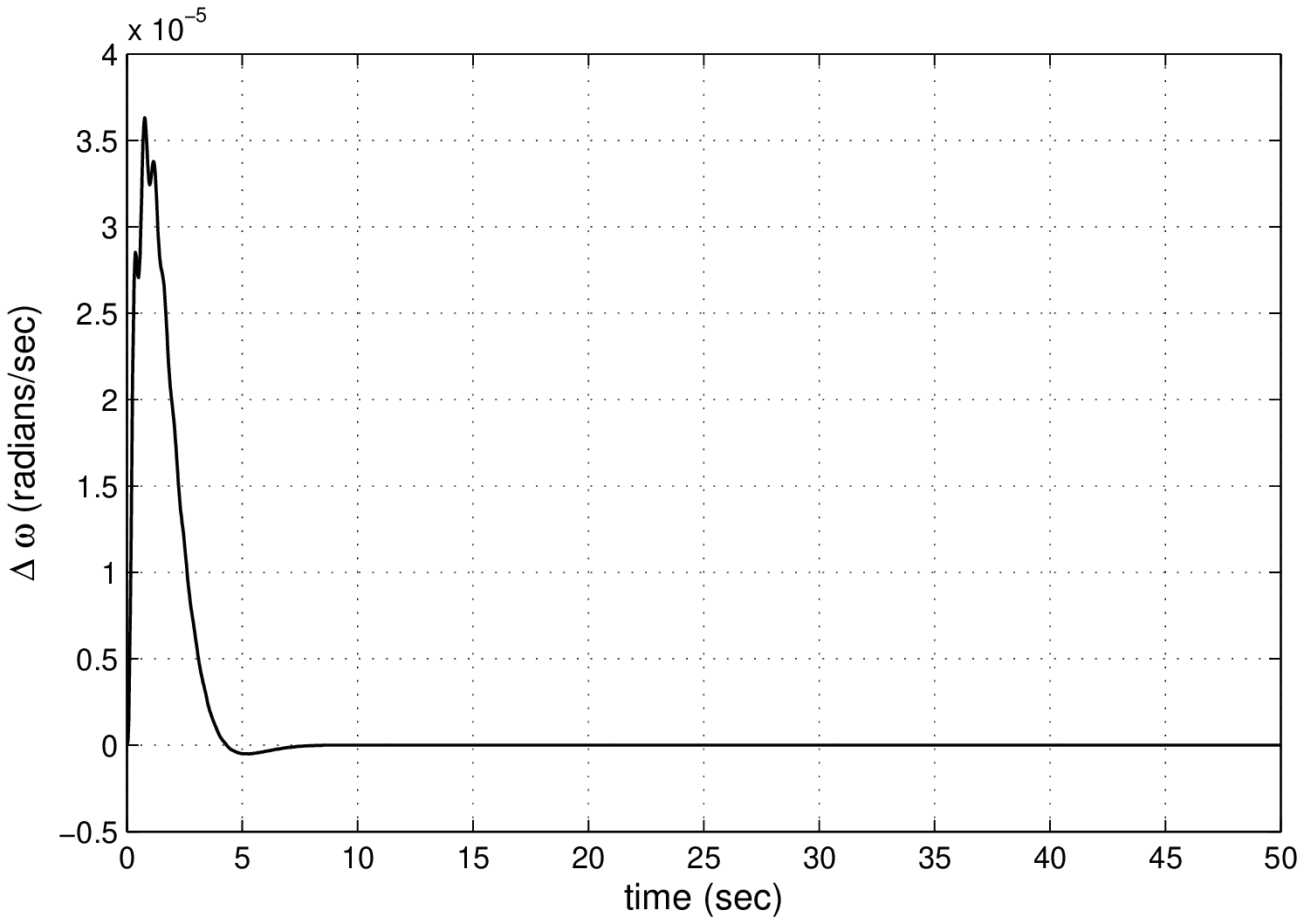}
          \caption{Plot of $\Delta \omega $(s) vs time for the PID compensated (LFC) SISO system}
          \label{fig:lfc9}
\end{figure}

\begin{figure}
          \centering
          \includegraphics[trim=0cm 0cm 0cm 0cm, clip=true, totalheight=0.26\textheight, width=0.52\textwidth]{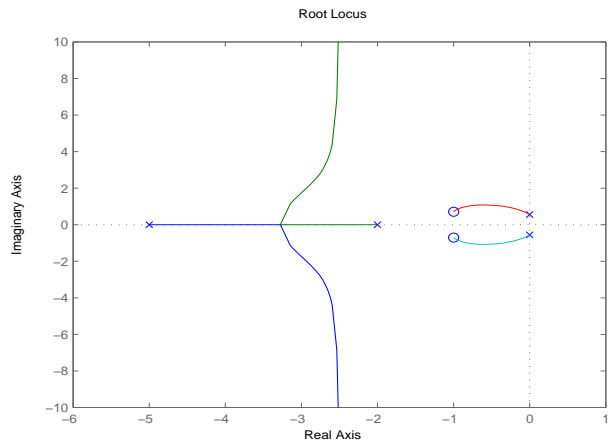}
          \caption{Root locus plot for the PID compensated (LFC) SISO system}
          \label{fig:lfc10}
\end{figure}

In order to stabilize this system we design a proportional-integral-derivative (PID) controller. The simulink model of the PID compensated (LFC) SISO system is given in \autoref{fig:lfc7}. The output of the PID controller $\Delta u_T(s)$ in the 
frequency domain is given by
\begin{equation}
          \Delta u_T(s)=K_{p1}\Delta e_T(s)+\frac{K_{i1}}{s}\Delta e_T(s)+K_{d1}s\Delta e_T(s)
 \label{eq:pid1}
\end{equation}
where $K_{p1}$ is the proportional gain, $K_{i1}$ is the integral gain, $K_{d1}$ is the derivative gain, and $\Delta e_T(s)$ is the error
signal.
Expressing \autoref{eq:pid1} as a transfer function
\begin{equation}
\begin{aligned}
         \frac{\Delta u_T(s)}{\Delta e_T(s)} &= K_{p1}+\frac{K_{i1}}{s}+K_{d1}s\\
                                             &= \frac{K_{d1}s^2+K_{p1}s+K_{i1}}{s}\\
 \end{aligned}                                            
 \label{eq:pid2}
\end{equation}         
From \autoref{fig:lfc7} the open loop transfer function of the PID compensated (LFC) SISO system can be written as
\begin{equation}
\begin{aligned}
          G_{PID1}(s) &= \bigg{(}\frac{K_{d1}s^2+K_{p1}s+K_{i1}}{s}\bigg{)}\frac{G(s)}{1+G(s)H(s)}\\
                     &=\bigg{(}\frac{K_{d1}s^2+K_{p1}s+K_{i1}}{s}\bigg{)}\bigg{(} 
                       \frac{g_{55}f_{42}f_{28}s}{(s-f_{52})(s-f_{41})(s^2-f_{27}s-A_{23})-f_{51}f_{42}f_{28}s}\bigg{)}\\
          H_{PID1}(s) &= 1\\
  \therefore \   G_{PID1}(s)H_{PID1}(s) &= \bigg{(}\frac{K_{d1}s^2+K_{p1}s+K_{i1}}{s}\bigg{)}\bigg{(} 
                       \frac{g_{55}f_{42}f_{28}s}{(s-f_{52})(s-f_{41})(s^2-f_{27}s-A_{23})-f_{51}f_{42}f_{28}s}\bigg{)}\\   
\end{aligned}
\label{eq:pid3}
\end{equation}
The PID gains are tuned to $K_{p1}=200$, $K_{i1}=150$, and $K_{d1}=100$. Substituting these PID gains and 
numerical values for the constant coefficients in \autoref{eq:pid3} we get
\begin{equation}
          G_{PID1}(s)H_{PID1}(s) = \frac{52.74 s^3 + 105.5 s^2 + 79.11 s}{5 s^5 + 35 s^4 + 51.53 s^3 + 10.69 s^2 + 15.27 s}
\label{eq:pid4}
\end{equation}          
\autoref{fig:lfc8} and \autoref{fig:lfc9} show the plots for $\Delta \delta $(s) and $\Delta \omega $(s) vs time for the PID compensated (LFC) SISO system, respectively. Also, \autoref{fig:lfc10} shows the root locus plot for the PID compensated (LFC) SISO system. From this 
plot it is evident that the PID compensated (LFC) SISO system is stable since the root locus lies entirely in the left half s-plane.

\newpage
\subsection{AVR Dynamics}

\begin{figure}
          \centering
          \includegraphics[trim=9cm 3cm 9cm 3cm, clip=true, totalheight=0.7\textheight, width=0.13\textwidth, angle=90]{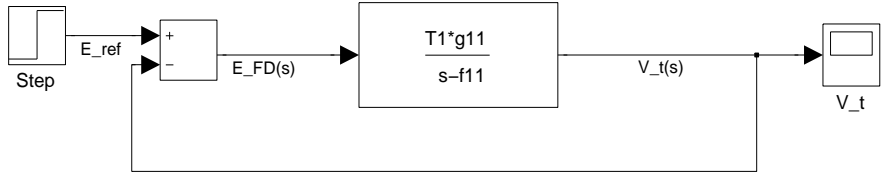}
          \caption{Simulink model of the uncompensated (AVR) SISO system for a step input}
          \label{fig:avr1}  
\end{figure}

\begin{figure}
          \centering      
          \includegraphics[trim=0cm 0cm 0cm 0cm, clip=true, totalheight=0.25\textheight, width=0.5\textwidth]{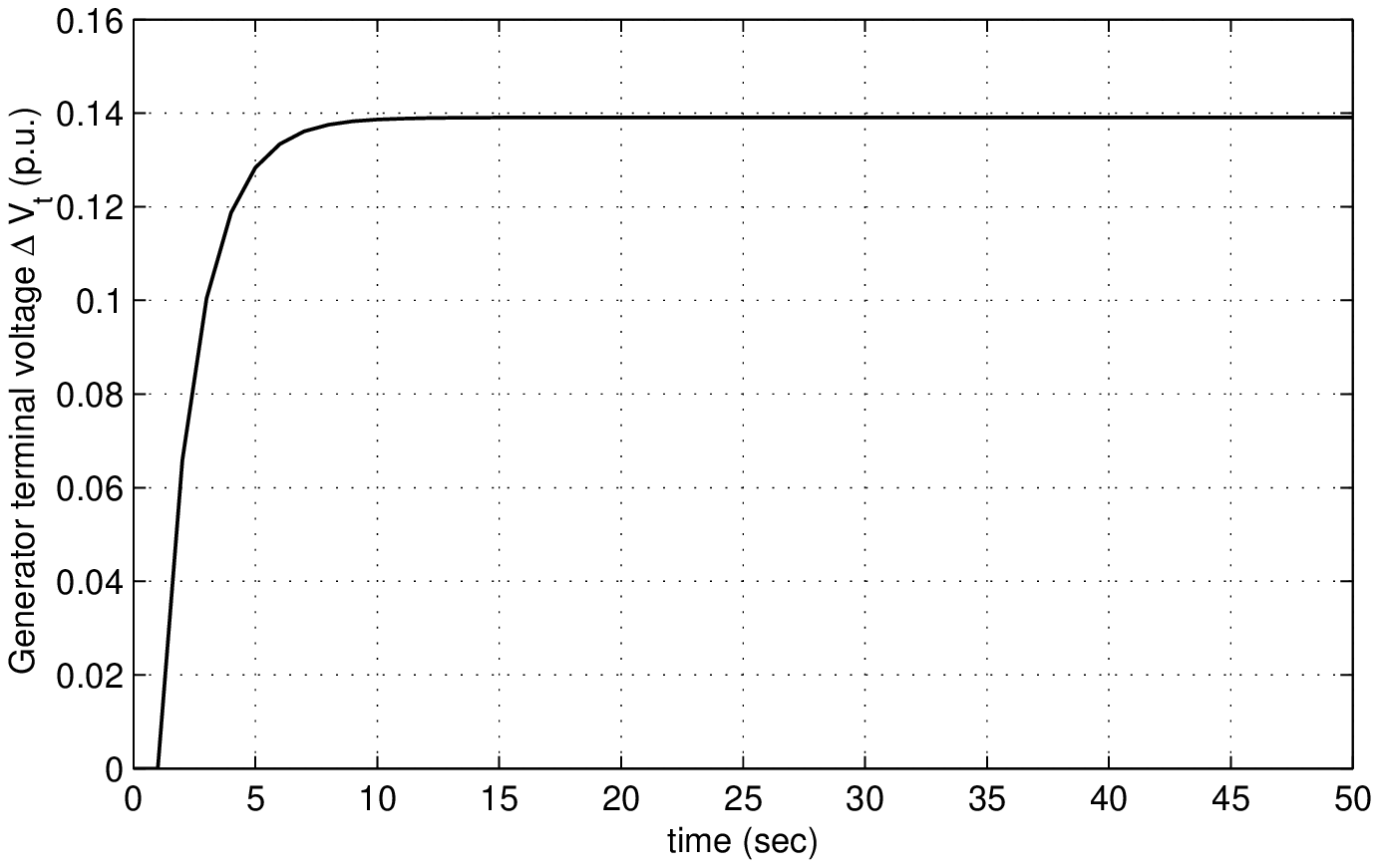}
          \caption{Plot of the terminal voltage $\Delta V_t$(s) vs time for the uncompensated (AVR) SISO system for a step input}
          \label{fig:avr2}          
          \includegraphics[trim=0cm 0cm 0cm 0cm, clip=true, totalheight=0.25\textheight, width=0.5\textwidth]{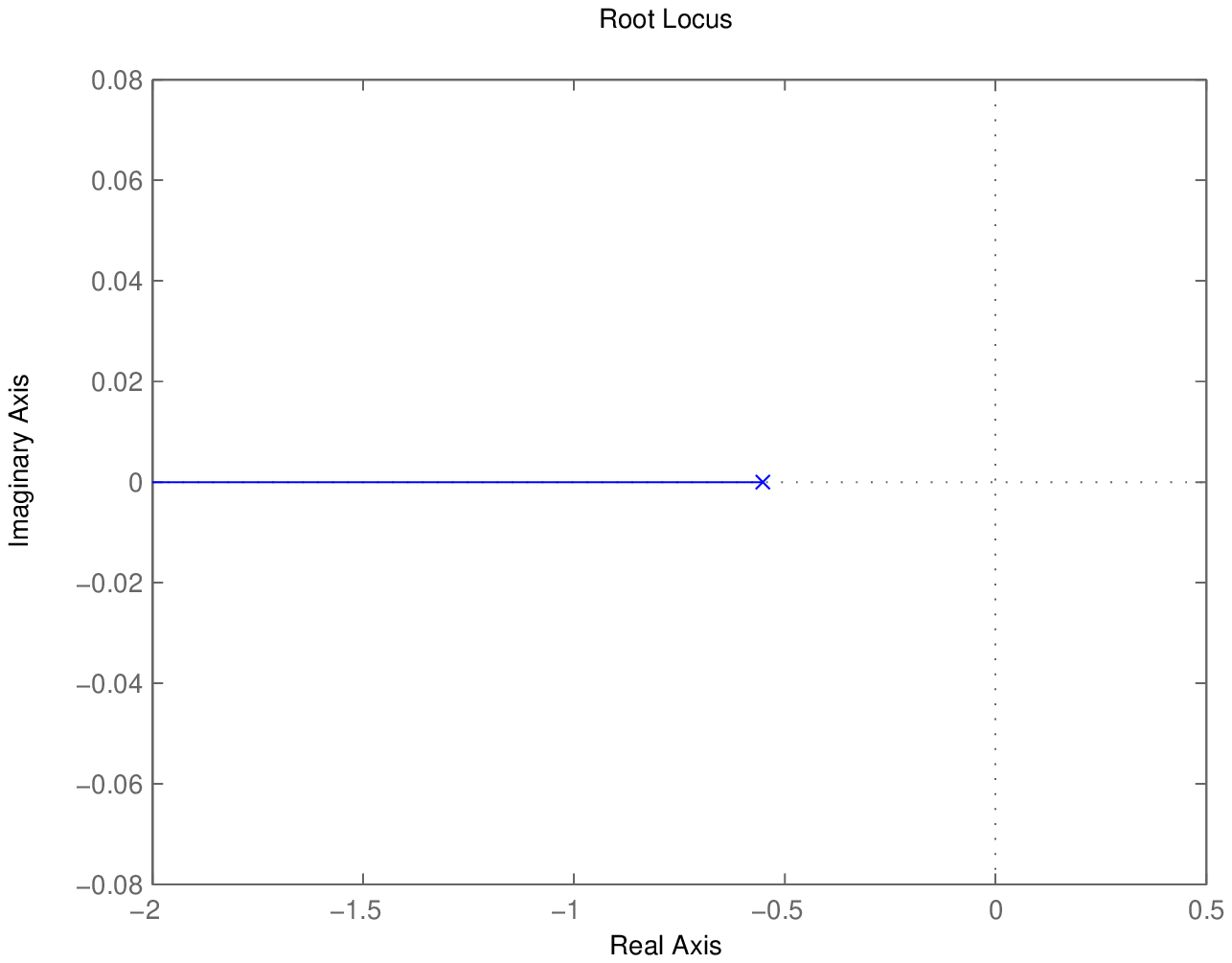}
          \caption{Root locus plot for the uncompensated (AVR) SISO system}
          \label{fig:avr3}   
\end{figure}

In this section we present the dynamics of the (AVR) SISO system with control input $\Delta E_{FD}$. In addition we also design a PID controller for the (AVR) SISO system. 
From \autoref{eq:decouple3} the dynamics of the (AVR) SISO system are
\begin{equation}
         \Delta \dot{E}'_q = f_{11}\Delta E'_q+g_{11}\Delta E_{FD}\\
         \label{eq:avr1}
\end{equation}
Taking the Laplace transform of \autoref{eq:avr1} assuming zero initial conditions we get
\begin{equation}
         s\Delta E'_q(s)=f_{11}\Delta E'_q(s)+g_{11}\Delta E_{FD}(s)
\label{eq:avr2}
\end{equation} 
Rearranging and expressing \autoref{eq:avr2} as a transfer function
\begin{equation}
          \frac{\Delta E'_q(s)}{\Delta E_{FD}(s)}=\frac{g_{11}}{s-f_{11}}
\label{eq:avr3}
\end{equation}
From \autoref{eq:gt17} the output terminal voltage $\Delta V_t$ is 
 \begin{equation}
                 \Delta V_t=T_1\Delta E'_q+T_2\Delta \delta
\label{eq:avr4}
\end{equation}
For the decoupled (AVR) SISO system with $\Delta E_{FD}$ as the input and $\Delta V_t$ as the output the coupling
$T_2\Delta \delta$ between the (LFC) SISO system and the (AVR) SISO system is neglected.
Therefore,
\begin{equation}
                 \Delta V_t=T_1\Delta E'_q
\label{eq:avr5}
\end{equation}
Taking the Laplace transform of \autoref{eq:avr5} 
\begin{equation}
                 \Delta V_t(s)=T_1\Delta E'_q(s)
\label{eq:avr6}
\end{equation} 
Combining \autoref{eq:avr3} and \autoref{eq:avr6} we get
\begin{equation}
                 \frac{\Delta V_t(s)}{\Delta E_{FD}(s)}=\frac{T_1g_{11}}{s-f_{11}}
\label{eq:avr7}
\end{equation} 
Substituting numerical values for the coefficients in \autoref{eq:avr7} evaluated at the nominal operating point we get
\begin{equation}
                 \frac{\Delta V_t(s)}{\Delta E_{FD}(s)}=\frac{0.08913}{s+0.5517}
\label{eq:avr8}
\end{equation}
\autoref{fig:avr1} shows the simulink model of the uncompensated (AVR) SISO system for a step input. In \autoref{fig:avr1} let 
\begin{equation}
       G_1(s)=\frac{T_1g_{11}}{s-f_{11}}
\label{eq:avrstep1}
\end{equation}
and
\begin{equation}
       H_1(s)=1
\label{eq:avrstep2}
\end{equation}       
Therefore, the open loop transfer function of the uncompensated (AVR) SISO system shown in \autoref{fig:avr1} is
\begin{equation}
       G_1(s)H_1(s)=\frac{T_1g_{11}}{s-f_{11}}
\label{eq:avrstep3}
\end{equation}
The closed loop transfer function of the uncompensated (AVR) SISO system shown in \autoref{fig:avr1} is
\begin{equation}
\begin{aligned}
        \frac{\Delta V_t(s)}{\Delta E_{ref}(s)} &=\frac{G_1(s)}{1+G_1(s)H_1(s)}=\frac{\frac{T_1g_{11}}{s-f_{11}}}
        {1+\frac{T_1g_{11}}{s-f_{11}}}\\
                &= \frac{T_1g_{11}}{s-f_{11}+T_1g_{11}}
  \end{aligned}              
\label{eq:avrstep4}
\end{equation}
For a step input, $\Delta E_{ref}(s)=\frac{1}{s}$. From the final value theorem, the steady state value of $\Delta V_t (s)$
is 
\begin{equation}
\begin{aligned}
                \Delta V_{t(ss)} &=  \lim_{s \to 0} s\Delta V_t (s)\\
                                    &= \lim_{s \to 0} s\Delta E_{ref}(s)\bigg{(}\frac{T_1g_{11}}{s-f_{11}+T_1g_{11}}\bigg{)}\\
                                    &= \frac{T_1g_{11}}{T_1g_{11}-f_{11}}
 \end{aligned}              
\label{eq:avrstep5}
\end{equation}
Substituting numerical values for the coefficients in \autoref{eq:avrstep5}
\begin{equation}
                \Delta V_{t(ss)}=0.1391
\label{eq:avrstep6}
\end{equation} 
\autoref{fig:avr2} shows the terminal voltage $\Delta V_t$(s) vs time plot for the uncompensated (AVR) SISO system for a step input.
From this plot we can see that $\Delta V_t$(s) settles to a steady state value of 0.1391 which is in agreement with the steady state
value calculated in \autoref{eq:avrstep6}. \autoref{fig:avr3} shows the root locus plot for the uncompensated (AVR) SISO system. 
From this plot it is evident that the uncompensated (AVR) SISO system is stable. 
 
\newpage 
\subsubsection{PID controller design for the AVR loop}

\begin{figure}
          \centering    
          \includegraphics[trim=9cm 3cm 9cm 3cm, clip=true, totalheight=0.7\textheight, width=0.13\textwidth, angle=90]{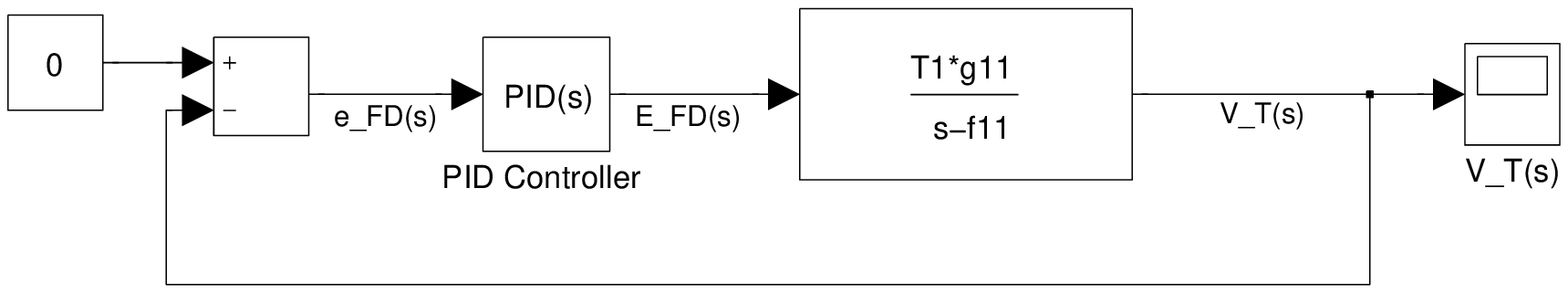}
          \caption{Simulink model of the PID compensated (AVR) SISO system}
          \label{fig:pidavr1} 
          \includegraphics[trim=0cm 0cm 0cm 0cm, clip=true, totalheight=0.2\textheight, width=0.4\textwidth]{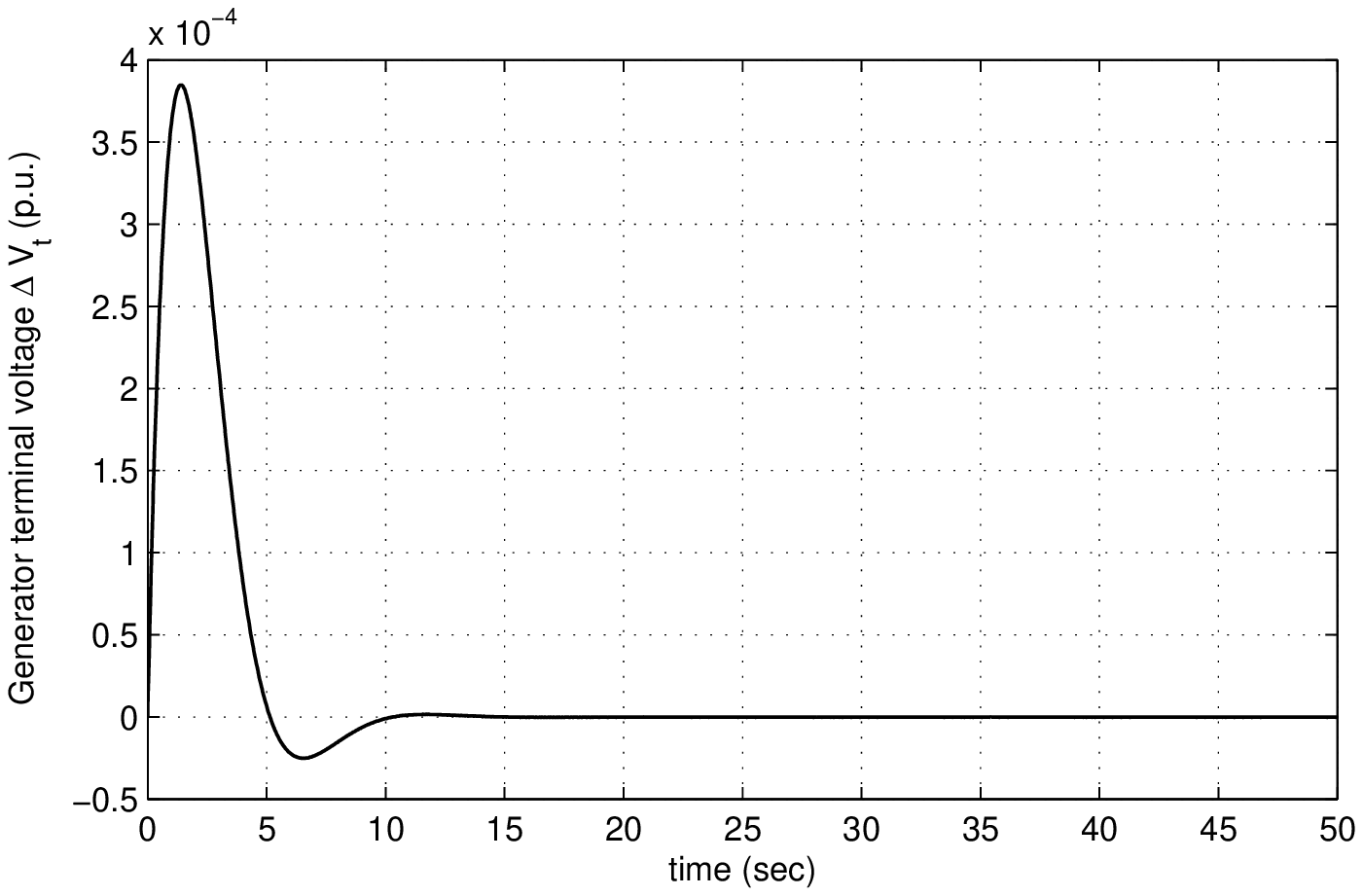}
          \caption{Plot of $\Delta V_t$(s) vs time for the PID compensated (AVR) SISO system}
          \label{fig:pidavr2}
          \includegraphics[trim=0cm 0cm 0cm 0cm, clip=true, totalheight=0.2\textheight, width=0.4\textwidth]{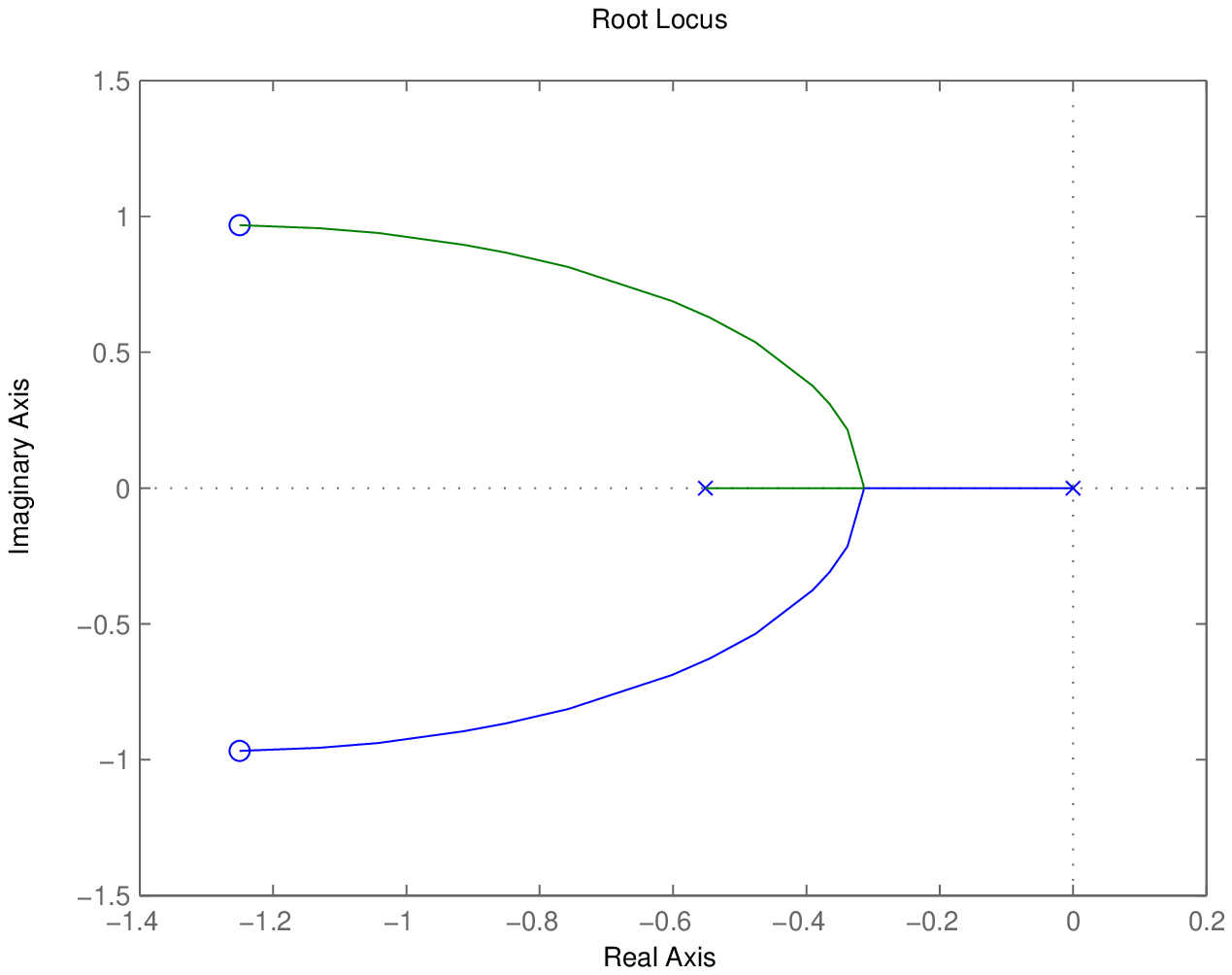}
          \caption{Root locus plot for the PID compensated (AVR) SISO system}
          \label{fig:pidavr3}
\end{figure}

We design a PID controller to improve the transient and steady state response of the (AVR) SISO system.
The output of this PID controller $\Delta E_{FD}(s)$ in the 
frequency domain is given by
\begin{equation}
          \Delta E_{FD}(s)=K_{p2}\Delta e_{FD}(s)+\frac{K_{i2}}{s}\Delta e_{FD}(s)+K_{d2}s\Delta e_{FD}(s)
 \label{eq:avrpid1}
\end{equation}
where $K_{p2}$ is the proportional gain, $K_{i2}$ is the integral gain, $K_{d2}$ is the derivative gain, and $\Delta e_{FD}(s)$ is the error signal.
Expressing \autoref{eq:avrpid1} as a transfer function
\begin{equation}
\begin{aligned}
         \frac{\Delta E_{FD}(s)}{\Delta e_{FD}(s)} &= K_{p2}+\frac{K_{i2}}{s}+K_{d2}s\\
                                             &= \frac{K_{d2}s^2+K_{p2}s+K_{i2}}{s}\\
 \end{aligned}                                            
 \label{eq:avrpid2}
\end{equation}
From \autoref{fig:pidavr1} the open loop transfer function of the PID compensated (AVR) SISO system can be written as
\begin{equation}
\begin{aligned}
      G_{PID2}(s) &= \bigg{(}\frac{K_{d2}s^2+K_{p2}s+K_{i2}}{s}\bigg{)}\bigg{(}\frac{T_1g_{11}}{s-f_{11}}\bigg{)}\\
      H_{PID2}(s) &= 1\\
  \therefore \  G_{PID2}(s)H_{PID2}(s) &= \bigg{(}\frac{K_{d2}s^2+K_{p2}s+K_{i2}}{s}\bigg{)}\bigg{(}\frac{T_1g_{11}}{s-f_{11}}\bigg{)}\\ 
\end{aligned}                                            
 \label{eq:avrpid3}
\end{equation}
The PID gains $K_{p2}=10$, $K_{i2}=10$, and $K_{d2}=4$ are tuned so as to get the best system response. Substituting these PID gains and 
numerical values for the constant coefficients in \autoref{eq:avrpid3} we get
\begin{equation}
          G_{PID2}(s)H_{PID2}(s) = \frac{0.3565 s^2 + 0.8913 s + 0.8913}{s^2 + 0.5517 s}
\label{eq:avrpid4}
\end{equation}
\autoref{fig:pidavr2} shows the generator terminal voltage $\Delta V_t$(s) vs time plot for the PID compensated (AVR) SISO system. 
From this plot we can see that $\Delta V_t$(s) attains its steady state value of 0 in approximately 10 seconds.
\autoref{fig:pidavr3} shows the root locus plot for the PID compensated (AVR) SISO system. The PID compensated (AVR) SISO system
is stable since the root locus lies in the left half s-plane.

\newpage
          
\subsection{LFC and AVR including coupling}

As there is weak coupling between the LFC and AVR subsystems, the rotor angle $\delta $ and thus the frequency $\omega $, and
the terminal voltage $V_t$ were controlled separately. In this section we study the effect of coupling between the 
LFC and the AVR system. In the previous section we saw that
the weak coupling $A_{21}\Delta E'_q$ between the LFC and the AVR loop which 
appeared in the coupled system in \autoref{eq:decouple1} was neglected in the LFC loop in \autoref{eq:decouple2}. Also,
 the weak coupling $A_{13}\Delta \delta$ between the LFC and the AVR loop which 
appeared in the coupled system in \autoref{eq:decouple1} was neglected in the AVR loop in \autoref{eq:decouple3}. In the 
coupled system including the LFC and AVR dynamics we do not neglect these terms.
Thus, from \autoref{eq:decouple1} we can write 

\begin{equation}
        \begin{aligned}
               &\Delta \dot{E}'_q = f_{11}\Delta E'_q+A_{13}\Delta \delta +g_{11}\Delta E_{FD}\\
               &\Delta \dot{\omega } = A_{21}\Delta E'_q+f_{27}\Delta \omega +
               A_{23}\Delta \delta+f_{28}\Delta T_m\\
               &\Delta \dot{\delta } = \Delta \omega \\
\end{aligned}
\label{eq:couple1}
\end{equation}
Taking the Laplace transform of the expression for $\Delta \dot{E}'_q$ in \autoref{eq:couple1} 
\begin{equation}
           s\Delta E'_q(s)=f_{11}\Delta E'_q(s)+A_{13}\Delta \delta (s)+g_{11}\Delta E_{FD}(s)
\label{eq:couple2}
\end{equation}
Rearranging \autoref{eq:couple2} we get
\begin{equation}
         \Delta E'_q(s)=\frac{A_{13}}{s-f_{11}}\Delta \delta (s)+\frac{g_{11}}{s-f_{11}}\Delta E_{FD}(s)
\label{eq:couple3}
\end{equation}
Also from \autoref{eq:gt17} the first output which is the terminal voltage $V_t$ of the synchronous generator for the coupled system including LFC and AVR dynamics is         
 \begin{equation}
                 \Delta  V_t=T_1\Delta  E'_q+T_2\Delta \delta
\label{eq:couple4}
\end{equation}
The Laplace transform of \autoref{eq:couple4} gives
\begin{equation}
                 \Delta  V_t(s)=T_1\Delta  E'_q(s)+T_2\Delta \delta (s)
\label{eq:couple4point1}
\end{equation}
Next, taking the derivative of $\Delta \dot{\omega }$ we get 
\begin{equation}
        \Delta \ddot{\omega }=A_{21}\Delta \dot{E}'_q+f_{27}\Delta \dot{\omega }+A_{23}\Delta \omega +f_{28}\Delta \dot{T}_m
\label{eq:couple5}
\end{equation} 
\begin{figure}
          \centering
          \includegraphics[trim=4.1cm 1.2cm 4.1cm 1.2cm, clip=true, totalheight=0.5\textheight, width=0.45\textwidth, angle=90]{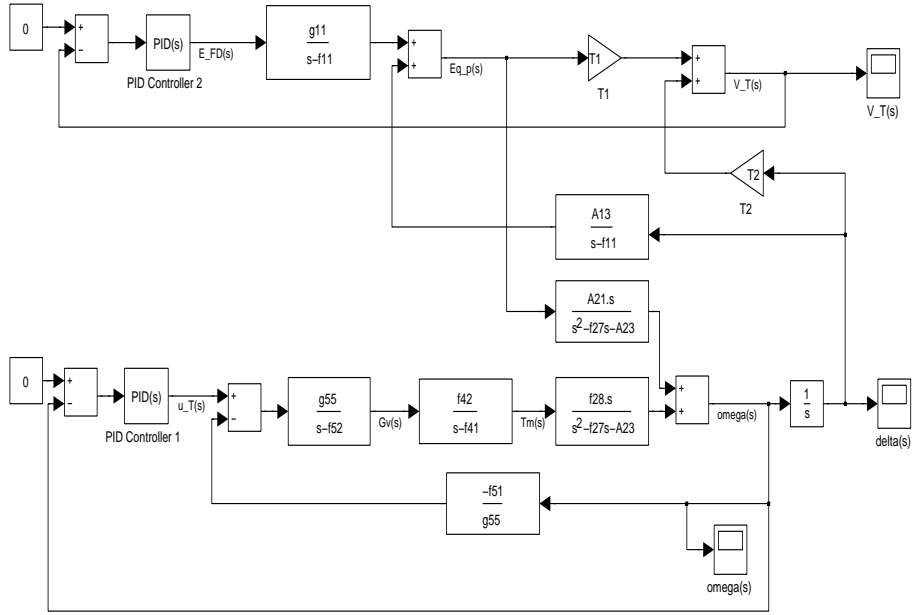}
          \caption{Simulink model of the PID compensated (LFC+AVR) MIMO system including coupling}
          \label{fig:pidlfcavr1}
\end{figure}
Taking the Laplace transform of \autoref{eq:couple5}
\begin{equation}
         s^2\Delta \omega (s)=A_{21}s\Delta E'_q(s)+f_{27}s\Delta \omega (s)+A_{23}\Delta \omega (s)+f_{28}s\Delta T_m(s)
\label{eq:couple6}
\end{equation} 
Rearranging \autoref{eq:couple6} we get
\begin{equation}
              \Delta \omega (s)=\frac{A_{21}s}{s^2-f_{27}s-A_{23}}\Delta E'_q(s)+\frac{f_{28}s}{s^2-f_{27}s-A_{23}}\Delta T_m(s)
\label{eq:couple7}
\end{equation} 
Also the Laplace transform of $\Delta \dot{\delta } = \Delta \omega $ is $s\Delta \delta (s)=\Delta \omega (s)$. 
Thus, the second output which is the rotor angle $\Delta \delta (s)$ of the synchronous generator
for the coupled system including the LFC and AVR dynamics can be written as    
\begin{equation}
              \Delta \delta (s)=\frac{A_{21}}{s^2-f_{27}s-A_{23}}\Delta E'_q(s)+\frac{f_{28}}{s^2-f_{27}s-A_{23}}\Delta T_m(s)
\label{eq:couple8}
\end{equation}

\section{PID Controller Design}

\subsection{PID Controller Design based on linear model}

\begin{figure}
          \centering
          \includegraphics[trim=0cm 0cm 0cm 0cm, clip=true, totalheight=0.26\textheight, width=0.52\textwidth]{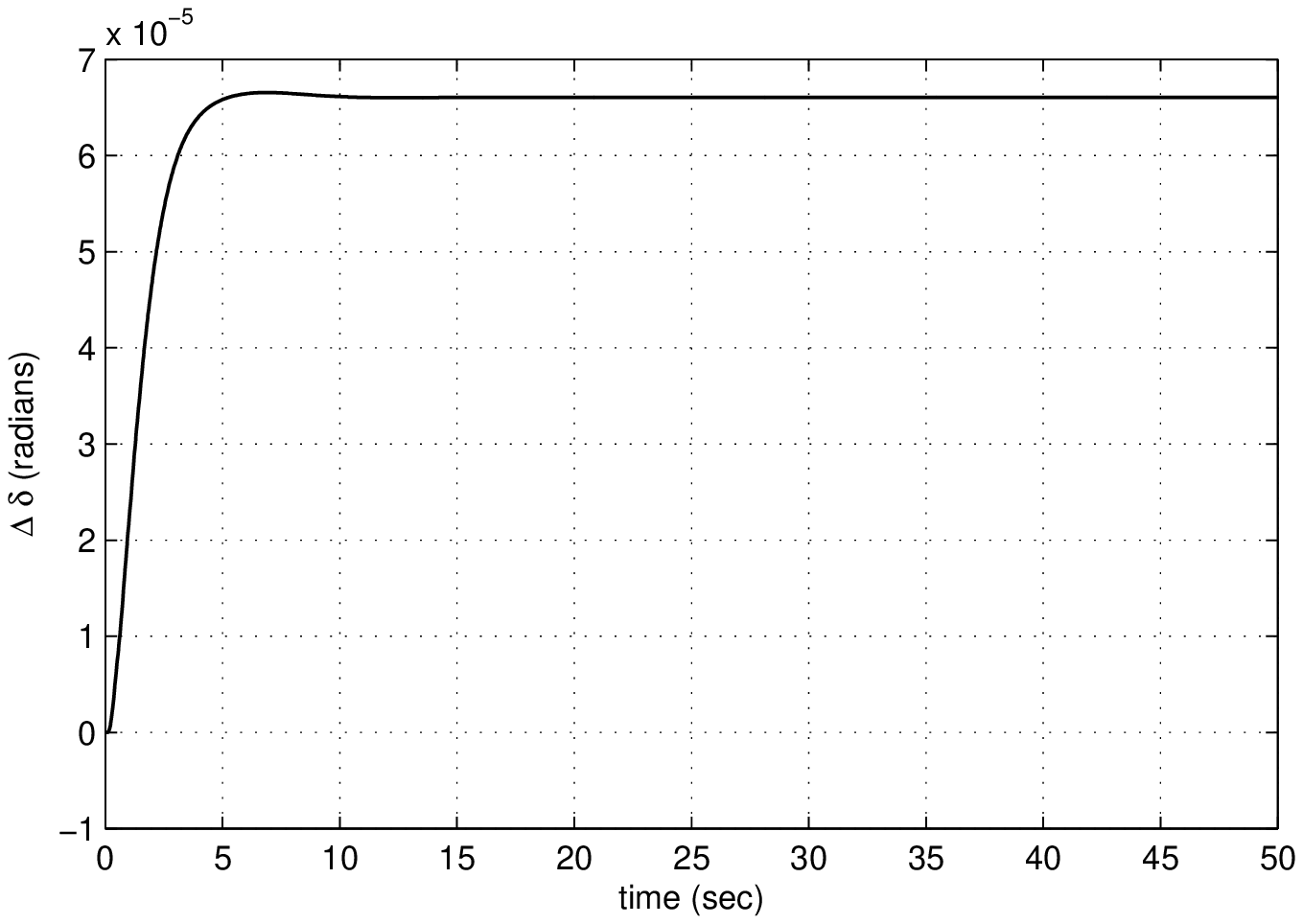}
          \caption{Plot of $\Delta \delta $(s) vs time for the PID compensated (LFC+AVR) MIMO system including coupling}
          \label{fig:pidlfcavr2}
          \includegraphics[trim=0cm 0cm 0cm 0cm, clip=true, totalheight=0.26\textheight, width=0.52\textwidth]{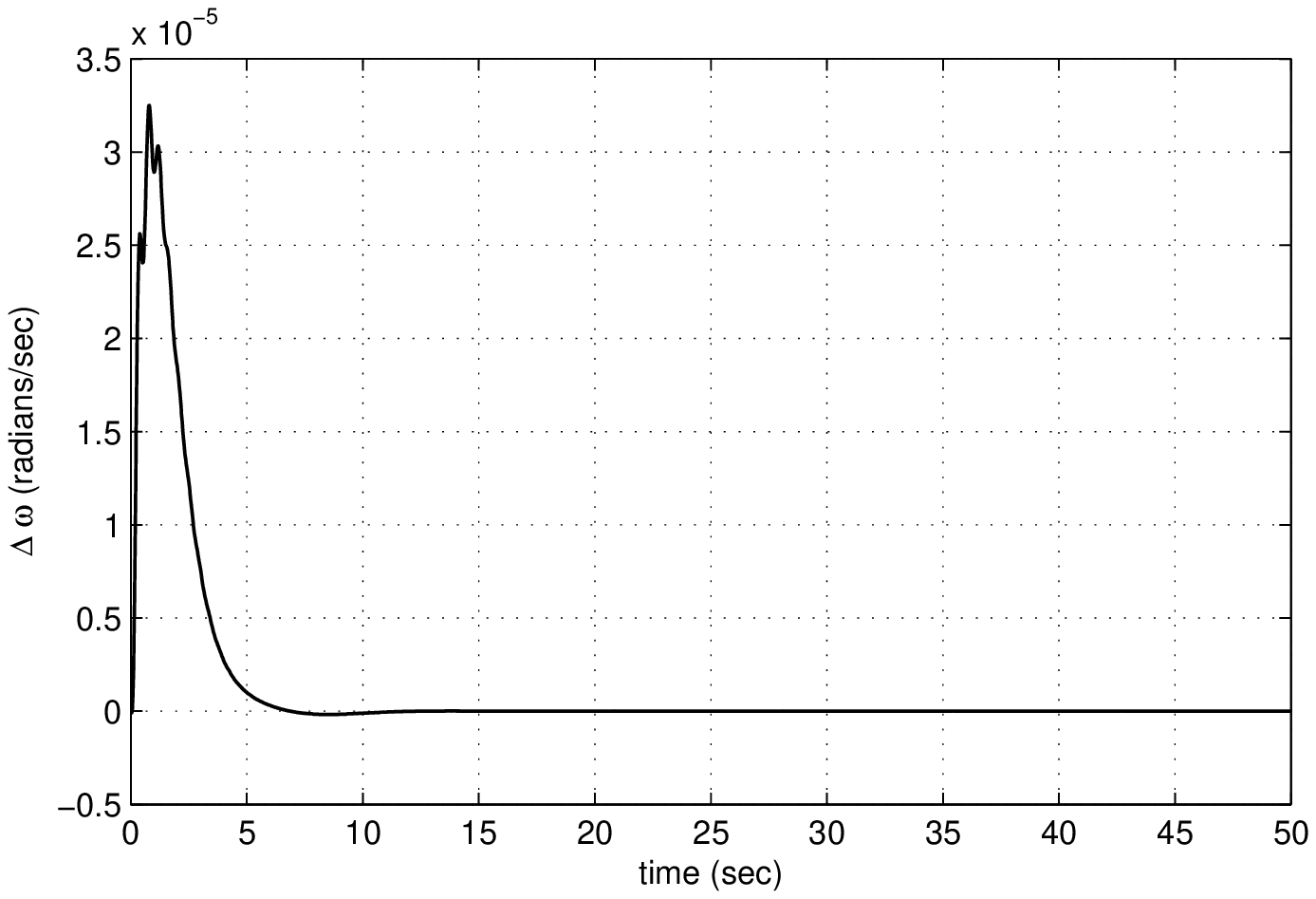}
          \caption{Plot of $\Delta \omega $(s) vs time for the PID compensated (LFC+AVR) MIMO system including coupling}
          \label{fig:pidlfcavr3}
          \includegraphics[trim=0cm 0cm 0cm 0cm, clip=true, totalheight=0.26\textheight, width=0.52\textwidth]{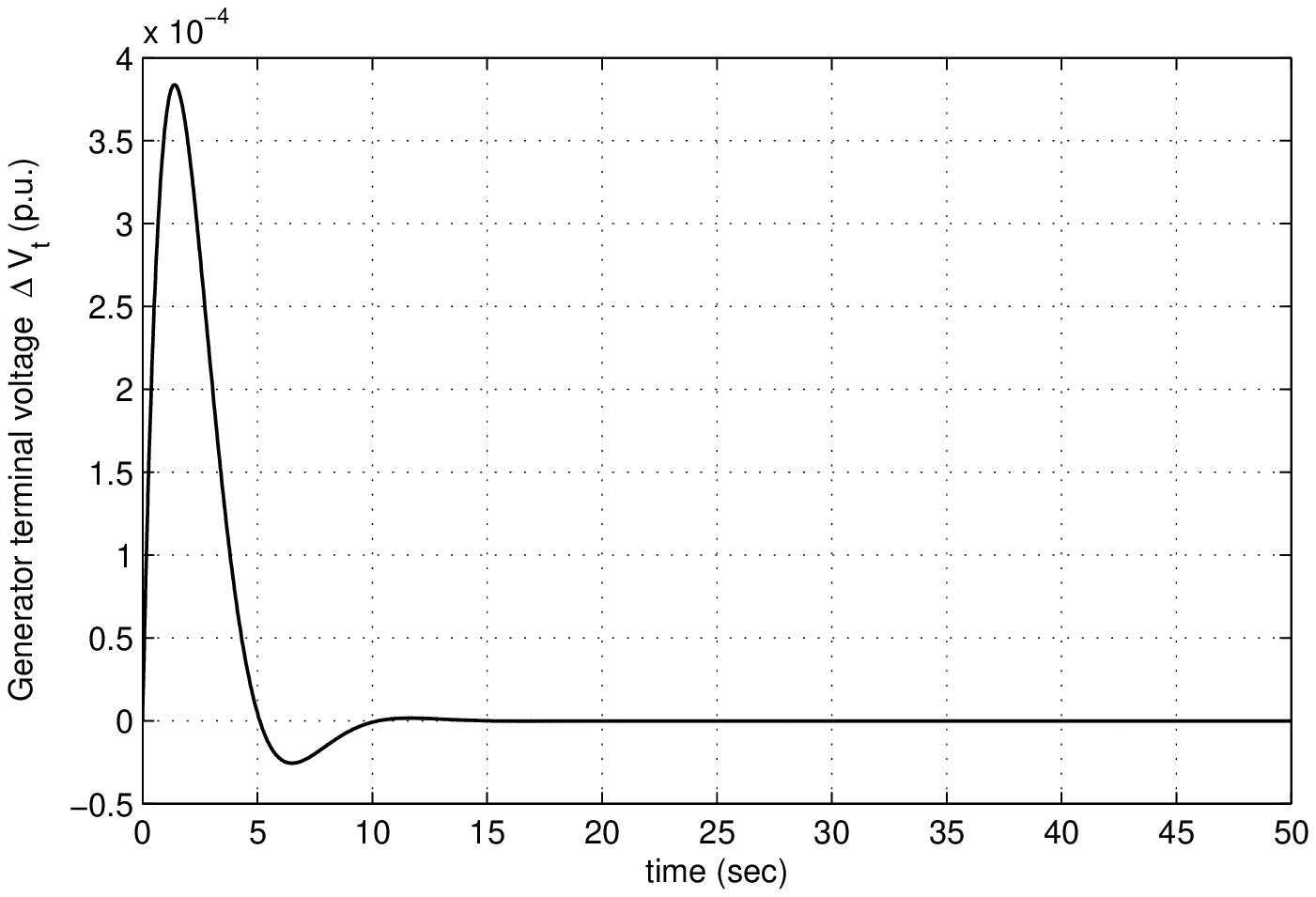}
          \caption{Plot of $\Delta V_t$(s) vs time for the PID compensated (LFC+AVR) MIMO system including coupling}
          \label{fig:pidlfcavr4}
\end{figure}

\begin{figure}
          \centering
          \includegraphics[trim=0cm 0cm 0cm 0cm, clip=true, totalheight=0.26\textheight, width=0.52\textwidth]{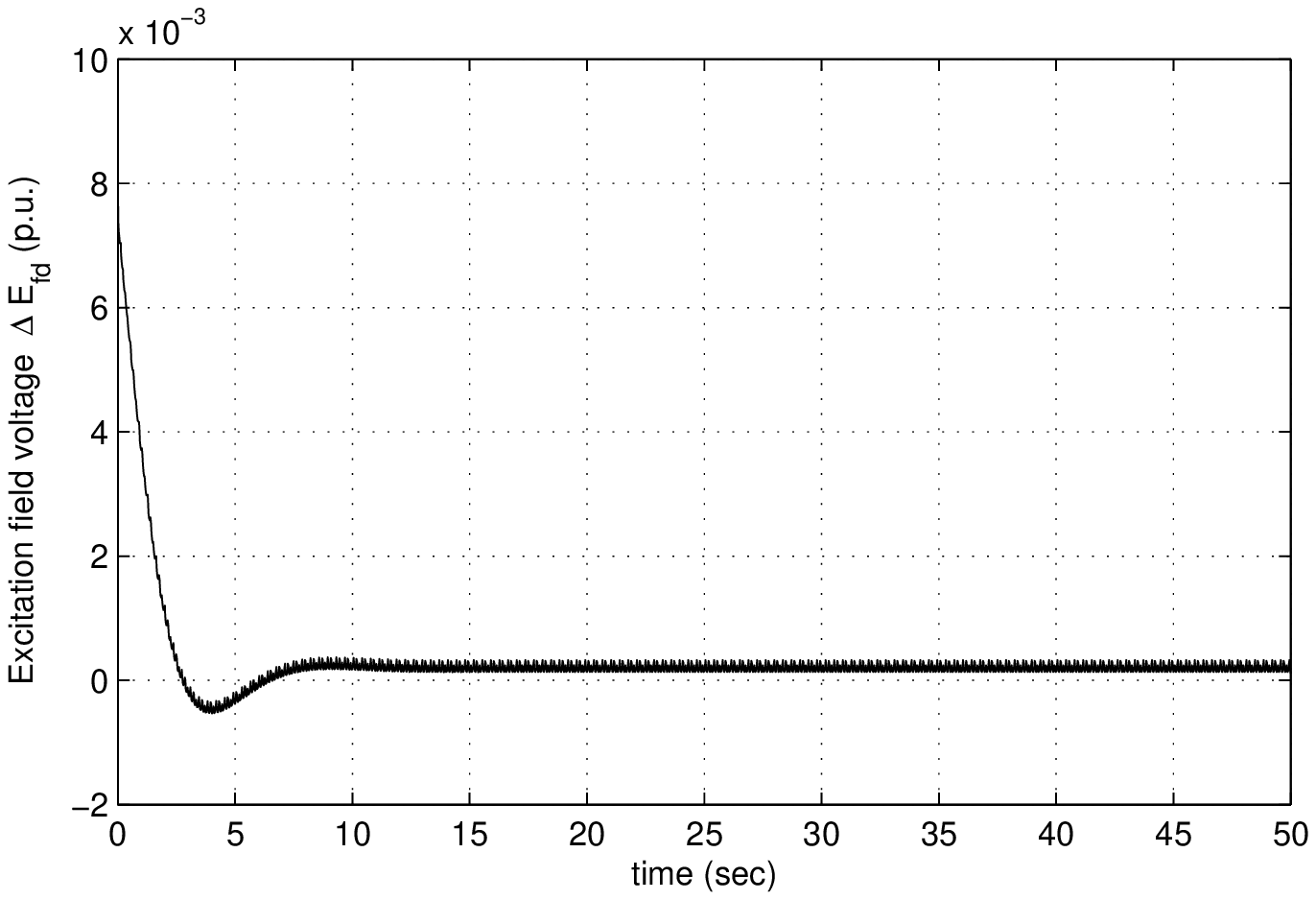}
          \caption{Plot of $\Delta E_{fd} $(s) vs time for the PID compensated (LFC+AVR) MIMO system including coupling}
          \label{fig:pidlfcavr5}
          \includegraphics[trim=0cm 0cm 0cm 0cm, clip=true, totalheight=0.26\textheight, width=0.52\textwidth]{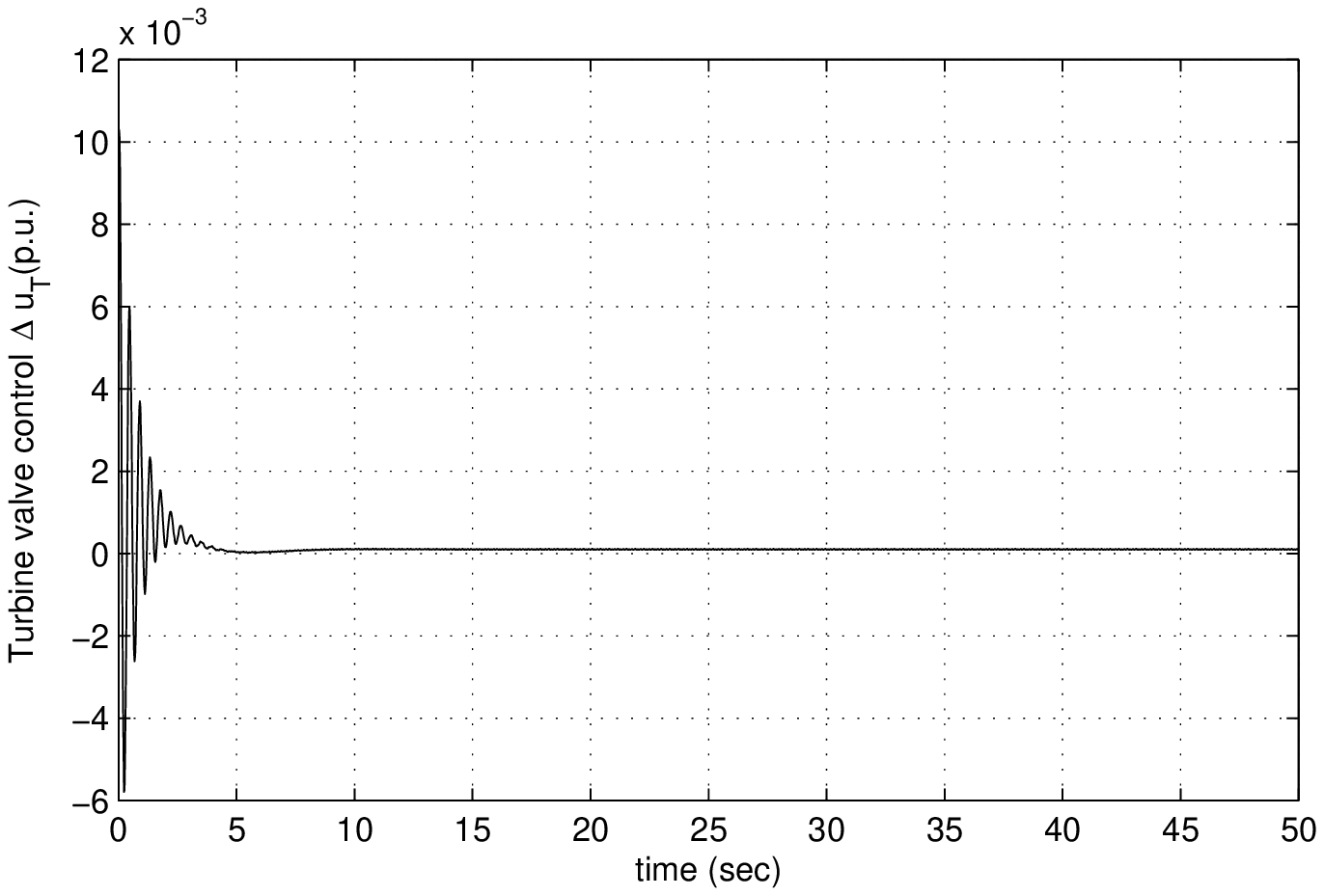}
          \caption{Plot of $\Delta u_T $(s) vs time for the PID compensated (LFC+AVR) MIMO system including coupling}
          \label{fig:pidlfcavr6}
\end{figure}

The two PID controllers that were designed for the LFC and the AVR SISO systems in the previous section, are now tested on the 
combined LFC and AVR system including coupling which is the same as the linear model. A simulink block diagram which consists of the two PID controllers and the combined LFC and AVR system including coupling is constructed in \autoref{fig:pidlfcavr1}. \autoref{fig:pidlfcavr2} to \autoref{fig:pidlfcavr6} show $\Delta \delta $(s), $\Delta \omega $(s), $\Delta V_t$(s), $\Delta E_{fd} $(s), and $\Delta u_T $(s) vs time plots for the PID compensated combined (LFC+AVR) MIMO system including coupling. 
The plots obtained when the coupling coefficients were set to zero i.e. in the decoupled LFC and AVR SISO systems, are
identical to the plots obtained for the combined (LFC+AVR) MIMO system including coupling. Thus, separate frequency and voltage control
of the synchronous generator and turbine-governor system connected to an infinite bus is justified.

\newpage
\subsection{Simulation results for the PID Controllers applied to the Reduced Order Nonlinear Model}

\begin{figure}
          \centering
          \includegraphics[trim=0cm 0cm 0cm 0cm, clip=true, totalheight=0.27\textheight, width=0.54
           \textwidth]  {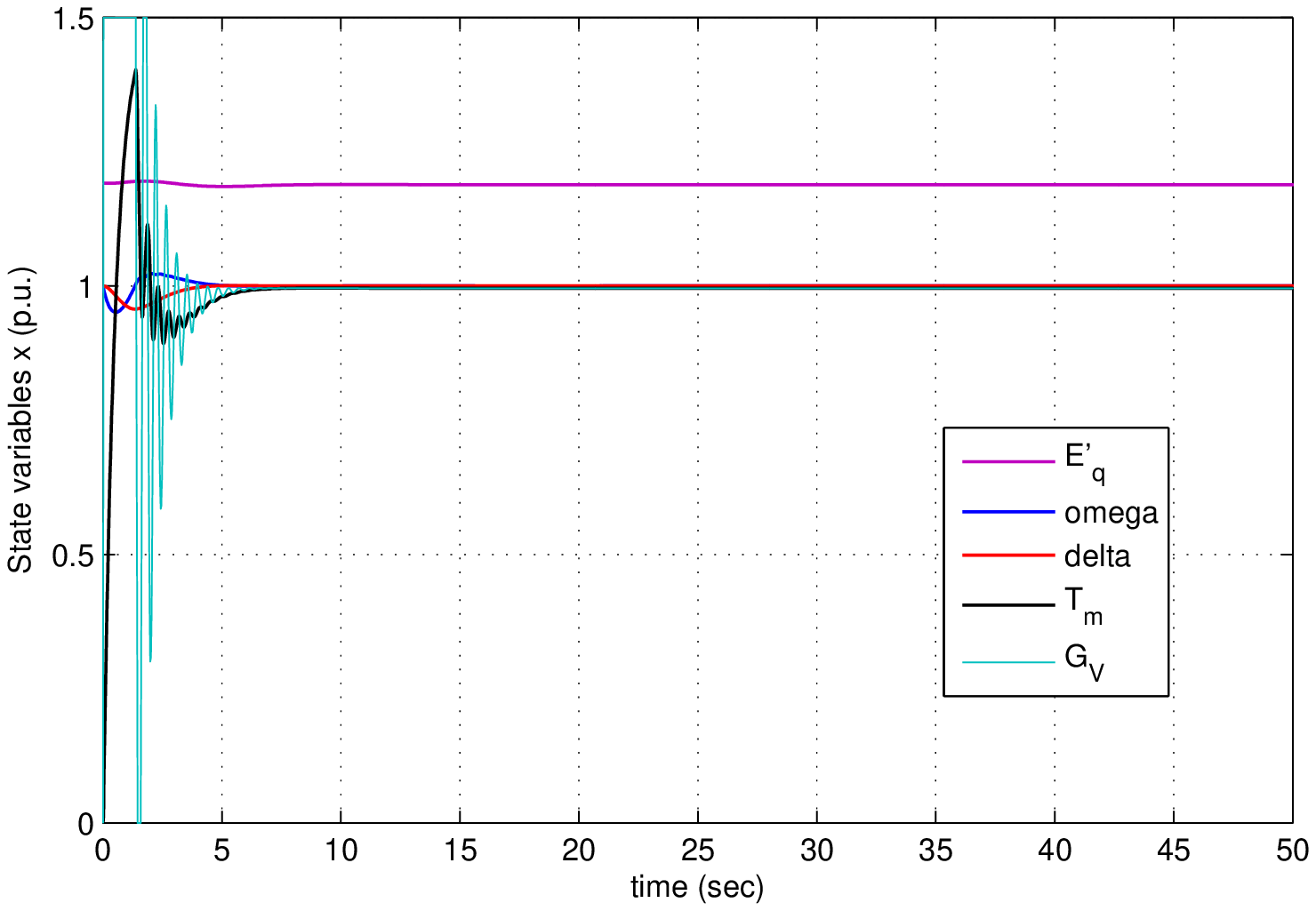}
          \caption{Plot of the state variables $E'_q$, $\omega $, $\delta $, 
           $T_m$, and $G_V$ vs time for the PID controller applied to the reduced order nonlinear model}
          \label{fig:pidnonlineartest1}
          \includegraphics[trim=0cm 0cm 0cm 0cm, clip=true, totalheight=0.27\textheight, 
           width=0.54\textwidth]   {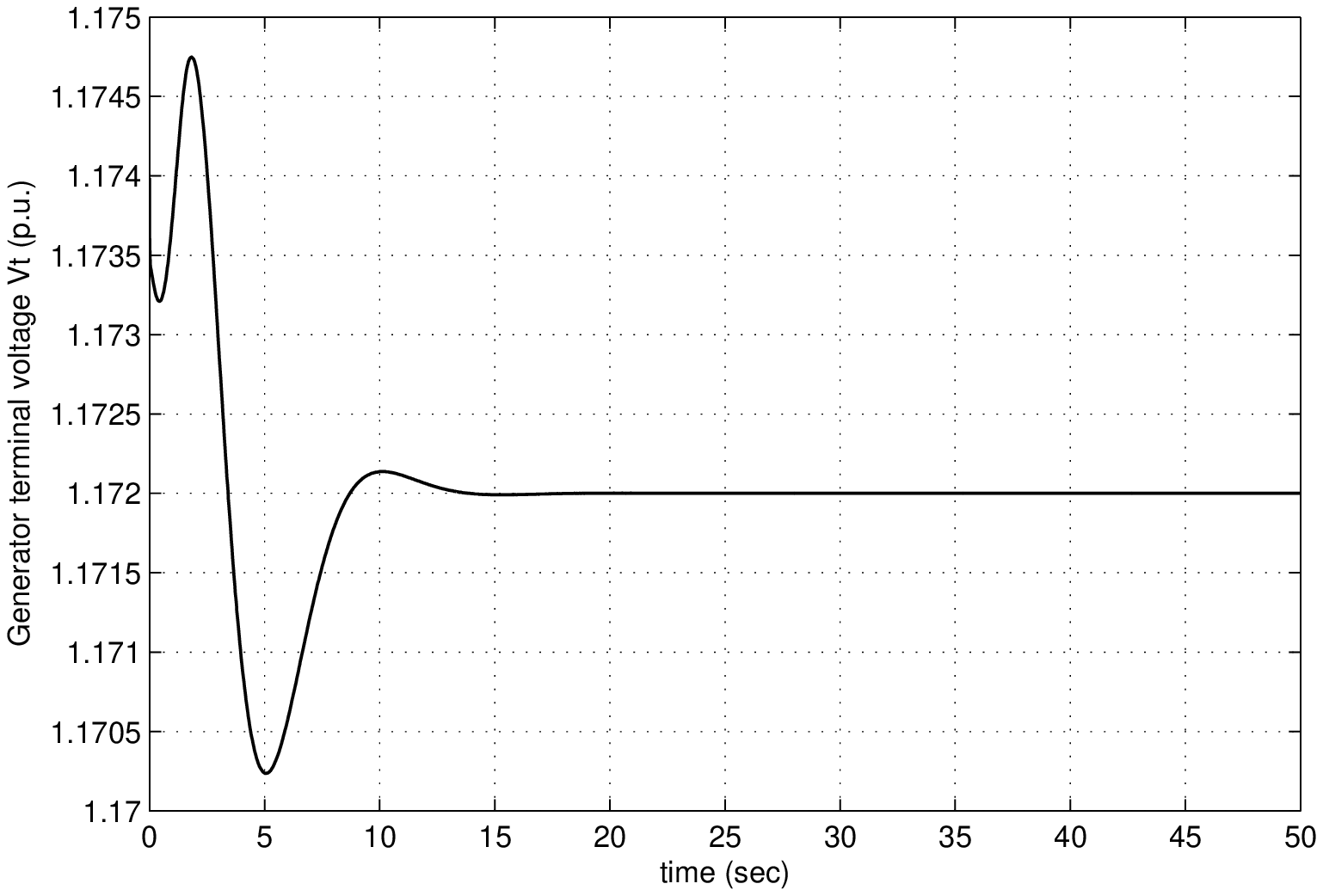}
          \caption{Plot of the generator terminal voltage $V_t$ vs time for the PID controller applied to the reduced 
           order nonlinear  model}
          \label{fig:pidnonlineartest2}
          \includegraphics[trim=0cm 0cm 0cm 0cm, clip=true, totalheight=0.27\textheight, 
           width=0.54\textwidth]{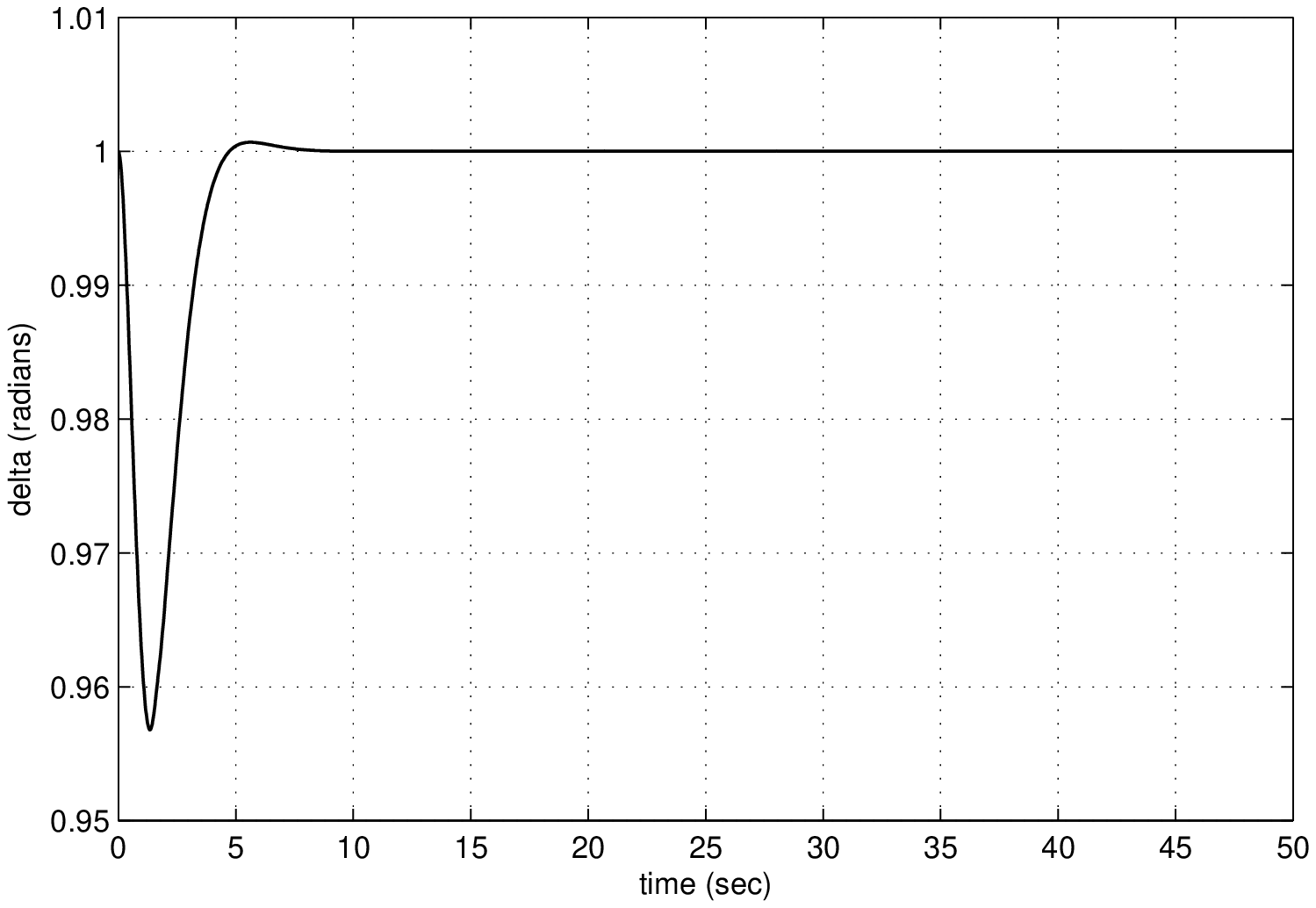}
          \caption{Plot of the rotor angle $\delta $ vs time for the PID controller applied to the reduced order nonlinear model}
          \label{fig:pidnonlineartest3}
\end{figure}

\begin{figure}
          \centering          
          \includegraphics[trim=0cm 0cm 0cm 0cm, clip=true, totalheight=0.27\textheight, 
          width=0.54\textwidth]{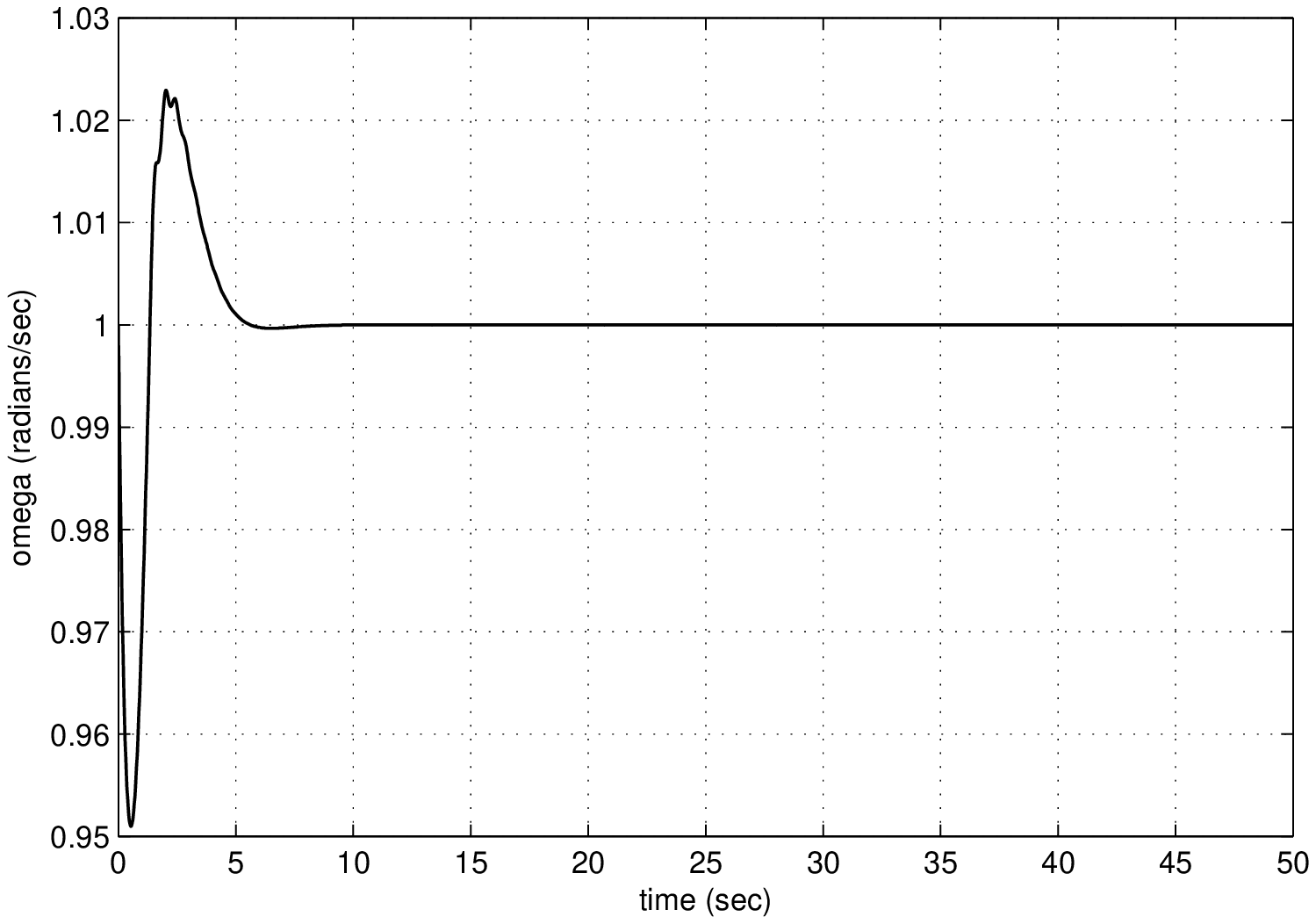}
          \caption{Plot of the frequency $\omega  $ vs time for the PID controller applied to the reduced order nonlinear model}
          \label{fig:pidnonlineartest4}
          \includegraphics[trim=0cm 0cm 0cm 0cm, clip=true, totalheight=0.27\textheight, 
          width=0.54\textwidth]{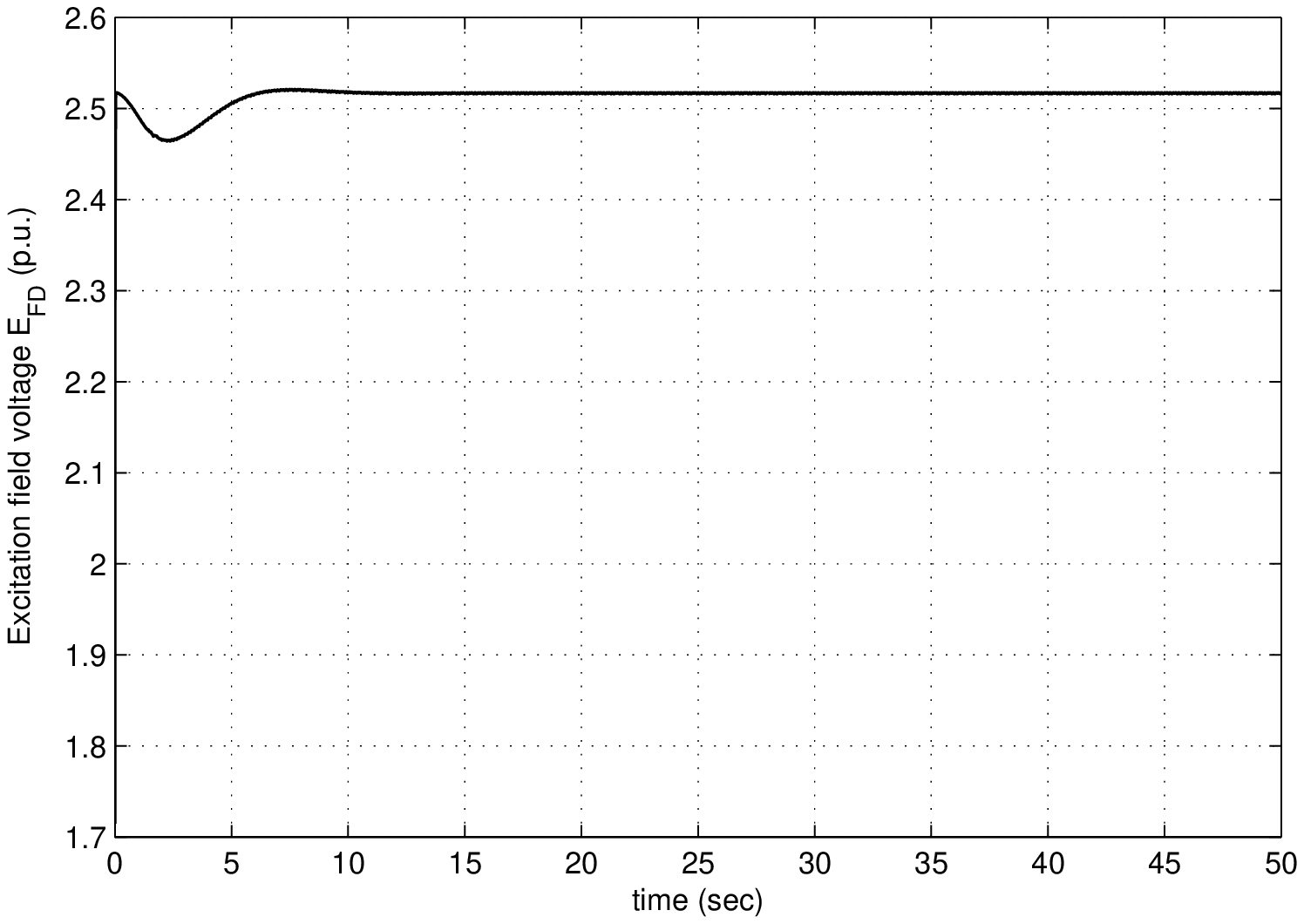}
          \caption{Plot of the generator excitation field control $E_{FD} $ vs time for the PID controller applied to 
          the reduced order nonlinear model}
          \label{fig:pidnonlineartest5}
          \includegraphics[trim=0cm 0cm 0cm 0cm, clip=true, totalheight=0.27\textheight, 
          width=0.54\textwidth]{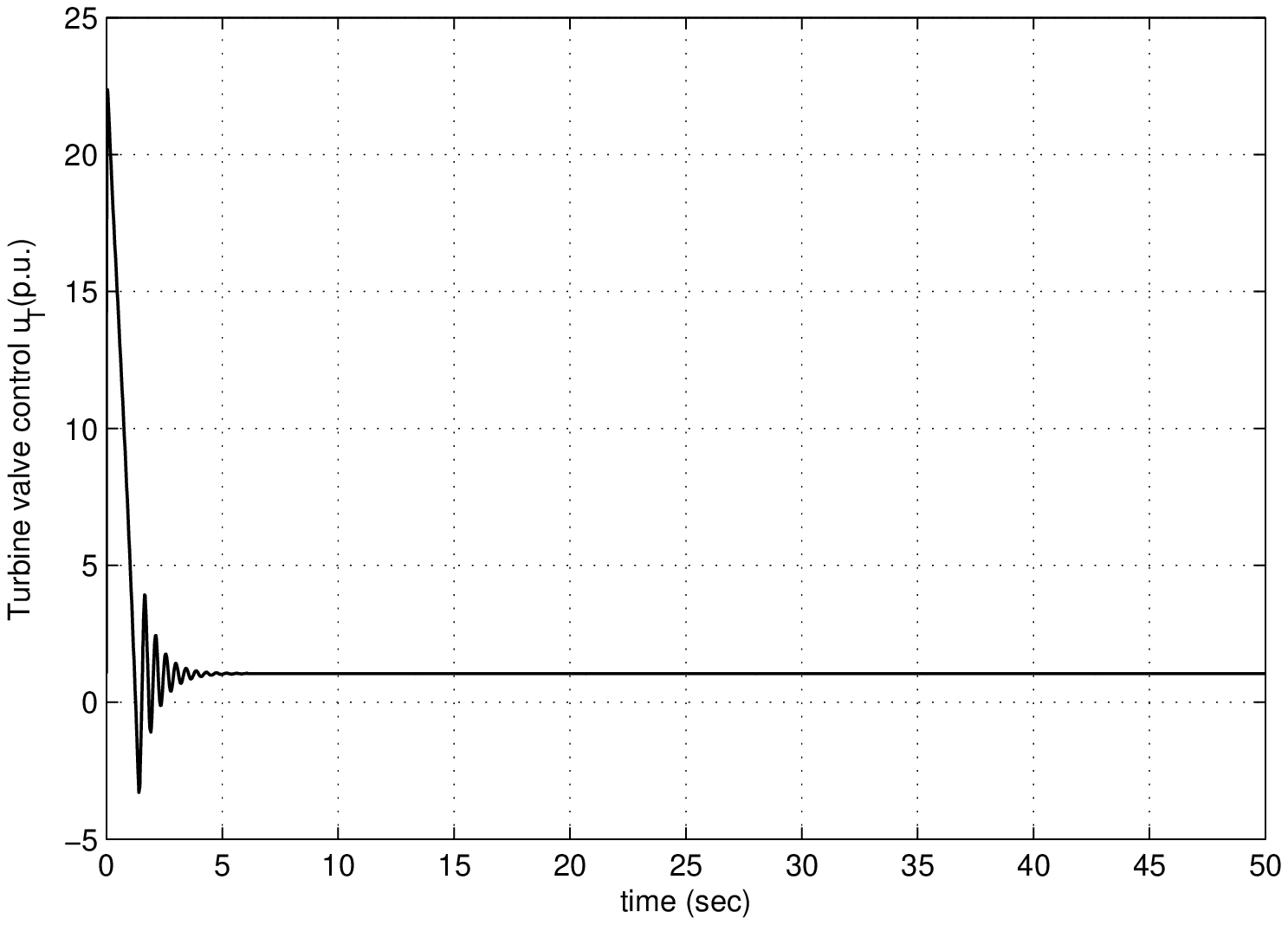}
          \caption{Plot of the turbine valve control $u_T$ vs time for the PID controller applied to the reduced order nonlinear model}
          \label{fig:pidnonlineartest6}
\end{figure}

The two PID controllers that were designed for the decoupled LFC and AVR SISO systems and then tested on the (LFC+AVR) MIMO system including coupling are now tested on the reduced order nonlinear model.
\autoref{fig:pidnonlineartest1} to \autoref{fig:pidnonlineartest6} 
show simulation results for the two PID controllers that are tested on the reduced order nonlinear model. From these
results we can see that all the state variables, outputs, and the control inputs attain their respective steady state 
values in approximately 10 seconds. 

\newpage
\subsection{Simulation results for the PID Controllers applied to the Truth Model}

\begin{figure}
          \centering
          \includegraphics[trim=0cm 0cm 0cm 0cm, clip=true, totalheight=0.27\textheight, width=0.54
           \textwidth]  {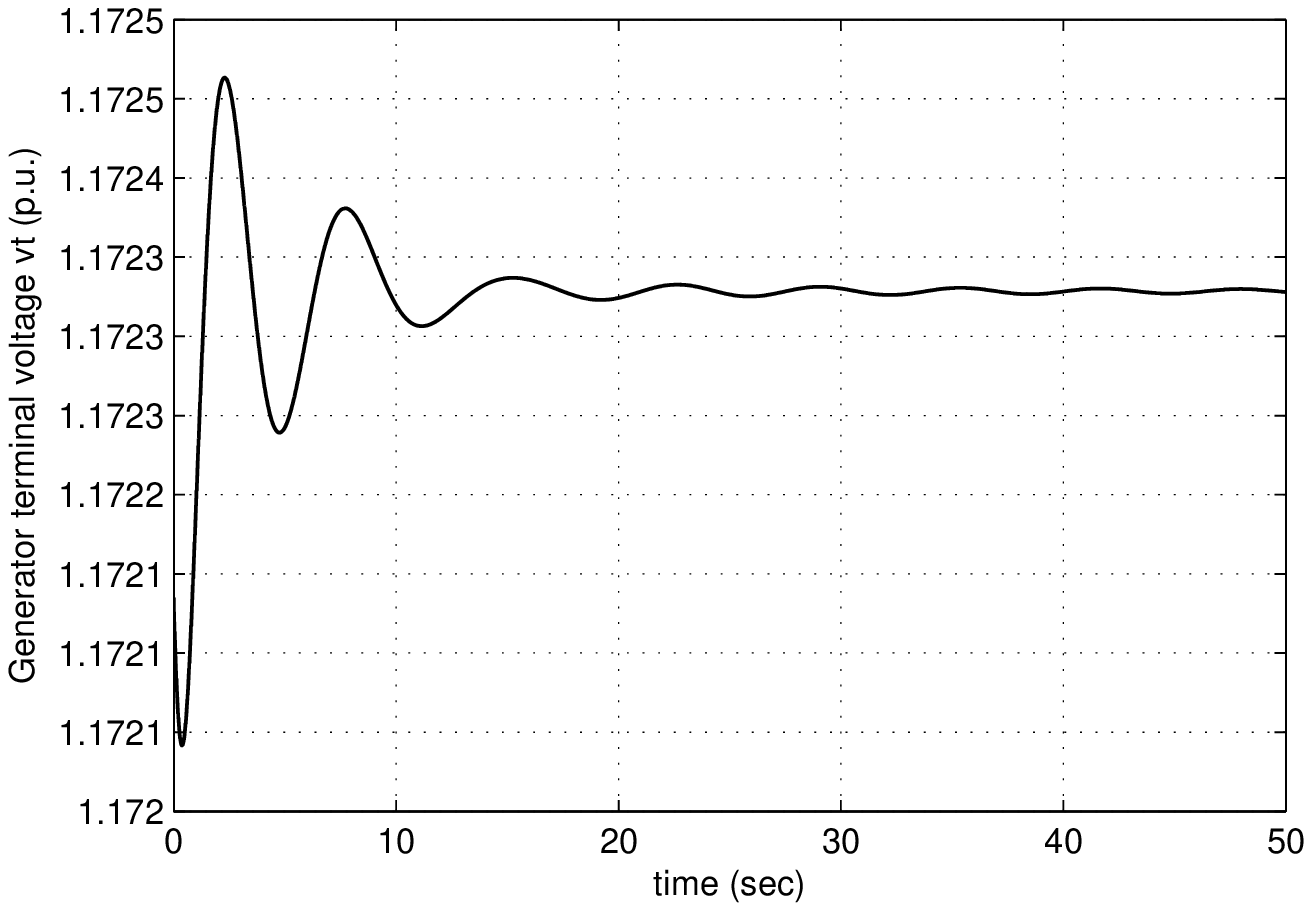}
          \caption{Plot of the generator terminal voltage $V_t$ vs time for the PID controllers applied to the truth model}
          \label{fig:pidtruth1}
          \includegraphics[trim=0cm 0cm 0cm 0cm, clip=true, totalheight=0.27\textheight, width=0.54\textwidth]{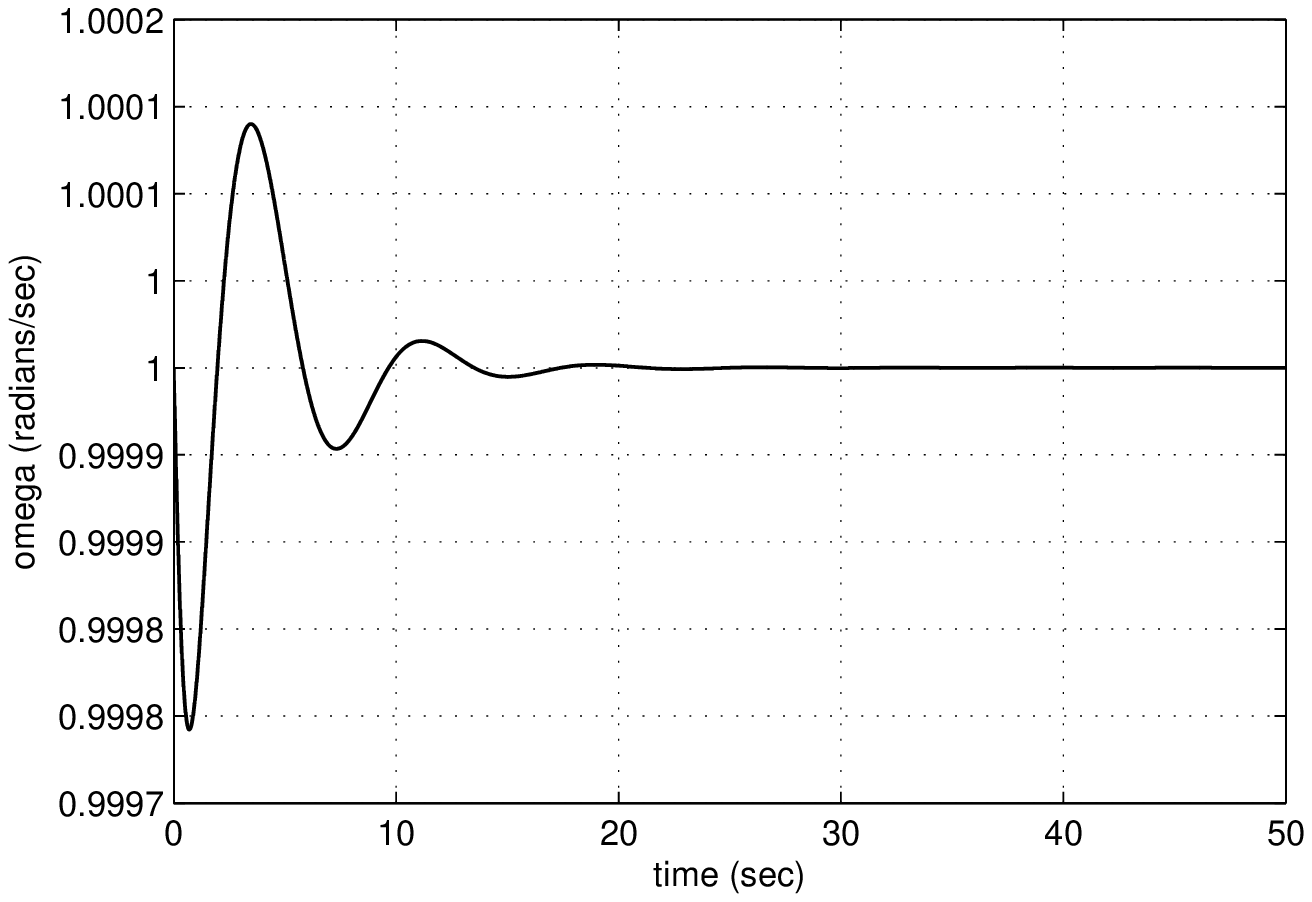}
          \caption{Plot of the angular velocity $\omega $ vs time for the PID controllers applied to the truth model}
          \label{fig:pidtruth2}
          \includegraphics[trim=0cm 0cm 0cm 0cm, clip=true, totalheight=0.27\textheight, width=0.54\textwidth]{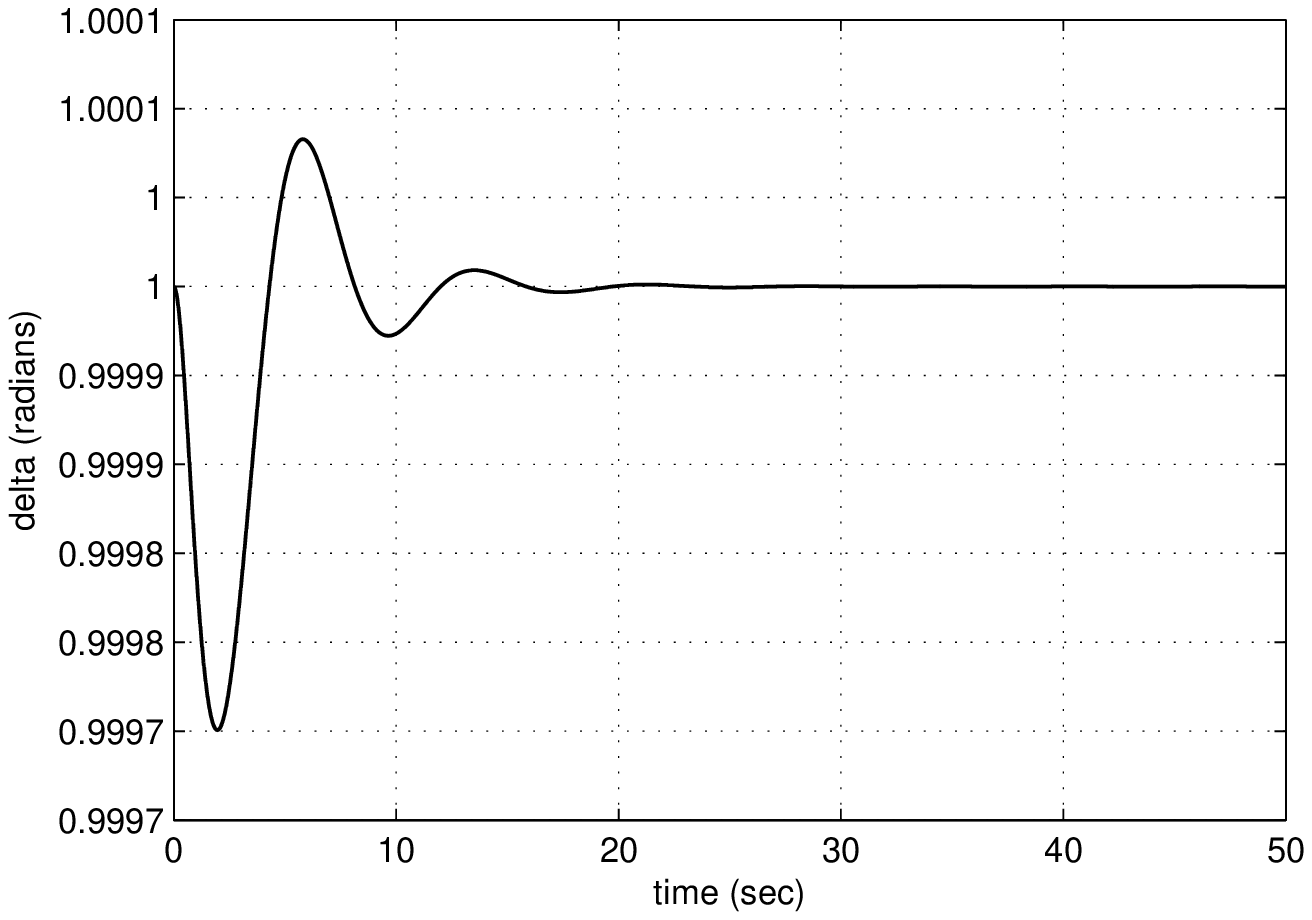}
          \caption{Plot of the rotor angle $\delta $ vs time for the PID controllers applied to the truth model}
          \label{fig:pidtruth3}
\end{figure}

\begin{figure}
          \centering          
          \includegraphics[trim=0cm 0cm 0cm 0cm, clip=true, totalheight=0.27\textheight, width=0.54\textwidth]{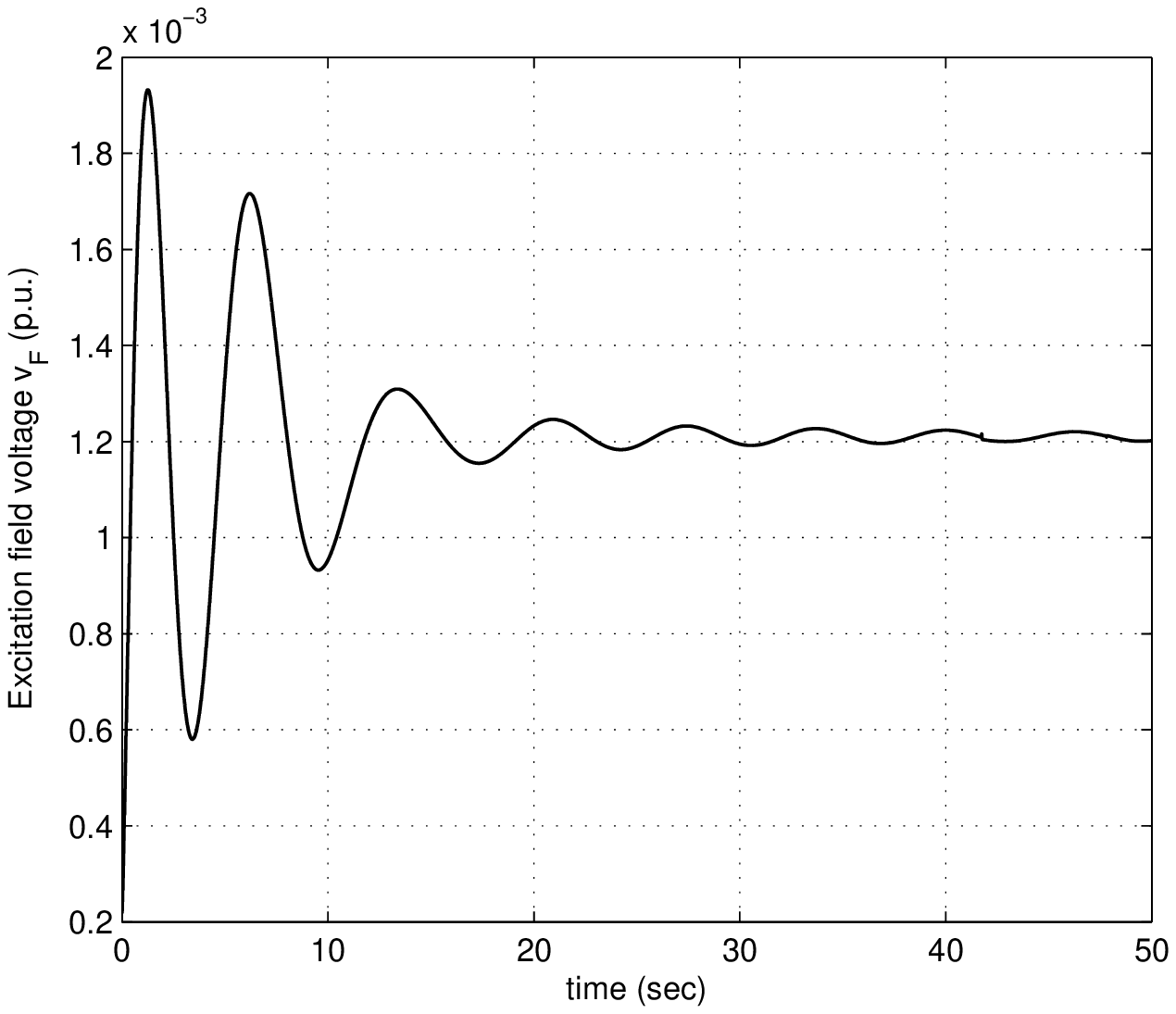}
          \caption{Plot of the control input $V_F$ vs time for the PID controllers applied to the truth model}
          \label{fig:pidtruth4}
          \includegraphics[trim=0cm 0cm 0cm 0cm, clip=true, totalheight=0.27\textheight, width=0.54\textwidth]{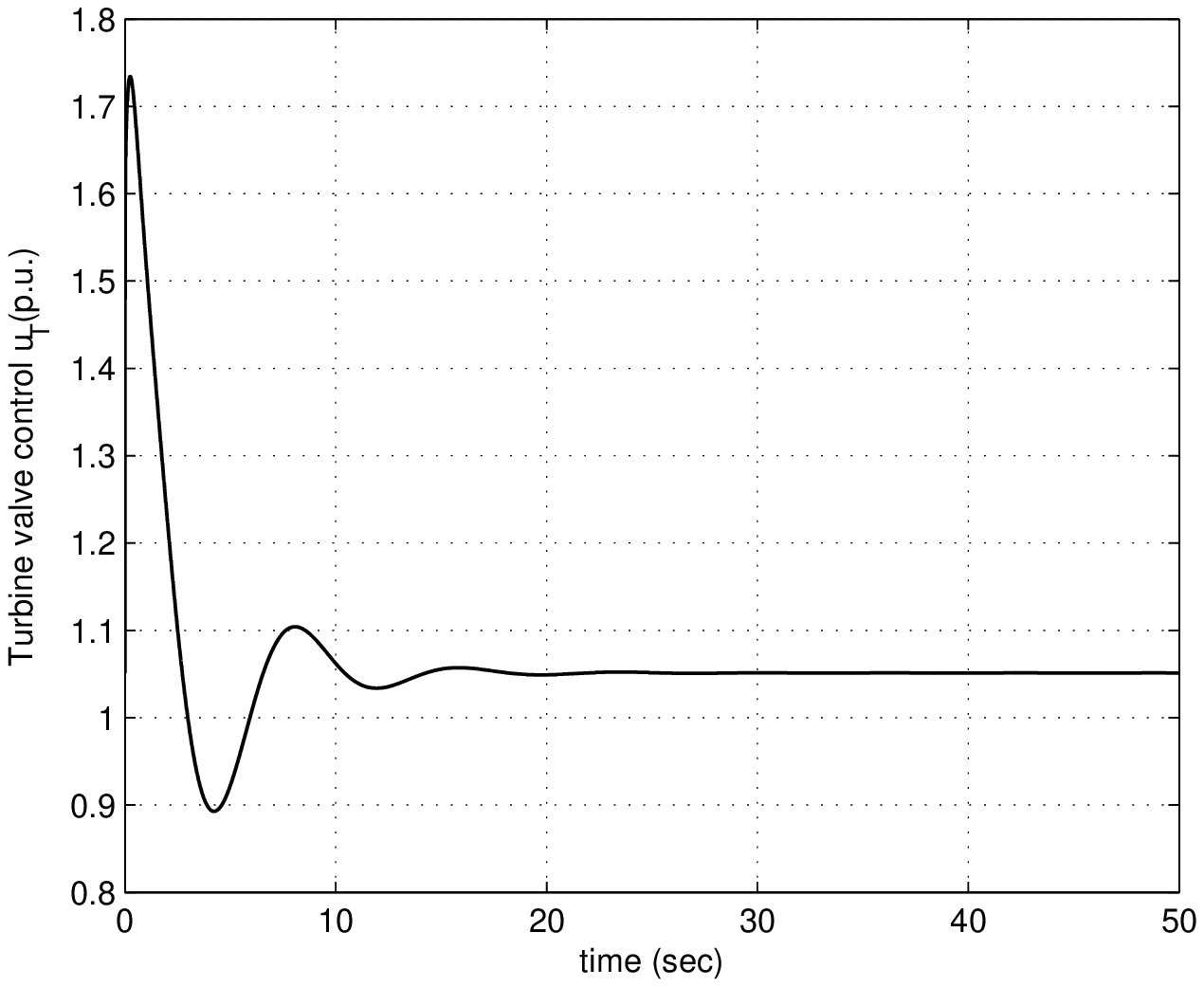}
          \caption{Plot of the control input $u_T$ vs time for the PID controllers applied to the truth model}
          \label{fig:pidtruth5}          
\end{figure}

 The two PID controllers that were designed for the decoupled reduced order LFC and AVR SISO systems, and then 
tested on the (LFC+AVR) MIMO system including coupling and the reduced order nonlinear model, are now tested on the truth model. The PID controller which controls the LFC loop is re-tuned to $K_{p1}=2000$, $K_{i1}=1500$, and $K_{d1}=1000$. Whereas, the PID controller which controls the AVR loop is re-tuned to $K_{p2}=2000$, $K_{i2}=15000$, and $K_{d2}=4$. \autoref{fig:pidtruth1} to \autoref{fig:pidtruth5} show simulation results for the two PID controllers that are tested on the truth model. From these simulation results we can see that the generator terminal voltage $V_t$, angular velocity $\omega $, rotor angle $\delta $, and the two control inputs $V_F$ and $u_T$,
settle to their respective steady state values.

\newpage                                                                            
\section{Linear State-Space Controller Design}

In this section we design linear state-space controllers for the synchronous generator and turbine
connected to an infinite bus. 

\subsection{State Feedback Controller design using LQR methodology}

\subsubsection{LQR Design based on linear model}

\begin{figure}
          \centering
          \includegraphics[trim=0cm 0cm 0cm 0cm, clip=true, totalheight=0.27\textheight, width=0.54\textwidth]{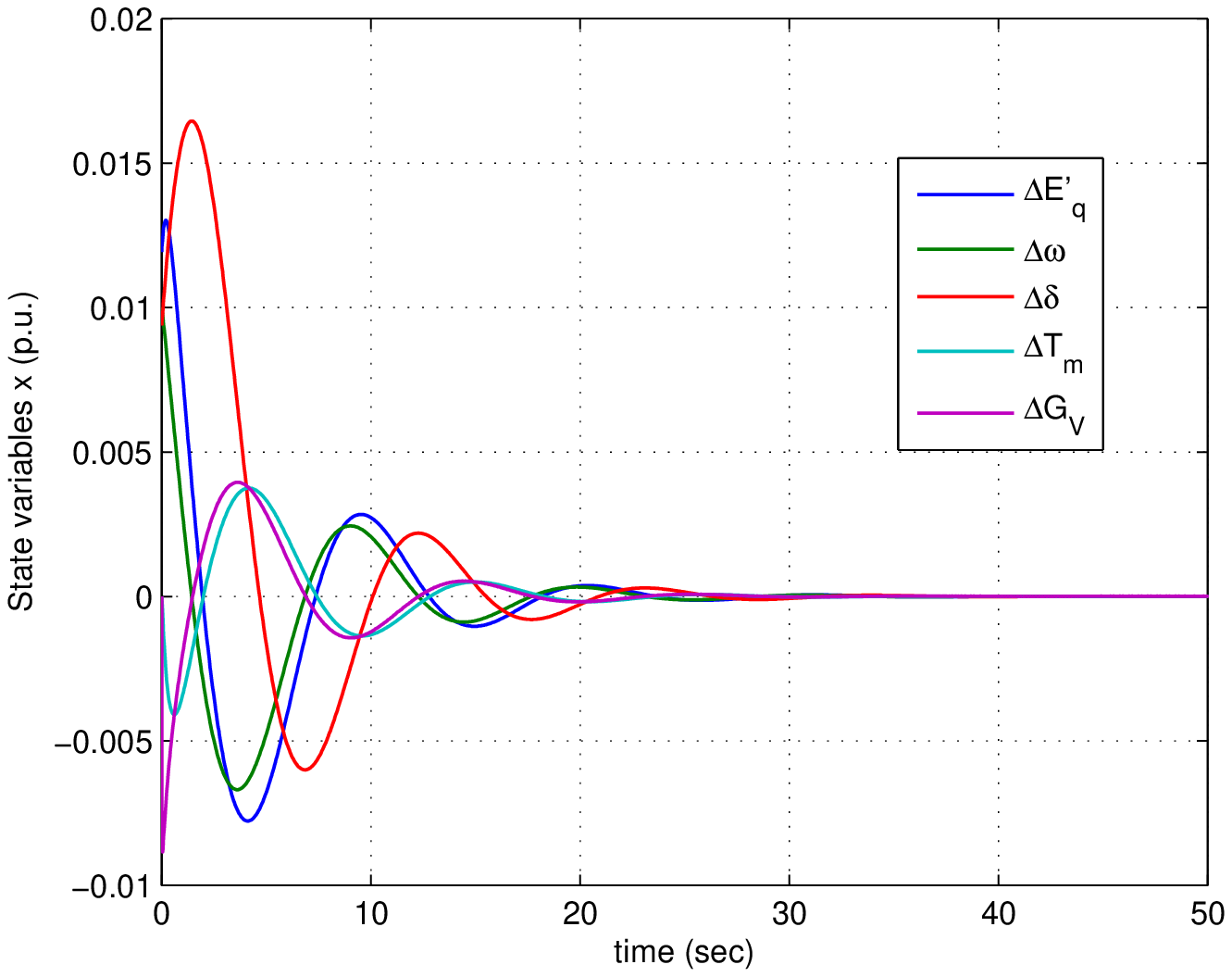}
          \caption{Plot of the state variables $\Delta E'_q$, $\Delta \omega $, $\Delta \delta $, $\Delta T_m$, and 
                   $\Delta G_V$ vs time for the full-state feedback LQR applied to the reduced order linear model}
          \label{fig:lqr1}
          \includegraphics[trim=0cm 0cm 0cm 0cm, clip=true, totalheight=0.27\textheight, width=0.54\textwidth]{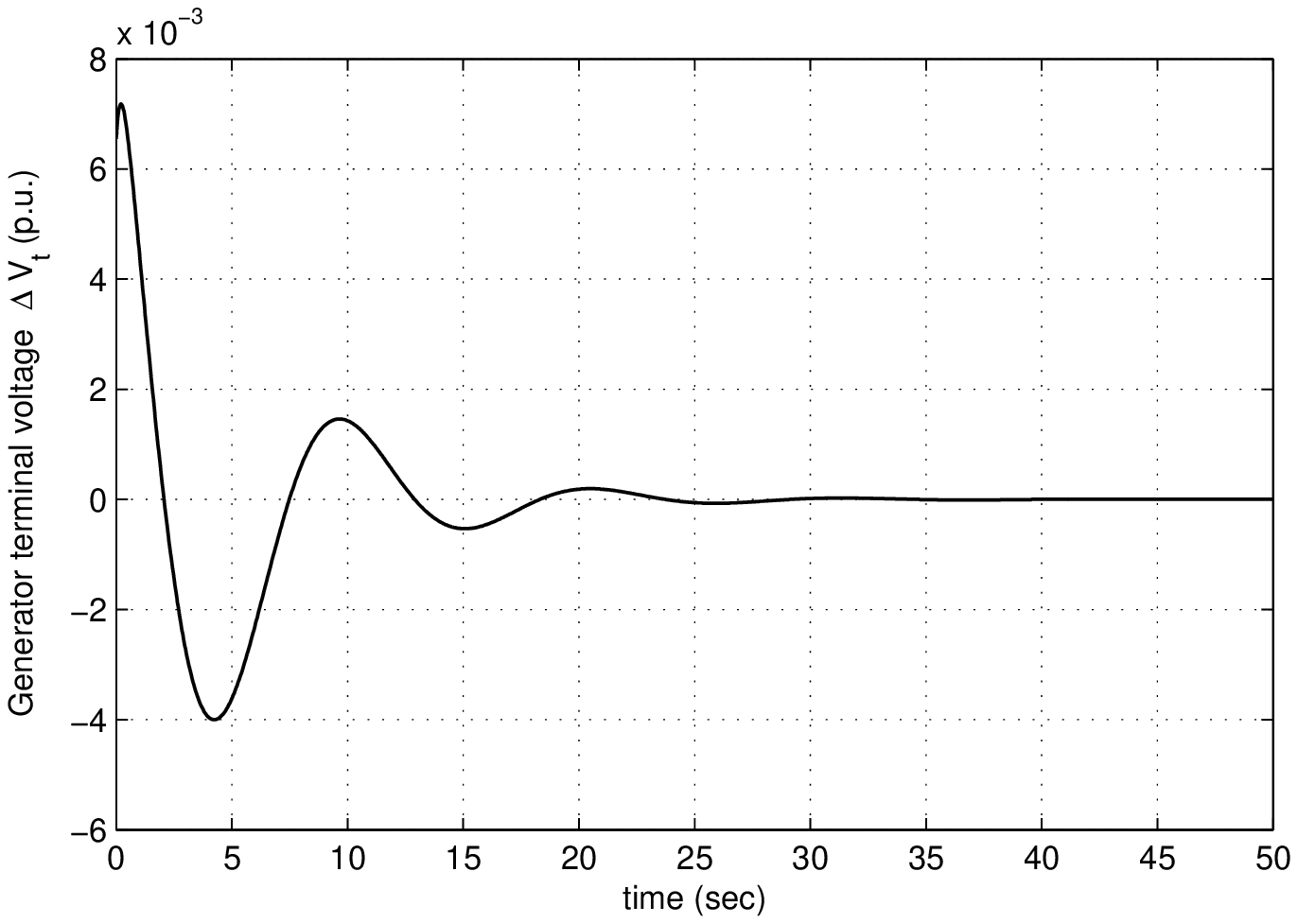}
          \caption{Plot of the generator terminal voltage $\Delta V_t$ vs time for the full-state feedback 
          LQR applied to the reduced order linear model}
          \label{fig:lqr2}
          \includegraphics[trim=0cm 0cm 0cm 0cm, clip=true, totalheight=0.27\textheight, width=0.54\textwidth]{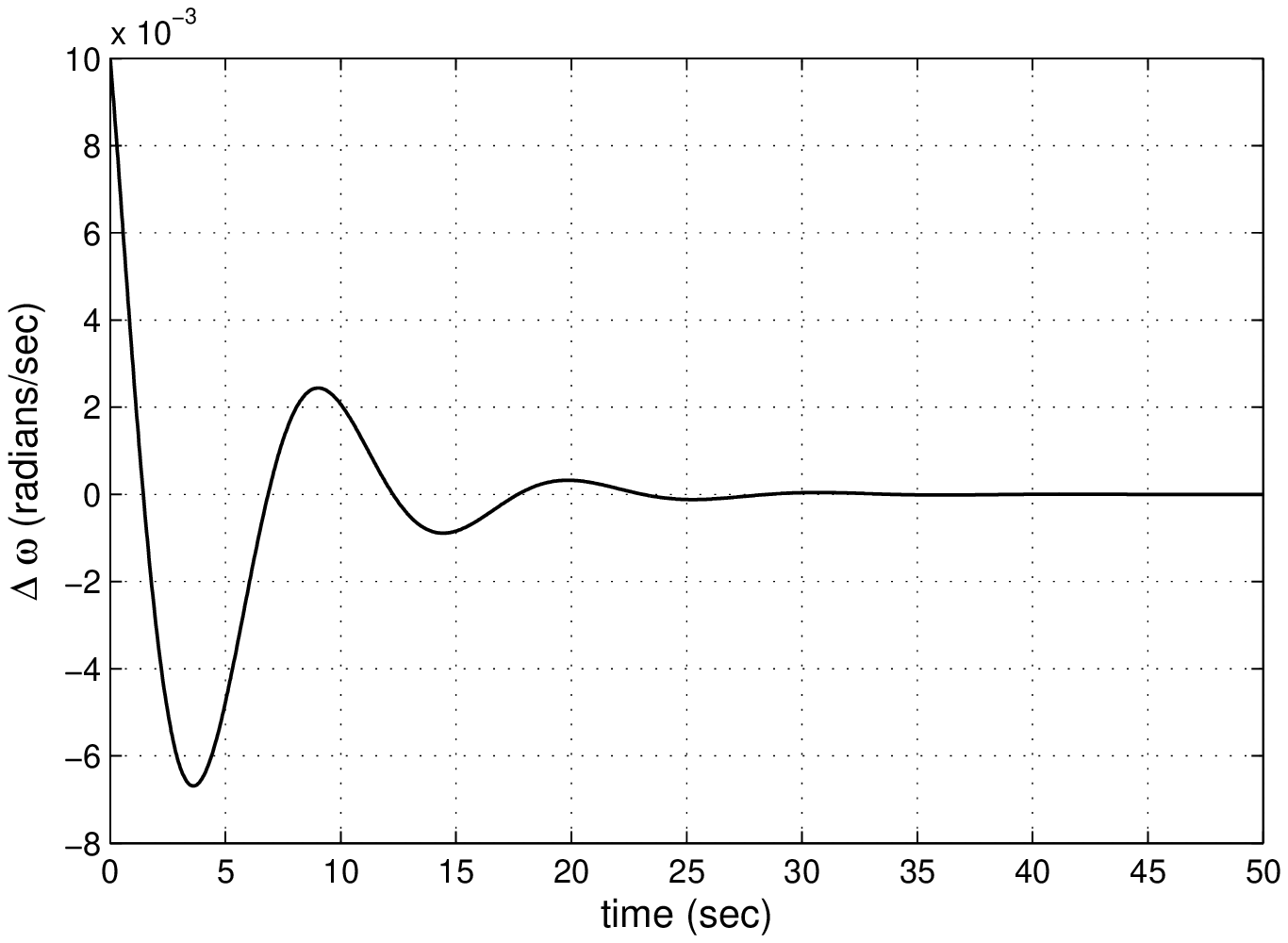}
          \caption{Plot of $\Delta \omega $ vs time for the full-state feedback LQR applied to the reduced order linear model}
          \label{fig:lqr3}
\end{figure}

\begin{figure}
          \centering
          \includegraphics[trim=0cm 0cm 0cm 0cm, clip=true, totalheight=0.27\textheight, width=0.54\textwidth]{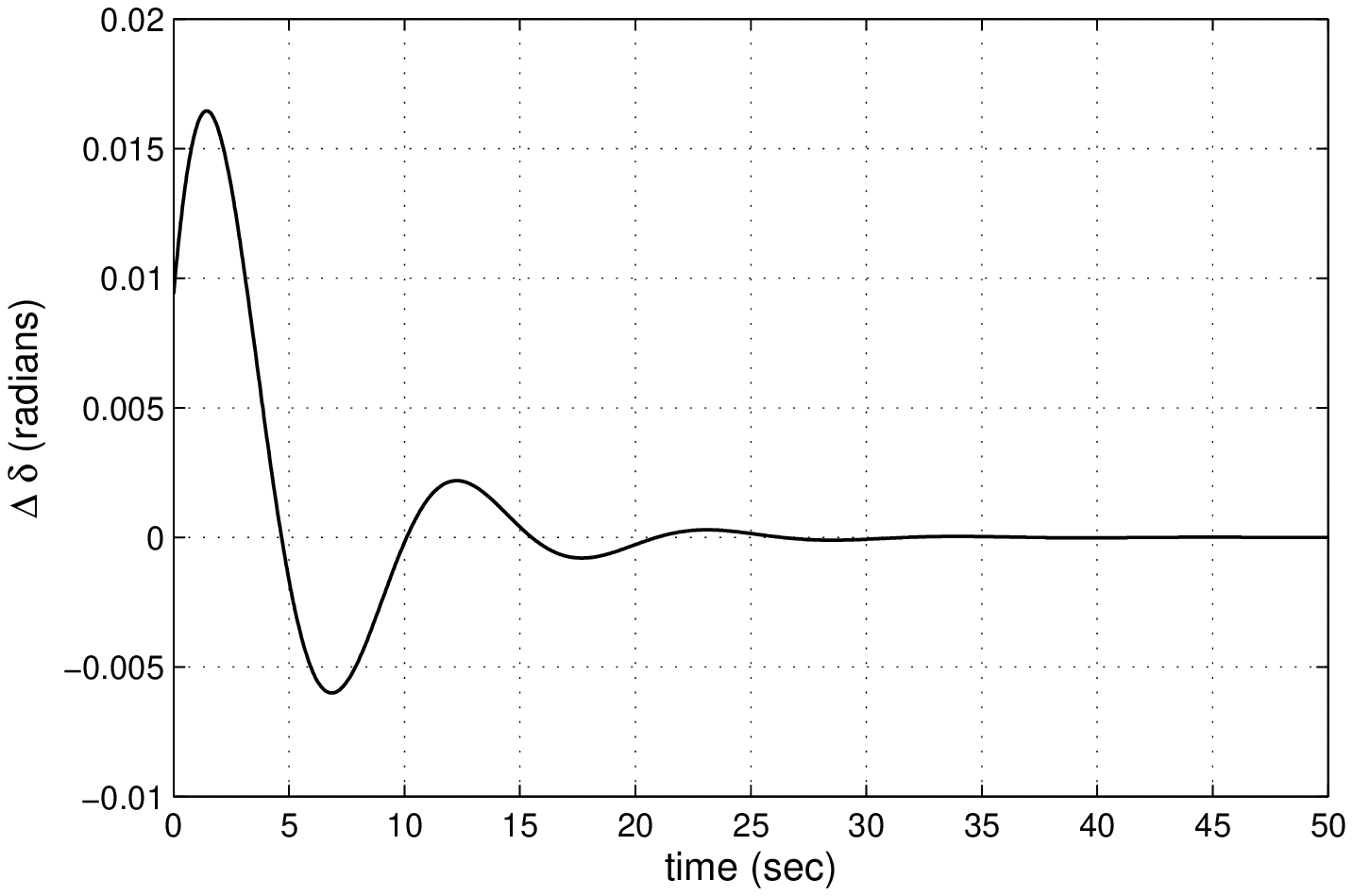}
          \caption{Plot of $\Delta \delta $ vs time for the full-state feedback LQR applied to the reduced order linear model}
          \label{fig:lqr4}
          \includegraphics[trim=0cm 0cm 0cm 0cm, clip=true, totalheight=0.27\textheight, width=0.54\textwidth]{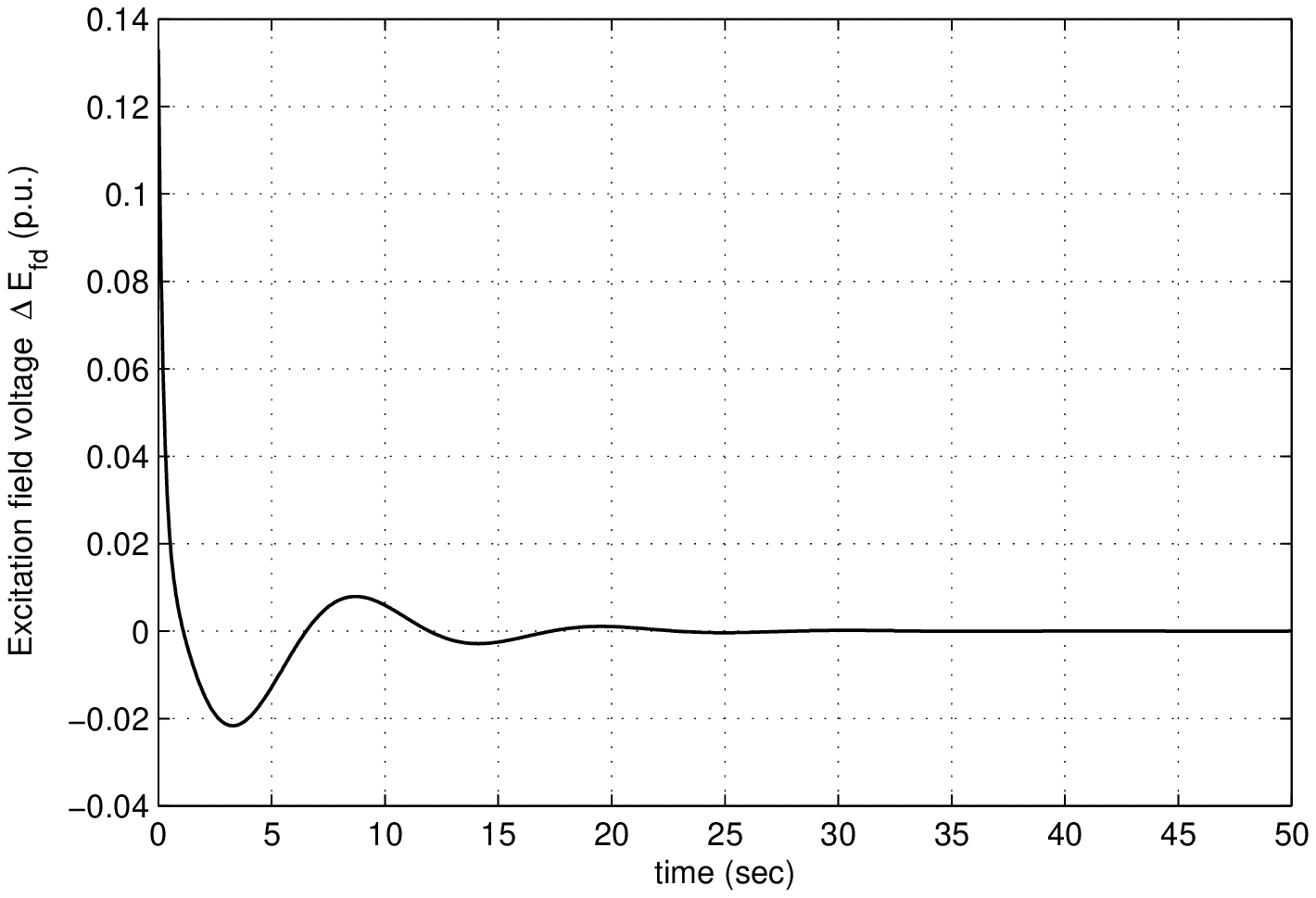}
          \caption{Plot of the control input $\Delta E_{fd}$ vs time for the full-state feedback LQR applied to 
          the reduced order linear model}
          \label{fig:lqr5}
          \includegraphics[trim=0cm 0cm 0cm 0cm, clip=true, totalheight=0.27\textheight, width=0.54\textwidth]{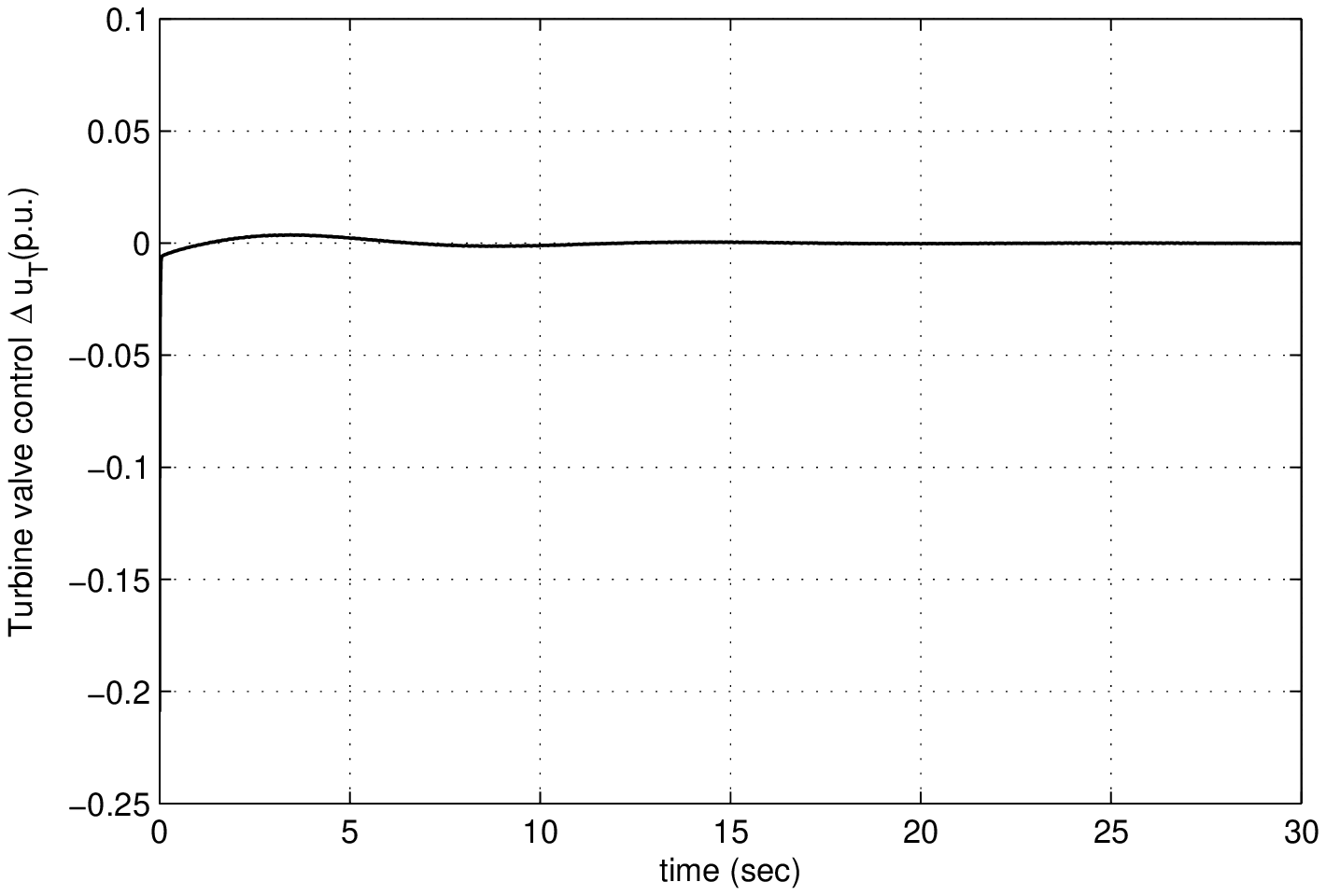}
          \caption{Plot of the control input $\Delta u_T$ vs time for the full-state feedback 
          LQR applied to the reduced order linear model}
          \label{fig:lqr6}          
\end{figure}

We first design a linear-quadratic regulator (LQR) for the linearized model of the synchronous generator and turbine
connected to an infinite bus. Let us assume that we have sensors to measure all the states and we use a full-state feedback controller (regulator)
of the form
\begin{equation}
          \mathbf{u}=-K\mathbf{x}
\label{eq:lc1}
\end{equation}
that seeks to drive the states to zero. Since, the system is MIMO with $2$ inputs and $5$ states, LQR controller design entails finding the $2\times 5$ gain vector $K$. This can be done by directly using the $lqr$ command in MATLAB. 
For this, letting 
\begin{equation}
         J=\int_0^\infty (\mathbf{x}^TQ\mathbf{x}+\mathbf{u}^TR\mathbf{u}) \ dt 
\label{eq:lc2}
\end{equation} 
we seek to find the gain vector $K$ to minimize the cost function $J$. Minimization of $J$ results in driving
$\mathbf{x}(t)$ to zero with as little control energy and state deviations as possible, with the balance between control
energy and state deviations specified via the $Q$ and $R$ matrices. Here, assume the $5\times 5$ $Q$ matrix is diagonal
with diagonal elements $q_i \geq  0$ (each providing a weight for a different element of the deviation of the state)
and the $2\times 2$ $R$ matrix is diagonal with diagonal elements $r_i >  0$ (each providing a weight for the deviation
of the two control inputs. The values for $Q$ and $R$ are used as design parameters.\\
A methodology to tune the $Q$ and $R$ matrices is given as follows: If all the $q_i = 0$, then the excursions of the states 
are high while the control input tries to force the state to zero. 
High values of $q_i$ relative to $r_i$ mean that you are willing to use lots of control energy to keep 
state excursions small while driving it to zero. Clearly, you cannot pick $r_i = 0$
as this results in allowing infinite control energy to force the state to zero, typically then very fast.
Finding the gain $K$ to minimize $J$ involves solving the $Riccati$ equation $A^TP+PA-PBR^{-1}B^TP+Q=0$, where $K=R^{-1}B^TP$. We use the Matlab $lqr$
command to directly solve for the gain vector $K$ given $A$, $B$, $Q$, and $R$.
Thus, by using
the weighting matrices
\begin{equation}
          Q=\bbm 300 & 0  & 0  & 0  & 0\\ 0 &  250 &  0  & 0  & 0\\ 0 &  0 &  200 &  0 &  0\\ 
            0 &  0  & 0  & 200  & 0\\ 0  & 0  & 0 &  0  & 250 \ebm
 \label{eq:lc4}
\end{equation} 
and           
\begin{equation}
        R=\bbm 0.5 & 0\\ 0 & 0.5\ebm
\label{eq:lc5}
\end{equation} 
and the state space matrices $(A, B)$ as given in \autoref{eq:linear25} and \autoref{eq:linear26} the 
control gain $K$ is found to be 
  \begin{equation}
        K=\bbm 23.7240 & -36.3457 & -5.5938 & -2.5612 & -0.0454\\ -1.3381 & 21.0340 & 1.5703 & 9.0242 & 21.5437\ebm
\label{eq:lc6}
\end{equation}
\autoref{fig:lqr1} shows time history of the state variables ($\Delta E'_q$, $\Delta \omega $, $\Delta \delta $, $\Delta T_m$, and 
                   $\Delta G_V$) for the full-state feedback LQR controller applied to the reduced order linear model. From this plot we can see that the LQR
drives all the states to $0$ in approximately 20 to 25 seconds. Similarly, from \autoref{fig:lqr2} to \autoref{fig:lqr6} we can
see that $\Delta V_t$, $\Delta \omega $, $\Delta \delta $, $\Delta E_{fd}$, and $\Delta u_T$ converge to zero in approximately 25 seconds.

\subsubsection{Simulation results for the Full-State Feedback LQR applied to the Reduced Order Nonlinear Model}

\begin{figure}
          \centering
          \includegraphics[trim=0cm 0cm 0cm 0cm, clip=true, totalheight=0.27\textheight, 
           width=0.54\textwidth]   {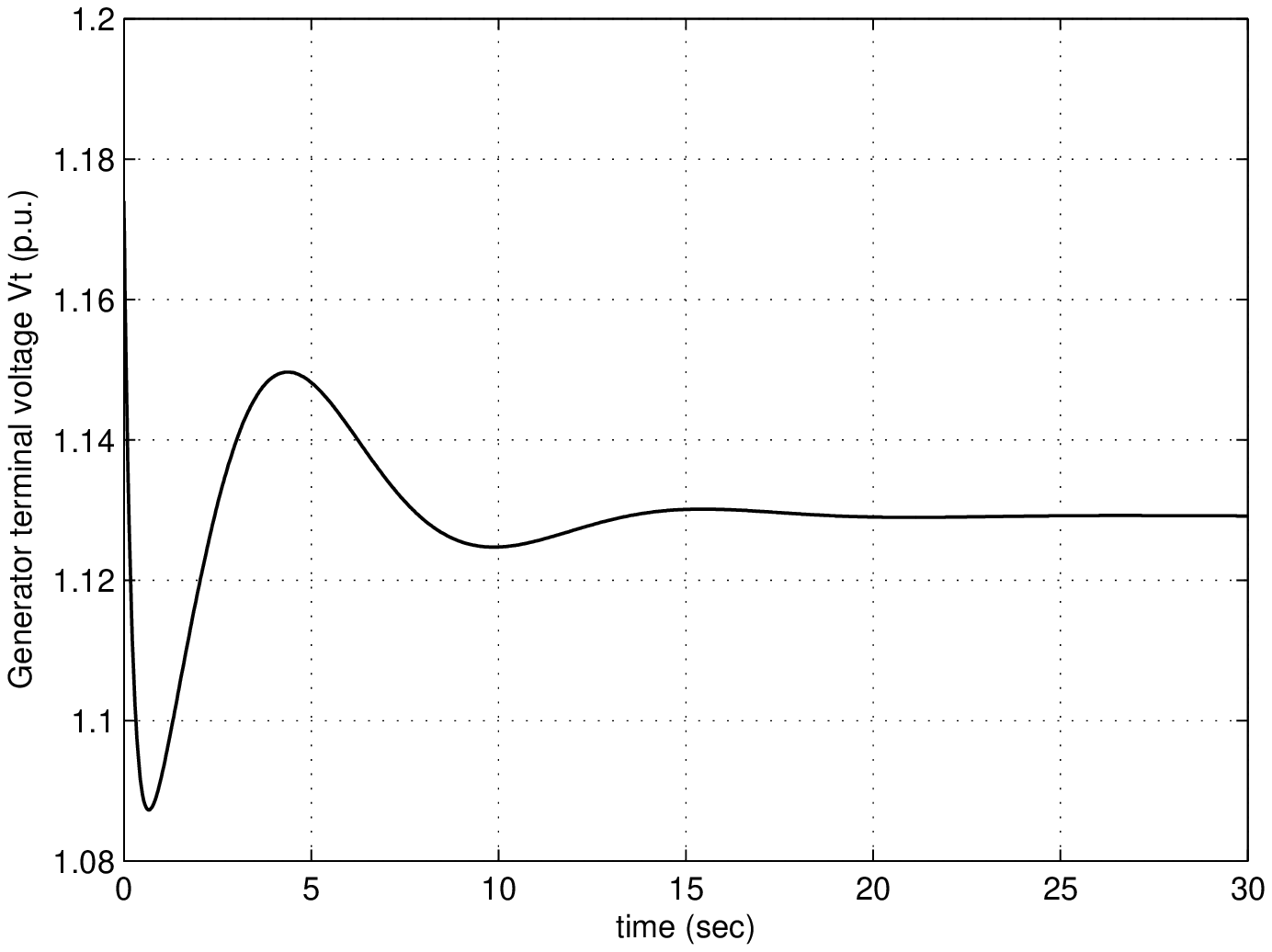}
          \caption{Plot of the generator terminal voltage $V_t$ vs time for the LQR-based full-state feedback 
           controller with the original gains of subsection 7.1.1 applied to the reduced order nonlinear  model }
          \label{fig:lqrnonlineartest}
          \includegraphics[trim=0cm 0cm 0cm 0cm, clip=true, totalheight=0.27\textheight, 
           width=0.54\textwidth]{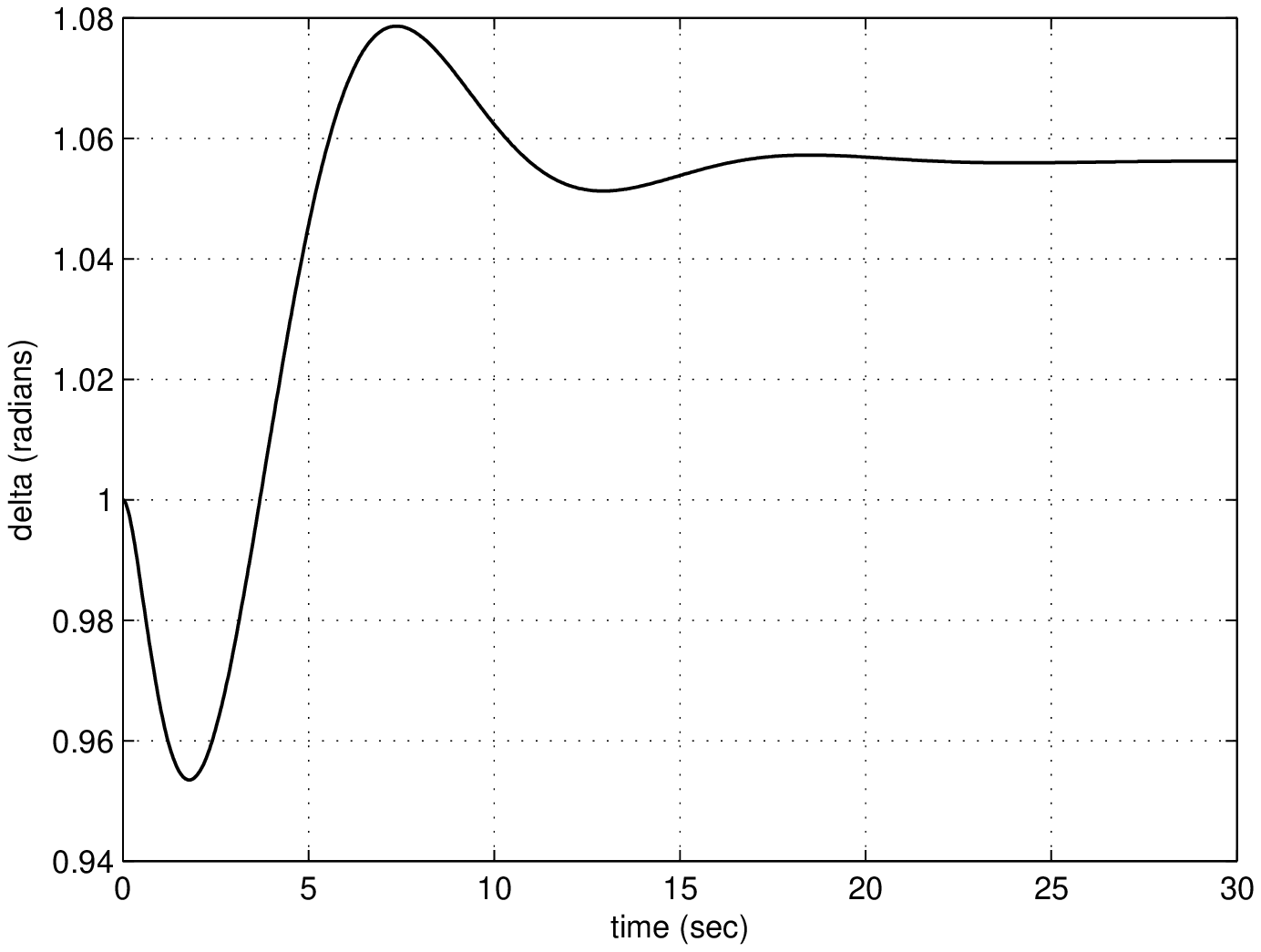}
          \caption{Plot of the rotor angle $\delta $ vs time for the LQR-based full-state feedback controller with the 
          original gains of subsection 7.1.1 applied to 
          the reduced order nonlinear model}
          \label{fig:lqrnonlineartest0}
          \includegraphics[trim=0cm 0cm 0cm 0cm, clip=true, totalheight=0.27\textheight, width=0.54
           \textwidth]  {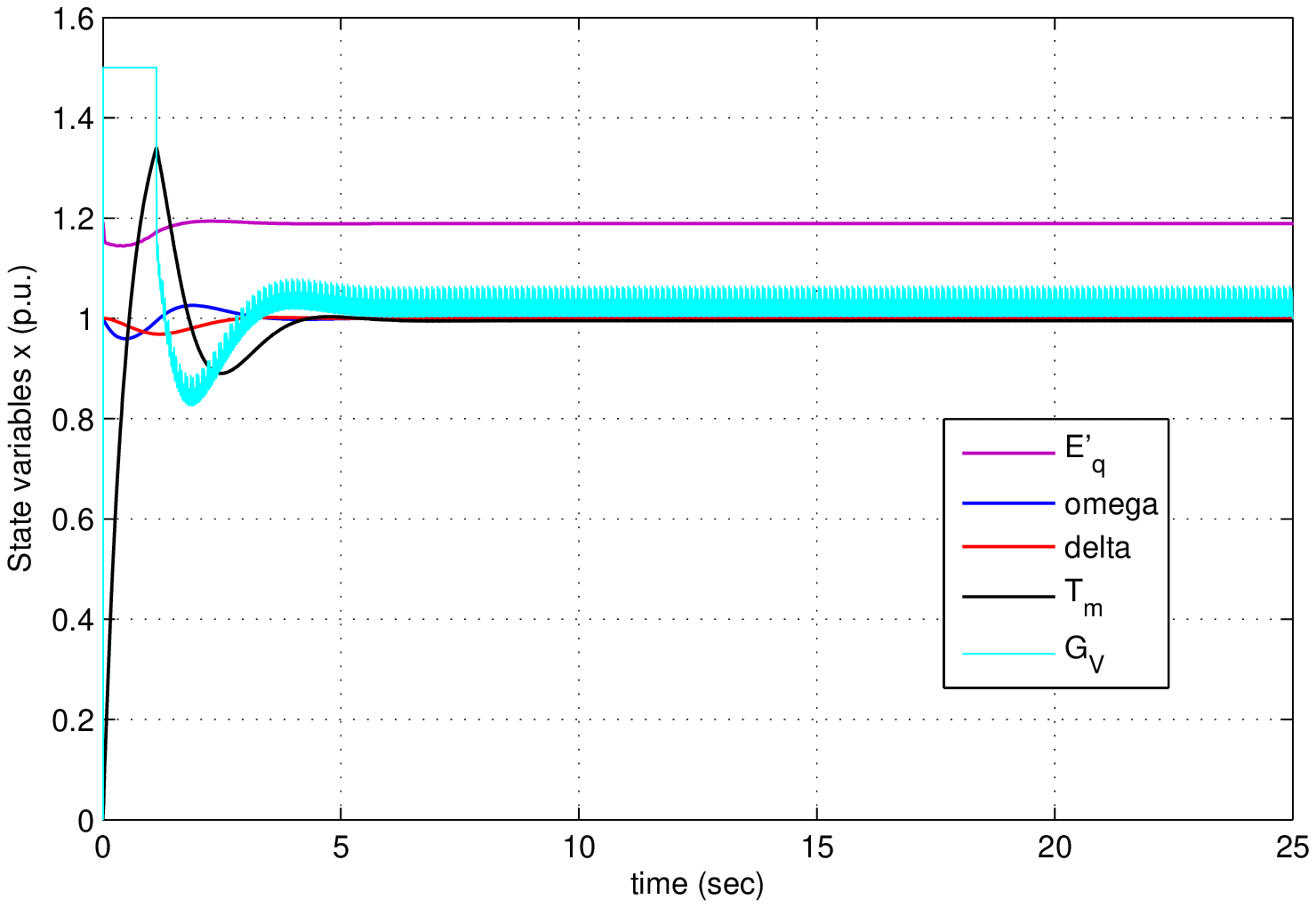}
          \caption{Plot of the state variables $E'_q$, $\omega $, $\delta $, 
           $T_m$, and $G_V$ vs time for the re-tuned LQR-based full-state feedback controller applied to the reduced order nonlinear model}
          \label{fig:lqrnonlineartest1}
\end{figure}

\begin{figure}
          \centering                           
          \includegraphics[trim=0cm 0cm 0cm 0cm, clip=true, totalheight=0.27\textheight, 
           width=0.54\textwidth]   {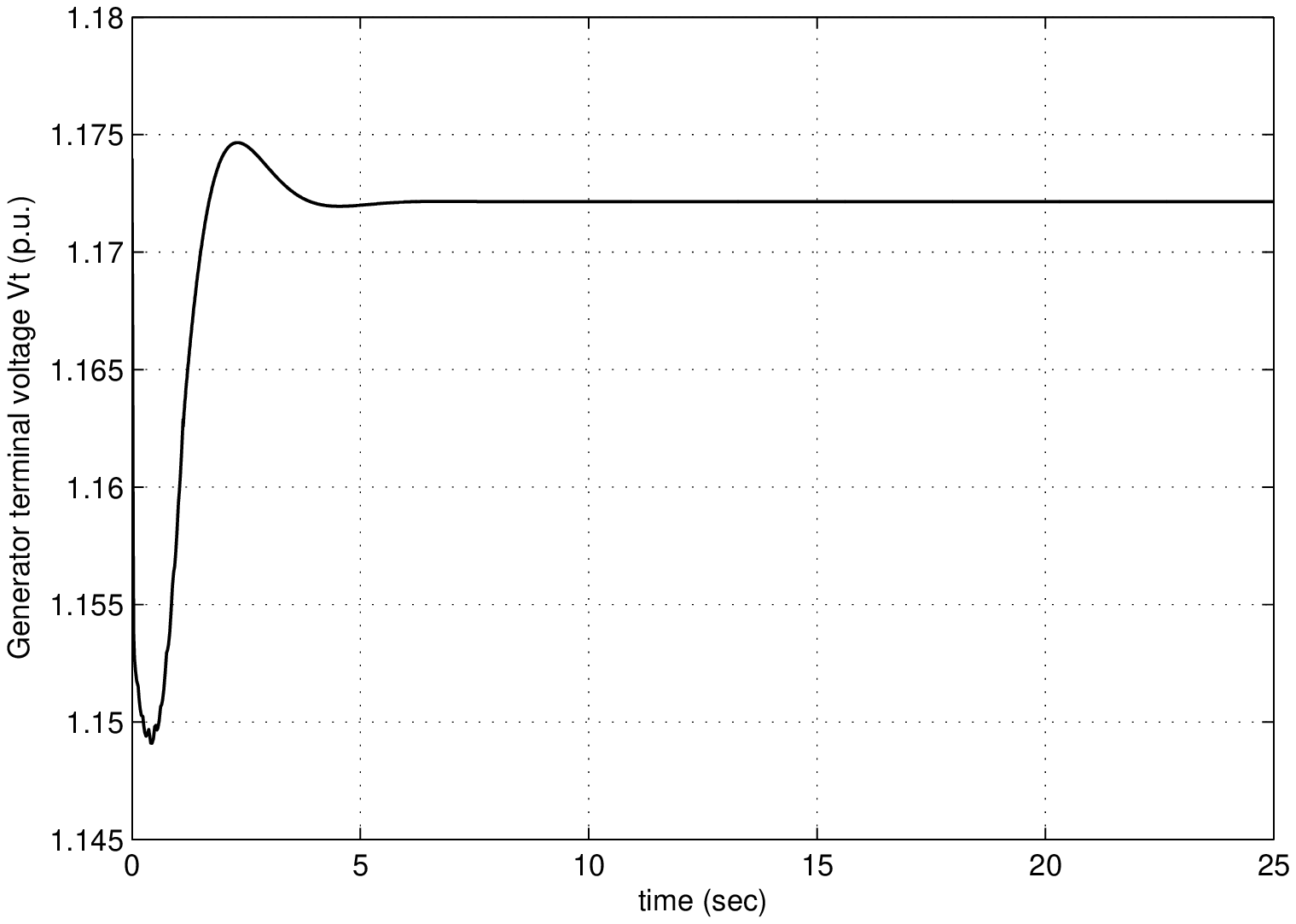}
          \caption{Plot of the generator terminal voltage $V_t$ vs time for the re-tuned LQR-based full-state feedback 
           controller applied to the reduced 
           order nonlinear  model}
          \label{fig:lqrnonlineartest2}
          \includegraphics[trim=0cm 0cm 0cm 0cm, clip=true, totalheight=0.27\textheight, 
           width=0.54\textwidth]{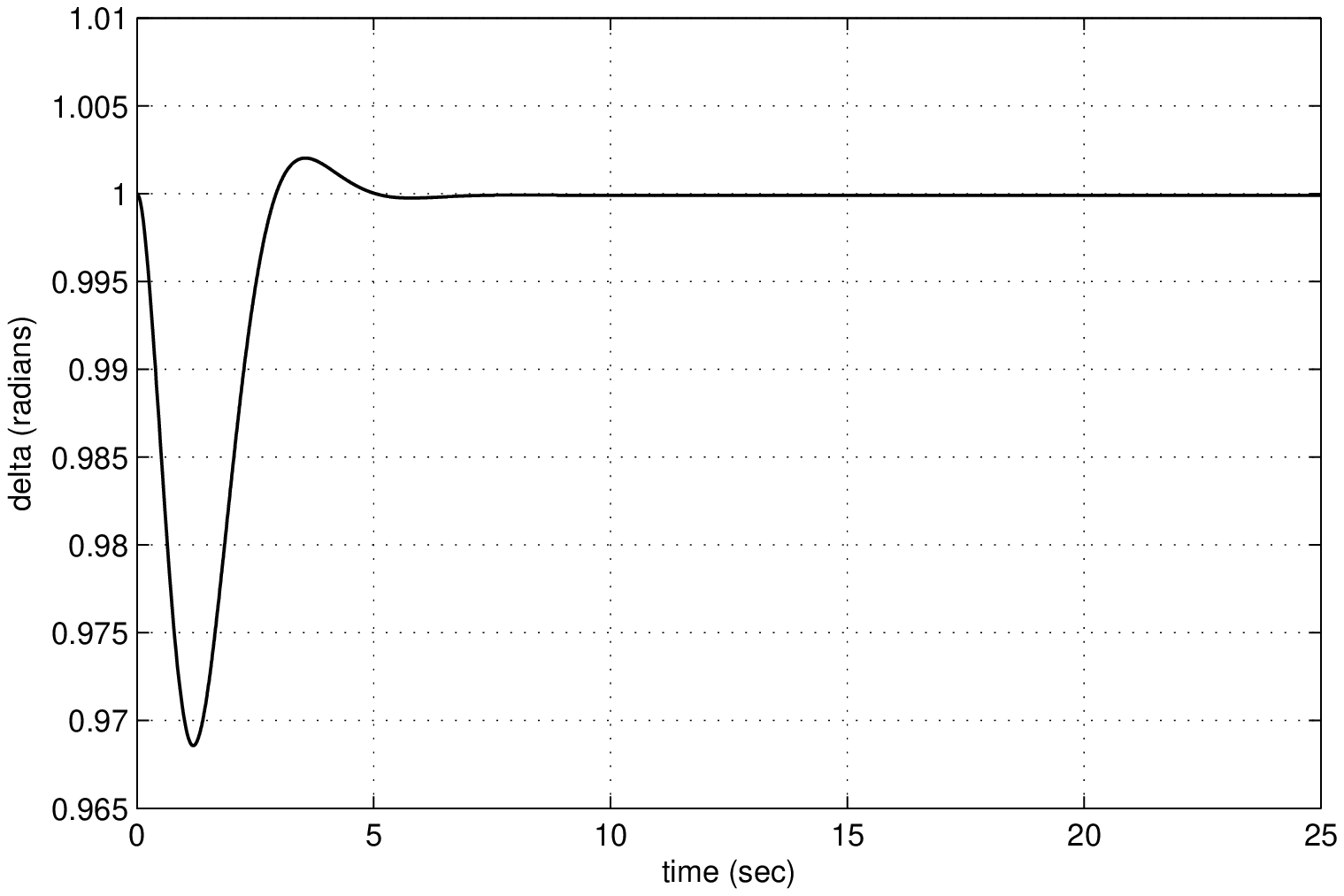}
          \caption{Plot of the rotor angle $\delta $ vs time for the re-tuned LQR-based full-state feedback controller applied to 
          the reduced order nonlinear model}
          \label{fig:lqrnonlineartest3}
          \includegraphics[trim=0cm 0cm 0cm 0cm, clip=true, totalheight=0.27\textheight, 
          width=0.54\textwidth]{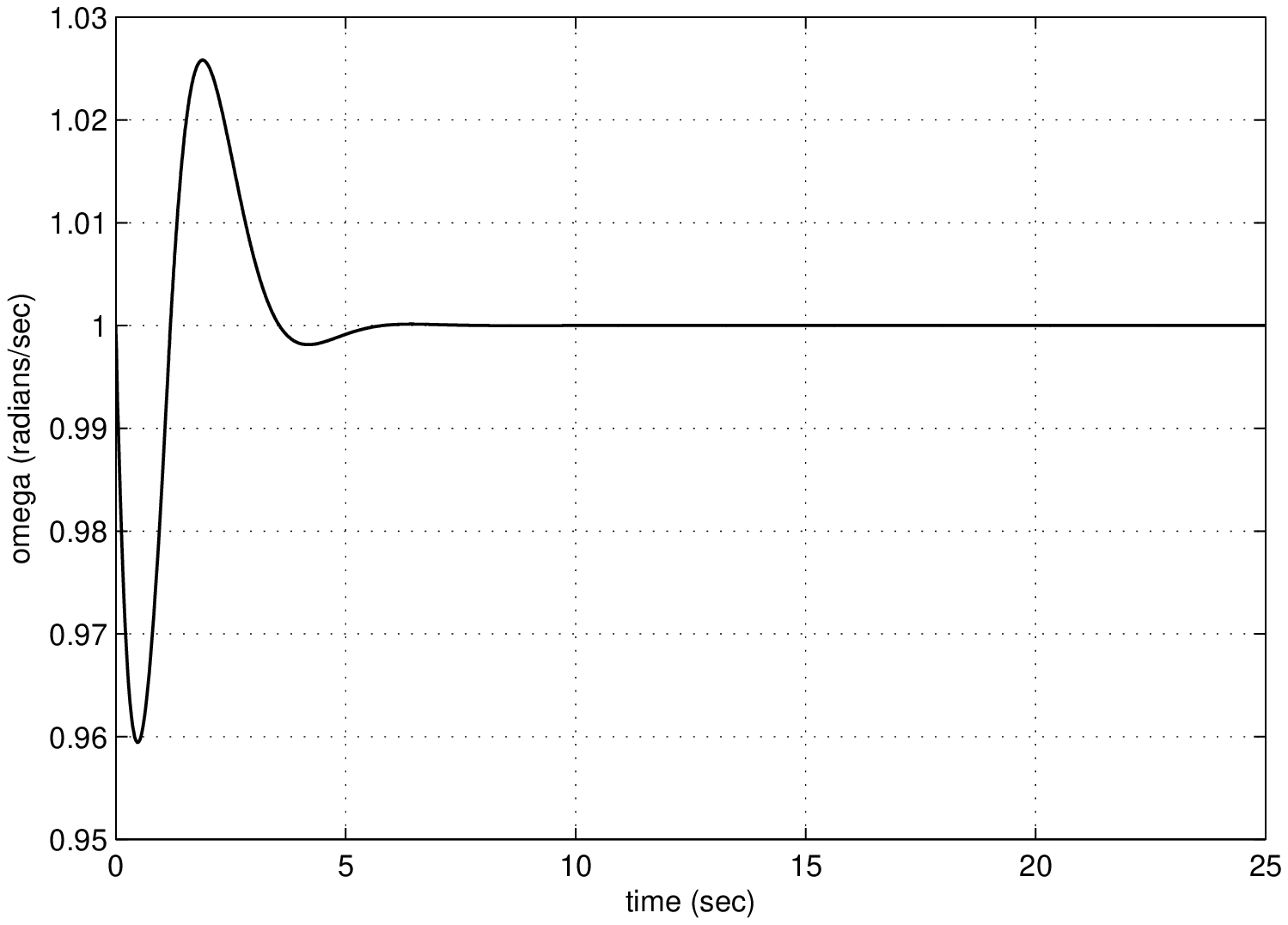}
          \caption{Plot of the frequency $\omega  $ vs time for the re-tuned LQR-based full-state feedback controller applied to the 
          reduced order nonlinear model}
          \label{fig:lqrnonlineartest4}
\end{figure}

\begin{figure}
          \centering          
          \includegraphics[trim=0cm 0cm 0cm 0cm, clip=true, totalheight=0.27\textheight, width=0.54\textwidth]{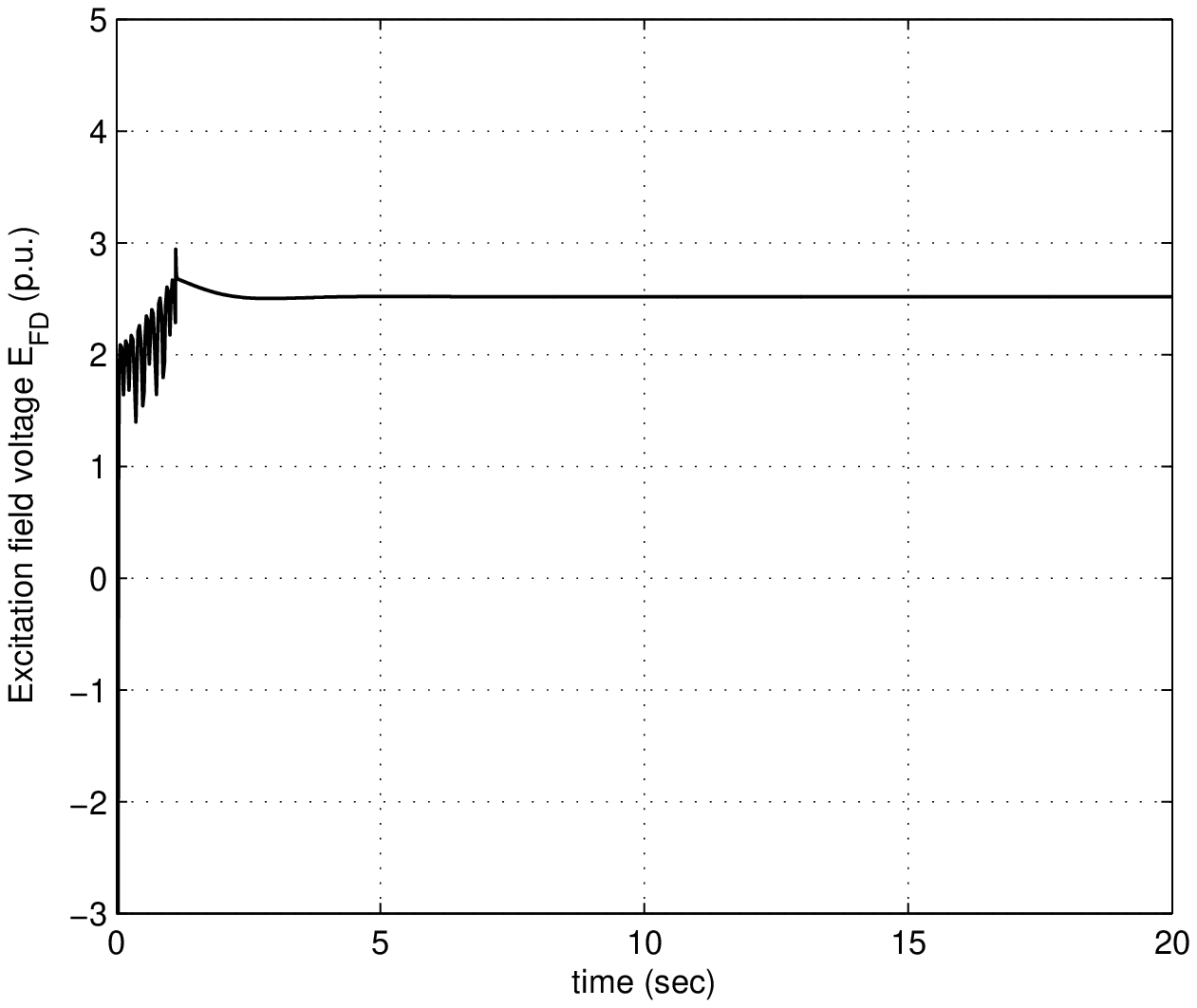}
          \caption{Plot of the control input $E_{fd}$ vs time for the full-state feedback LQR applied to 
          the reduced order nonlinear model}
          \label{fig:lqrnonlineartest5}         
          \includegraphics[trim=0cm 0cm 0cm 0cm, clip=true, totalheight=0.27\textheight, width=0.54\textwidth]{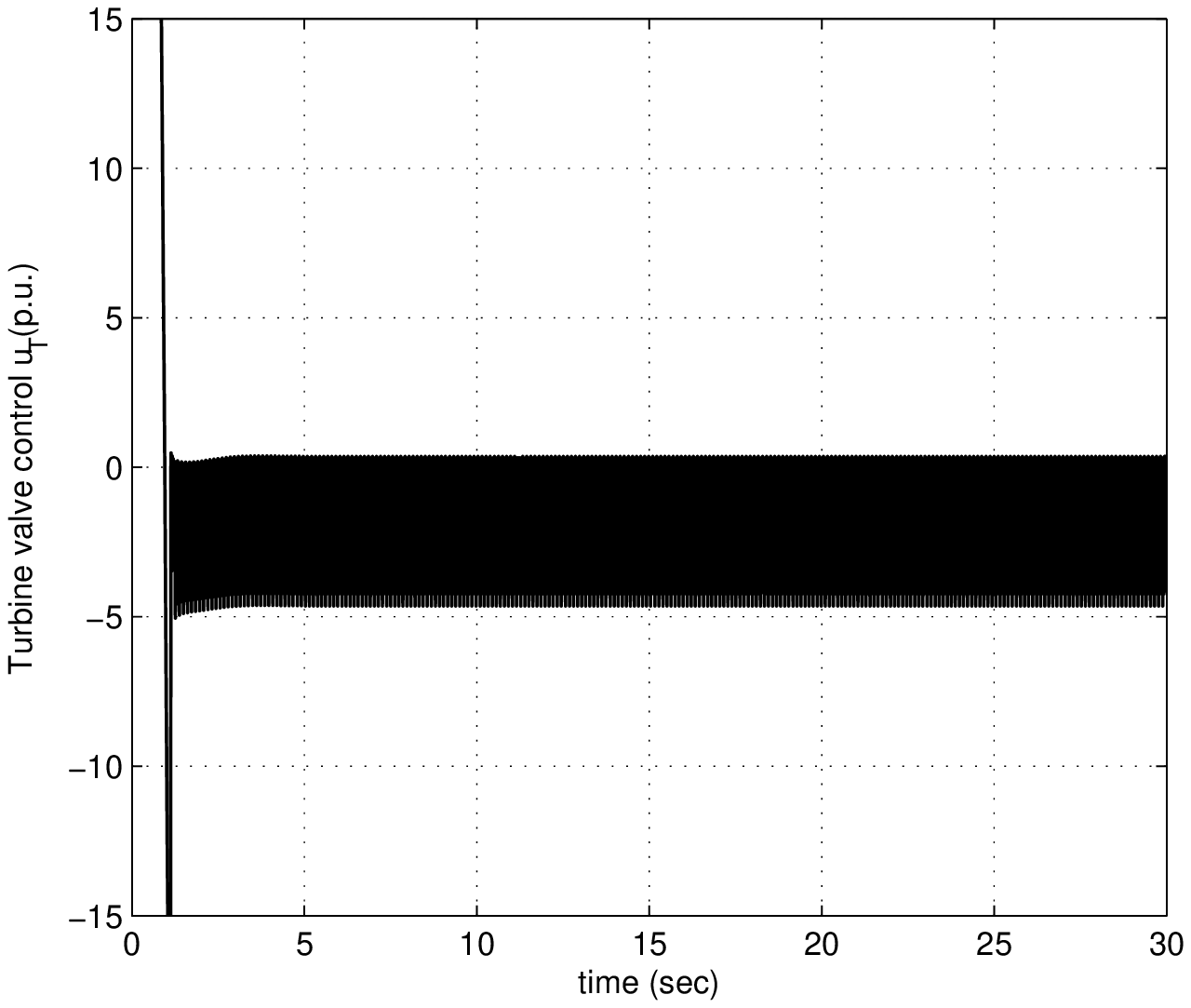}
          \caption{Plot of the control input $u_T$ vs time for the full-state feedback 
          LQR applied to the reduced order nonlinear model}
          \label{fig:lqrnonlineartest6}          
\end{figure}

The LQR based full-state feedback controller that was earlier designed for the
reduced order linear model is now tested on the reduced order nonlinear model. 
\autoref{fig:lqrnonlineartest} and \autoref{fig:lqrnonlineartest0} show plots for the generator terminal voltage
$V_t$, and rotor angle $\delta $, respectively. The generator terminal voltage
$V_t$ settles to a new steady state value of 1.13 p.u. which deviates from the desired steady state value of 1.1723 p.u. by an amount
of 0.0423 p.u. Similarly, the rotor angle $\delta $ settles to a new steady state value of 1.056 p.u. which deviates from the desired steady state value of 1 p.u. by an amount of 0.056 p.u. We observe a steady state error when the LQR with original gains of subsection 7.1.1
is applied to the reduced order nonlinear model.
Therefore, the gains of the controller are tuned once
again so that the controller works efficiently on the reduced order nonlinear model.
We use the Matlab $lqr$ command to directly solve for the gain vector $K$ given $A, B, Q,$ and $R$. The system matrices 
$(A, B)$ are evaluated at the nominal operating point, and $Q$ and $R$ matrices are appropriately
tuned as per the procedure explained in the LQR design section.
Thus, by using the feedback law,
\begin{equation}
          \mathbf{u}=-K\mathbf{x}
\label{eq:lqrtest1}
\end{equation}
the weighting matrices
\begin{equation}
          Q=\bbm 40000 & 0  & 0  & 0  & 0\\ 0 &  10000 &  0  & 0  & 0\\ 0 &  0 &  250000 &  0 &  0\\ 
            0 &  0  & 0  & 500  & 0\\ 0  & 0  & 0 &  0  & 500 \ebm
 \label{eq:lqrtest2}
\end{equation} 
and           
\begin{equation}
        R=\bbm 0.07 & 0\\ 0 & 0.07\ebm
\label{eq:lqrtest3}
\end{equation} 
and the state space matrices $(A, B)$ as given in \autoref{eq:linear25} and \autoref{eq:linear26} the 
control gain $K$ is found to be 
  \begin{equation}
        K=\bbm 753.9172 & -575.5829 & -610.0649 & -27.6436 &  -0.1301\\ -3.8375 & 1782.567 & 1474.307 &  128.4938  & 84.1272\ebm
\label{eq:lqrtest4}
\end{equation}
\autoref{fig:lqrnonlineartest1} to \autoref{fig:lqrnonlineartest6} 
show simulation results for the LQR-based full-state feedback controller applied to the reduced order nonlinear model with the gains re-tuned. From these
results we can see that all the state variables, and outputs, attain their respective steady state 
values in approximately 8-10 seconds. \autoref{fig:lqrnonlineartest5} and \autoref{fig:lqrnonlineartest6} show plots of the two control
inputs, $E_{fd}$ and $u_T$, respectively.

\newpage          
\subsubsection{Simulation results for the Full-State Feedback LQR applied to the Truth Model}

\begin{figure}
          \centering
          \includegraphics[trim=0cm 0cm 0cm 0cm, clip=true, totalheight=0.27\textheight, width=0.54
           \textwidth]  {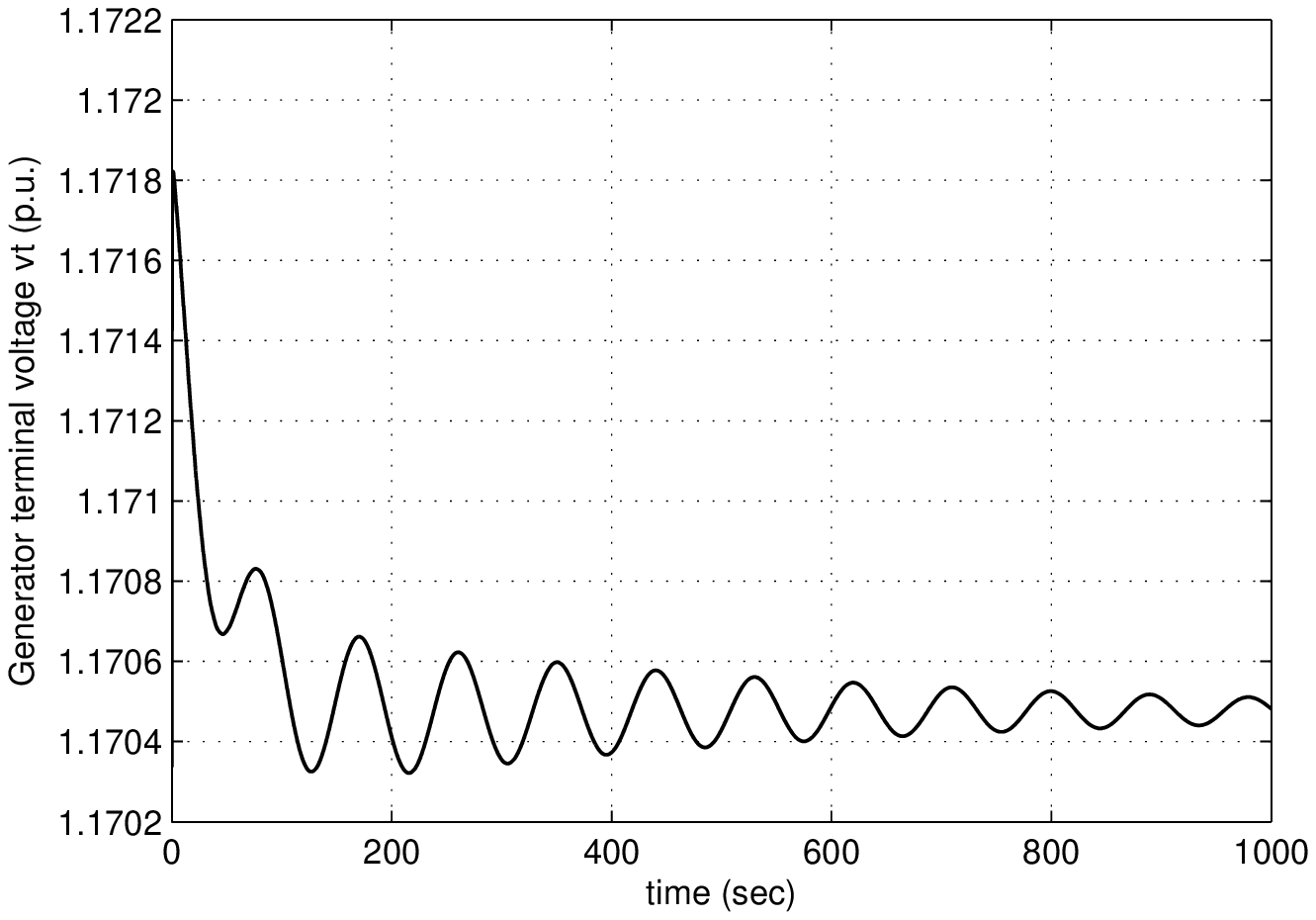}
          \caption{Plot of the generator terminal voltage $V_t$ vs time for the LQR-based full-state feedback controller
          applied to the truth model}
          \label{fig:lqrtruth1}
          \includegraphics[trim=0cm 0cm 0cm 0cm, clip=true, totalheight=0.27\textheight, width=0.54\textwidth]{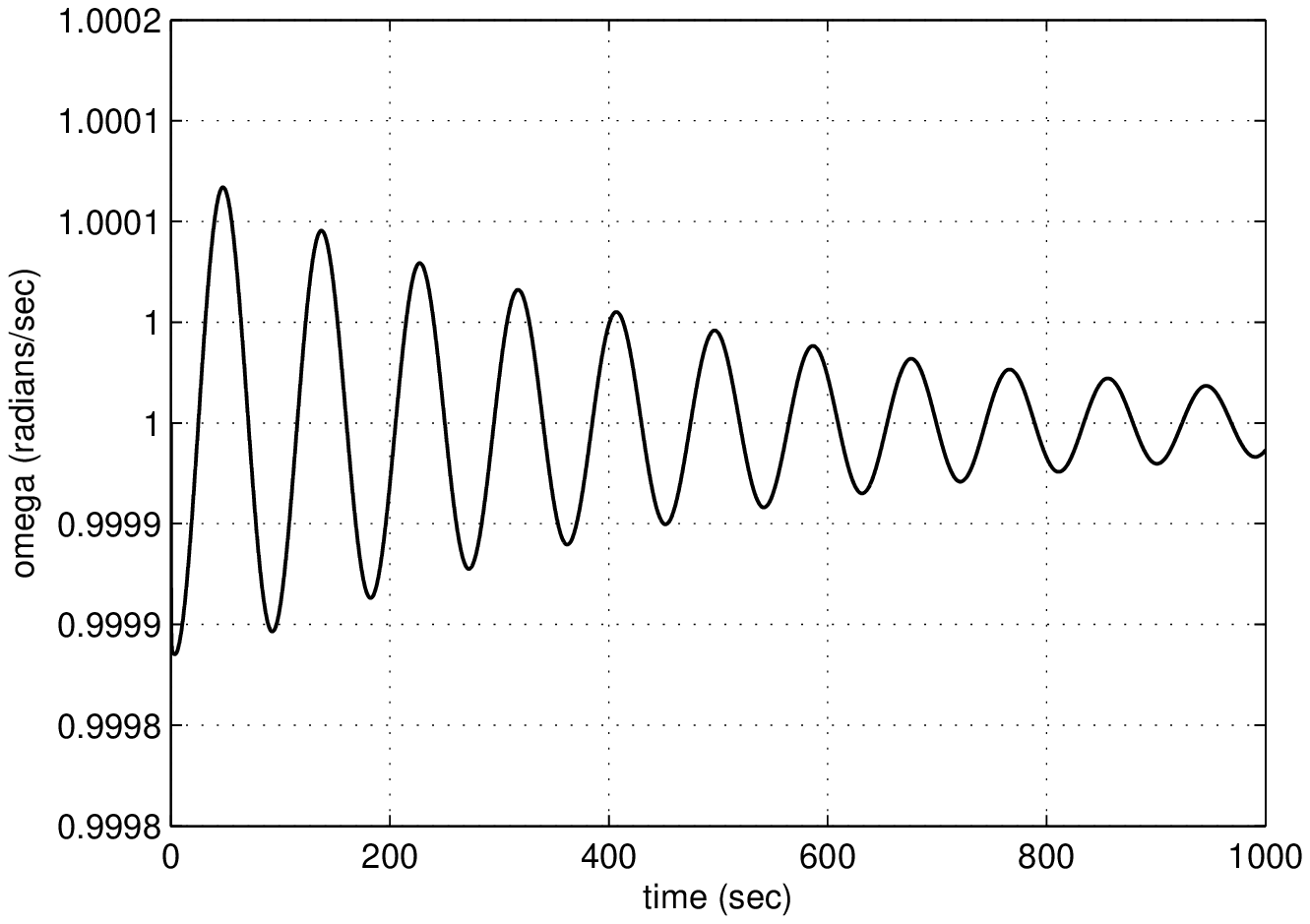}
          \caption{Plot of the angular velocity $\omega $ vs time for the LQR-based full-state feedback controller
          applied to the truth model}
          \label{fig:lqrtruth2}
          \includegraphics[trim=0cm 0cm 0cm 0cm, clip=true, totalheight=0.27\textheight, width=0.54\textwidth]{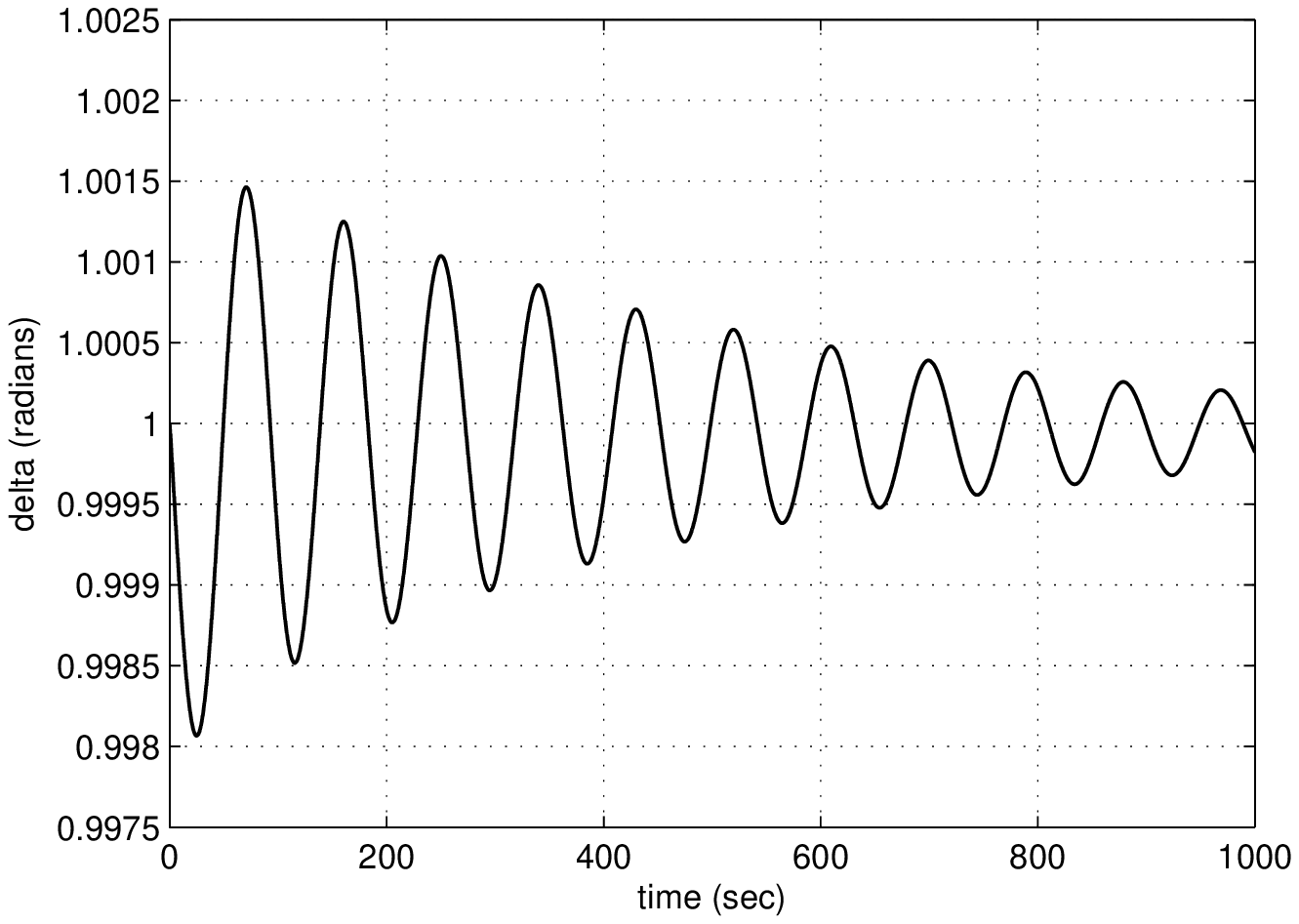}
          \caption{Plot of the rotor angle $\delta $ vs time for the LQR-based full-state feedback controller applied to the truth model}
          \label{fig:lqrtruth3}
\end{figure}

\begin{figure}
          \centering          
          \includegraphics[trim=0cm 0cm 0cm 0cm, clip=true, totalheight=0.27\textheight, width=0.54\textwidth]{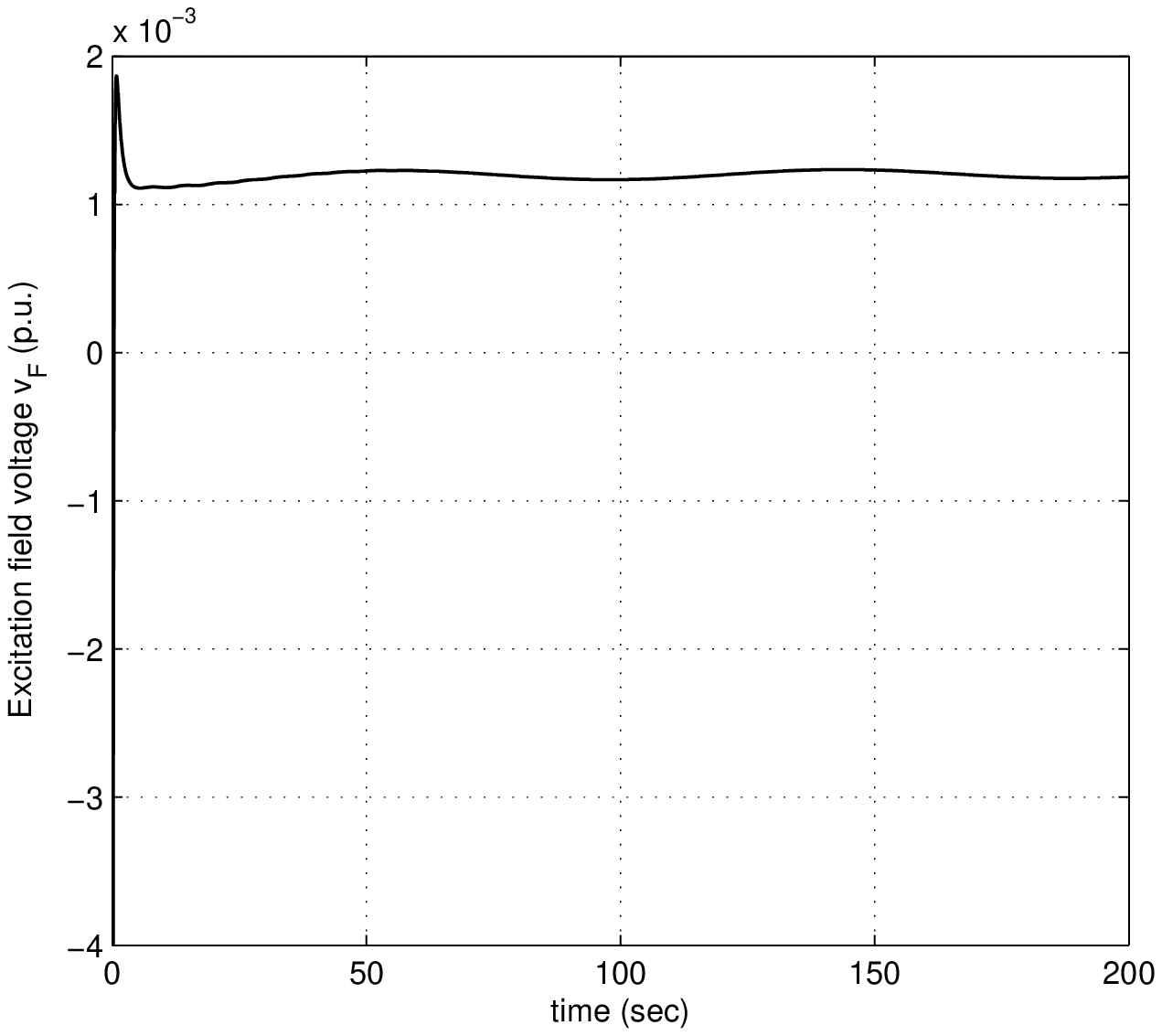}
          \caption{Plot of the control input $V_F$ vs time for the full-state feedback LQR applied to 
          the truth model}
          \label{fig:lqrtruth4}
          \includegraphics[trim=0cm 0cm 0cm 0cm, clip=true, totalheight=0.27\textheight, width=0.54\textwidth]{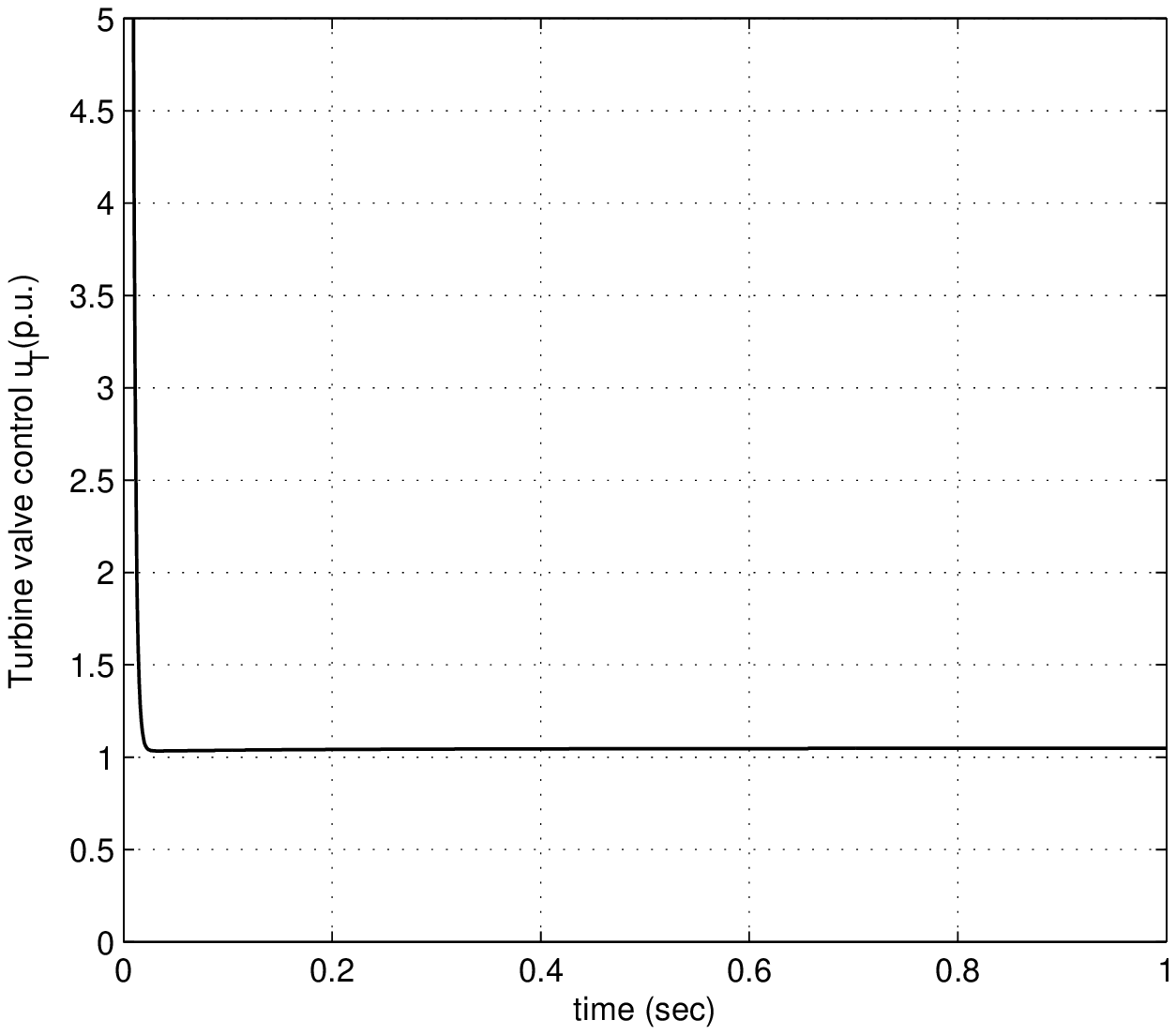}
          \caption{Plot of the control input $u_T$ vs time for the full-state feedback 
          LQR applied to the truth model}
          \label{fig:lqrtruth5}          
\end{figure}

The LQR-based full-state feedback controller that was designed for the reduced order linear model and then re-tuned and tested on the reduced order nonlinear model, is now tested
on the truth model. There is a non physical state $E'_q$ in the reduced order model that needs to be reconstructed from the states of the truth model. The LQR-based full-state feedback controller can be implemented on the truth model by using either \autoref{eq:nonlinearflctruth120} or \autoref{eq:nonlinearflctruth130} to express $E'_q$ as a function of state variables of the
truth model. The gains of this controller are the same as the LQR controller that was re-tuned and tested on the reduced order nonlinear model. \autoref{fig:lqrtruth1}
to \autoref{fig:lqrtruth5} show simulation results for the LQR based full state feedback controller applied to the truth model. From these simulation results we can see that the generator terminal voltage $V_t$ oscillates about a steady state value of 1.1705 p.u. which deviates from the desired steady state value of 1.1723 p.u. by an amount of 0.0018 p.u. Angular velocity $\omega $, oscillates about its desired steady state value of 1 p.u., and the rotor angle $\delta $, oscillates about its desired state value of 1 p.u. These oscillations decay with time. Also, the generator 
excitation voltage $V_F$ settles to its steady state value of 0.00121 p.u. and the turbine valve control settles to its
steady state value of 1.0512 p.u.
\begin{equation}
          E'_q=\frac{e_{14}}{e_{11}}I_F+\frac{e_{12}}{e_{11}}\cos(\delta -\alpha )
            +\frac{e_{13}}{e_{11}}\sin(\delta -\alpha ) 
\label{eq:nonlinearflctruth120}
\end{equation}

\begin{equation}
          E'_q=e_{14}I_F+L_2I_d 
\label{eq:nonlinearflctruth130}
\end{equation}
    
\newpage

\subsection{State Feedback Controller design using Pole Placement Technique}

\subsubsection{Pole Placement Design based on linear model}

\begin{figure}
          \centering
          \includegraphics[trim=0cm 0cm 0cm 0cm, clip=true, totalheight=0.27\textheight, width=0.54\textwidth]{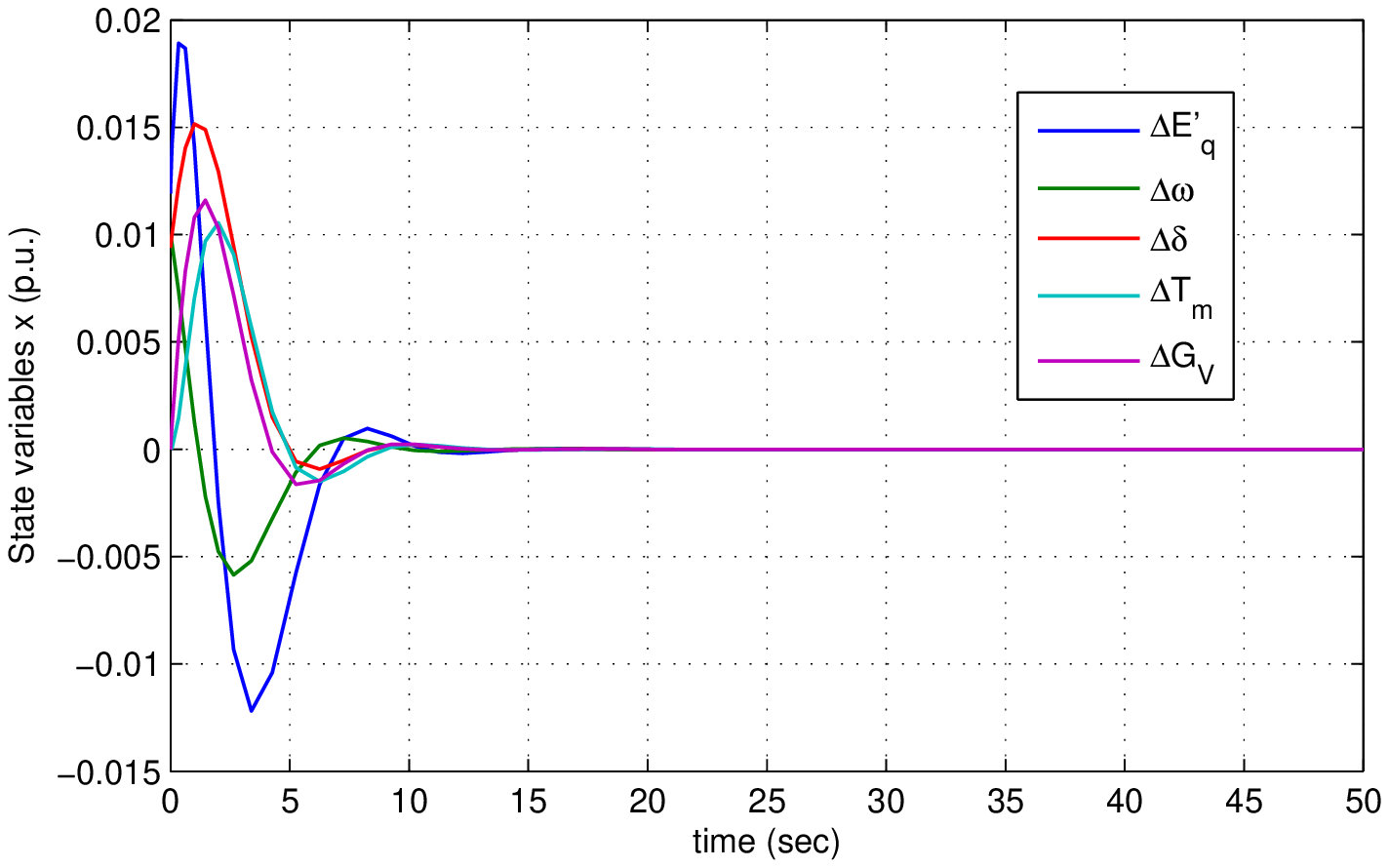}
          \caption{Plot of the state variables $\Delta E'_q$, $\Delta \omega $, $\Delta \delta $, $\Delta T_m$, and 
                   $\Delta G_V$ vs time for the pole placement based full-state feedback controller 
          applied to the reduced order linear model}
          \label{fig:poleplace1}
          \includegraphics[trim=0cm 0cm 0cm 0cm, clip=true, totalheight=0.27\textheight, width=0.54\textwidth]{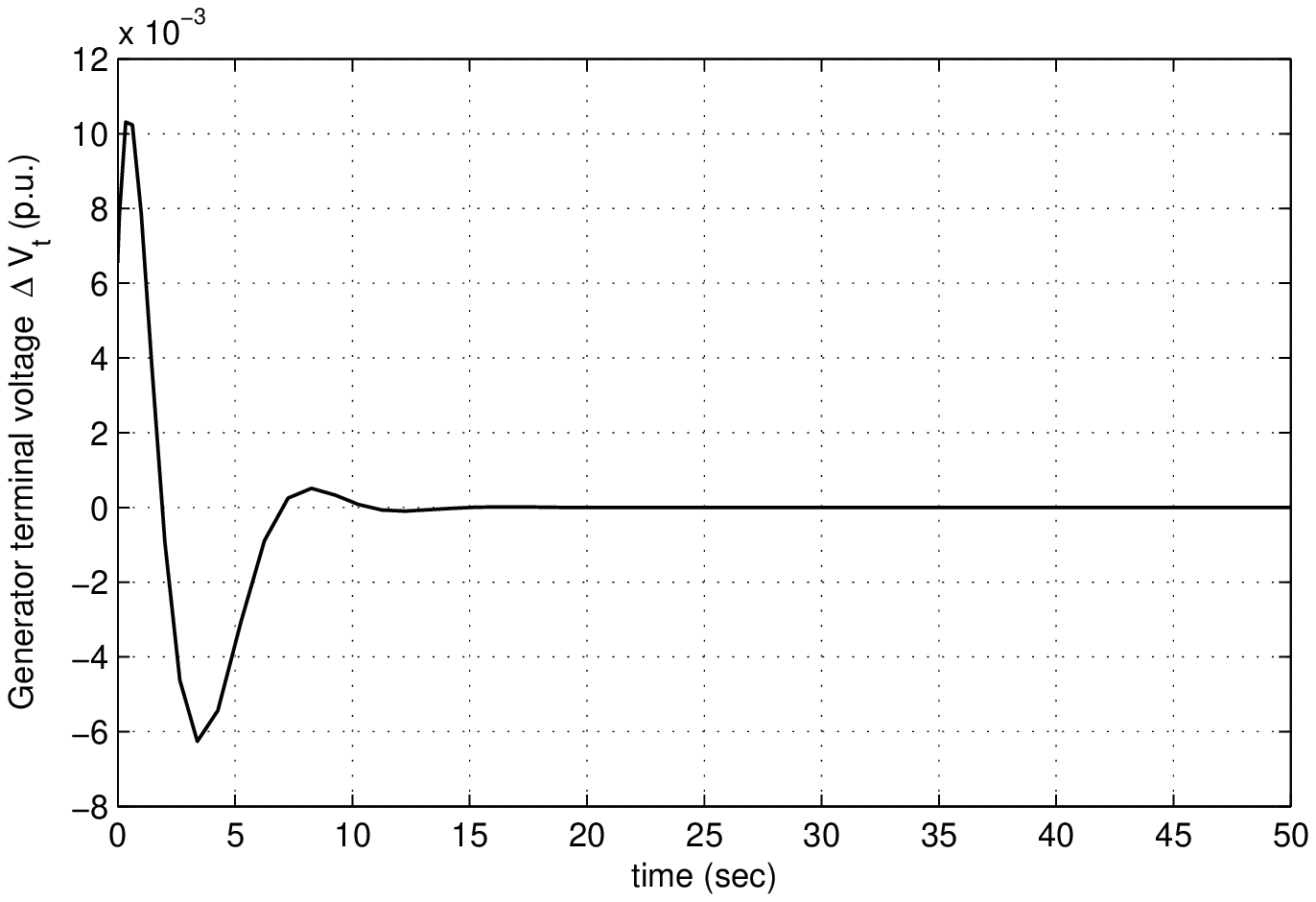}
          \caption{Plot of the generator terminal voltage $\Delta V_t$ vs time for 
          the pole placement based full-state feedback controller 
          applied to the reduced order linear model}
          \label{fig:poleplace2}
          \includegraphics[trim=0cm 0cm 0cm 0cm, clip=true, totalheight=0.27\textheight, width=0.54\textwidth]{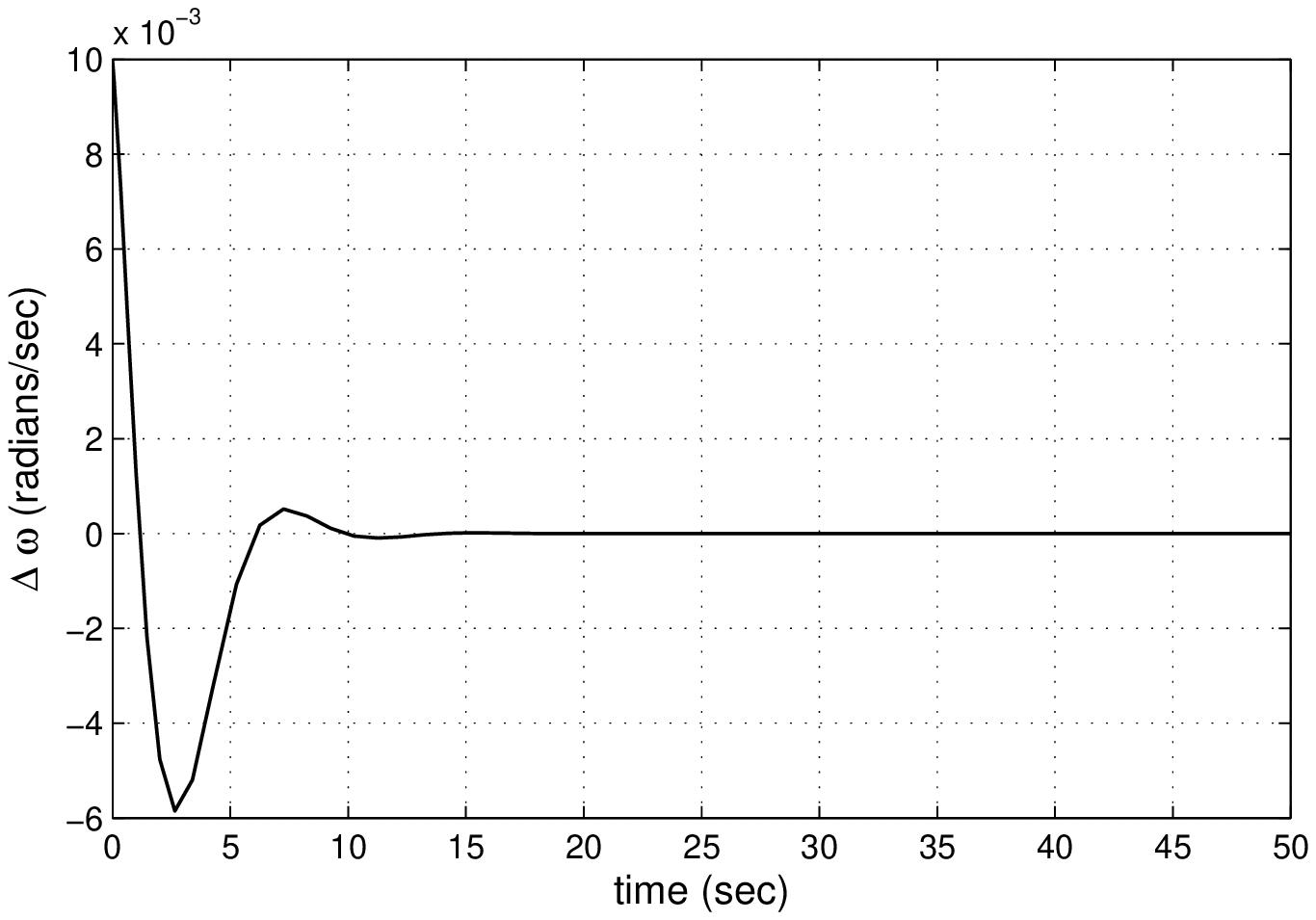}
          \caption{Plot of $\Delta \omega $ vs time for the pole placement based full-state feedback controller 
          applied to the reduced order linear model}
          \label{fig:poleplace3}
\end{figure}

\begin{figure}
          \centering
          \includegraphics[trim=0cm 0cm 0cm 0cm, clip=true, totalheight=0.27\textheight, width=0.54\textwidth]{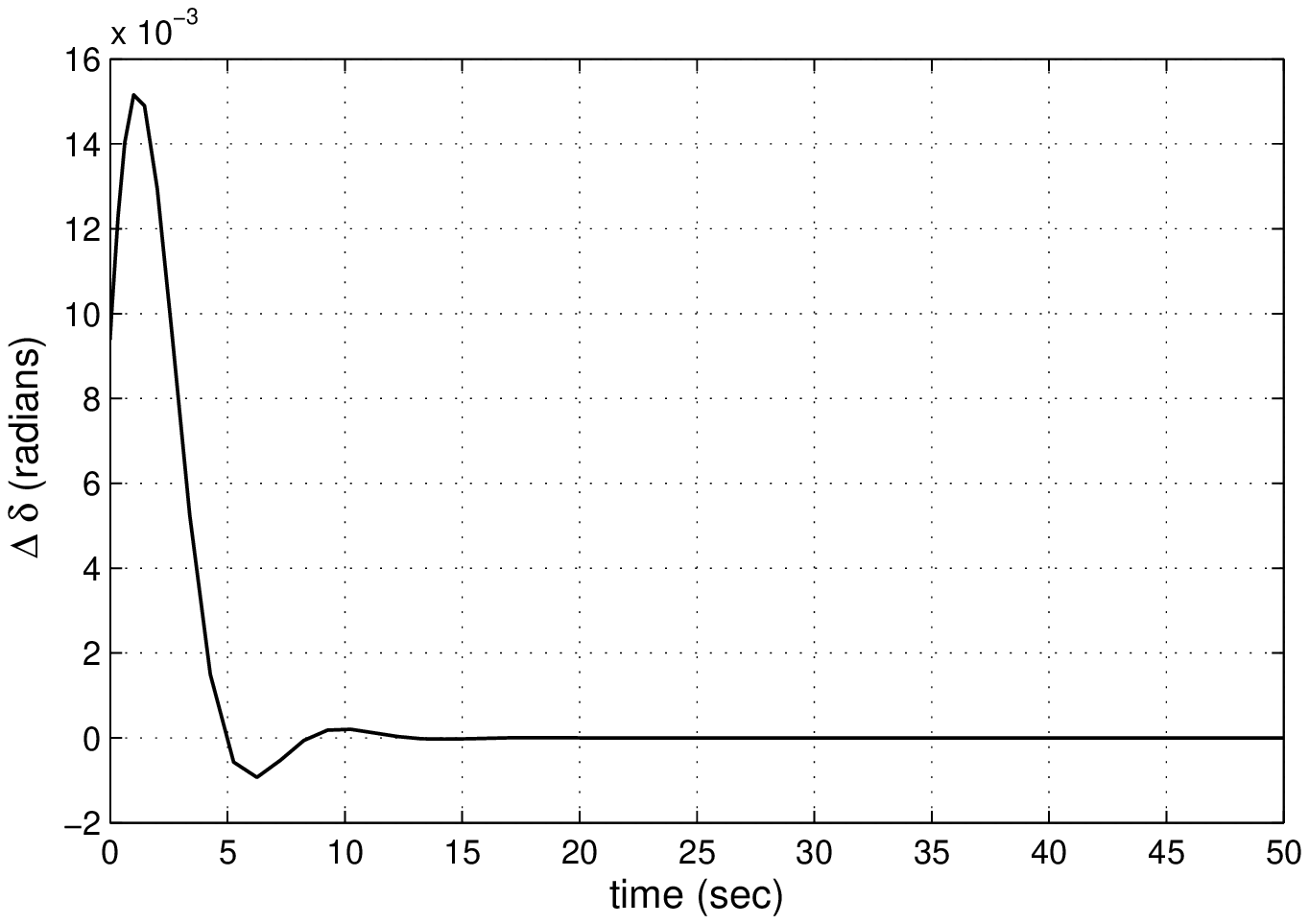}
          \caption{Plot of $\Delta \delta $ vs time for the pole placement based full-state feedback controller 
          applied to the reduced order linear model}
          \label{fig:poleplace4}
          \includegraphics[trim=0cm 0cm 0cm 0cm, clip=true, totalheight=0.27\textheight, width=0.54\textwidth]{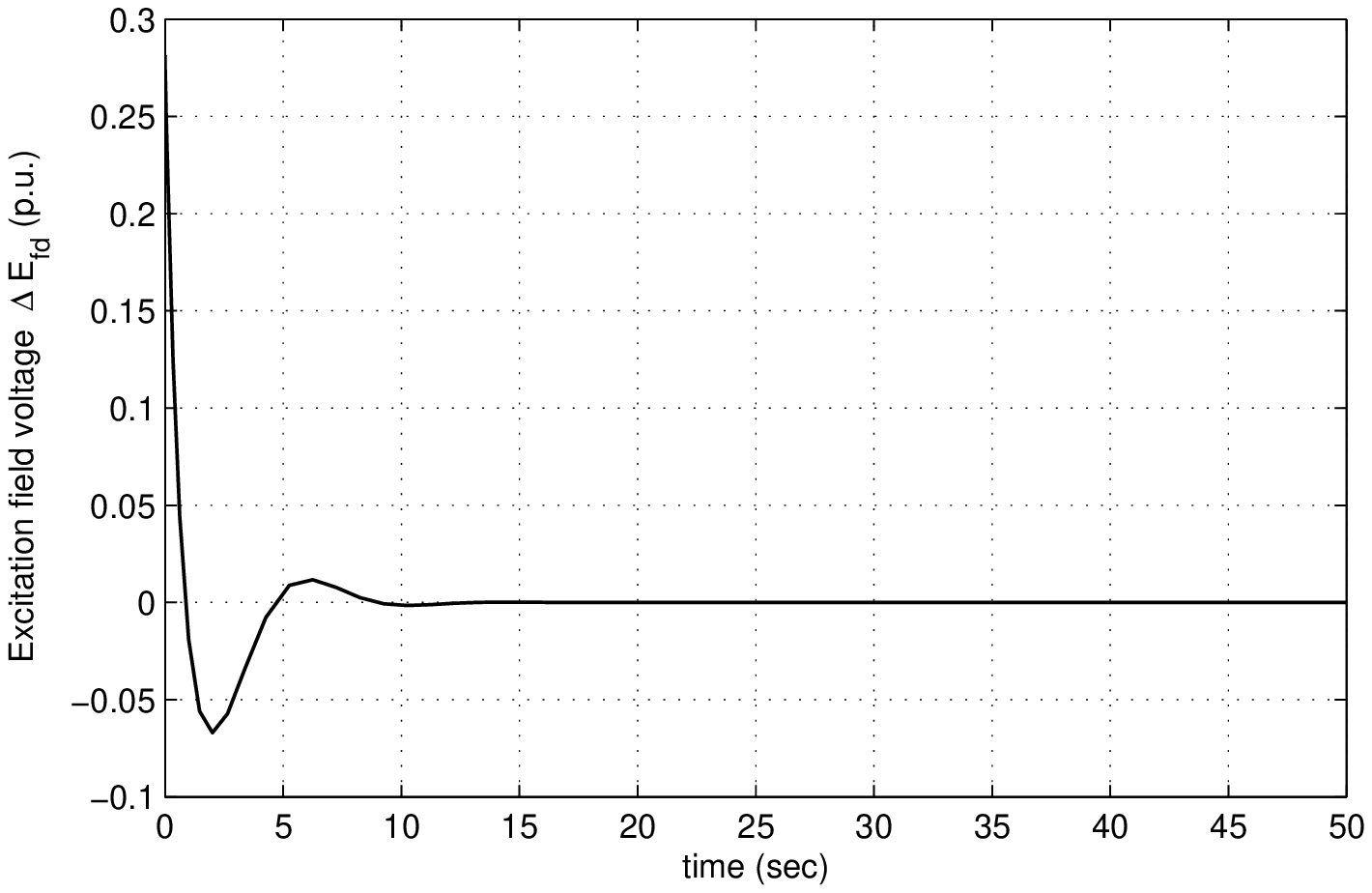}
          \caption{Plot of the control input $\Delta E_{fd}$ vs time for 
          the pole placement based full-state feedback controller 
          applied to the reduced order linear model}
          \label{fig:poleplace5}
          \includegraphics[trim=0cm 0cm 0cm 0cm, clip=true, totalheight=0.27\textheight, width=0.54\textwidth]{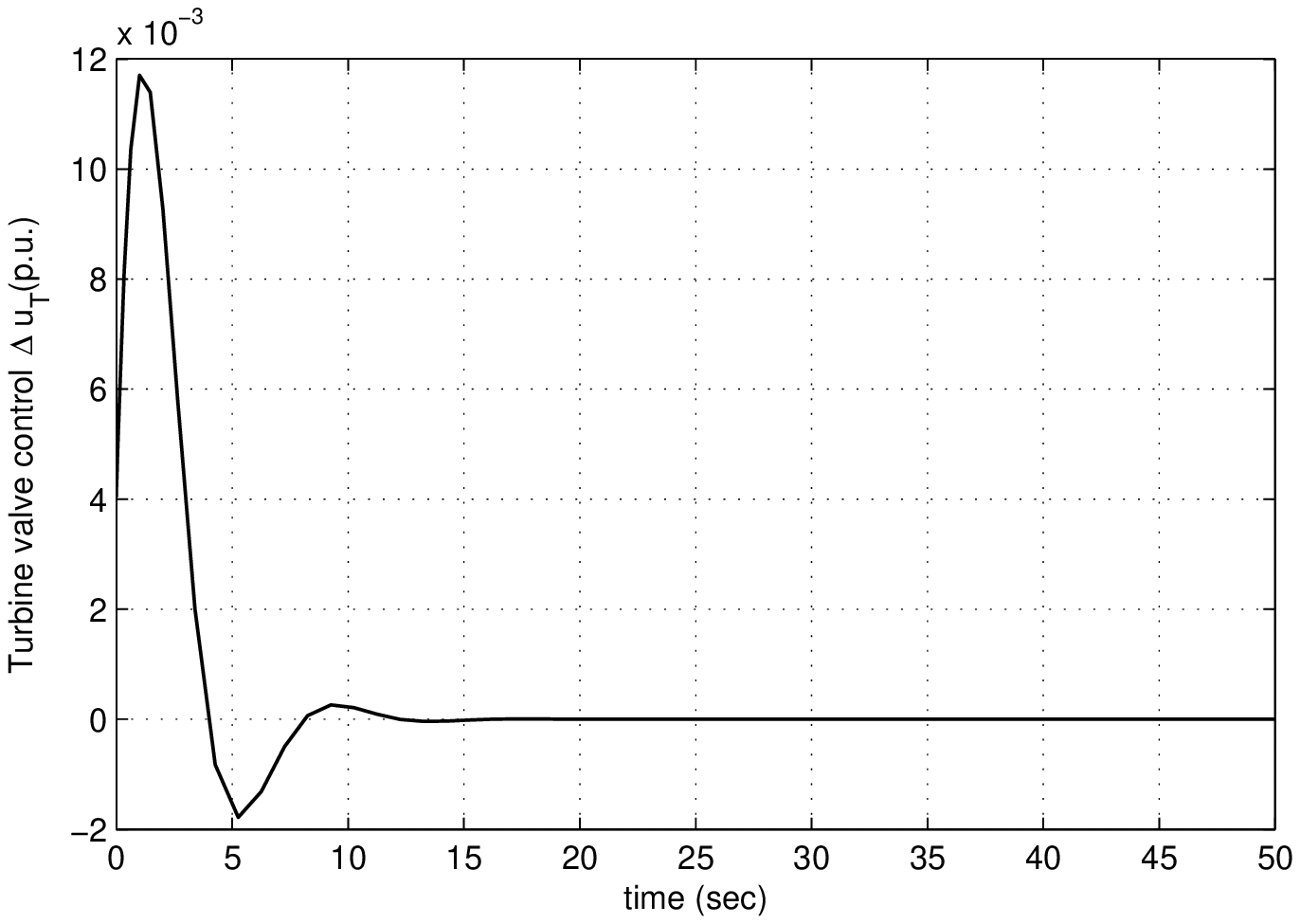}
          \caption{Plot of the control input $\Delta u_T$ vs time for the pole placement based full-state feedback controller 
          applied to the reduced order linear model}
          \label{fig:poleplace6}
\end{figure}

 The characteristic equation of a compensated linear system with controller $\mathbf{u}=-K\mathbf{x}$ is
$|s\mathbf{I}-\mathbf{A}+\mathbf{B}K| =0$. If the system is in the phase-variable canonical form
then the characteristic equation of the compensated linear system is
\begin{equation}
|s\mathbf{I}-\mathbf{A}+\mathbf{B}K| =s^n+(a_{n-1}+k_n)+\cdot \cdot \cdot \cdot +(a_1+k_2)s+(a_0+k_1)=0
\label{eq:pole1}
\end{equation}
where $a_{n-1}$, $a_{n-2}$ $\cdot \cdot \cdot $ $a_1$, and $a_0$ are the coefficients of the characteristic equation 
$|s\mathbf{I}-\mathbf{A}|$ = 0 and $k_n$ $\cdot \cdot \cdot $ $k_1$, are gains of the control matrix $K$. For the specified
closed-loop pole locations $-\lambda _1$, $\cdot \cdot \cdot $ $-\lambda _n$ the desired characteristic equation is

\begin{equation}
      \alpha _c(s)=(s+\lambda _1)(s+\lambda _2)\cdot \cdot \cdot (s+\lambda _n)= s^n+\alpha _{n-1}s^{n-1}+\cdot \cdot \cdot \cdot +\alpha _1s+\alpha _0=0
\label{eq:pole2}
\end{equation} 
The design objective is to find the gain matrix $K$ such that the characteristic equation
for the controlled system is identical to the desired characteristic equation. Thus, the gain vector $K$ is obtained by equating coefficients of \autoref{eq:pole1} and \autoref{eq:pole2} and for the $i^{th}$ coefficient we get 
\begin{equation}
         k_i=\alpha _i-a_i
\label{eq:pole3}
\end{equation} 
If the state model is not in the phase-variable canonical form, we can use the transformation
technique to transform the given state model to the phase-variable canonical
form which results in the following formula, known as Ackermann's formula \cite{RTWQ}. 
\begin{equation}
     K=[0 \ \ 0 \ \ \cdot \cdot \cdot \cdot \ \ 0 \ \ 1]S^{-1}\alpha _c(A)
\label{eq:pole4}
\end{equation}     
where the matrix $S$ is given by
\begin{equation}
           S=[B \ \ AB \ \ A^2B \ \ \cdot \cdot \cdot \cdot \ \ A^{n-1}B]
\label{eq:pole5}
\end{equation} 
and  $\alpha _c(A)$ is given by
\begin{equation}
             \alpha _c(A)=A^n+\alpha _{n-1}A^{n-1}+\cdot \cdot \cdot \cdot +\alpha _1A+\alpha _0I
\label{eq:pole6}
\end{equation}  
The MATLAB function $K$=place$(A,B,p)$ can be used to design the controller gain matrix $K$, where $p$ is a row vector
containing the desired closed-loop poles. The closed-loop poles are selected such that all the state variables remain within a specific limit and converge to zero in minimum time. The closed-loop poles cannot be placed too close to the imaginary axis as the relative stability of the system decreases and the system oscillations increase. Also if they are placed far away from the imaginary axis the settling time of the state variables and hence the output decreases, but there is no guarantee that the state variables and the control inputs will remain within a reasonable
physical limit. After some trial and error the desired closed-loop poles of the generator-turbine system were selected as
\begin{equation}
        p=[-0.8, -0.9, -0.7, -1.1, -1]
\label{eq:pole7}
\end{equation}  
The controller gain matrix $K$ corresponding to these closed loop poles is 
\begin{equation}      
K =\bbm 12.6444 & -44.1610  &  0.9926  & -3.1472  & -8.9082\\
    0.0535  & -0.3867 &  -0.0861 &   0.0788  &  -1.0390\ebm
\label{eq:pole8}
\end{equation}      
\autoref{fig:poleplace1} to \autoref{fig:poleplace4} show that the state variables 
$\Delta E'_q$, $\Delta \omega $, $\Delta \delta $, $\Delta T_m$, $\Delta G_V$, and the outputs $\Delta V_t$,  $\Delta \omega $ and $\Delta \delta $ settle to their steady state value of zero in approximately 10 seconds. \autoref{fig:poleplace5} and \autoref{fig:poleplace6} show plots for the two control inputs $\Delta E_{fd}$, and $\Delta u_T$ respectively.

\newpage

\subsubsection{Simulation Results for the Full-State Feedback Pole Placement controller applied to the Reduced Order Nonlinear
Model}

\begin{figure}
          \centering      
          \includegraphics[trim=0cm 0cm 0cm 0cm, clip=true, totalheight=0.27\textheight, 
           width=0.54\textwidth]   {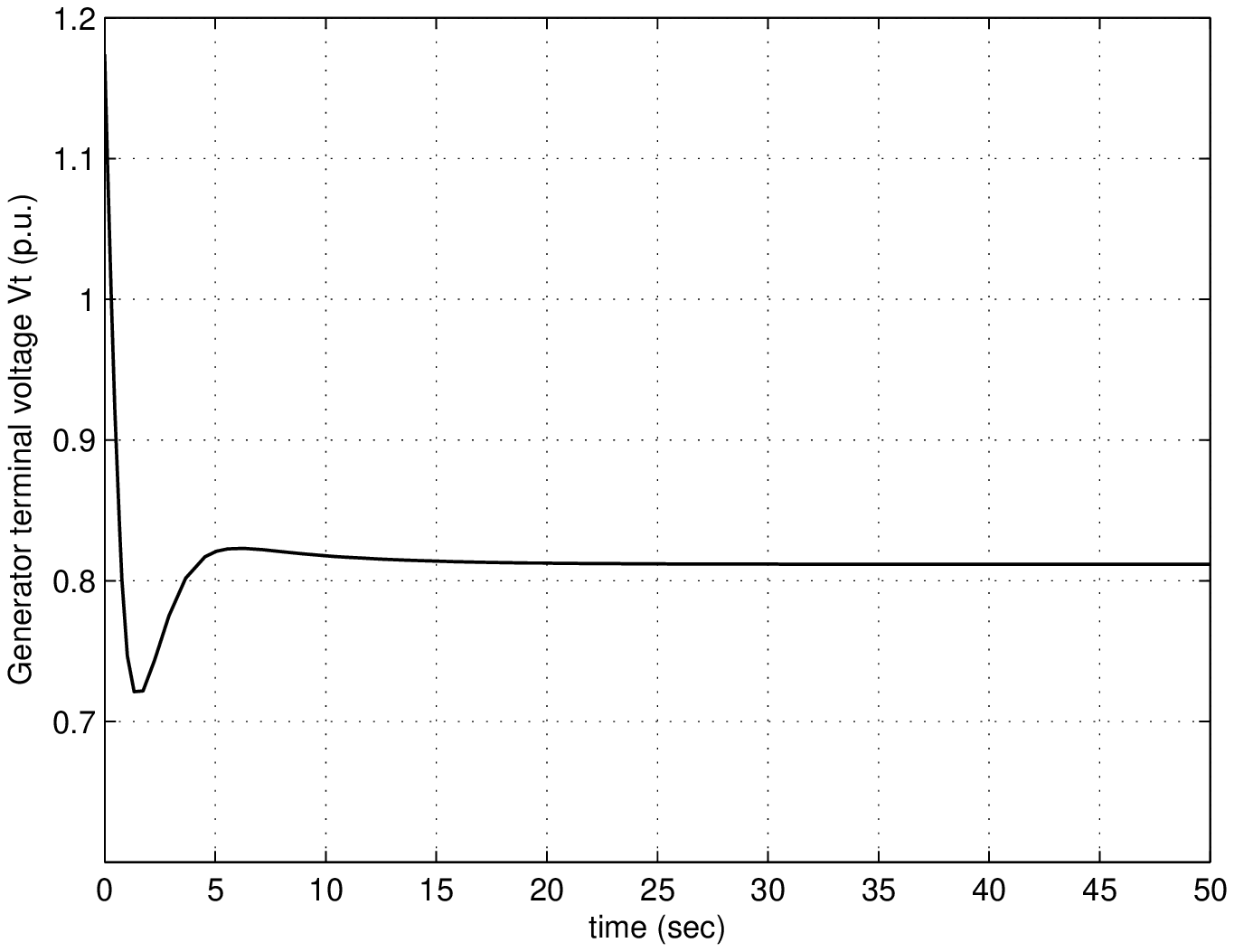}
          \caption{Plot of the generator terminal voltage $V_t$ vs time for the pole placement based full-state feedback 
           controller with the original gains of subsection 7.2.1 applied to the reduced 
           order nonlinear  model}
          \label{fig:polenonlineartest}
          \includegraphics[trim=0cm 0cm 0cm 0cm, clip=true, totalheight=0.27\textheight, 
           width=0.54\textwidth]{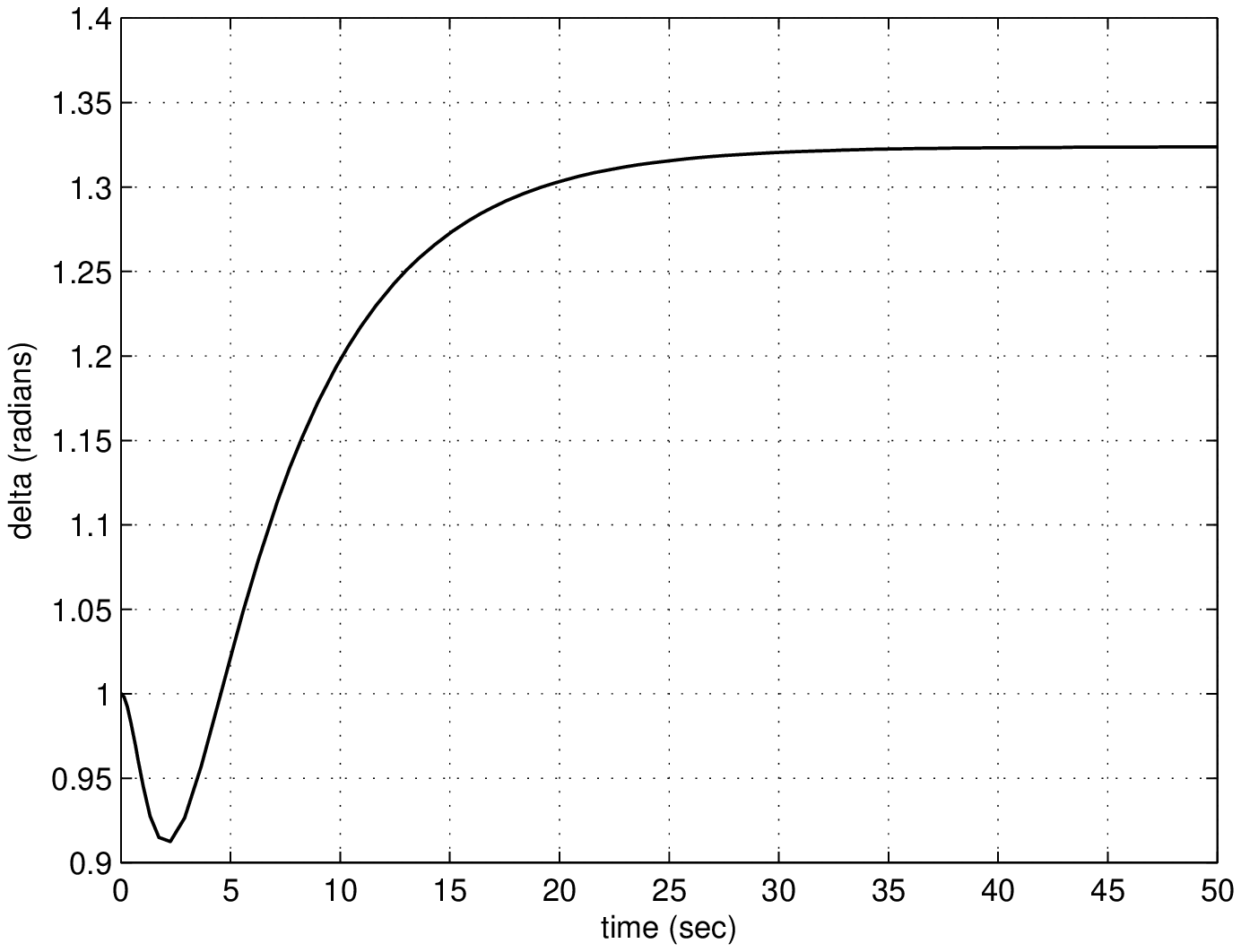}
          \caption{Plot of the rotor angle $\delta $ vs time for the pole placement based full-state feedback controller with the original gains of subsection 7.2.1 applied to 
          the reduced order nonlinear model}
          \label{fig:polenonlineartest0}
          \includegraphics[trim=0cm 0cm 0cm 0cm, clip=true, totalheight=0.27\textheight, width=0.54
           \textwidth]  {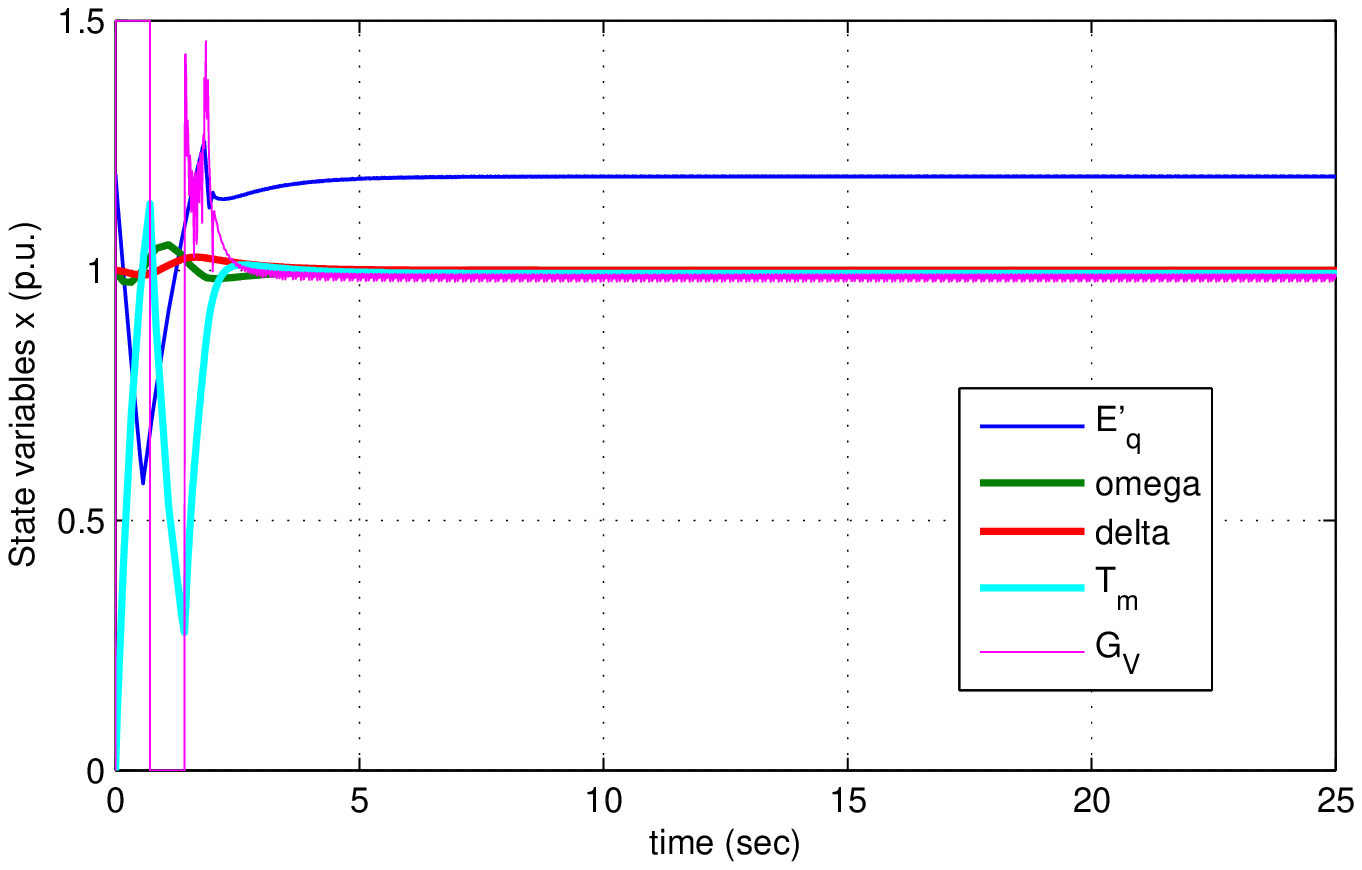}
          \caption{Plot of the state variables $E'_q$, $\omega $, $\delta $, 
           $T_m$, and $G_V$ vs time for the pole placement based full-state feedback controller applied to the 
          reduced order nonlinear model, with the gains re-tuned}
          \label{fig:polenonlineartest1}
\end{figure}

\begin{figure}
          \centering    
          \includegraphics[trim=0cm 0cm 0cm 0cm, clip=true, totalheight=0.27\textheight, 
           width=0.54\textwidth]{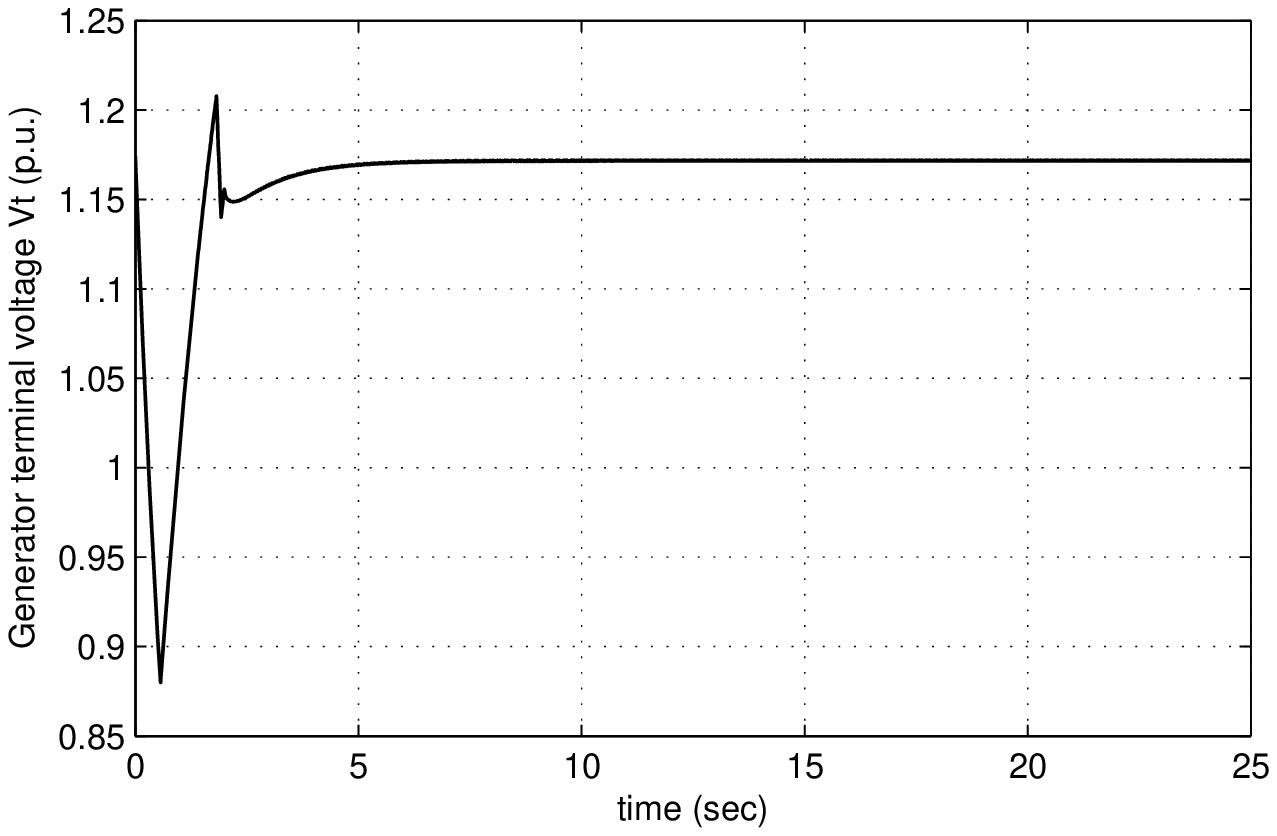}
          \caption{Plot of the generator terminal voltage $V_t$ vs time for the pole placement based full-state feedback 
           controller applied to the reduced 
           order nonlinear  model, with the gains re-tuned}
          \label{fig:polenonlineartest2}
          \includegraphics[trim=0cm 0cm 0cm 0cm, clip=true, totalheight=0.27\textheight, 
           width=0.54\textwidth]{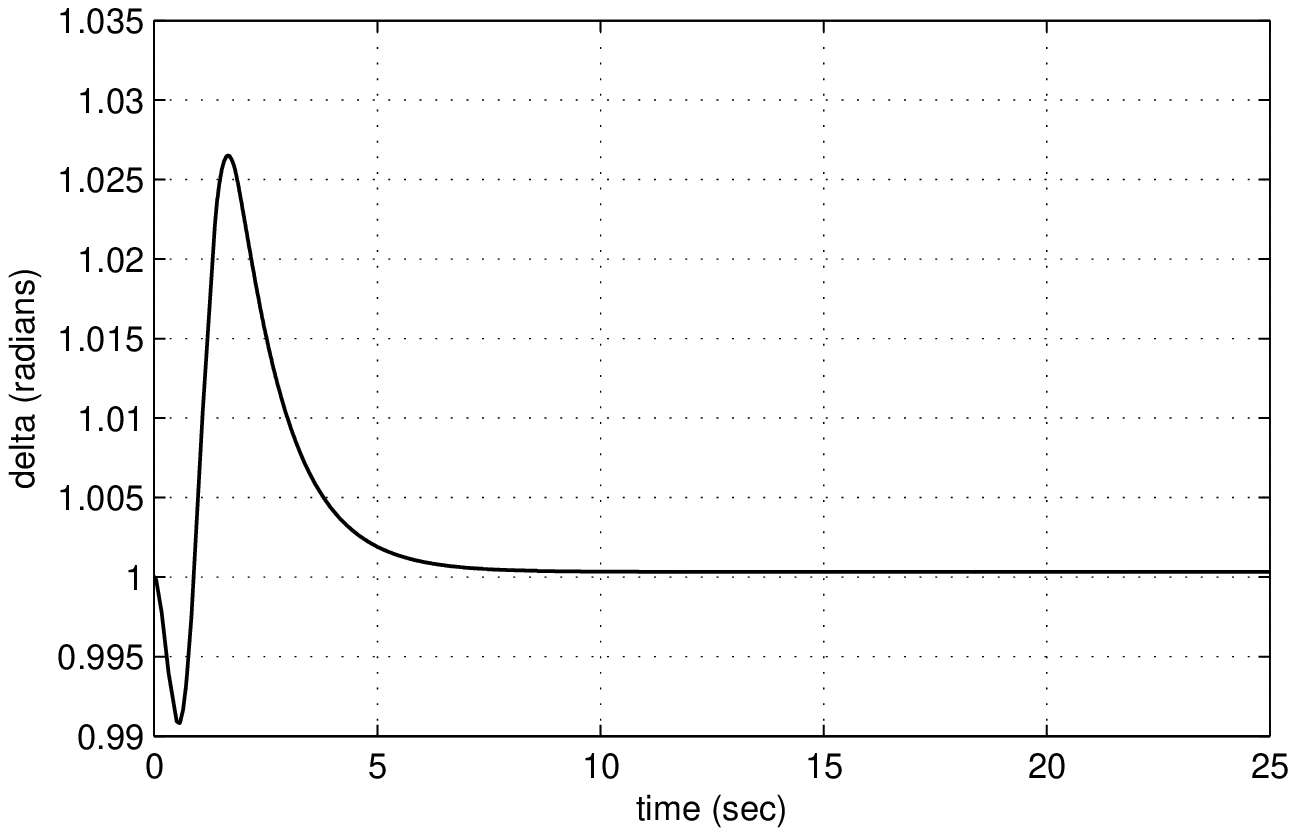}
          \caption{Plot of the rotor angle $\delta $ vs time for the pole placement based full-state feedback controller applied to 
          the reduced order nonlinear model, with the gains re-tuned}
          \label{fig:polenonlineartest3}
          \includegraphics[trim=0cm 0cm 0cm 0cm, clip=true, totalheight=0.27\textheight, 
          width=0.54\textwidth]{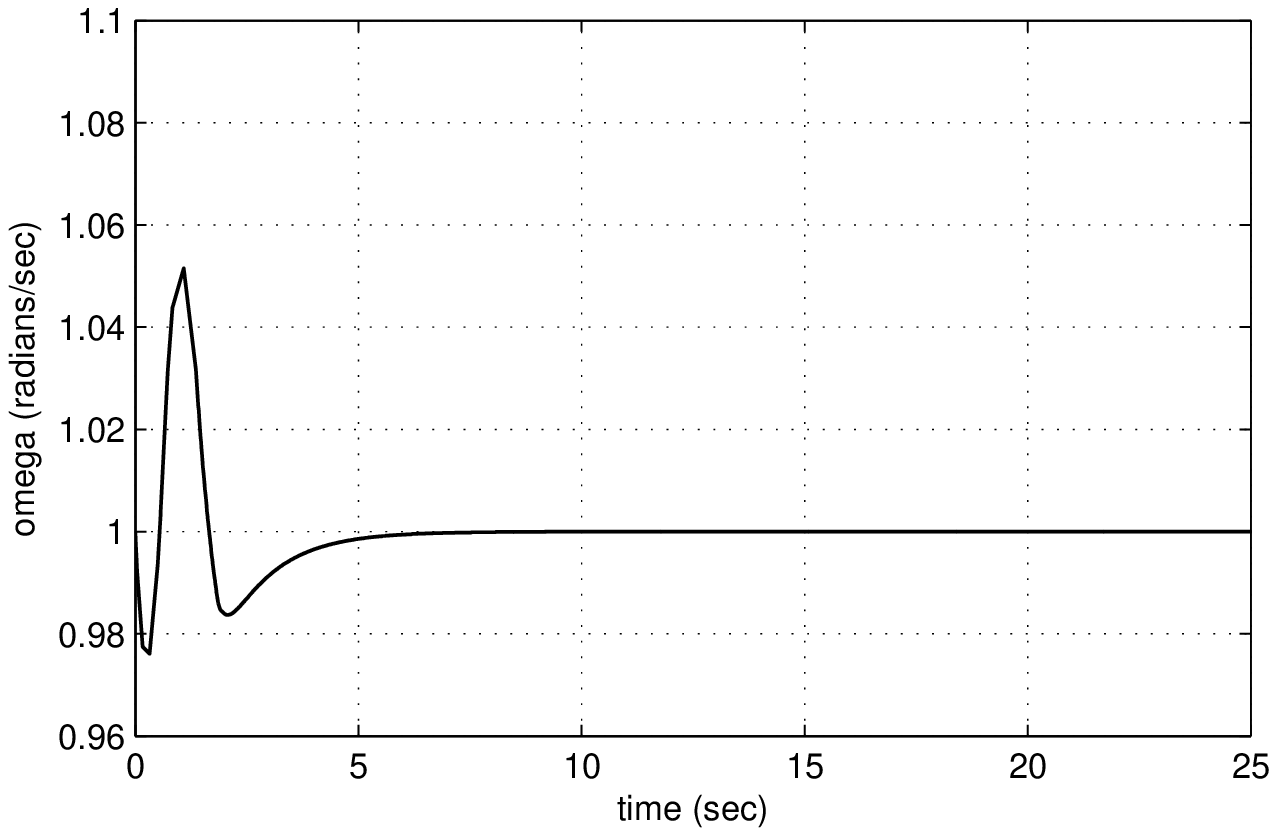}
          \caption{Plot of the frequency $\omega  $ vs time for the pole placement based full-state feedback controller 
          applied to the reduced order nonlinear model, with the gains re-tuned}
          \label{fig:polenonlineartest4}          
\end{figure}

We now test the pole placement based full-state feedback controller that was designed earlier for 
the reduced order linear model, on the reduced order nonlinear model. \autoref{fig:polenonlineartest} and \autoref{fig:polenonlineartest0} show plots for the generator terminal voltage
$V_t$, and rotor angle $\delta $, respectively. The generator terminal voltage
$V_t$ settles to a new steady state value of 0.8116 p.u. which deviates from the desired steady state value of 1.1723 p.u. by an amount
of 0.3607 p.u. Similarly the rotor angle $\delta $ settles to a new steady state value of 1.324 p.u. which deviates from the desired steady state value of 1 p.u. by an amount of 0.324 p.u. We observe a large steady state error when the pole placement controller with original gains of subsection 7.2.1
is applied to the reduced order nonlinear model.
Therefore, the gains of the controller are once again tuned by
appropriately choosing the desired pole locations, to get satisfactory performance.
The MATLAB function $K$=place$(A,B,p)$ can be used to design the controller gain matrix $K$, where $p$ is a row vector
containing the desired closed-loop poles. After some trial and error the desired closed-loop poles of the 
generator-turbine system were selected as
\begin{equation}
        p=[-300, -0.9, -280, -5, -70]
\label{eq:poletest1}
\end{equation} 
We see here that poles p(1), p(3), and p(5) are placed far away from the imaginary axis so that the state variables $E'_q$, $\delta $,
and $G_V$ attain their desired steady state values in less time. At the same time care is taken that these state variables
stay within a safe physical limit. 
The controller gain matrix $K$ corresponding to these closed loop poles is 
\begin{equation}      
K =\bbm 2020 & -376860 &  -304950  & -18250  & -130\\
    -10  &  4840  &  3770  &  330  &  60\ebm
\label{eq:poletest2}
\end{equation}      
\autoref{fig:polenonlineartest1}-\autoref{fig:polenonlineartest4} show that the state variables 
$E'_q$, $\omega $, $\delta $, $T_m$, $G_V$, and the outputs $V_t$ and $\delta $
settle to their respective steady state values in approximately 5 to 8 seconds.
  
\newpage
\subsubsection{Simulation Results for the Full-State Feedback Pole Placement controller applied to the Truth
Model}

\begin{figure}
          \centering
          \includegraphics[trim=0cm 0cm 0cm 0cm, clip=true, totalheight=0.27\textheight, width=0.54\textwidth]{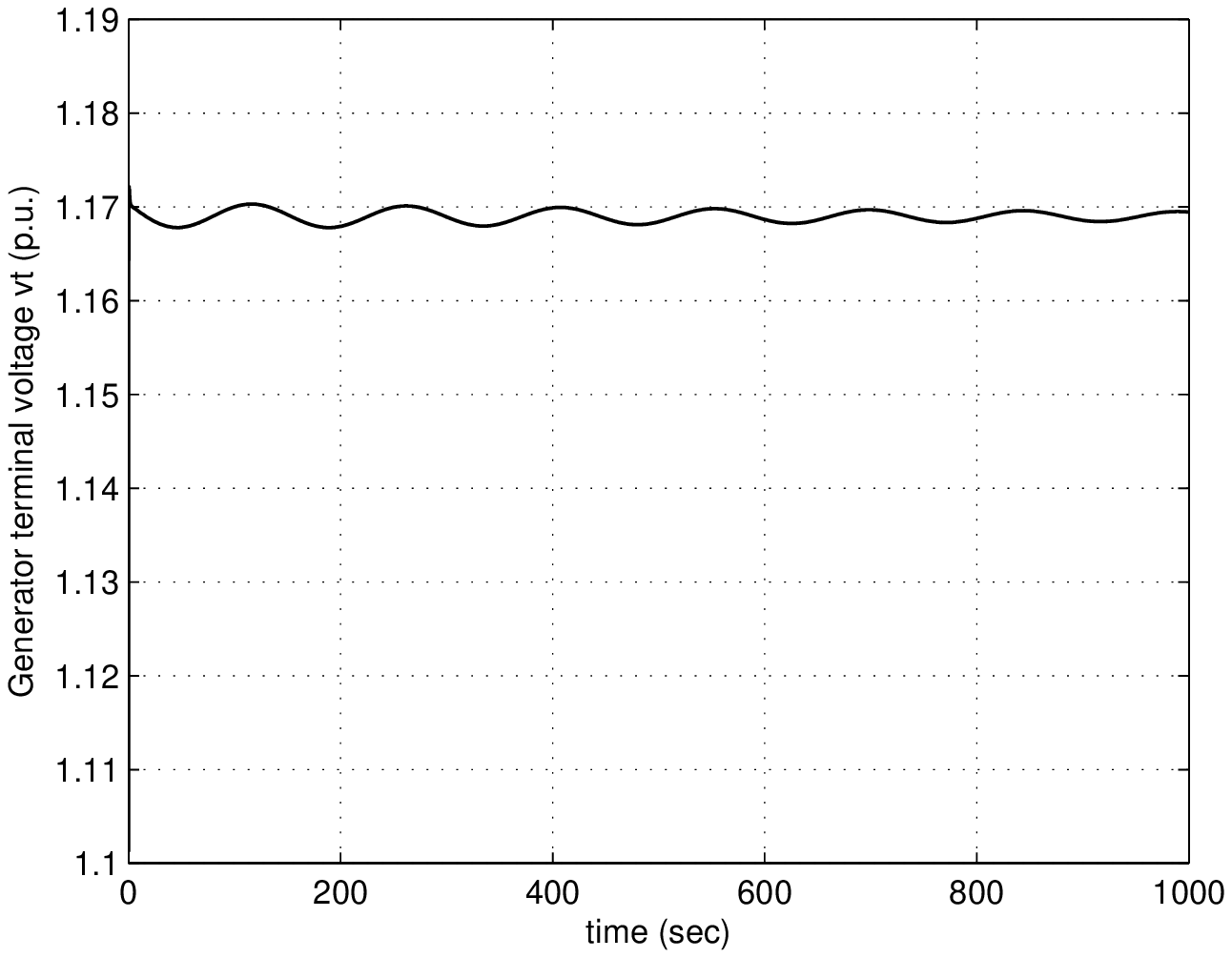}
          \caption{Plot of the generator terminal voltage $V_t$ vs time for 
          the pole placement based full-state feedback controller 
          applied to the truth model}
          \label{fig:poleplacetruth1}
          \includegraphics[trim=0cm 0cm 0cm 0cm, clip=true, totalheight=0.27\textheight, width=0.54\textwidth]{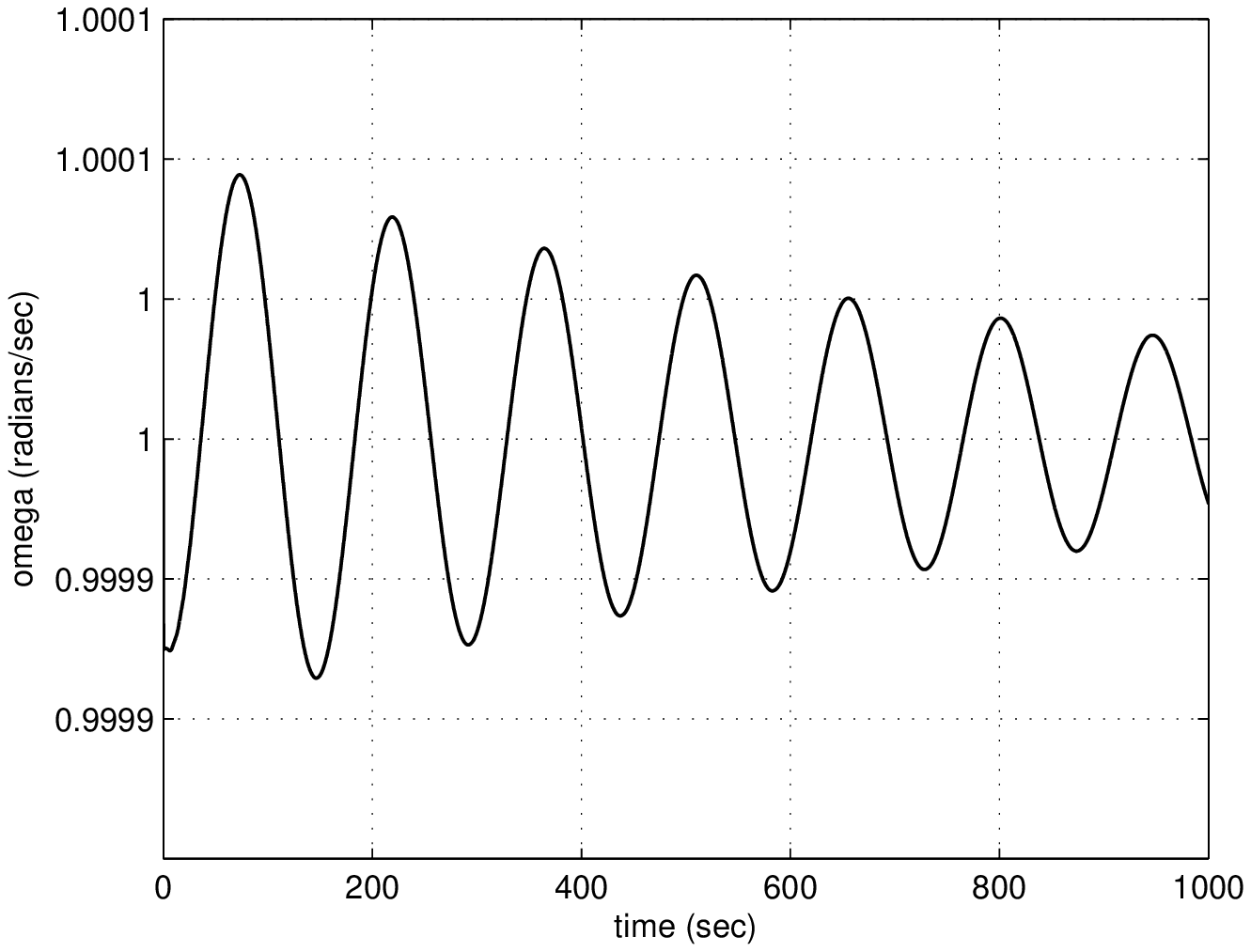}
          \caption{Plot of $\omega $ vs time for the pole placement based full-state feedback controller 
          applied to the truth model}
          \label{fig:poleplacetruth2}
          \includegraphics[trim=0cm 0cm 0cm 0cm, clip=true, totalheight=0.27\textheight, width=0.54\textwidth]{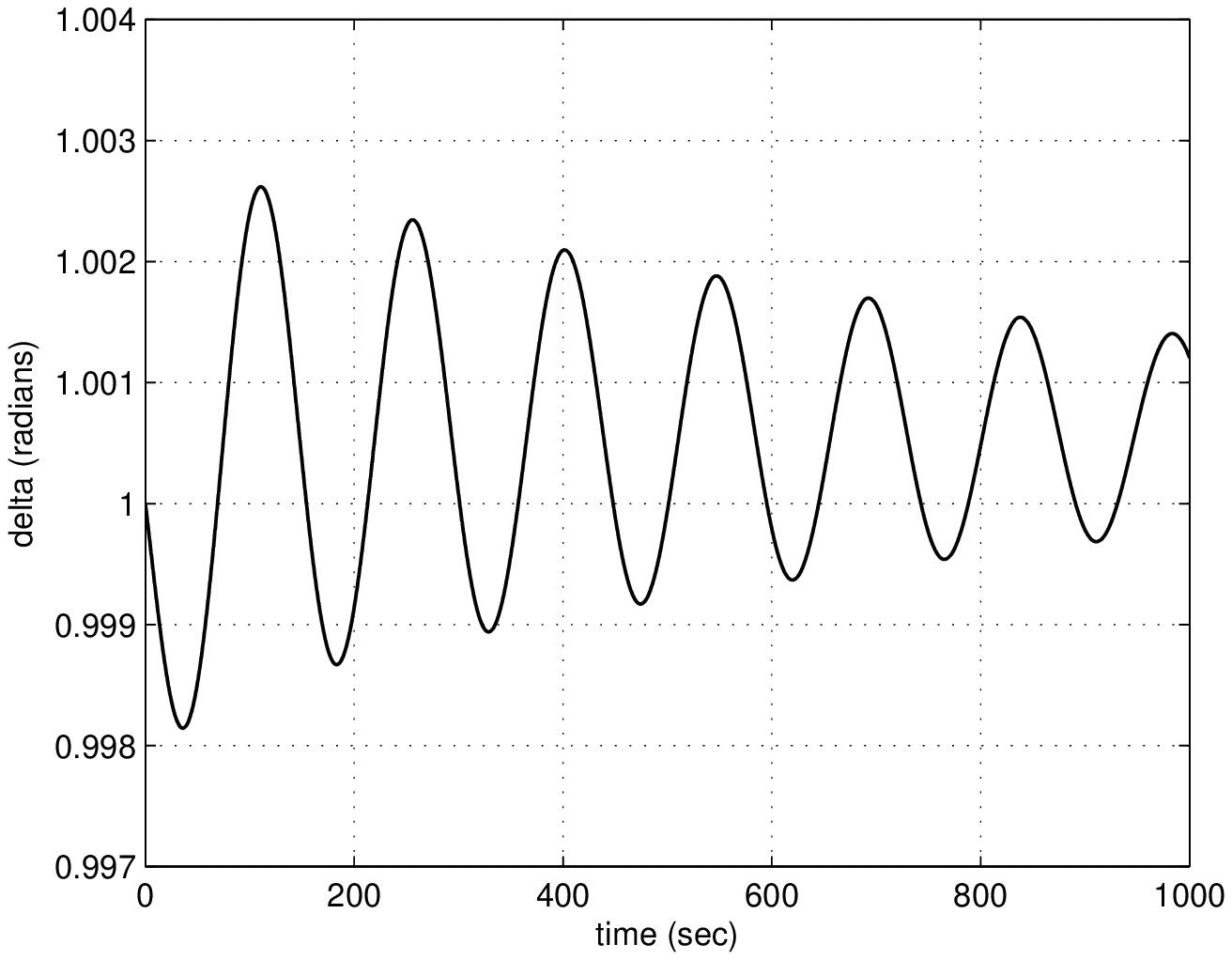}
          \caption{Plot of $\delta $ vs time for the pole placement based full-state feedback controller 
          applied to the truth model}
          \label{fig:poleplacetruth3}
\end{figure}

\begin{figure}
          \centering          
          \includegraphics[trim=0cm 0cm 0cm 0cm, clip=true, totalheight=0.27\textheight, width=0.54\textwidth]{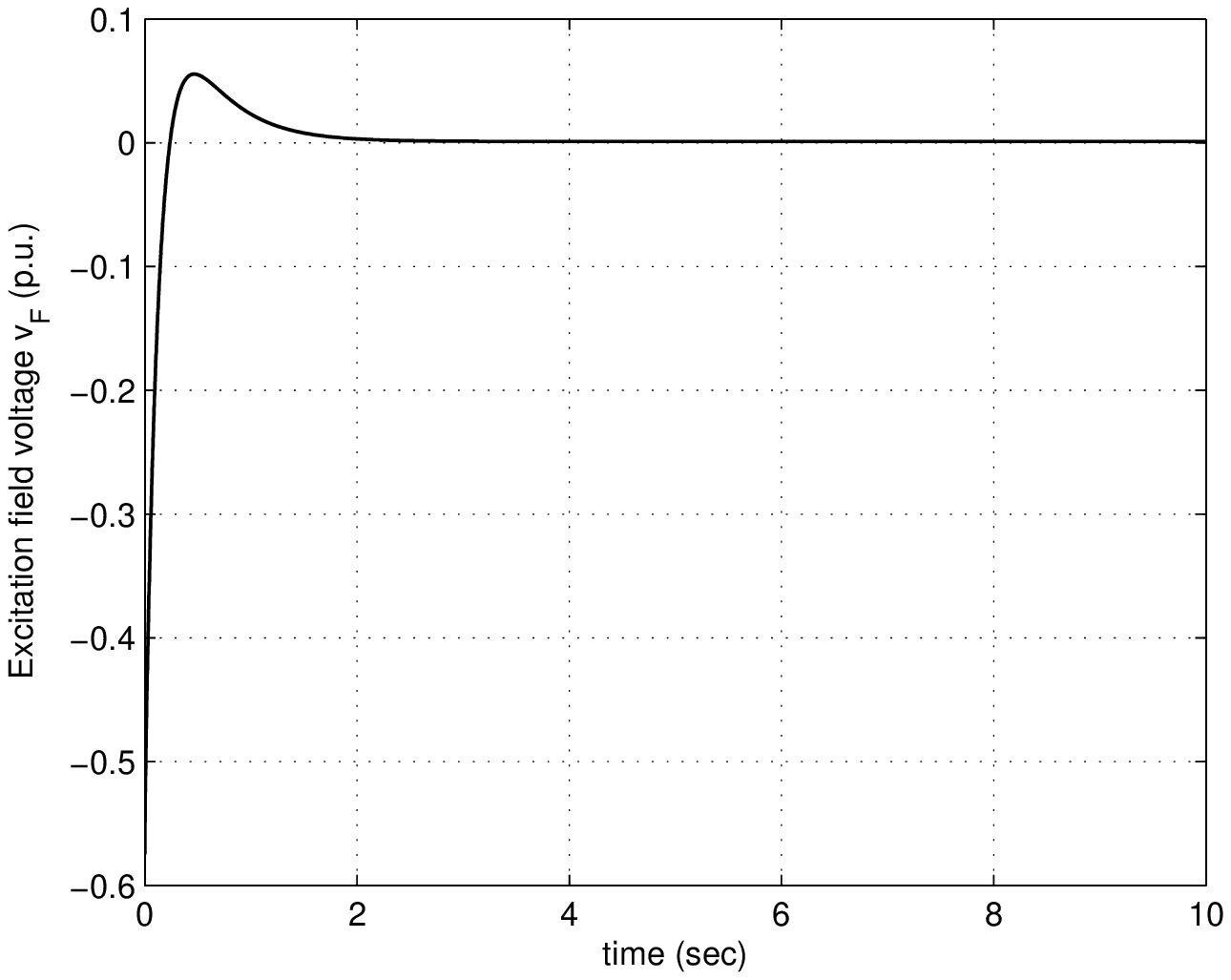}
          \caption{Plot of the control input $V_F$ vs time for 
          the pole placement based full-state feedback controller 
          applied to the truth model}
          \label{fig:poleplacetruth4}
          \includegraphics[trim=0cm 0cm 0cm 0cm, clip=true, totalheight=0.27\textheight, width=0.54\textwidth]{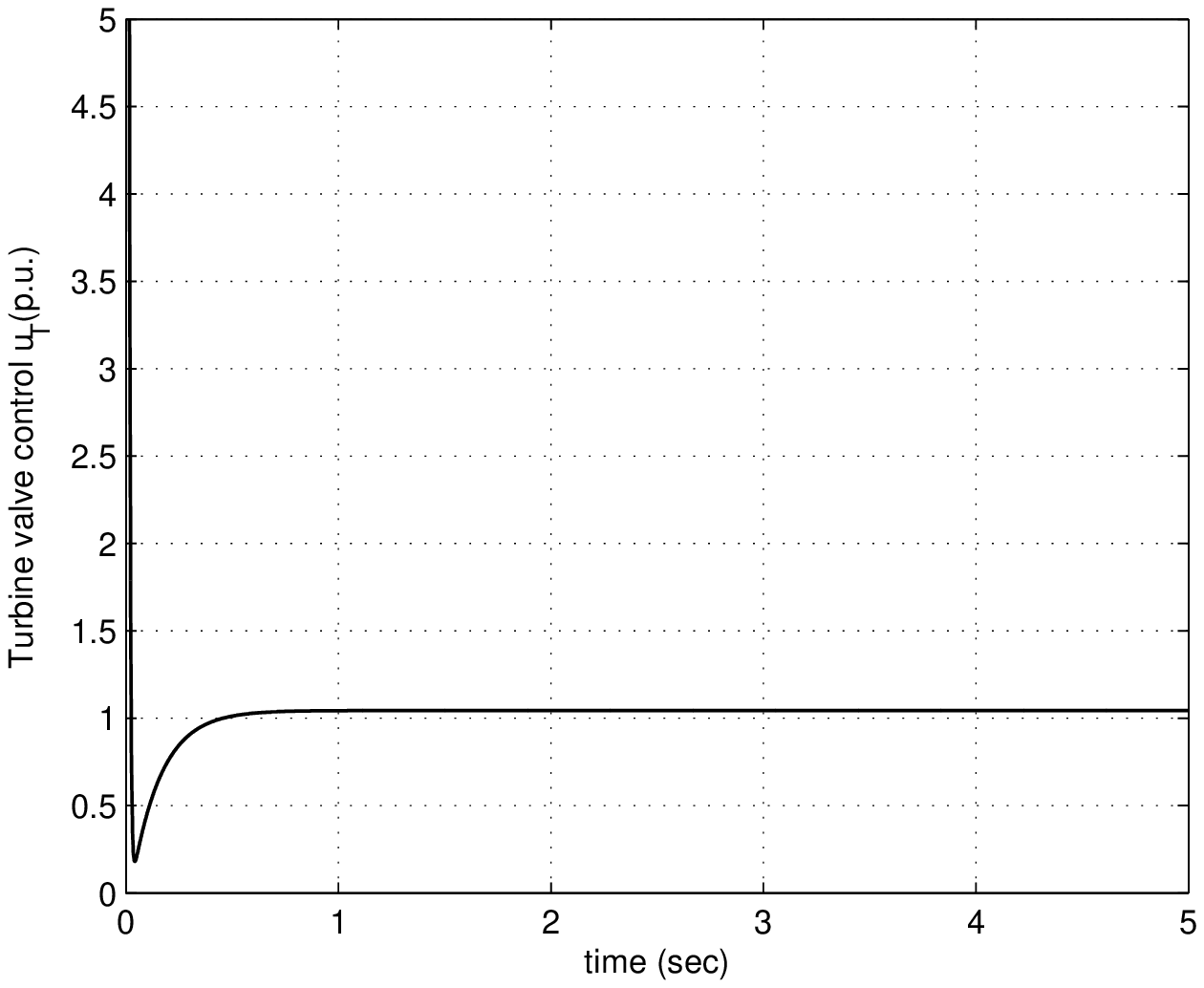}
          \caption{Plot of the control input $u_T$ vs time for the pole placement based full-state feedback controller 
          applied to the truth model}
          \label{fig:poleplacetruth5}
\end{figure}

We now test the pole placement based full-state feedback controller on the truth model. 
The gains of the controller are re-tuned by
appropriately choosing the desired pole locations, to get satisfactory performance.
The MATLAB function $K$=place$(A,B,p)$ is used to design the controller gain matrix $K$, where $p$ is a row vector
containing the desired closed-loop poles. After some trial and error the desired closed-loop poles of the 
generator-turbine system were selected as
\begin{equation}
        p=[-8.00+j0.05, -8.00-j0.05, -200, -250, -0.1]
\label{eq:poletruthtest1}
\end{equation} 
The controller gain matrix $K$ corresponding to these closed loop poles is 
\begin{equation}      
K =\bbm 1526.38 & -48186.24 &  -3062.64  & -1166.75 & -32.7\\
    -0.9566  &  1321.51 &  49.346  &  103.52  &  39.965\ebm
\label{eq:poletruthtest2}
\end{equation}   
\autoref{fig:poleplacetruth1} to \autoref{fig:poleplacetruth5} show simulation results
for the pole placement based full-state feedback controller applied to the truth model. From these results we can see that $V_t$ 
oscillates about a steady state value of 1.17 p.u. which deviates from the desired steady state value of 1.1723 p.u. by an amount of 0.0023 p.u., $\omega $ oscillates about its desired steady state value of 1 p.u., and $\delta $ oscillates about a steady state value of 1.0005 p.u. which deviates from the desired steady state value of 1 p.u. by an amount of 0.0005 p.u.. Small
oscillations which decay with time are seen for $\omega $, and $\delta $. \autoref{fig:poleplacetruth4} and 
\autoref{fig:poleplacetruth5} show plots for the two control inputs $V_F$ and $u_T$ respectively. As seen from these plots, the generator 
excitation voltage $V_F$ settles to its steady state value of 0.00121 p.u. and the turbine valve control settles to its
steady state value of 1.0512 p.u. 
   
\newpage
\subsection{Output Feedback Controller Design}

\subsubsection{Observer-based LQR Design based on linear model}

\begin{figure}
          \centering
          \includegraphics[trim=0cm 0cm 0cm 0cm, clip=true, totalheight=0.275\textheight, width=0.58\textwidth]{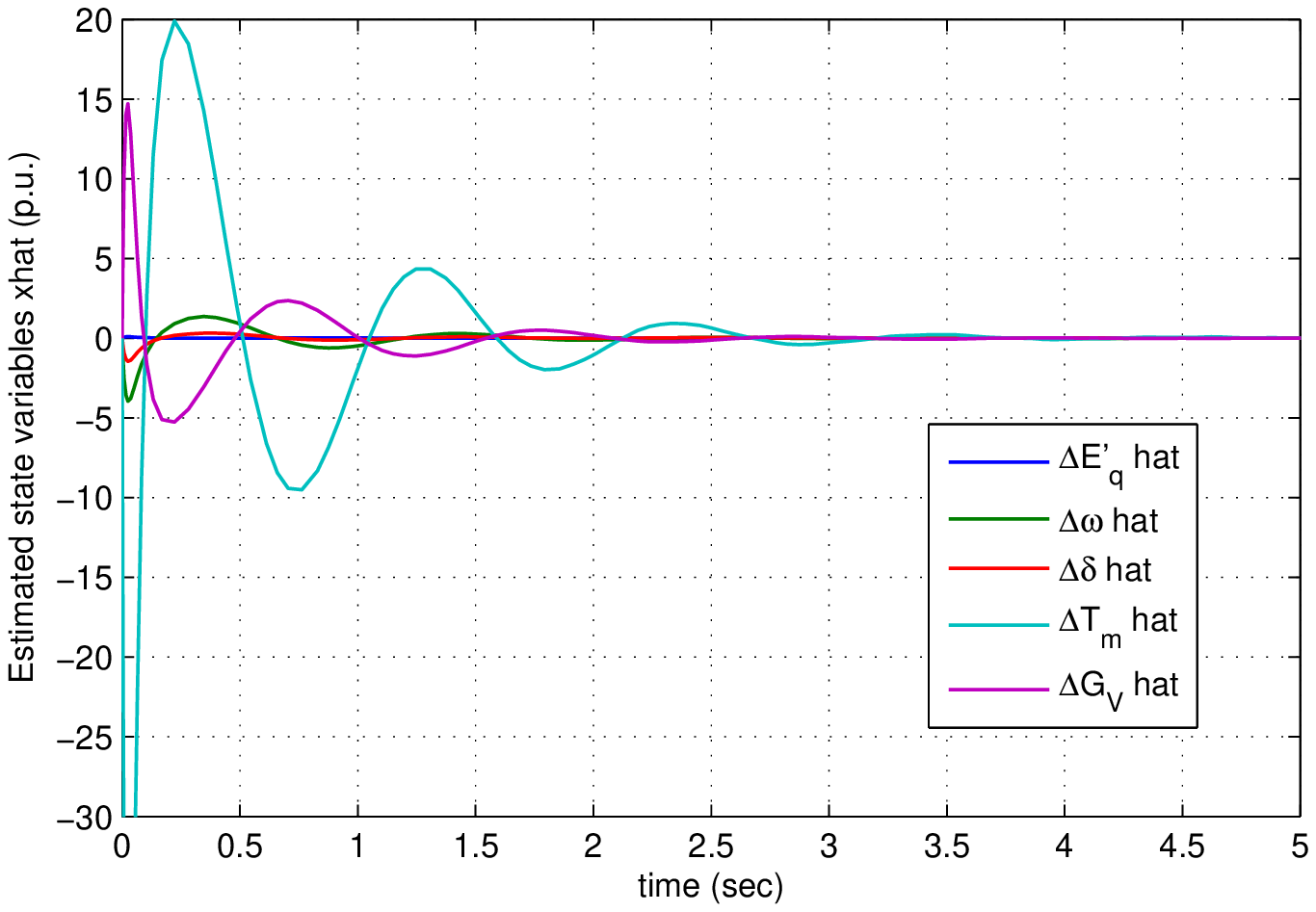}
          \caption{Plot of the estimated state variables $\Delta \hat{E'}_q$, $\Delta \hat{\omega }$, $\Delta \hat{\delta }$, 
           $\Delta \hat{T}_m$, and $\Delta \hat{G}_V$ vs time for the observer-based LQR 
           applied to the reduced order linear model with output $\Delta V_t$ measured}
          \label{fig:lqrobs1}
          \includegraphics[trim=0cm 0cm 0cm 0cm, clip=true, totalheight=0.275\textheight, width=0.58\textwidth]{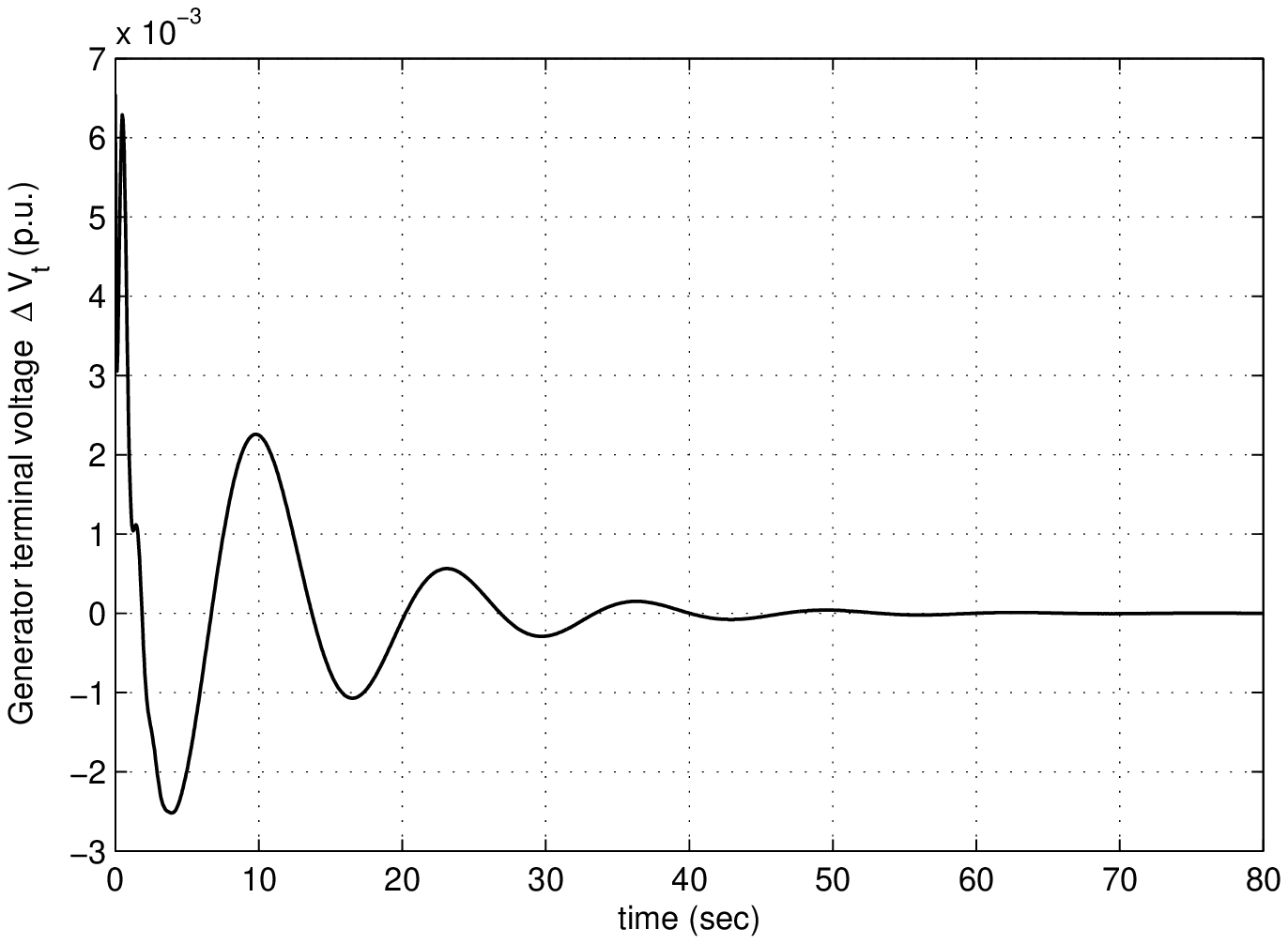}
          \caption{Plot of the generator terminal voltage $\Delta V_t$ vs time for the observer-based LQR  
          applied to the reduced order linear model with output $\Delta V_t$ measured}
          \label{fig:lqrobs2}
          \includegraphics[trim=0cm 0cm 0cm 0cm, clip=true, totalheight=0.275\textheight, width=0.58\textwidth]{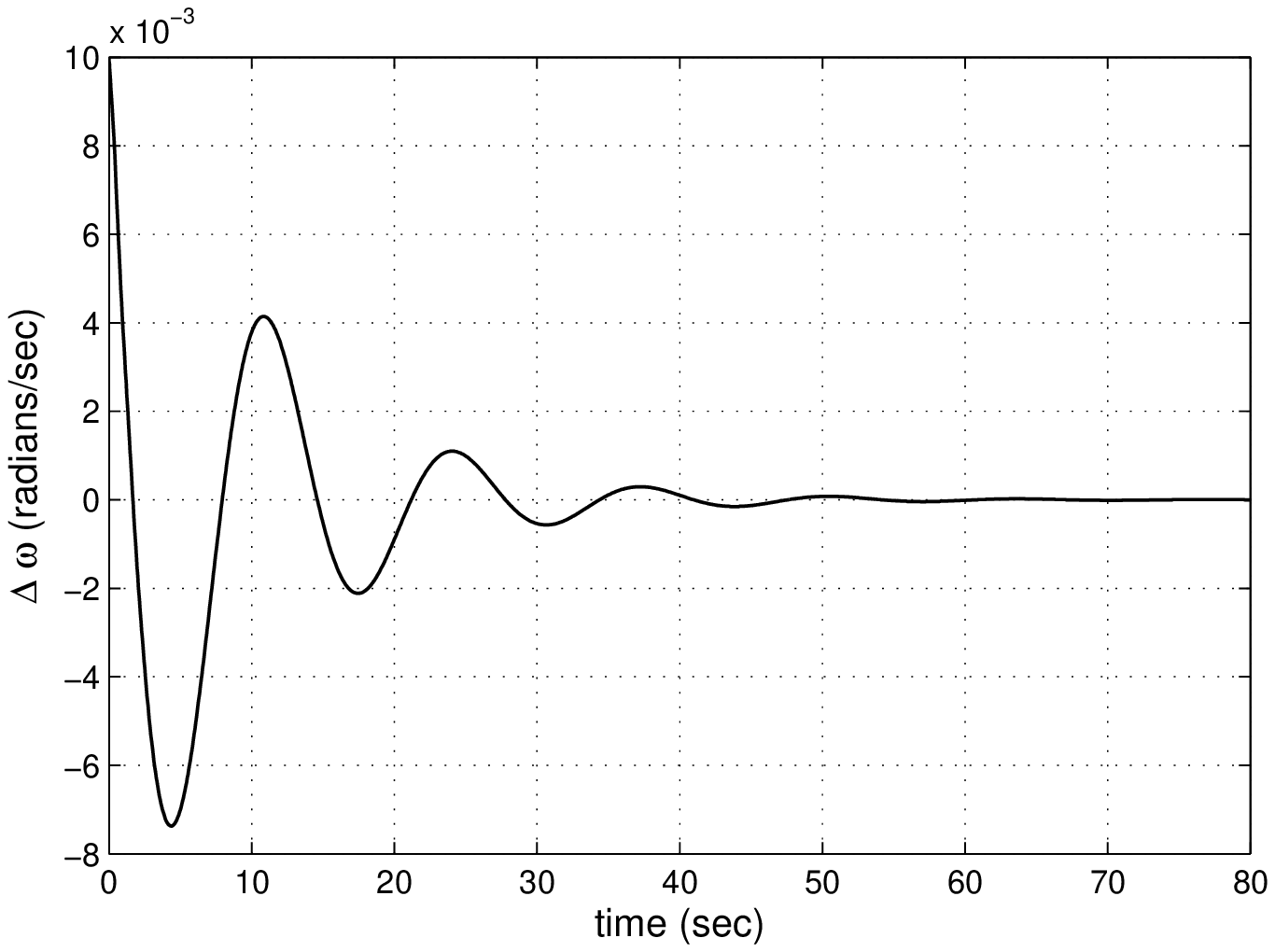}
          \caption{Plot of $\Delta \omega  $ vs time for the observer-based LQR applied to the reduced order 
          linear model with output $\Delta V_t$ measured}
          \label{fig:lqrobs3}
\end{figure}

\begin{figure}
          \centering
          \includegraphics[trim=0cm 0cm 0cm 0cm, clip=true, totalheight=0.275\textheight, width=0.58\textwidth]{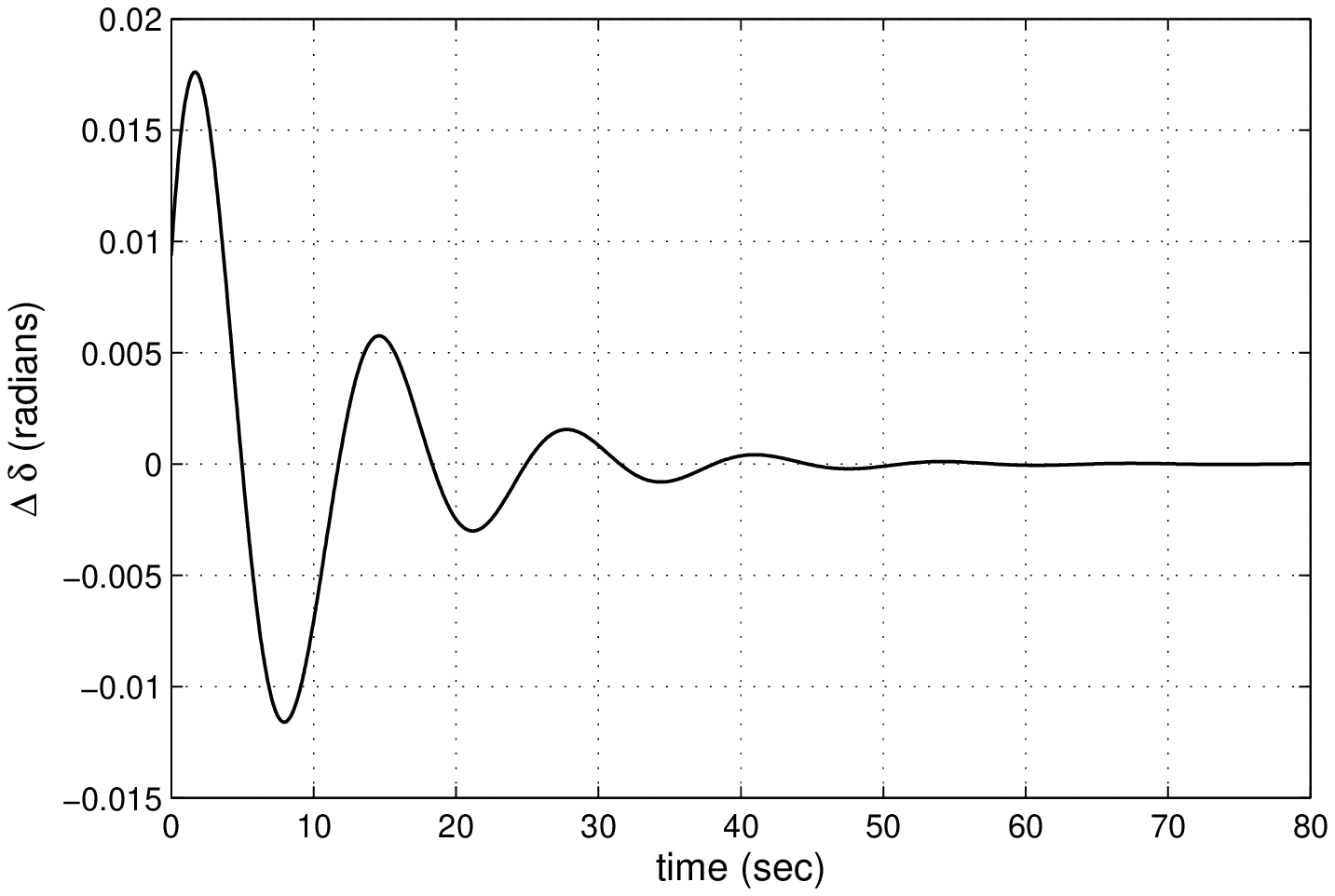}
          \caption{Plot of the rotor angle $\Delta \delta $ vs time for the observer-based LQR 
           applied to the reduced order linear model with output $\Delta V_t$ measured}
          \label{fig:lqrobs4}
          \includegraphics[trim=0cm 0cm 0cm 0cm, clip=true, totalheight=0.275\textheight, width=0.58\textwidth]{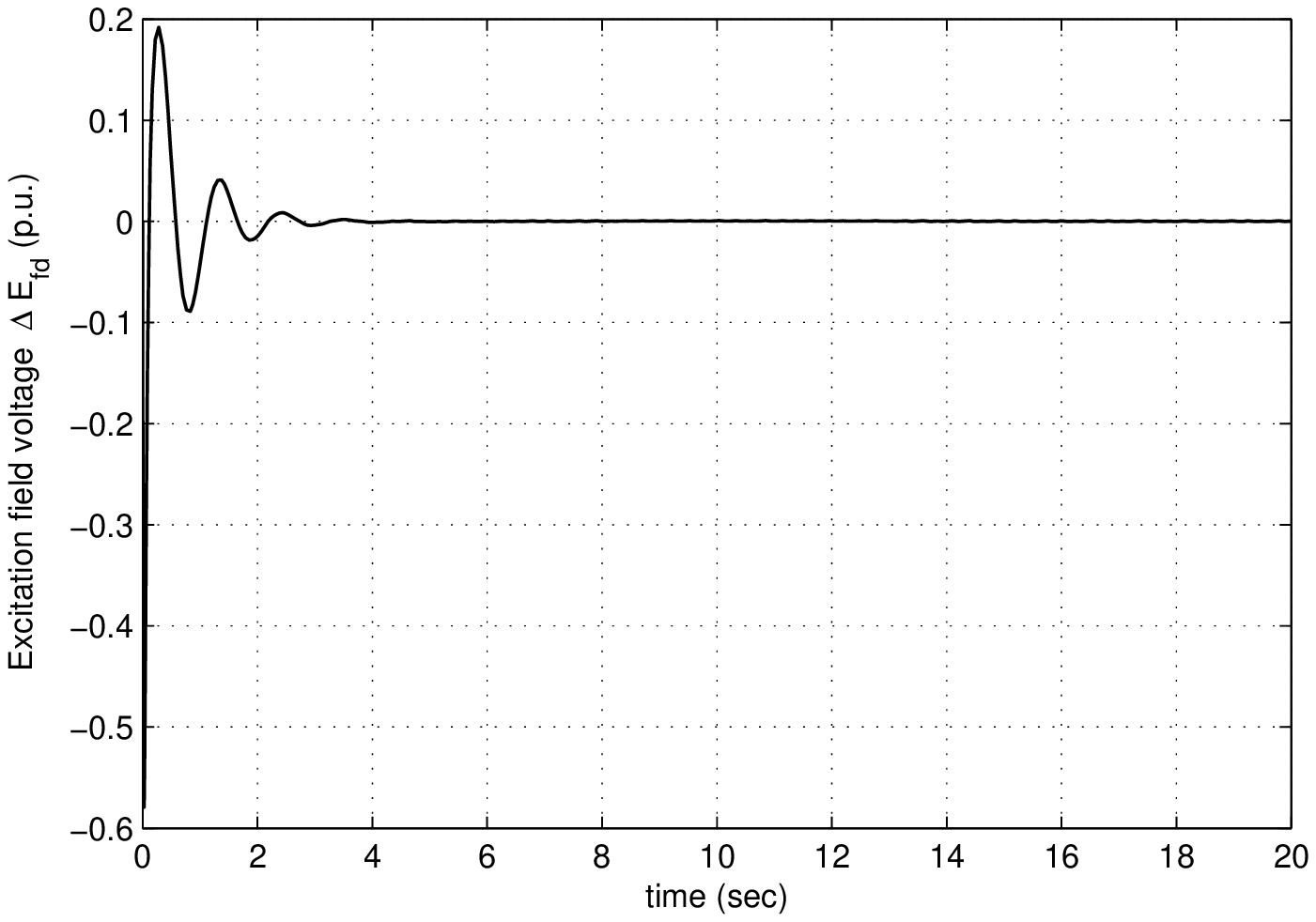}
          \caption{Plot of the control input $\Delta E_{fd}$ vs time for the observer-based LQR  
          applied to the reduced order linear model with output $\Delta V_t$ measured}
          \label{fig:lqrobs5}
          \includegraphics[trim=0cm 0cm 0cm 0cm, clip=true, totalheight=0.275\textheight, width=0.58\textwidth]{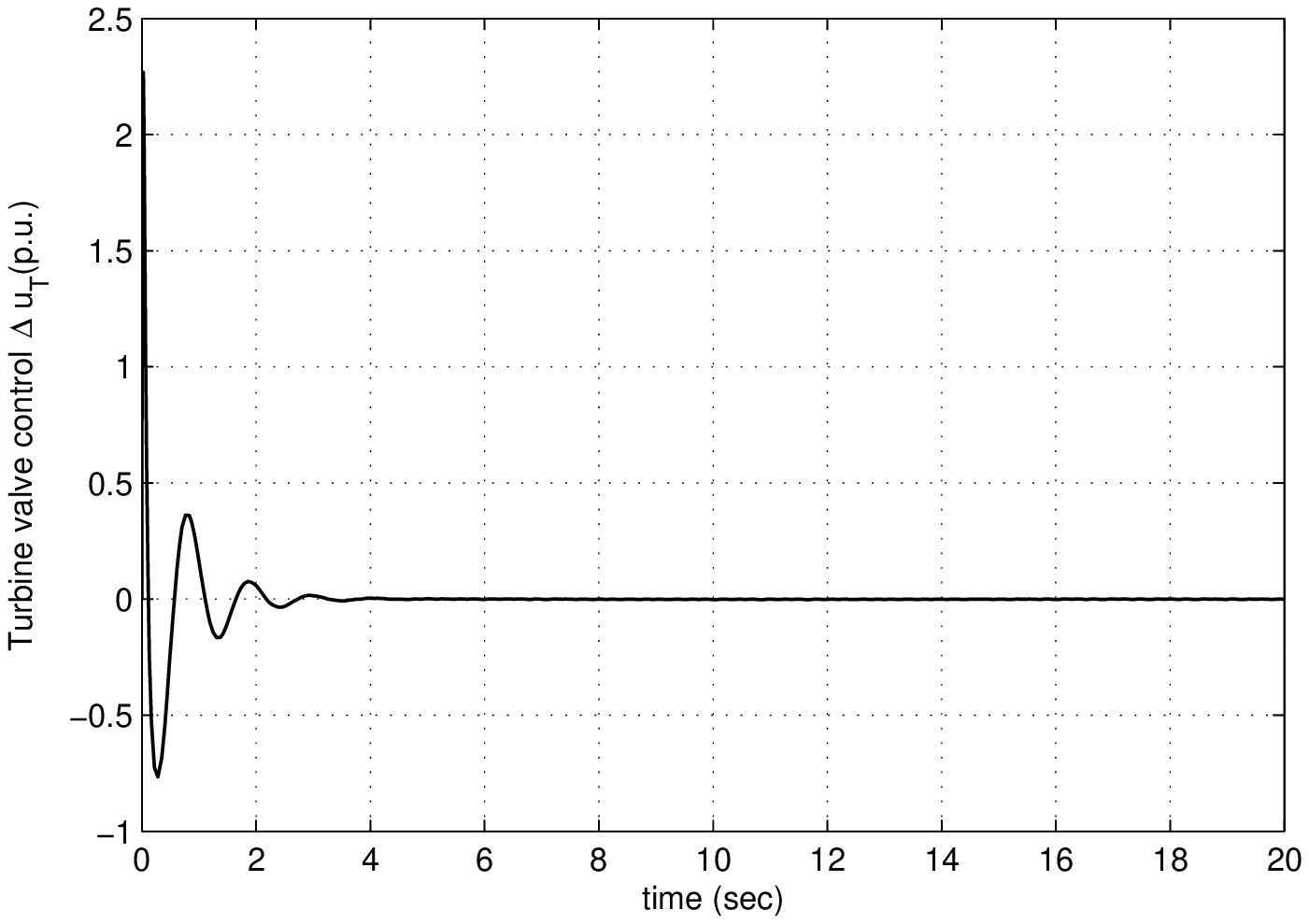}
          \caption{Plot of the control input $\Delta u_T$ vs time for the observer-based LQR applied to the reduced order 
          linear model with output $\Delta V_t$ measured}
          \label{fig:lqrobs6}
\end{figure}

In a practical generator-turbine system connected to an infinite bus not
 all the states are available for measurement. For example the 
second output of the system which is the rotor angle cannot be measured using a sensor. Thus, we need to design an observer
to estimate the rotor angle and the remaining unmeasured states. Once the estimator is designed, using the separation principle 
state estimates are used as the input to the state feedback controller to obtain a regulator for the system. We first
use only one sensor to measure only the first output $\Delta V_t$. Later we use two sensors to measure both the outputs $\Delta V_t$,
and $\Delta \omega $.
Using the $A$ and $B$ matrices as found earlier and 
\begin{equation}
\begin{aligned}
               \mathbf{y}&= C_1\mathbf{x}=\bbm T_1 & 0 & T_2 & 0 & 0\ebm \mathbf{x}\\
        \mathrm{i.e.} \ \   \Delta V_t &= \bbm 0.5258 & 0 & 0.0294 & 0 & 0\ebm \mathbf{x} \\
\end{aligned}                         
\label{eq:ob1}
\end{equation}                         
 where we use only the first output i.e. the generator terminal voltage $\Delta V_t$ measurement for the observer design.
Using MATLAB we can verify that the system is observable using the generator terminal voltage $\Delta V_t$ i.e. $C_1$ as given in 
\autoref{eq:ob1}. The observer that produces an estimate of the state, $\hat{x}$, is of the form 
\begin{equation}
\begin{aligned}
                 \dot{\hat{x}} &= A\hat{x}+Bu+L(y-\hat{y})\\
                        \hat{y} &= C_1\hat{x}
\end{aligned}                         
\label{eq:ob2}
\end{equation}
The observer gain matrix $L$ in the above equation is designed such that the error dynamics of the estimator decay faster than
the remaining dynamics. Let us denote $e=x-\hat{x}$ as the estimation error. Then, the error dynamics of the estimator are given by
\begin{equation}
        \dot{e}=(A-LC_1)e
\label{eq:ob3}
\end{equation}
The procedure to design the observer gain matrix $L$ is as follows: We first calculate the eigenvalues of the closed loop system, i.e.
eigenvalues of the closed loop $Acp$ matrix where $Acp=A-BK$.
The real parts of the estimator poles are selected 10-15 times to the left of the closed loop poles of the system $Acp$ matrix,
so that the error dynamics of the estimator decay to zero faster. This can be achieved by appropriately selecting the 
scaling factor $\rho $ which places the estimator poles to the left of the closed loop poles of the system. The following MATLAB command can be used to design the observer gain matrix\\
\\
$poles=eig(Acp);$\\
$p=rho*[poles(1), poles(2), poles(3), poles(4), poles(5)];$\\
$Ltilde = place(A',C_1',p)$;\\
$L=Ltilde';$ \\
\\
The estimator gain matrix $L$ for $\rho=12$ is
\begin{equation}
     L=\bbm 1510.4\\-65508.0\\-24107.6\\-1004232.6\\276190.3\ebm
\label{eq:ob4}
\end{equation} 
 Next, the estimated states are fed back to the feedback controller to obtain a regulator for the system
\begin{equation}
         u=-K\hat{x}
\label{eq:ob5}
\end{equation}
Choosing the weighting matrices
\begin{equation}
          Q=\bbm 1 & 0  & 0  & 0  & 0\\ 0 &  1 &  0  & 0  & 0\\ 0 &  0 &  1 &  0 &  0\\ 
            0 &  0  & 0  & 1  & 0\\ 0  & 0  & 0 &  0  & 1 \ebm
 \label{eq:obvt1}
\end{equation} 
and           
\begin{equation}
        R=\bbm 20 & 0\\ 0 & 20\ebm
\label{eq:obvt2}
\end{equation} 
and the state space matrices $(A, B)$ as given in \autoref{eq:linear25} and \autoref{eq:linear26} the 
control gain $K$ for the observer-based LQR with the output $V_t$ measured is found to be 
  \begin{equation}
        K=\bbm 0.0384  & -0.0635  &  0.0126  & -0.0070  & -0.0027\\  -0.0811  &  0.1973 &  -0.0236  &  0.0382  &  0.0395\ebm
\label{eq:obvt3}
\end{equation} 
The $Q$ and $R$ matrices are designed such that all the estimated state variables and the control signals stay within
a specific limit. From the choice of the $Q$ and $R$ matrices for the estimator-based LQR with output $\Delta V_t$ measured, 
we can see that the
weights of the $Q$ matrix are reduced and weights of the $R$ matrix are increased compared to the full state feedback LQR.
In other words the amount of control energy that can be provided such that all the state variables and control inputs remain within a limit
is significantly reduced. This is allowed by the separation principle where the controller and the observer gains are designed independently. Thus, the time taken by the estimated states and hence the outputs to reach the steady state value of 0
is larger than the state feedback case. This is evident from \autoref{fig:lqrobs1} to \autoref{fig:lqrobs4} where the estimated states
$\Delta \hat{E'}_q$, $\Delta \hat{\omega }$, $\Delta \hat{\delta }$, 
           $\Delta \hat{T}_m$, $\Delta \hat{G}_V$, and the outputs $\Delta V_t$, $\Delta \omega $, and $\Delta \delta $ take more time
           to converge to zero as compared to the LQR with full state feedback. \autoref{fig:lqrobs5} and \autoref{fig:lqrobs6} show
plots for the two control inputs $\Delta E_{fd}$, and $\Delta u_T$ respectively. \\

\begin{figure}
          \centering
          \includegraphics[trim=0cm 0cm 0cm 0cm, clip=true, totalheight=0.27\textheight, width=0.54\textwidth]{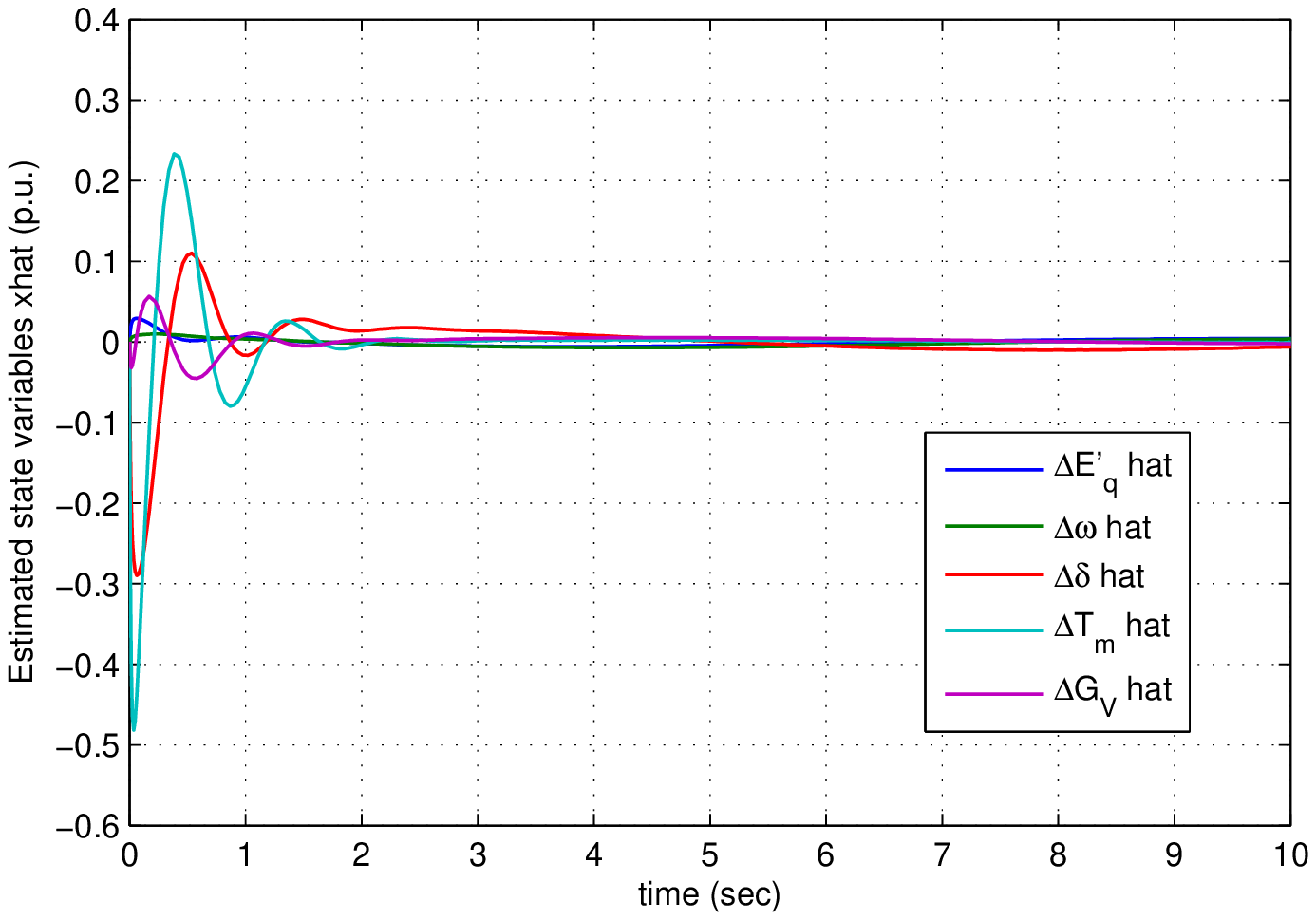}
          \caption{Plot of the estimated state variables $\Delta \hat{E'}_q$, $\Delta \hat{\omega }$, $\Delta \hat{\delta }$, 
           $\Delta \hat{T}_m$, and $\Delta \hat{G}_V$ vs time for the observer-based LQR 
           applied to the reduced order linear model with outputs $\Delta V_t$ and $\Delta \omega $ measured}
           \label{fig:lqrobsomega1}
          \includegraphics[trim=0cm 0cm 0cm 0cm, clip=true, totalheight=0.27\textheight, width=0.54\textwidth]{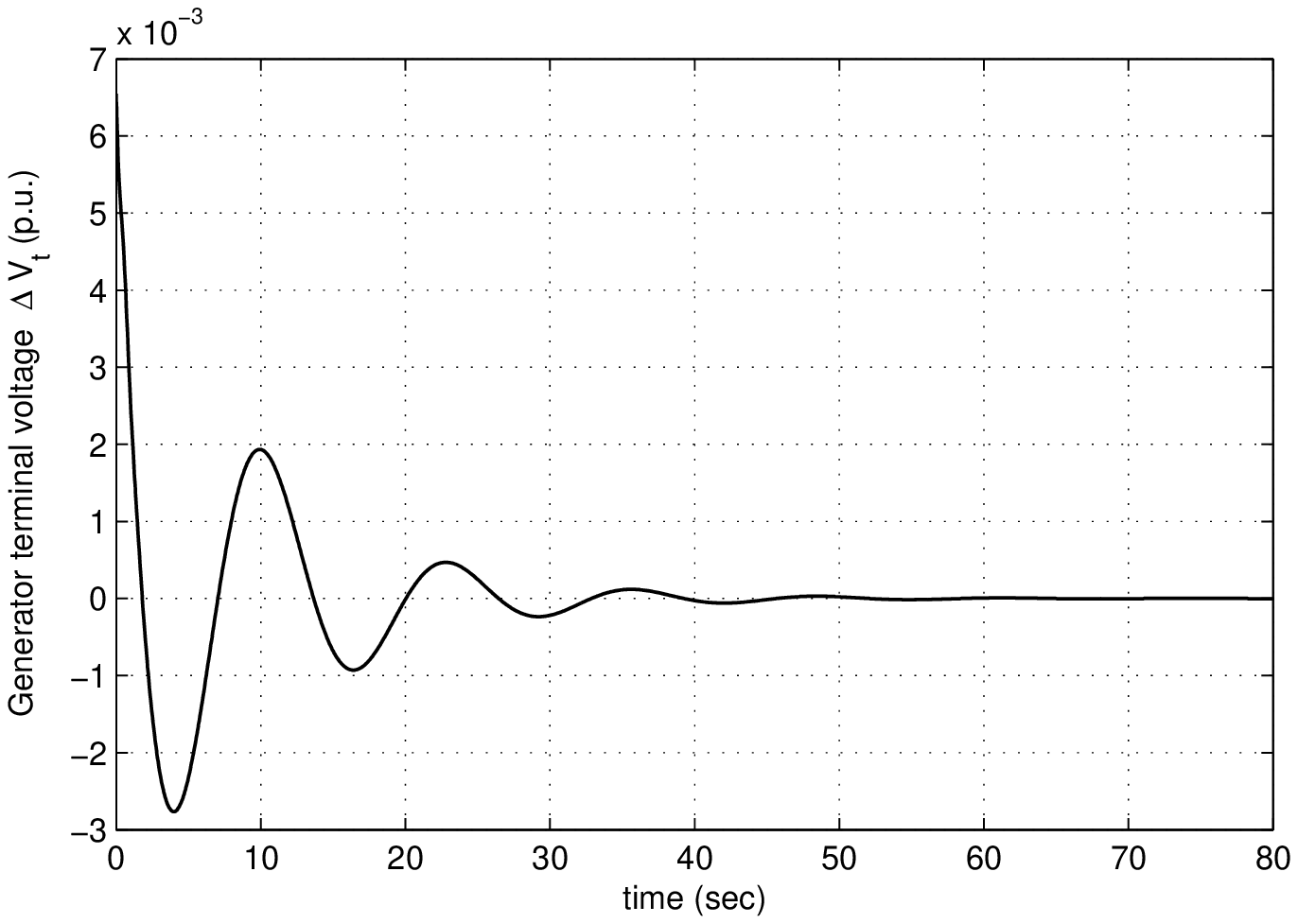}
          \caption{Plot of the generator terminal voltage $\Delta V_t$ vs time for the observer-based LQR 
          applied to the reduced order linear model with outputs $\Delta V_t$ and $\Delta \omega $ measured}
          \label{fig:lqrobsomega2}
          \includegraphics[trim=0cm 0cm 0cm 0cm, clip=true, totalheight=0.27\textheight, width=0.54\textwidth]{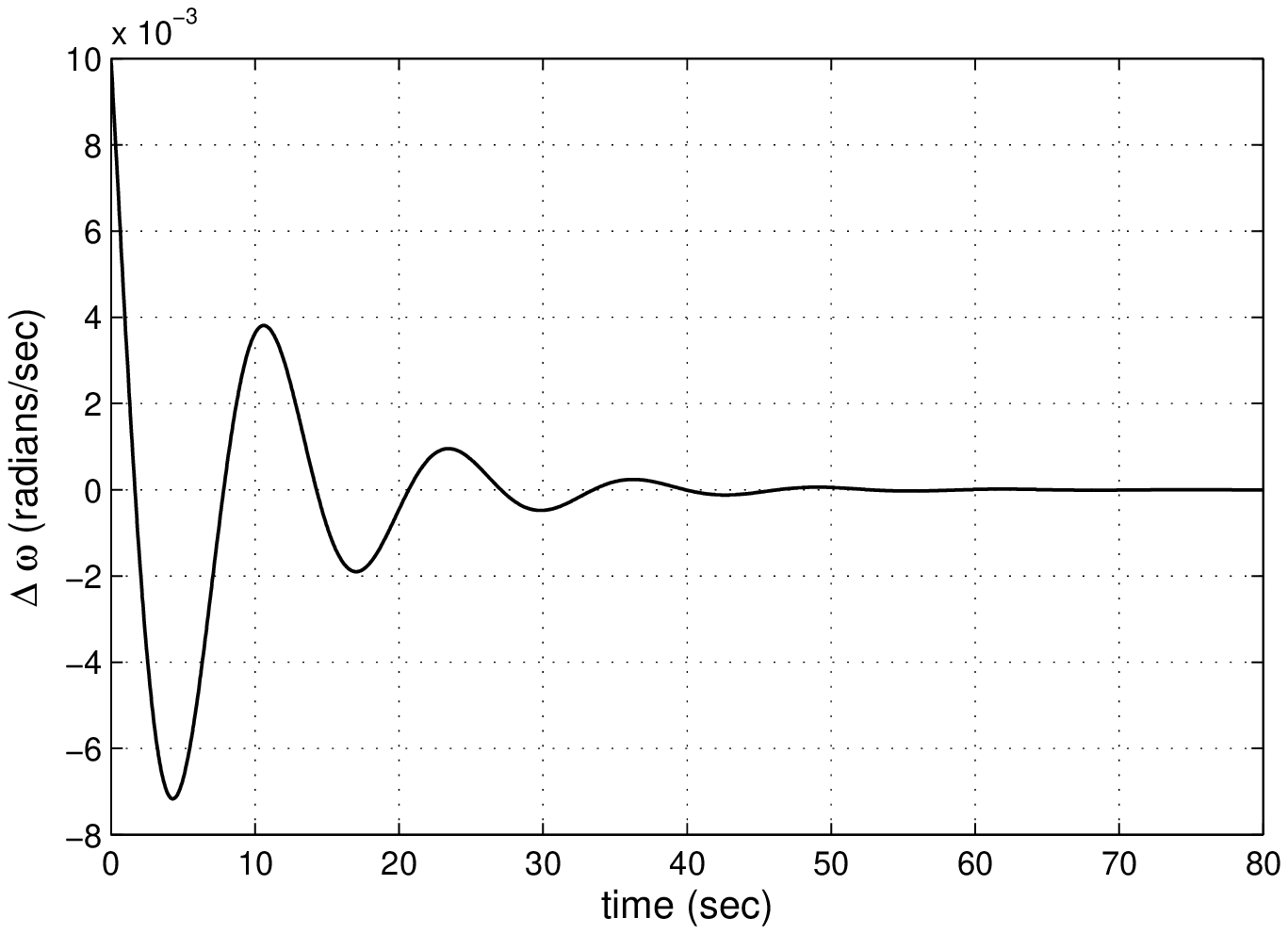}
          \caption{Plot of $\Delta \omega $ vs time for the observer-based LQR applied to the reduced order 
          linear model with outputs $\Delta V_t$ and $\Delta \omega $  measured}
          \label{fig:lqrobsomega3}       
\end{figure}

\begin{figure}
          \centering          
          \includegraphics[trim=0cm 0cm 0cm 0cm, clip=true, totalheight=0.27\textheight, width=0.54\textwidth]{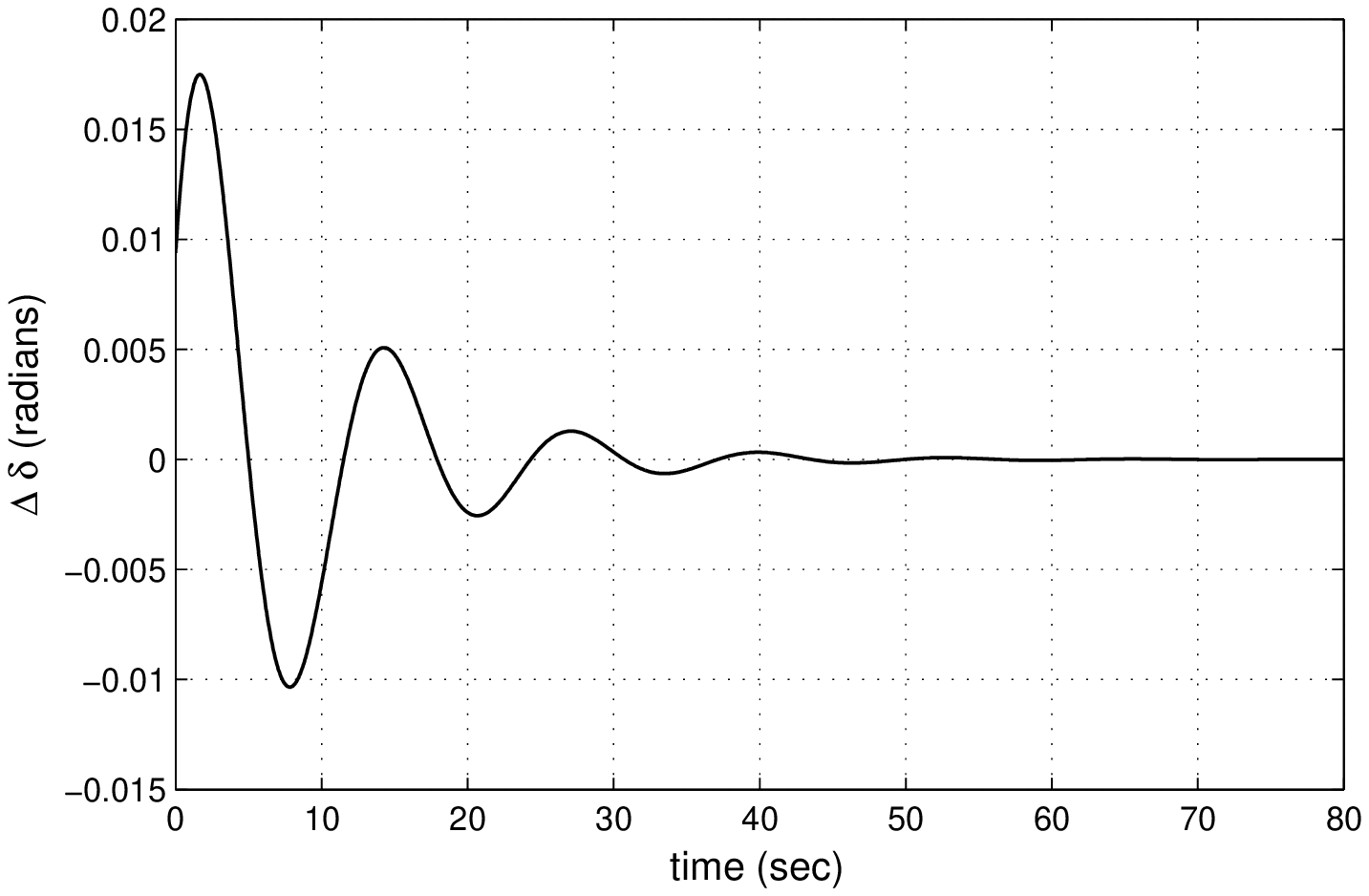}
          \caption{Plot of $\Delta \delta $ vs time for the observer-based LQR applied to the reduced order 
          linear model with outputs $\Delta V_t$ and $\Delta \omega $  measured}
          \label{fig:lqrobsomega4}
          \includegraphics[trim=0cm 0cm 0cm 0cm, clip=true, totalheight=0.27\textheight, width=0.54
           \textwidth]  {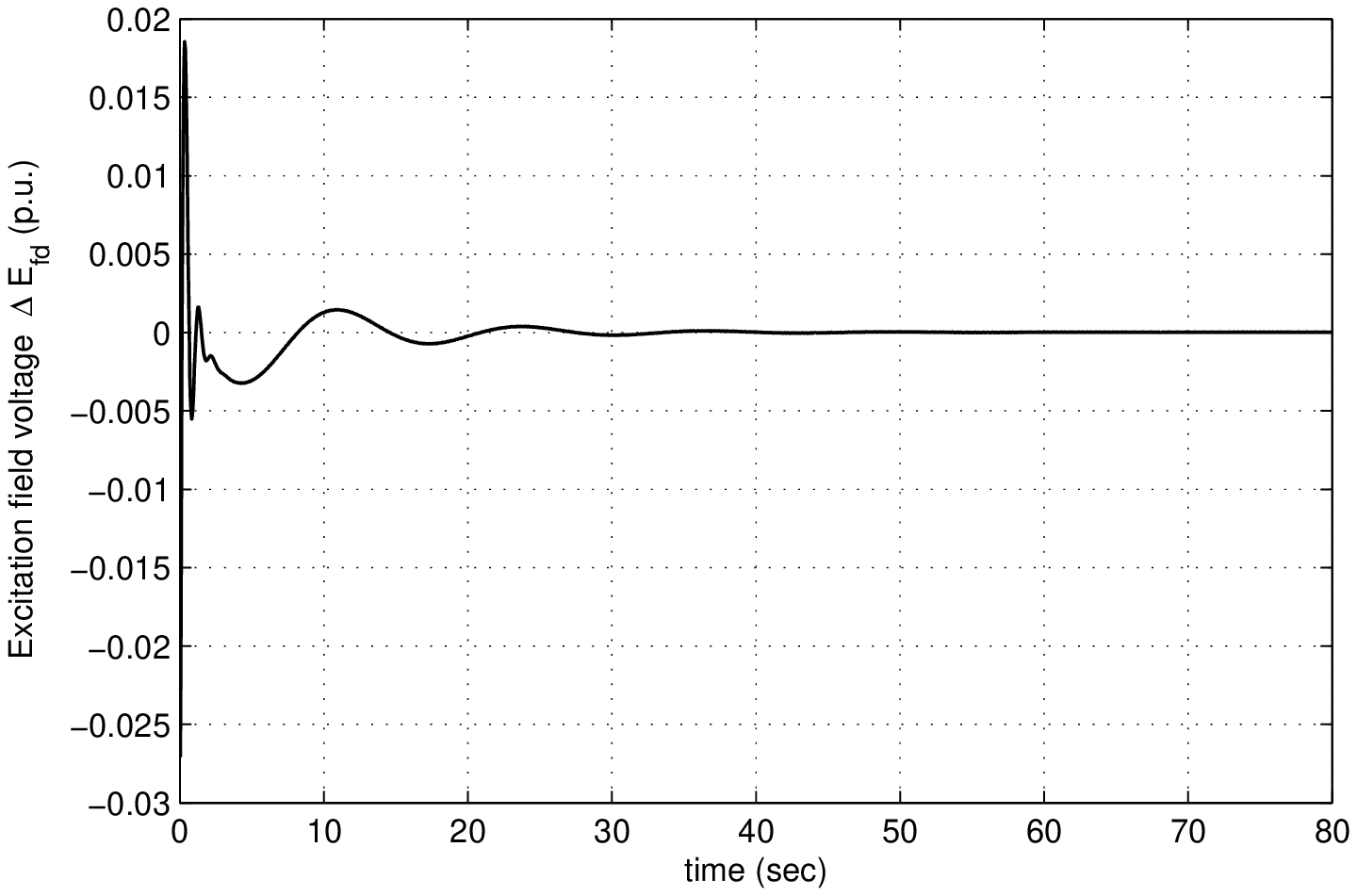}
          \caption{Plot of the control input $\Delta E_{fd}$ vs time for the observer-based LQR applied to the 
           reduced order linear model with outputs $\Delta V_t$ and $\Delta \omega $ measured}
          \label{fig:lqrobsomega5}
          \includegraphics[trim=0cm 0cm 0cm 0cm, clip=true, totalheight=0.27\textheight, width=0.54\textwidth]{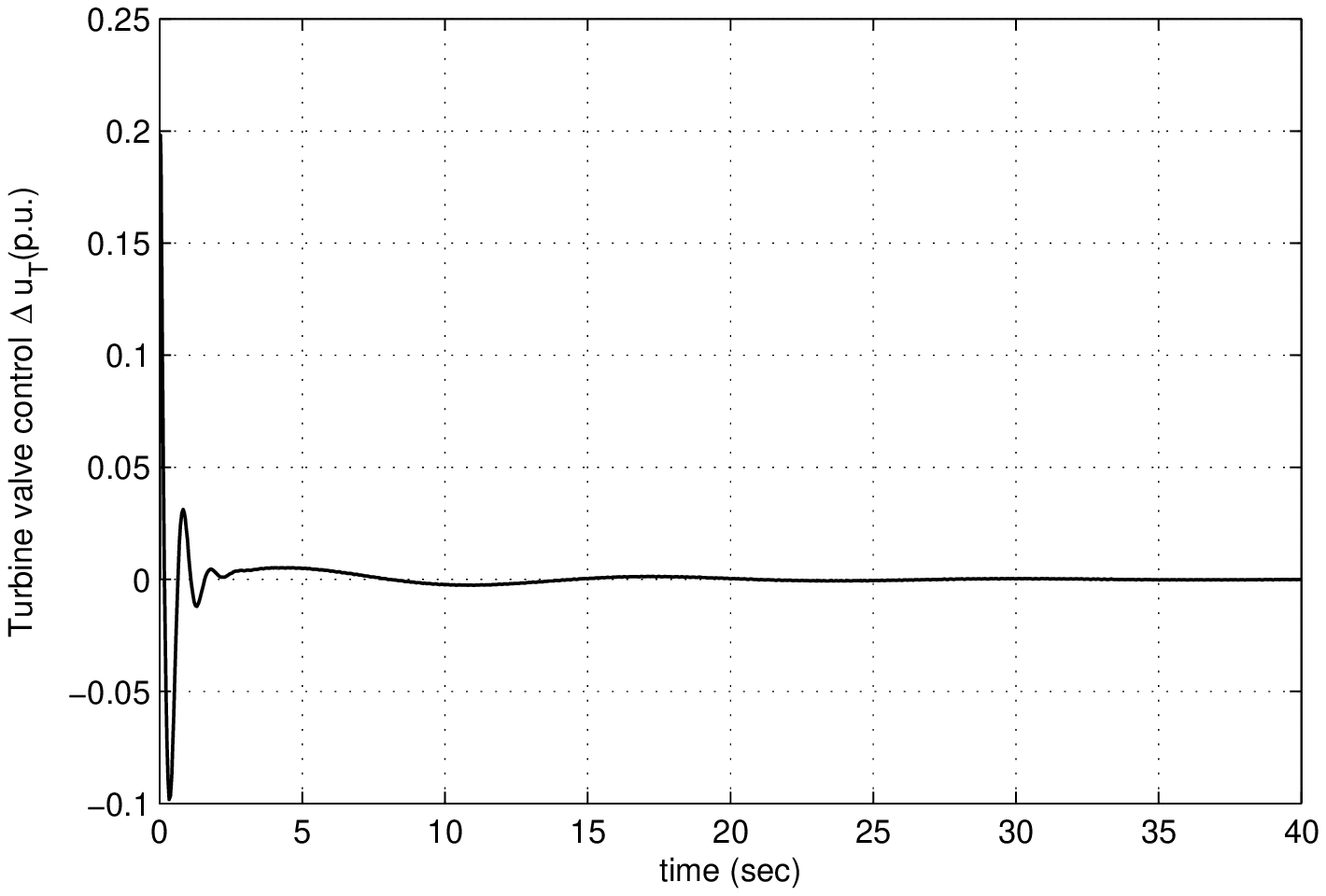}
          \caption{Plot of the control input $\Delta u_T$ vs time for the observer-based LQR 
          applied to the reduced order linear model with outputs $\Delta V_t$ and $\Delta \omega $ measured}
          \label{fig:lqrobsomega6}  
\end{figure}   

A better estimate of the states can be obtained by also using the angular velocity or frequency $\Delta \omega $ of 
the synchronous generator which can be measured using a sensor, in addition to the generator terminal voltage $\Delta V_t$. 
In this case we have 
\begin{equation}
\begin{aligned}
               \mathbf{y}&= C_2\mathbf{x}=\bbm T_1 & 0 & T_2 & 0 & 0\\ 
                  0 & 1 & 0 & 0 & 0 \ebm \mathbf{x}\\
              \mathrm{i.e.} \ \        \bbm \Delta V_t\\ \Delta \omega \ebm &= \bbm 0.5258 & 0 & 0.0294 & 0 & 0\\ 
                  0 & 1 & 0 & 0 & 0 \ebm \mathbf{x} \\
\end{aligned}                         
\label{eq:ob6} 
\end{equation} 
Using MATLAB we can verify that the system is observable using the generator terminal voltage $\Delta V_t$ and the frequency
$\Delta \omega $ i.e. $C_2$ as given in \autoref{eq:ob6}.
The estimator gain matrix $L$ for the output matrix $C_2$ as given in \autoref{eq:ob6} and $\rho=12$ is
\begin{equation}
     L=\bbm 256.459 & 71.2024\\20.6146 & 34.7633\\-1846.26 & -922.386 \\-5032.595 & -1309.90\\-531.7120 & -135.4256\ebm
\label{eq:ob7}
\end{equation} 
Choosing the weighting matrices
\begin{equation}
          Q=\bbm 1 & 0  & 0  & 0  & 0\\ 0 &  1 &  0  & 0  & 0\\ 0 &  0 &  1 &  0 &  0\\ 
            0 &  0  & 0  & 1  & 0\\ 0  & 0  & 0 &  0  & 1 \ebm
 \label{eq:obvtomega1}
\end{equation} 
and           
\begin{equation}
        R=\bbm 1 & 0\\ 0 & 1\ebm
\label{eq:obvtomega2}
\end{equation} 
and the state space matrices $(A, B)$ as given in \autoref{eq:linear25} and \autoref{eq:linear26} the 
control gain $K$ for the observer-based LQR with the outputs $\Delta V_t$ and $\Delta \omega $ measured is found to be 
  \begin{equation}
        K=\bbm 0.4722 &  -0.8024  &  0.0599 & -0.0726  & -0.0195\\  -0.5758  &  1.6563  & -0.0271 &  0.3948  &  0.5217 \ebm
\label{eq:obvtomega3}
\end{equation}
From the choice of the $Q$ and $R$ matrices for the second observer when the frequency $\Delta \omega $ is measured in addition to 
the output $\Delta V_t$, we can see that the weights of the $R$ matrix are reduced compared to the case when only $\Delta V_t$  is measured. Thus, the amount of control effort that can be applied to the system such that all the states and the control
inputs remain within a specific bound is more, and hence the state variables and the outputs converge to zero faster compared to the first observer with only $\Delta V_t$
measured, but not as fast as the full state
feedback LQR controller. This is evident from \autoref{fig:lqrobsomega1} to \autoref{fig:lqrobsomega4}
where the estimated states
$\Delta \hat{E'}_q$, $\Delta \hat{\omega }$, $\Delta \hat{\delta }$, 
           $\Delta \hat{T}_m$, $\Delta \hat{G}_V$, and the outputs $\Delta V_t$, $\Delta \omega $, and $\Delta \delta $ of the observer based LQR with $\Delta V_t$ and $\Delta \omega $ measured take less time
           to converge to zero as compared to the observer-based LQR with only $\Delta V_t$ measured. However, the time taken is larger
           compared to the LQR with full state feedback, which gives the best results. \autoref{fig:lqrobsomega5} and \autoref{fig:lqrobsomega6} show
plots for the two control inputs $\Delta E_{fd}$, and $\Delta u_T$ respectively.

\newpage
\subsubsection{Observer-Based Pole Placement Controller Design based on linear model}

\begin{figure}
          \centering
          \includegraphics[trim=0cm 0cm 0cm 0cm, clip=true, totalheight=0.27\textheight, width=0.54
           \textwidth]  {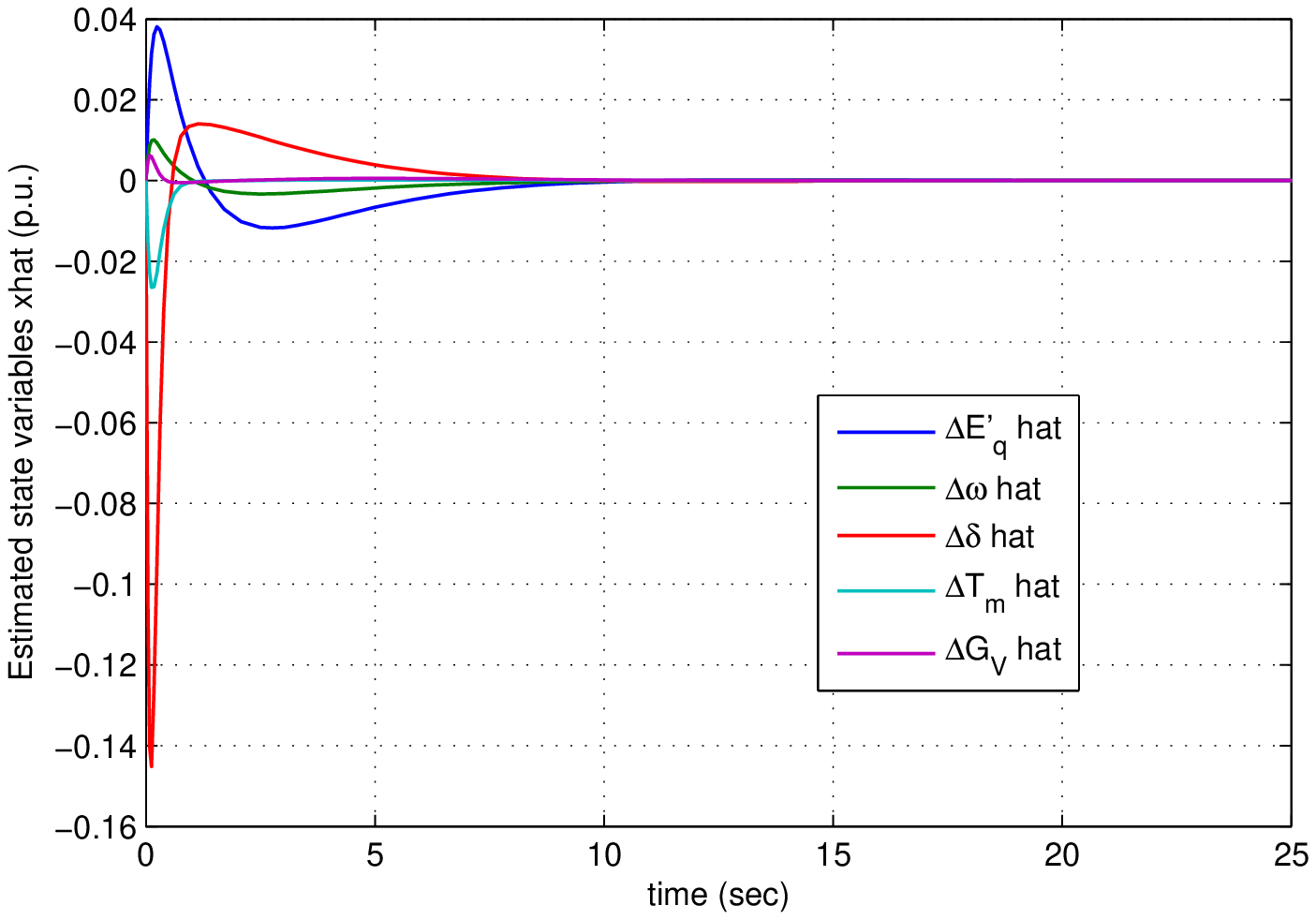}
          \caption{Plot of the estimated state variables $\Delta \hat{E'}_q$, $\Delta \hat{\omega }$, $\Delta \hat{\delta }$, 
           $\Delta \hat{T}_m$, and $\Delta \hat{G}_V$ vs time for the observer-based pole placement controller applied to 
           the reduced order linear model with 
           outputs $\Delta V_t$ and $\Delta \omega $ measured}
          \label{fig:poleobsomega1}
          \includegraphics[trim=0cm 0cm 0cm 0cm, clip=true, totalheight=0.27\textheight, width=0.54\textwidth]{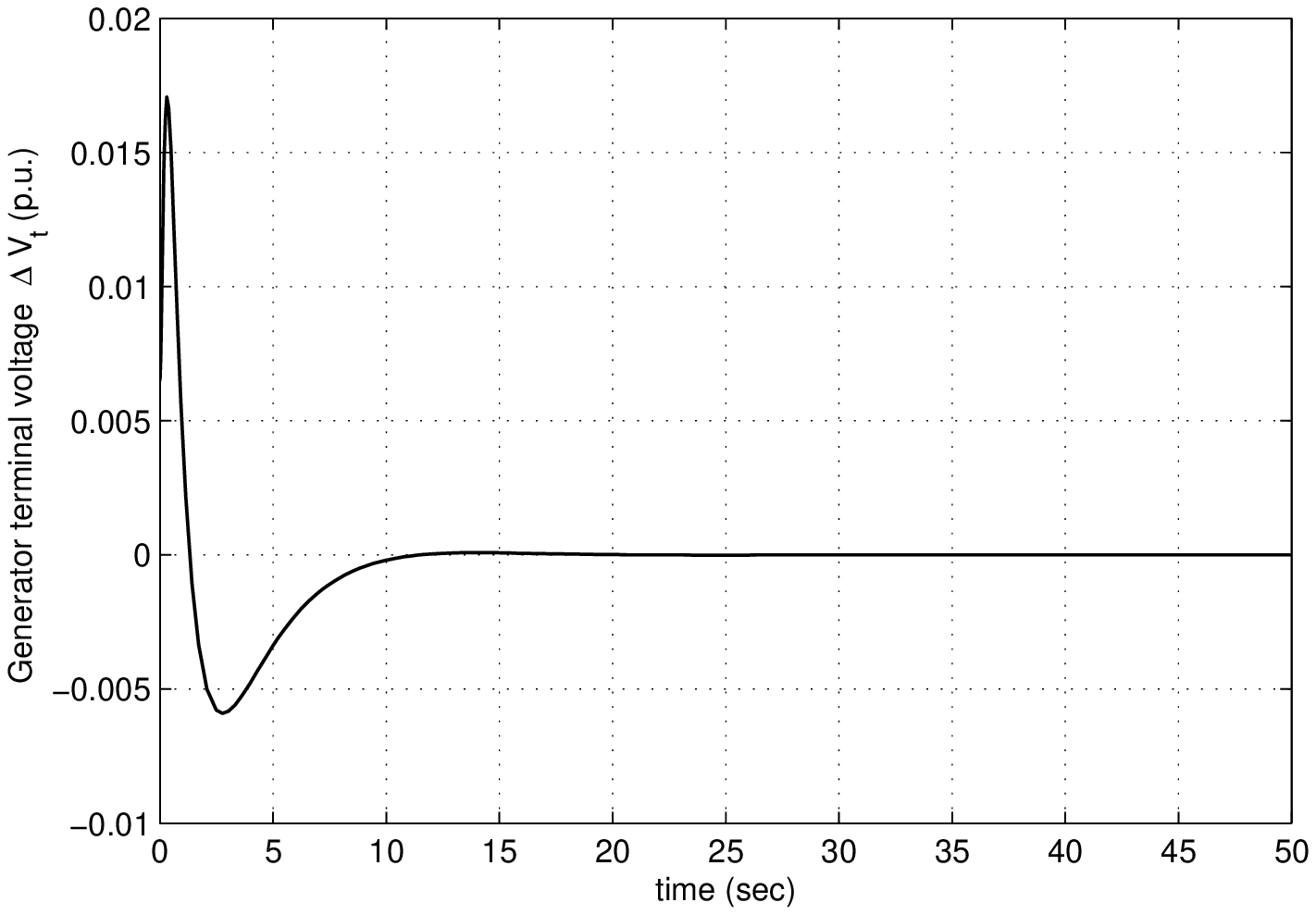}
          \caption{Plot of the generator terminal voltage $\Delta V_t$ vs time for the observer-based pole placement controller 
          applied to the reduced order linear model with outputs $\Delta V_t$ and $\Delta \omega $ measured}
          \label{fig:poleobsomega2}
          \includegraphics[trim=0cm 0cm 0cm 0cm, clip=true, totalheight=0.27\textheight, width=0.54\textwidth]{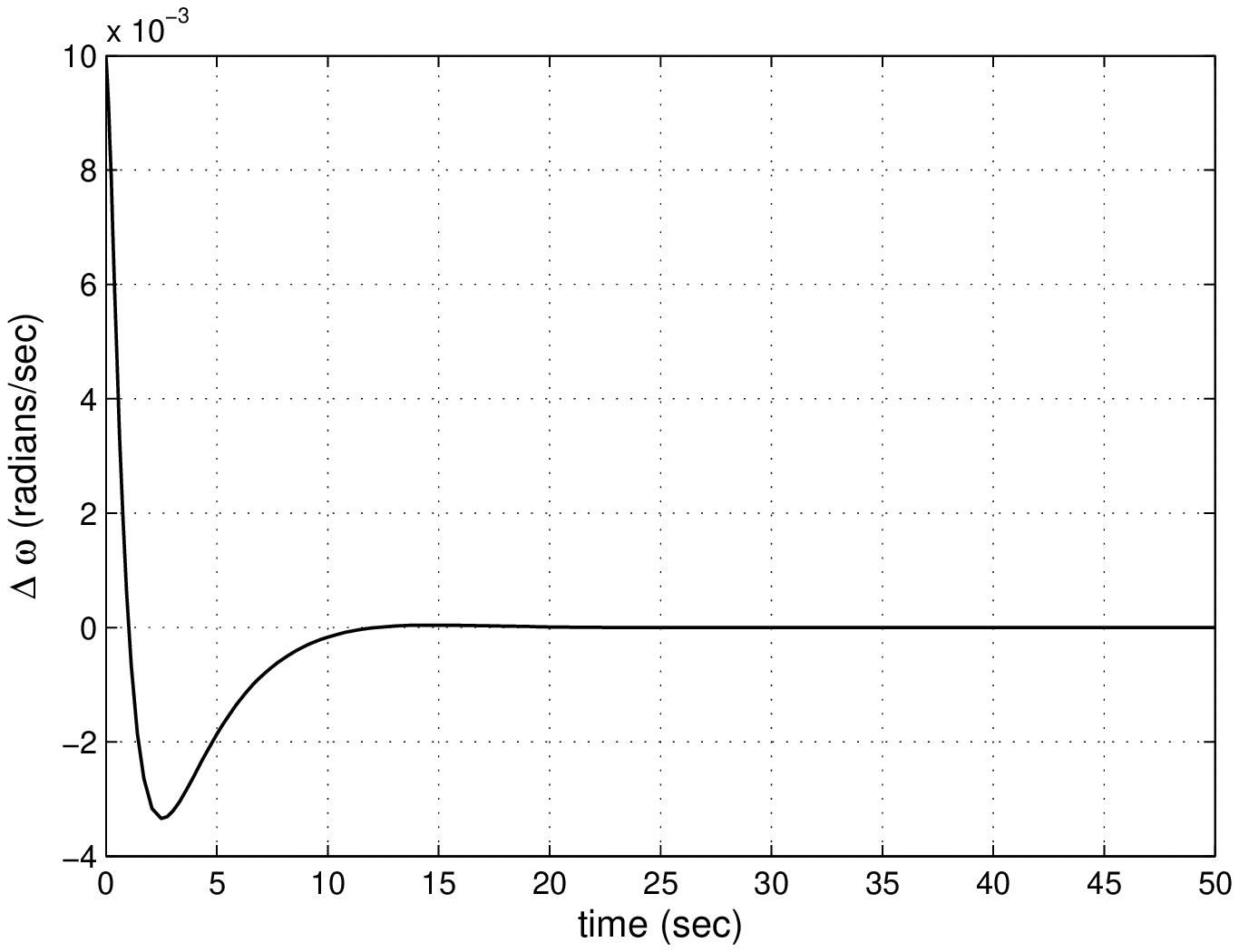}
          \caption{Plot of $\Delta \omega $ vs time for the observer-based pole placement controller applied to the 
          reduced order linear model with outputs $\Delta V_t$ and 
          $\Delta \omega $  measured}
          \label{fig:poleobsomega3}
\end{figure}

\begin{figure}
          \centering
          \includegraphics[trim=0cm 0cm 0cm 0cm, clip=true, totalheight=0.27\textheight, width=0.54
           \textwidth]  {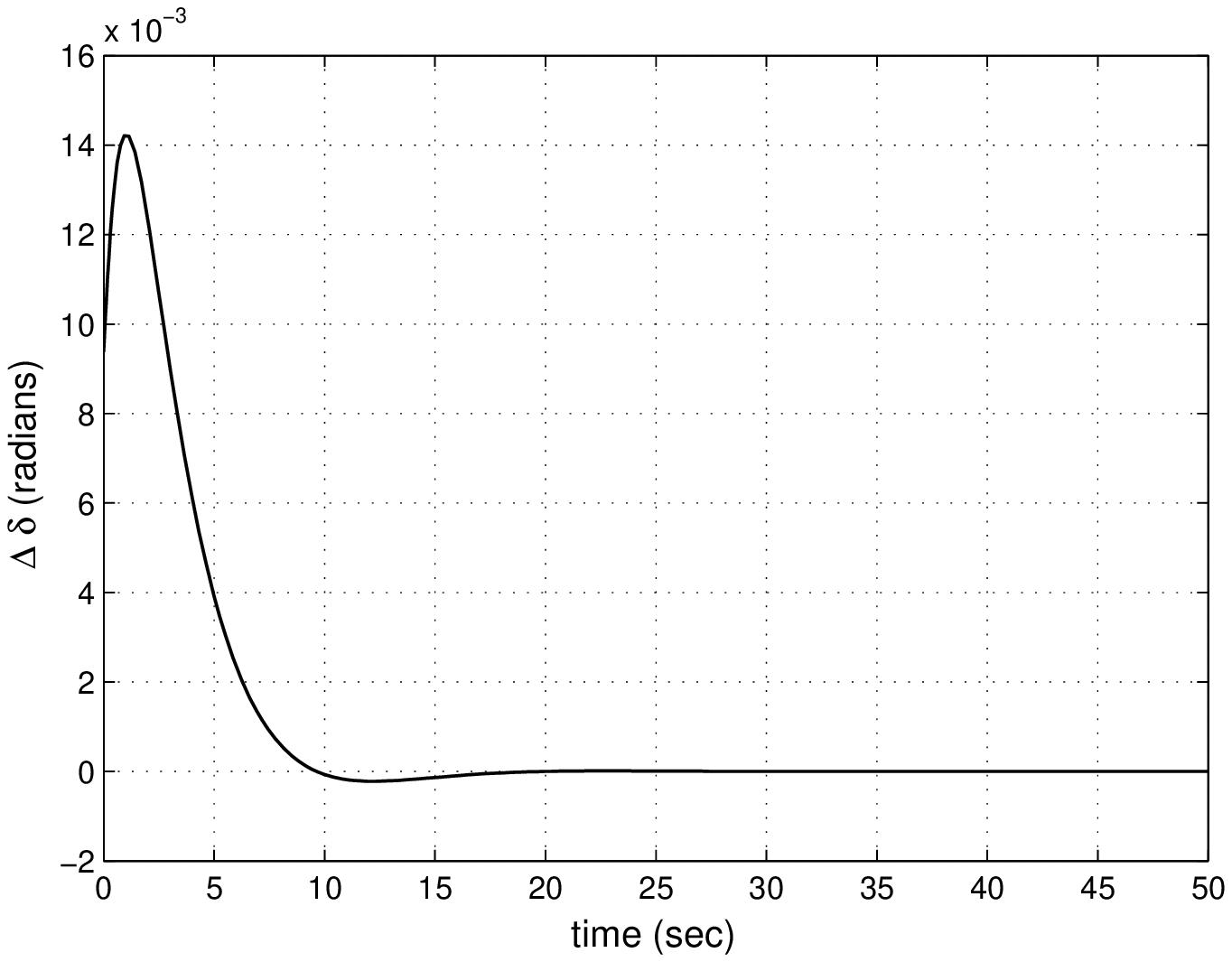}
          \caption{Plot of $\Delta \delta $ 
           vs time for the observer-based pole placement controller applied to 
           the reduced order linear model with 
           outputs $\Delta V_t$ and $\Delta \omega $ measured}
          \label{fig:poleobsomega4}
          \includegraphics[trim=0cm 0cm 0cm 0cm, clip=true, totalheight=0.27\textheight, width=0.54\textwidth]{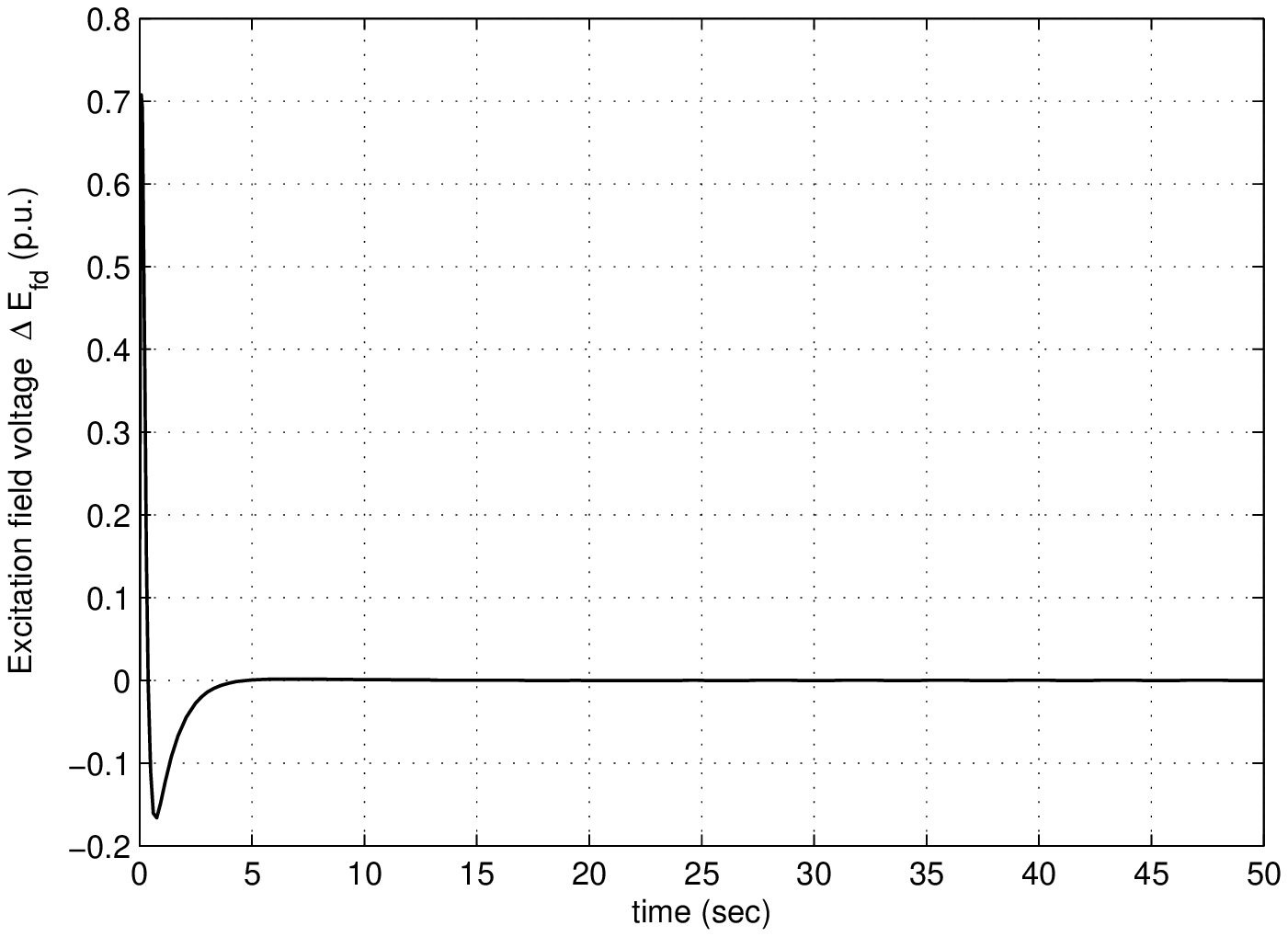}
          \caption{Plot of the control input $\Delta E_{fd}$ vs time for the observer-based pole placement controller 
          applied to the reduced order linear model with outputs $\Delta V_t$ and $\Delta \omega $ measured}
          \label{fig:poleobsomega5}
          \includegraphics[trim=0cm 0cm 0cm 0cm, clip=true, totalheight=0.27\textheight, width=0.54\textwidth]{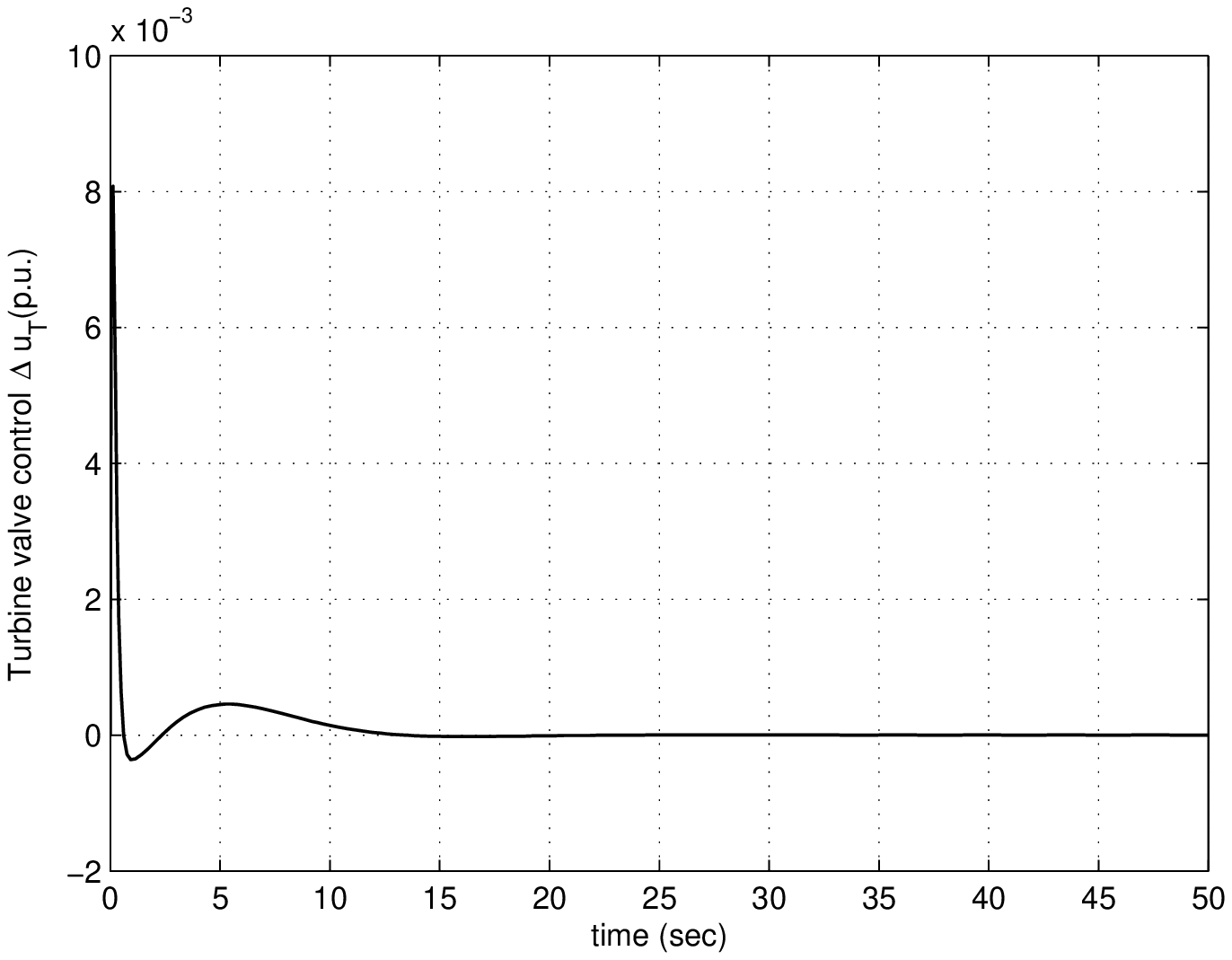}
          \caption{Plot of the control input $\Delta u_T$ vs time for the observer-based pole placement controller applied to the 
          reduced order linear model with outputs $\Delta V_t$ and 
          $\Delta \omega $  measured}
          \label{fig:poleobsomega6}
\end{figure}

In this section we design an observer-based pole placement controller for the reduced order linear model of the 
generator-turbine system connected to an infinite bus. The procedure used to design an observer for the pole placement controller is similar to the one used earlier for the observer-based LQR. The two outputs, generator terminal voltage $\Delta V_t$ and frequency $\Delta \omega $, that can be measured
using sensors are used for the observer design. Thus we have 
\begin{equation}
\begin{aligned}
               \mathbf{y}&= C_2\mathbf{x}=\bbm T_1 & 0 & T_2 & 0 & 0\\ 
                  0 & 1 & 0 & 0 & 0 \ebm \mathbf{x}\\
              \mathrm{i.e.} \ \        \bbm \Delta V_t\\ \Delta \omega \ebm &= \bbm 0.5258 & 0 & 0.0294 & 0 & 0\\ 
                  0 & 1 & 0 & 0 & 0 \ebm \mathbf{x} \\
\end{aligned}                         
\label{eq:obpole1} 
\end{equation}
The estimator gain matrix $L$ for the output matrix $C_2$ as given in \autoref{eq:ob6} and $\rho=12$ is
\begin{equation}
     L=\bbm 68.7022 &  -2.8308\\0.2306  &  18.6252\\-608.8626  & 34.4711\\-661.5803 & 387.1455\\-61.4420  & 58.6700\ebm
\label{eq:obpole2}
\end{equation}
The desired closed-loop poles of the generator-turbine system for the observer-based pole placement controller were selected as
\begin{equation}
        p=[-0.7, -0.8, -0.5, -0.9, -0.8]
\label{eq:obpole3}
\end{equation}  
The controller gain matrix $K$ corresponding to these closed loop poles is 
\begin{equation}      
K =\bbm 9.7459 & -26.9495  &  4.8337 &  -0.8674  & -9.0425\\
   0.0299  & -0.2030 &  -0.0207  &  0.1337 &  -1.1007 \ebm   
\label{eq:obpole4}
\end{equation}
From \autoref{fig:poleobsomega1} to \autoref{fig:poleobsomega4} we can see that for the observer-based pole placement controller the estimated state variables $\Delta \hat{E'}_q$, $\Delta \hat{\omega }$, $\Delta \hat{\delta }$, 
           $\Delta \hat{T}_m$, $\Delta \hat{G}_V$, and the outputs $\Delta V_t$, $\Delta \omega $, and $\Delta \delta $ 
           converge to zero in approximately 12 seconds. \autoref{fig:poleobsomega5} and \autoref{fig:poleobsomega6} show plots for the two control inputs $\Delta E_{fd}$, and $\Delta u_T$ respectively. The performance of the observer-based pole placement 
           controller is comparable to the performance of the full-state feedback pole-placement controller.


\newpage
\subsubsection{Simulation Results for the Observer-based LQR applied to the Reduced Order Nonlinear Model}

\begin{figure}
          \centering    
          \includegraphics[trim=0cm 0cm 0cm 0cm, clip=true, totalheight=0.27\textheight, 
           width=0.54\textwidth]   {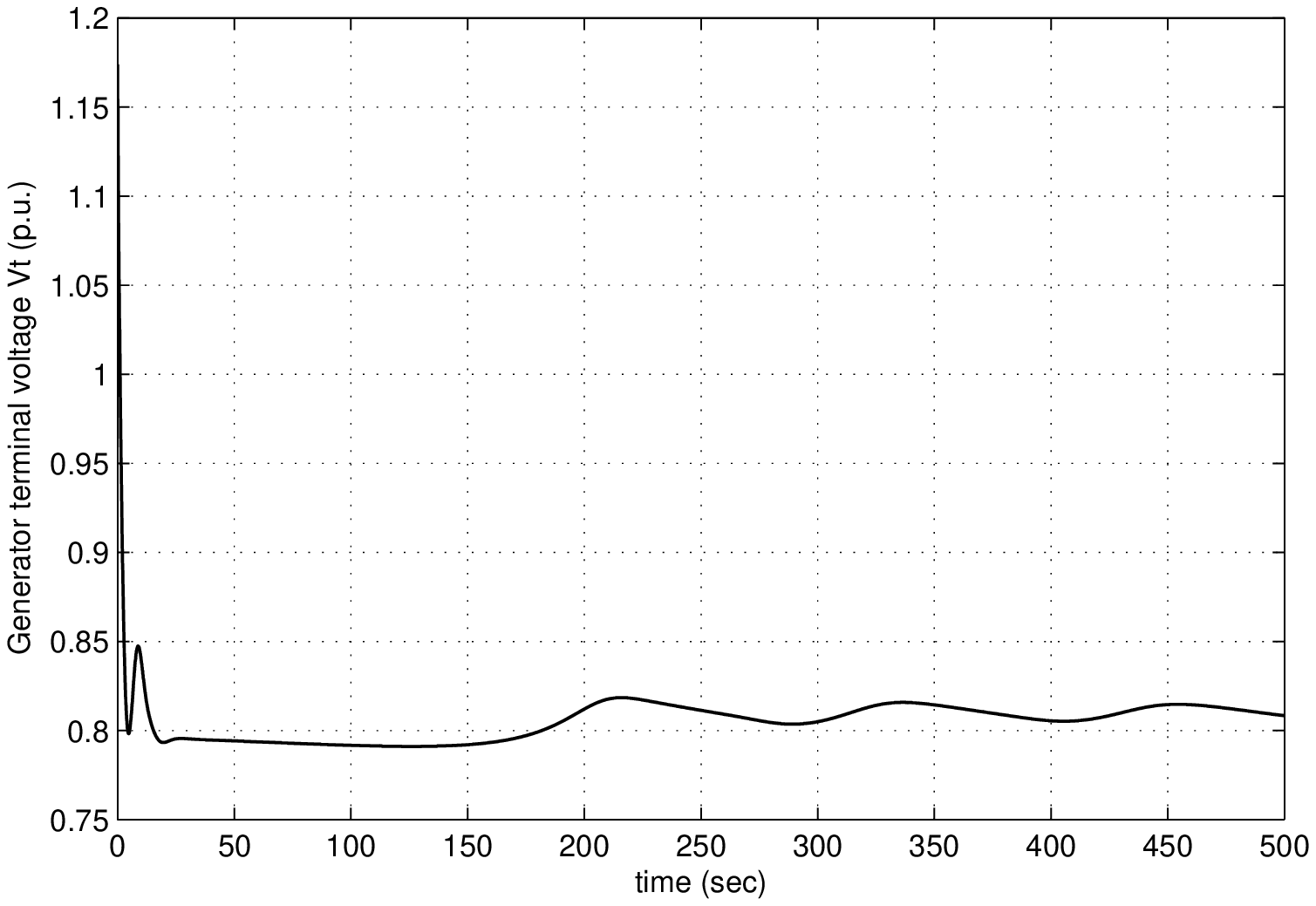}
          \caption{Plot of the generator terminal voltage $V_t$ vs time for the observer-based LQR applied to the reduced 
           order nonlinear  model}
          \label{fig:LQRobsnonlineartest1}
          \includegraphics[trim=0cm 0cm 0cm 0cm, clip=true, totalheight=0.27\textheight, 
           width=0.54\textwidth]{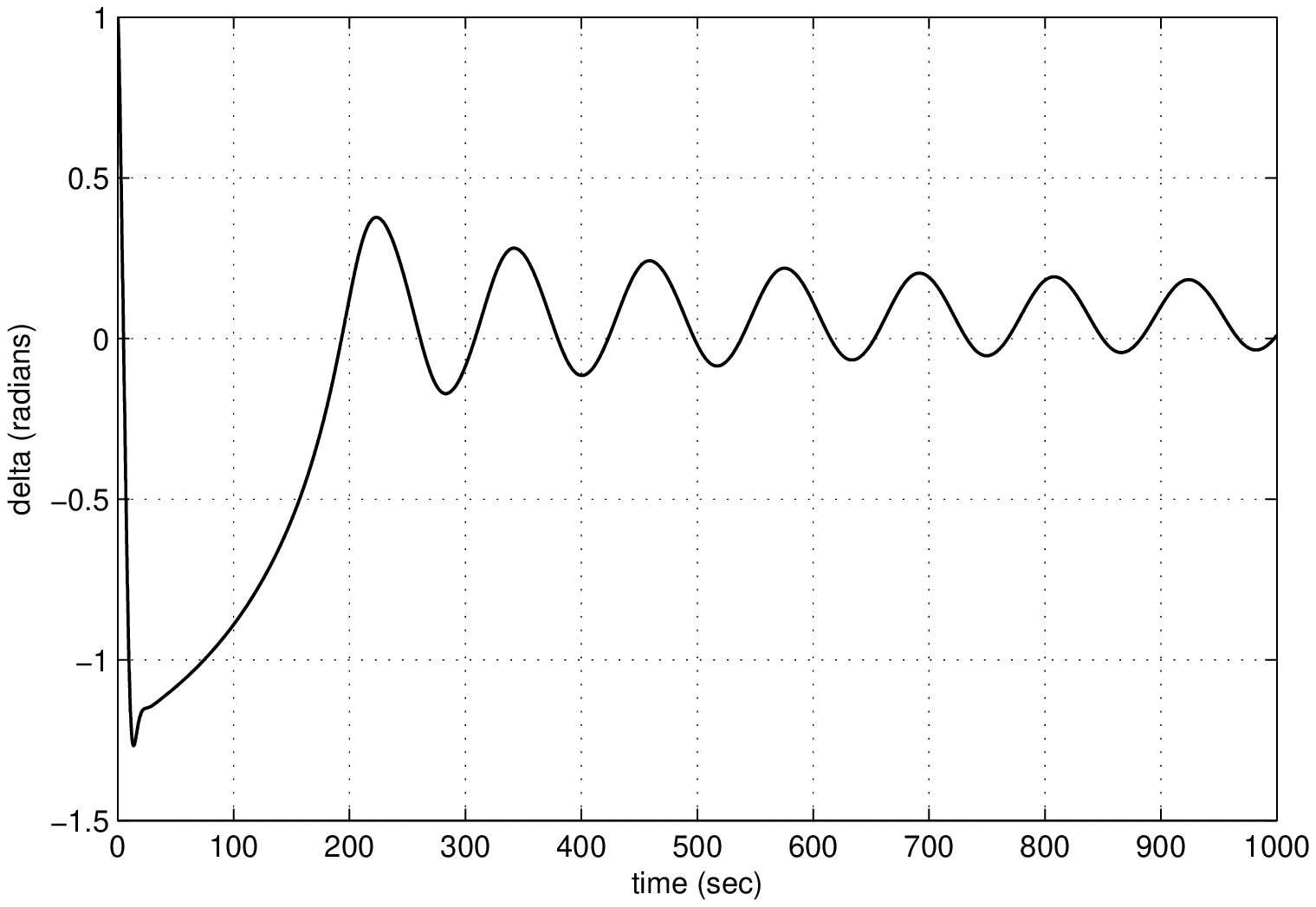}
          \caption{Plot of the rotor angle $\delta $ vs time for the observer-based LQR applied to 
          the reduced order nonlinear model}
          \label{fig:LQRobsnonlineartest2}
          \includegraphics[trim=0cm 0cm 0cm 0cm, clip=true, totalheight=0.27\textheight, 
          width=0.54\textwidth]{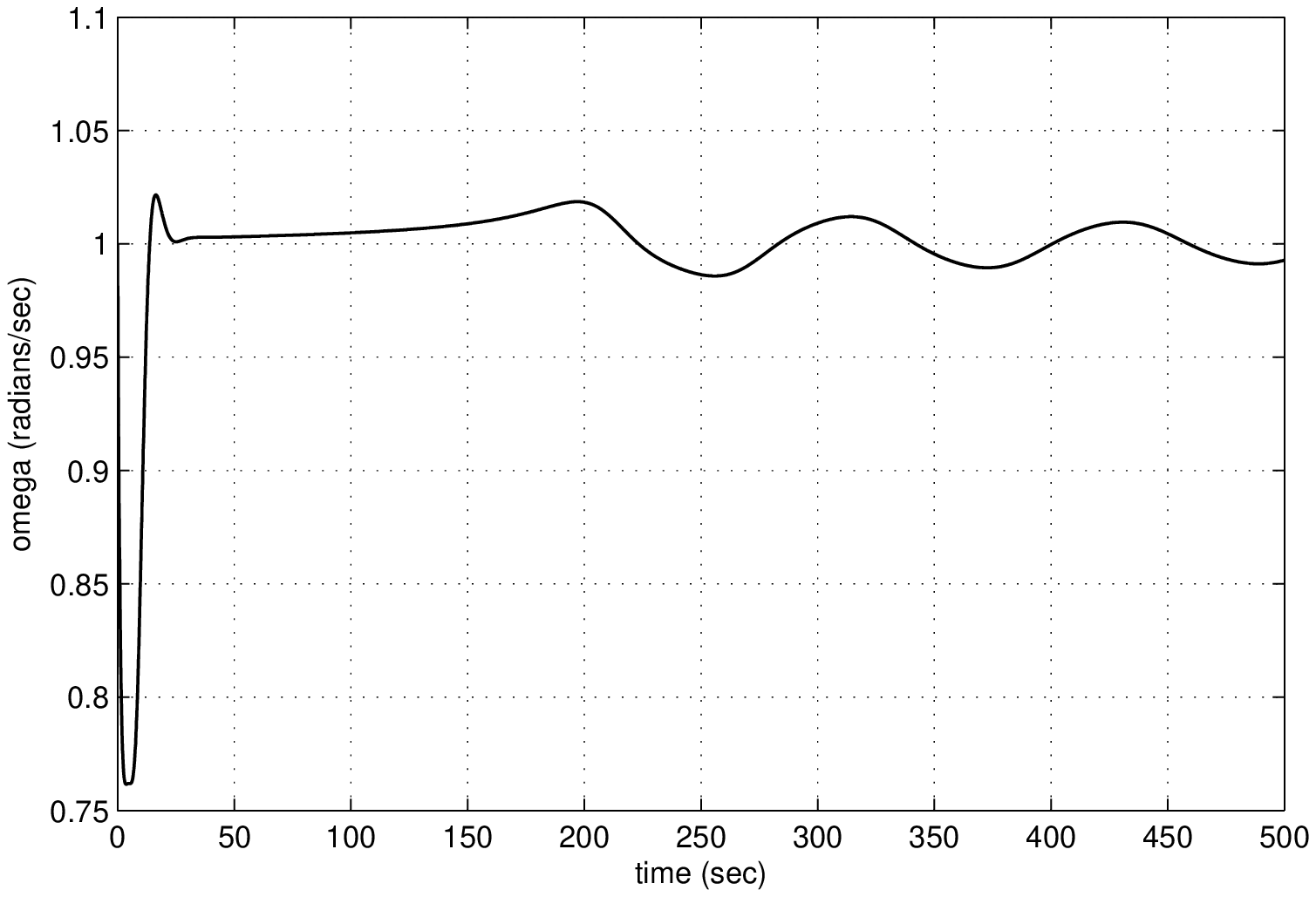}
          \caption{Plot of the frequency $\omega  $ vs time for the observer-based LQR applied to the reduced order nonlinear model}
          \label{fig:LQRobsnonlineartest3}          
\end{figure}

We now test the observer-based LQR on the reduced order nonlinear model. The generator terminal voltage
$V_t$ and the angular velocity $\omega $ are measured using sensors. The results obtained by directly applying the observer-based
LQR that was designed for the reduced order linear model in subsection 7.3.1 are poor. Therefore, the gains of the controller and observer poles are adjusted so that the controller works efficiently on the reduced order nonlinear model. 
Using the angular velocity or frequency $\omega $ of 
the synchronous generator which can be measured using a sensor, in addition to the generator terminal voltage $V_t$ 
the output equation can be written as
\begin{equation}
\begin{aligned}
               \mathbf{y}&= C_2\mathbf{x}=\bbm T_1 & 0 & T_2 & 0 & 0\\ 
                  0 & 1 & 0 & 0 & 0 \ebm \mathbf{x}\\
              \mathrm{i.e.} \ \        \bbm \Delta V_t\\ \Delta \omega \ebm &= \bbm 0.5258 & 0 & 0.0294 & 0 & 0\\ 
                  0 & 1 & 0 & 0 & 0 \ebm \mathbf{x} \\
\end{aligned}                         
\label{eq:oblq1} 
\end{equation} 
The estimator gain matrix $L$ for the output matrix $C_2$ as given in \autoref{eq:oblq1} and $\rho=12$ is
\begin{equation}
     L=\bbm 145.5739 & 29.0262\\-9.1126 & 22.2527\\-528.9546 & -442.3042 \\-2676.907 & -678.8487\\834.9962 & 219.7840\ebm
\label{eq:oblq2}
\end{equation} 
Choosing the weighting matrices
\begin{equation}
          Q=\bbm 5 & 0  & 0  & 0  & 0\\ 0 &  5 &  0  & 0  & 0\\ 0 &  0 &  0.5 &  0 &  0\\ 
            0 &  0  & 0  & 0.05  & 0\\ 0  & 0  & 0 &  0  & 5 \ebm
 \label{eq:oblq3}
\end{equation} 
and           
\begin{equation}
        R=\bbm 1000 & 0\\ 0 & 1000\ebm
\label{eq:oblq4}
\end{equation} 
and the state space matrices $(A, B)$ as given in \autoref{eq:linear25} and \autoref{eq:linear26} the 
control gain $K$ for the observer-based LQR with the outputs $V_t$ and $\omega $ measured is found to be 
  \begin{equation}
        K=\bbm 0.0021 & -0.0027 & 0.0005 & -0.0003 & -0.0001 \\ -0.0036 & 0.0079 & -0.0014 & 0.0009 & 0.0028  \ebm
\label{eq:oblq5}
\end{equation}
\autoref{fig:LQRobsnonlineartest1}, \autoref{fig:LQRobsnonlineartest2}, and \autoref{fig:LQRobsnonlineartest3} show  
simulation results for the re-tuned observer-based LQR applied to the reduced order nonlinear model. From these simulation results we can see 
that the generator terminal voltage $V_t$ oscillates between 0.8 and 0.82 p.u. which is different from the desired steady state
value of 1.172 p.u. by an amount of 0.372 p.u., the rotor angle $\delta $ oscillates between 0 and 0.4 p.u. with the amplitude of oscillations decreasing with time, which is also different from the desired steady state value of 1 p.u., and the frequency $\omega $ oscillates about the desired steady state value of 1 p.u. Even after re-tuning the gains of the controller and the observer poles
it is not possible to improve the results too much. A significant amount of steady state error is observed. The linear
observer is not able to accurately estimate the states of the reduced order nonlinear model, i.e. it is not robust to
uncertainties or state deviations between the reduced order nonlinear and the reduced order linear model. 
To overcome this difficulty i.e. to
increase the robustness of the observer we design a linear quadratic gaussian (LQG) controller, in which the gains of the Kalman 
filter are tuned using loop transfer recover (LTR) procedure, in the next subsection.

\newpage
\subsubsection{LTR-based LQG Controller applied to the Reduced Order Nonlinear Model}

In this section we present the design of a linear quadratic Gaussian (LQG) controller, where a design or gain adjustment procedure
 in the time domain is used which is analogous to loop shaping in the frequency domain, to adjust the gains of the Kalman filter. 
 This gain adjustment procedure not only improves the robustness of the observer but also asymptotically achieves the same
loop transfer function as a full-state feedback controller \citep{CDS4}. 
A basic requirement for every point of an adjustment trajectory is stability of the observer error dynamics. If this
requirement is not satisfied then closed-loop stability of the system is also lost. One way to assure stable error dynamics is to restrict the observer to be a Kalman filter for some set of noise parameters or covariance matrices $V_1$ and $V_2$, where $V_1$ is 
a $5\times 5$ matrix of intensities related to the plant disturbances or process noise, and $V_2$ is 
a $2\times 2$ matrix of intensities related to the measurement or sensor noise. Thus let the observer gain $H(q)$ be given by
the Kalman filter expression
\begin{equation}
           H(q)=\Sigma (q)C^TV^{-1}_2
\label{eq:lqg_ltr5}
\end{equation}           
with $\Sigma (q)$ defined by the Riccati equation
\begin{equation}
           A\Sigma (q)+\Sigma (q)A^T+V_1(q)-\Sigma (q)C^TV^{-1}_2C\Sigma (q)=0
\label{eq:lqg_ltr6}
\end{equation}
We select $V_1=V^T_1>0$ and $V_2=V^T_2>0$ with $(A,V^{\frac{1}{2}}_1)$ and $(C,A)$ stabilizable and observable, respectively.
$V_1(q)$ which is a function of $q$ is designed as
\begin{equation}
\begin{aligned}
                 V_1(q)&=V_{10}+q^2BVB^T\\
                 V_2(q)&=V_{20}\\
   \end{aligned}              
\label{eq:lqg_ltr7}
\end{equation} 
where $V_{10}$ and $V_{20}$ are noise intensities appropriate for the nominal plant, and $V$ is any positive definite
symmetric matrix, which are chosen as
\begin{equation}
\begin{aligned}
              V_{10}&=\bbm 1 & 0 & 0 & 0 & 0\\0 & 1 & 0 & 0 & 0\\0 & 0 & 1 & 0 & 0\\0 & 0 & 0 & 1 & 0\\0 & 0 & 0 & 0 & 1\ebm\\
              V_{20}&=\bbm 1 & 0\\0 & 1\ebm\\
              V&=\bbm 1 & 0\\0 & 1\ebm\\
 \end{aligned}             
\label{eq:lqg_ltr8}
\end{equation}              
With these selections, the observer gain for $q=0$ corresponds to the nominal Kalman filter gain. However as $q$ approaches
infinity, the gains are seen from \autoref{eq:lqg_ltr6} to satisfy 
\begin{equation}
                   \frac{HV_2H^T}{q^2}\rightarrow BVB^T
\label{eq:lqg_ltr9}
\end{equation}  
By substituting \autoref{eq:lqg_ltr7} in \autoref{eq:lqg_ltr6} and dividing \autoref{eq:lqg_ltr6} by $q^2$ we get
\begin{equation}
             A\bigg{(}\frac{\Sigma (q)}{q^2}\bigg{)}+\bigg{(}\frac{\Sigma (q)}{q^2}\bigg{)}A^T+\frac{V_{10}}{q^2}+BVB^T
               -q^2\bigg{(}\frac{\Sigma (q)}{q^2}\bigg{)}C^TV^{-1}_2C\bigg{(}\frac{\Sigma (q)}{q^2}\bigg{)}=0
\label{eq:lqg_ltr10}
\end{equation}
From \citep{CDS4} it can be seen that
\begin{equation}
               \frac{\Sigma (q)}{q^2}\rightarrow 0 \ \ \ \ \mathrm{as} \ q\rightarrow \infty
\label{eq:lqg_ltr11}
\end{equation} 
whenever the transfer function $C(sI-A)^{-1}B$ has no right half plane zeros.
Consequently,
\begin{equation}
               q^2\bigg{(}\frac{\Sigma (q)}{q^2}\bigg{)}C^TV^{-1}_2C\bigg{(}\frac{\Sigma (q)}{q^2}\bigg{)}\rightarrow BVB^T
\label{eq:lqg_ltr12}
\end{equation}
and thus \autoref{eq:lqg_ltr9} is established. Solutions of \autoref{eq:lqg_ltr9} must necessarily be of the form
\begin{equation}
               \frac{H(q)}{q}\rightarrow BV^{\frac{1}{2}}(V^{\frac{1}{2}}_2)^{-1}
\label{eq:lqg_ltr13}
\end{equation} 
where $V^{\frac{1}{2}}$ denotes some square root of $V$, i.e. $(V^{\frac{1}{2}})^T V^{\frac{1}{2}}=V$, and similarly
$V^{\frac{1}{2}}_2$ is some square root of $V_2$. 
The design adjustment procedure defined by \autoref{eq:lqg_ltr5} to \autoref{eq:lqg_ltr7} will achieve the desired 
robustness improvement objective. We can see that higher the value of $q$, higher will be the observer gain $H$ 
and thus higher will be the robustness of the system which is maximum at $q=\infty $. Our objective is to
apply the LTR-based LQG controller on the reduced order nonlinear model. From simulations it is verified that for 
very large values of $q$ the system goes unstable. Thus the choice of $q$ is very important and it cannot be
arbitrarily selected to a large value, when we are applying
this design adjustment procedure on a nonlinear model. The design parameter $q$ is suitable selected to be equal to 
$9.0005$ to get satisfactory closed loop performance of the system. From the separation principle we know that
the gains of the Kalman filter $H(q)$ and the controller gains $K$ can be deigned independently. We use the LQR algorithm  
to design the controller gains. Thus by using the feedback law,
\begin{equation}
          \mathbf{u}=-K\mathbf{\hat{x}}
\label{eq:ltr1}
\end{equation}
\begin{figure}
          \centering
          \includegraphics[trim=0cm 0cm 0cm 0cm, clip=true, totalheight=0.27\textheight, width=0.54
           \textwidth]  {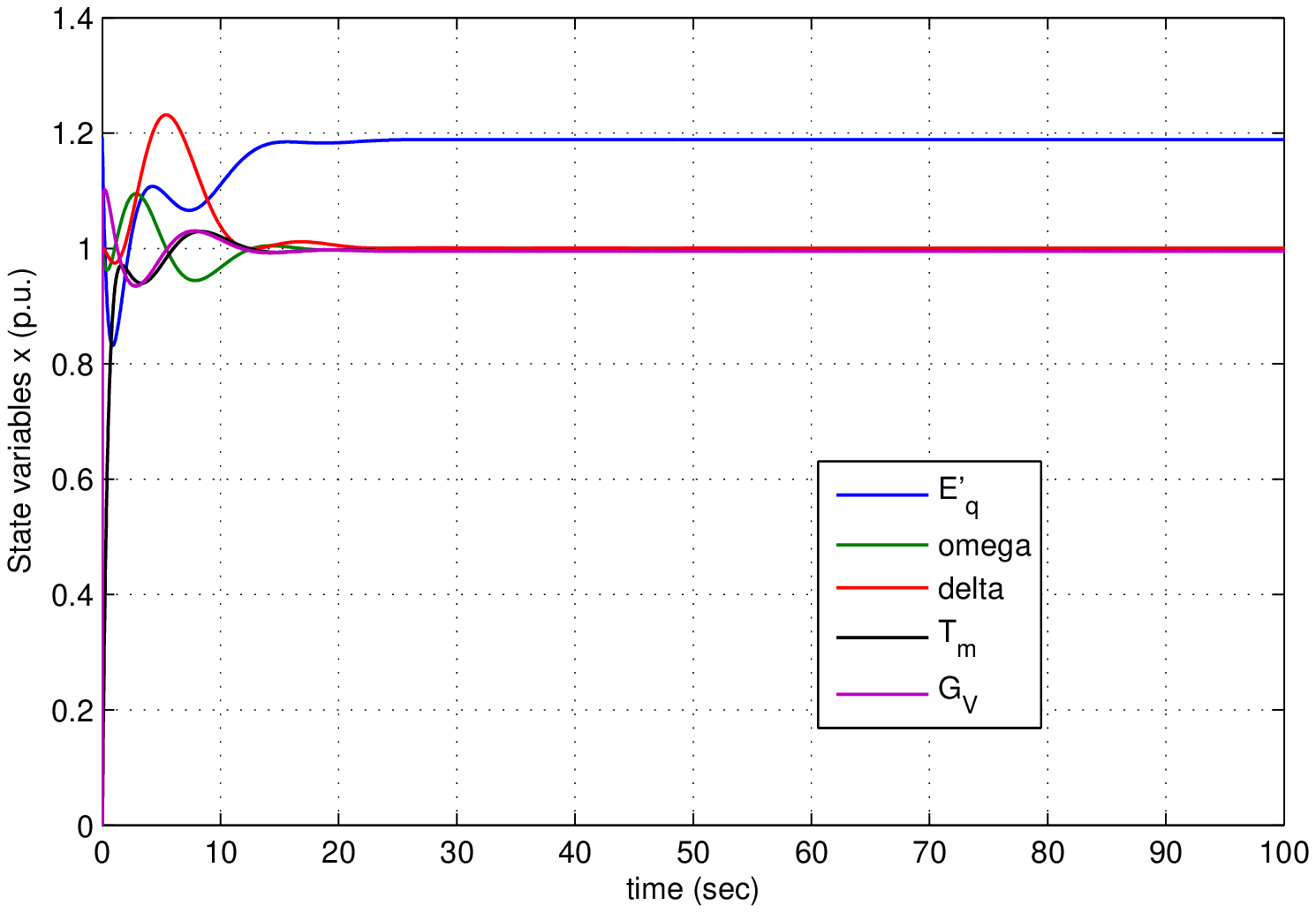}
          \caption{Plot of the state variables $E'_q$, $\omega $, $\delta $, 
           $T_m$, and $G_V$ vs time for the LTR-based LQG controller applied to the reduced order nonlinear model}
          \label{fig:ltrnonlinear1}
          \includegraphics[trim=0cm 0cm 0cm 0cm, clip=true, totalheight=0.27\textheight, 
           width=0.54\textwidth]   {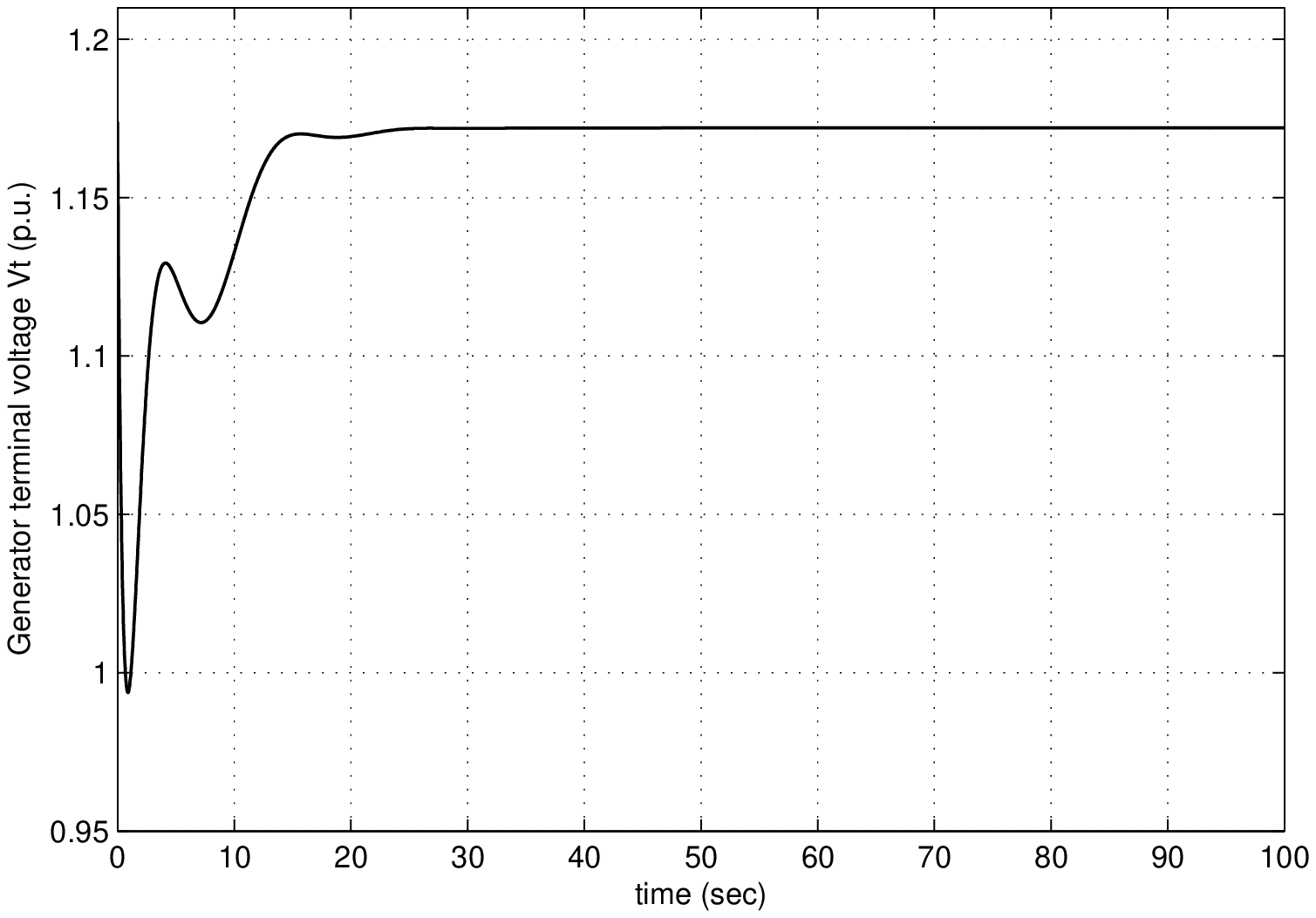}
          \caption{Plot of the generator terminal voltage $V_t$ vs time for the LTR-based LQG controller applied to the reduced 
           order nonlinear  model}
          \label{fig:ltrnonlinear2}
          \includegraphics[trim=0cm 0cm 0cm 0cm, clip=true, totalheight=0.27\textheight, 
           width=0.54\textwidth]{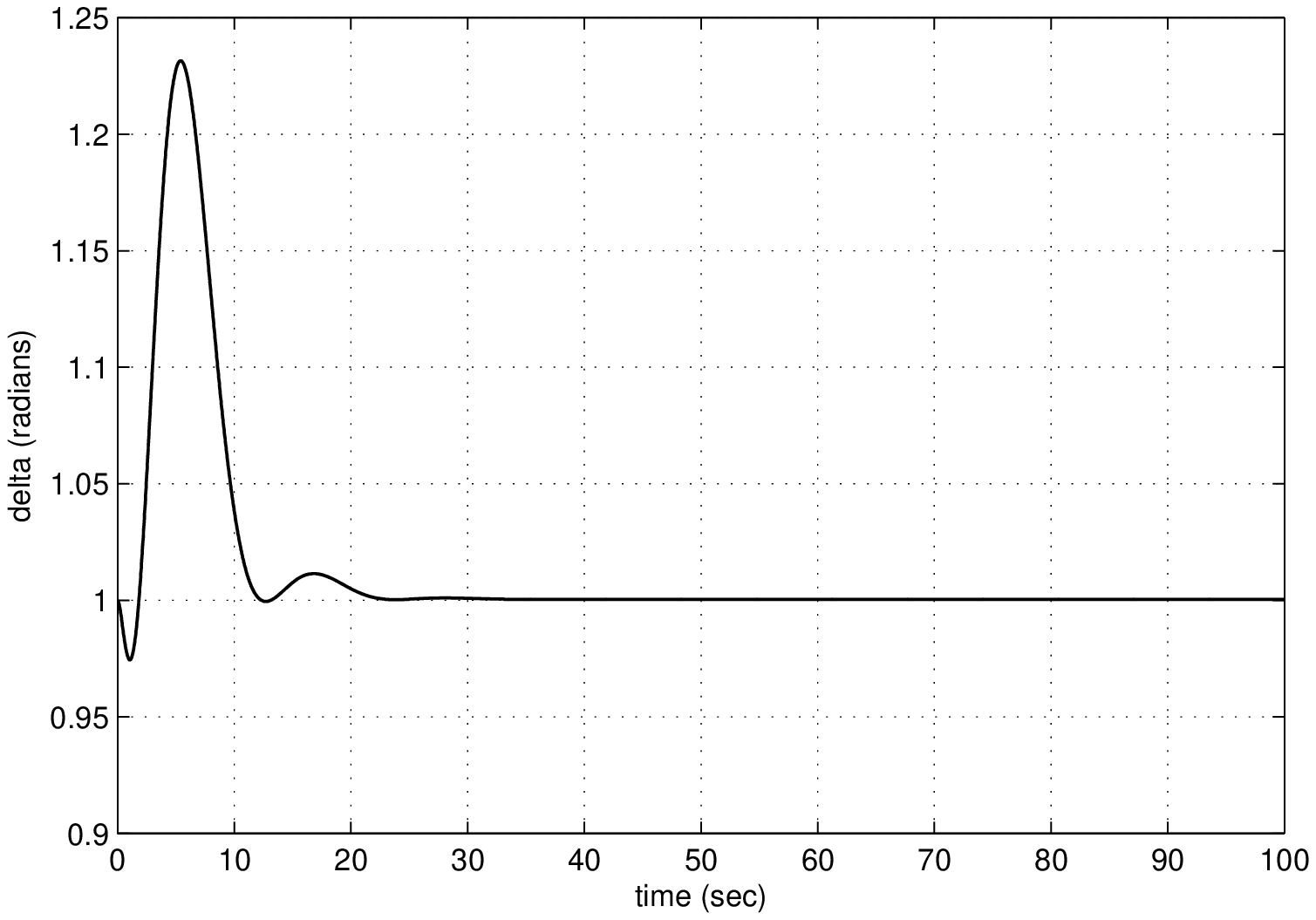}
          \caption{Plot of the rotor angle $\delta $ vs time for the LTR-based LQG controller applied to 
          the reduced order nonlinear model}
          \label{fig:ltrlinear3}
\end{figure}
          
\begin{figure}
          \centering          
          \includegraphics[trim=0cm 0cm 0cm 0cm, clip=true, totalheight=0.27\textheight, 
          width=0.54\textwidth]{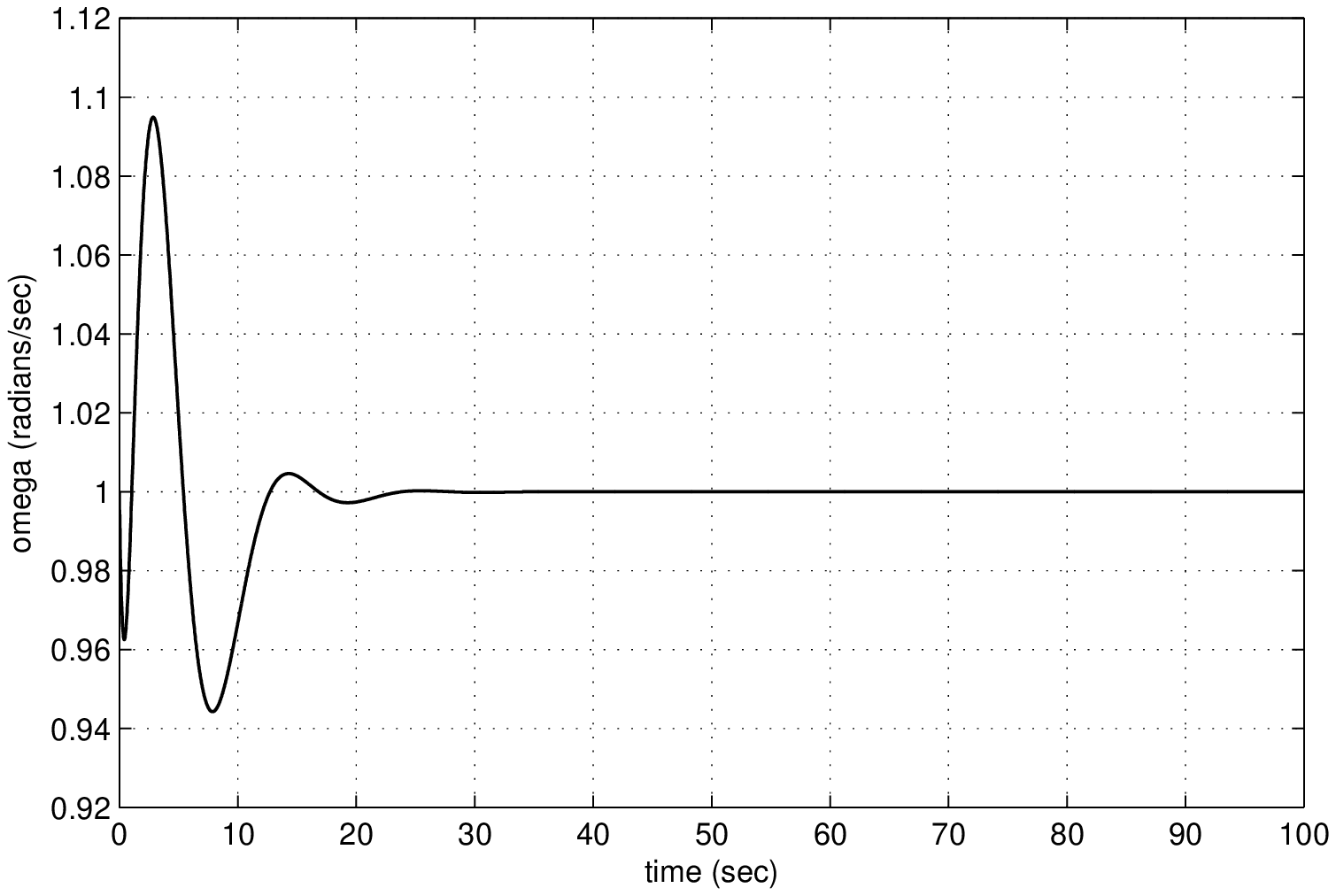}
          \caption{Plot of the frequency $\omega  $ vs time for the LTR-based LQG controller
          applied to the reduced order nonlinear model}
          \label{fig:ltrnonlinear4}
          \includegraphics[trim=0cm 0cm 0cm 0cm, clip=true, totalheight=0.27\textheight, 
          width=0.54\textwidth]{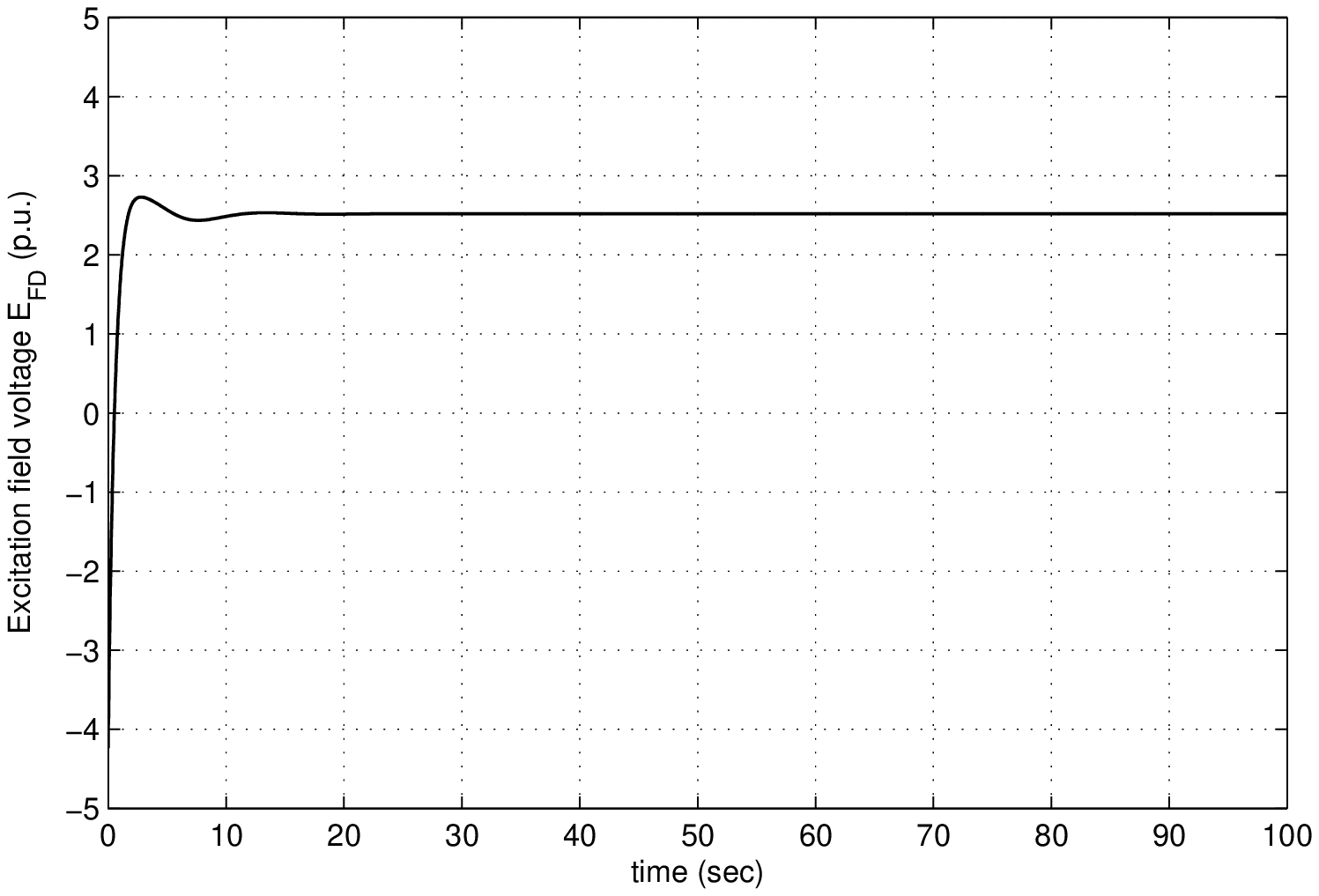}
          \caption{Plot of the generator excitation field control $E_{FD} $ vs time for the
           LTR-based LQG controller applied to 
          the reduced order nonlinear model}
          \label{fig:ltrnonlinear5}
          \includegraphics[trim=0cm 0cm 0cm 0cm, clip=true, totalheight=0.27\textheight, 
          width=0.54\textwidth]{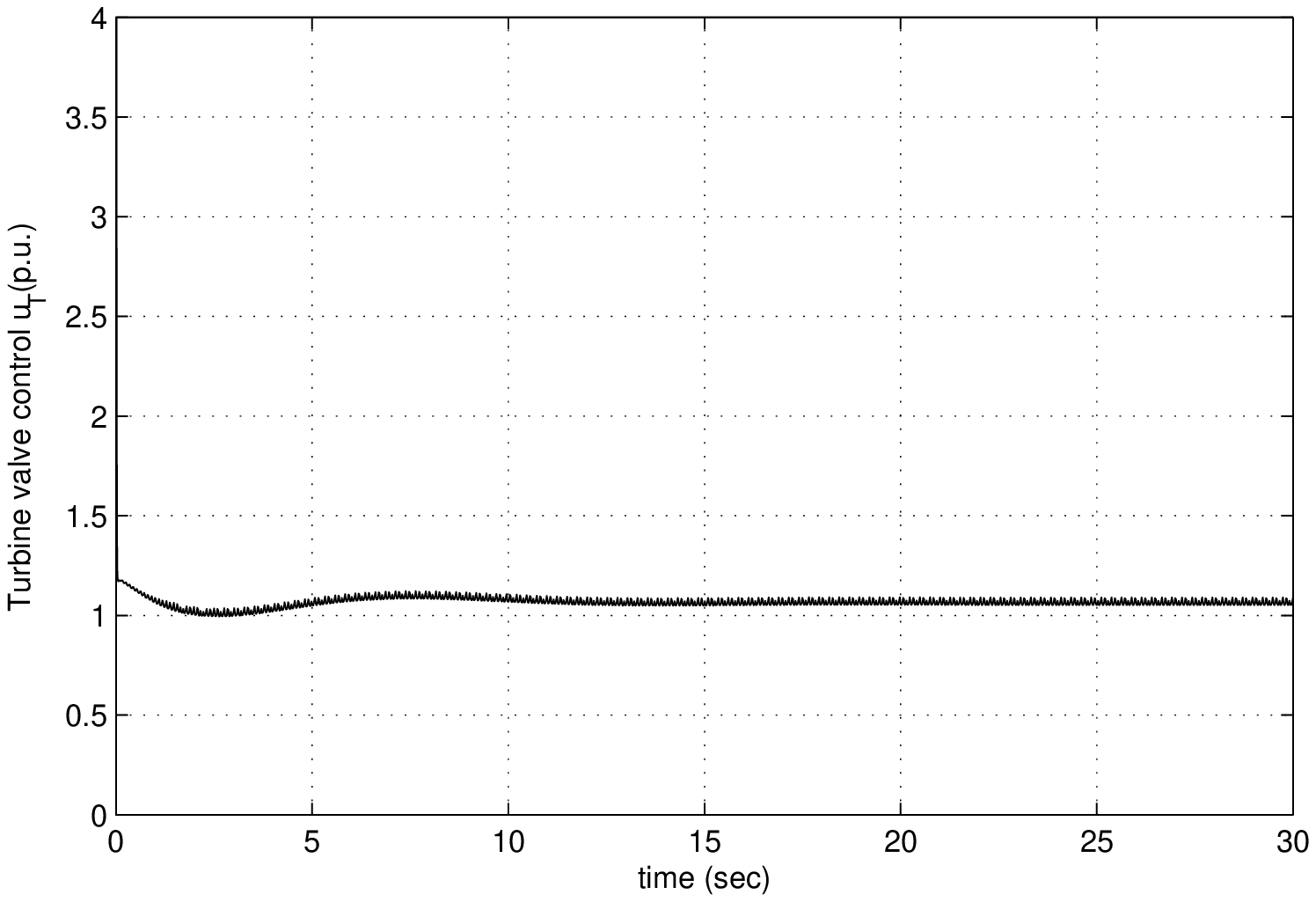}
          \caption{Plot of the turbine valve control $u_T$ vs time for the LTR-based LQG controller
         applied to the reduced order nonlinear model}
          \label{fig:ltrnonlinear6}
\end{figure}
where $\mathbf{\hat{x}}$ are the states estimated by the Kalman filter with loop shaping,
the weighting matrices
\begin{equation}
          Q=\bbm 1254.75 & 0  & 0  & 0  & 0\\ 0 &  1500 &  0  & 0  & 0\\ 0 &  0 &  544.5 &  0 &  0\\ 
            0 &  0  & 0  & 142.5  & 0\\ 0  & 0  & 0 &  0  & 1500 \ebm
 \label{eq:ltr2}
\end{equation} 
and           
\begin{equation}
        R=\bbm 1 & 0\\ 0 & 1\ebm
\label{eq:ltr3}
\end{equation} 
and the state space matrices $(A, B)$ as given in \autoref{eq:linear30} and \autoref{eq:linear31} the 
control gain $K$ is found to be 
  \begin{equation}
        K=\bbm 34.5065 & -49.5197 & -5.6995 & -4.1955 & -0.0432\\ -1.2745 & 28.0922 & -0.0281 & 4.3245 & 37.7873\ebm
\label{eq:ltr4}
\end{equation} 
\autoref{fig:ltrnonlinear1} to \autoref{fig:ltrnonlinear6} show simulation results for the LTR-based LQG 
controller applied to the reduced order nonlinear model. From these results we can see that all the state variables, outputs,
and control inputs, attain their desired steady state values respectively. Thus a significant improvement in the robustness
is observed when the Kalman filter gains are tuned using the loop-shaping design adjustment procedure explained above. The
performance of the LTR-based LQG controller is comparable to that of the LQR-based full-state feedback controller. 

\newpage
\subsubsection{LTR-based LQG Controller applied to the Truth Model}

It is not always possible to measure the rotor angle $\delta $ of the synchronous generator. Thus in cases where
the rotor angle cannot be measured, state feedback based control techniques are not applicable to the truth model. Hence it is necessary 
to design an observer-based output feedback controller which can be applied to the truth model.   

\begin{figure}
          \centering
          \includegraphics[trim=0cm 0cm 0cm 0cm, clip=true, totalheight=0.27\textheight, width=0.54
           \textwidth]  {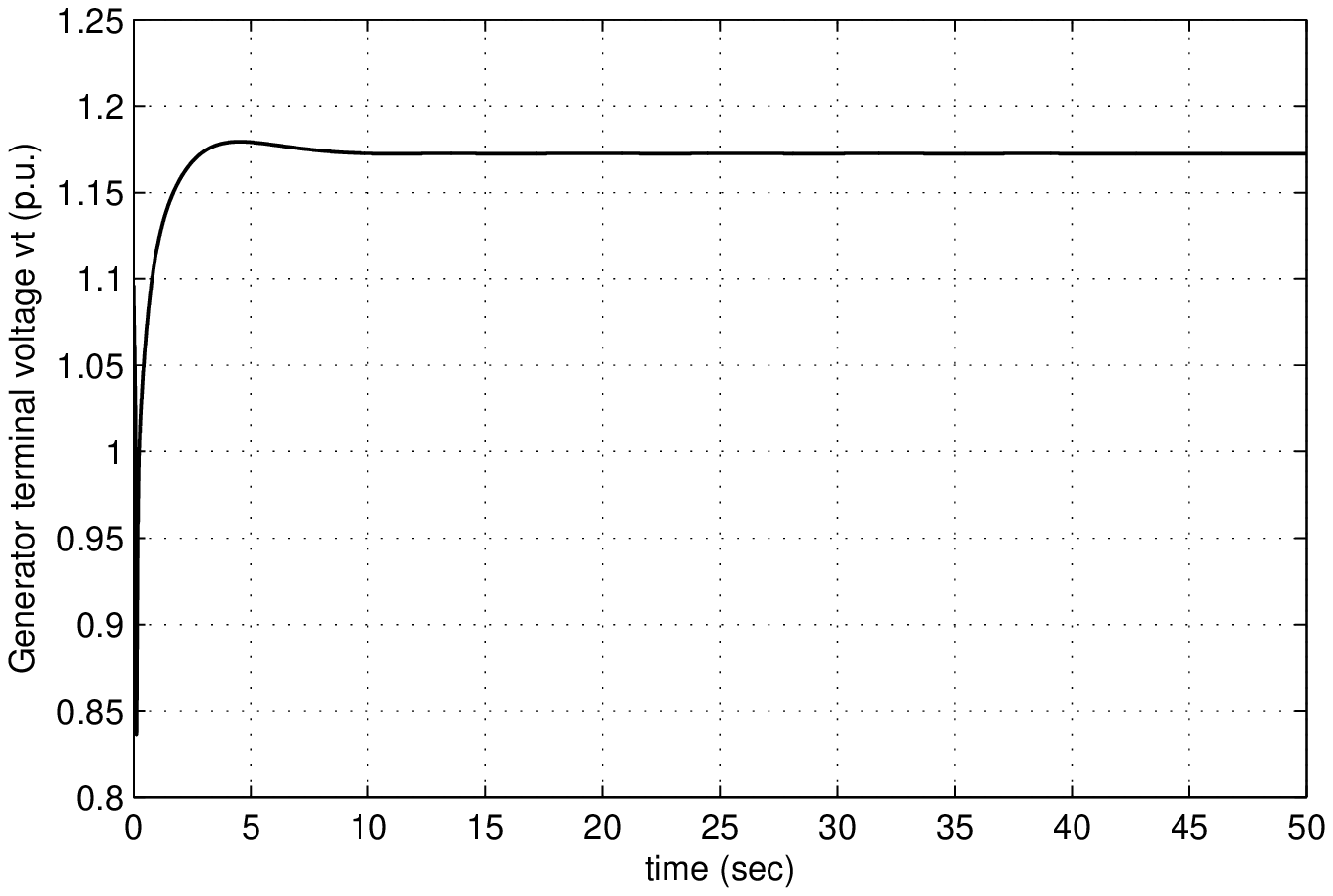}
          \caption{Plot of the generator terminal voltage $V_t$ vs time for the LTR-based LQG controller
          applied to the truth model}
          \label{fig:ltrtruth1}
          \includegraphics[trim=0cm 0cm 0cm 0cm, clip=true, totalheight=0.27\textheight, width=0.54\textwidth]{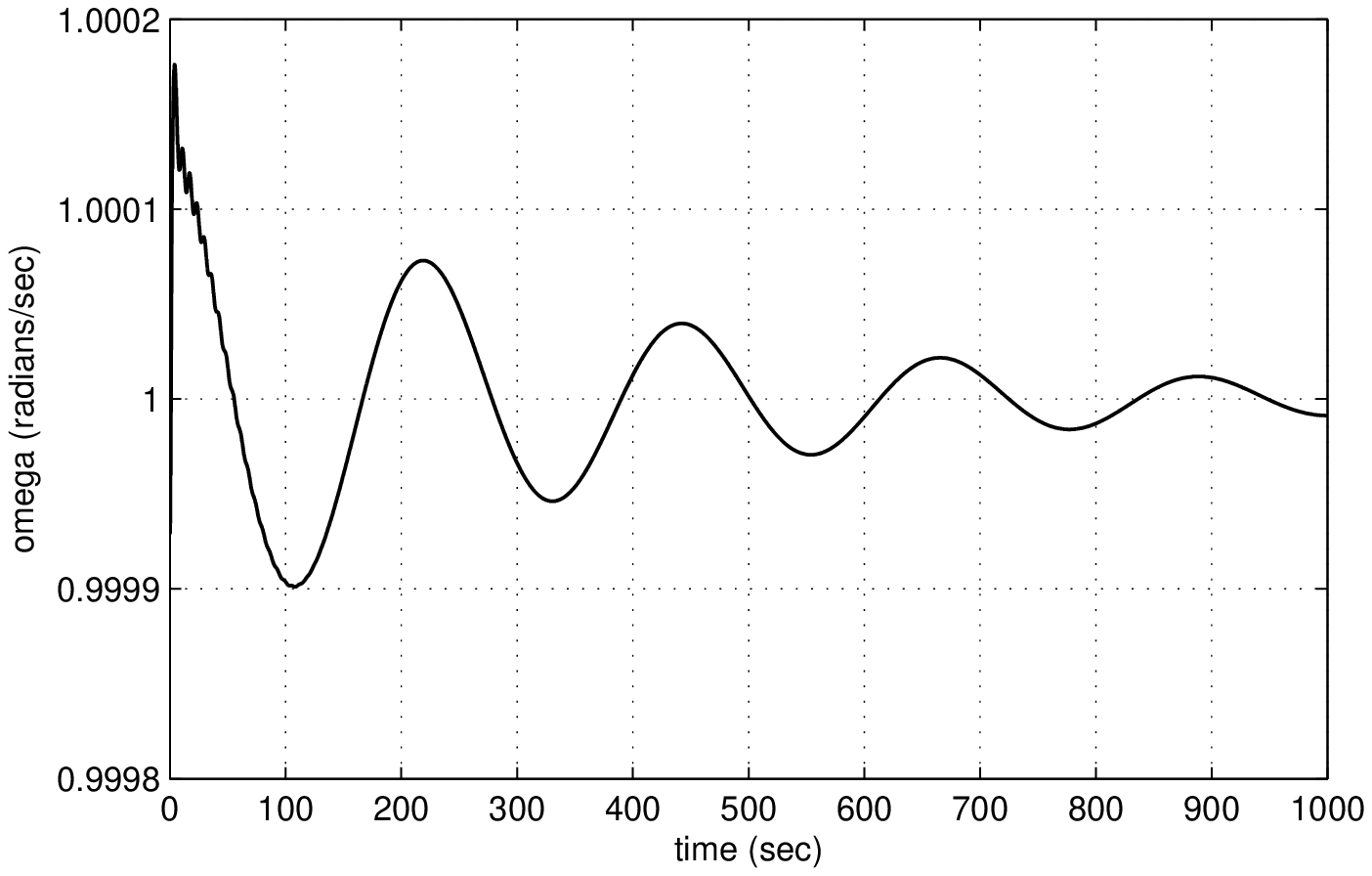}
          \caption{Plot of the angular velocity $\omega $ vs time for the LTR-based LQG controller
          applied to the truth model}
          \label{fig:ltrtruth2}
          \includegraphics[trim=0cm 0cm 0cm 0cm, clip=true, totalheight=0.27\textheight, width=0.54\textwidth]{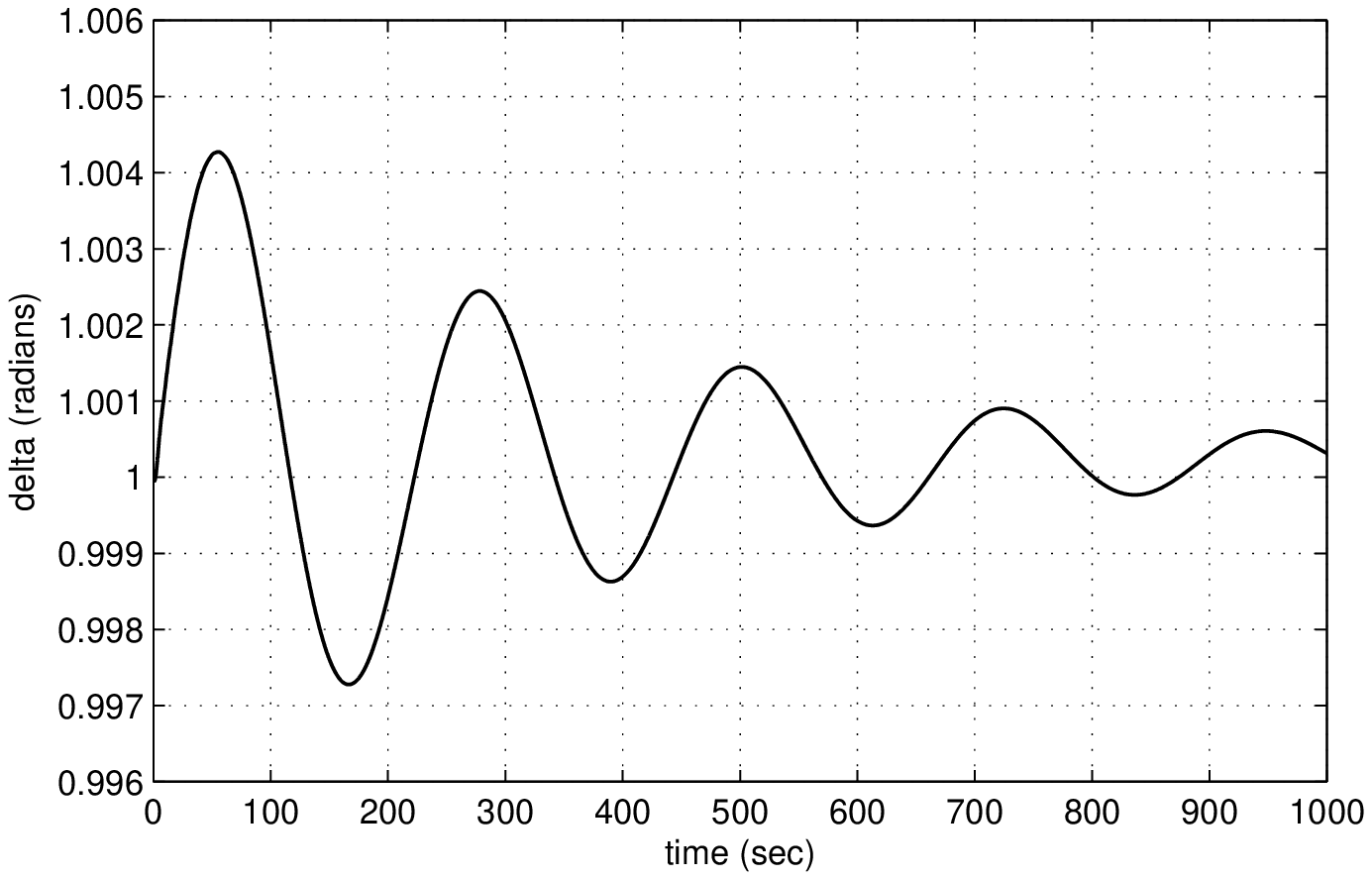}
          \caption{Plot of the rotor angle $\delta $ vs time for the LTR-based LQG controller applied to the truth model}
          \label{fig:ltrtruth3}
\end{figure}

\begin{figure}
          \centering          
          \includegraphics[trim=0cm 0cm 0cm 0cm, clip=true, totalheight=0.27\textheight, 
          width=0.54\textwidth]{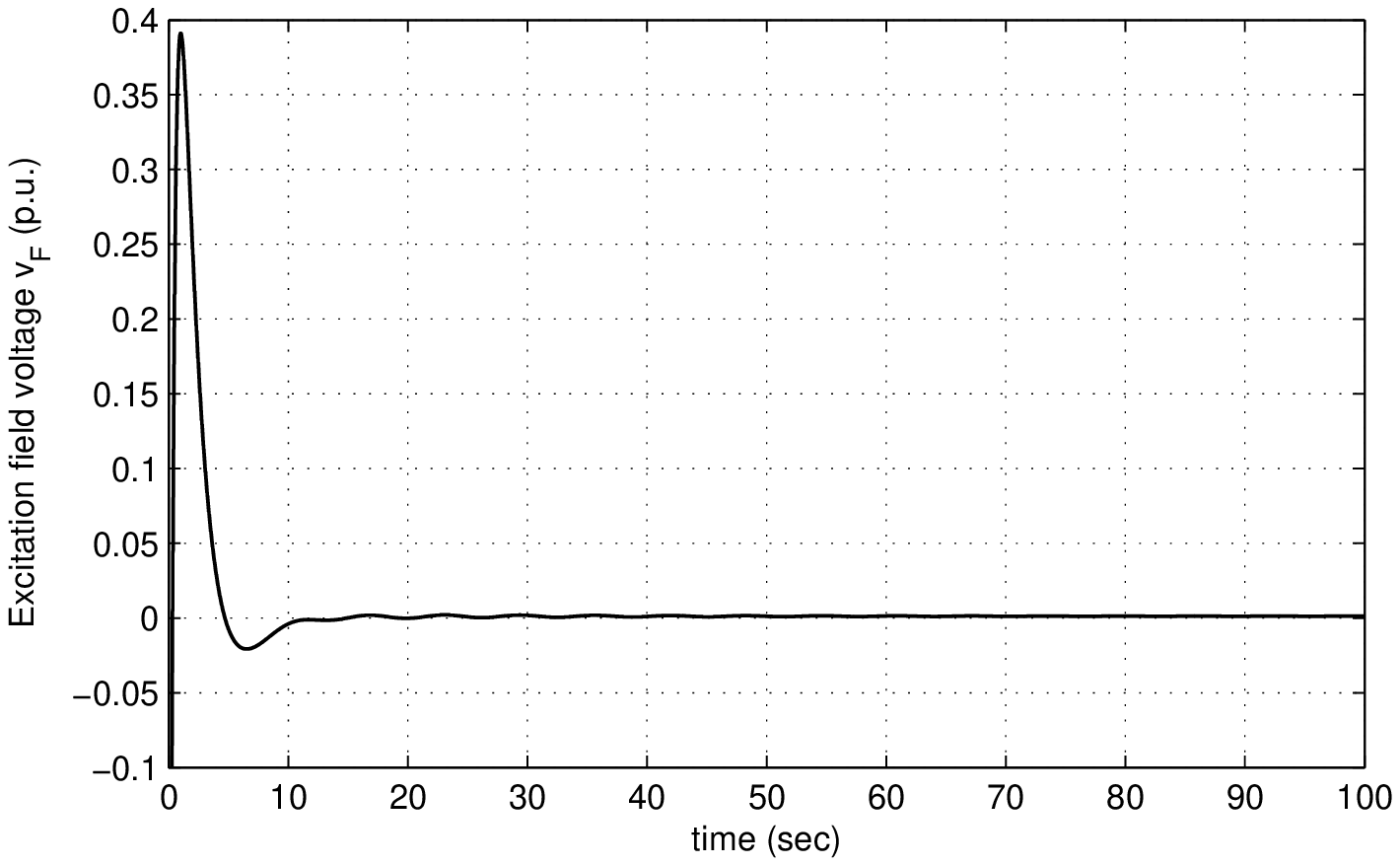}
          \caption{Plot of the generator excitation voltage $V_F$ vs time for the 
          LTR-based LQG controller applied to the truth model }
          \label{fig:ltrtruth4}
          \includegraphics[trim=0cm 0cm 0cm 0cm, clip=true, totalheight=0.27\textheight, 
          width=0.54\textwidth]{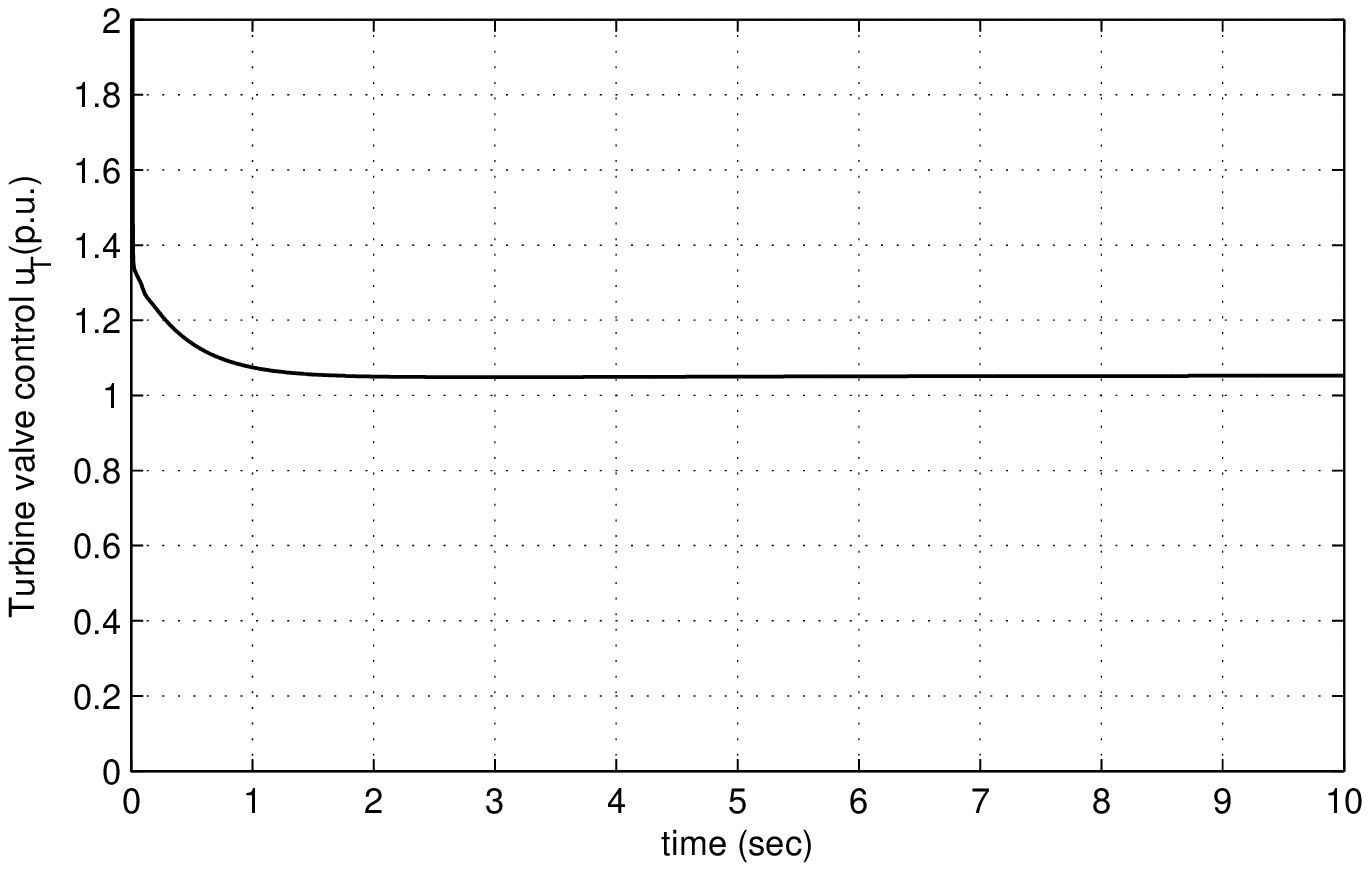}
          \caption{Plot of the turbine valve control $u_T$ vs time for the LTR-based LQG controller applied to the truth model}
          \label{fig:ltrtruth5}
\end{figure}

The loop transfer recovery (LTR)-based linear quadratic gaussian (LQG) controller that was tested on
the reduced order nonlinear model, is now tested on the truth model. The covariance matrices $V_1$, $V_2$, that are used 
in the design of the Kalman filter, the loop shaping parameter $q$, and the controller gains are tuned once again so that the LTR-based LQG controller works efficiently on the truth model.
We select $V_1=V^T_1>0$ and $V_2=V^T_2>0$ with $(A,V^{\frac{1}{2}}_1)$ and $(C,A)$ stabilizable and observable, respectively.
$V_1(q)$ which is a function of $q$ is designed as
\begin{equation}
\begin{aligned}
                 V_1(q)&=V_{10}+q^2BVB^T\\
                 V_2(q)&=V_{20}\\
   \end{aligned}              
\label{eq:ltrtruth1}
\end{equation}
$V_{10}$ and $V_{20}$ are noise intensities appropriate for the nominal plant, and $V$ is any positive definite
symmetric matrix, which are chosen as
\begin{equation}
\begin{aligned}
              V_{10}&=\bbm 1 & 0 & 0 & 0 & 0\\0 & 1 & 0 & 0 & 0\\0 & 0 & 1 & 0 & 0\\0 & 0 & 0 & 1 & 0\\0 & 0 & 0 & 0 & 1\ebm\\
              V_{20}&=\bbm 0.65 & 0\\0 & 0.65\ebm\\
              V&=\bbm 1 & 0\\0 & 1\ebm\\
 \end{aligned}             
\label{eq:ltrtruth2}
\end{equation} 
The loop shaping parameter $q$ is tuned to 5.25. We use the LQR algorithm  
to design the controller gains. Thus by using the feedback law,
\begin{equation}
          \mathbf{u}=-K\mathbf{\hat{x}}
\label{eq:ltrtruth3}
\end{equation}
where $\mathbf{\hat{x}}$ are the states estimated by the Kalman filter with loop shaping,
the weighting matrices
\begin{equation}
          Q=\bbm 7500 & 0  & 0  & 0  & 0\\ 0 &  15000 &  0  & 0  & 0\\ 0 &  0 &  16500 &  0 &  0\\ 
            0 &  0  & 0  & 7500  & 0\\ 0  & 0  & 0 &  0  & 7500 \ebm
 \label{eq:ltrtruth4}
\end{equation} 
and           
\begin{equation}
        R=\bbm 1 & 0\\ 0 & 1\ebm
\label{eq:ltrtruth5}
\end{equation} 
and the state space matrices $(A, B)$ as given in \autoref{eq:linear30} and \autoref{eq:linear31} the 
control gain $K$ is found to be 
  \begin{equation}
        K=\bbm 87.3944 & -216.7677 & -60.7947 & -13.4353 & -0.0618\\ -1.8244 & 98.0650 & 17.7303 & 42.1399 & 85.8027\ebm
\label{eq:ltrtruth6}
\end{equation} \autoref{fig:ltrtruth1} to \autoref{fig:ltrtruth5} show simulation results
for the LTR-based LQG controller applied to the truth model. From \autoref{fig:ltrtruth1}, \autoref{fig:ltrtruth2}, and
\autoref{fig:ltrtruth3} we can see that $V_t$, $\omega $, and $\delta $ settle to their respective steady state values. Small
oscillations which decay with time are seen for $\omega $, and $\delta $. \autoref{fig:ltrtruth4} and 
\autoref{fig:ltrtruth5} show the plots for the two control inputs $V_F$ and $u_T$ respectively. The generator 
excitation voltage $V_F$ settles to its steady state value of 0.00121 p.u. and the turbine valve control settles to its
steady state value of 1.0512 p.u.


\newpage

\section{Nonlinear Feedback Linearizing Controller Design}

We present the design of a nonlinear input-state feedback linearizing controller for the nonlinear model 
of the synchronous generator and turbine connected to an infinite bus. Before we proceed to the nonlinear controller
design, we give the fundamentals of full-state feedback linearization for single-input, single-output
(SISO) nonlinear control systems \citep{ISID, HKL}, which are affine in the control input, that is, systems of the form
\begin{equation}
\begin{aligned}
              \dot{x} &= f(x)+g(x)u\\
                   y  &= h(x)\\
\end{aligned}                         
\label{eq:nfc1} 
\end{equation} 
where $x\in \mathbb{R}^n$, $u\in\mathbb{R}$, $y\in\mathbb{R}$. We assume that the vector fields $f:D \rightarrow \mathbb{R}^n$, 
$g:D \rightarrow \mathbb{R}^n$ and the readout map $h:D \rightarrow \mathbb{R}$ are smooth in the domain $D\subset \mathbb{R}^n$,
that is their partial derivatives with respect to $x$ of any order exist and are continuous in $D$. Our objective is to 
find a smooth full state feedback control law of the form 
 \begin{equation}
          u=\alpha (x)+\beta (x)v
\label{eq:nfc2}           
\end{equation}
and a change of variables $z=T(x)$ that transform the nonlinear system into an equivalent linear system, i.e.
the closed loop system
\begin{equation}
 \begin{aligned}
              \dot{x} &= f(x)+g(x)\alpha (x)+g(x)\beta (x)v\\
                   y  &= h(x)\\
\end{aligned}                         
\label{eq:nfc3} 
\end{equation}
is completely linearized. The ability to use state feedback control to convert a nonlinear state
equation into a controllable linear state equation by canceling nonlinearities requires the nonlinear
state equation to have the structure
\begin{equation}
          \dot{x}=Ax+B\gamma (x)[u-\alpha (x)]
\label{eq:nfc4} 
\end{equation}
where $A$ is $n\times n$, $B$ is $n\times p$, the pair $(A,B)$ is controllable, the functions
$\alpha :\mathbb{R}^n\rightarrow \mathbb{R}^p$ and $\gamma :\mathbb{R}^n\rightarrow \mathbb{R}^{p\times p}$ are
defined in a domain $D\subset \mathbb{R}^n$ that contains the origin, and the matrix $\gamma (x)$ is nonsingular
for every $x\in D$. If the state equation takes the form \autoref{eq:nfc4}, then we can linearize it via the state 
feedback 
\begin{equation}
          u=\alpha (x)+\beta (x)v
\label{eq:nfc5}           
\end{equation}
where $\beta (x)=\gamma ^{-1}(x)$, to obtain the linear state equation
\begin{equation}
           \dot{x}=Ax+Bv
\label{eq:nfc6}           
\end{equation} 
For stabilization, we design $v=-Kx$ such that $A-BK$ is Hurwitz. The overall nonlinear stabilizing state feedback
control is           
\begin{equation}
          u=\alpha (x)-\beta (x)Kx
\label{eq:nfc7}           
\end{equation}

\subsection*{Example 1 \citep{HKL}}

To introduce the idea of exact feedback linearization, let us consider the pendulum
stabilization problem. Inspection of the state equation
\begin{equation}
\begin{aligned}
             &\dot{x}_1=x_2\\
             &\dot{x}_2=-a[\sin (x_1+\delta )-\sin \delta ]-bx_2+cu\\
 \end{aligned}            
\label{eq:nfl1}           
\end{equation}              
shows that we can choose $u$ as 
\begin{equation}
                u=\bigg{(}\frac{a}{c}\bigg{)}[\sin (x_1+\delta )-\sin \delta ]+\frac{v}{c}
\label{eq:nfl2}           
\end{equation}
to cancel the nonlinear term $a[\sin (x_1+\delta )-\sin \delta ]$. This cancellation results in the
linear system
\begin{equation}
\begin{aligned}
             &\dot{x}_1=x_2\\
             &\dot{x}_2=-bx_2+v\\
 \end{aligned}            
\label{eq:nfl3}           
\end{equation} 
Thus, the stabilization problem for the nonlinear system has been reduced to a
stabilization problem for a controllable linear system. We can proceed to design a
stabilizing linear state feedback control
\begin{equation}
                v=k_1x_1+k_2x_2
\label{eq:nfl4}           
\end{equation}
to locate the eigenvalues of the closed-loop system
\begin{equation}
\begin{aligned}
             &\dot{x}_1=x_2\\
             &\dot{x}_2=k_1x_1+(k_2-b)x_2\\
 \end{aligned}            
\label{eq:nfl5}           
\end{equation}
in the open left-half plane. The overall state feedback control law is given by
\begin{equation}
                u=\bigg{(}\frac{a}{c}\bigg{)}[\sin (x_1+\delta )-\sin \delta ]+\frac{1}{c}(k_1x_1+k_2x_2)
\label{eq:nfl6}           
\end{equation}

However, if the nonlinear state equation does not have the structure of \autoref{eq:nfc4} for one choice of state variables,
it does not mean that we cannot linearize the system via feedback. A co-ordinate transformation which transforms the 
old states into new state variables, can be used to convert the nonlinear state equation in the original states into a structure
of \autoref{eq:nfc4} in the new state variables. 

\subsection*{Example 2 \citep{HKL}}

Consider the system
\begin{equation}
\begin{aligned}
             &\dot{x}_1=a\sin x_2\\
             &\dot{x}_2=-x^2_1+u\\
 \end{aligned}            
\label{eq:nfl7}           
\end{equation}
We cannot simply choose $u$ to cancel the nonlinear term $a\sin x_2$. However, if we
first change the variables by the transformation
\begin{equation}
\begin{aligned}
             &z_1=x_1\\
             &z_2=a\sin x_2=\dot{x}_1\\
 \end{aligned}            
\label{eq:nfl8}           
\end{equation}
then $z_1$ and $z_2$ satisfy
\begin{equation}
\begin{aligned}
             &\dot{z}_1=z_2\\
             &\dot{z}_2=a\cos x_2(-x^2_1+u)\\
 \end{aligned}            
\label{eq:nfl9}           
\end{equation}
and the nonlinearities can be canceled by the control
\begin{equation}
             u= x^2_1+\frac{1}{a\cos x_2}v        
\label{eq:nfl10}           
\end{equation}
which is well defined for $-\frac{\pi }{2} < x_2 < \frac{\pi }{2}$. The state equation in the new coordinates
($z_1$, $z_2$) can be found by inverting the transformation to express ($x_1$, $x_2$) in
terms of ($z_1$, $z_2$); that is,
\begin{equation}
\begin{aligned}
             &x_1=z_1\\
             &x_2=\sin ^{-1}\bigg{(}\frac{z_2}{a}\bigg{)}\\
 \end{aligned}            
\label{eq:nfl11}           
\end{equation}
which is well defined for $-a < z_2 < a$. The transformed state equation is given by
\begin{equation}
\begin{aligned}
             &\dot{z}_1=z_2\\
             &\dot{z}_2=a\cos\bigg{(}\sin ^{-1}\bigg{(}\frac{z_2}{a}\bigg{)}\bigg{)}(-z^2_1+u)\\
 \end{aligned}            
\label{eq:nfl12}           
\end{equation}

When a change of variables $z=T(x)$ is used to transform the state equation
from the $x$-coordinates to the $z$-coordinates, the map $T$ must be invertible; that is it must have an inverse map
$T^{-1}(\cdot )$ such that $x=T^{-1}(z)$ for all $z\in T(D)$, where $D$ is the domain of $T$. Moreover, because the derivatives
of $z$ and $x$ should be continuous, we require both $T(\cdot)$ and $T^{-1}(\cdot )$ to be continuously differentiable.
A continuously differentiable map with a continuously differentiable inverse is known as diffeomorphism \citep{ISID, HKL}. Thus, we can define a 
feedback linearizable system as\\
\\
Definition \cite{HKL}:
A nonlinear system
\begin{equation}
\begin{aligned}
              \dot{x} &= f(x)+g(x)u\\
                   y  &= h(x)\\
\end{aligned}                         
\label{eq:nfc8} 
\end{equation}
where $f:D \rightarrow \mathbb{R}^n$ and $g:D \rightarrow \mathbb{R}^{n\times p}$ are sufficiently smooth
on a domain $D\subset \mathbb{R}^n$, is said to be feedback linearizable (or input-state linearizable) if there exists
a diffeomorphism $T:D \rightarrow \mathbb{R}^n$ such that $D_z=T(D)$ contains the origin and the change of variables
$z=T(x)$ transforms the system into the form
\begin{equation}
          \dot{z}=Az+B\gamma (x)[u-\alpha (x)]
\label{eq:nfc9} 
\end{equation}
with $(A,B)$ controllable and $\gamma (x)$ nonsingular for all $x\in D$\\

\subsection{Nonlinear Feedback Linearizing Controller Design for the Reduced Order Model}

We now present the design of an input-state feedback linearizing controller for the reduced order model of the 
synchronous generator and turbine-governor system connected to an infinite bus. The fifth order nonlinear model of the 
system in the original coordinates is
\begin{equation}
\begin{aligned}
      &\dot{E}'_q = f_{11}E'_q+f_{12}\cos(\delta -\alpha)
             +f_{13}\sin(\delta -\alpha)+g_{11}E_{FD}\\
       &\dot{\omega } = f_{21}E'^2_q+
       f_{22}E'_q\cos(\delta -\alpha)
      + f_{23}E'_q\sin(\delta -\alpha)
      + f_{24}\sin(\delta -\alpha)\cos(\delta -\alpha)\\
      & \ \ \ \ \ +f_{25}\cos^2(\delta -\alpha)
      + f_{26}\sin^2(\delta -\alpha)+f_{27}\omega +f_{28}T_m\\      
       &\dot{\delta } = \omega -1\\
       &\dot{T}_{m} = f_{41}T_{m}+f_{42}G_{V}\\
       &\dot{G}_V = f_{51}\omega +f_{52}G_V+g_{55}u_T\\
       \end{aligned}
 \label{eq:nfc10}
\end{equation}
where $x = [E'_q, \omega, \delta, T_m, G_V]^\mathrm{T}$ are the original state variables 
and $u = [u_1, u_2]^\mathrm{T} = [ E_{FD}, u_T]^\mathrm{T}$ are the two control inputs.
We now define a transformation $z=T(x)$ that transforms the nonlinear state equations into a structure of \autoref{eq:nfc4}.
The new state variables are chosen as
\begin{equation}
\begin{aligned}
              z_1 &= \delta \\
              z_2 &= \dot{z}_1=\dot{\delta }=\omega -1\\
              z_3 &= \dot{z_2}=\ddot{\delta }=\dot{\omega }=f_{21}E'^2_q+
       f_{22}E'_q\cos(\delta -\alpha)
      + f_{23}E'_q\sin(\delta -\alpha)
      + f_{24}\sin(\delta -\alpha)\cos(\delta -\alpha)\\
      & \ \ \ \ \ \ \ \ \ \ \ \ \ \ \ \ \ \ \ \ \ +f_{25}\cos^2(\delta -\alpha)
      + f_{26}\sin^2(\delta -\alpha)+f_{27}\omega +f_{28}T_m\\\\
              z_4 &= T_m\\
              z_5 &=\dot{z}_4= \dot{T}_m=f_{41}T_{m}+f_{42}G_{V}\\
\end{aligned}
 \label{eq:nfc11}
\end{equation}
Thus, we have
\begin{equation}
\begin{aligned}
           \dot{z}_1 &= \dot{\delta }=\omega -1=z_2\\
           \dot{z}_2 &= \dot{\omega }=\ddot{\delta }=z_3\\ 
           \dot{z}_3 &= \ddot{\omega }=\dddot{\delta }=\sigma  _1(x)+\gamma  _1(x)E_{FD}\\
           \dot{z}_4 &= \dot{T}_m=z_5\\
           \dot{z}_5 &= \ddot{T}_m=\sigma  _2(x)+\gamma  _2(x)u_T\\
\end{aligned}
 \label{eq:nfc12}
\end{equation}
The state transformations given above are invertible and exist throughout the domain of stable operation
$0<\delta <180^\circ $. 
From \autoref{eq:nfc10} and \autoref{eq:nfc12}, i.e. by taking the derivative of $\dot{\omega}$ we have
\begin{equation}
\begin{aligned}
          \dot{z}_3 &= \ddot{\omega }=2f_{21}E'_q\dot{E}'_q+f_{22}\dot{E}'_q\cos(\delta -\alpha )
          -f_{22}E'_q\sin(\delta -\alpha )\dot{\delta }+f_{23}\dot{E}'_q\sin(\delta -\alpha )\\
                & \ \ +f_{23}E'_q\cos(\delta -\alpha )\dot{\delta }+f_{24}\cos^2(\delta -\alpha )\dot{\delta }
           -f_{24}\sin^2(\delta -\alpha )\dot{\delta }
             -2f_{25}\cos(\delta -\alpha )\sin(\delta -\alpha )\dot{\delta }\\
          & \ \ +2f_{26}\cos(\delta -\alpha )\sin(\delta -\alpha )\dot{\delta }+f_{27}\dot{\omega }+f_{28}\dot{T}_m\\
\end{aligned}
 \label{eq:nfc13}
\end{equation}
Substituting $\dot{\delta }$, $\dot{E}'_q$, $\dot{\omega }$, and $\dot{T}_m$ from \autoref{eq:nfc10} in \autoref{eq:nfc13} we get
\begin{equation}
\begin{aligned}
            \dot{z}_3 &= 2f_{21}E'_q(f_{11}E'_q+f_{12}\cos(\delta -\alpha)
             +f_{13}\sin(\delta -\alpha )+g_{11}E_{FD})\\
             & \ \ +f_{22}(f_{11}E'_q+f_{12}\cos(\delta -\alpha )
             +f_{13}\sin(\delta -\alpha )+g_{11}E_{FD})\cos(\delta -\alpha )\\
             & \ \ -f_{22}E'_q\sin(\delta -\alpha )(\omega -1)+f_{23}(f_{11}E'_q+f_{12}\cos(\delta -\alpha )
             +f_{13}\sin(\delta -\alpha )+g_{11}E_{FD})\sin(\delta -\alpha )\\
             & \ \ +f_{23}E'_q\cos(\delta -\alpha )(\omega -1)
             +f_{24}\cos^2(\delta -\alpha )(\omega -1)-f_{24}\sin^2(\delta -\alpha )(\omega -1)\\
             & \ \ -2f_{25}\cos(\delta -\alpha )\sin(\delta -\alpha )(\omega -1)
             +2f_{26}\cos(\delta -\alpha )\sin(\delta -\alpha )(\omega -1)\\
             & \ \ +f_{27}(f_{21}E'^2_q+f_{22}E'_q\cos(\delta -\alpha )
             + f_{23}E'_q\sin(\delta -\alpha )
             + f_{24}\sin(\delta -\alpha )\cos(\delta -\alpha )\\
             & \ \ +f_{25}\cos^2(\delta -\alpha )
             + f_{26}\sin^2(\delta -\alpha )+f_{27}\omega +f_{28}T_m)\\
             & \ \ +f_{28}(f_{41}T_{m}+f_{42}G_{V})\\
\end{aligned}
 \label{eq:nfc14}
\end{equation}
On rearranging and simplifying \autoref{eq:nfc14} we get
\begin{equation}
\begin{aligned}
            \dot{z}_3 &= (2f_{11}f_{21}+f_{27}f_{21}){E'_q}^2\\
            & \ \ +(2f_{21}f_{12}+f_{22}f_{11}-f_{23}+f_{27}f_{22})E'_q\cos(\delta -\alpha )\\
            & \ \ +(2f_{21}f_{13}+f_{22}+f_{23}f_{11}+f_{27}f_{23})E'_q\sin(\delta -\alpha )\\
            & \ \ +(f_{22}f_{12}-f_{24}+f_{27}f_{25})\cos^2(\delta -\alpha )\\
            & \ \ +(f_{23}f_{13}+f_{24}+f_{27}f_{26})\sin^2(\delta -\alpha )\\
            & \ \ +(f_{22}f_{13}+f_{23}f_{12}+2f_{25}-2f_{26}+f_{27}f_{24})\sin(\delta -\alpha)\cos(\delta -\alpha )\\
            & \ \ +f^2_{27}\omega +(f_{27}f_{28}+f_{28}f_{41})T_m+f_{28}f_{42}G_v\\
            & \ \ +(f_{23})E'_q\omega \cos(\delta -\alpha )\\
            & \ \ +(-f_{22})E'_q\omega \sin(\delta -\alpha )\\
            & \ \ +(f_{24})\omega \cos^2(\delta -\alpha )\\
            & \ \ +(-f_{24})\omega \sin^2(\delta -\alpha )\\
            & \ \ +(-2f_{25}+2f_{26})\omega \sin(\delta -\alpha)\cos(\delta -\alpha )\\
            & \ \ +(2f_{21}g_{11}E'_q+f_{22}g_{11}\cos(\delta -\alpha )+f_{23}g_{11}\sin(\delta -\alpha ))E_{FD}\\
\end{aligned}
 \label{eq:nfc15}
\end{equation} 
For simplification let us denote  
\begin{equation}
\begin{aligned}
               & 2f_{11}f_{21}+f_{27}f_{21}=p_{31},\\ 
               & 2f_{21}f_{12}+f_{22}f_{11}-f_{23}+f_{27}f_{22}=p_{32},\\
               & 2f_{21}f_{13}+f_{22}+f_{23}f_{11}+f_{27}f_{23}=p_{33},\\
               & f_{22}f_{12}-f_{24}+f_{27}f_{25}=p_{34},\\
               & f_{23}f_{13}+f_{24}+f_{27}f_{26}=p_{35},\\
               & f_{22}f_{13}+f_{23}f_{12}+2f_{25}-2f_{26}+f_{27}f_{24}=p_{36},\\
               & f^2_{27}=p_{37},\ \ f_{27}f_{28}+f_{28}f_{41}=p_{38},\\
               & f_{28}f_{42}=p_{39},\ \ f_{23}=q_{31}, \ \ -f_{22}=q_{32},\\ 
               & f_{24}=q_{33}, \ \ -f_{24}=q_{34},\ \ -2f_{25}+2f_{26}=q_{35},\\
               & 2f_{21}g_{11}=r_{31}, \ \ f_{22}g_{11}=r_{32}, \ \ f_{23}g_{11}=r_{33}\\
\end{aligned}
 \label{eq:nfc16}
\end{equation} 
On substituting \autoref{eq:nfc16} in \autoref{eq:nfc15} we get
\begin{equation}
\begin{aligned}
                \dot{z}_3 &= p_{31}{E'_q}^2+p_{32}E'_q\cos(\delta -\alpha )
             +p_{33}E'_q\sin(\delta -\alpha )
             +p_{34}\cos^2(\delta -\alpha )+p_{35}\sin^2(\delta -\alpha )\\
            & \ \ +p_{36}\sin(\delta -\alpha)\cos(\delta -\alpha )
             +p_{37}\omega +p_{38}T_m+p_{39}G_v\\
            & \ \ +q_{31}E'_q\omega \cos(\delta -\alpha )+q_{32}E'_q\omega \sin(\delta -\alpha )
               +q_{33}\omega \cos^2(\delta -\alpha )\\
            & \ \ +q_{34}\omega \sin^2(\delta -\alpha )+q_{35}\omega \sin(\delta -\alpha)\cos(\delta -\alpha )\\
            & \ \ +(r_{31}E'_q+r_{32}\cos(\delta -\alpha )+r_{33}\sin(\delta -\alpha ))E_{FD}\\
\end{aligned}
 \label{eq:nfc17}
\end{equation}
From \autoref{eq:nfc12} we have
\begin{equation}
        \dot{z}_3 =\sigma  _1(x)+\gamma  _1(x)E_{FD}=v_1
\label{eq:nfc18}
\end{equation}
Thus, equating \autoref{eq:nfc17} and \autoref{eq:nfc18} we get
\begin{equation}
\begin{aligned}
              \sigma  _1(x) &= p_{31}{E'_q}^2+p_{32}E'_q\cos(\delta -\alpha )
             +p_{33}E'_q\sin(\delta -\alpha )
             +p_{34}\cos^2(\delta -\alpha )+p_{35}\sin^2(\delta -\alpha )\\
            & \ \ +p_{36}\sin(\delta -\alpha)\cos(\delta -\alpha )
             +p_{37}\omega +p_{38}T_m+p_{39}G_v\\
              & \ \ +q_{31}E'_q\omega \cos(\delta -\alpha )+q_{32}E'_q\omega \sin(\delta -\alpha )
               +q_{33}\omega \cos^2(\delta -\alpha )\\
            & \ \ +q_{34}\omega \sin^2(\delta -\alpha )+q_{35}\omega \sin(\delta -\alpha)\cos(\delta -\alpha )\\
       \mathrm{i.e} \ \ \      \sigma  _1(x) &= p_{31}{x_1}^2+p_{32}x_1\cos(x_3 -\alpha )
             +p_{33}x_1\sin(x_3 -\alpha )
             +p_{34}\cos^2(x_3 -\alpha )+p_{35}\sin^2(x_3 -\alpha )\\
            & \ \ +p_{36}\sin(x_3 -\alpha)\cos(x_3 -\alpha )
             +p_{37}x_2 +p_{38}x_4+p_{39}x_5\\
              & \ \ +q_{31}x_1x_2\cos(x_3 -\alpha )+q_{32}x_1x_2\sin(x_3 -\alpha )
               +q_{33}x_2\cos^2(x_3 -\alpha )\\
            & \ \ +q_{34}x_2 \sin^2(x_3 -\alpha )+q_{35}x_2 \sin(x_3 -\alpha)\cos(x_3 -\alpha )\\
             \\
             \gamma  _1(x) &= r_{31}E'_q+r_{32}\cos(\delta -\alpha )+r_{33}\sin(\delta -\alpha )\\
  \mathrm{i.e} \ \ \           \gamma  _1(x) &= r_{31}x_1+r_{32}\cos(x_3 -\alpha )+r_{33}\sin(x_3 -\alpha )\\
\end{aligned}
 \label{eq:nfc19}
\end{equation}
Thus, using \autoref{eq:nfc18} and \autoref{eq:nfc19} we can compute the excitation field EMF $E_{FD}$ as
\begin{equation}
\begin{aligned}
          E_{FD} &= \gamma  ^{-1}_1(x)(v_1-\sigma  _1(x))\\
          E_{FD} &= \alpha _1(x)+\beta _1(x)v_1\\
\end{aligned}          
\label{eq:nfc20}
\end{equation}
where $\alpha _1(x)=-\gamma  ^{-1}_1(x)\sigma  _1(x)$ and $\beta _1(x)=\gamma  ^{-1}_1(x)$.             
From \autoref{eq:nfc10} and \autoref{eq:nfc12}, i.e. by taking the derivative of $\dot{T}_m$ we have
\begin{equation}
\begin{aligned} 
           \dot{z}_5 = \ddot{T}_m &= f_{41}\dot{T}_m+f_{42}\dot{G}_v\\
                     &= f_{41}(f_{41}T_m+f_{42}G_v)+f_{42}(f_{51}\omega +f_{52}G_v+g_{55}u_T)\\
                     &= f_{42}f_{51}\omega +f^2_{41}T_m+(f_{41}f_{42}+f_{42}f_{52})G_v+f_{42}g_{55}u_T\\
\end{aligned}
 \label{eq:nfc21}
\end{equation}
Let us denote $f_{42}f_{51}=p_{51}$, $f^2_{41}=p_{52}$, $f_{41}f_{42}+f_{42}f_{52}=p_{53}$, and $f_{42}g_{55}=r_{51}$.
Thus, \autoref{eq:nfc21} can be simplified as
\begin{equation}
                \dot{z}_5 = \ddot{T}_m = p_{51}\omega +p_{52}T_m+p_{53}G_v+r_{51}u_T
\label{eq:nfc22}
\end{equation}          
From \autoref{eq:nfc12} we have
\begin{equation}
        \dot{z}_5 =\sigma _2(x)+\gamma _2(x)u_T=v_2
\label{eq:nfc23}
\end{equation}
Thus, equating \autoref{eq:nfc22} and \autoref{eq:nfc23} we get
\begin{equation}
\begin{aligned}
          \sigma  _2(x) &= p_{51}\omega +p_{52}T_m+p_{53}G_v\\
   \mathrm{i.e} \ \ \        \sigma  _2(x) &= p_{51}x_2 +p_{52}x_4+p_{53}x_5\\
          \\
          \gamma  _2(x) &= r_{51}
\end{aligned}
 \label{eq:nfc24}
\end{equation}
Thus, using \autoref{eq:nfc23} and \autoref{eq:nfc24} we can compute the turbine valve control $u_T$ as
\begin{equation}
\begin{aligned}
          u_T &= \gamma  ^{-1}_2(x)(v_2- \sigma _2(x))\\
           u_T &= \alpha _2(x)+\beta _2(x)v_2\\
 \end{aligned}       
\label{eq:nfc25}
\end{equation}
where $\alpha _2(x)=-\gamma  ^{-1}_2(x)\sigma  _2(x)$ and $\beta _2(x)=\gamma  ^{-1}_2(x)$.\\ 
\\
We next design the linear controllers $v_1$ and $v_2$. The purpose of the linear controllers $v_1$ and $v_2$ is to regulate the state
variables $\delta $ and $T_m$ to their set points. The set points for the remaining state variables are found through the equilibrium condition $\delta (t)=1$ radian, and $T_m(t)=1.0012$ p.u.. Application of the state feedback control 
$u_1=E_{FD}=\alpha _1(x)+\beta _1(x)v_1$, $u_2=u_T=\alpha _2(x)+\beta _2(x)v_2$, and the state transformation $z=T(x)$ to the nonlinear model of the system, results in a linear model of the system in the new coordinates as
\begin{equation}
   \dot{\mathbf{z}}=\mathbf{A}\mathbf{z}+\mathbf{B}\mathbf{v}
 \label{eq:nfc26}
\end{equation}
where
\begin{equation}
\begin{aligned}
          \mathbf{z}^T &= \bbm z_1 & z_2 & z_3 & z_4 & z_5\ebm\\
          \mathbf{v} &= \bbm v_1\\ v_2\ebm
\end{aligned}
 \label{eq:nfc27}
\end{equation}

\begin{equation}
     \mathbf{A}=\bbm 0 & 1 & 0 & 0 & 0\\0 & 0 & 1 & 0 & 0\\0 & 0 & 0 & 0 & 0\\
                0 & 0 & 0 & 0 & 1\\0 & 0 & 0 & 0 & 0\ebm
\label{eq:nfc28}
\end{equation}                
\begin{equation}
         \mathbf{B}=\bbm 0 & 0\\0 & 0\\1 & 0\\0 & 0\\0 & 1\ebm
\label{eq:nfc29}
\end{equation}
 The linear controller $\mathbf{v}=-K\mathbf{z}$ can be designed either by pole placement or LQR technique 
such that $A-BK$ is Hurwitz. We use the LQR technique to design the controller gain matrix $K$.
Thus, by using the feedback law,
\begin{equation}
          \mathbf{v}=-K\mathbf{z}
\label{eq:nfc30}
\end{equation}
the weighting matrices
\begin{equation}
          Q=\bbm 300 & 0  & 0  & 0  & 0\\ 0 &  250 &  0  & 0  & 0\\ 0 &  0 &  200 &  0 &  0\\ 
            0 &  0  & 0  & 200  & 0\\ 0  & 0  & 0 &  0  & 250 \ebm
 \label{eq:nfc31}
\end{equation} 
and           
\begin{equation}
        R=\bbm 0.07 & 0\\ 0 & 0.07\ebm
\label{eq:nfc32}
\end{equation} 
and the state space matrices $(A, B)$ as given in \autoref{eq:nfc28} and \autoref{eq:nfc29} the 
control gain $K$ is found to be 
  \begin{equation}
        K=\bbm 65.4654 & 104.0206 & 55.3641 & 0 & 0\\ 0 & 0 & 0 & 53.4522 & 60.6493\ebm
\label{eq:nfc33}
\end{equation}  
Thus, we have
\begin{equation}
\begin{aligned}
            v_1 &= -K_{11}(z_1-z_{1d})-K_{12}(z_2-z_{2d})-K_{13}(z_3-z_{3d})\\
                &= -K_{11}(\delta -\delta _d)-K_{12}(\dot{\delta }-\dot{\delta }_d)-K_{13}(\ddot{\delta }-\ddot{\delta }_d)\\
 \end{aligned}               
\label{eq:nfc34}
\end{equation}
and 
\begin{equation}
\begin{aligned}
            v_2 &= -K_{24}(z_4-z_{4d})-K_{25}(z_5-z_{5d})\\
                &= -K_{24}(T_m - T_{md})-K_{25}(\dot{T}_m-\dot{T}_{md})\\
 \end{aligned}               
\label{eq:nfc35}
\end{equation}
where $z_{1d}$, $z_{2d}$, $z_{3d}$, $z_{4d}$, $z_{5d}$ are the desired values of the new state variables. Since we want the set point
to be an equilibrium, all the derivatives of $z_1$ and $z_4$ have to be zero. Thus 
$z_{2d}=z_{3d}=z_{5d}=0$. Also $z_{1d}=1$ radian, and $z_{4d}=1.0012$ p.u. are appropriately 
chosen within a reasonable physical limit of the system. The limits of the excitation field voltage of the 
generator are given by $E_{FD(max)}$ = 5 p.u. and $E_{FD(min)}$ = -5 p.u. The limits of the turbine gate opening are
given by $G_{V(max)}$ = 1.5 p.u. and $G_{V(min)}$ = 0 p.u. The operating conditions of the system are
as given in example 3.4 with  
          $I_{F0} = 1.6315$,  
          $I_{q0} = 0.4047$,
          $I_{d0} = -0.9185$,
          $V_{q0} = 0.9670$,
          $V_{d0} = -0.6628$, 
          $V_\infty = 1.00$,
          $V_{t0} = 1.172$,
          $\delta _0-\alpha = 53.736^\circ $, 
          $E'_{q0} = 1.1925$, and 
          $\tau '_{d0} = 5.90$

\begin{figure}
          \centering
          \includegraphics[trim=0cm 0cm 0cm 0cm, clip=true, totalheight=0.27\textheight, width=0.54
           \textwidth]  {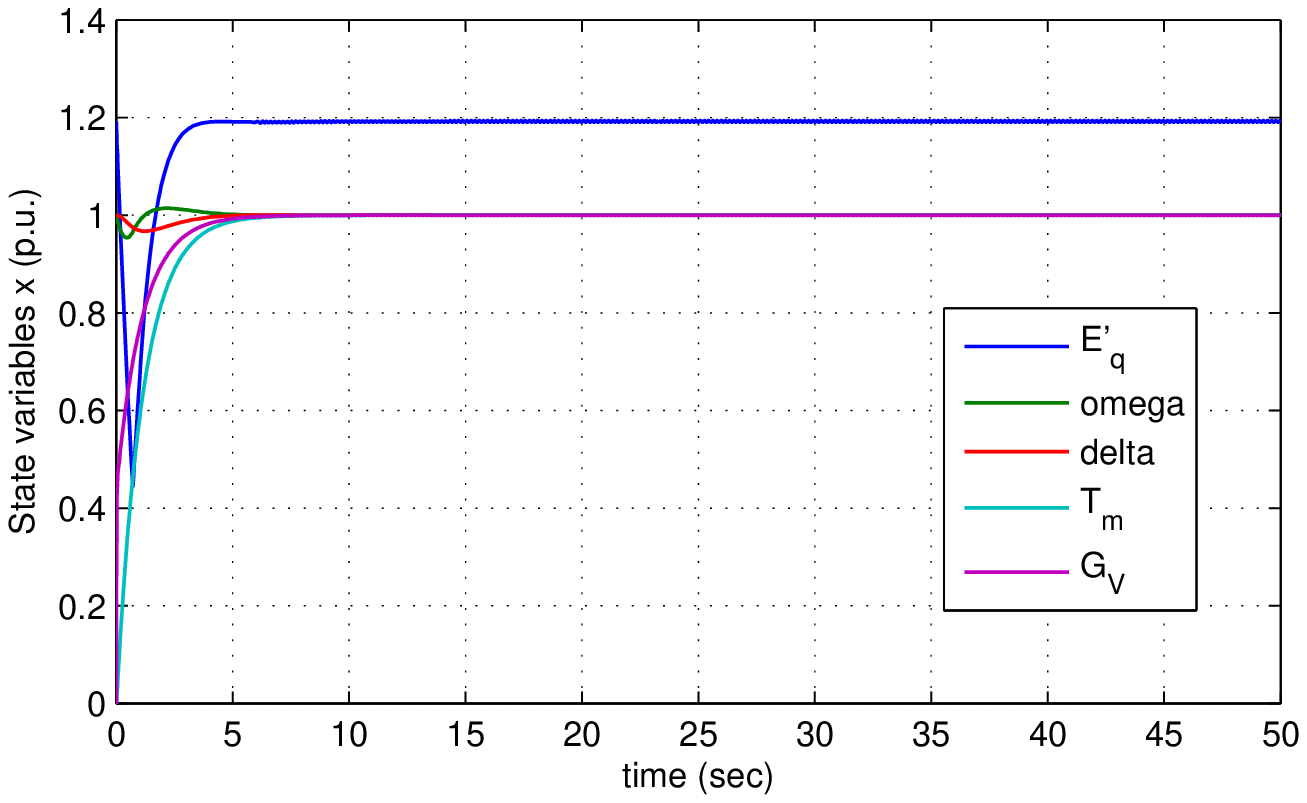}
          \caption{Plot of the state variables $E'_q$, $\omega $, $\delta $, 
           $T_m$, and $G_V$ vs time for the input-state nonlinear feedback linearizing controller applied to the reduced order model}
          \label{fig:isnonlinear1}
          \includegraphics[trim=0cm 0cm 0cm 0cm, clip=true, totalheight=0.27\textheight, width=0.54\textwidth]{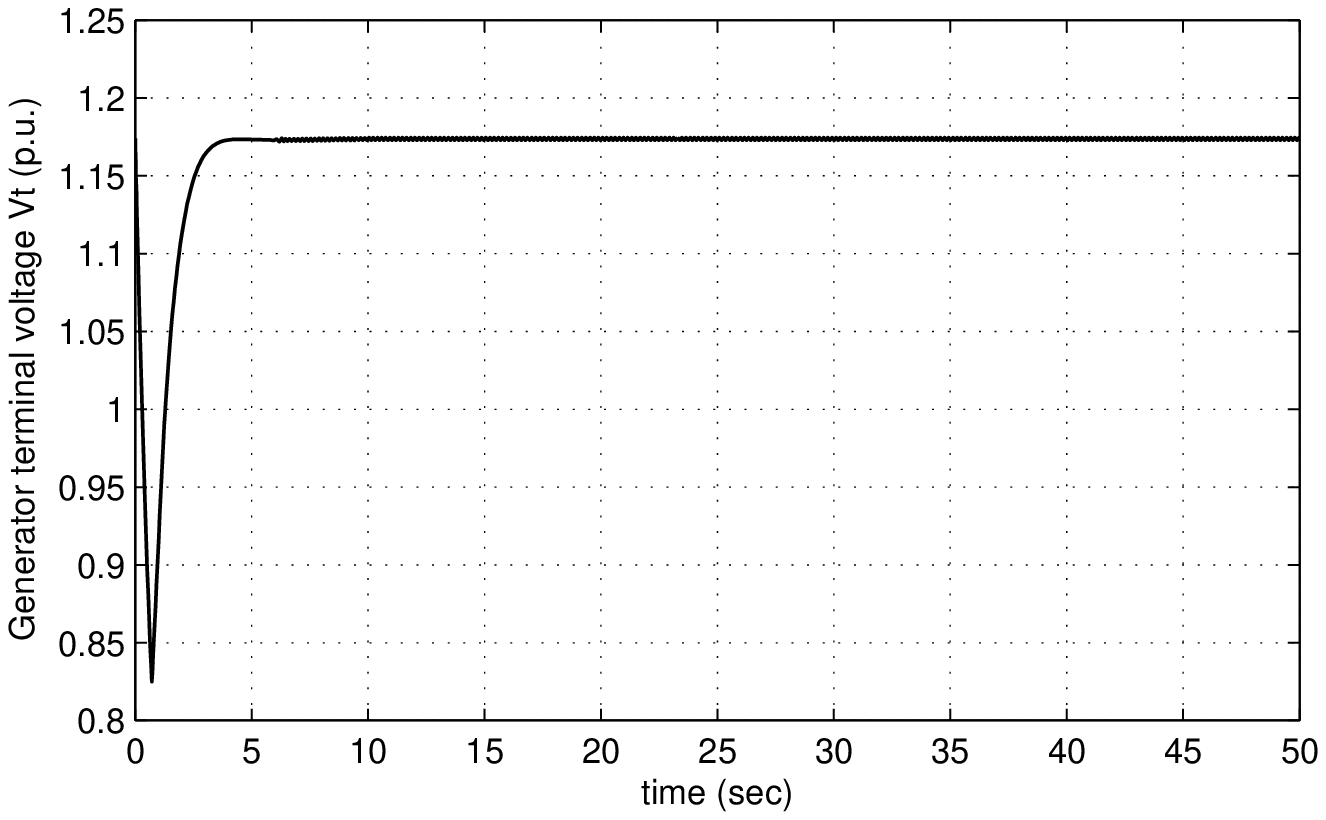}
          \caption{Plot of the generator terminal voltage $V_t$ vs time for the input-state nonlinear feedback linearizing 
          controller applied to the reduced order model}
          \label{fig:isnonlinear2}
          \includegraphics[trim=0cm 0cm 0cm 0cm, clip=true, totalheight=0.27\textheight, width=0.54\textwidth]{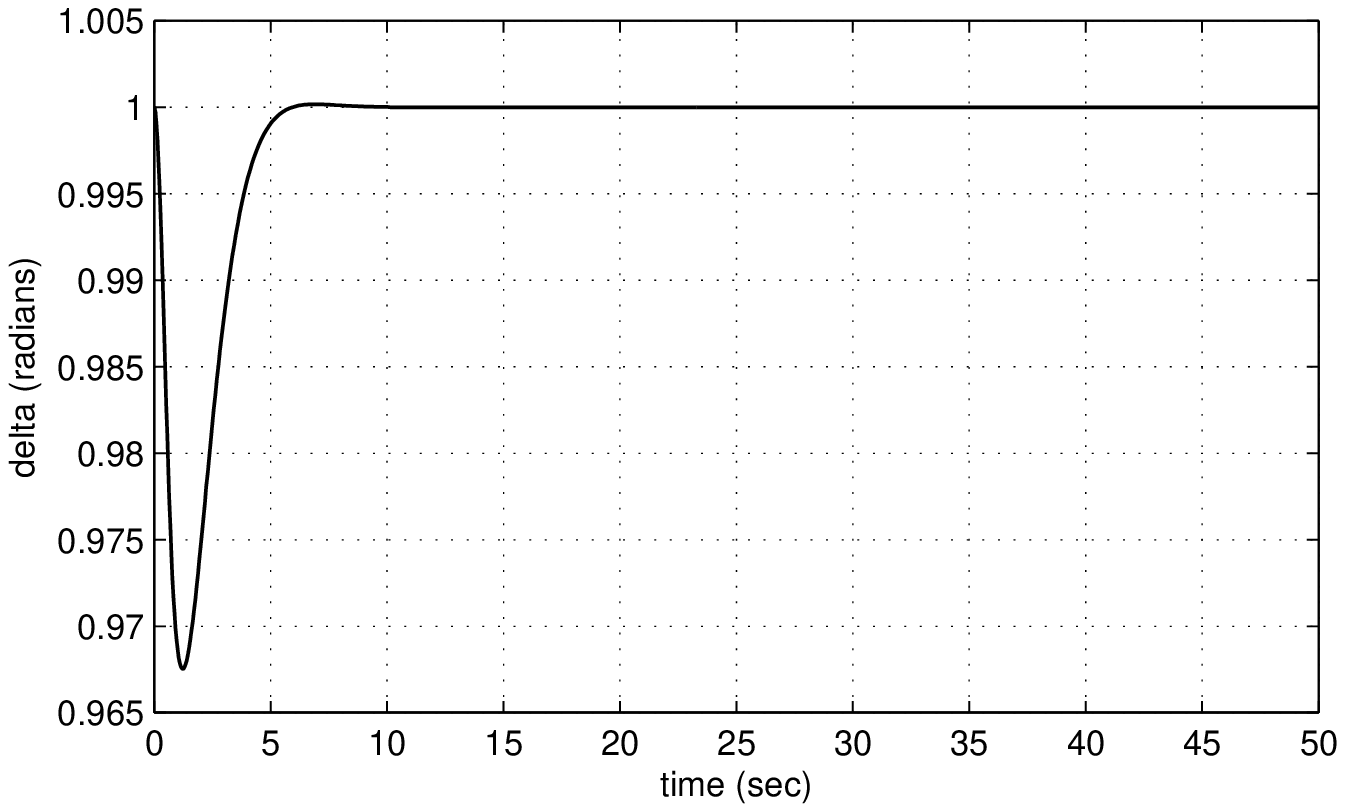}
          \caption{Plot of $\delta $ vs time for the input-state nonlinear feedback linearizing controller 
          applied to the reduced order model}
          \label{fig:isnonlinear3}
\end{figure}

\begin{figure}
          \centering          
          \includegraphics[trim=0cm 0cm 0cm 0cm, clip=true, totalheight=0.27\textheight, width=0.54\textwidth]{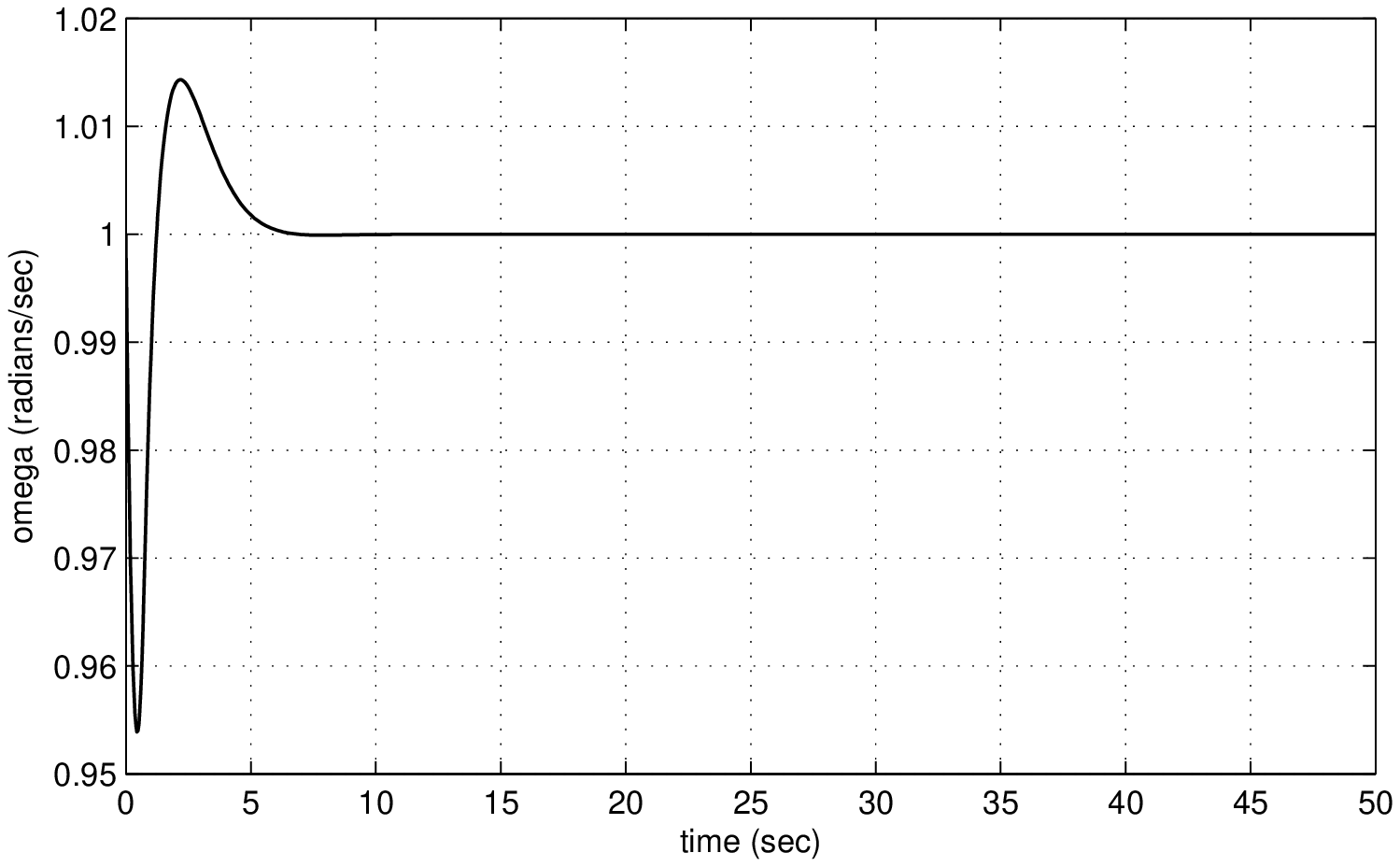}
          \caption{Plot of $\omega  $ vs time for the input-state nonlinear feedback linearizing controller applied to 
          the reduced order model}
          \label{fig:isnonlinear4}
          \includegraphics[trim=0cm 0cm 0cm 0cm, clip=true, totalheight=0.27\textheight, width=0.54\textwidth]{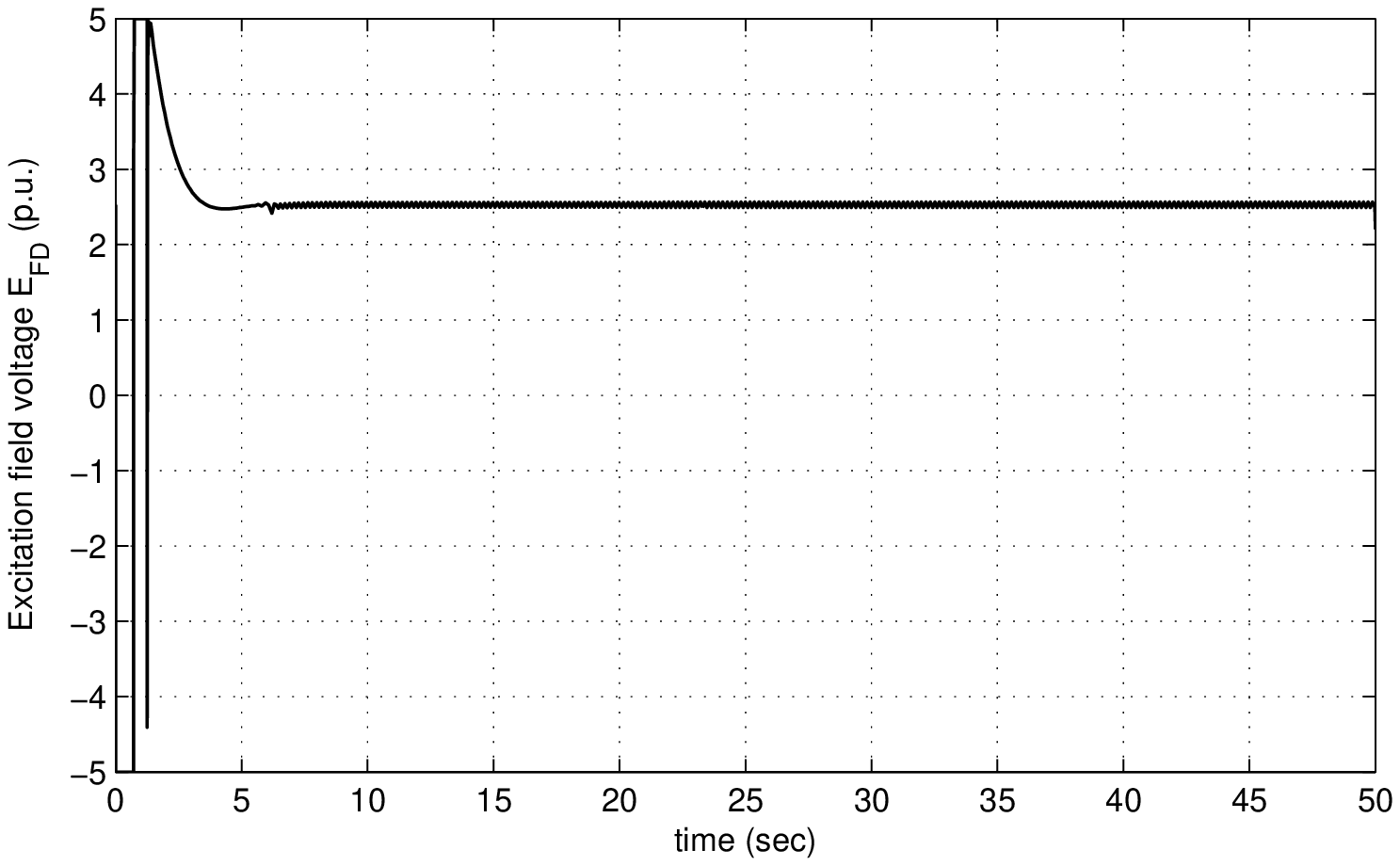}
          \caption{Plot of the generator excitation field control $E_{FD} $ vs time for the input-state nonlinear 
          feedback linearizing controller applied to the reduced order model}
          \label{fig:isnonlinear5}
          \includegraphics[trim=0cm 0cm 0cm 0cm, clip=true, totalheight=0.27\textheight, width=0.54\textwidth]{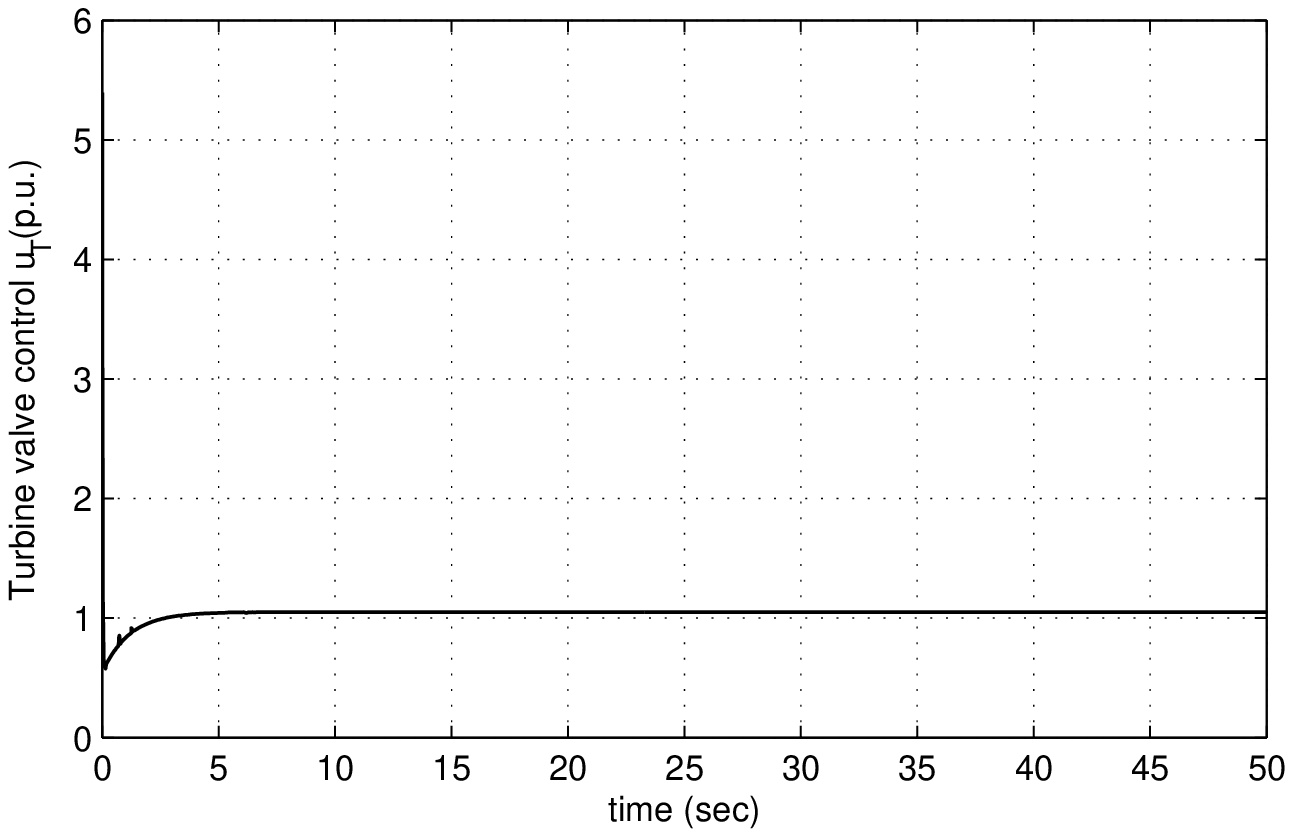}
          \caption{Plot of the turbine valve control $u_T$ vs time for the input-state nonlinear 
          feedback linearizing controller applied to the reduced order model}
          \label{fig:isnonlinear6}
\end{figure}

From \autoref{fig:isnonlinear1} we can see that the state variables $E'_q$, $\omega $, $\delta $, 
           $T_m$, and $G_V$ attain their respective steady state values in approximately 5 to 7 seconds. \autoref{fig:isnonlinear2}
           shows that the generator terminal voltage $V_t$ reaches a steady state value of 1.172 p.u. which is
           equal to the desired steady state operating point $V_{t0} = 1.172$, in approximately 5 seconds.
           \autoref{fig:isnonlinear3} shows
           that the rotor angle $\delta $ settles to a steady state value of approximately 1 radians, which is
           equal to the steady state operating condition $\delta _{0}=1$ radian, in approximately 7 seconds. 
           The angular velocity or the frequency $\omega $ 
           of the synchronous generator settles to a steady state value of 1 rad/sec which is equal to the desired operating condition
           $\omega _{0}$= 1 rad/sec which can be verified from \autoref{fig:isnonlinear4}. Also, from \autoref{fig:isnonlinear5} and
           \autoref{fig:isnonlinear6} we can see that the control inputs, $E_{FD}$ and $u_T$, settle to their steady state values of
           2.529 p.u. and 1.0512 p.u. respectively. To verify our simulation results we analytically compute 
           the steady state equilibrium values of the original state variables and the control inputs.
           The steady state equilibrium in the original co-ordinates can be computed by solving the differential
           equation, $\dot{x}=f(x)+g(x)u=0$. Thus using \autoref{eq:nfc10} we can write     
\begin{equation}
\begin{aligned}
      &\dot{E}'_q = f_{11}E'_q+f_{12}\cos(\delta -\alpha)
             +f_{13}\sin(\delta -\alpha)+g_{11}E_{FD}=0\\
       &\dot{\omega } = f_{21}E'^2_q+
       f_{22}E'_q\cos(\delta -\alpha)
      + f_{23}E'_q\sin(\delta -\alpha)
      + f_{24}\sin(\delta -\alpha)\cos(\delta -\alpha)\\
      & \ \ \ \ \ +f_{25}\cos^2(\delta -\alpha)
      + f_{26}\sin^2(\delta -\alpha)+f_{27}\omega +f_{28}T_m=0\\      
       &\dot{\delta } = \omega -1=0\\
       &\dot{T}_{m} = f_{41}T_{m}+f_{42}G_{V}=0\\
       &\dot{G}_V = f_{51}\omega +f_{52}G_V+g_{55}u_T=0\\
       \end{aligned}
 \label{eq:steady}
\end{equation}
The steady state equilibrium in the original co-ordinates $E'_q$, $\omega $, $\delta $, 
           $T_m$, and $G_V$ can be found by using $z_{1d}=\delta =1$ radian, and $z_{4d}=T_m=1.0012$ p.u., and solving the differential equation
\begin{equation}
\begin{aligned}
       &\dot{\omega } = f_{21}E'^2_q+
       f_{22}E'_q\cos(\delta -\alpha)
      + f_{23}E'_q\sin(\delta -\alpha)
      + f_{24}\sin(\delta -\alpha)\cos(\delta -\alpha)\\
      & \ \ \ \ \ +f_{25}\cos^2(\delta -\alpha)
      + f_{26}\sin^2(\delta -\alpha)+f_{27}\omega +f_{28}T_m=0\\      
       &\dot{\delta } = \omega -1=0\\
       &\dot{T}_{m} = f_{41}T_{m}+f_{42}G_{V}=0\\       
       \end{aligned}
 \label{eq:steady1}
\end{equation}
Also the steady state equilibrium of the control inputs $E_{FD}$ and $u_T$ can be found by using $\dot{E}'_q=0$, and $\dot{G}_V=0$, from \autoref{eq:steady}, and the steady state equilibrium values of the original state variables $E'_{q(ss)}$, $\omega _{(ss)}$, $\delta _{(ss)}$, $T_{m(ss)}$, and $G_{V(ss)}$ obtained by solving \autoref{eq:steady1} i.e.
\begin{equation}
\begin{aligned}
      0 &= f_{11}E'_{q0}+f_{12}\cos(\delta _0-\alpha )
             +f_{13}\sin(\delta _0-\alpha )+g_{11}E_{FD(ss)}\\
             &= -0.5517\times 1.1925+0.3822\times 0.5915+0.0037\times 0.8063+0.1695\times E_{FD(ss)}\\
       0 &= f_{51}\omega _0+f_{52}G_{V0}+g_{55}u_{T(ss)}\\
       &= -0.25\times 1-5\times 1.0012+5\times u_{T(ss)}
       \end{aligned}
 \label{eq:steady2}
\end{equation} 
Solving \autoref{eq:steady2} we get $E_{FD(ss)}=2.529$ and $u_{T(ss)}=1.0512$, which are in agreement with our simulation results.
       
\subsection{Simulation Results for the Nonlinear Feedback Linearizing Controller applied to the Truth Model}

For the reduced order nonlinear model the state variables  are $x = [E'_q, \omega, \delta, T_m, G_V]^\mathrm{T}$, the two control inputs are $u = [ E_{FD}, u_T]^\mathrm{T}$, and the two regulated outputs are $y=[V_t, \omega ]^\mathrm{T}$. Whereas, for the truth model $x = [I_d, I_F, I_D, I_q, I_Q, \omega , \delta , T_m, G_V]^\mathrm{T}$ is the vector of state variables, 
$u = [ V_F, u_T]^\mathrm{T}$ is vector of control inputs, and $y = [ V_t, \omega ]^\mathrm{T}$ is the vector of outputs. Thus we can
see that the first control input $V_F$ of the truth model and the first control input $E_{FD}$ of the reduced order model are different.
As given in
\autoref{eq:pu6} the excitation field emf, $E_{FD}$, which is the first control input for the reduced order nonlinear model is
related to field voltage, $V_F$, which is the first control input for the truth model by the following expression 
\begin{equation}
         \bigg{(}\frac{V_F}{r_F}\bigg{)}\omega _RkM_F=E_{FD}
\label{eq:nonlinearflctruth3}
\end{equation}
\autoref{eq:nonlinearflctruth3} can be rearranged to get the following expression 
\begin{equation}
\begin{aligned}
         V_F=&\bigg{(}\frac{r_F}{\omega _RkM_F}\bigg{)}E_{FD}\\
             =&e_{15}E_{FD}\\
\end{aligned}             
\label{eq:EFD_VF}
\end{equation}
In the above expression, $\omega _R=1$ p.u., and $e_{15}=(\frac{r_F}{\omega _RkM_F})$. 
Also there is a non physical state $E'_q$ in the reduced order model that needs to be reconstructed from the states of the truth model.

\begin{figure}
          \centering
          \includegraphics[trim=0cm 0cm 0cm 0cm, clip=true, totalheight=0.27\textheight, width=0.54
           \textwidth]  {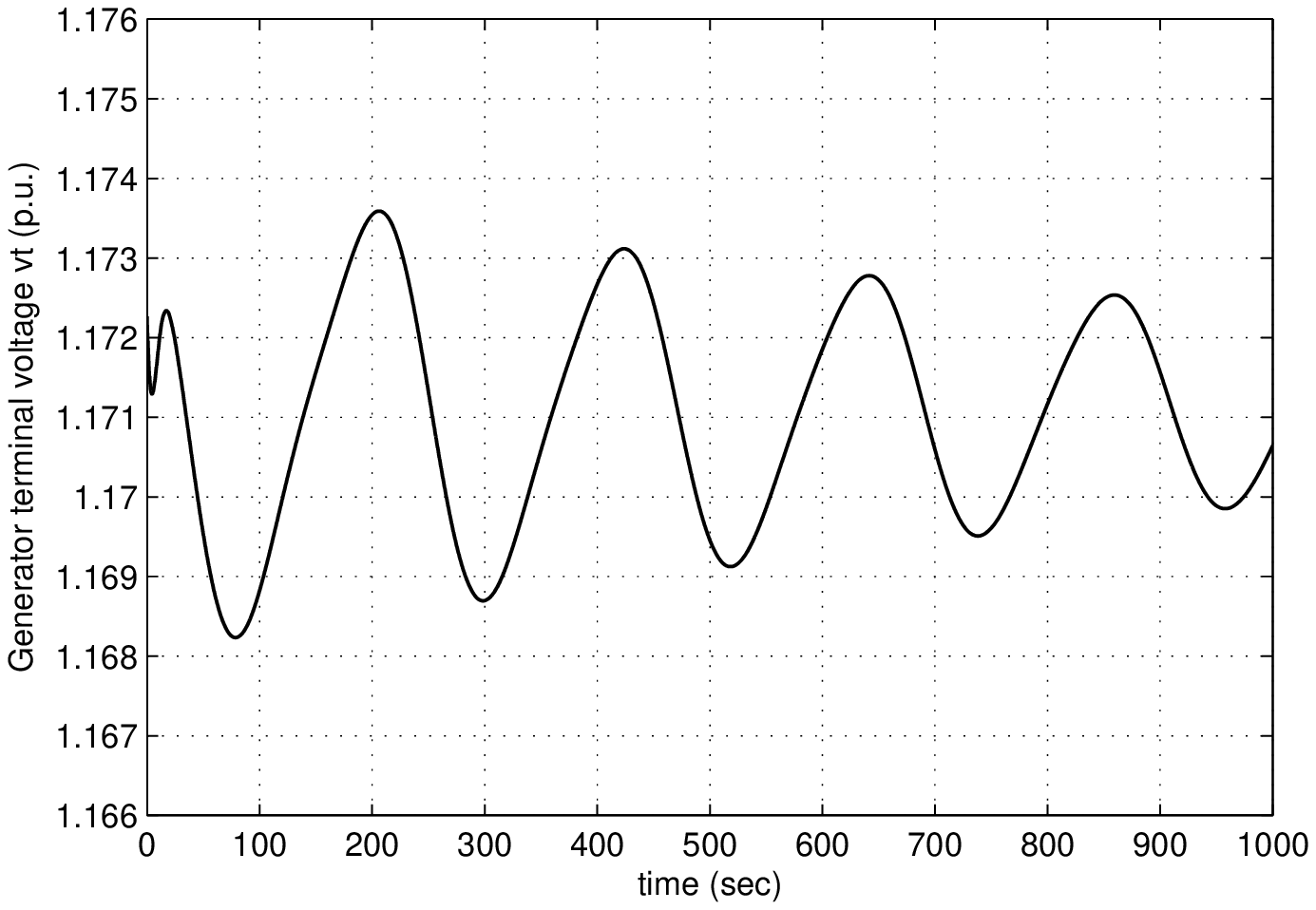}
          \caption{Plot of the generator terminal voltage $V_t$ vs time for the nonlinear feedback linearizing controller
          applied to the truth model}
          \label{fig:nonlineartruth1}
          \includegraphics[trim=0cm 0cm 0cm 0cm, clip=true, totalheight=0.27\textheight, width=0.54\textwidth]{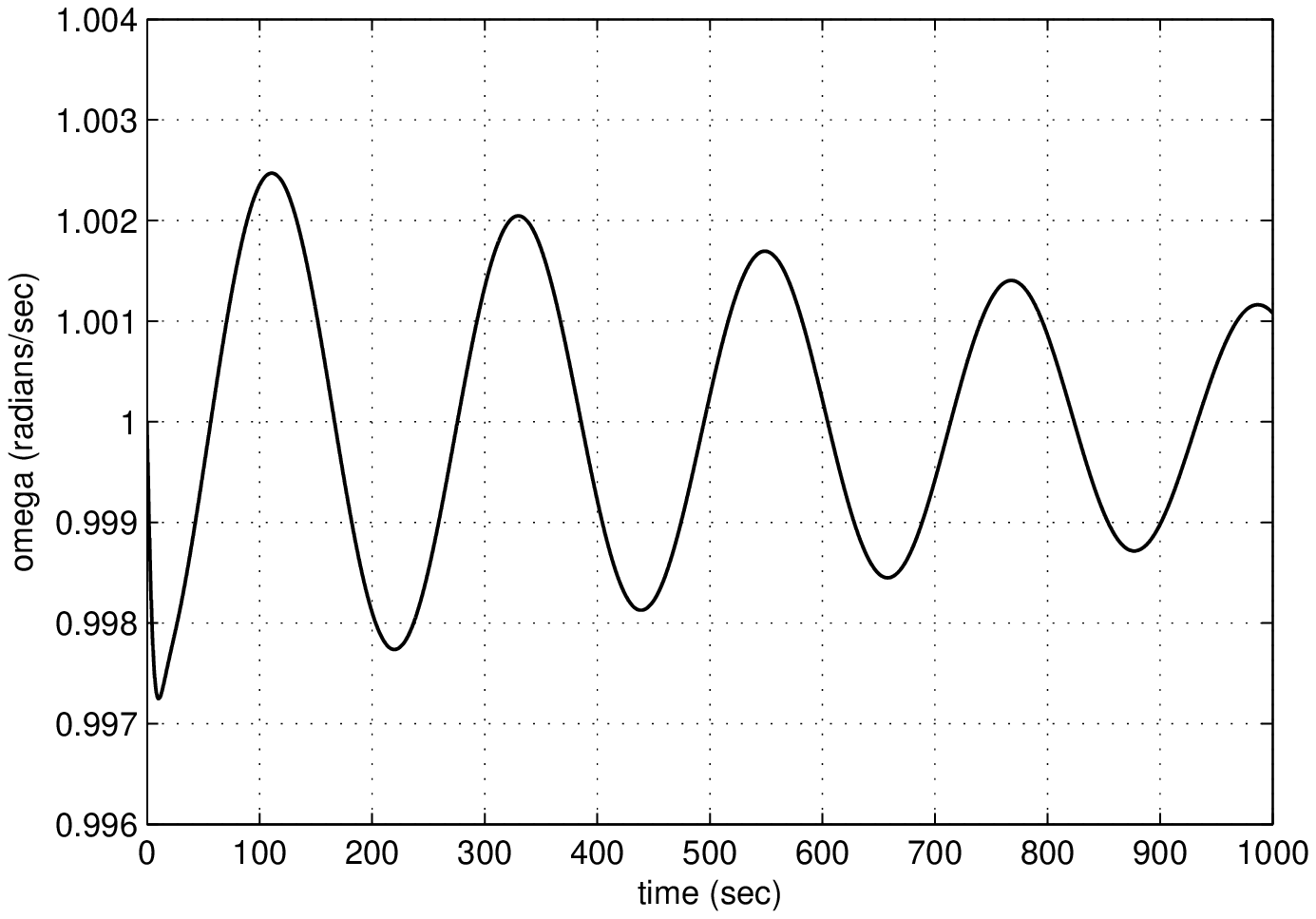}
          \caption{Plot of the angular velocity $\omega $ vs time for the nonlinear feedback linearizing controller
          applied to the truth model}
          \label{fig:nonlineartruth2}
          \includegraphics[trim=0cm 0cm 0cm 0cm, clip=true, totalheight=0.27\textheight, width=0.54\textwidth]{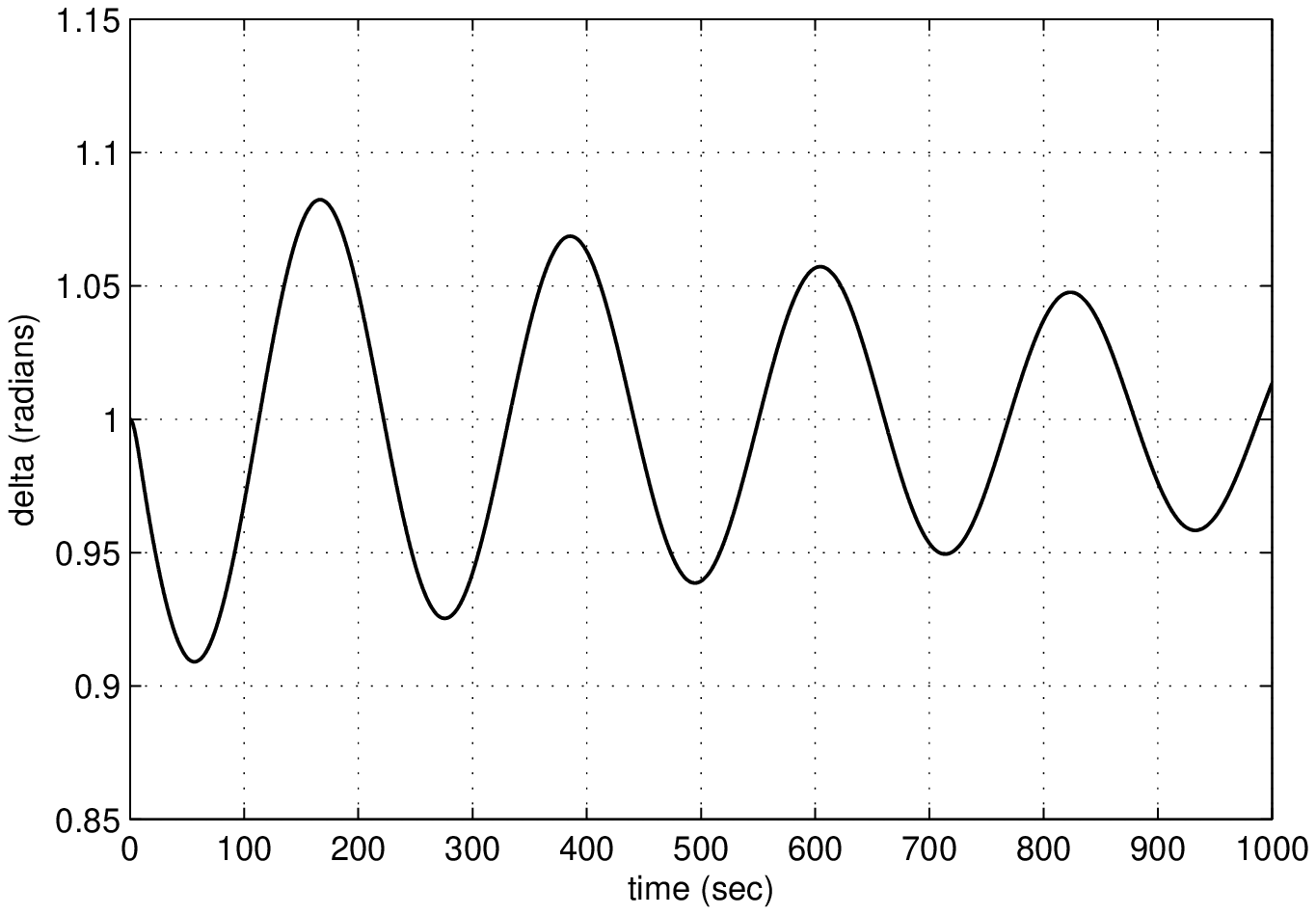}
          \caption{Plot of the rotor angle $\delta $ vs time for the nonlinear feedback linearizing controller applied to the truth model}
          \label{fig:nonlineartruth3}
\end{figure}

\begin{figure}
          \centering          
          \includegraphics[trim=0cm 0cm 0cm 0cm, clip=true, totalheight=0.27\textheight, 
          width=0.54\textwidth]{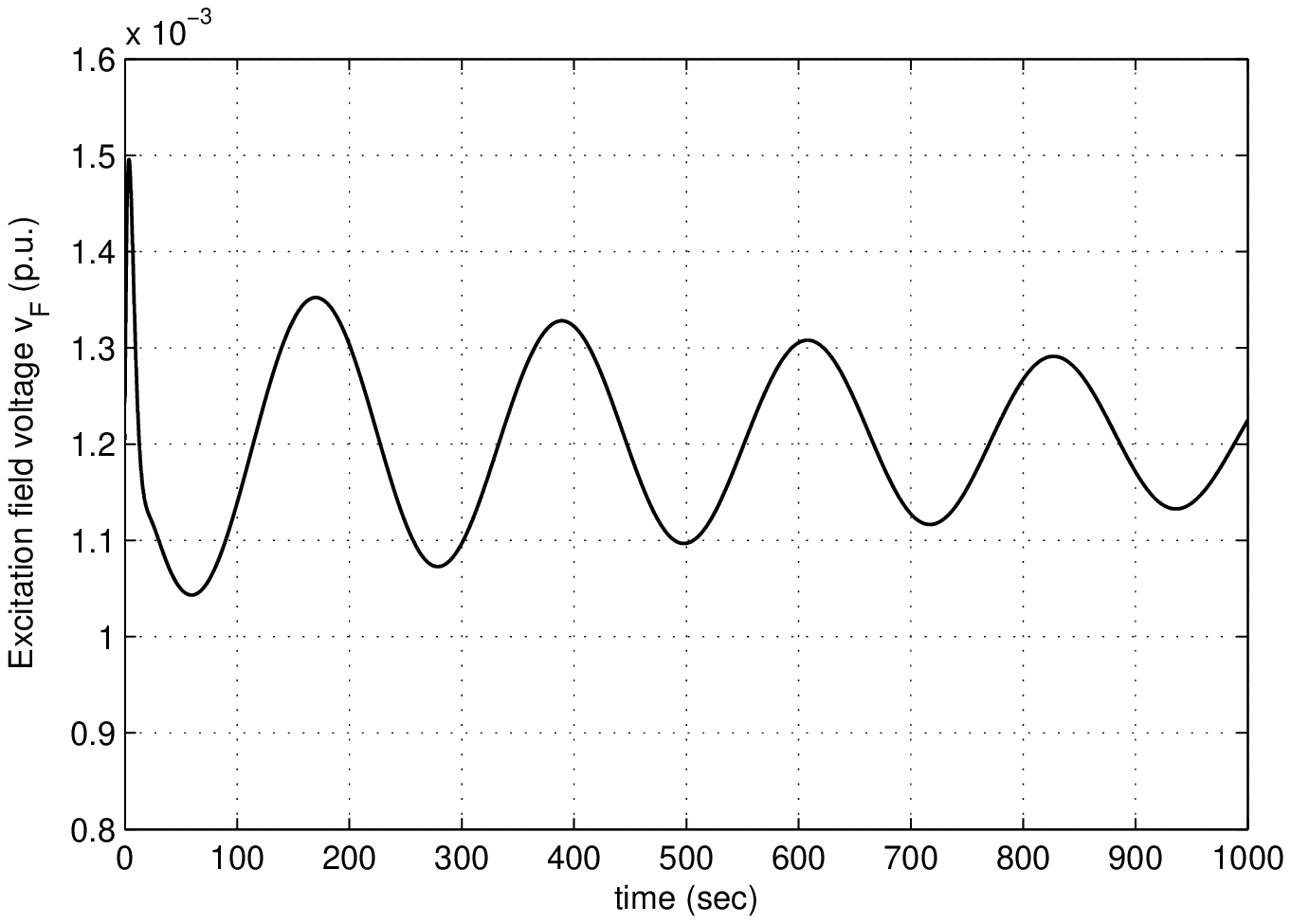}
          \caption{Plot of the generator excitation voltage $V_F$ vs time for the 
          nonlinear feedback linearizing controller applied to the truth model }
          \label{fig:nonlineartruth4}
          \includegraphics[trim=0cm 0cm 0cm 0cm, clip=true, totalheight=0.27\textheight, 
          width=0.54\textwidth]{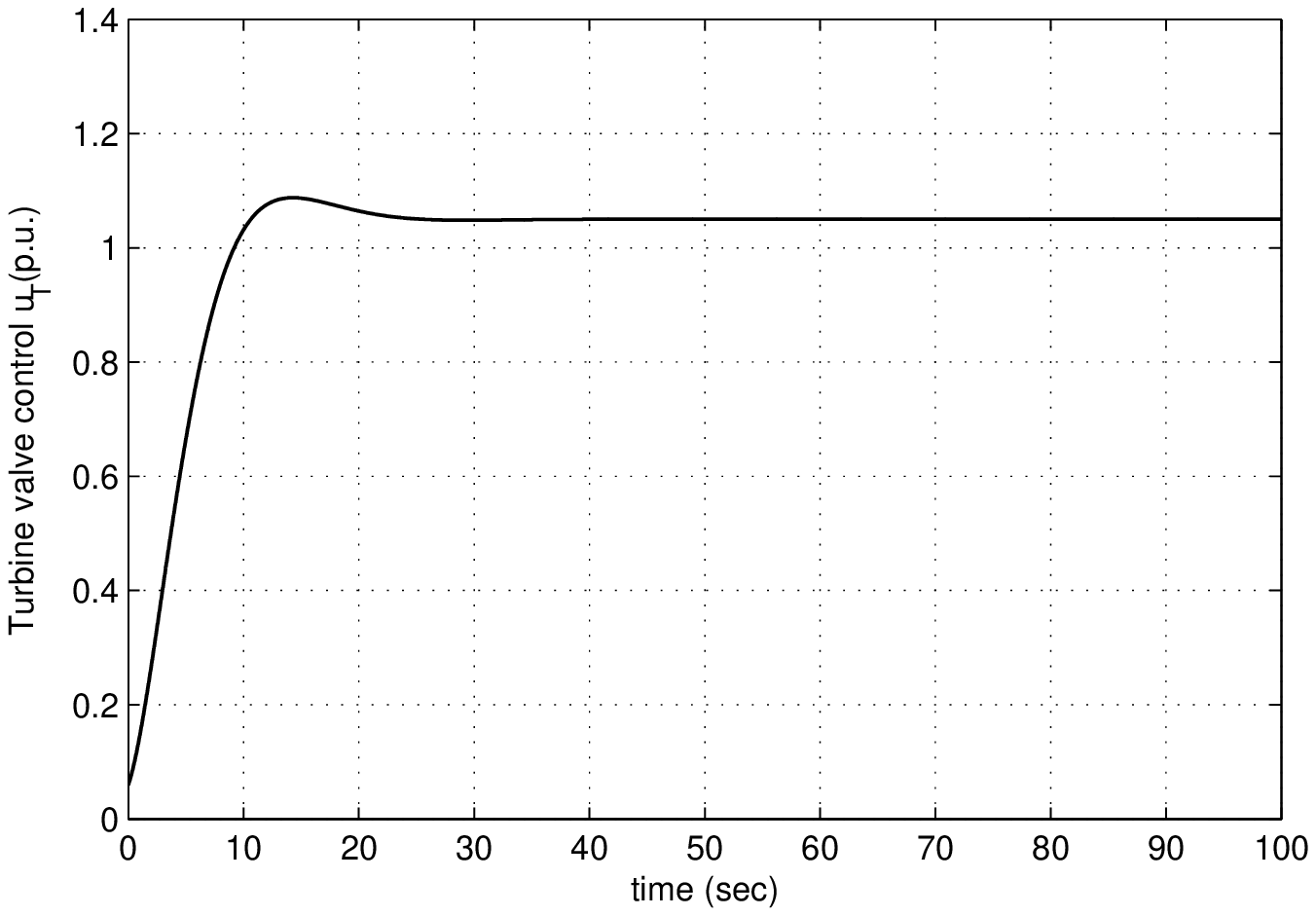}
          \caption{Plot of the turbine valve control $u_T$ vs time for the nonlinear feedback 
          linearizing controller applied to the truth model}
          \label{fig:nonlineartruth5}
\end{figure}

The nonlinear feedback linearizing controller that was designed for the reduced order nonlinear model is now tested on the 
truth model. From \autoref{eq:nfc20} the excitation field EMF, $E_{FD}$, is computed as
\begin{equation}
\begin{aligned}
          E_{FD} &= \gamma  ^{-1}_1(x)(v_1-\sigma  _1(x))\\
          E_{FD} &= \alpha _1(x)+\beta _1(x)v_1\\
\end{aligned}          
\label{eq:nonlinearflctruth1}
\end{equation}
where $\alpha _1(x)=-\gamma  ^{-1}_1(x)\sigma  _1(x)$; $\beta _1(x)=\gamma  ^{-1}_1(x)$ and
\begin{equation}
\begin{aligned}
              \sigma  _1(x) &= p_{31}{E'_q}^2+p_{32}E'_q\cos(\delta -\alpha )
             +p_{33}E'_q\sin(\delta -\alpha )
             +p_{34}\cos^2(\delta -\alpha )+p_{35}\sin^2(\delta -\alpha )\\
            & \ \ +p_{36}\sin(\delta -\alpha)\cos(\delta -\alpha )
             +p_{37}\omega +p_{38}T_m+p_{39}G_v\\
              & \ \ +q_{31}E'_q\omega \cos(\delta -\alpha )+q_{32}E'_q\omega \sin(\delta -\alpha )
               +q_{33}\omega \cos^2(\delta -\alpha )\\
            & \ \ +q_{34}\omega \sin^2(\delta -\alpha )+q_{35}\omega \sin(\delta -\alpha)\cos(\delta -\alpha )\\
       \mathrm{i.e} \ \ \      \sigma  _1(x) &= p_{31}{x_1}^2+p_{32}x_1\cos(x_3 -\alpha )
             +p_{33}x_1\sin(x_3 -\alpha )
             +p_{34}\cos^2(x_3 -\alpha )+p_{35}\sin^2(x_3 -\alpha )\\
            & \ \ +p_{36}\sin(x_3 -\alpha)\cos(x_3 -\alpha )
             +p_{37}x_2 +p_{38}x_4+p_{39}x_5\\
              & \ \ +q_{31}x_1x_2\cos(x_3 -\alpha )+q_{32}x_1x_2\sin(x_3 -\alpha )
               +q_{33}x_2\cos^2(x_3 -\alpha )\\
            & \ \ +q_{34}x_2 \sin^2(x_3 -\alpha )+q_{35}x_2 \sin(x_3 -\alpha)\cos(x_3 -\alpha )\\
             \gamma  _1(x) &= r_{31}E'_q+r_{32}\cos(\delta -\alpha )+r_{33}\sin(\delta -\alpha )\\
  \mathrm{i.e} \ \ \           \gamma  _1(x) &= r_{31}x_1+r_{32}\cos(x_3 -\alpha )+r_{33}\sin(x_3 -\alpha )\\
\end{aligned}
 \label{eq:nonlinearflctruth2}
\end{equation}
From \autoref{eq:EFD_VF} the field voltage $V_F$ can be written as
\begin{equation}
             V_F=\bigg{(}\frac{r_F}{\omega _RkM_F}\bigg{)}E_{FD}=\bigg{(}\frac{r_F}{\omega _RkM_F}\bigg{)}
                     \gamma  ^{-1}_1(x)(v_1-\sigma  _1(x))
                                                 =e_{15}(\alpha _1(x)+\beta _1(x)v_1)
\label{eq:nonlinearflctruth4}
\end{equation}
In \autoref{eq:nonlinearflctruth4}, $V_F$ depends on  $\sigma  _1(x)$ and $\gamma  _1(x)$ which depend on the
fictitious state variable $E'_q$, which is not a physical quantity that can be measured or a state variable of the truth model. 
Since $E'_q$ cannot be measured
using sensors, we cannot apply the field voltage $V_F$ directly on the truth model
without eliminating the state variable $E'_q$ in the above expression. This problem can be solved by 
expressing $E'_q$ as a function of any of the measurable states of the truth model. From \autoref{eq:two27} as given 
in the derivation of the reduced order model in the appendix we have
\begin{equation}
          E=E'_q-(L_d-L'_d)I_d 
\label{eq:nonlinearflctruth5}
\end{equation}
Substituting $I_d$ from \autoref{eq:simplified7} in the above equation
\begin{equation}
          E=E'_q-(L_d-L'_d)\bigg{(}\frac{-(E'_q-V_{\infty q})(L_q+L_e)-V_{\infty d}(r+R_e)}{(r+R_e)^2+(L'_d+L_e)(L_q+L_e)}\bigg{)} 
\label{eq:nonlinearflctruth6}
\end{equation}
For simplification of the above expression let us denote
$L_d-L'_d=L_2$, $L_q+L_e=L_1$, $r+R_e=R_1$, $(r+R_e)^2+(L'_d+L_e)(L_q+L_e)=M_1$.
Thus \autoref{eq:nonlinearflctruth6} can be written as
\begin{equation}
          E=E'_q-L_2\bigg{(}\frac{-(E'_q-V_{\infty q})L_1-V_{\infty d}R_1}{M_1}\bigg{)} 
\label{eq:nonlinearflctruth7}
\end{equation}
The above expression can be simplified to get
\begin{equation}
          E=\bigg{(}1+\frac{L_1L_2}{M_1}\bigg{)}E'_q-\bigg{(}\frac{L_1L_2V_\infty }{M_1}\bigg{)}\cos(\delta -\alpha )
            -\bigg{(}\frac{R_1L_2V_\infty }{M_1}\bigg{)}\sin(\delta -\alpha ) 
\label{eq:nonlinearflctruth8}
\end{equation}
Let us denote 
\begin{equation}
\begin{aligned}
             &\bigg{(}1+\frac{L_1L_2}{M_1}\bigg{)}=e_{11}\\
             &\bigg{(}\frac{L_1L_2V_\infty }{M_1}\bigg{)}=e_{12}\\
             &\bigg{(}\frac{R_1L_2V_\infty }{M_1}\bigg{)}=e_{13}\\
  \end{aligned}           
\label{eq:nonlinearflctruth9}
\end{equation}             
Substituting \autoref{eq:nonlinearflctruth9} in \autoref{eq:nonlinearflctruth8} and rearranging, $E'_q$ can be written as
\begin{equation}
          E'_q=\frac{1}{e_{11}}E+\frac{e_{12}}{e_{11}}\cos(\delta -\alpha )
            +\frac{e_{13}}{e_{11}}\sin(\delta -\alpha ) 
\label{eq:nonlinearflctruth10}
\end{equation}
From \autoref{eq:pu4} the excitation field emf, $E$, is related to the field current $I_F$ as
\begin{equation}
       E=\omega _RkM_FI_F =e_{14}I_F        
\label{eq:nonlinearflctruth11}
\end{equation}
Substituting \autoref{eq:nonlinearflctruth11} in \autoref{eq:nonlinearflctruth10}
\begin{equation}
          E'_q=\frac{e_{14}}{e_{11}}I_F+\frac{e_{12}}{e_{11}}\cos(\delta -\alpha )
            +\frac{e_{13}}{e_{11}}\sin(\delta -\alpha ) 
\label{eq:nonlinearflctruth12}
\end{equation}
$E'_q$ is now expressed as a function of the field current $I_F$, and rotor angle $\delta $, which are state variables
of the truth model. While the field current $I_F$ can be measured using a sensor, it is not always possible to measure
the rotor angle $\delta $, using a sensor. In this case we assume that $\delta $ can be measured.
Using \autoref{eq:nonlinearflctruth5} and \autoref{eq:nonlinearflctruth11}, $E'_q$ can also be written as
\begin{equation}
          E'_q=e_{14}I_F+L_2I_d 
\label{eq:nonlinearflctruth13}
\end{equation}
where $I_F$ and $I_d$ are both state variables of the truth model which can be measured using senors. 
Thus the first control input for the truth model which is the field voltage, $V_F$, as given in \autoref{eq:nonlinearflctruth4}
can be implemented on the truth model by using either \autoref{eq:nonlinearflctruth12} or \autoref{eq:nonlinearflctruth13} to eliminate $E'_q$ in the expression for $V_F$.
We prefer \autoref{eq:nonlinearflctruth12} instead of \autoref{eq:nonlinearflctruth13} since the rotor angle $\delta $,
is a state variable in both the reduced order and the truth model, whereas the direct axis current $I_d$, is not a state variable
in the reduced order model.
From \autoref{eq:nfc25} the second control input for the truth model which is the turbine valve control $u_T$, is given by
\begin{equation}
\begin{aligned}
          u_T &= \gamma  ^{-1}_2(x)(v_2- \sigma _2(x))\\
           u_T &= \alpha _2(x)+\beta _2(x)v_2\\
 \end{aligned}       
\label{eq:nonlinearflctruth14}
\end{equation}
 where $\alpha _2(x)=-\gamma  ^{-1}_2(x)\sigma  _2(x)$, $\beta _2(x)=\gamma  ^{-1}_2(x)$ and
\begin{equation}
\begin{aligned}
          \sigma  _2(x) &= p_{51}\omega +p_{52}T_m+p_{53}G_v\\
   \mathrm{i.e} \ \ \        \sigma  _2(x) &= p_{51}x_2 +p_{52}x_4+p_{53}x_5\\
          \\
          \gamma  _2(x) &= r_{51}
\end{aligned}
 \label{eq:nonlinearflctruth15}
\end{equation}
As seen earlier, in the section for nonlinear feedback linearizing controller design for the reduced order model,
 application of the state feedback control, 
$u_1=E_{FD}=\alpha _1(x)+\beta _1(x)v_1$, $u_2=u_T=\alpha _2(x)+\beta _2(x)v_2$, and the state transformation $z=T(x)$ to the 
reduced order nonlinear model of the system, results in a linear model of the system in the new coordinates. However, this is true only
for the reduced order model, but not true for the truth model. The resulting closed loop system in the case of the truth model is not linear at all.
The linear controllers $v_1$ and $v_2$ were designed by using the LQR technique for the reduced order nonlinear model.
The gains of this linear LQR controller are tuned once again so that the nonlinear controller works efficiently on the truth model.
Thus, by using the feedback law,
\begin{equation}
          \mathbf{v}=-K\mathbf{z}
\label{eq:nonlinearflctruth20}
\end{equation}
the weighting matrices
\begin{equation}
          Q=\bbm 250 & 0  & 0  & 0  & 0\\ 0 &  250 &  0  & 0  & 0\\ 0 &  0 &  250 &  0 &  0\\ 
            0 &  0  & 0  & 250  & 0\\ 0  & 0  & 0 &  0  & 250 \ebm
 \label{eq:nonlinearflctruth21}
\end{equation} 
and           
\begin{equation}
        R=\bbm 30000 & 0\\ 0 & 30000\ebm
\label{eq:nonlinearflctruth22}
\end{equation} 
and the state space matrices $(A, B)$ as given in \autoref{eq:nfc28}and \autoref{eq:nfc29} the 
control gain $K$ is found to be 
\begin{equation}
        K=\bbm 0.0913 & 0.4201 & 0.9212 & 0 & 0\\ 0 & 0 & 0 & 0.0913 & 0.4369\ebm
\label{eq:nonlinearflctruth23}
\end{equation} 
Thus, we have
\begin{equation}
\begin{aligned}
            v_1 &= -K_{11}(z_1-z_{1d})-K_{12}(z_2-z_{2d})-K_{13}(z_3-z_{3d})\\
                &= -K_{11}(\delta -\delta _d)-K_{12}(\dot{\delta }-\dot{\delta }_d)-K_{13}(\ddot{\delta }-\ddot{\delta }_d)\\
 \end{aligned}               
\label{eq:ntruth1}
\end{equation}
and 
\begin{equation}
\begin{aligned}
            v_2 &= -K_{24}(z_4-z_{4d})-K_{25}(z_5-z_{5d})\\
                &= -K_{24}(T_m - T_{md})-K_{25}(\dot{T}_m-\dot{T}_{md})\\
 \end{aligned}               
\label{eq:ntruth2}
\end{equation}
where $z_{1d}$, $z_{2d}$, $z_{3d}$, $z_{4d}$, $z_{5d}$ are the desired values of the new state variables.
\autoref{fig:nonlineartruth1} to \autoref{fig:nonlineartruth5} show simulation results
for the nonlinear feedback linearizing controller applied to the truth model. From \autoref{fig:nonlineartruth1}, \autoref{fig:nonlineartruth2}, and \autoref{fig:nonlineartruth3}, we can see that $V_t$, $\omega $, and $\delta $ oscillate about their respective steady state values, where the oscillations slowly decay with time. \autoref{fig:nonlineartruth4} and 
\autoref{fig:nonlineartruth5} show the plots for the two control inputs, $V_F$ and $u_T$, respectively. The generator 
excitation voltage $V_F$ oscillates about its steady state value of 0.00121 p.u. and the turbine valve control settles to its
steady state value of 1.0512 p.u.


\newpage
\section{Simulation results for the Controllers at different Operating Points}

In this section we test the LTR-based LQG, nonlinear feedback linearizing controller, and the LQR-based full-state feedback 
controllers, which were tested on the truth model in the previous section, at new operating points. The operating points at which the controllers are tested are given in \autoref{tab:operating_points}.
Operating point I, is the original desired operating point about which the reduced order model was linearized, and all the controllers
were designed and tested in the previous sections. In this section we will test the controllers 
that we tested on the truth model, in the previous section at Operating Point I, at new operating points II and III respectively, without making any change to the controller gains, i.e. the controller gains that were computed for Operating Point I, are unchanged.
Only parameters in the controllers that change with a change in the operating condition are the desired reference values
of the state variables and the outputs.
In all three operating conditions the desired steady state operating value of the frequency $\omega $ is 1 p.u., which is
evident from the differential equation, $\dot{\delta }=\omega -1$. Also the infinite bus voltage $V_\infty $ is 1 p.u. at all three
operating conditions. At Operating Condition I, machine loading or the real power $P$ generated by the synchronous generator
is 1 p.u. at 0.85 lagging power factor conditions. At Operating Point II, the real power $P$ generated by the synchronous
generator is 0.6368 p.u. at 0.9892 lagging power factor conditions. At Operating Point III, the real power $P$ generated by the synchronous generator is 1.3466 p.u. at 0.652 lagging power factor conditions. Thus, from Operating Point II we can see that as the power factor is increased from 0.85 to 0.9892 the real power generated by the synchronous generator decreases from 1 p.u. to 0.6368 p.u. Similarly, from Operating Condition III we can see that as the power factor is reduced to 0.652 the real power generated by the synchronous generator or the machine loading increases to 1.3466 p.u. Also the stator current $I_a$ of the synchronous generator which is equal to 1.0037 p.u. at operating condition I, is reduced to 0.6323 p.u., when the machine loading is reduced at Operating Point II, and the stator current increases to 1.4764 p.u., when the load on the synchronous generator is increased at Operating Condition III. Thus, by varying the machine loading i.e. by increasing or decreasing the load on the generator, the operating
conditions are varied.

\begin{centergroup}
	\captionof{table}{Operating Points of the SMIB}
	\begin{tabular}{|c|c|c|c|}
		\hline
		 & Operating Point I & Operating Point II & Operating Point III\\
		\hline
	        Variables (in p.u.) & & & \\ 
		\hline
		    $I_{d0}$ & -0.9185 & -0.4818 & -1.4281\\
		\hline
		    $I_{F0}$ & 1.6315 & 1.0228 & 2.37786\\
		\hline
		   $I_{D0}$ & $-4.6204\times 10^{-6}$ & 0 & 0\\
		\hline   
		   $I_{q0}$ & 0.4047 & 0.4094 & 0.37472\\
		 \hline  
          $I_{Q0}$ & $5.9539\times 10^{-5}$ & 0 & 0\\
         \hline 
          $\omega _0$ & 1 & 1 & 1\\
         \hline 
          $\delta _0$ & 1 & 1.0325 & 0.88676\\
         \hline 
          $T_{m0}$ & 1.0012 & 0.6373 & 1.34899\\
         \hline 
          $G_{V0}$ & 1.0012 & 0.6373 & 1.34899\\
          \hline
          $V_{q0}$ & 0.9670 & 0.7659 & 1.2575\\
          \hline
          $V_{d0}$ & -0.6628 & -0.6710 & -0.6130\\ 
          \hline
          $V_{t0}$ & 1.172 & 1.0182 & 1.3990\\ 
          \hline
          Stator current $I_{a0}$ & 1.0037 & 0.6323 & 1.4764\\
          \hline
          $V_\infty $ & 1.00 & 1.00 & 1.00\\
          \hline
          $\alpha $ & $3.5598^\circ $ & $3.5598^\circ $ & $3.5598^\circ $\\
          \hline
          $E'_{q0}$ & 1.1925 & 0.8844 & 1.6078\\ 
          \hline
          $\tau '_{d0}$ & 5.90 & 5.90 & 5.90\\
          \hline
          Real power ($P=V_tI_a\cos \phi $) & 1.00 & 0.6368 & 1.3466\\
          \hline
          power factor ($PF=\cos \phi $) & 0.85 & 0.9892 & 0.652\\
          \hline
	\end{tabular}
	\label{tab:operating_points}
\end{centergroup}

\newpage
\subsection{Simulation results for the Controllers at Operating Point II}

\begin{figure}
          \centering
          \includegraphics[trim=0cm 0cm 0cm 0cm, clip=true, totalheight=0.27\textheight, width=0.54
           \textwidth]  {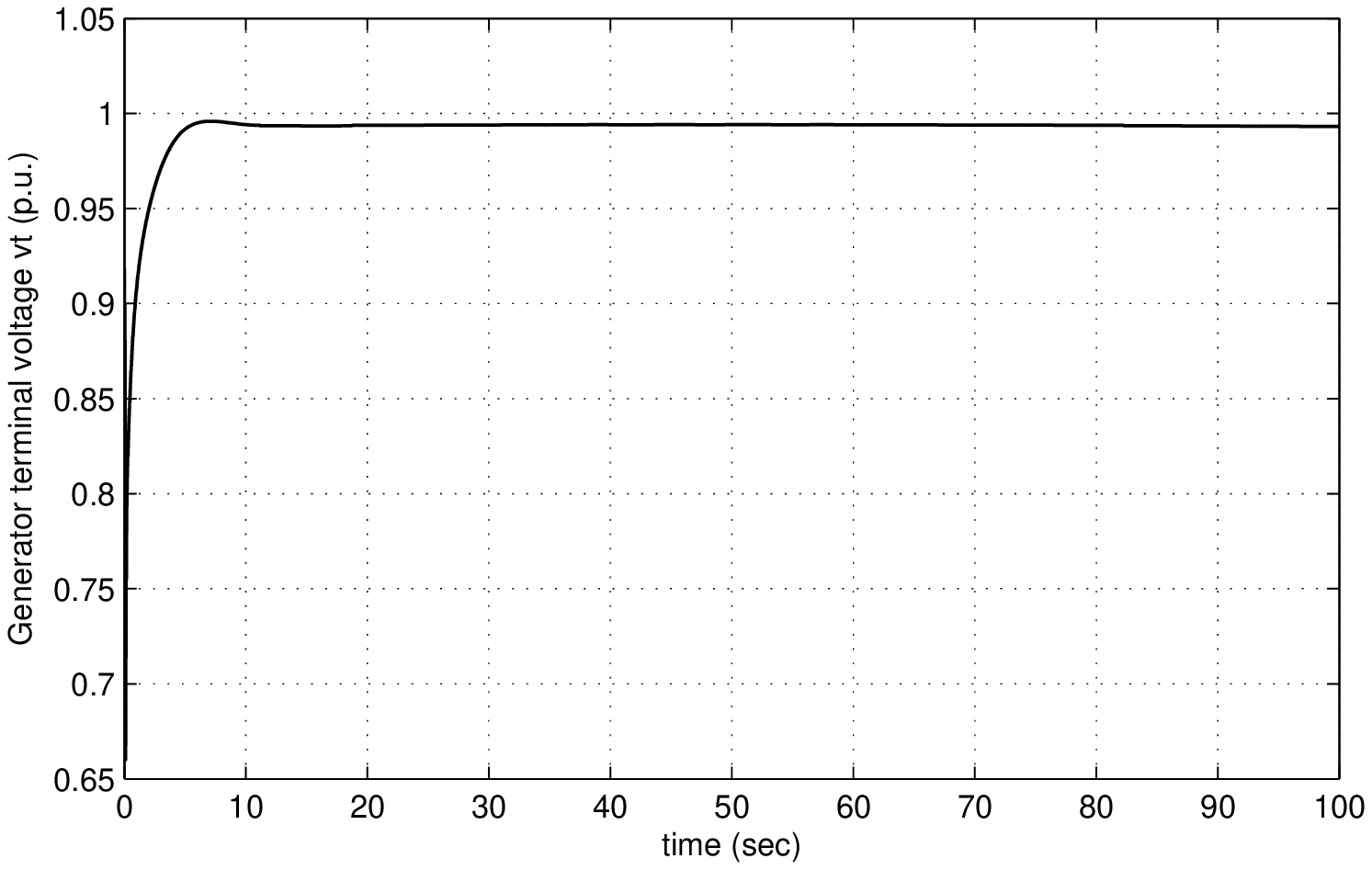}
          \caption{Plot of the generator terminal voltage $V_t$ vs time for the LTR-based LQG controller
          applied to the truth model (Operating Point II)}
          \label{fig:ltrtruth1op2}
          \includegraphics[trim=0cm 0cm 0cm 0cm, clip=true, totalheight=0.27\textheight, width=0.54\textwidth]{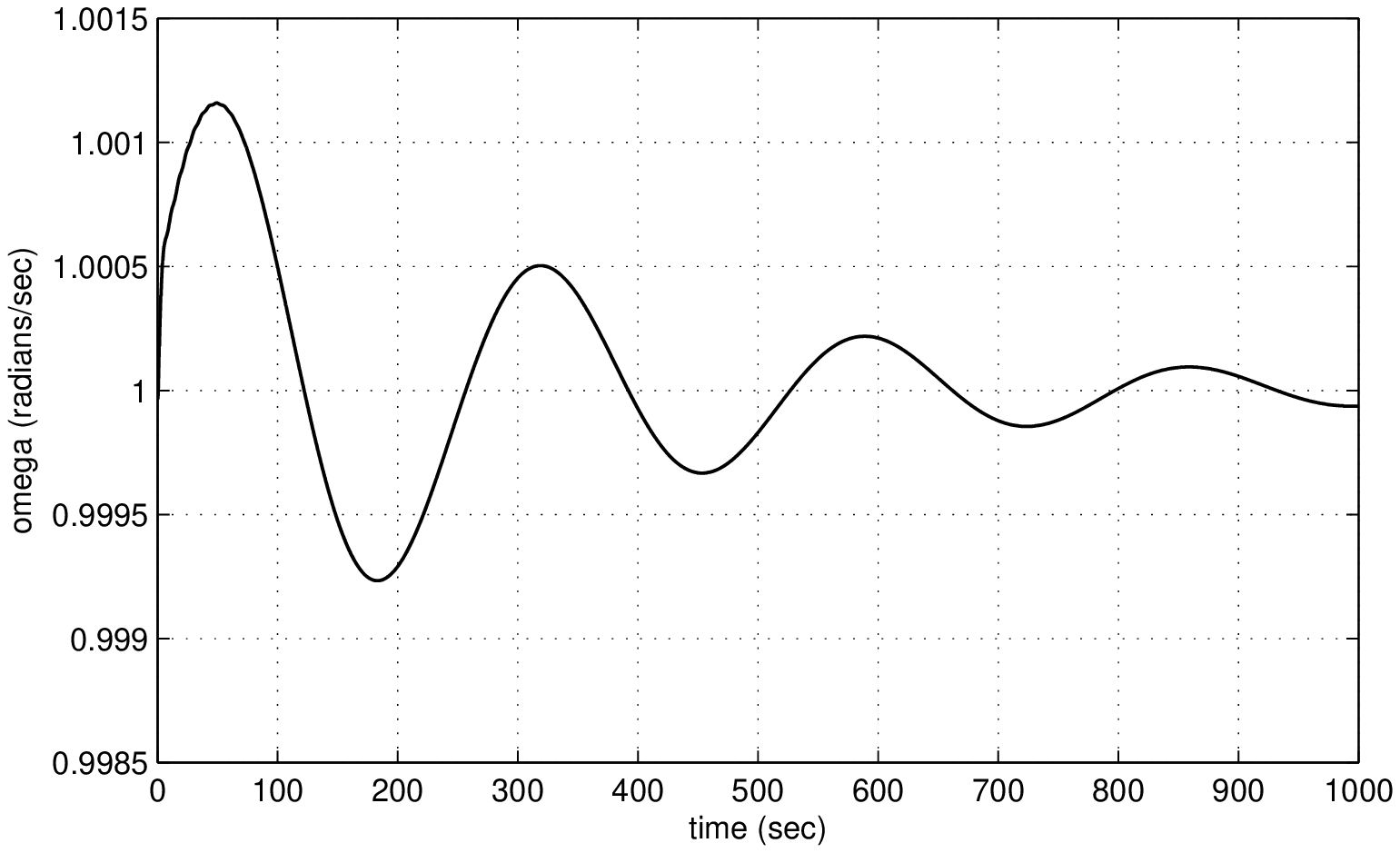}
          \caption{Plot of the angular velocity $\omega $ vs time for the LTR-based LQG controller
          applied to the truth model (Operating Point II)}
          \label{fig:ltrtruth2op2}
          \includegraphics[trim=0cm 0cm 0cm 0cm, clip=true, totalheight=0.27\textheight, width=0.54\textwidth]{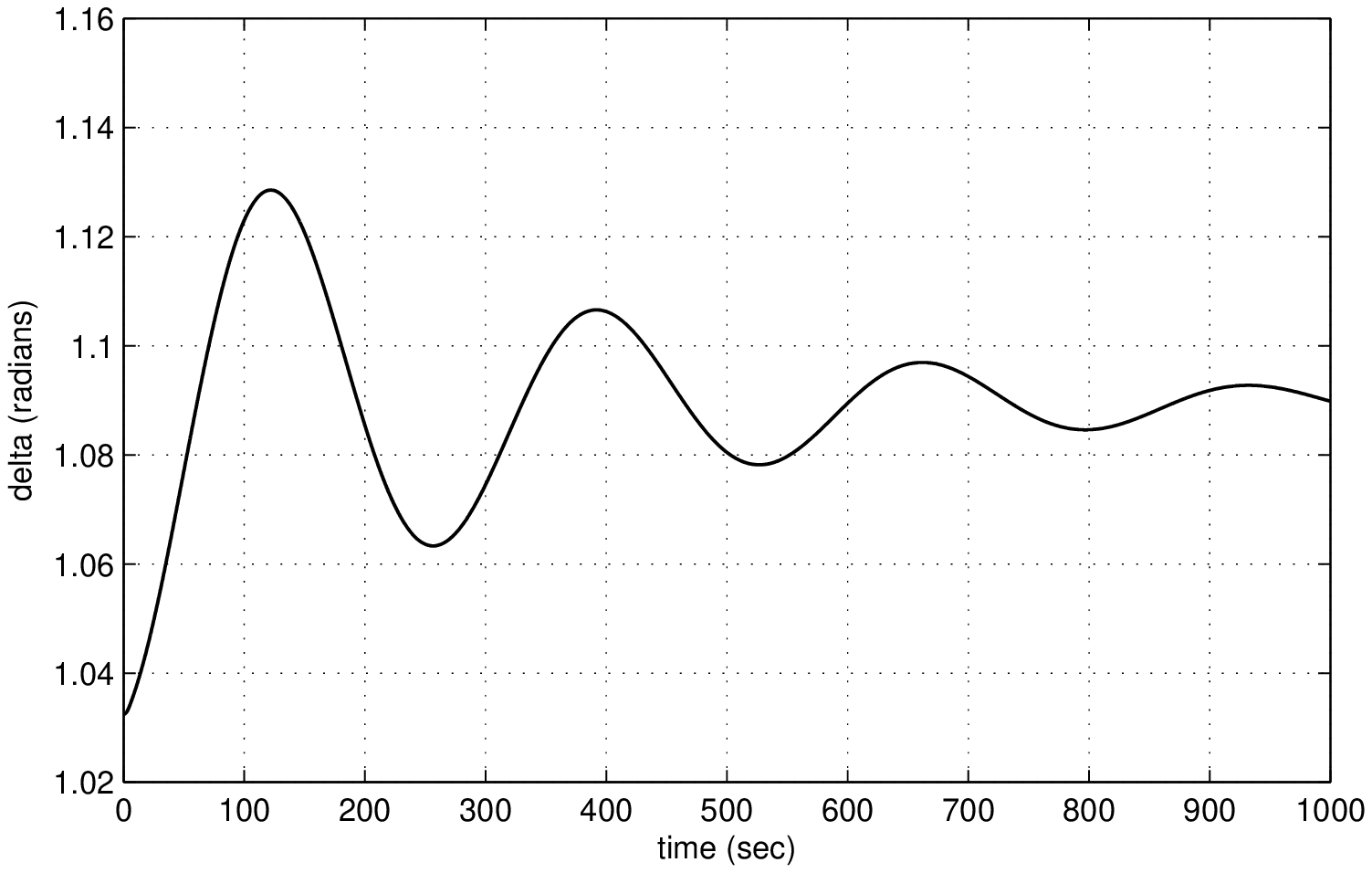}
          \caption{Plot of the rotor angle $\delta $ vs time for the LTR-based LQG controller applied to the truth model
          (Operating Point II) }
          \label{fig:ltrtruth3op2}
\end{figure}

\autoref{fig:ltrtruth1op2}, \autoref{fig:ltrtruth2op2}, and \autoref{fig:ltrtruth3op2} show the simulation results for
the LTR-based LQG controller 
 applied to the truth model at Operating Point II. From these results we can see that
the generator terminal voltage $V_t$ settles to a steady state value of 0.9923 p.u. which is different from the desired steady state
value of 1.0182 p.u., with a steady state error of 0.0259 p.u. The angular velocity $\omega $ oscillates about the desired steady state value of 1 p.u., where the  oscillations decay with time. 
 Also the rotor angle $\delta $ oscillates about a new steady state value of 1.09 p.u., which deviates from the desired steady state value of 1.0325 p.u., with a steady state error of 0.0575 p.u. Thus, we observe a small steady state error, when the LTR-based LQG controller is applied to the truth model at Operating Point II.

\begin{figure}
          \centering
          \includegraphics[trim=0cm 0cm 0cm 0cm, clip=true, totalheight=0.27\textheight, width=0.54
           \textwidth]  {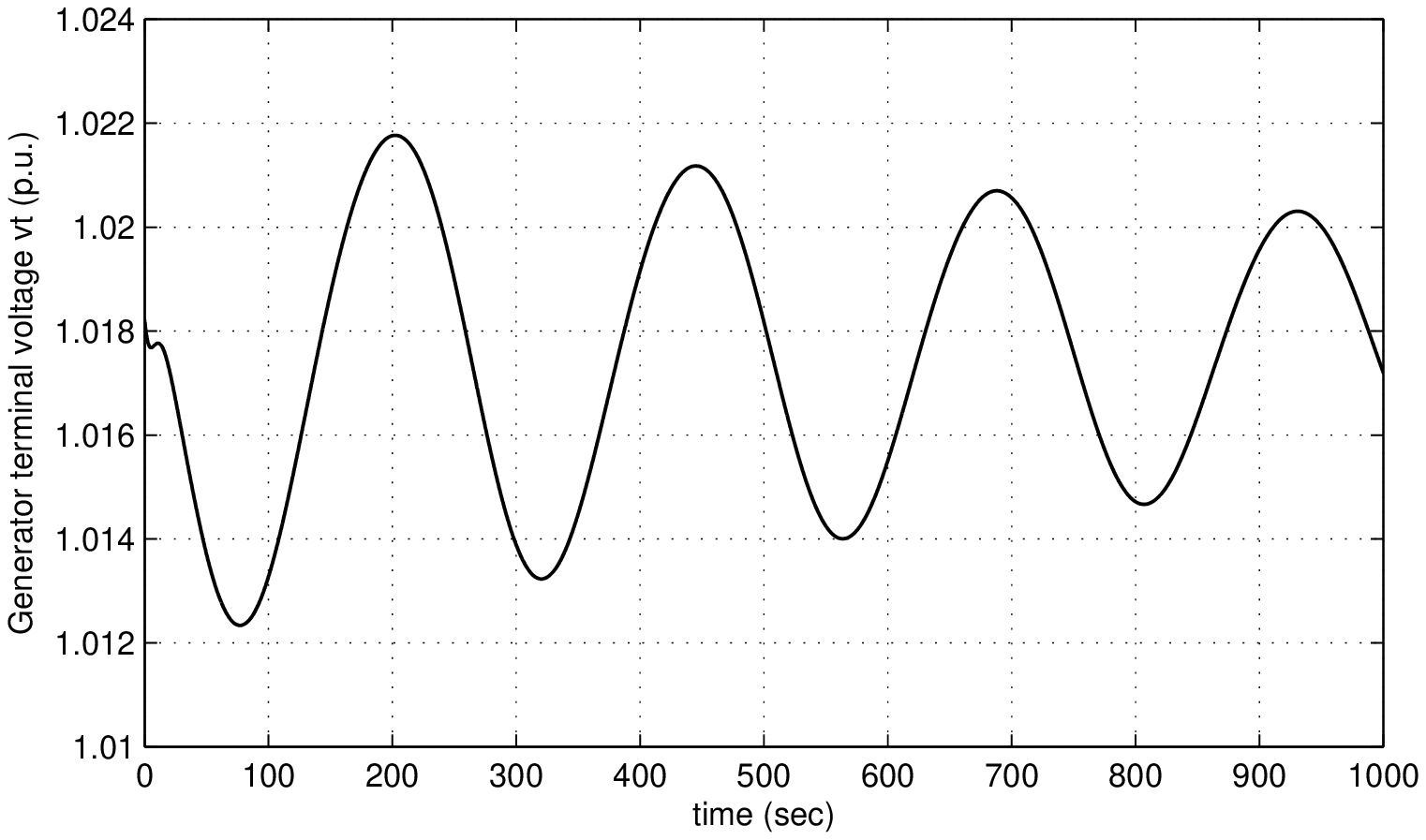}
          \caption{Plot of the generator terminal voltage $V_t$ vs time for the nonlinear feedback linearizing controller
          applied to the truth model (Operating Point II)}
          \label{fig:nonlineartruth1op2}
          \includegraphics[trim=0cm 0cm 0cm 0cm, clip=true, totalheight=0.27\textheight, width=0.54\textwidth]  {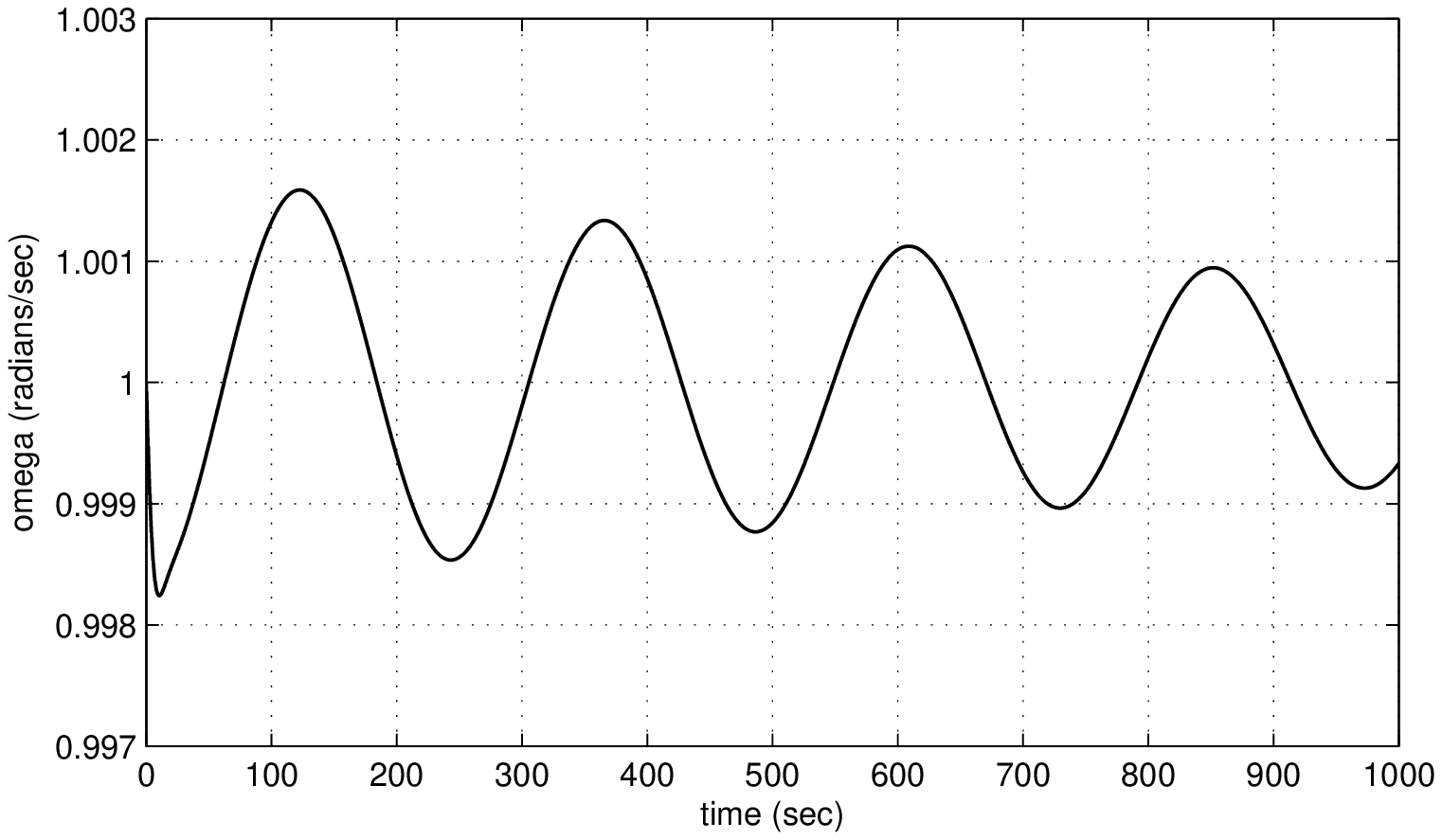}
          \caption{Plot of the angular velocity $\omega $ vs time for the nonlinear feedback linearizing controller
          applied to the truth model (Operating Point II)}
          \label{fig:nonlineartruth2op2}
          \includegraphics[trim=0cm 0cm 0cm 0cm, clip=true, totalheight=0.27\textheight, width=0.54\textwidth]{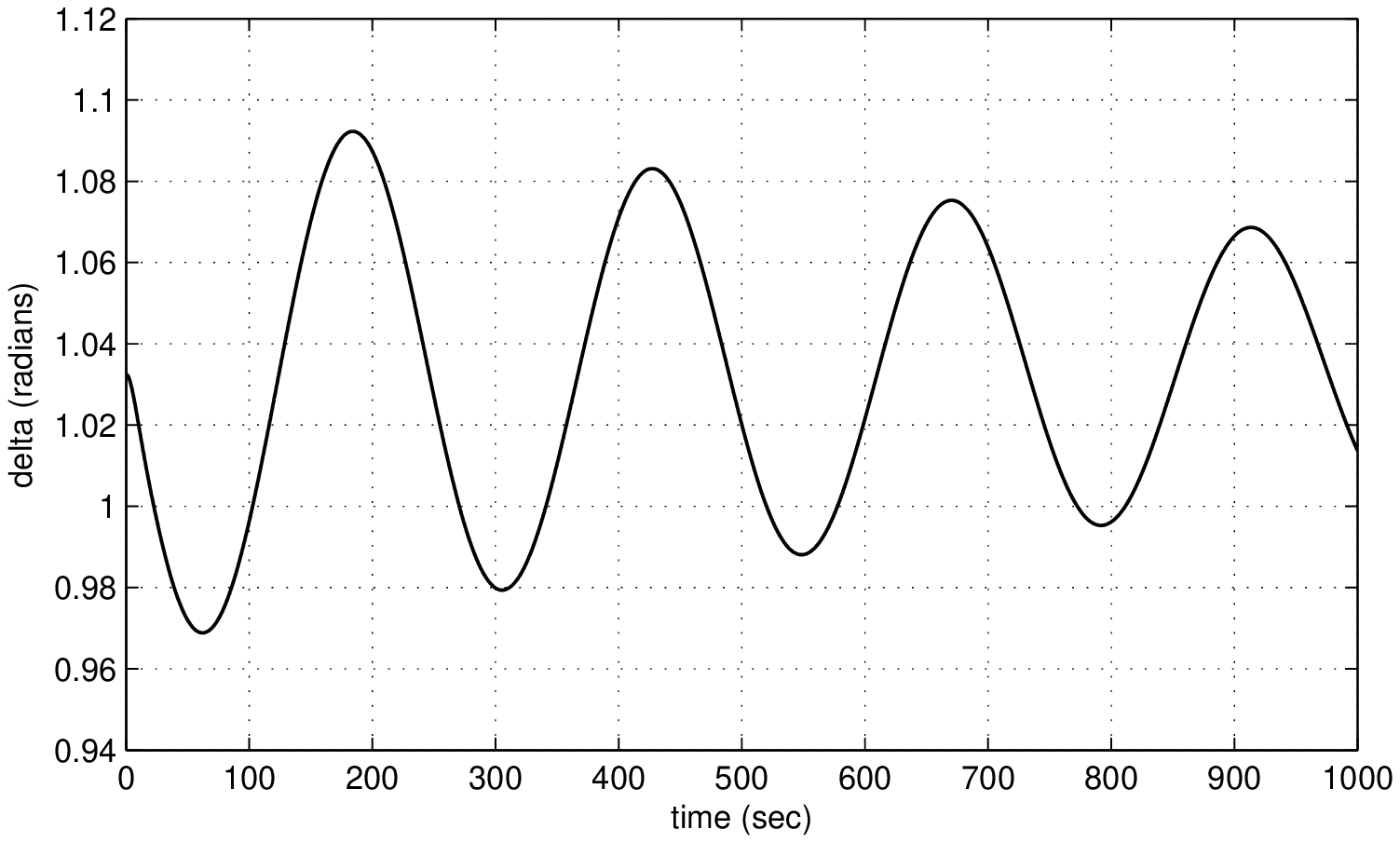}
          \caption{Plot of the rotor angle $\delta $ vs time for the nonlinear feedback linearizing controller 
           applied to the truth model (Operating Point II)}
          \label{fig:nonlineartruth3op2}
\end{figure}

\autoref{fig:nonlineartruth1op2}, \autoref{fig:nonlineartruth2op2}, and \autoref{fig:nonlineartruth3op2} show the simulation results for
the nonlinear feedback linearizing controller applied to the truth model at Operating Point II. The generator terminal
voltage $V_t$ oscillates about the desired steady state value of 1.0182 p.u., the angular velocity $\omega $ oscillates about the
desired steady state value of 1 p.u., and the rotor angle $\delta $ oscillates about the
desired steady state value of 1.0325 p.u. The magnitude of these oscillations are more than that seen for the LTR-based LQG controller. The transient response of the nonlinear feedback linearizing controller is not as good as the
LTR-based LQG controller, but the steady state error seen in the linear controller is significantly reduced in the nonlinear controller
at Operating Point II.

\begin{figure}
          \centering
          \includegraphics[trim=0cm 0cm 0cm 0cm, clip=true, totalheight=0.27\textheight, width=0.54
           \textwidth]  {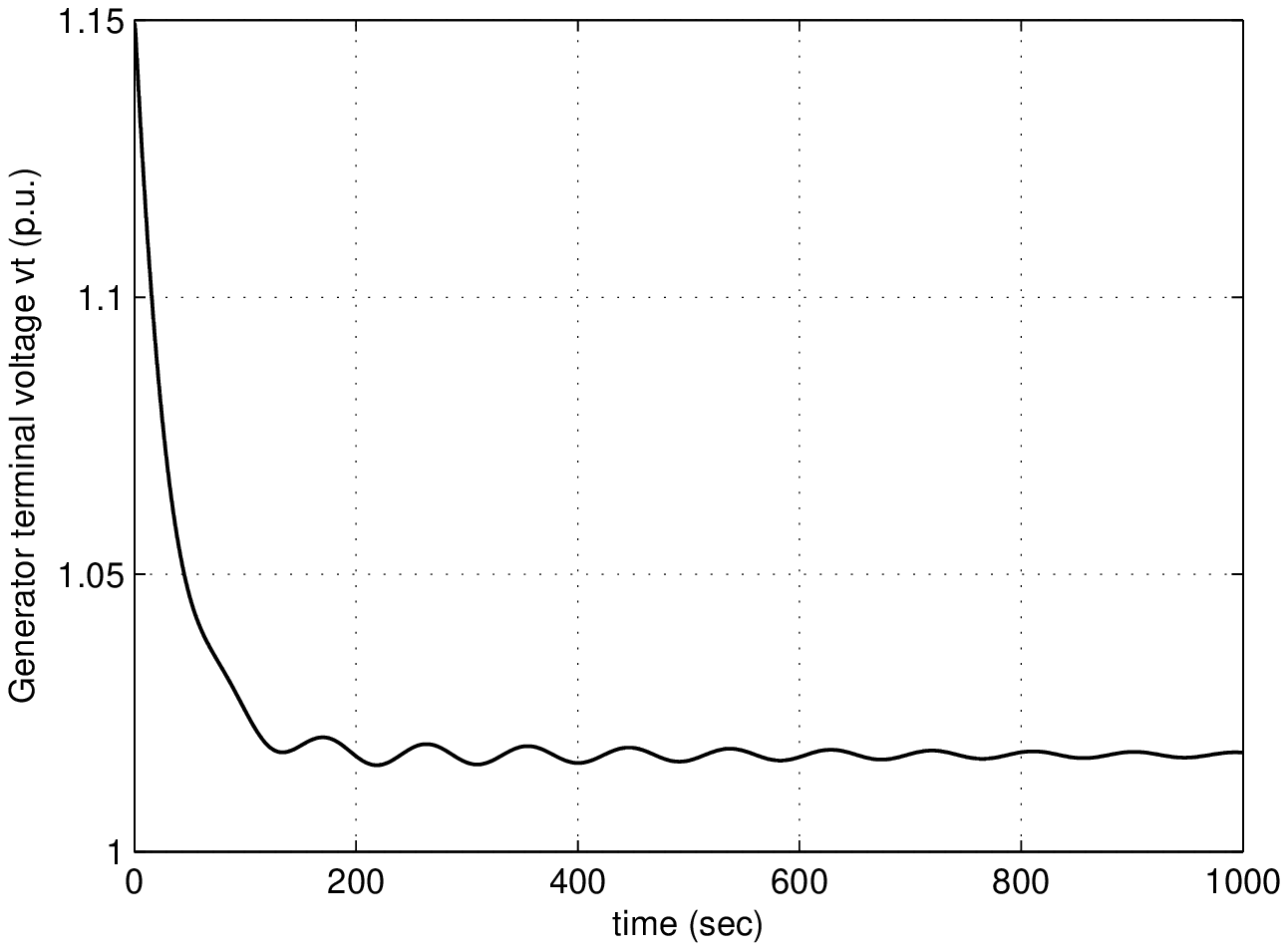}
          \caption{Plot of the generator terminal voltage $V_t$ vs time for the LQR-based full-state feedback controller
          applied to the truth model (Operating Point II)}
          \label{fig:lqrtruth1op2}
          \includegraphics[trim=0cm 0cm 0cm 0cm, clip=true, totalheight=0.27\textheight, width=0.54\textwidth]{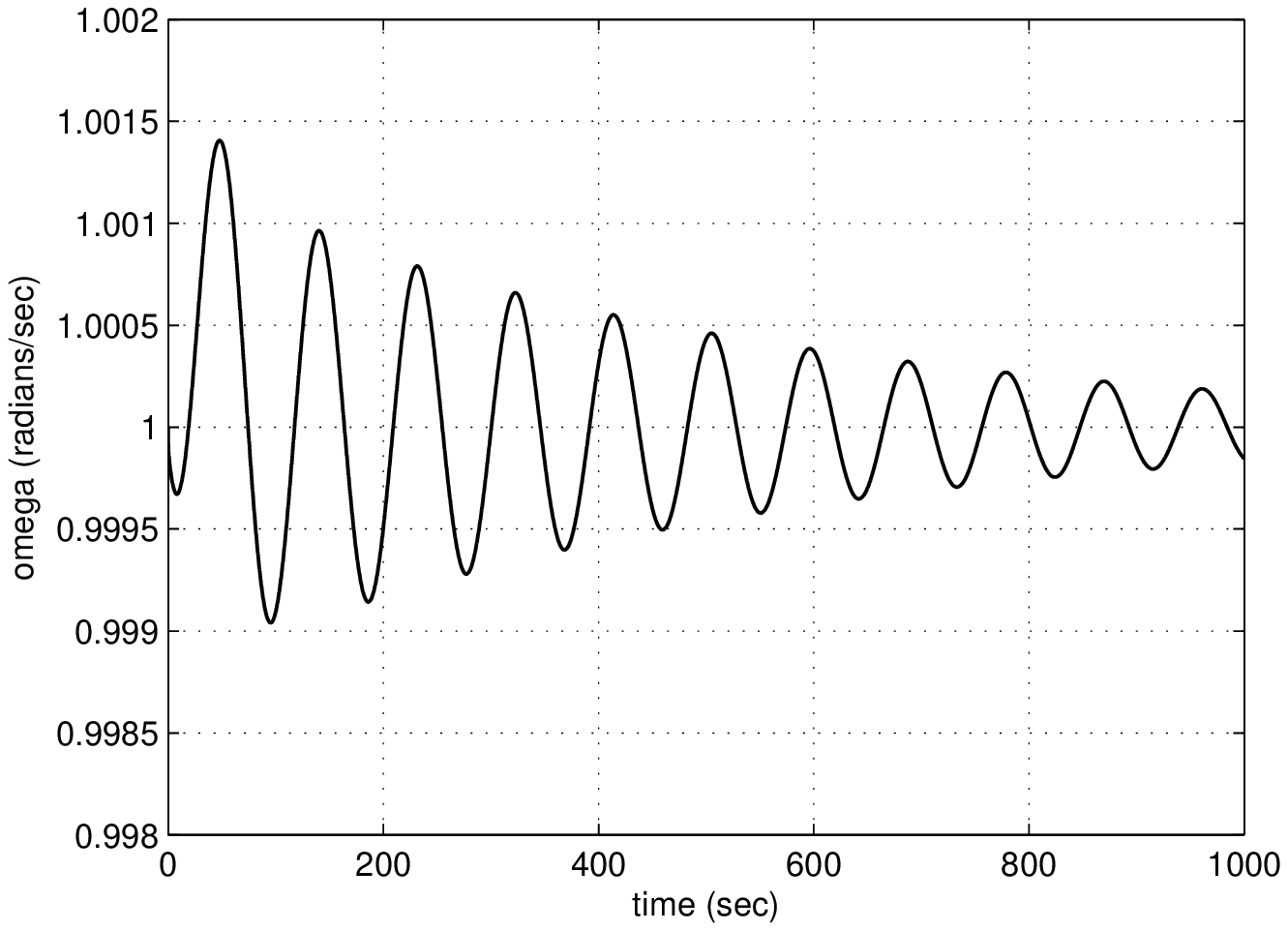}
          \caption{Plot of the angular velocity $\omega $ vs time for the LQR-based full-state feedback controller
          applied to the truth model (Operating Point II)}
          \label{fig:lqrtruth2op2}
          \includegraphics[trim=0cm 0cm 0cm 0cm, clip=true, totalheight=0.27\textheight, width=0.54\textwidth]{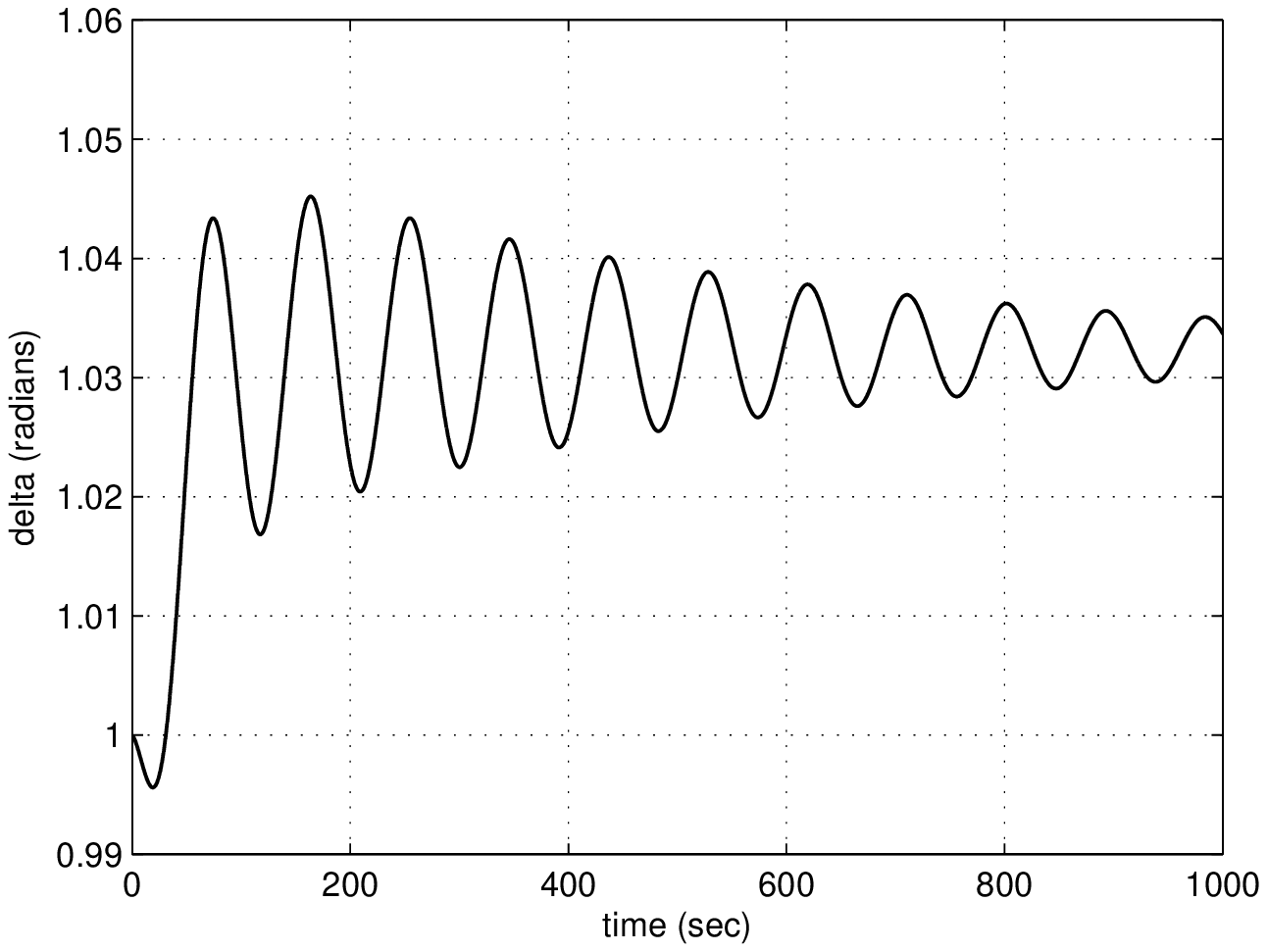}
          \caption{Plot of the rotor angle $\delta $ vs time for the LQR-based full-state feedback controller applied 
          to the truth model (Operating Point II)}
          \label{fig:lqrtruth3op2}
\end{figure}

\autoref{fig:lqrtruth1op2}, \autoref{fig:lqrtruth2op2}, and \autoref{fig:lqrtruth3op2} show the simulation results for
the LQR-based full-state feedback controller applied to the truth model at Operating Point II. 
The generator terminal voltage $V_t$ settles to the desired steady state
value of 1.0182 p.u. The angular velocity $\omega $ oscillates about the desired steady state value of 1 p.u., where the oscillations decay with time. Also the rotor angle $\delta $, oscillates about the desired steady state value of 1.0325 p.u. 
The steady state response of the LQR-based full-state feedback controller is much better than the LTR-based LQG controller at
Operating Point II. Also, the transient response of the LQR-based 
full-state feedback controller is better than the transient response of the nonlinear feedback linearizing controller at operating 
condition II.

\newpage
\subsection{Simulation results for the Controllers at Operating Point III}

\begin{figure}
          \centering
          \includegraphics[trim=0cm 0cm 0cm 0cm, clip=true, totalheight=0.27\textheight, width=0.54
           \textwidth]  {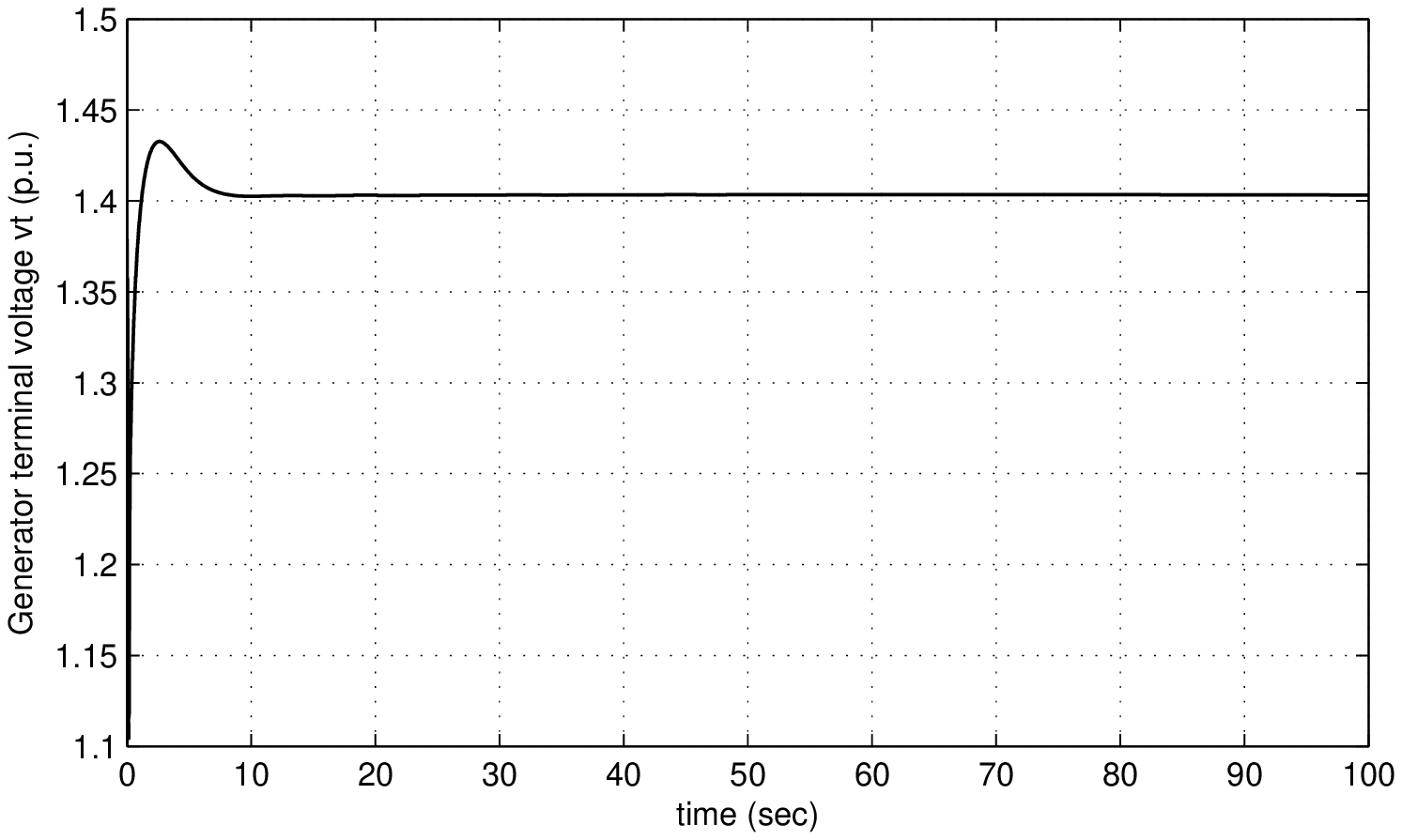}
          \caption{Plot of the generator terminal voltage $V_t$ vs time for the LTR-based LQG controller
          applied to the truth model (Operating Point III)}
          \label{fig:ltrtruth1op3}
          \includegraphics[trim=0cm 0cm 0cm 0cm, clip=true, totalheight=0.27\textheight, width=0.54\textwidth]{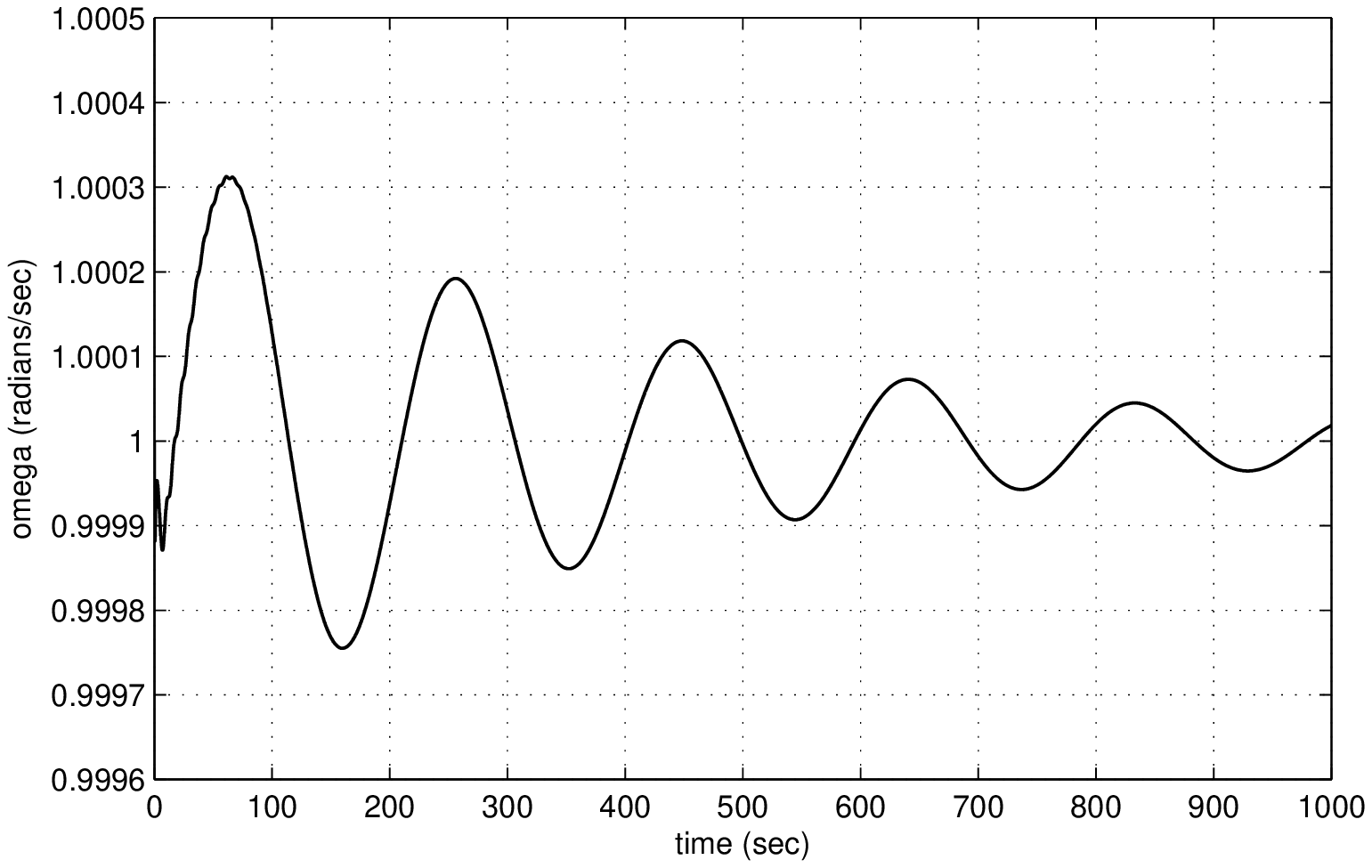}
          \caption{Plot of the angular velocity $\omega $ vs time for the LTR-based LQG controller
          applied to the truth model (Operating Point III)}
          \label{fig:ltrtruth2op3}
          \includegraphics[trim=0cm 0cm 0cm 0cm, clip=true, totalheight=0.27\textheight, width=0.54\textwidth]{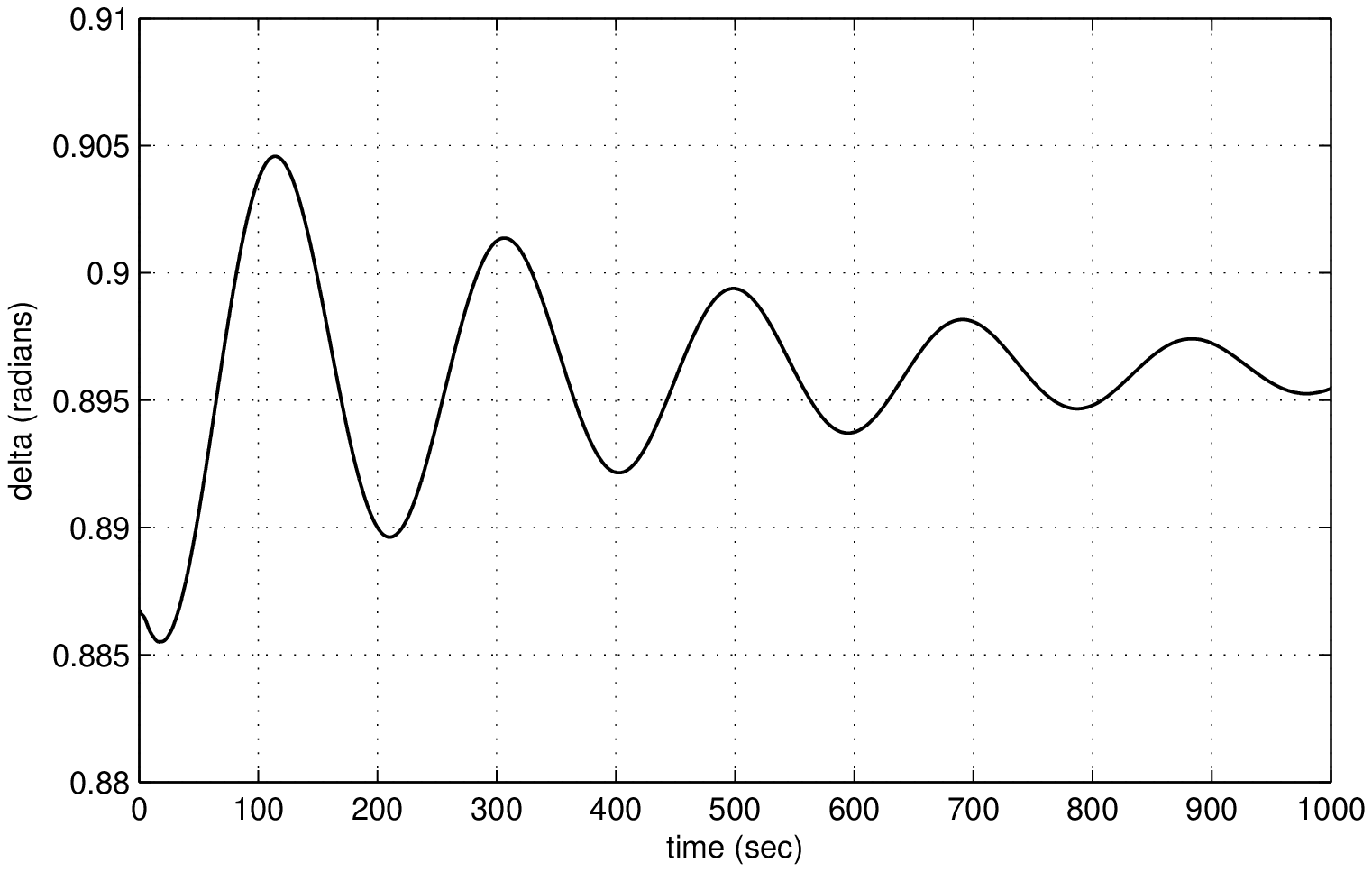}
          \caption{Plot of the rotor angle $\delta $ vs time for the LTR-based LQG controller applied to the truth model
          (Operating Point III) }
          \label{fig:ltrtruth3op3}
\end{figure}

\autoref{fig:ltrtruth1op3}, \autoref{fig:ltrtruth2op3}, and \autoref{fig:ltrtruth3op3} show the simulation results for
the LTR-based LQG controller applied to the truth model at Operating Point III. From these results we can see that
the generator terminal voltage $V_t$ settles to a steady state value of 1.403 p.u. which is different from the desired steady state
value of 1.3990 p.u., with a steady state error of 0.004 p.u. The angular velocity $\omega $ oscillates about the desired steady state value of 1 p.u., where the  oscillations decay with time. Also the rotor angle $\delta $, oscillates about a new steady state value of 
0.896 p.u. which deviates from the desired steady state value of 0.88676 p.u., with a steady state error of 0.00924 p.u. Thus, we observe a small steady state error, when the LTR-based LQG controller is applied to the truth model at Operating Point III.

\begin{figure}
          \centering
          \includegraphics[trim=0cm 0cm 0cm 0cm, clip=true, totalheight=0.27\textheight, width=0.54
           \textwidth]  {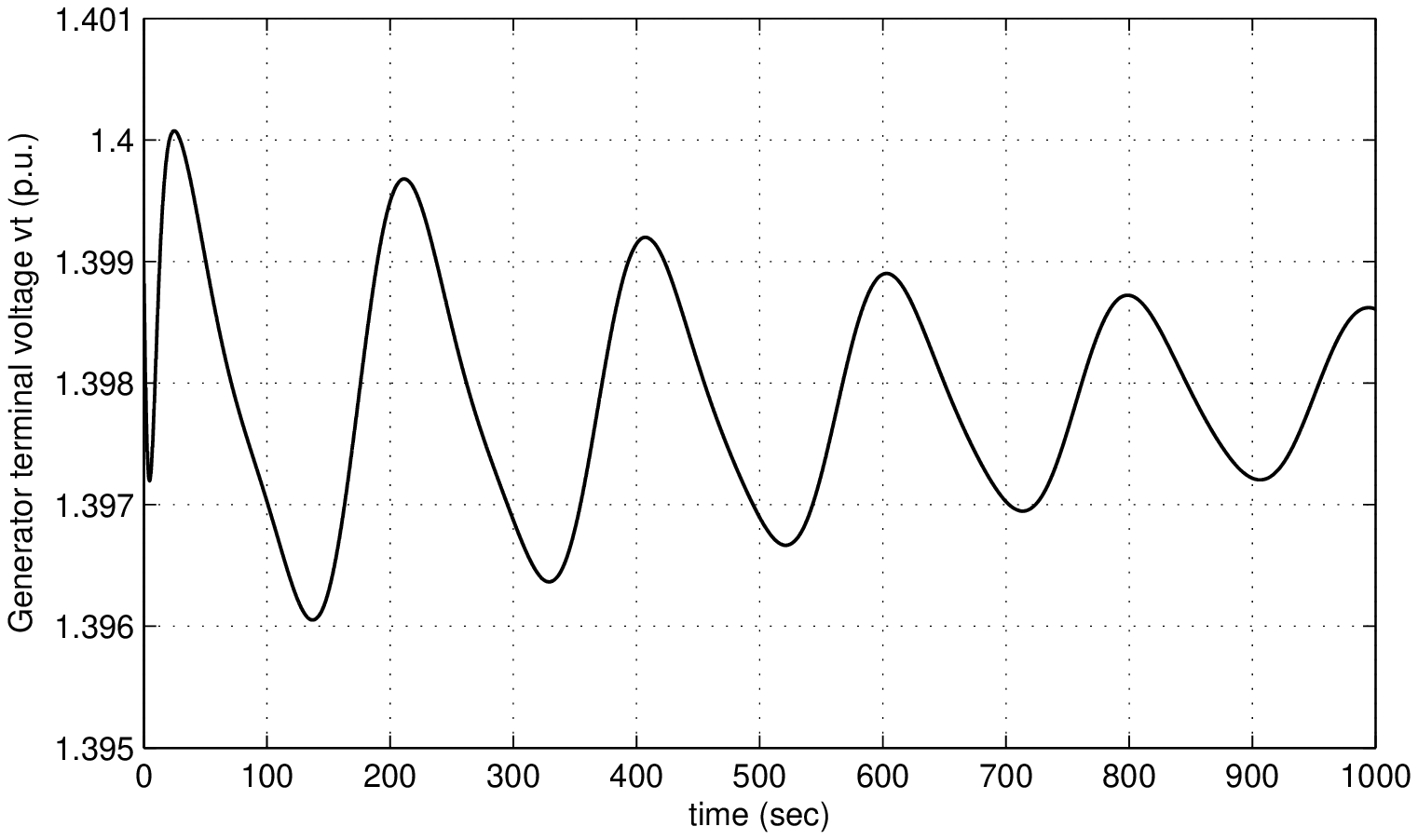}
          \caption{Plot of the generator terminal voltage $V_t$ vs time for the nonlinear feedback linearizing controller
          applied to the truth model (Operating Point III)}
          \label{fig:nonlineartruth1op3}
          \includegraphics[trim=0cm 0cm 0cm 0cm, clip=true, totalheight=0.27\textheight, width=0.54\textwidth]  {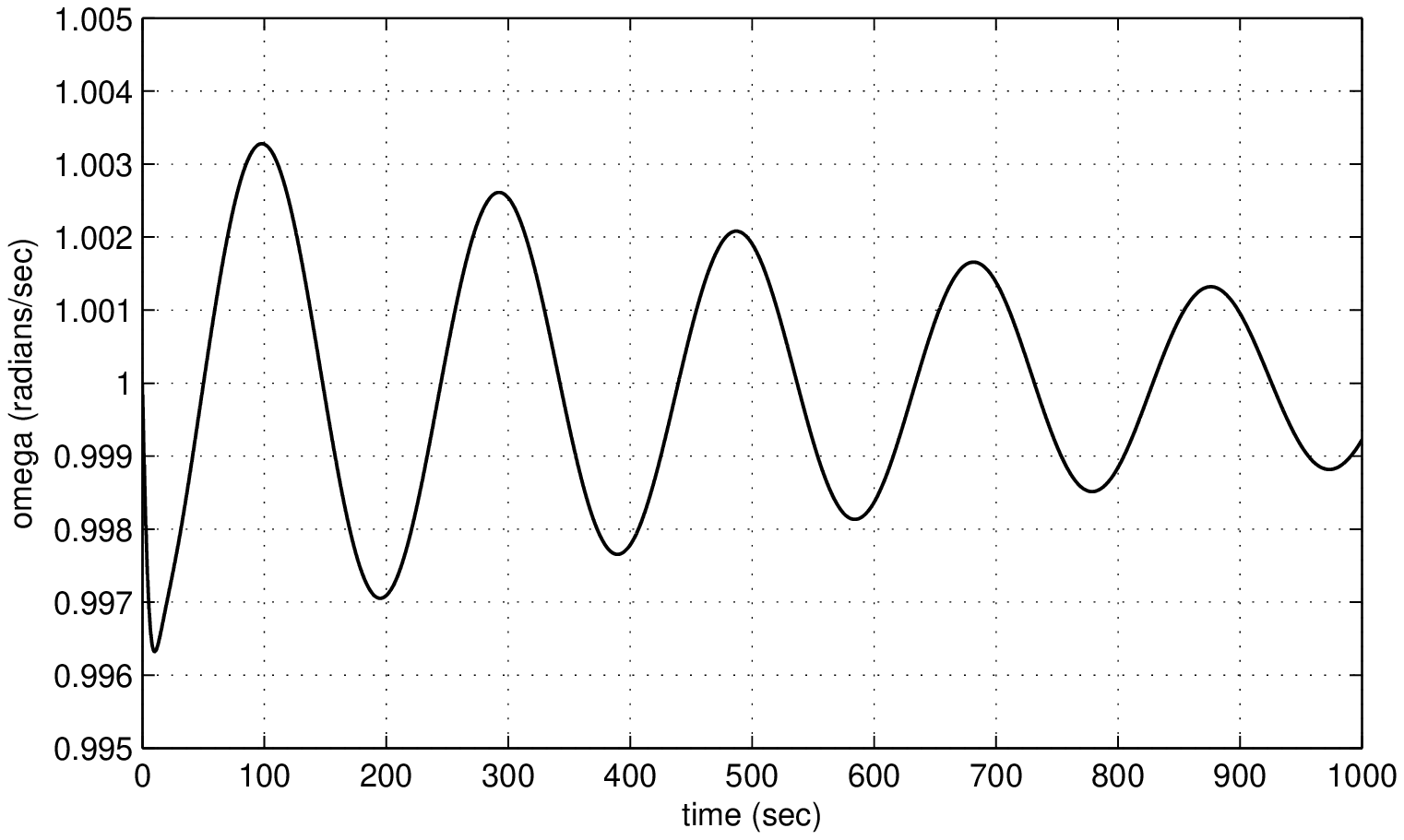}
          \caption{Plot of the angular velocity $\omega $ vs time for the nonlinear feedback linearizing controller
          applied to the truth model (Operating Point III)}
          \label{fig:nonlineartruth2op3}
          \includegraphics[trim=0cm 0cm 0cm 0cm, clip=true, totalheight=0.27\textheight, width=0.54\textwidth]{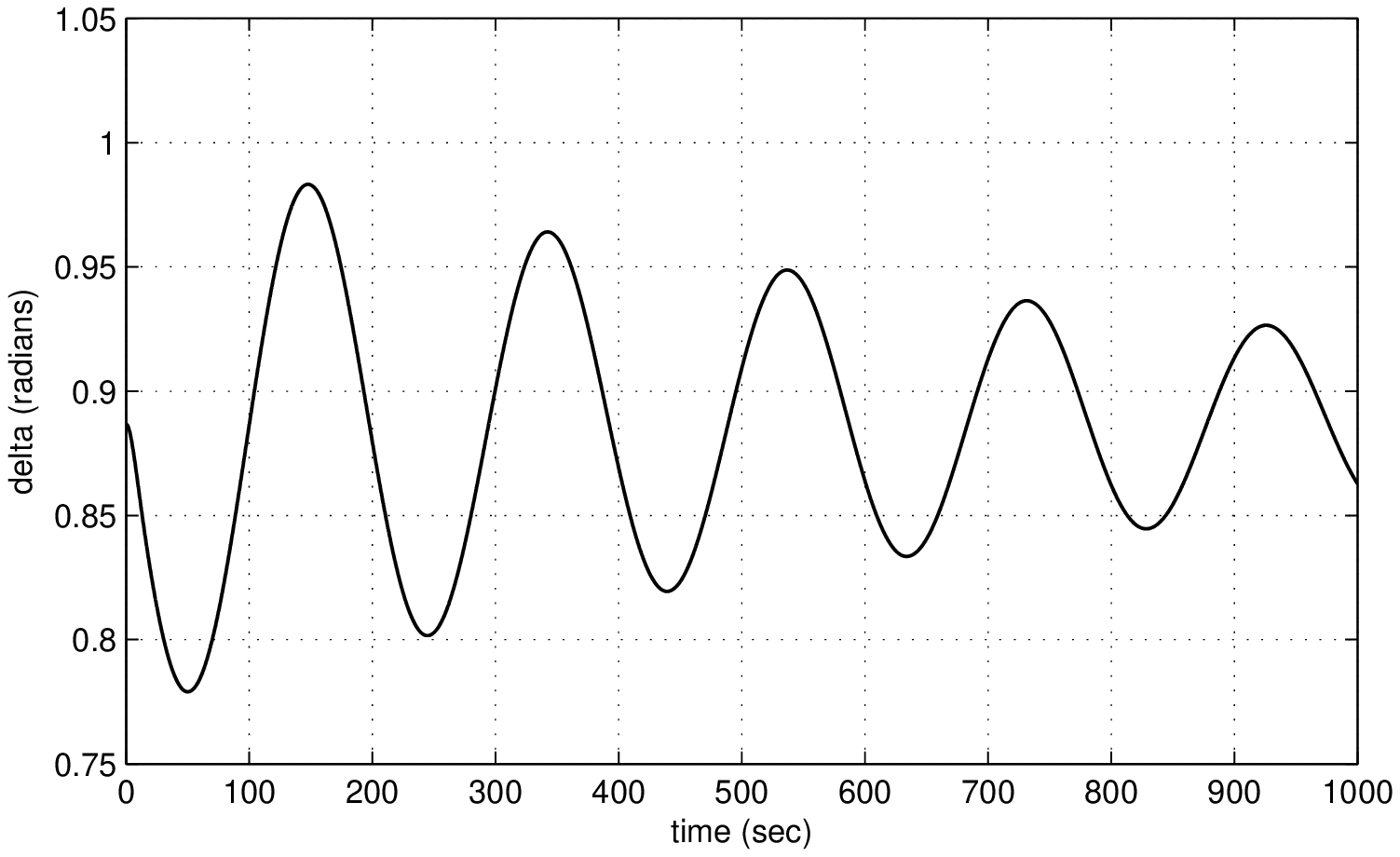}
          \caption{Plot of the rotor angle $\delta $ vs time for the nonlinear feedback linearizing controller 
           applied to the truth model (Operating Point III)}
          \label{fig:nonlineartruth3op3}
\end{figure}

\autoref{fig:nonlineartruth1op3}, \autoref{fig:nonlineartruth2op3}, and \autoref{fig:nonlineartruth3op3} show the simulation results for
the nonlinear feedback linearizing controller applied to the truth model at Operating Point III. The generator terminal
voltage $V_t$ oscillates about the desired steady state value of 1.3990 p.u., the angular velocity $\omega $ oscillates about the
desired steady state value of 1 p.u., and the rotor angle $\delta $ oscillates about the
desired steady state value of 0.88676 p.u. The magnitude of these oscillations are more than that seen for the LTR-based LQG controller. The transient response of the LTR-based LQG controller is better than that of the nonlinear feedback linearizing controller, but the steady state error seen in the linear controller is significantly reduced in the nonlinear controller
at operating point III.

\begin{figure}
          \centering
          \includegraphics[trim=0cm 0cm 0cm 0cm, clip=true, totalheight=0.27\textheight, width=0.54
           \textwidth]  {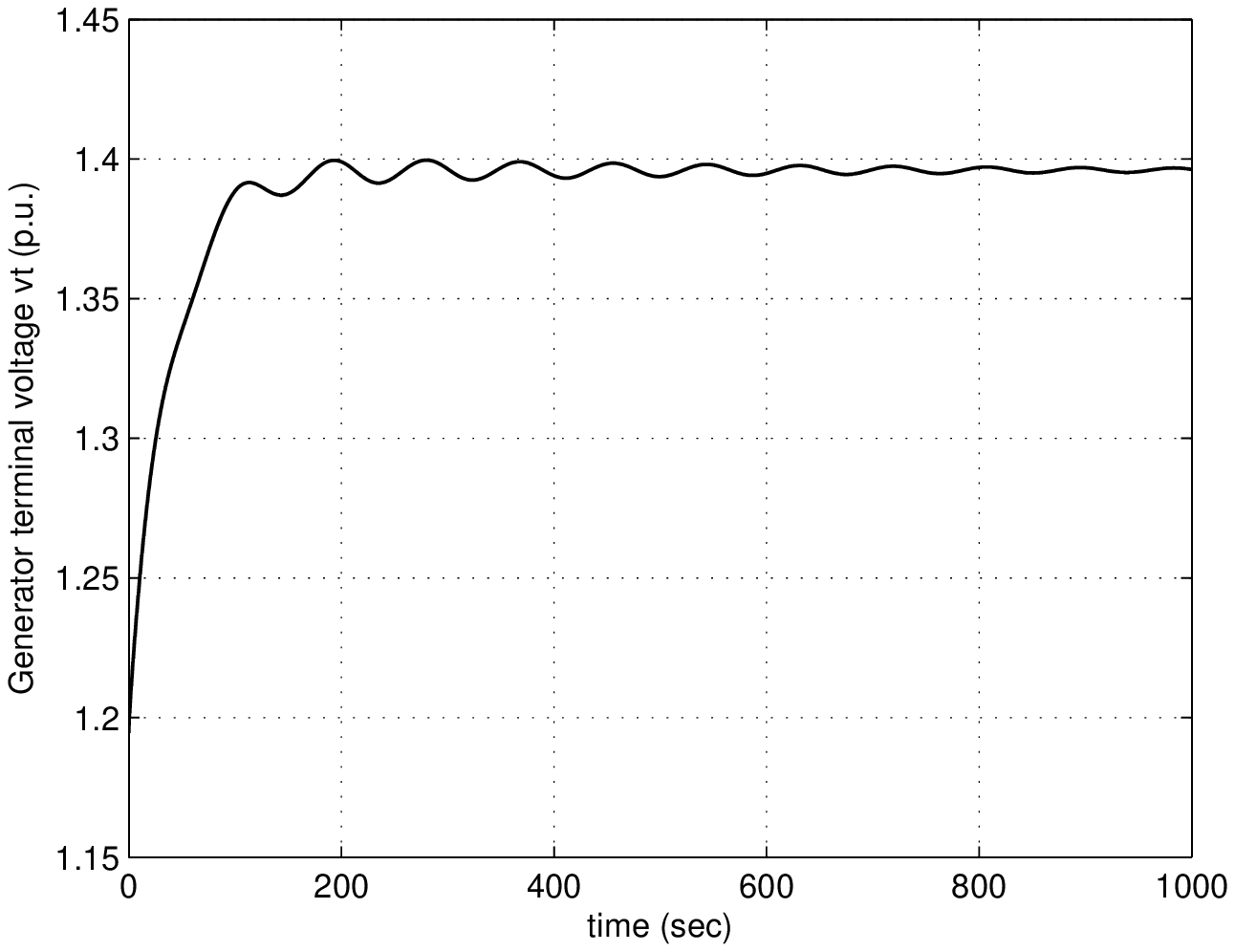}
          \caption{Plot of the generator terminal voltage $V_t$ vs time for the LQR-based full-state feedback controller
          applied to the truth model (Operating Point III)}
          \label{fig:lqrtruth1op3}
          \includegraphics[trim=0cm 0cm 0cm 0cm, clip=true, totalheight=0.27\textheight, width=0.54\textwidth]{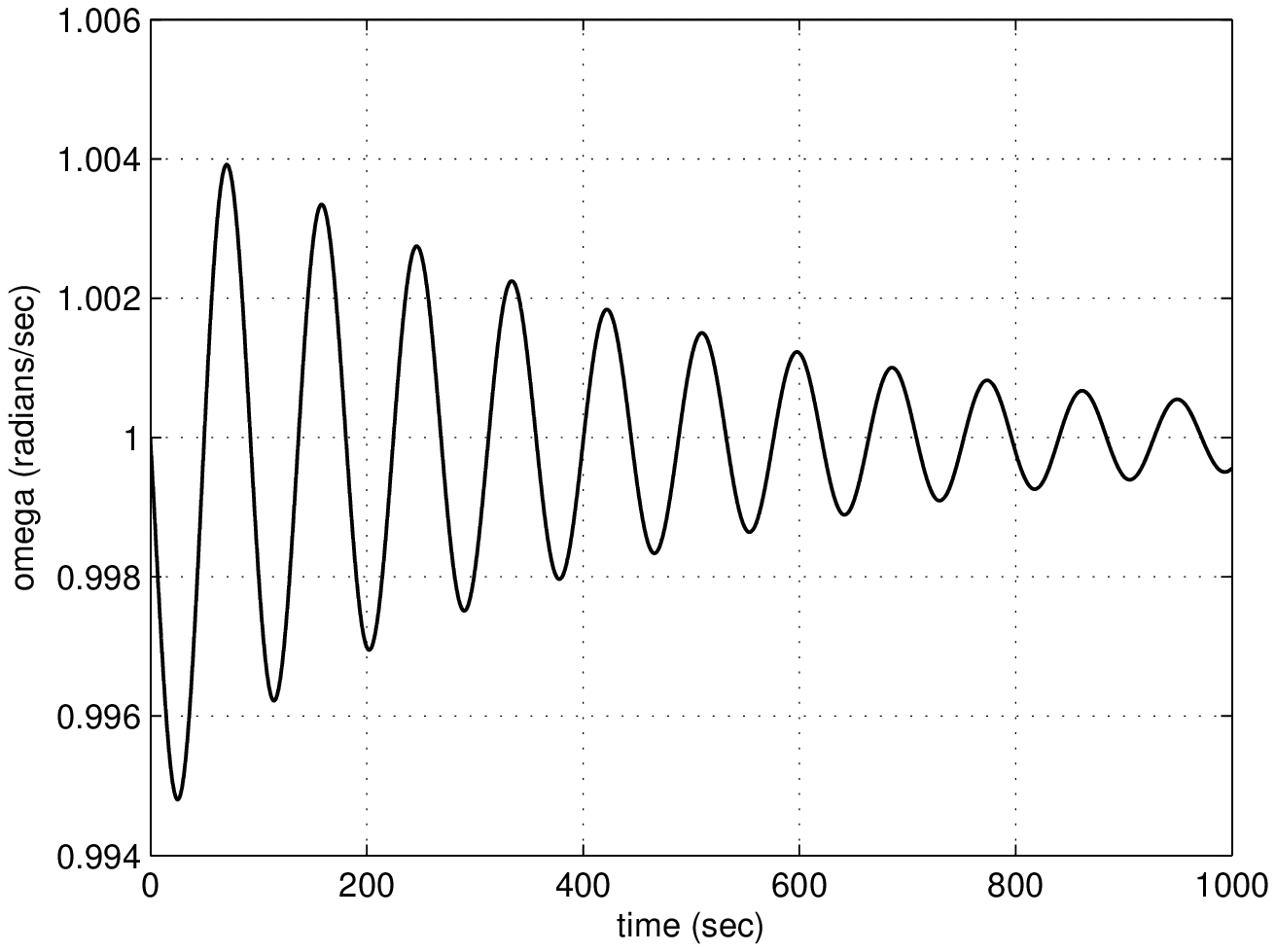}
          \caption{Plot of the angular velocity $\omega $ vs time for the LQR-based full-state feedback controller
          applied to the truth model (Operating Point III)}
          \label{fig:lqrtruth2op3}
          \includegraphics[trim=0cm 0cm 0cm 0cm, clip=true, totalheight=0.27\textheight, width=0.54\textwidth]{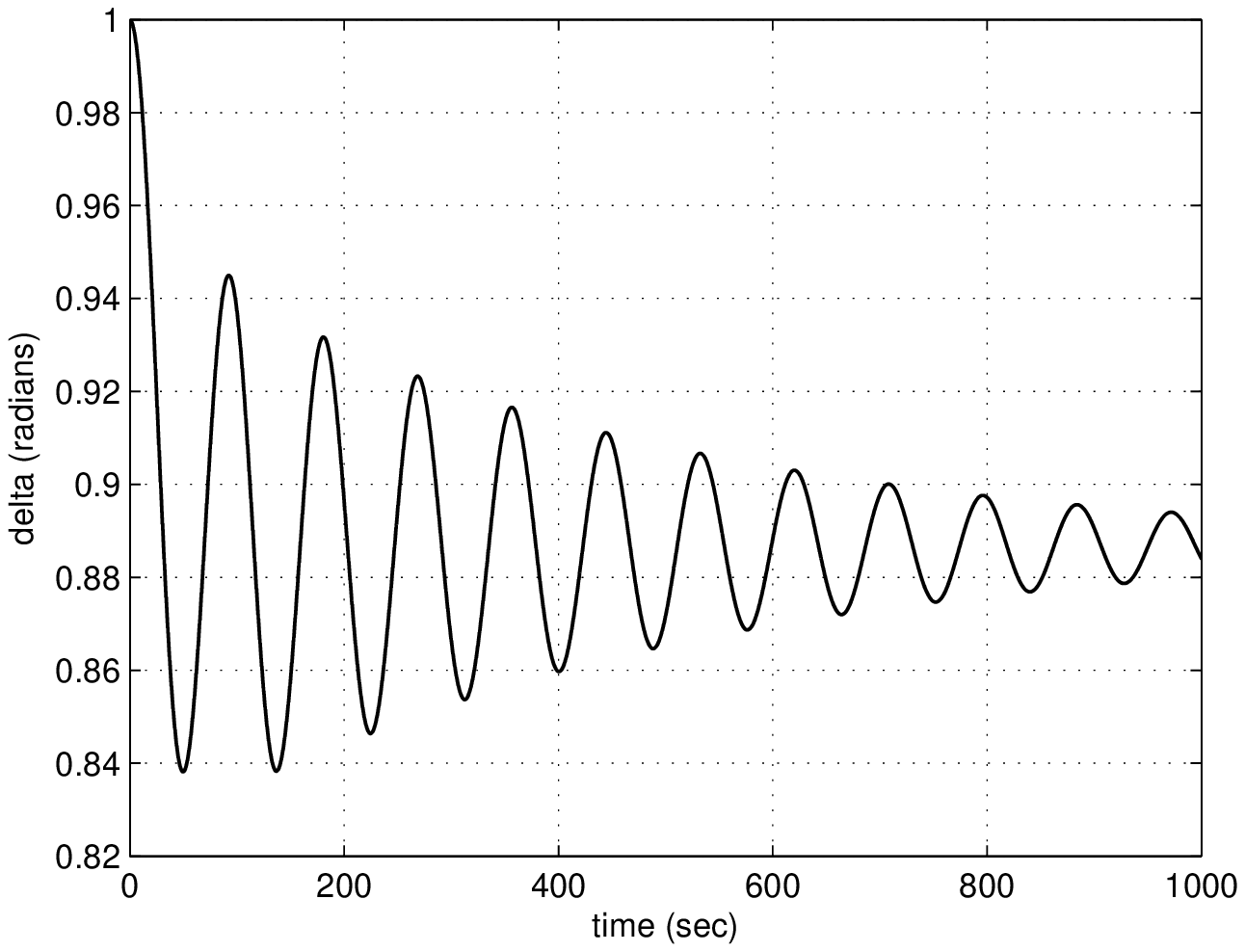}
          \caption{Plot of the rotor angle $\delta $ vs time for the LQR-based full-state feedback controller applied 
          to the truth model (Operating Point III)}
          \label{fig:lqrtruth3op3}
\end{figure}

\autoref{fig:lqrtruth1op3}, \autoref{fig:lqrtruth2op3}, and \autoref{fig:lqrtruth3op3} show the simulation results
for the LQR-based full-state feedback controller applied to the truth model at Operating Point III. 
The generator terminal voltage $V_t$ settles to a steady state value of 1.3964 p.u. which is different from the desired steady state
value of 1.3990 p.u., with a steady state error of 0.0026 p.u. The angular velocity $\omega $ oscillates about the desired steady state value of 1 p.u., where the oscillations decay with time. Also the rotor angle $\delta $ oscillates about a steady state value of 
0.885 p.u. which deviates from the desired steady state value of 0.88676 p.u., with a steady state error of 0.00176 p.u. A negligible
steady state error is observed, when the LQR-based full-state feedback controller is applied to the truth model at Operating Point III.
The steady state error is slightly less than that of the LTR-based LQG controller. Also the transient response of the LQR-based 
full-state feedback controller is better than that of the nonlinear feedback linearizing controller at operating 
condition III.

From these simulation results we can conclude that for small variations in the operating conditions, i.e. when
the rotor angle $\delta $, is varied by a small amount from $\delta _0=1$ at operating point I, to $\delta _0=1.0325$ at 
operating point II, and $\delta _0=0.88676$ at operating point III, the transient response of the two linear controllers
is better than that of the nonlinear controller, but the steady state error is the least for the nonlinear controller, followed
by the LQR-based full-state feedback controller which also has a negligible steady state error compared to the LTR-based LQG controller.
The controllers were also tested for large variations in the operating conditions, especially when the rotor angle is varied
by a large amount from $\delta _0=1$ at operating point I. It is observed that as the new operating point is moved further 
away from operating point I, the LTR-based LQG controller applied to the truth model goes unstable, whereas the nonlinear feedback
linearizing controller and the LQR-based full-state feedback controller applied to the truth model remain stable. The LQR-based full-state feedback controller 
shows a small steady state error, whereas the performance of the nonlinear feedback linearizing
controller is independent of the operating condition.

\clearpage
\bibliography{power_course}

\clearpage
 \section{Appendix}
\label{sec:appendix}

\subsection{Derivation of Park's Transformation}

In this section we first present the stator and rotor inductances of a synchronous generator and then 
derive the voltage equation of the synchronous generator in the static frame of reference \cite{BV00}. Park's
transformation is then used to transform the quantities in the static frame to the rotating frame of reference. Next, we derive
the voltage equation of the synchronous generator in the rotating frame of reference.

\subsubsection{Stator and Rotor Inductances}

Here, we present expressions for the self and mutual inductances of the stator and the rotor of a synchronous 
generator. We relate the flux linkage $\mbox{\boldmath$\lambda$}$ to the current 
$\mathbf{i}$ through $\mbox{\boldmath$\lambda$} = \mathbf{L}(\theta)\mathbf{i}$ assuming a linear
relationship, where $\mathbf{L}(\theta)$ is a $6\times6$ inductance matrix relating six flux linkages to
the six currents. The element $L_{ij}$ is calculated by finding the flux linkages of the
$i$th coil after setting all currents equal to zero except the current in the $j$th coil.
To simplify the analysis, we only consider the DC and the fundamental harmonic term.

The self-inductance of the stator coils are given by \cite{BV00}
\begin{equation}
	\begin{aligned}
		L_{aa} &= \frac{\lambda_{aa'}}{i_a} = L_s + L_m \cos 2\theta, \ L_s > L_m
		\geq 0\\
		L_{bb} &= \frac{\lambda_{bb'}}{i_b} = L_s + L_m \cos 2\bigg(\theta -
		          \frac{2\pi}{3}\bigg) \\
	        L_{cc} &= \frac{\lambda_{cc'}}{i_c} = L_s + L_m \cos 2\bigg(\theta +
		          \frac{2\pi}{3}\bigg) \\
	\end{aligned}
	\label{eq:ssi1}
\end{equation}
where $L_{aa}$, $L_{bb}$, and $L_{cc}$ denote the self-inductances of the $a$, $b$, and $c$ axes stator coils respectively.

\psset{framesep=1.5pt}
	\begin{centergroup}
		\begin{pspicture}(-5,-5)(5,5)
		\pscircle(0,0){3.8}

		\multido{\i=4+60,\n=56+60}{6}{\psarc(0,0){2.5}{\i}{\n}}

		\SpecialCoor
		\multido{\i=4+60,\n=-4+60}{6}{\psline(2.5;\i)(2.8;\i)(2.8;\n)(2.5;\n)}
		
		\multido{\i=0+60}{6}{\pscircle(2.65;\i){0.15}}
		\multido{\i=60+120}{3}{\pscircle[fillstyle=solid,
		    fillcolor=black](2.65;\i){0.05}}
		\uput{0}[0](2.5;0){$\times$} 
		\uput{0}[120](2.5;120){$\times$}
		\uput{0}[240](2.5;240){$\times$}

		\psarc(0,0){2.3}{80}{190}\psarc(0,0){2.3}{260}{370}
		\psline(2.3;80)(1.73;95)
		\psline(2.3;190)(1.73;175)
		\psline(2.3;260)(1.73;275)
		\psline(2.3;370)(1.73;355)
		\psline(1.73;95)(1.73;355)\psline(1.73;175)(1.73;275)

		\pscircle(2.1;135){0.15}\pscircle(2.1;315){0.15}
		\pscircle[fillstyle=solid,fillcolor=black](2.1;135){0.05}
		\uput{0}[315](1.90;315){$\times$}

		\pscircle(1.4;45){0.15}\pscircle(1.8;45){0.15}
		\pscircle(1.4;225){0.15}\pscircle(1.8;225){0.15}
		\pscircle[fillstyle=solid,fillcolor=black](1.4;225){0.05}
		\pscircle[fillstyle=solid,fillcolor=black](1.8;225){0.05}
		\uput{0}[45](1.2;45){$\times$}\uput{0}[45](1.6;45){$\times$}

		\psline[linestyle=dashed]{->}(0;0)(5;45)
		\psline[linestyle=dashed]{->}(0;0)(5;90)
		\psline[linestyle=dashed]{->}(0;0)(5;135)
		\psline[linestyle=dashed]{->}(0;0)(5;210)
		\psline[linestyle=dashed]{->}(0;0)(5;330)
		\psline[linestyle=dashed](0;0)(2.4;0)
		\psline[linestyle=dashed](0;0)(2.4;180)
		\psarc{->}{4.5}{90}{135}

		\uput{0}[112.5](4.6;112.5){$\theta$}
		\uput{0.1}[0](4.7;90){Reference}\uput{0.5}[0](4.3;90){axis}
	        \uput{0.1}[180](4.7;135){Direct}\uput{0.7}[180](4.2;135){axis}
		\uput{0.2}[0](4.7;45){Quadrature}\uput{0.8}[0](4.2;45){axis} 

		\uput{0}[0](2.6;-10){$a'$}\uput{0}[0](2.6;10){$i_a$}
		\uput{0}[0](2.7;70){$c$}\uput{0}[0](3;70){$i_c$}
		\uput{0}[120](2.9;120){$i_b$}\uput{0}[0](2.8;110){$b'$}
                \uput{0}[0](2.85;170){$i_a$}\uput{0}[0](2.8;190){$a$}
		\uput{0}[0](3.1;240){$c'$}
		\uput{0}[0](2.8;290){$i_b$}\uput{0}[300](2.9;300){$b$}
		\uput{0.1}[0](1.8;40){$D$}\uput{0.1}[0](1.3;38){$F$}
		\uput{0.1}[0](1.8;65){$i_D$}\uput{0.1}[0](1.3;70){$i_F$}
		\uput{0.1}[0](2.1;230){$D'$}\uput{0.1}[0](1.6;230){$F'$}

		\uput{0.1}[0](1.8;295){$i_Q$}\uput{0.1}[0](1.8;325){$Q$}
		\uput{0.1}[0](2.2;125){$Q'$}
		
		\pscircle[linestyle=dashed](2.85;45){0.15}\pscircle[linestyle=dashed](2.85;225){0.15}
		\pscircle[fillstyle=solid,fillcolor=black](2.85;225){0.05}
		\uput{0}[45](2.65;45){$\times$}
		\uput{0.3}[0](2.85;45){$d$}
		\uput{0.3}[0](3.05;225){$d'$}
		\uput{0.3}[90](2.85;45){$i_d$}
		
		\pscircle[linestyle=dashed](3.05;135){0.15}\pscircle[linestyle=dashed](3.05;315){0.15}
		\pscircle[fillstyle=solid,fillcolor=black](3.05;135){0.05}
		\uput{0}[315](2.85;315){$\times$}
		\uput{0.3}[90](3.05;135){$q'$}
		\uput{0.3}[0](3.05;315){$q$}
		\uput{0.3}[180](3.05;135){$i_q$}
		
                \end{pspicture}
		\captionof{figure}{Machine schematic.}
		\label{fig:electrical_d}
		\end{centergroup}
Referring to \autoref{fig:electrical_d}, the magnetomotive force (mmf), (physical
driving motive force that produces magnetic flux) due to the current in coil $aa'$ is
effective in the vertical direction. The resulting flux, with centerline in the vertical
direction, is maximum when $\theta = 0$ or $\pi$, and minimum when $\theta = \pi/2$ or
$3\pi/2$. The variation of $L_{aa}(\theta)$ is $\pi$-periodic. In \autoref{eq:ssi1}, $L_s$
is the DC term and $L_m$ is the magnitude of the fundamental harmonic term. Similar
analysis is applied for $L_{bb}$ and $L_{cc}$, but phase shifts of $2\pi/3$ and $-2\pi/3$
need to be included for $L_{bb}$ and $L_{cc}$, respectively. Likewise, mutual inductances
between the stator coils are given as follows,
\begin{equation}
	\begin{aligned}
		L_{ab} &= \frac{\lambda_{aa'}}{i_b} = -\bigg[M_s + L_m\cos2\bigg(\theta +
		\frac{\pi}{6}\bigg)\bigg],\ M_s > L_m \geq 0 \\
		L_{bc} &= \frac{\lambda_{bb'}}{i_c} = -\bigg[M_s + L_m\cos2\bigg(\theta -
		\frac{\pi}{2}\bigg)\bigg] \\
		L_{ca} &= \frac{\lambda_{cc'}}{i_a} = -\bigg[M_s + L_m\cos2\bigg(\theta +
		\frac{5\pi}{6}\bigg)\bigg] \\
	\end{aligned}
\end{equation}
where $L_{ab}$ is the mutual inductance between the $a$ axis and the $b$ axis stator coils,
$L_{bc}$ is the mutual inductance between the $b$ axis and the $c$ axis stator coils,
$L_{ca}$ is the mutual inductance between the $c$ axis and the $a$ axis stator coils, and 
$M_s$ is the DC term of the mutual inductances. Self-inductances of the rotor
coils are given by
\begin{equation}
	\begin{aligned}
		L_{FF} &= \frac{\lambda_{FF'}}{i_F} = L_F,\ L_F > 0 \\
		L_{DD} &= \frac{\lambda_{DD'}}{i_D} = L_D,\ L_D > 0 \\
		L_{QQ} &= \frac{\lambda_{QQ'}}{i_Q} = L_Q,\ L_Q > 0 \\
	\end{aligned}
\end{equation}
where $L_{FF}$, $L_{DD}$, and $L_{QQ}$ are self-inductances of the field axis, the direct axis, and the quadrature axis 
rotor coils respectively.
Also, mutual inductances between the stator coils and the rotor coils are given
as
\begin{equation}
	\begin{aligned}
		L_{aF} &= \frac{\lambda_{aa'}}{i_F} = M_F\cos \theta,\ M_F>0 \\
		L_{aD} &= \frac{\lambda_{aa'}}{i_D} = M_D\cos \theta,\ M_D>0 \\
		L_{aQ} &= \frac{\lambda_{bb'}}{i_Q} = M_Q\sin \theta,\ M_Q>0 \\
		L_{FD} &= \frac{\lambda_{FF'}}{i_D} = M_R \\
		L_{FQ} &= \frac{\lambda_{FF'}}{i_Q} = 0 \\
        L_{DQ} &= \frac{\lambda_{DD'}}{i_Q} = 0 \\ 
	\end{aligned}
	\label{eq:other_mutual_inductance}
\end{equation}
where $L_{aF}$, $L_{aD}$, and $L_{aQ}$ are mutual inductances between the $a$ axis stator coil and the field $F$ axis, the direct $D$ axis, and the quadrature $Q$ axis rotor coils respectively. Also $L_{FD}$, $L_{FQ}$, and $L_{DQ}$ are mutual inductances between 
the rotor coils.
Like the mutual inductances between stator coils, the mutual inductances between the
stator coils and the rotor coils for phase $b$ and $c$ are similar to the ones for phase
$a$ which are expressed in \autoref{eq:other_mutual_inductance}.
Thus, because of symmetry the $6\times6$ inductance matrix is partitioned into $4$
submatrices, i.e.,
\begin{equation}
	\mathbf{L}(\theta) = \bbm \mathbf{L}_{11}(\theta) & \mathbf{L}_{12}(\theta)\\
	                          \mathbf{L}_{21}(\theta) & \mathbf{L}_{22}(\theta) \\
				  \ebm
\label{eq:im1}				  
\end{equation}
where each submatrix is expressed as 
\begin{equation}
	\begin{aligned}
		\mathbf{L}_{11} &= 
		        \bbm L_s+L_m\cos2\theta &
		             -M_s-L_m\cos2\bigg(\theta+\frac{\pi}{6}\bigg) &
		             -M_s-L_m\cos2\bigg(\theta+\frac{5\pi}{6}\bigg) \\
			     -M_s-L_m\cos2\bigg(\theta+\frac{\pi}{6}\bigg) &
			     L_s + L_m\cos2\bigg(\theta-\frac{2\pi}{3}\bigg) &
			     -M_s-L_m\cos2\bigg(\theta-\frac{\pi}{2}\bigg) \\
			     -M_s-L_m\cos2\bigg(\theta+\frac{5\pi}{6}\bigg) &
			     -M_s-L_m\cos2\bigg(\theta-\frac{\pi}{2}\bigg) &
			     L_s + L_m\cos2\bigg(\theta + \frac{2\pi}{3}\bigg) \\
			\ebm \\
		\mathbf{L}_{12} &= \mathbf{L}_{21}^\mathrm{T} = 
		        \bbm M_F\cos\theta & M_D\cos\theta & M_Q\sin\theta \\
			     M_F\cos\bigg(\theta - \frac{2\pi}{3}\bigg) &
			     M_D\cos\bigg(\theta - \frac{2\pi}{3}\bigg) &
			     M_Q\sin\bigg(\theta - \frac{2\pi}{3}\bigg) \\
			     M_F\cos\bigg(\theta + \frac{2\pi}{3}\bigg) &
			     M_D\cos\bigg(\theta + \frac{2\pi}{3}\bigg) &
			     M_Q\sin\bigg(\theta + \frac{2\pi}{3}\bigg) \\ \ebm \\
	        \mathbf{L}_{22} &= \bbm L_F & M_R & 0 \\ M_R & L_D & 0 \\ 0 & 0 & L_Q \\ 
		                   \ebm
		\\
	\end{aligned}
	\label{eq:im2}
\end{equation}
From \autoref{eq:basic}, using $\mathbf{\lambda}=\mathbf{L}(\theta)\mathbf{i}$, we get
\begin{equation}
	\mathbf{v} = -\mathbf{R}\mathbf{i} - \frac{d\mathbf{L}(\theta)}{dt}\mathbf{i} -
	\mathbf{L}(\theta)\frac{d\mathbf{i}}{dt}
	\label{eq:vol_static}
\end{equation}
\autoref{eq:vol_static} gives the voltage equation of the synchronous generator in the static frame.

\subsubsection{Park's Transformation}

To simplify the equations and in some important cases obtain linear time-invariant
equations, we introduce the \emph{zero-direct-quadrature}
transformation (also called Park's transformation \cite{BV00}, shown in \autoref{fig:park_trans}). By
using Park's transformation, three AC quantities in the static $abc$ frame are converted into two DC
quantities in the rotating $0dq$ frame. The rotating frame which is represented by the direct and the 
quadrature axis in \autoref{fig:park_trans} is rotating in the same direction and has the same frequency as 
that of the rotor.
\begin{centergroup}
	\begin{pspicture}(-5,-5)(5,5)
		\SpecialCoor
		\psline[linestyle=dashed]{->}(0;0)(5.5;45)
		\psline[linestyle=dashed](0;0)(5.5;225)
		\psline[linestyle=dashed]{->}(0;0)(5;90)
		\psline[linestyle=dashed]{->}(0;0)(5.5;135)
		\psline[linestyle=dashed](0;0)(5.5;315)
		\psline[linestyle=dashed]{->}(0;0)(5;210)
		\psline[linestyle=dashed]{->}(0;0)(5;330)
		\uput{0.1}[180](4.7;135){Direct}\uput{0.7}[180](4.2;135){axis}
		\uput{0.2}[0](4.7;45){Quadrature}\uput{0.8}[0](4.2;45){axis} 
		\uput{0.1}[0](4.7;90){a}\uput{0.1}[0](4.7;205){b}\uput{0}[0](4.7;335){c}
		\psarc{->}{4.5}{90}{135}\uput{0}[112.5](4.6;112.5){$\theta$}
                \uput{0.1}[0](3;90){$i_a$}
		\uput{0.1}[0](2.9;200){$i_b$}
		\uput{0}[0](2.7;335){$i_c$}
		\uput{0}[0](2.6;143){$i_d$}\uput{0}[0](2.5;40){$i_q$}

		\psline[linewidth=2pt]{->}(0;0)(3;90)
		\psline[linewidth=2pt]{->}(0;0)(3;210)
		\psline[linewidth=2pt]{->}(0;0)(3;330)

		\psline[linewidth=2pt]{->}(0;0)(2.12;135)
		\psline[linewidth=2pt]{->}(0;0)(2.12;45)
		\psline[linestyle=dashed](2.12;135)(3;90)(2.12;45)

		\psline[linewidth=2pt]{->}(0;0)(2.9;315)
		\psline[linewidth=2pt]{->}(0;0)(0.78;45)
		\psline[linestyle=dashed](2.9;315)(3;330)(0.78;45)

		\psline[linewidth=2pt]{->}(0;0)(2.9;225)
		\psline[linewidth=2pt]{->}(0;0)(0.78;135)
		\psline[linestyle=dashed](2.9;225)(3;210)(0.78;135)

	\end{pspicture}
	
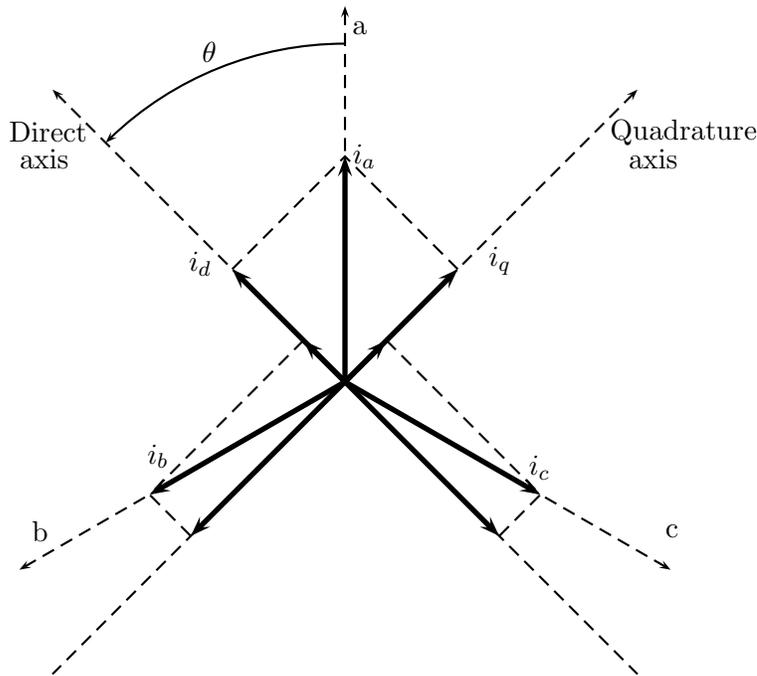
\captionof{figure}{Park's Transformation}
	\label{fig:park_trans}
\end{centergroup}
We use the current variables to show Park's Transformation. As shown in 
\autoref{fig:park_trans}, we define the current on the direct axis as $i_d$ and the
current on the quadrature axis as $i_q$. The Park's transformation is given by
\begin{equation}
	\begin{aligned}
	     i_0 &=\sqrt{\frac{2}{3}}\bigg(\frac{1}{\sqrt{2}}i_a + \frac{1}{\sqrt{2}}
                      i_b + \frac{1}{\sqrt{2}}i_c\bigg) \\
		i_d &=\sqrt{\frac{2}{3}}\bigg( i_a \cos\theta +
		i_b\cos(\theta-\frac{2\pi}{3})+ i_c\cos(\theta+\frac{2\pi}{3}) \bigg) \\
		i_q &=\sqrt{\frac{2}{3}}\bigg( i_a\sin\theta +
		i_b\sin(\theta-\frac{2\pi}{3})+ i_c\sin(\theta+\frac{2\pi}{3})\bigg) \\
        \end{aligned}
	\label{eq:park}
\end{equation}
where the coefficient $\sqrt{\frac{2}{3}}$ is introduced to satisfy the power
conservation between the $abc$ frame and the $0dq$ frame. The
zero-sequence component $i_0$, indicated by the subscript 0, is also included. 

 The zero sequence component is required to yield a unique transformation of the three stator-phase
quantities; it corresponds to components of armature current which produce no net air-gap
flux and hence no net flux linking the rotor circuits \cite{FKU03}. Under
balanced-three-phase conditions, there are no zero-sequence components (zero-sequence
components equal 0). The Park's transformation is usually expressed in matrix form
\begin{equation}
	\bbm  i_0 \\ i_d \\ i_q \ebm = \sqrt{\frac{2}{3}} 
    \bbm \frac{1}{\sqrt{2}} & \frac{1}{\sqrt{2}} & \frac{1}{\sqrt{2}}\\ \cos\theta & \cos(\theta -
	\frac{2\pi}{3}) & \cos(\theta + \frac{2\pi}{3}) \\
	\sin\theta & \sin(\theta - \frac{2\pi}{3}) & \sin(\theta + \frac{2\pi}{3})\ebm 
	 \bbm i_a \\ i_b \\ i_c \ebm
	\label{eq:park_matrix}
\end{equation}
We denote
\begin{equation*}
	\mathbf{P} = \sqrt{\frac{2}{3}} 
    \bbm \frac{1}{\sqrt{2}} & \frac{1}{\sqrt{2}} & \frac{1}{\sqrt{2}}\\ \cos\theta & \cos(\theta -
	\frac{2\pi}{3}) & \cos(\theta + \frac{2\pi}{3}) \\
	\sin\theta & \sin(\theta - \frac{2\pi}{3}) & \sin(\theta + \frac{2\pi}{3})\ebm 
\end{equation*}
It is easy to show that $\mathbf{P}$ is nonsingular and $\mathbf{P}^{-1} =
\mathbf{P}^\mathrm{T}$. 

\subsubsection{Voltage Equation in the Rotating Frame}

We first presented expressions for the self and mutual inductances of the synchronous generator and then 
derived the voltage equation of the synchronous generator in the static frame of reference. Then we used Park's transformation
to convert the AC quantities in the static frame to the rotating frame of reference. Now we present the voltage equations of the synchronous generator in the rotating frame of reference.

If we express \autoref{eq:park} in matrix form, we have
\begin{equation}
	\mathbf{i}_{0dq} = \mathbf{P}\mathbf{i}_{abc}
	\label{eq:park_current}
\end{equation}
Similarly, for voltages and flux linkages, we get
\begin{equation}
	\begin{aligned}
		\mathbf{v}_{0dq} &= \mathbf{P}\mathbf{v}_{abc} \\
		\mbox{\boldmath$\lambda$}_{0dq} &=
		\mathbf{P}\mbox{\boldmath$\lambda$}_{abc} \\
	\end{aligned}
	\label{eq:park_v_flux}
\end{equation}
where the $0dq$ terms are the ones in the rotating frame and $abc$ terms are the ones in
the static $abc$ frame. We need to consider all six components of each current,
voltage, or flux linkage vector; and we want to transform the stator-based ($abc$)
variables into rotor-based ($0dq$) variables while keeping the original rotor quantities
unaffected. So, we define
\begin{equation}
	\mathbf{i}_B \triangleq \left[ 
	\begin{array}{c} 
		i_0 \\ i_d \\ i_q \\ \hline i_F \\ i_D \\ i_Q 
	\end{array} \right] = \left[
	\begin{array}{c|c}
		& \\ \mathbf{P} & \mathbf{0} \\ & \\ \hline
		& \\ \mathbf{0} & \mathbf{I} \\ & \\
	\end{array} \right] \left[ 
	\begin{array}{c}
		i_a \\ i_b \\ i_c \\ \hline i_F \\ i_D \\ i_Q 
	\end{array} \right] = \mathbf{B}\mathbf{i}
	\label{eq:i_park}
\end{equation}
where $\mathbf{I}$ is the $3\times3$ identity matrix and $\mathbf{0}$ is the $3\times3$
zero matrix. Similarly, we define 
\begin{equation}
	\begin{aligned}
		\mathbf{v}_B &= \mathbf{Bv} \\
		\mbox{\boldmath$\lambda$}_B &= \mathbf{B}\mbox{\boldmath$\lambda$} \\
	\end{aligned}
	\label{eq:v_flux_park}
\end{equation}
Substituting \autoref{eq:i_park} and \autoref{eq:v_flux_park} into
$\mbox{\boldmath$\lambda$} = \mathbf{L}\mathbf{i}$ we obtain
\begin{equation}
	\begin{aligned}
		\mathbf{B}^{-1}\mbox{\boldmath$\lambda$}_B &=
		\mathbf{L}\mathbf{B}^{-1}\mathbf{i}_B \\
		\mbox{\boldmath$\lambda$}_B &= \mathbf{BLB}^{-1}\mathbf{i}_B =
		\mathbf{L}_B\mathbf{i}_B \\
	\end{aligned}
	\label{eq:flux_current_park}
\end{equation}
where $\mathbf{L}_B \triangleq \mathbf{B}\mathbf{L}\mathbf{B}^{-1}$. Using the rule for finding the 
transpose of a product of matrices, we find that $\mathbf{L}^\mathrm{T}_B=(\mathbf{BL}\mathbf{B}^\mathrm{T})^\mathrm{T}=\mathbf{B}\mathbf{L}^\mathrm{T}
\mathbf{B}^\mathrm{T}=\mathbf{B}\mathbf{L}\mathbf{B}^\mathrm{T}=\mathbf{L}_B$. Thus $\mathbf{L}_B$ is 
symmetric because $\mathbf{L}$ is symmetric.
By premultiplying by $\mathbf{B}^{-1}$ and postmultiplying by $\mathbf{B}$ we also find the relationship
$\mathbf{L}=\mathbf{B}^{-1}\mathbf{L}_B\mathbf{B}$. It is easy to check
that $\mathbf{B}^{-1} = \mathbf{B}^\mathrm{T} = \bbm
\mathbf{P}^\mathrm{T} &  \mathbf{0} \\ \mathbf{0} & \mathbf{I} \ebm $. Thus we can easily calculate 
$\mathbf{L}_B$ as follows:
\begin{equation}
\begin{aligned}
        \mathbf{L}_B &= \bbm \mathbf{P} & \mathbf{0} \\ \mathbf{0} & \mathbf{1} \ebm 
                        \bbm \mathbf{L}_{11} & \mathbf{L}_{12} \\ \mathbf{L}_{21} & \mathbf{L}_{22} \ebm 
                        \bbm \mathbf{P}^\mathrm{T} & \mathbf{0} \\ \mathbf{0} & \mathbf{1} \ebm\\ 
                     &= \bbm \mathbf{P}\mathbf{L}_{11}\mathbf{P}^\mathrm{T} & \mathbf{P}\mathbf{L}_{12} \\
                        \mathbf{L}_{21}\mathbf{P}^\mathrm{T} & \mathbf{L}_{22} \ebm\\
 \end{aligned}
\label{eq:plp}
\end{equation}
Consider now the terms in \autoref{eq:plp}. As expected, $\mathbf{L}_{22}$ is unchanged. By a tedious but
straightforward calculation, we can show that
\begin{equation}
               \mathbf{P}\mathbf{L}_{11}\mathbf{P}^\mathrm{T}=\bbm L_0 & 0 & 0 \\ 0 & L_d & 0 \\ 0 & 0 & L_q \ebm
\label{eq:plp1}
\end{equation}
where
\begin{equation}
\begin{aligned}
                        L_0 &\triangleq L_s-2M_s\\
                        L_d &\triangleq L_s+M_s+\frac{3}{2}L_m\\
                        L_0 &\triangleq L_s+M_s-\frac{3}{2}L_m\\
\end{aligned}
\label{eq:plp2}
\end{equation} 
With a knowledge of linear algebra we can show that this simplification occurs because the columns
of $\mathbf{P}^\mathrm{T}$ are orthonormal eigenvectors of $\mathbf{L}_{11}$, and hence the similarity
transformation yields a diagonal matrix of eigenvalues. \\
\\
Consider next the off-diagonal submatrix $\mathbf{L}_{21}\mathbf{P}^\mathrm{T}=(\mathbf{P}\mathbf{L}_{12})^\mathrm{T}.$
Using \autoref{eq:im2},
\\
\begin{equation}
    \mathbf{L}_{21}\mathbf{P}^\mathrm{T}=\bbm M_Fcos\theta  &  M_Fcos(\theta -\frac{2\pi }{3}) &  
    M_Fcos(\theta +\frac{2\pi }{3}) \\  M_Dcos\theta  &  M_Dcos(\theta -\frac{2\pi }{3}) &  
    M_Dcos(\theta +\frac{2\pi }{3}) \\  M_Qsin\theta  &  M_Qsin(\theta -\frac{2\pi }{3}) &  
    M_Qsin(\theta +\frac{2\pi }{3}) \ebm \sqrt {\frac{2}{3}} \bbm \frac{1}{\sqrt{2}} & cos\theta  & sin\theta \\
    \frac{1}{\sqrt{2}} & cos(\theta -\frac{2\pi }{3})  & sin(\theta -\frac{2\pi }{3}) \\ 
    \frac{1}{\sqrt{2}} & cos(\theta +\frac{2\pi }{3})  & sin(\theta +\frac{2\pi }{3}) \ebm
\label{eq:plp3}
\end{equation}
\\
Here we recognize that the first two rows of $\mathbf{L}_{21}$  are proportional to the second column of 
$\mathbf{P}^\mathrm{T}$, which we previously indicated was orthogonal to the remaining two columns of $\mathbf{P}^\mathrm{T}$.
The third row of $\mathbf{L}_{21}$ is proportional to the third column of $\mathbf{P}^\mathrm{T}$. Thus, multiplying
rows into columns, it is easy to evaluate the product of the two matrices.
We get
\\
\begin{equation}
    \mathbf{L}_{21}\mathbf{P}^\mathrm{T}=\bbm 0 & \sqrt {\frac{3}{2}}M_F & 0 \\ 0 & \sqrt {\frac{3}{2}}M_D & 0\\
    0 & 0 & \sqrt {\frac{3}{2}}M_Q \ebm
\label{eq:plp4}
\end{equation}
\\    
To simplify the notation, let $k \triangleq \sqrt{\frac{3}{2}}$; then, from \autoref{eq:im2},
\autoref{eq:plp1}, \autoref{eq:plp2}, and \autoref{eq:plp4}, we get
\\
\begin{equation}
\mathbf{L}_B=\left[ \begin{array}{ccc|ccc} L_0 & 0 & 0 & 0 & 0 & 0\\
             0 & L_d & 0 & kM_F & kM_D & 0\\ 0 & 0 & L_q & 0 & 0 & kM_Q\\ \hline
             0 & kM_F & 0 & L_F & M_R & 0\\ 0 & kM_D & 0 & M_R & L_D & 0\\
             0 & 0 & kM_Q & 0 & 0 & L_Q\end{array}\right]
\label{eq:plp5}
\end{equation}
\\            
Note that the matrix $\mathbf{L}_B$ is simple, sparse, symmetric, and constant.
\\ 
We next derive the voltage-current relations using Park's variables. Starting with \autoref{eq:basic},
which is repeated here,
\begin{equation}
	\mathbf{v} = -\mathbf{R}\mathbf{i} - \frac{d\mbox{\boldmath$\lambda$}}{dt}
	\label{eq:pve1}
\end{equation}
and using \autoref{eq:i_park} and \autoref{eq:v_flux_park}, we find that
\begin{equation}
       \mathbf{B}^{-1}\mathbf{v}_B=-\mathbf{R}\mathbf{B}^{-1}\mathbf{i}_B-\frac{d}{dt}(\mathbf{B}^{-1}\mbox{\boldmath$\lambda$} _B)
\label{eq:pve2}
\end{equation}
Premultiplying on the left by $\mathbf{B}$, we get
\begin{equation}
       \mathbf{v}_B=-\mathbf{B}\mathbf{R}\mathbf{B}^{-1}\mathbf{i}_B-\mathbf{B}\frac{d}{dt}(\mathbf{B}^{-1}\mbox{\boldmath$\lambda$} _B)
\label{eq:pve3}
\end{equation}
\\ 
Using the identity
\begin{equation}
     \mathbf{B}\mathbf{R}\mathbf{B}^{-1}=\left[ \begin{array}{c|c}
     & \\ \mathbf{P} & \mathbf{0}\\ & \\ \hline & \\ \mathbf{0} & \mathbf{I} \\ & \end{array}\right]
     \left[ \begin{array}{ccc|ccc} r & & & & & \\ & r & & & \mathbf{0} & \\ 
     & & r & & & \\ \hline & & & r_F & & \\ & \mathbf{0} & & & r_D & \\
     & & & & & r_Q\end{array}\right]\left[ \begin{array}{c|c}
     & \\ \mathbf{P}^{-1} & \mathbf{0}\\ & \\ \hline & \\ \mathbf{0} & \mathbf{I} \\ & \end{array}\right]=\mathbf{R}
\label{eq:pve4}
\end{equation}     
\\
We can simplify \autoref{eq:pve3} by replacing $\mathbf{B}\mathbf{R}\mathbf{B}^{-1}$ by $\mathbf{R}$. 
Continuing, and using the rule for differentiating a product, we get
\begin{equation}
       \mathbf{v}_B=-\mathbf{R}\mathbf{i}_B-\mathbf{B}\frac{d\mathbf{B}^{-1}}{dt}\mbox{\boldmath$\lambda$} _B
       -\frac{d\mbox{\boldmath$\lambda$} _B}{dt}
\label{eq:pve5}
\end{equation}
\\
Next, we wish to obtain a more explicit expression for the matrix $\mathbf{B}\frac{d\mathbf{B}^{-1}}{dt}$ in 
\autoref{eq:pve5}. We first calculate $\mathbf{B}\frac{d\mathbf{B}^{-1}}{d\theta }$. Using \autoref{eq:i_park} and 
$\mathbf{B}^{-1} = \mathbf{B}^\mathrm{T} = \bbm \mathbf{P}^\mathrm{T} &  \mathbf{0} \\ \mathbf{0} & \mathbf{I} \ebm $,
we get
\begin{equation}
       \mathbf{B}\frac{d\mathbf{B}^{-1}}{d\theta }=\bbm \mathbf{P} & \mathbf{0}\\ \mathbf{0} & \mathbf{I}\ebm
       \bbm \frac{d\mathbf{P}^{-1}}{d\theta } & \mathbf{0}\\ \mathbf{0} & \mathbf{0}\ebm
       =\bbm \mathbf{P}\frac{d\mathbf{P}^{-1}}{d\theta } & \mathbf{0}\\ \mathbf{0} & \mathbf{0}\ebm
\label{eq:pve6}
\end{equation}
\\       
where
\begin{equation}
         \mathbf{P}\frac{d\mathbf{P}^{-1}}{d\theta }=\frac{2}{3}\bbm 0 & 0 & 0\\ 0 & 0 & \frac{3}{2}\\
         0 & -\frac{3}{2} & 0 \ebm = \bbm 0 & 0 & 0\\ 0 & 0 & 1\\ 0 & -1 & 0\ebm
\label{eq:pve7}
\end{equation}
\\ 
Then using \autoref{eq:pve7} in \autoref{eq:pve6}, we get
\begin{equation}
      \mathbf{B}\frac{d\mathbf{B}^{-1}}{d\theta }=\left[\begin{array}{ccc|c}
      0 & 0 & 0 & \\ 0 & 0 & 1 & \mathbf{0}\\ 0 & -1 & 0 & \\ \hline
      & \mathbf{0} & & \mathbf{0}\end{array}\right]
\label{eq:pve8}
\end{equation}
\\ 
a $6\times 6$ matrix with only two nonzero elements. Finally noting that $\mathbf{B}\frac{d\mathbf{B}^{-1}}{dt}=
 \mathbf{B}\frac{d\mathbf{B}^{-1}}{d\theta }\frac{d\theta }{dt}$ and substituting \autoref{eq:pve8} in \autoref{eq:pve5},
 we get
\begin{equation}
       \mathbf{v}_B=-\mathbf{R}\mathbf{i}_B-\dot{\theta }\bbm 0 \\ \lambda _q \\ -\lambda _d \\ 0 \\ 0 \\ 0 \ebm
       -\frac{d\mbox{\boldmath$\lambda$} _B}{dt}
\label{eq:pve9}
\end{equation}
\\ 
Note 1: If the shaft position is uniform (i.e. $\dot{\theta }$ = $\frac{d\theta }{dt}$ = $\omega $ = constant)
\autoref{eq:pve9} is linear and time invariant. Very often as a good approximation we can assume
that $\dot{\theta }$ = constant.\\
Note 2: Although, superficially \autoref{eq:pve9} looks very much like \autoref{eq:basic}, it is important
to note that \autoref{eq:pve9} is basically much simpler. In \autoref{eq:pve9}, 
\begin{equation}
\mbox{\boldmath$\lambda$} _B =\mathbf{L}_B\mathbf{i}_B
\end{equation}
where $\mathbf{L}_B$ 
is the constant matrix as given in
\autoref{eq:plp5}, while in \autoref{eq:basic}, 
\begin{equation}
\mbox{\boldmath$\lambda$}=\mathbf{L}\mathbf{i}
\end{equation}
 where 
$\mathbf{L}=\mathbf{L}(\theta )$
is a very complicated matrix as given in \autoref{eq:im1} and \autoref{eq:im2}. 

\subsection{Per Unit Conversion}
 
A per-unit system is the expression of system quantities as fractions of a defined base unit quantity. The 
stator voltages of a synchronous generator are in the kilovolt range and the field voltage is at a much lower level \cite{AF03}. 
This problem can be solved
by using the per unit conversion i.e. normalizing the equations to a convenient base value and expressing all voltages in
p.u. (or percent) of base.
Calculations are simplified because quantities expressed as per-unit are the same regardless of the voltage level. 
Generally base values of power and voltage are chosen. The base power may be the rating of a 
single piece of apparatus such as a motor or generator. If a system is being studied, the base power is usually 
chosen as a convenient round number such as 10 MVA or 100 MVA. The base voltage is chosen as the 
nominal rated voltage of the system. All other base quantities are derived from these two base quantities.

Let us consider an example.
For a single phase system, suppose we select the base quantities of voltage [V] and apparent power [VA] as
$V_B$ and $S_B$ respectively, then we can find the base quantities of current $I_B$ and impedance $Z_B$ as follows.
\begin{equation}
\begin{aligned}
            I_B &= \frac{S_B}{V_B}\\
            Z_B &= \frac{V_B}{I_B}=\frac{V^2_B}{S_B}\\
\end{aligned}
\end{equation}
 All other base quantities can be derived in a similar manner using simple laws of physics.            
Note: Apparent power $|S|$ is the absolute value of complex power $S$, where $S=P+jQ$, $P$ = real power in watt [W],
and $Q$ = reactive power in volt-ampere reactive [Var].
 
\subsubsection{Choosing a base for stator quantities}

\ \ \ The variables $v_d$, $v_q$, $i_d$, $i_q$, $\lambda _d$, and $\lambda _q$ are stator quantities because they relate directly
to the $abc$ phase quantities through Park's transformation. Using the subscript B to indicate
$base$ and R to indicate $rated$, we choose the following stator base quantities \cite{AF03}.
\begin{itemize}
\item $S_B=S_R$ = stator rated apparent power VA/phase, VA rms
\item $V_B=V_R$ = stator rated line-to-neutral voltage, V rms
\item $\omega _B =\omega _R$ = generator rated speed, elec rad/s
\end{itemize}
Note: The RMS value of a set of values (or a continuous-time waveform) is the square root of the 
arithmetic mean (average) of the squares of the original values (or the square of the function that 
defines the continuous waveform).
We now develop the relations for the various base quantities. From the stator base quantities 
we compute the following:
\begin{itemize}
        \item $I_B=\frac{S_B}{V_B}=\frac{S_R}{V_R}$ \ A rms
        \item $t_B=\frac{1}{\omega _B}=\frac{1}{\omega _R}$ \ s
        \item $\lambda _B=V_Bt_B=\frac{V_R}{\omega _R}=L_BI_B$ \ Wb turn
        \item $R_B=\frac{V_B}{I_B}=\frac{V_R}{I_R}$ \ $\Omega$ 
        \item $L_B=\frac{V_Bt_B}{I_B}=\frac{V_R}{I_R\omega _R}$ \ H
\end{itemize}
Thus by choosing the three base quantities $S_B$, $V_B$, and $t_B$, we can compute base
values for all quantities of interest.

To normalize any quantity or to find its per unit value, it is divided by the base quantity of the same dimension.
For example, for currents we write
$$i_u=\frac{i(A)}{I_B(A)}\ \mathrm{p.u.}$$
where we use the subscript $u$ to indicate p.u. Later, when there is no danger of ambiguity
in the notation, this subscript is omitted.

\subsubsection{Choosing a base for rotor quantities}

\ \ \ While choosing a base for rotor quantities, the choice of equal time base throughout all parts
of a circuit with mutual coupling is the important constraint \cite{AF03} . It can be shown that the choice
of a common time base $t$, forces the VA base to be equal in all circuit parts and also forces 
the base mutual inductance to be the geometric mean of the base self-inductances if equal p.u. mutuals 
are to result; i.e., $M_{12B}=(L_{1B}L_{2B})^{1/2}$.

For the synchronous machine the choice of $S_B$ is based on the rating of the stator,
and the time base is fixed by the rated radian frequency. These base quantities must be
the same for the rotor circuits as well. It should be remembered, however, that the
stator VA base is much larger than the VA rating of the rotor (field) circuits \cite{AF03}. Hence
some rotor base quantities are bound to be very large, making the corresponding p.u.
rotor quantities appear numerically small. Therefore, care should be exercised in the
choice of the remaining free rotor base term, since all other rotor base quantities will
then be automatically determined. There is a choice of quantities, but the question is,
Which is more convenient?

To illustrate the above, consider a machine having a stator rating of $100 \times  10^6$ VA/
phase. Assume that its exciter has a rating of 250 V and l,000 A. If, for example, we
choose $I_{RB}$ = 1000 A, $V_{RB}$ will then be 100,000 V; and if we choose $V_{RB}$ = 250 V, then
$I_{RB}$ will be 4,00,000 A.

Is one choice more convenient than the other? Are there other more desirable
choices? The answer lies in the nature of the coupling between the rotor and the stator
circuits. It would seem desirable to choose some base quantity in the rotor to give the
correct base quantity in the stator. For example, we can choose the base rotor current
to give, through the magnetic coupling, the correct base stator flux linkage or open
circuit voltage. Even then there is some latitude in the choice of the base rotor current,
depending on the condition of the magnetic circuit.

The choice made here for the free rotor base quantity is based on the concept of
equal mutual flux linkages. This means that base field current or base $d$ axis amortisseur
current will produce the same space fundamental of air gap flux as produced by base
stator current acting in the fictitious $d$ winding.
Referring to the flux linkage equations derived in the previous section
let $i_d = I_B$, $i_F = I_{FB}$, and $i_D = I_{DB}$ be applied one by one with other currents set
to zero. If we denote the magnetizing inductances ( $l$ = leakage inductances) as
\begin{equation}
\begin{aligned}
L_{md} &\triangleq L_d-l_d\ \mathrm{H} \\
L_{mq} &\triangleq L_q-l_q\ \mathrm{H} \\
L_{mF} &\triangleq L_F-l_F\ \mathrm{H} \\
L_{mQ} &\triangleq L_Q-l_Q\ \mathrm{H} \\
L_{mD} &\triangleq L_D-l_D\ \mathrm{H} \\
\end{aligned}
\end{equation}
and equate the mutual flux linkages in each winding,
\begin{equation}
\begin{aligned}
\lambda_{md} &= L_{md}I_B=kM_FI_{FB}=kM_DI_{DB}\ \mathrm{Wb} \\
\lambda_{mq} &= L_{mq}I_B=kM_QI_{QB}\ \mathrm{Wb} \\
\lambda_{mF} &= kM_FI_B=L_{mF}I_{FB}=M_RI_{DB}\ \mathrm{Wb}  \\
\lambda_{mQ} &= k M_QI_B=L_{mQ}I_{QB}\ \mathrm{Wb} \\
\lambda_{mD} &= kM_DI_B=M_RI_{FB}=L_{mD}I_{DB}\ \mathrm{Wb} \\
\end{aligned}
\end{equation}
Then we can show that,
\begin{equation}
\begin{aligned}
      L_{md}I^2_B &= L_{mF}I^2_{FB}=L_{mD}I^2_{DB}=kM_FI_BI_{FB}=kM_DI_BI_{DB}=M_RI_{FB}I_{DB}\\
      L_{mq}I^2_B &= kM_QI_BI_{QB}=L_{mQ}I^2_{QB}\\
\end{aligned}
\label{eq:pu0}
\end{equation}
and this is the fundamental constraint among base currents.
From the previous equation and the requirement for equal $S_B$, we compute
\begin{equation}
\begin{aligned}
\frac{V_{FB}}{V_B} &= \frac{I_B}{I_{FB}}=\bigg{(}\frac{L_{mF}}{L_{md}}\bigg{)}^{1/2}=\frac{kM_F}{L_{md}}=\frac{L_{mF}}{kM_F}
=\frac{M_R}{kM_D}\triangleq k_F\\
\frac{V_{DB}}{V_B} &= \frac{I_B}{I_{DB}}=\bigg{(}\frac{L_{mD}}{L_{md}}\bigg{)}^{1/2}=\frac{kM_D}{L_{md}}=\frac{L_{mD}}{kM_D}
=\frac{M_R}{kM_F}\triangleq k_D\\
\frac{V_{QB}}{V_B} &= \frac{I_B}{I_{QB}}=\bigg{(}\frac{L_{mQ}}{L_{mq}}\bigg{)}^{1/2}=\frac{kM_Q}{L_{mq}}=\frac{L_{mQ}}{kM_Q}
\triangleq k_Q\\
\end{aligned}
\label{eq:pu1}
\end{equation}
These basic constraints permit us to compute
\begin{equation}
\begin{aligned}
              R_{FB} &= k^2_FR_B \ \ \mathrm{\Omega } \\  
              R_{DB} &= k^2_DR_B \ \ \mathrm{\Omega } \\  
              R_{QB} &= k^2_QR_B \ \ \mathrm{\Omega } \\
              L_{FB} &= k^2_FL_B \ \ \mathrm{H} \\ 
              L_{DB} &= k^2_DL_B \ \ \mathrm{H} \\ 
              L_{QB} &= k^2_QL_B \ \ \mathrm{H} \\
\end{aligned}
\label{eq:pu2}
\end{equation}
and since the base mutuals must be the geometric mean of the base self-inductances
\begin{equation}
\begin{aligned}
             M_{FB} &= k_FL_B \ \ \mathrm{H} \\ 
             M_{DB} &= k_DL_B \ \ \mathrm{H} \\ 
             M_{QB} &= k_QL_B \ \ \mathrm{H} \\ 
             M_{RB} &= k_Fk_DL_B \ \ \mathrm{H} \\
\end{aligned}
\label{eq:pu3}
\end{equation}

\subsubsection{The correspondence of per unit stator EMF to rotor quantities}

\ \ \ The particular choice of base quantities used here gives p.u. values
of $d$ and $q$ axis stator currents and voltages that are $\sqrt {3}$ times the rms values \cite{AF03}. We also
note that the coupling between the $d$ axis rotor and stator involves the factor $k =\sqrt \frac{3}{2}$,
and similarly for the q axis. For example, the contribution to the $d$ axis stator flux linkage
$\lambda _d$ due to the field current $i_F$ is $kM_Fi_F$ and so on. In synchronous machine
equations it is often desirable to convert a rotor current, flux linkage, or voltage to an
equivalent stator EMF. These expressions are developed in this section.

The basis for converting a field quantity to an equivalent stator EMF is that at open
circuit a field current $i_F$ A corresponds to an EMF of $i_F\omega _RM_F$ V peak. If the rms
value of this EMF is $E$, then in MKS units we have
\begin{equation}
\begin{aligned}
       &i_F\omega _RM_F  = \sqrt {2}E\\ 
       &i_F\omega _RkM_F = \sqrt {3}E\\
\end{aligned}
\label{eq:pu4}
\end{equation}       
Since $M_F$, and $\omega _R$  are known constants for a given machine, the field current corresponds
to a given EMF by a simple scaling factor. Thus, $E$ is the stator air gap rms voltage in
pu corresponding to the field current $i_F$ in pu.
We can also convert a field flux linkage $\lambda _F$, to a corresponding stator EMF. At
steady-state open circuit conditions $\lambda _F= L_Fi_F$, and this value of field current $i_F$, when
multiplied by $\omega _RM_F$, gives a peak stator voltage the rms value of which is denoted by
$E'_q$. We can show that the d axis stator EMF corresponding to the field flux linkage $\lambda _F$,
is given by
\begin{equation}
         \lambda _F\frac{\omega _RkM_F}{L_F}=\sqrt{3}E'_q
\label{eq:pu5}
\end{equation}
By the same reasoning a field voltage $v_F$, corresponds (at steady state) to a field current
$\frac{v_F}{r_F}$. This in turn corresponds to a peak stator EMF $(\frac{v_F}{r_F})\omega _RM_F$. If the rms
value of this EMF is denoted by $E_{FD}$, the d axis stator EMF corresponds to a field voltage $v_F$, or
\begin{equation}
         \bigg{(}\frac{v_F}{r_F}\bigg{)}\omega _RkM_F=\sqrt {3}E_{FD}
\label{eq:pu6}
\end{equation}

\subsection{Sub-transient and Transient Inductances and Time Constants}

Before proceeding to the derivation of the simplified model of a synchronous generator we present and define some important terms. 
In this section we define sub-transient and transient inductances and the corresponding time constants \cite{AF03}
which are used in the derivation of a simplified model of a synchronous generator.

\subsubsection{Sub-transient and Transient Inductances}

If all the rotor circuits are short circuited and balanced three-phase voltages are suddenly impressed
upon the stator terminals, the flux linking the $d$ axis circuit will depend initially on the 
sub-transient inductances, and after a few cycles on the transient inductances \cite{AF03}.\\
Let the phase voltages suddenly applied to the stator be given by
\begin{equation}
\bbm v_a \\ v_b \\ v_c \ebm =\sqrt {2}V\bbm cos\theta \\ cos(\theta -120) \\ cos(\theta +120) \ebm u(t)
\label{eq:st1}
\end{equation}
where $u(t)$ is a unit step function and $V$ is the rms phase voltage. Then using Park's transformation
we can show that
\begin{equation}
\bbm v_0 \\ v_d \\ v_q \ebm =\bbm 0 \\ \sqrt {3}Vu(t) \\ 0 \ebm
\label{eq:st2}
\end{equation}
Immediately after the voltage is applied, the flux linkages $\lambda _F$ and $\lambda _D$
are still zero, since they cannot change instantly. Thus at $t=0^+$ from \autoref{eq:direct_axis} we have
\begin{equation}
\begin{aligned}
         \lambda _F &= 0=kM_Fi_d+L_Fi_F+M_Ri_D\\
         \lambda _D &= 0=kM_Di_d+L_Di_D+M_Ri_F\\
\end{aligned}
\label{eq:st3}
\end{equation}
Therefore
\begin{equation}
\begin{aligned}
        i_F &= -\frac{kM_FL_D-kM_DM_R}{L_FL_D-M^2_R}i_d\\
        i_D &= -\frac{kM_DL_F-kM_FM_R}{L_FL_D-M^2_R}i_d\\       
\end{aligned}
\label{eq:st4}
\end{equation}        
Substituting in \autoref{eq:direct_axis} for $\lambda _d$ , we get (at $t=0^+$) 
\begin{equation}
          \lambda _d= \bigg{(}L_d-\frac{k^2M^2_FL_D+L_Fk^2M^2_D-2kM_FkM_DM_R}{L_FL_D-M^2_R}\bigg{)}i_d
\label{eq:st5}          
\end{equation} 
The sub-transient inductance is defined as the initial stator flux linkage per unit of stator current, 
with all the rotor circuits shorted and previously unenergized. Thus by definition
\begin{equation}
    \lambda _d \triangleq L''_di_d
    \label{eq:st6}
\end{equation}
where $L''_d$ is the d axis sub-transient inductance. From \autoref{eq:st5} and \autoref{eq:st6}
\begin{equation}
\begin{aligned}
          L''_d &= L_d-\frac{k^2M^2_FL_D+L_Fk^2M^2_D-2kM_FkM_DM_R}{L_FL_D-M^2_R}\\
          L''_d &= L_d-\frac{L_D+L_F-2L_{AD}}{\frac{L_FL_D}{L^2_{AD}}-1}
\end{aligned}          
\label{eq:st7}          
\end{equation}
where $L_{AD}=L_D-l_D=L_F-L_f=L_d-L_d=kM_F=kM_D=M_R$.\\
 $l_D$, $l_d$, and $l_F$ are the leakage inductances
of the $d$, $F$, and $D$ circuits respectively.\\
If the balanced voltages described by \autoref{eq:st1} are suddenly applied to a machine with no damper winding,
the same procedure will yield (at $t=0^+$)
\begin{equation}
\begin{aligned}
       i_F &= -\frac{kM_F}{L_F}i_d\\
       \lambda _d &= \bigg{[}L_d-\frac{(kM_F)^2}{L_F} \bigg{]}i_d=L'_di_d\\
\end{aligned}          
\label{eq:st8}          
\end{equation}
where $L'_d$ is the $d$ axis transient inductance; i.e.,
\begin{equation}
       L'_d= \bigg{[}L_d-\frac{(kM_F)^2}{L_F} \bigg{]}=L_d-\frac{L^2_{AD}}{L_F}          
\label{eq:st9}
\end{equation}  
In a machine with damper windings, after a few cycles from the start of the transient described in this section,
the damper winding current decays rapidly to zero and the effective stator inductance is the transient inductance.

For a salient pole machine with amortisseur windings a $q$ axis damper circuit exists, but there is no other
$q$ axis rotor winding. For such a machine the stator flux linkage after the initial sub-transient dies out is determined
by essentially the same circuit as that of the steady-state $q$ axis flux linkage. Thus for a salient pole machine
it is customary to consider the $q$ axis transient inductance to be the same as the $q$ axis synchronous inductance.
Repeating the previous procedure for the $q$ axis circuits of a salient pole machine,
\begin{equation}
         \lambda _Q=0=kM_Qi_q+L_Qi_Q
\label{eq:st10}
\end{equation} 
or       
\begin{equation}
         i_Q=-\frac{kM_Q}{L_Q}i_q
\label{eq:st11}
\end{equation}
Substituting in the equation for $\lambda _q$,
\begin{equation}
         \lambda _q=L_qi_q+kM_Qi_Q
\label{eq:st12}
\end{equation}    
or
\begin{equation}
        \lambda _q=\bigg{[}L_q-\frac{(kM_Q)^2}{L_Q}\bigg{]}i_q \triangleq L''_qi_q
 \label{eq:st13}
\end{equation}
where $L''_q$ is the $q$ axis sub-transient inductance
\begin{equation}
        L''_q=L_q-\frac{(kM_Q)^2}{L_Q}=L_q-\frac{L^2_{AQ}}{L_Q}
 \label{eq:st14}
\end{equation} 
Also, for a salient pole machine the $q$ axis transient inductance $L'_q$ is approximately equal to the $q$ axis synchronous inductance
which is the same as $q$ axis sub-transient inductance $L''_q$.
\begin{equation}
        L''_q=L_q-\frac{(kM_Q)^2}{L_Q}=L'_q
 \label{eq:st14_1}
\end{equation}       
 We can also see that when $i_Q$ decays to zero after a few cycles, the $q$ axis effective
inductance in the transient period is the same as $L_q$. Thus for this type of machine
\begin{equation}
           L'_q=L_q
\label{eq:st15}
\end{equation}
Since the reactance is the product of the rated angular speed and the inductance, and since
in p.u. $\omega _R=1$, the sub-transient and transient reactances are numerically equal to the corresponding
values of inductances in (p.u.). It is important to note that for a round rotor machine $L''_q < L'_q < L_q$.

\subsubsection{Time constants}

\ \ \ We start with the stator circuits open circuited \cite{AF03}. Consider a step change in the field voltage;
i.e., $v_F=V_Fu(t)$. From \autoref{eq:dq_eq1} the voltage equations can be written as
\begin{equation}
\begin{aligned}
             r_Fi_F+\dot{\lambda }_F &= V_Fu(t)\\
             r_Di_D+\dot{\lambda }_D &= 0\\
 \end{aligned}            
 \label{eq:tc1}
\end{equation}
and from \autoref{eq:direct_axis} the flux linkages are given by (note that $i_d$ = 0)
\begin{equation}
\begin{aligned}
             \lambda _D &= L_Di_D+M_Ri_F\\
             \lambda _F &= L_Fi_F+M_Ri_D\\
 \end{aligned}            
 \label{eq:tc2}
\end{equation}            
Again at $t=0^+$, $\lambda _D=0$, which gives for that instant
\begin{equation}
         i_F=-\bigg{(}\frac{L_D}{M_R}\bigg{)}i_D
 \label{eq:tc3}
\end{equation}
Substituting for the flux linkages using \autoref{eq:tc2} in \autoref{eq:tc1} we get,
\begin{equation}
\begin{aligned}
             \frac{V_F}{L_F} &= \bigg{(}\frac{r_F}{L_F}\bigg{)}i_F+\dot{i}_F+\bigg{(}\frac{M_R}{L_F}\bigg{)}\dot{i}_D\\
             0 &= \bigg{(}\frac{r_D}{M_R}\bigg{)}i_D+\dot{i}_F+\bigg{(}\frac{L_D}{M_R}\bigg{)}\dot{i}_D\\
 \end{aligned}            
 \label{eq:tc4}
\end{equation}
Subtracting and substituting for $i_F$ using \autoref{eq:tc3},
\begin{equation}
            \dot{i}_D+\bigg{(}\frac{r_DL_F+r_FL_D}{L_FL_D-M^2_R}\bigg{)}i_D=-V_F\bigg{(}\frac{M_R}{L_FL_D-M^2_R}\bigg{)}
\label{eq:tc5}
\end{equation}
Usually in pu $r_D \gg r_F$, while $L_D$ and $L_F$ are of similar magnitude. Therefore we can write, 
approximately,
\begin{equation}
            \dot{i}_D+\bigg{(}\frac{r_D}{L_D-\frac{M^2_R}{L_F}}\bigg{)}i_D=
            -V_F\bigg{(}\frac{\frac{M_R}{L_F}}{L_D-\frac{M^2_R}{L_F}}\bigg{)}
\label{eq:tc6}
\end{equation}           
\autoref{eq:tc6} shows that $i_D$ decays with a time constant
\begin{equation}
            \tau ''_{d0}=\frac{L_D-\frac{M^2_R}{L_F}}{r_D}
\label{eq:tc7}
\end{equation}
This is the $d$ axis open circuit subtransient time constant. It is denoted open circuit because
by definition the stator circuits are open.
When the damper winding is not available or after the decay of the subtransient current, we can show that the
field current is affected only by the parameters of the field circuit; i.e.,
\begin{equation}
          r_Fi_F+L_F\dot{i}_F=V_Fu(t)
\label{eq:tc8}
\end{equation}
The time constant of this transient is the $d$ axis transient open circuit time constant $\tau '_{d0}$, where
\begin{equation}
             \tau '_{d0}=\frac{L_F}{r_F}
 \label{eq:tc9}
\end{equation}
When the stator is short circuited, the corresponding $d$ axis time constants are given by
\begin{equation}
\begin{aligned}
             \tau ''_{d} &= \tau ''_{d0}\frac{L''_d}{L'_d}\\
             \tau '_{d} &= \tau '_{d0}\frac{L'_d}{L_d}\\
 \end{aligned}            
 \label{eq:tc10}
\end{equation}      
A similar analysis of the transient in the $q$ axis circuits of a salient pole machine shows that the 
time constants are given by  
\begin{equation}
\begin{aligned}
             \tau ''_{q0} &= \frac{L_Q}{r_Q}\\
             \tau ''_{q} &= \tau ''_{q0}\frac{L''_q}{L_q}\\
 \end{aligned}            
 \label{eq:tc11}
\end{equation}

\subsection{Simplified Model of the Synchronous Generator}

The truth model of the synchronous generator, takes into account the various effects introduced by 
different rotor circuits, i.e., both field  effects and damper-winding effects. This 
model includes seven nonlinear differential equations for a single synchronous generator.
In addition to these, other equations describing the load constraints, the excitation system, and the 
mechanical torque of the turbine-speed governor system must be included in the mathematical model.
Thus the complete mathematical description of a large power system is exceedingly complex, and simplifications 
are often used in modeling the system \cite{AF03}.

In a stability study the response of a large number of synchronous machines to a given disturbance is 
investigated. The complete mathematical description of the system will therefore be very complicated unless some 
simplifications are used. Often only a few machines are modeled in detail, usually those nearest the disturbance,
while others are described by simpler models.

In this section we first derive a two-axis simplified model of the synchronous generator and then proceed to the 
third order simplified model sometimes referred to in literature as the one-axis model of the synchronous generator.

\subsubsection{The two-axis model}

In the two-axis model the following assumptions are made \cite{AF03}:
\begin{itemize}
\item Transient effects are accounted for, while the sub-transient effects are 
neglected. The transient effects are dominated by the rotor circuits, which are the field circuit in the $d$ axis
and an equivalent circuit in the $q$ axis formed by the solid rotor. 
\end{itemize}
\begin{itemize}
\item In the stator voltage equations the terms $\dot{\lambda}_d$ and $\dot{\lambda}_q$ are negligible
compared to the speed voltage terms and that $\omega \cong \omega _R$ = 1 p.u.
\end{itemize}
\begin{itemize}
\item Amortisseur or damper winding effects are neglected
\end{itemize}
The machine will thus have two stator circuits and two rotor circuits. However, the number of differential equations
describing these circuits is reduced by two since $\dot{\lambda}_d$ and $\dot{\lambda}_q$ are neglected in the stator
voltage equations (the stator voltage equations are now algebraic equations).

The stator transient flux linkages are defined by
\begin{equation}
\begin{aligned}
          \lambda '_d &\triangleq \lambda _d-L'_di_d\\
          \lambda '_q &\triangleq \lambda _q-L'_qi_q\\
 \end{aligned}
\label{eq:two1}
\end{equation}
and the corresponding transient stator voltages are defined by
\begin{equation}
\begin{aligned}
          e'_d &\triangleq -\omega \lambda '_q=-\omega _R\lambda '_q\\
          e'_q &\triangleq \omega \lambda '_d=\omega _R\lambda '_d\\
 \end{aligned}
\label{eq:two2}
\end{equation}       
From \autoref{eq:dq_eq1} the stator voltages $v_d$ and $v_q$ for the truth model of the synchronous generator are given by
\begin{equation}
	\begin{aligned}
		v_d &= -ri_d - \omega\lambda_q - \frac{d\lambda_d}{dt} \\
		v_q &= -ri_q + \omega\lambda_d - \frac{d\lambda_q}{dt} \\
	\end{aligned}
	\label{eq:twopoint1}
\end{equation}
In the two-axis model we assume that in the stator voltage equations, the terms $\dot{\lambda}_d$ 
and $\dot{\lambda}_q$ are negligible compared to the speed voltage terms $\omega\lambda_q$ and 
$\omega\lambda_d$, and that $\omega \cong \omega _R$ = 1 p.u. Thus, neglecting $\frac{d\lambda_d}{dt}$ and $\frac{d\lambda_q}{dt}$
in \autoref{eq:twopoint1} and substituting $\omega \cong \omega _R$ we get

\begin{equation}
\begin{aligned}
          v_d &= -ri_d-\omega _R\lambda_q\\
          v_q &= -ri_q+\omega _R\lambda_d\\
\end{aligned}
\label{eq:two3}
\end{equation}
Substituting $\lambda _d$ and $\lambda _q$ from \autoref{eq:two1} in \autoref{eq:two3} we get
\begin{equation}
\begin{aligned}
          v_d &= -ri_d-\omega _R(\lambda' _q+L'_qi_q)\\
              &= -ri_d-\omega _R\lambda' _q-\omega _RL'_qi_q\\
          v_q &= -ri_q+\omega _R(\lambda' _d+L'_di_d)\\
              &= -ri_q+\omega _R\lambda' _d+\omega _RL'_di_d\\
\end{aligned}
\label{eq:two3point1}
\end{equation}
Also substituting $e'_d \triangleq -\omega \lambda '_q=-\omega _R\lambda '_q$, and
$e'_q \triangleq \omega \lambda '_d=\omega _R\lambda '_d$ as given in \autoref{eq:two2} into \autoref{eq:two3point1}
we get
\begin{equation}
\begin{aligned}
          v_d &= -ri_d-\omega _RL'_qi_q+e'_d\\
          v_q &= -ri_q+\omega _RL'_di_d+e'_q\\
\end{aligned}
\label{eq:two4}
\end{equation}
substituting $\omega _R$ = 1 p.u. in \autoref{eq:two4} and rearranging we get
\begin{equation}
\begin{aligned}
          e'_d &= v_d+ri_d+L'_qi_q\\
               &= v_d+ri_d+L'_di_q+(L'_q-L'_d)i_q\\
          e'_q &= v_q+ri_q-L'_di_d\\
\end{aligned}
\label{eq:two5}
\end{equation}          
The above equation is in the per unit system. Since the term $(L'_q-L'_d)i_q$ is usually small, 
it can be neglected in \autoref{eq:two5}. 
Thus we can write, approximately,
\begin{equation}
          e'_d \cong  v_d+ri_d+L'_di_q
 \label{eq:two6}
\end{equation} 

The voltages $e'_q$ and $e'_d$ are the $q$ and $d$ components of a voltage $e'$ behind the transient reactance.
It is interesting to note that since $e'_d$ and $e'_q$ are $d$ and $q$ axis stator voltages, they represent
$\sqrt {3}$ times the equivalent stator rms voltages. For example, $e'_d=\sqrt{3}E'_d$, and $e'_q=\sqrt{3}E'_q$. 
Also, in this model the voltage $e'$, which corresponds to the transient flux linkages in the machine, is not a constant.
Rather it will change due to the changes in the flux linkage of the $d$ and $q$ axis rotor circuits.

We now develop the differential equations for the voltages $e'_d$ and $e'_q$ for the two axis model. 
Note, all the equations given below are in per unit. The $d$ axis flux linkage equations for the truth model of 
a synchronous generator as given in \autoref{eq:direct_axis} are

\begin{equation}
	\bbm \lambda_d \\ \lambda_F \\ \lambda_D \ebm = 
	    \bbm L_d & kM_F & kM_D \\ kM_F & L_F & M_R \\ kM_D & M_R & L_D\\ \ebm 
	    \bbm i_d \\ i_F \\ i_D \ebm
	\label{eq:dqaxis}
\end{equation}
i.e.
\begin{equation}
\begin{aligned}
                 \lambda _d &= L_di_d+kM_Fi_F+kM_Di_D\\
                 \lambda _F &= kM_Fi_d+L_Fi_F+M_Ri_D\\
\end{aligned}                 
\label{eq:dqaxis1}
\end{equation}
In the two axis model the damper winding effects are neglected. Thus, the expression for $\lambda_D$ and the terms $kM_Di_D$,
and $M_Ri_D$ in \autoref{eq:dqaxis} and \autoref{eq:dqaxis1}, which are related to the direct axis damper winding are neglected. 
Neglecting the above terms, the $d$ axis flux linkage equations for the two-axis model of a synchronous generator are
\begin{equation}
\begin{aligned}
            \lambda _d &= L_di_d+kM_Fi_F\\
            \lambda _F &= kM_Fi_d+L_Fi_F\\
 \end{aligned}
\label{eq:two7}
\end{equation}
The previous two equations can be solved simultaneously to compute $i_F$
\begin{equation}
                 i_F=\frac{\lambda _F}{L_F}-\frac{kM_F}{L_F}i_d
\label{eq:two7point1}
\end{equation}
Substituting $i_F$ from \autoref{eq:two7point1} in the $\lambda _d$ expression of \autoref{eq:two7} we have
\begin{equation}
\begin{aligned}
           \lambda _d &= L_di_d+kM_F\bigg{(}\frac{\lambda _F}{L_F}-\frac{kM_F}{L_F}i_d\bigg{)}\\
                      &= L_di_d-\frac{(kM_F)^2}{L_F}i_d+kM_F\frac{\lambda _F}{L_F}\\
                      &= \bigg{(}L_d-\frac{(kM_F)^2}{L_F}\bigg{)}i_d+kM_F\frac{\lambda _F}{L_F}\\
\end{aligned}
\label{eq:two7point2}
\end{equation} 
From \autoref{eq:st9} and \autoref{eq:pu4} we have                    
\begin{equation}
       L'_d= \bigg{(}L_d-\frac{(kM_F)^2}{L_F} \bigg{)}        
\label{eq:two7point3}
\end{equation} 
\begin{equation}
         \lambda _F\frac{\omega _RkM_F}{L_F}=\sqrt{3}E'_q
\label{eq:two7point4}
\end{equation}
Since all the equations are in per unit we substitute $\omega _R$ = 1 p.u. in the above expression.
i.e.
\begin{equation}
         kM_F\frac{\lambda _F}{L_F}=\sqrt{3}E'_q
\label{eq:two7point5}
\end{equation}
Substituting \autoref{eq:two7point3} and \autoref{eq:two7point5} in \autoref{eq:two7point2} we get
\begin{equation}
          \lambda _d=L'_di_d+\sqrt{3}E'_q
\label{eq:two7point6}
\end{equation}
i.e.
\begin{equation}
          \lambda _d-L'_di_d=\sqrt{3}E'_q
\label{eq:two7point6.5}
\end{equation}
Using \autoref{eq:two1} and \autoref{eq:two2} we can verify that
\begin{equation}
          \lambda '_d \triangleq \lambda _d-L'_di_d=\omega _R\lambda '_d=e'_q=\sqrt{3}E'_q
\label{eq:two7point7}
\end{equation}
The $q$ axis flux linkage equations of the synchronous generator as given in \autoref{eq:quadrature_axis} are

\begin{equation}
	\bbm \lambda_q \\ \lambda_Q \ebm = \bbm L_q & kM_Q \\ kM_Q & L_Q \ebm 
	\bbm i_q \\ i_Q \ebm
	\label{eq:two7point8}
\end{equation}
i.e.
\begin{equation}
\begin{aligned}
	\lambda_q &= L_qi_q+kM_Qi_Q\\
	\lambda_Q &= kM_Qi_q+L_Qi_Q\\
	\end{aligned}
	\label{eq:two7point9}
\end{equation}
The previous two equations can be solved simultaneously to compute $i_Q$
\begin{equation}
         i_Q=\frac{\lambda _Q}{L_Q}-\frac{kM_Q}{L_Q}i_q
\label{eq:two7point10}
\end{equation}         
Substituting $i_Q$ from \autoref{eq:two7point10} in the $\lambda_q$ expression of \autoref{eq:two7point9} we have
\begin{equation}
\begin{aligned}
	\lambda_q &= L_qi_q+kM_Q\bigg{(}\frac{\lambda _Q}{L_Q}-\frac{kM_Q}{L_Q}i_q\bigg{)}\\
	          &= L_qi_q-\frac{(kM_Q)^2}{L_Q}i_q+kM_Q\frac{\lambda _Q}{L_Q}\\
              &= \bigg{(}L_q-\frac{(kM_Q)^2}{L_Q}\bigg{)}i_q+kM_Q\frac{\lambda _Q}{L_Q}\\
\end{aligned}
	\label{eq:two7point11}
\end{equation} 
From \autoref{eq:st14_1} we have
 \begin{equation}
        L'_q=L_q-\frac{(kM_Q)^2}{L_Q}
 \label{eq:two7point12}
\end{equation}
Substituting \autoref{eq:two7point12} in \autoref{eq:two7point11} we get 
\begin{equation}
           \lambda _q= L'_qi_q+kM_Q\frac{\lambda _Q}{L_Q}
\label{eq:two7point13}
\end{equation} 
i.e.
\begin{equation}
           \lambda _q-L'_qi_q=kM_Q\frac{\lambda _Q}{L_Q}
\label{eq:two7point14}
\end{equation}
Substituting $\lambda '_q \triangleq \lambda _q-L'_qi_q$ as given in \autoref{eq:two1}
and $e'_d \triangleq -\lambda '_q=-\omega _R\lambda '_q$ as given in \autoref{eq:two2} in \autoref{eq:two7point14}
we can verify that
 \begin{equation}
          \lambda '_q \triangleq \lambda _q-L'_qi_q=kM_Q\frac{\lambda _Q}{L_Q}=\omega _R\lambda '_q=-e'_d
\label{eq:two7point15}
\end{equation}
Therefore,
\begin{equation}
         e'_d \triangleq \sqrt{3}E'_d= -\frac{kM_Q}{L_Q}\lambda _Q 
  \label{eq:two7point16}
\end{equation}   
We define
\begin{equation}
\begin{aligned}
            \sqrt{3}E &=e_q \triangleq kM_Fi_F \\
            \sqrt{3}E_d &=e_d \triangleq -kM_Qi_Q \\
 \end{aligned}
\label{eq:two15}
\end{equation}
From \autoref{eq:two7point15} we have
 \begin{equation}
          \lambda _q-L'_qi_q=kM_Q\frac{\lambda _Q}{L_Q}=-e'_d
\label{eq:two15point1}
\end{equation}
Substituting $\lambda_q = L_qi_q+kM_Qi_Q$ from \autoref{eq:two7point9} in \autoref{eq:two15point1} we get
\begin{equation}
             L_qi_q+kM_Qi_Q-L'_qi_q=-e'_d
\label{eq:two15point2}
\end{equation} 
Substituting  $\sqrt{3}E_d =e_d \triangleq -kM_Qi_Q $ from \autoref{eq:two15} in \autoref{eq:two15point2} we get
\begin{equation}
           L_qi_q-\sqrt{3}E_d-L'_qi_q=-e'_d
\label{eq:two15point3}
\end{equation} 
Dividing \autoref{eq:two15point3} by $\sqrt{3}$ we get
\begin{equation}
           L_q\frac{i_q}{\sqrt{3}}-E_d-L'_q\frac{i_q}{\sqrt{3}}=-\frac{e'_d}{\sqrt{3}}
\label{eq:two15point4}
\end{equation}
Converting all quantities to rms values, i.e. by substituting $\frac{i_q}{\sqrt{3}}=I_q$, and $\frac{e'_d}{\sqrt{3}}=E'_d$
where $I_q$ and $E'_d$ are the rms quantities, \autoref{eq:two15point4} can be written as
\begin{equation}
           L_qI_q-E_d-L'_qI_q=-E'_d
\label{eq:two15point5}
\end{equation}
i.e.
\begin{equation}
           E'_d=E_d-(L_q-L'_q)I_q
\label{eq:two15point6}
\end{equation}          
Since the damper windings are short circuited, from the $Q$ circuit voltage equation as given in \autoref{eq:dq_eq1} we have
\begin{equation}
        v_Q=r_Qi_Q+\frac{d\lambda _Q}{dt}=0
\label{eq:two15point7}
\end{equation}
Substituting $\lambda_Q = kM_Qi_q+L_Qi_Q$ from \autoref{eq:two7point9} in \autoref{eq:two15point7}
\begin{equation}
        v_Q=r_Qi_Q+kM_Q\dot{i}_q+L_Q\dot{i}_Q=0
\label{eq:two15point8}
\end{equation}
Dividing \autoref{eq:two15point8} by $\sqrt{3}$ we have
\begin{equation}
        v_Q=r_Q\frac{i_Q}{\sqrt{3}}+kM_Q\frac{\dot{i}_q}{\sqrt{3}}+L_Q\frac{\dot{i}_Q}{\sqrt{3}}=0
\label{eq:two15point9}
\end{equation} 
Writing $I_Q=\frac{i_Q}{\sqrt{3}}$, where $I_Q$ is the rms value of $i_Q$
\begin{equation}
        v_Q=r_QI_Q+kM_Q\dot{I}_q+L_Q\dot{I}_Q=0
\label{eq:two15point10}
\end{equation}
Rearranging \autoref{eq:two15point10} we get
\begin{equation}
             -\dot{I}_Q=\frac{r_Q}{L_Q}I_Q+\frac{kM_Q}{L_Q}\dot{I}_q
\label{eq:two15point11}
\end{equation} 
From \autoref{eq:two15} we have            
$\sqrt{3}E_d =e_d \triangleq -kM_Qi_Q$. Expressing all terms as rms quantities we write
\begin{equation}
       E_d =-kM_QI_Q
\label{eq:two15point12}
\end{equation}
Differentiating \autoref{eq:two15point12} we get
\begin{equation}
       \dot{E}_d =-kM_Q\dot{I}_Q
\label{eq:two15point13}
\end{equation}
Substituting $-\dot{I}_Q$ from \autoref{eq:two15point11} in \autoref{eq:two15point12}
\begin{equation}
         \dot{E}_d=kM_Q\frac{r_Q}{L_Q}I_Q+\frac{(kM_Q)^2}{L_Q}\dot{I}_q
\label{eq:two15point14}
\end{equation} 
Differentiating \autoref{eq:two15point6} we get        
\begin{equation}
           \dot{E}'_d=\dot{E}_d-(L_q-L'_q)\dot{I}_q
\label{eq:two16}
\end{equation}
Substituting $\dot{E}_d$ from \autoref{eq:two15point14} in \autoref{eq:two16}
\begin{equation}
\begin{aligned}
           \dot{E}'_d &= kM_Q\frac{r_Q}{L_Q}I_Q+\frac{(kM_Q)^2}{L_Q}\dot{I}_q-(L_q-L'_q)\dot{I}_q\\
                      &= \frac{r_Q}{L_Q}kM_QI_Q+\bigg{(}\frac{(kM_Q)^2}{L_Q}-L_q+L'_q\bigg{)}\dot{I}_q\\
\end{aligned}                      
\label{eq:two17}
\end{equation}
Substituting $\frac{L_Q}{r_Q}=\tau '_{q0}$, $kM_QI_Q=-E_d$ from \autoref{eq:two15}, and 
 $\frac{(kM_Q)^2}{L_Q}-L_q=-L'_q$ from \autoref{eq:st14_1} in \autoref{eq:two17} we get
\begin{equation}
        \dot{E}'_d = -\frac{1}{\tau '_{q0}}E_d
\label{eq:two18}
\end{equation} 
From \autoref{eq:two15point6} we have $E'_d=E_d-(L_q-L'_q)I_q$. Thus, substituting $-E_d=-E'_d-(L_q-L'_q)I_q$ 
 in \autoref{eq:two18}     
\begin{equation}
      \dot{E}'_d = \frac{1}{\tau '_{q0}}[-E'_d-(L_q-L'_q)I_q]
\label{eq:two19}
\end{equation}
The field voltage equation for the truth model of the synchronous generator as given in \autoref{eq:dq_eq1} is
\begin{equation}
		v_F = r_Fi_F + \frac{d\lambda_F}{dt} 
	\label{eq:eqprime}
\end{equation}
Substituting 
$\lambda _F = kM_Fi_d+L_Fi_F$ from \autoref{eq:two7} in \autoref{eq:eqprime} 

\begin{equation}
		v_F = r_Fi_F + kM_F\dot{i}_d+L_F\dot{i}_F
	\label{eq:eqprime1}
\end{equation}
From \autoref{eq:pu4}, \autoref{eq:pu5}, and \autoref{eq:pu6} we have

\begin{equation}
\begin{aligned}
       &i_F\omega _RkM_F = \sqrt {3}E\\
       &\lambda _F\frac{\omega _RkM_F}{L_F}=\sqrt{3}E'_q\\
       &\bigg{(}\frac{v_F}{r_F}\bigg{)}\omega _RkM_F=\sqrt {3}E_{FD}\\
\end{aligned}
\label{eq:eqprime2}
\end{equation}
Since all equations are in per unit, $\omega _R$ = $1$. 
Substituting 
$\lambda _F = kM_Fi_d+L_Fi_F$ from \autoref{eq:two7} in the $\sqrt{3}E'_q$ expression of \autoref{eq:eqprime2}
 
\begin{equation}
\begin{aligned}
              & (kM_Fi_d+L_Fi_F)\frac{\omega _RkM_F}{L_F}=\sqrt{3}E'_q\\
              & \frac{\omega _R(kM_F)^2}{L_F}i_d+i_F\omega _RkM_F=\sqrt{3}E'_q\\             
\end{aligned}
\label{eq:eqprime3}
\end{equation}
Substituting $i_F\omega _RkM_F = \sqrt {3}E$ from \autoref{eq:eqprime2} and $\frac{(kM_F)^2}{L_F}=L_d-L'_d$ from
\autoref{eq:st9} in \autoref{eq:eqprime3} we get
\begin{equation}
           (L_d-L'_d)i_d+\sqrt {3}E=\sqrt{3}E'_q
\label{eq:eqprime4}
\end{equation}           
i.e.
\begin{equation}
           \sqrt {3}E+L_di_d=\sqrt{3}E'_q+L'_di_d
\label{eq:eqprime5}
\end{equation}             
Multiplying \autoref{eq:eqprime1} by $\frac{kM_F}{r_F}$ we get
\begin{equation}
		\frac{v_F}{r_F}kM_F = i_FkM_F + \frac{(kM_F)^2}{r_F}\dot{i}_d+\frac{L_F}{r_F}kM_F\dot{i}_F
	\label{eq:eqprime6}
\end{equation}              
Now substituting $\frac{v_F}{r_F}kM_F=\sqrt {3}E_{FD}$, $i_FkM_F=\sqrt {3}E$, and $kM_F\dot{i}_F=\sqrt{3}\dot{E}$
from \autoref{eq:eqprime2} in \autoref{eq:eqprime6} we get
\begin{equation}
		\sqrt {3}E_{FD}= \sqrt {3}E + \frac{(kM_F)^2}{r_F}\dot{i}_d+\frac{L_F}{r_F}\sqrt{3}\dot{E}
	\label{eq:eqprime7}
\end{equation} 
On differentiating \autoref{eq:eqprime5} 
\begin{equation}
           \sqrt {3}\dot{E}+L_d\dot{i}_d=\sqrt{3}\dot{E}'_q+L'_d\dot{i}_d
\label{eq:eqprime8}
\end{equation}
Substituting $\sqrt {3}\dot{E}$ from \autoref{eq:eqprime8} in \autoref{eq:eqprime7} we get
\begin{equation}
\begin{aligned}
		\sqrt {3}E_{FD} &= \sqrt {3}E + \frac{(kM_F)^2}{r_F}\dot{i}_d+\frac{L_F}{r_F}(\sqrt{3}\dot{E}'_q-L_d\dot{i}_d+L'_d\dot{i}_d)\\
		                &= \sqrt {3}E+\frac{L_F}{r_F}\sqrt{3}\dot{E}'_q+
                           \bigg{(}\frac{(kM_F)^2}{r_F}+\frac{L_F}{r_F}(L'_d-L_d)\bigg{)}\dot{i}_d\\
  \end{aligned}                         
	\label{eq:eqprime9}
\end{equation}
Substituting $\frac{(kM_F)^2}{L_F}=L_d-L'_d$ from \autoref{eq:st9} in \autoref{eq:eqprime9}

\begin{equation}
\begin{aligned}
		\sqrt {3}E_{FD} &= \sqrt {3}E+\frac{L_F}{r_F}\sqrt{3}\dot{E}'_q+
                           \bigg{(}\frac{L_F}{r_F}(L_d-L'_d)+\frac{L_F}{r_F}(L'_d-L_d)\bigg{)}\dot{i}_d\\
                        &= \sqrt {3}E+\frac{L_F}{r_F}\sqrt{3}\dot{E}'_q+0\\
\end{aligned}   
	\label{eq:eqprime10}
\end{equation}
Dividing \autoref{eq:eqprime10} by $\sqrt {3}$ and rearranging we get
\begin{equation}
    \dot{E}'_q = \frac{1}{\tau '_{d0}}(E_{FD}-E)
 \label{eq:eqprime11}
\end{equation}   
where $\tau '_{d0}=\frac{L_F}{r_F}$, and $\sqrt {3}E+L_di_d=\sqrt{3}E'_q+L'_di_d$ i.e 
\begin{equation}
   E=E'_q-(L_d-L'_d)I_d
 \label{eq:eqprime12}
\end{equation}
here $I_d$ is the rms quantity which is given by $I_d=\frac{i_d}{\sqrt{3}}$.\\   
To complete the description of the system, the electrical torque is given by 
$T_{e\phi }=\lambda _di_q-\lambda _qi_d$. Substituting $\lambda _d=\lambda '_d+L'_di_d$, and $\lambda _q=\lambda '_q+L'_qi_q$
from \autoref{eq:two1} we compute
\begin{equation}
           T_{e\phi }=\lambda '_di_q+L'_di_di_q-\lambda '_qi_d-L'_qi_qi_d
\label{eq:te1}
\end{equation} 
Substituting $e'_d=-\omega _R\lambda '_q$, and  $e'_q=\omega _R\lambda '_d$ from \autoref{eq:two2} 
in \autoref{eq:te1}
\begin{equation}
           T_{e\phi }=e'_di_d+e'_qi_q+L'_di_di_q-L'_qi_qi_d
\label{eq:te2}
\end{equation}
Converting all terms to rms quantities by dividing \autoref{eq:te2} by $3$ 
\begin{equation}
           \frac{T_{e\phi }}{3}=\frac{e'_d}{\sqrt{3}}\frac{i_d}{\sqrt{3}}+\frac{e'_q}{\sqrt{3}}\frac{i_q}{\sqrt{3}}
             +L'_d\frac{i_d}{\sqrt{3}}\frac{i_q}{\sqrt{3}}-L'_q\frac{i_q}{\sqrt{3}}\frac{i_d}{\sqrt{3}}
\label{eq:te3}
\end{equation}
 By using $I_q=\frac{i_q}{\sqrt{3}}$, $I_d=\frac{i_d}{\sqrt{3}}$, $E'_d=\frac{e'_d}{\sqrt{3}}$, $E'_q=\frac{e'_q}{\sqrt{3}}$,
and $T_e=\frac{T_{e\phi }}{\sqrt{3}}$ where $I_q$, $I_d$, $E'_d$, $E'_q$ are the rms quantities, $T_{e\phi }$ is the 
per unit generator electromagnetic torque defined on a per phase VA base, and $T_e$ is
the per unit generator electromagnetic torque defined on a three-phase $(3\phi )$ VA base.
Thus, the electromagnetic torque equation of the synchronous generator is
\begin{equation}          
           T_e=E'_dI_d+E'_qI_q-(L'_q-L'_d)I_dI_q
\label{eq:te4}
\end{equation}
From \autoref{eq:swing8}, \autoref{eq:te_pu}, and \autoref{eq:omega_pu}
we can write the mechanical dynamics for the two-axis model of the synchronous generator in the p.u.\ system as
\begin{equation}
\begin{aligned} 
            \tau _j\dot{\omega } &= T_m-D\omega -(E'_dI_d+E'_qI_q-(L'_q-L'_d)I_dI_q)\\            
            \dot{\delta } &= \omega -1\\
\end{aligned}
\label{eq:two24}
\end{equation} 
where $\tau _j$ is the time constant in per unit and is given by $\tau _j=2H\omega _R=2H\omega _B$, where $\omega _B=\omega _R$ = rated or base
angular velocity of the synchronous generator.
Thus, the system equations for the two-axis model of the synchronous generator consist of four differential equations
which are summarized below,
\begin{equation}
\begin{aligned} 
               \dot{E}'_d &= \frac{1}{\tau '_{q0}}[-E'_d-(L_q-L'_q)I_q]\\
               \dot{E}'_q &= \frac{1}{\tau '_{d0}}(E_{FD}-E)\\
               \dot{\omega } &= \frac{1}{\tau _j}[T_m-D\omega -(E'_dI_d+E'_qI_q-(L'_q-L'_d)I_dI_q)]\\            
               \dot{\delta } &= \omega -1\\
\end{aligned}
\label{eq:two25}
\end{equation}  
              
\subsubsection{The one-axis or the third-order simplified model}

 \ \ \ \ To obtain the third order simplified model which is also referred to in literature 
as the one-axis model the following assumptions are made:
\begin{itemize}
\item Amortisseur or damper winding effects are neglected.
\end{itemize}
\begin{itemize}
\item The $\dot{\lambda}_d$ and $\dot{\lambda}_q$ terms in the stator and load voltage equations are neglected compared to the speed 
voltage terms $\omega\lambda_q$ and $\omega\lambda_d$.
\end{itemize}
\begin{itemize}
\item The terms $\omega\lambda$ in the stator and load voltage equations are assumed to be approximately equal to $\omega_R\lambda$.
\end{itemize}
\begin{itemize}
\item The effect of the $Q$ circuit i.e. the differential equation for $E'_d$ or $e'_d$ which is a function of the current $i_Q$ is neglected.
\end{itemize}
It is similar to the two-axis model presented in the previous section except that the absence of the $Q$
circuit eliminates the differential equation for $E'_d$ or $e'_d$ which is a function of the current $i_Q$.
The voltage behind the transient reactance $e'$ has only the component $e'_q$ changing by the field effects
according to \autoref{eq:eqprime11}. The component $e'_d$ is completely determined from the currents and $v_d$.
Thus, the system equations in per unit are
 \begin{equation}
            \dot{E}'_q = \frac{1}{\tau '_{d0}}(E_{FD}-E)   
 \label{eq:two26}
\end{equation}
where
\begin{equation}
          E=E'_q-(L_d-L'_d)I_d 
\label{eq:two27}
\end{equation}
as given in \autoref{eq:eqprime12}.  
Since the damper winding effects are neglected substituting 
$\dot{\lambda }_d=0$ in \autoref{eq:twopoint1}, and using \autoref{eq:two1} and 
\autoref{eq:two2} we obtain the equation for $E'_d$ as
\begin{equation}
             E'_d=V_d+L'_qI_q+rI_d 
\label{eq:two28}
\end{equation}
From \autoref{eq:two25} the $\dot{E}'_d$ expression for the two-axis model is
\begin{equation}
\dot{E}'_d = \frac{1}{\tau '_{q0}}[-E'_d-(L_q-L'_q)I_q]
\label{eq:two29}
\end{equation}
However, in the one-axis or the third order model an additional assumption is that the differential equation for 
$E'_d$ or $e'_d$ which is a function of the current $i_Q$ is neglected. Thus, substituting
$\dot{E}'_d$ = $0$ in \autoref{eq:two29} we get
\begin{equation}
            E'_d=-(L_q-L'_q)I_q
\label{eq:two30}
\end{equation}
\autoref{eq:two30} gives the expression for $E'_d$ for the one-axis model, and it is treated as an algebraic constraint.
Thus, the basic difference between the two-axis and the one-axis model is that in the two-axis model we have a differential 
equation for $E'_d$ whereas in the one-axis model we have an algebraic expression for $E'_d$.
The electromagnetic torque equation for the two-axis model as given in \autoref{eq:te4} is
\begin{equation}
     T_e=E'_dI_d+E'_qI_q-(L'_q-L'_d)I_dI_q
 \label{eq:two31}
\end{equation}
Now substituting the algebraic constraint $E'_d=-(L_q-L'_q)I_q$ from \autoref{eq:two30} in 
\autoref{eq:two31} we obtain the electromagnetic torque equation for the one-axis model 
\begin{equation}
\begin{aligned}
           T_e &= -(L_q-L'_q)I_qI_d+E'_qI_q-(L'_q-L'_d)I_dI_q\\
               &= E'_qI_q+(-L_q+L'_q-L'_q+L'_d)I_dI_q\\
               &= E'_qI_q-(L_q-L'_d)I_dI_q\\
   \end{aligned}
 \label{eq:two32}
\end{equation}
The mechanical dynamics for the one-axis model of the synchronous generator in the per unit system are
\begin{equation}
\begin{aligned}
        \tau _j\dot{\omega } &= T_m-D\omega -T_e\\
        \tau _j\dot{\omega } &= T_m-D\omega -[E'_qI_q-(L_q-L'_d)I_dI_q] \\
        \dot{\delta } &= \omega -1\\
 \end{aligned}
\label{eq:two33}
\end{equation} 
The system equations for the one-axis or the third order simplified model of the synchronous generator 
are summarized below, 
\begin{equation}
\begin{aligned}
            \dot{E}'_q &= \frac{1}{\tau '_{d0}}(E_{FD}-E) \\
            \dot{\omega } &= \frac{1}{\tau _j}[T_m-D\omega -(E'_qI_q-(L_q-L'_d)I_dI_q)]\\
            \dot{\delta } &= \omega -1\\
\end{aligned}
\label{eq:two34}
\end{equation}

\end{document}